\documentclass[a4paper,10pt,twoside,openright]{book}

\usepackage[spanish,USenglish]{babel}
\usepackage[utf8]{inputenc}
\usepackage[usenames, dvipsnames]{color}
\usepackage{graphicx}
\usepackage{enumerate}
\usepackage{multirow}
\usepackage{graphics}
\usepackage{appendix}
\usepackage[nottoc,numbib]{tocbibind}
\usepackage{verbatim} 

\usepackage{tikz}
\usetikzlibrary{shapes,arrows}
\usepackage{subcaption}
\usepackage{amsmath,amsthm,amssymb,mathrsfs}
\usepackage[final]{pdfpages}
\usepackage[]{placeins,flafter}
\usepackage[none]{hyphenat} 
\usepackage{xcolor}
\usepackage{adjustbox}
\usepackage{lscape} 
\usepackage[breaklinks=true]{hyperref} 

\usepackage{subfiles} 
\graphicspath{{Imagenes/}{../Imagenes/}{../../Imagenes/}} 

\usepackage[numbers]{natbib} 
\usepackage{csquotes} 

\usepackage{pdftexcmds}

\usepackage{makeidx}
\makeindex
\index{key}

\usepackage{emptypage} 
\usepackage{minitoc} 

\begin{document}


\selectlanguage{USenglish}

\begin{titlepage}

\vspace*{-4mm}
\begin{figure}[!h]
  \centering
	\includegraphics[width=50mm]{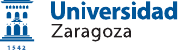}
\end{figure}

\vspace*{10mm}

\fontsize{12pt}{12pt}\selectfont
\begin{center}
Doctoral Thesis
\end{center}

\fontsize{20.74}{20.74}\selectfont
\begin{center}
Development of large Micromegas readout planes for experiments searching for rare events
\end{center}

\vspace*{1cm} 
\baselineskip 36pt
\begin{center}
\fontsize{12pt}{12pt}\selectfont
\center{\rm  Author}

\vspace*{0mm} 
\fontsize{17.28}{17.28}\selectfont
\center{Héctor Mirallas Sánchez}
\vspace*{1cm}
\baselineskip 36pt
\fontsize{12pt}{12pt}\selectfont
\center{\rm  Directors}
\vspace*{0mm}
\fontsize{14pt}{14pt}\selectfont
\center{Igor García Irastorza}
\center{Theopisti Dafní}
\end{center}

\setcounter{footnote}{1}

\vspace*{16.45mm}
\fontsize{12pt}{12pt}\selectfont
\begin{center}
Laboratorio de Física Nuclear y Astropartículas \\
Área de Física Atómica, Molecular y Nuclear \\
Departamento de Física Teórica \\
Universidad de Zaragoza \\
\vspace*{4mm} 
May 2024
\end{center}

\renewcommand{\thefootnote}{\arabic{footnote}}
\pagenumbering{gobble}
\end{titlepage}
\newpage

\title{Title of the Master's Final Project}
\author{Héctor Mirallas Sánchez}

\pagebreak
\cleardoublepage
\baselineskip 19pt

\renewcommand{\labelitemi}{$-$}
\renewcommand{\tablename}{Table}

\renewcommand{\appendixname}{Appendices}
\renewcommand{\appendixtocname}{Appendices}
\renewcommand{\appendixpagename}{Appendices}

\pagenumbering{Roman}





\pagenumbering{arabic}

\cleardoublepage
\thispagestyle{empty}

\setlength{\baselineskip}{12pt} 

\begin{center}
{\Large \bfseries Preface}
\vspace{2.5cm}
\end{center}

\addcontentsline{toc}{chapter}{Preface}

\pagestyle{myheadings}
\markboth{PREFACE}{PREFACE}

The research report presented below has been carried out under the umbrella of the University of Zaragoza, in collaboration with Jiao Tong University in Shanghai, specifically regarding the detectors for double beta decay, and continuously with other organizations such as CERN and CEA Irfu in Paris.

My affiliation with the University of Zaragoza during this period since 2015 has been through the doctoral program, in addition to the last 5 years working as technician for research support, with the main objective of developing low background detectors for the axion observatories CAST and IAXO.

My integration into the research group in which this thesis is developed occurred through several studies, all of which involved a high level of dedication within the group: postgraduate 'Introduction to Research' (2012), collaboration scholarship (2012), and 'Master in Physics and Physical Technologies' (2013). During this time, before starting the doctoral program in 2015, I had the opportunity to carry out my first development of a TPC-based detector for the CAST experiment.

After enrolling in the doctoral program in 2015, I had the opportunity to be part of the team for the PandaX-III experiment, focusing on the development of a tessellated readout plane for this detector, thus elaborating the first part of this thesis. Furthermore, starting in 2018, I also actively participated in the commissioning of the TREX-DM experiment, which was installed in the LSC (Laboratorio Subterráneo de Canfranc) during that year, forming the second part of this thesis. From 2019 onwards, concluding my collaboration with the PandaX-III project and having recently been employed as technician for research support at the University of Zaragoza, I balanced my dedication to the TREX-DM experiment with the work on detectors development in the IAXO project. The latter being outside the scope of this report.

Due to my previous studies in industrial engineering, throughout these years of research, my area of specialization has been the most technical and instrumental part of the experiments. Therefore, this thesis centers on the design, commissioning, and optimization of these experiments, including the engineering processes necessary to carry them out, which occupy a significant part of this report. Despite its technical nature, each of the parts worked on has been introduced and motivated within the framework of fundamental physics research, in which this type of experiment is typically classified.

During this work periods of design and development have alternated with periods of commissioning, interventions for corrections in the detectors, and data taking. Therefore, we decided that the structure of this report would follow a chronological order of these works, a natural order of events derived from continuous consensual decisions, always following a scientific interest.

\cleardoublepage
\thispagestyle{empty}

\setlength{\baselineskip}{12pt} 

\begin{center}
{\Large \bfseries Acknowledgments}
\vspace{2.5cm}
\end{center}

\addcontentsline{toc}{chapter}{Acknowledgments}

\pagestyle{myheadings}
\markboth{ACKNOWLEDGMENTS}{ACKNOWLEDGMENTS}

Any acknowledgments section worthy of being part of a research thesis, something that requires a great deal of dedication, will contain an almost infinite list of people who have been part of the life of the unfortunate researcher. They are the ones who have endured the countless scientific setbacks and have enjoyed the scarce achievements.

It is almost obligatory to begin by thanking family. And so I will, first thanking my family for bringing me into this world, of course. And second, and more importantly, for bringing me into this world and also giving me love, affection, and a happy childhood. The opportunities came later, due to the luck of belonging to a society that cares for the individual.

I considered not referring to anyone in particular in these acknowledgments, not naming anyone, if only to be contrary and not comply with the standard. But it is impossible, and at the very least, I want to address two very, very special people: my daughters, Lucía and Candela. They fill my mornings with joy and make them brighter, until they go to the school, at which point my daily 'party' begins. In the afternoons, they are usually with María, my wife, who supports me, tolerates me, and keeps me sane. How lucky I am to have you by my side!

Starting the mornings this way, one can face any laboratory that comes their way. And if not, you can always count on the wildcard of one of the colleagues who comes to help you. Because one of the fortunes I have had while working in this group is that the human quality is exceptional, as exceptional as the academic records of all of them, which will never cease to intimidate this humble and eternal student. Congratulations to all for your careers, but above all, and I say this from the heart, for being such 'cool' people.

I thank you for the care and politeness in the treatment you have directed towards me, the rigor of your advice, the professionalism of your actions. Even the patience you have had with me on many occasions.

Going up to Canfranc has always been a motivational boost, although lately things have not been going so well up there. Deep conversations in the car, during meals, in good company. Or those trips to CERN, exhausting but exciting, how many discussions and laughs they brought us. Also those trips to Shanghai or Paris, to meet with the rest of the collaboration teams. Let it be known, there are good people outside of Spain as well.

In short: to Theopisti, Juan, Igor, Gloria, Juanan, Alfonso, Xavi, Paco, Elisa, Patricia, Diana, Laura, Clara, Diego, Alicia, María (x3), Susana, Marisa, Javier (x3), Xinglong, Carmen, José Manuel, Miguel Ángel, Iván, Cristina, Oscar, David (x2), Luis, Pedro, Ángel (x3), Jaime, Julia, Konrad, Yikun, Álvaro, Jorge (x2), Itxaso, Fran. And Silvia, Fredi, Alberto, Sergio, Esther, Eduardo, Héctor, Thomas, Rui, Ke-Han, Shaobo, and so many others... In the end, you have made me write the names, I am sure I have left someone out, my apologies.

Something not very common in acknowledgments, as it goes against the spirit of this section, is to express disacknowledgments. And here I will not name names; I will only say that I estimate the proportion of these in relation to the acknowledgments to be barely~5\%. It is a fortune.

Ah! And something else not very common. I want to thank myself, why not, trying to detach myself as much as possible from any hint of narcissism, but with pride. Because I do not know if this work is deserving of the degree that its memory pursues, but I am sure it is deserving of recognition for the effort and dedication. Motivated by love for physics, passion for technology, and the excitement of working with the human group I have just named. Thank you very much to all.

\cleardoublepage
\renewcommand{\contentsname}{Contents}
\setlength{\baselineskip}{12pt} 
\tableofcontents

\setlength{\baselineskip}{12pt} 
\selectlanguage{USenglish}
\graphicspath{{../Imagenes/}{../../Imagenes/}} 

\renewcommand{\listtablename}{Table of contents}
\renewcommand{\tablename}{Table} 

\chapter{TPCs and Micromegas, an optimal combination}
\label{ch:Introduction}


\section{Time Projection Chambers}

Gas particle detectors are instruments that produce signals due to the interaction of particles passing through the gas contained in the detector. One type of gas detector, the Time Projection Chamber (TPC), is well known in the field of particle physics research since its invention in the late 1970s by Marx and Nygren~\cite{Marx1978}, resulting from the evolution of the Multi-Wire Proportional Chamber (MWPC)~\cite{Charpak1968}.

Multi-wire Proportional Chambers were widely used since their invention in 1968, as they represented a significant advancement in measuring the trajectories of particles. The MWPC consisted of a plane formed by independent and parallel wires, separated by a minimal distance ($\sim$~0.1~cm), and placed between two flat electrodes~\cite{Charpak1968}. When an electron or cloud of electrons, resulting from the ionization of the gas by a particle traversing the region between the electrodes, was attracted to one of the wires, a multiplication of these electrons occurred through the process of electron avalanche, due to the strong electric field applied in the area. Despite their limitations, mainly in resolution and the rate of interactions they could measure, they have been used in numerous applications, such as in medical imaging, X-ray and ultraviolet detection, neutron diffraction, and astronomy, among others.

As a result of the evolution of the MWPC, Drift Chambers with multi-wires appeared in 1971~\cite{Walenta1971}, in which the arrival time of the electrons at the wires was measured, thus allowing for improved spatial resolution perpendicular to them. Inheriting this idea and as a strategy for constructing detectors for the PEP (Positron-Electron Project) and PETRA (Positron-Electron Tandem Ring Accelerator) accelerators, TPCs emerged in 1978. They were formed by a volume of gas that was ionized when an ionizing particle traversed it, and in which the electrons produced by the ionization were directed towards a readout plane formed by multi-wires. The measurement of the arrival time of the electrons allowed, as in the case of drift chambers, the reconstruction of the spatial component along the axis perpendicular to the readout plane.


TPCs have undergone significant evolution since their invention. In addition to gas TPCs, today there are liquid-phase TPCs and even those with two phases (liquid-gas) within the same volume. Similarly, the readout planes have also undergone significant evolution, from MPGD (Micropattern Gaseous Detectors), an evolution of the multi-wire planes, to readout planes developed for detecting scintillation photons produced during ionization.

\subsection{Principle of operation of a gas TPC with charge readout}

\begin{figure}[!ht]
    \centering
    \includegraphics[width=0.75\textwidth]{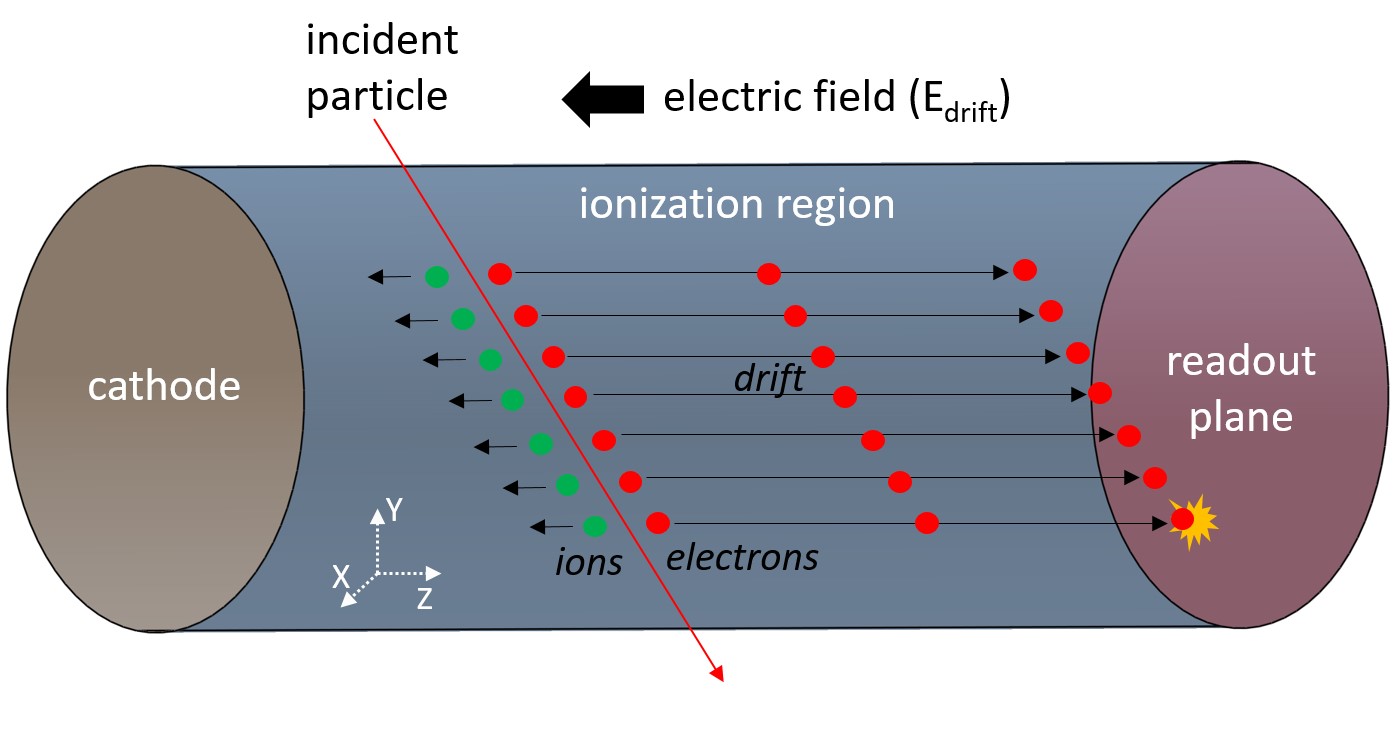}
    \caption{Principle of operation of a gas TPC with a charge readout plane: a particle traverses the ionization region and ionizes some of the gas atoms; the free electrons after ionization drift towards the readout plane due to the electric field~$E_{drift}$ applied in this region. Each cloud of electrons that reaches the readout plane generates a signal that is read and stored. The segmentation of the readout plane allows for independent reading of each area, thus enabling the localization of the event in the XY plane. The differential measurement of the arrival time of each signal allows for the reconstruction of the Z coordinate (perpendicular to the readout plane) relative to the event, knowing the speed of the electrons during their drift towards the readout plane.}
    \label{fig:Introduction_TPCWorkingPrinciple}
\end{figure}

When an ionizing particle traverses a gas TPC with a charge readout plane, it ionizes some of the gas atoms. The free electrons generated from the ionization drift towards the readout plane due to the application of an electric field. In this way, each cloud of electrons that reaches the readout plane generates a signal that can be read and stored (figure~\ref{fig:Introduction_TPCWorkingPrinciple}).

If the readout plane is divided into zones and each cloud of electrons that reaches a zone generates a signal that can be read independently, it is possible to reconstruct the trajectory of the ionizing particle in the XY plane of the readout plane. Furthermore, and this is what defines a TPC, if the time difference with which the primary electrons arrive at the readout plane is measured, it will be possible to reconstruct the trace in the XZ and YZ planes, with Z being the coordinate axis perpendicular to the readout plane and the direction followed by the primary electrons during their drift towards it.

In the case where a gas TPC with charge readout has a readout plane where the electrons can be multiplied, thanks to the effect of electronic avalanche, the signal obtained is also multiplied. The signal multiplication factor is proportional to the multiplication factor of the primary electrons, although its exact relationship will be determined by the properties of the readout plane (geometry, electric field distribution, materials, etc.)

\subsubsection{Interaction of Particles with Gaseous Detectors}

Particles that interact with a gaseous detector can primarily be divided into two groups: photons and charged particles.

The main interaction mechanisms of photons with gaseous detectors are the photoelectric effect, Compton scattering, and pair production.

A beam of photons is characterized by the fact that, although its intensity decreases as it passes through a material, the energy of the photons that are not eliminated from the beam does not degrade. However, when a charged particle interacts with a gaseous medium, it suffers a reduction in its energy, and depending on whether they are light or heavy, they behave differently: light particles, such as electrons, experience deviations from their trajectory, while heavy particles, such as $\alpha$ particles, do not change their direction.

The main interaction mechanisms of charged particles with gaseous detectors are inelastic collisions, in which Coulomb interactions occur with the atomic electrons of the medium, and Bremsstrahlung, in which they emit electromagnetic radiation during their deceleration that can ionize the medium.

\begin{figure}[!ht]
    \centering
    \includegraphics[width=0.6\textwidth]{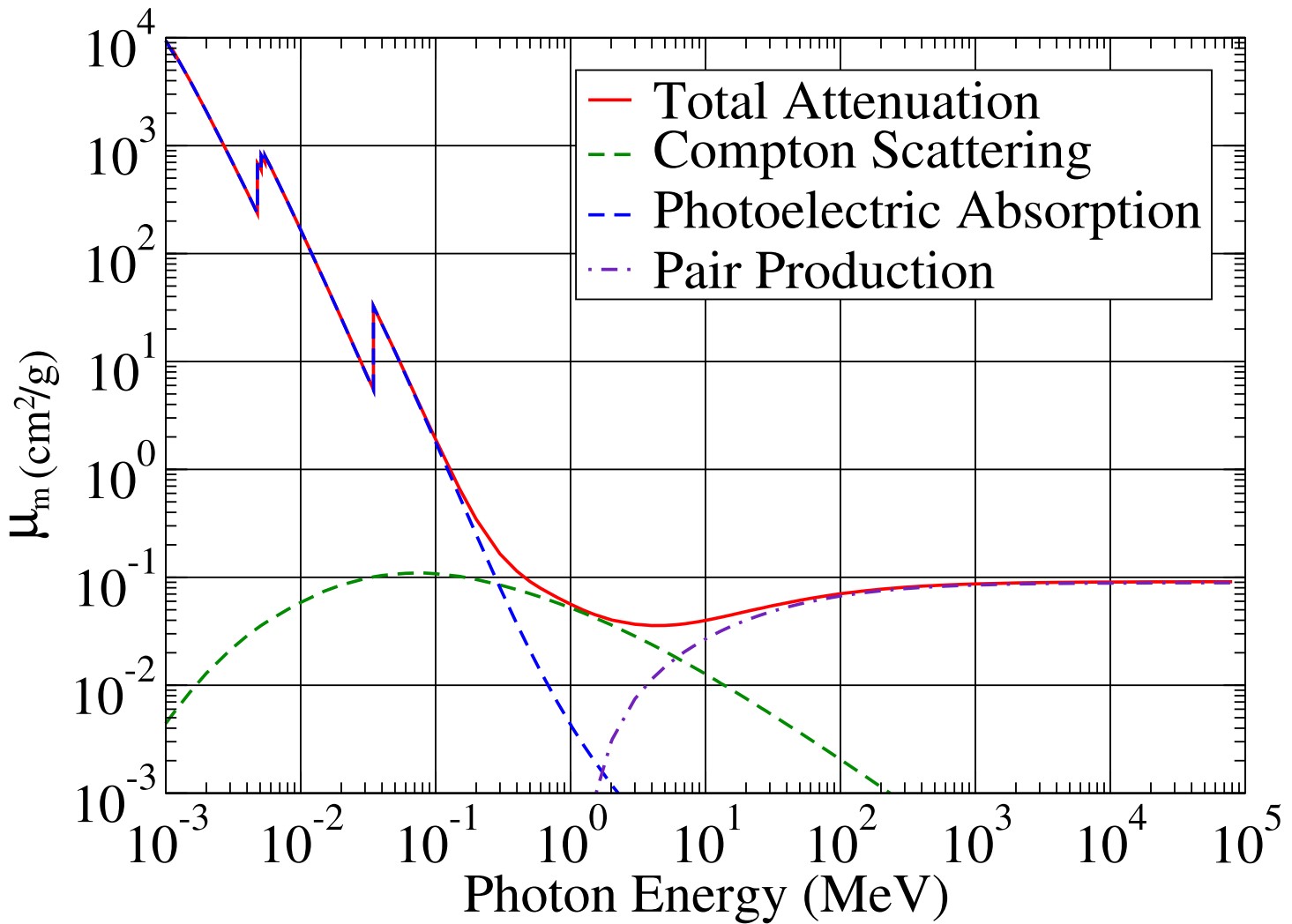}
    \caption{\cite{DianaCarolinaHerreraMunoz} Effective cross-section of photon interaction in Xenon as a function of its energy. It is observed that at low energies ($<$1~MeV) the photoelectric effect predominates over the other mechanisms. Between 1~MeV and 10~MeV, Compton scattering is predominant, while above 10~MeV, the predominant mechanism is pair production.}
    \label{fig:Introduction_AtenuacionFotonesXenon}
\end{figure}

\begin{figure}[!ht]
    \centering
    \includegraphics[width=1.0\textwidth]{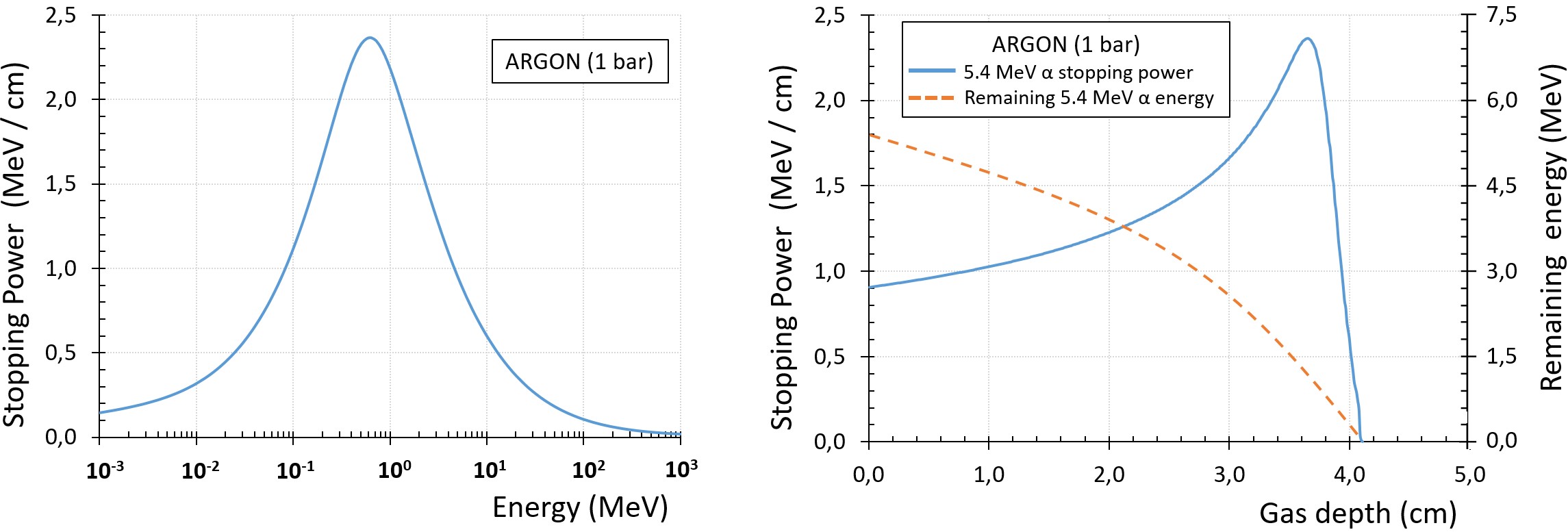}
    \caption{$\bullet$~Left:~Stopping power of $\alpha$ particles as a function of particle energy, for Argon at 1~bar pressure. Data to generate the curve collected from~\cite{NIST-ASTAR}. $\bullet$~Right:~Ionization density and remaining energy of $\alpha$ particles of 5.4~MeV (initial energy) as a function of their path in the gas, for Argon at 1~bar pressure, calculated from the stopping power. Note the Bragg peak at the end of the trajectory, where the ionization density increases.}
    \label{fig:Intro_StoppingPower}
\end{figure}

The interaction of photons with a given medium is defined by their effective cross-section (figure~\ref{fig:Introduction_AtenuacionFotonesXenon}). Meanwhile, the energy loss of a charged particle as it traverses the medium is known as stopping power (figure~\ref{fig:Intro_StoppingPower}).

\begin{figure}[!ht]
    \centering
    \includegraphics[width=0.5\textwidth]{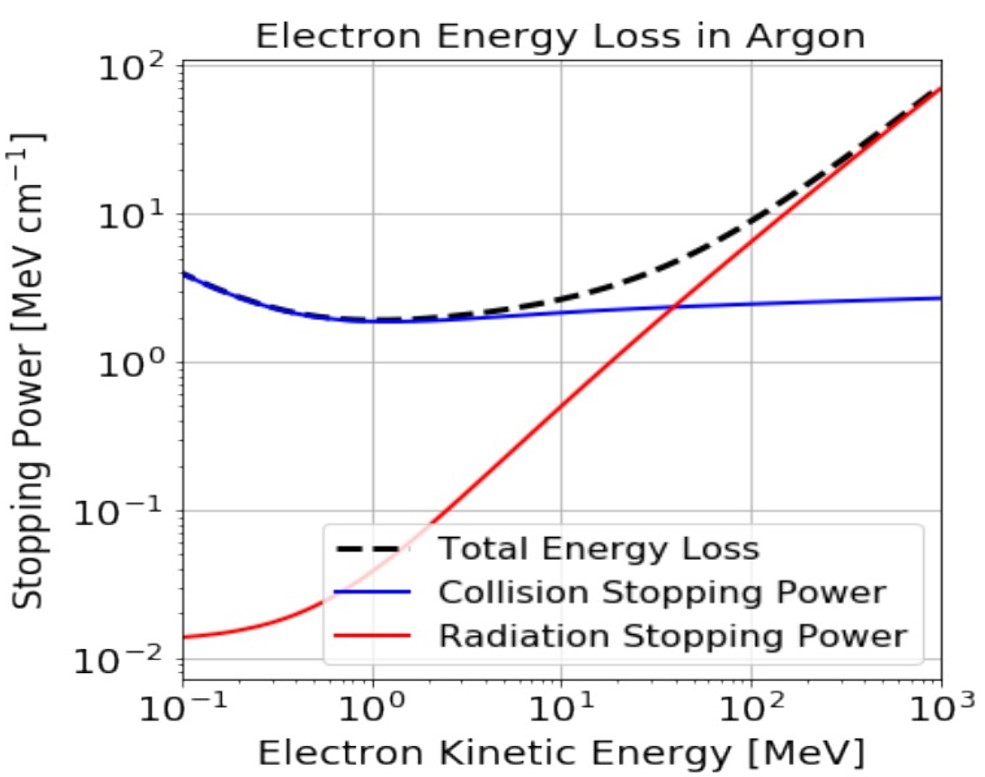}
    \caption{\cite{Adams2020b}~Stopping power of electrons in Argon.}
    \label{fig:Introduction_PoderFrenadoElectronsArgon}
\end{figure}

In the case of low-mass charged particles, such as electrons, it is necessary to take relativistic effects into account when calculating energy loss due to inelastic collisions, which alters the stopping power compared to heavy charged particles. Figure~\ref{fig:Introduction_PoderFrenadoElectronsArgon} shows an example of the stopping power of electrons in Argon, where it is observed that energy loss due to collisions is continuous and predominant at low energies ($<$~10~MeV), while at higher energies ($>$~10~MeV), the loss occurs mainly due to Bremsstrahlung.

\subsubsection{Ionization}

Depending on the processes by which the ionization of the gas occurs due to a particle passing through it, one can differentiate between primary and secondary ionization. In primary ionization, the particle traversing the gas directly generates ion-electron pairs through collision, while secondary ionization typically occurs when electrons resulting from primary ionization are ejected with sufficient energy to ionize the gas again. Penning transfer can also contribute to secondary ionization and occurs when excited gas atoms can ionize the molecules of another gas added to the mixture (Penning mixtures).

Although not all the energy~$E_0$ transferred by the particle passing through the gas is capable of ionizing the molecules within it, the number of ion-electron pairs~$N_e$ that can be produced is proportional to this energy:

\begin{equation} 
N_e=\frac{E_0}{W}
\label{eq:PairProduction}
\end{equation}

\noindent where~$W$ is the mean energy required to create an ion-electron pair in a given gas. It has been experimentally verified that in noble gases, $W$ is practically the same for $\alpha$ particles as for electrons, but there are differences for other complex molecules, such as hydrocarbons. It has also been experimentally confirmed that~$W$ is independent of the deposited energy.

However, two identical charged particles depositing an energy $E_0$ in a gas will not necessarily produce the same number of electron-ion pairs, as the inelastic collisions that occur are governed by probabilities, leading to a variance in the measurements of the number of primary electrons. This variance~$\sigma_e^2$ is quantified through the Fano factor~$F$ (in the range of 0.17-0.23 for Argon and 0.13-0.17 for Xenon), such that ~$\sigma_e^2=FN_e$. Therefore, there exists a limit on the resolution~$R$, defined as the Full Width at Half Maximum (\%FWHM), that can be achieved in a gaseous detector due to this statistical behavior:

\begin{equation} 
R=2.35\sqrt{\dfrac{W}{E_0}F}
\label{eq:ResolutionFanoFactor}
\end{equation}

\subsubsection{Drift of Electrons Towards the Readout Plane}

A free electron in a gas subjected to an electric field experiences an acceleration due to the applied field until a collision occurs with a gas molecule, during which it loses the acquired velocity, only to be accelerated again by the field. This succession of accelerations and collisions of the electron results in an average drift velocity $v_d$ defined as

\begin{equation} 
v_d=\dfrac{eE}{m}\tau
\label{eq:VelocidadDeriva}
\end{equation}

\noindent where~$\tau$ is the mean time between collisions.

Due to scattering with gas molecules, electrons can change direction, although on average they will follow the field lines. The process by which a cloud of electrons, originating from primary ionization, spreads as it moves through the TPC due to the electric field is called diffusion. Diffusion can be divided into transverse, when it occurs in a direction perpendicular to the drift direction, and longitudinal, when it occurs in the drift direction. The root mean square deviation $\sigma_d$ of the electron cloud after a traveled length $L$ can be expressed as

\begin{equation} 
\sigma_d=D\sqrt{L}
\label{eq:Difusion}
\end{equation}

\noindent where~$D$ is the diffusion coefficient.

\subsubsection{Charge Readout Planes}

The readout planes commonly used in the early TPCs employed techniques inherited from multi-wire proportional chambers (MWPC)~\cite{Charpak1968}, and since the wires forming the readout planes could not be installed either too close to each other or with high precision, their spatial resolution was limited. Therefore, a significant step in the evolution of these detectors was the use of micro-electronics and photolithography techniques to manufacture readout planes with greater precision and the capability to perform an initial stage of electron amplification, thanks to the avalanche effect~\cite{Oed1988}, an effect that was difficult to achieve in multi-wire chambers with high wire density. This marked the beginning of a new type of readout planes known as MPGD (Micropattern Gaseous Detectors).

The MPGD detectors that have become most established since then are primarily of two types: GEM (Gas Electron Multiplier)~\cite{Sauli1997} and Micromegas (MICRO-MEsh GAseous Structure)~\cite{Giomataris1996}. The latter technology will be the one used in the development of the large readout planes throughout this thesis.

\section{Micromegas Readout Planes}

\begin{figure}[!ht]
    \centering
    \includegraphics[width=0.95\textwidth]{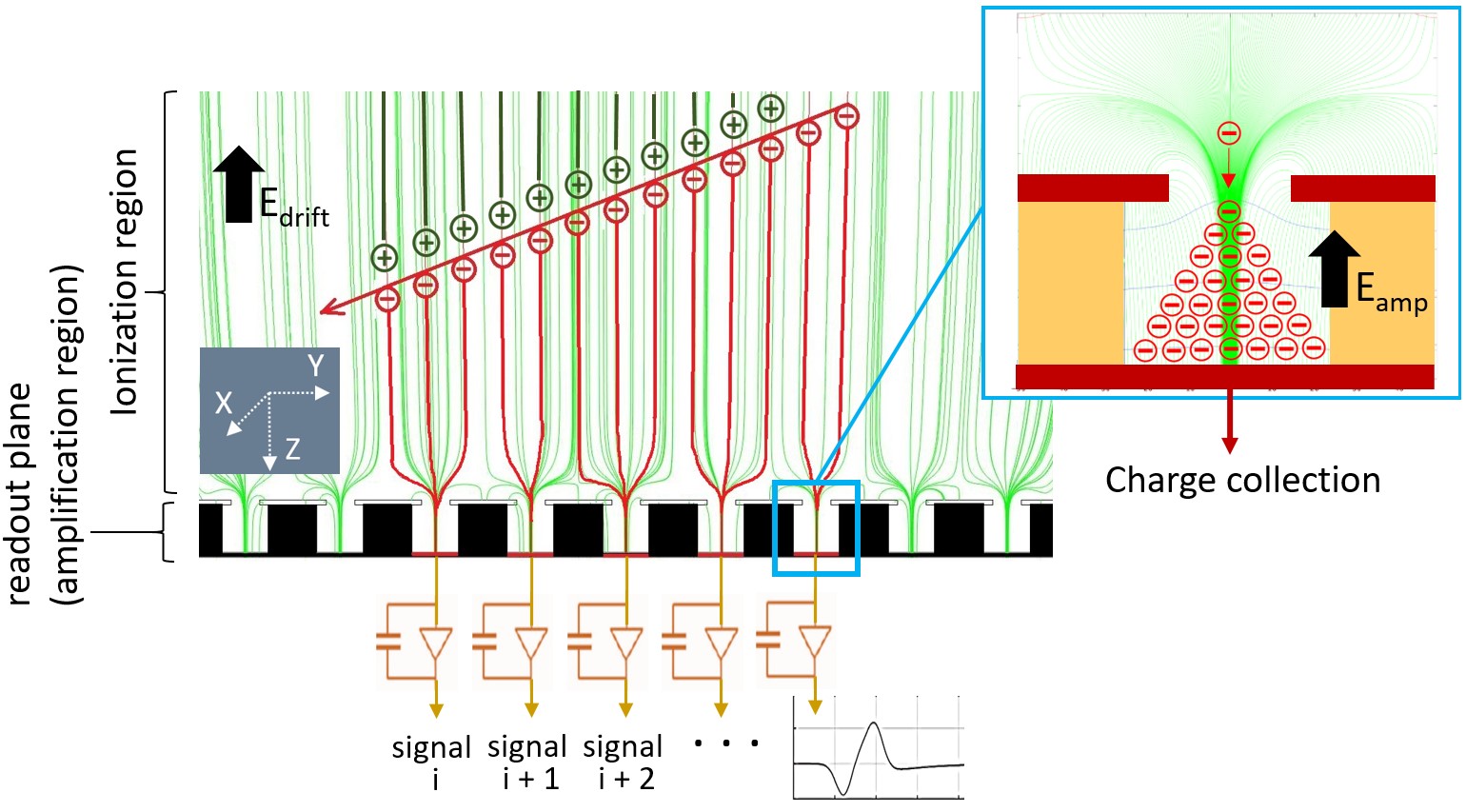}
    \caption{Principle of operation of a Micromegas (the diagram refers to a microbulk Micromegas): When the primary electrons resulting from the ionization of a particle (see figure~\ref{fig:Introduction_TPCWorkingPrinciple}) enter the amplification region, a multiplication of these electrons occurs through the avalanche effect, inducing a signal that is read and amplified by the acquisition electronics. The segmentation of the lower plane of the Micromegas allows for independent reading of each area, thus enabling the localization of the event in the XY plane of the Micromegas. The differential measurement of the arrival time of each signal allows for the reconstruction of the Z coordinate (perpendicular to the readout plane) relative to the event, knowing the drift velocity of the electrons in the conversion area.}
    \label{fig:Introduction_MicromegasWorkingPrinciple}
\end{figure}

Micromegas readout planes are structures designed for the amplification of electrons resulting from the ionization of the atoms that compose the gas, typically in a TPC, as a particle passes through it. Micromegas essentially consist of an upper plane with openings and a lower blind plane, separated by a small distance (figure~\ref{fig:Introduction_MicromegasWorkingPrinciple}). The high electric field applied between the two planes ($\sim$kV/mm) induces the avalanche effect: when one or more electrons penetrate through the openings of the upper plane of the Micromegas, this effect results in a multiplication of these electrons. The total charge resulting from this multiplication alters the electric field~$E_{amp}$, thereby inducing a signal between the two planes of the Micromegas, due to the electromagnetic coupling between them through the dielectric material that separates them (air, kapton, etc.). The signal can be read out either from the lower plane (anode readout, or pixels/strips if the anode is segmented) or from the upper plane (cathode readout, also known as mesh).

The techniques used for the production of Micromegas readout planes have been refined since their invention, and currently, there are two of them using semi-automatic processes, allowing for their production in larger quantities and with better quality: bulk technology~\cite{Giomataris2006} and microbulk technology~\cite{Andriamonje2010}.

In bulk technology, the distance between the upper and lower planes typically ranges from 128 to 256~$\mu$m, with a high gain (multiplication factor of the electrons entering the region with $E_{amp}$) around $\approx$~2·10$^4$ in Ar~+~5\% isobutane. It exhibits good uniformity in gain across the entire surface and achieves good energy resolutions, around 20\% FWHM for 5.9~keV photons in Ar~+~5\% isobutane.

In microbulk technology (figure~\ref{fig:Introduction_MicromegasWorkingPrinciple}), which is used in the development of the readout planes in this thesis, the distance between the upper and lower planes typically ranges from 25 to 100~$\mu$m. Similar gains to those of bulk technology are achieved, but with greater precision and uniformity, allowing for better energy resolutions close to the limit imposed by the Fano factor~\cite{Iguaz2012}, around 11\% FWHM for 5.9~keV photons in Ar~+~5\% isobutane.

In addition to greater uniformity, microbulk technology also has an intrinsic advantage over bulk technology: its radiopurity. Radiopurity can be defined as the ability of a material to minimize its contribution to the background of a detector, and is usually quantified in terms of radioactivity per unit mass or surface area of the radioactive isotopes contained in the material. Microbulk Micromegas are constructed from copper-kapton-copper laminates, materials that are well-known and very clean from a radiopurity standpoint. Only the chemical compounds used during their production, or exposure to an environment that causes contamination, can degrade their initial intrinsic cleanliness. This characteristic makes microbulk technology optimal for application in experiments searching for rare events and low background scenarios using gaseous TPCs, where the contribution to the radioactive background from any material forming the detector must be minimized.

\subsection{Need to Scale Micromegas Readout Planes}

The use of Time Projection Chambers (TPCs) in particle physics experiments has been growing since their invention, along with the development of their readout planes, reaching sizes equivalent to buildings (ALICE or ATLAS at CERN). However, their application in another class of experiments within this branch of physics, commonly referred to as rare event experiments, which operate at interaction rates and energies significantly lower than those typically encountered in particle physics experiments, has been more recent.

This work focuses on two such experiments aimed at searching for rare events: the search for neutrinoless double beta decay (PandaX-III experiment) and the search for WIMPs (TREX-DM experiment). Both utilize a large-sized gaseous TPC, necessitating that the corresponding readout plane is also large. Furthermore, the future of both types of experiments involves the construction of increasingly larger detectors, and consequently, the associated readout planes.

Aside from the specific interest for this work, in general, gaseous detectors in rare event experiments are reaching projected sizes that exceed a ton of gas as a target. Therefore, the work presented here also holds significance for the development of other readout planes as a possible alternative or support to other technologies used.

Both the PandaX-III and TREX-DM experiments employ microbulk Micromegas readout planes. This technology has become established in recent years and is continuously improving, although one of its limitations lies in the maximum sizes that can be manufactured. At the time this thesis began, the maximum size for the active area (effective electron collection area) was 20~x~20~cm$^2$, which was the limitation under which the readout plane for PandaX-III was developed, increasing to 25~x~25~cm$^2$ during the development of the plane for TREX-DM.

Due to this size limitation and the concurrent increase in the size of double beta experiments, there arose a need to develop a tiled readout plane, using a mosaic of modules with microbulk Micromegas. This work is carried out during the first part of this thesis in the context of the PandaX-III experiment. The second part, featuring a single readout plane of 25~x~25~cm$^2$ installed in TREX-DM, focuses on the specifications that define a Micromegas and its response in a high-pressure gaseous detector (10~bar).

Additionally, during the work on TREX-DM, the need to reduce the energy threshold of the experiment emerged, leading to the development of another composite readout plane, the GEM-Micromegas system, in which a GEM specifically manufactured for this system was installed above the Micromegas. Finally, leveraging the development of the Micromegas for TREX-DM, a new gaseous detector for the detection of surface $\alpha$ particles, called AlphaCAMM, is planned, designed, and implemented, addressing all the requirements of a low-background detector with a projection beyond the TREX-DM experiment.

\section[Sistemas de apoyo, adquisición de datos y análisis en TPCs con Micromegas]{Support systems, data acquisition, and analysis in TPCs with Micromegas \sectionmark{Support systems, data acquisition, ...}}
\sectionmark{Support systems, data acquisition, ...}

The complementary systems developed and implemented for the operation of the TPCs described in this thesis have been used and, in part, developed or modified. Although they are not the focus of this work, the most important and widely used ones are listed below:

\subsubsection{Slow control}

The \textit{slow control} monitors the state of the different subsystems that make up the experiment (gas system, acquisition system, calibration system, sensors, etc.) and acts on each of them in real time according to the programmed conditions.

The proper monitoring and storage of the variables managed by the \textit{slow control} allow for a greater or lesser understanding of the detector's state at any given moment and ultimately correlate these variables with the observables derived from the analysis of the data collected in the readout plane.

In the specific case of a TPC with Micromegas readout, it is important that the \textit{slow control} can monitor and control the parameters of the voltage sources used to produce the amplification field in the Micromegas (voltage, current, on, off, etc.). The state of these parameters is continuously stored, ensuring that the conditions under which the amplification of the electrons from the ionization region has occurred are known at all times.

The \textit{slow control} can also allow for the programming of actions on the voltage sources, or any other element controlled by it, in response to one or a series of events (reduction of the voltage in the Micromegas if a pressure loss or sparks occur, reduction of the drift field in the event of sparks at the cathode, blocking of the gas input and output ports in the event of uncontrolled pressure variations, etc.).

The level of implementation of the \textit{slow control} for the reading and control of the equipment, sensors, and actuators will determine whether the experiment operates in a manual, semi-automatic, or automatic mode.

\subsubsection{Gas System}

A gas TPC, by definition, contains a gas or a mixture of gases as a target for the particles to be detected. In a TPC with charge readout using Micromegas, the gas must be optimized for the three main processes that occur: ionization of the gas by the particle traversing the TPC, drift of the primary electrons towards the Micromegas, and the avalanche process within the Micromegas itself.

The gas system in a TPC supplies the appropriate mixture to the detection volume. Depending on the needs of the experiment, this system may be equipped with various sensors and actuators to ensure optimal conditions throughout the duration of the experiment.

In the case of a TPC with Micromegas, where both the drift process of the electrons towards the readout plane and the avalanche process are highly sensitive to impurities in the gas, a recirculation system with filters is typically installed, at a minimum for humidity and oxygen, to maintain the gas in good condition.

The proper implementation of the gas system in the \textit{slow control} will allow for the programming of appropriate conditions at each stage (filling cycle, emptying cycle, data acquisition mode, etc.).

\begin{figure}[!ht]
    \centering
    \includegraphics[width=1.0\textwidth]{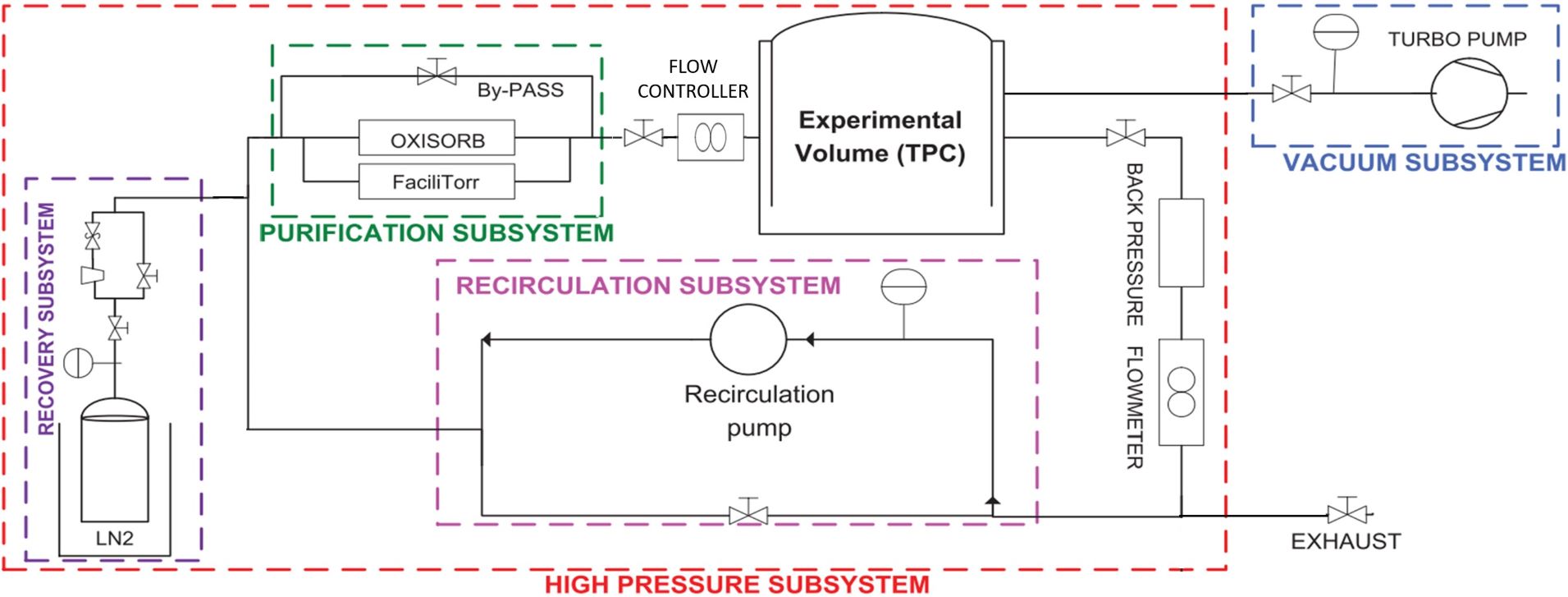}
    \caption{Scheme of the gas system used in the laboratories of the University of Zaragoza during the development of the readout plane for PandaX-III, similar to that used in the TREX-DM experiment.}
    \label{fig:Intro_GasSystem}
\end{figure}

Both the PandaX-III experiment and the TREX-DM experiment use a gas TPC at 10~bar pressure, with a mixture of Xenon and trimethylamine (TMA) in the case of PandaX-III and a mixture of Argon or Neon plus Isobutane (iC$_4$H$_{10}$) in the case of TREX-DM. The gas system in both cases is similar (figure~\ref{fig:Intro_GasSystem}).

To start the TPC, a vacuum is first created in the system using a turbo pump, and then the selected gas mixture for operation is injected from a reservoir where the mixture has been prepared beforehand. When the operating pressure in the TPC is reached, the recirculation system with pump and filters is activated. The filters are selected based on the mixture used and the type of impurities to be filtered, and in both cases, humidity and oxygen filters were used.

Various valves and sensors are distributed throughout the gas system to allow operation during the different modes of operation (vacuum, filling, recirculation, recovery of the mixture, etc.). A flow controller at the inlet of the chamber and a \textit{back pressure} at the outlet allow for automatic control of pressure and flow in the TPC during the different operating modes.

In the case of TREX-DM, the volume of gas used during measurements with the detector was not recovered, being vented to the atmosphere if a new emptying cycle was to be started to change the mixture, perform an intervention, or simply reduce the pressure in the TPC. However, in the case of PandaX-III, since the cost of the mixture was considerably higher, a manual cryogenic pumping system with liquid nitrogen was used to cool the reservoir and thus recover the gas from the TPC.

\subsubsection{Data Acquisition System}

The acquisition system or DAQ (\textit{Data Acquisition}) system is responsible for collecting, amplifying, shaping, digitizing, and storing the signals coming from the readout planes in a TPC. It generally consists of one or more electronic boards commonly called \textit{front-end}, located near the readout plane, which are responsible for amplifying the charge that reaches it, shaping it, and digitizing the generated signal; an electronic board commonly called \textit{back-end}, not necessarily located near the readout plane, which usually handles the management and synchronization of the \textit{front-end} boards; and a data storage system, usually a PC, that manages the reception and storage of the data coming from the acquisition electronics.

All these elements must be appropriately sized according to the needs of each TPC, depending on the number of channels to be read, the event rate produced, or the amount of information generated in each of them, among others. In the case of readout planes with charge collection, as occurs in Micromegas planes, the \textit{front-end} typically contains a charge amplifier (\textit{charge amplifier}) in the first stage, which integrates the current from the readout plane and produces an output voltage proportional to the integrated value of the input current.

Following the behavior of the charge when it is collected at the readout plane (charge density, arrival time, response of the readout plane), the \textit{front-end} can be adjusted so that the generated signal meets the requirements imposed by the type of analysis that will be executed subsequently.

\begin{figure}[!ht]
    \centering
    \includegraphics[width=0.7\textwidth]{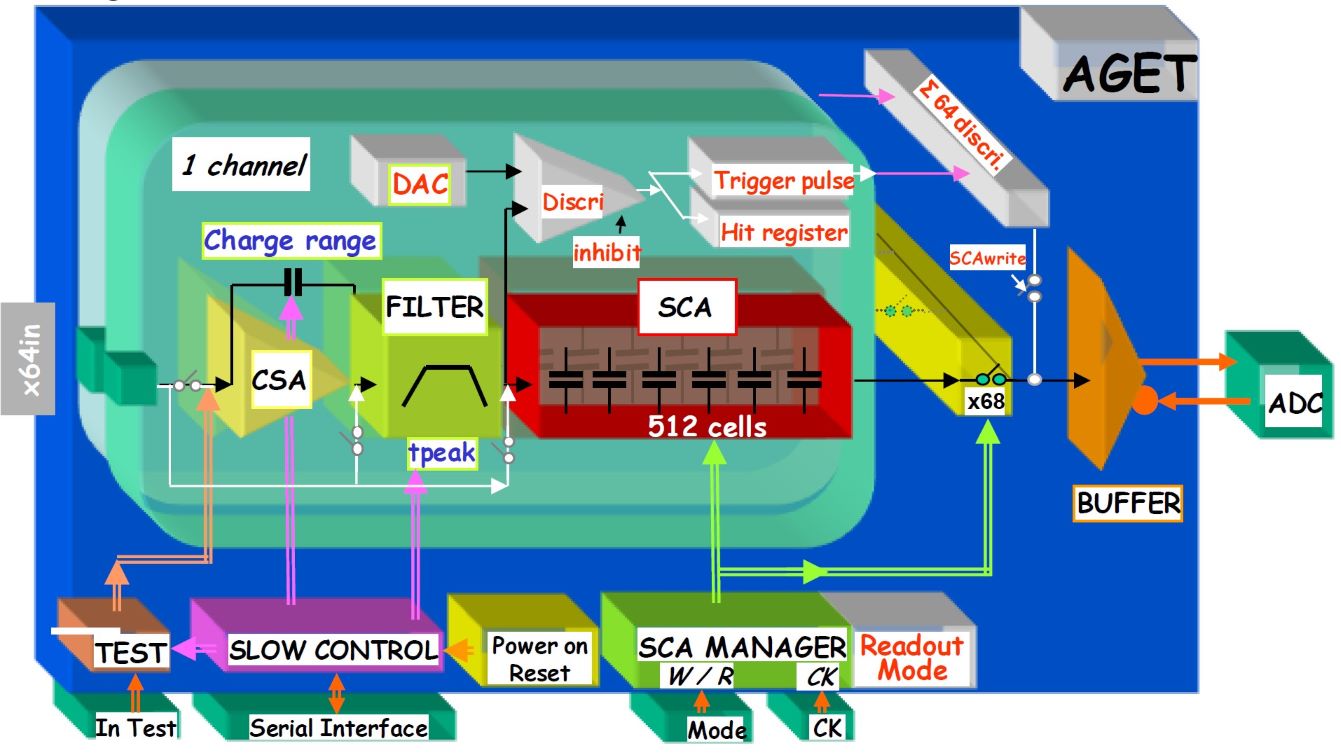}
    \caption{\cite{AGETforTPCs} Schematic of the architecture of the AGET chip.}
    \label{fig:Intro_EsquemaAGET}
\end{figure}

The data acquisition boards used in the work described in this thesis have been primarily developed at CEA Irfu in Paris, and most use the AGET chip~\cite{AGETforTPCs} (figure~\ref{fig:Intro_EsquemaAGET}) as a charge amplifier. This chip was manufactured with CMOS technology and driven by the strategies of the nuclear physics community for its use in TPCs.

Both in AlphaCAMM and in the first commissioning of TREX-DM, acquisition electronics based on the AFTER chip~\cite{Baron2008}, a predecessor of the AGET chip and very similar to it, have also been used. The main difference between the two is that the trigger in the AFTER chip must always be done externally, while in the AGET chip it can be managed by the chip itself, using the signal coming from the readout plane to which it is connected.

\subsubsection{REST-for-Physics, the framework for data analysis}

\begin{figure}[!ht]
    \centering
    \includegraphics[width=0.85\textwidth]{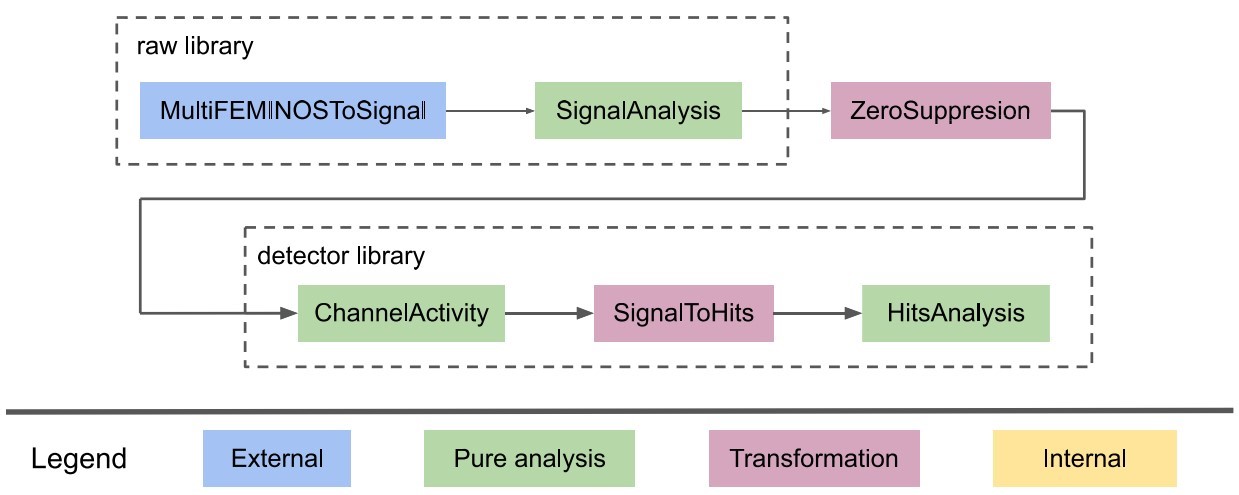}
    \caption{\cite{Altenmuller2022} Data flow during the analysis for each event recorded in a Micromegas by the FEC-Feminos acquisition electronics, used in several detectors in this thesis.}
    \label{fig:SR2MdataTaking_RESTflow}
\end{figure}

The analysis of the data obtained from a TPC must be treated and analyzed appropriately. In the case of the data collected from the detectors developed in this thesis, the environment or framework used for its analysis is the REST-for-Physics software~\cite{Altenmuller2022}, developed within the group over the past few years. Figure~\ref{fig:SR2MdataTaking_RESTflow} shows the sequence of processes in REST that comprise the analysis for each event recorded in a Micromegas.

\begin{figure}[!ht]
    \centering
    \includegraphics[width=1.0\textwidth]{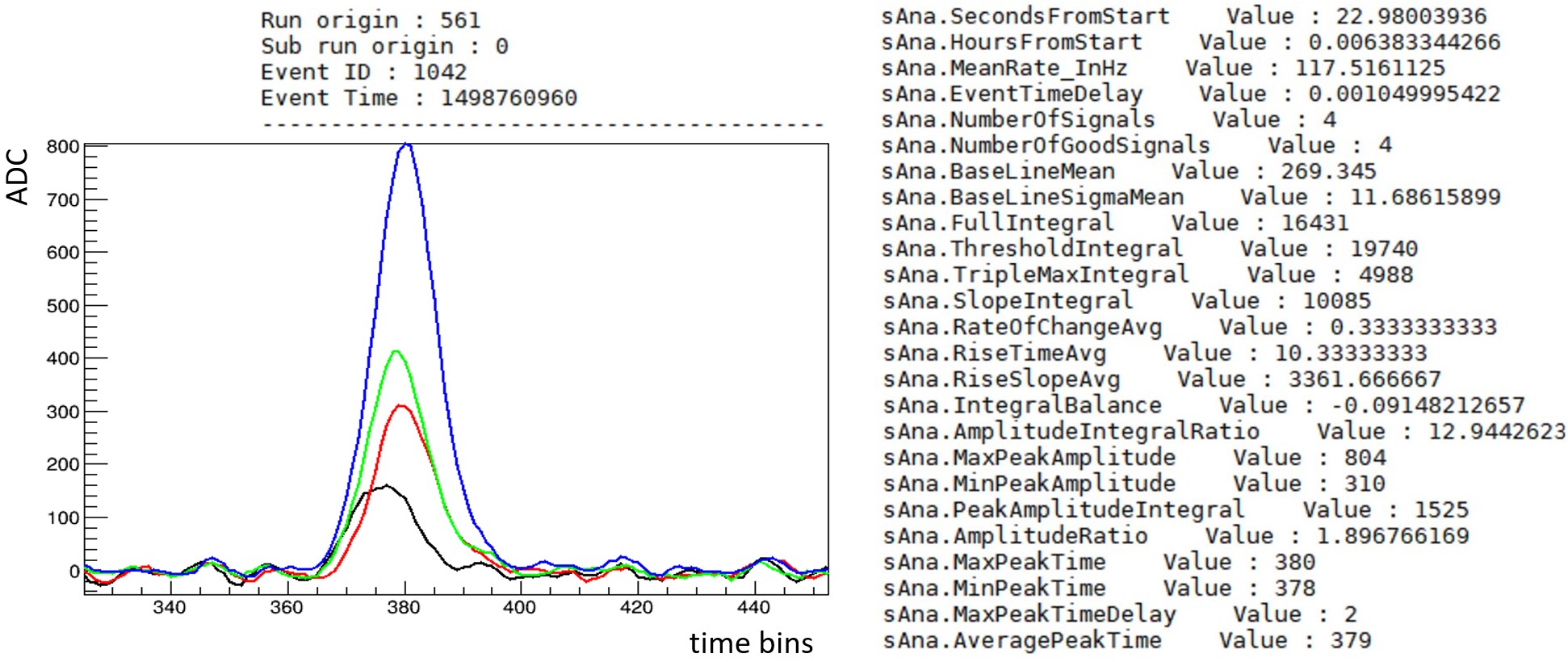}
    \caption{Signals collected by the acquisition system corresponding to an event generated in a TPC with Micromegas. $\bullet$~Left:~reconstruction of the signal from 512 values in ADC units for each channel read. Each color represents a different channel corresponding to a different area in the segmented Micromegas. $\bullet$~Right:~list of observables calculated by the SignalAnalysis process from the signals of the channels in the shown event.}
    \label{fig:Intro_SignalAndObservables}
\end{figure}

The first process transforms the TPC data collected into ROOT format~\cite{ROOT} as imposed by the acquisition system. This process must be generated specifically for each type of acquisition system used. In the case of Figure~\ref{fig:SR2MdataTaking_RESTflow}, it refers to the FEC-Feminos system with the AGET chip, along with the MultiFEMINOSToSignal process. The first analysis of the signal is performed in the SignalAnalysis process, calculating several observables (Figure~\ref{fig:Intro_SignalAndObservables}).

\begin{figure}[!ht]
    \centering
    \includegraphics[width=0.90\textwidth]{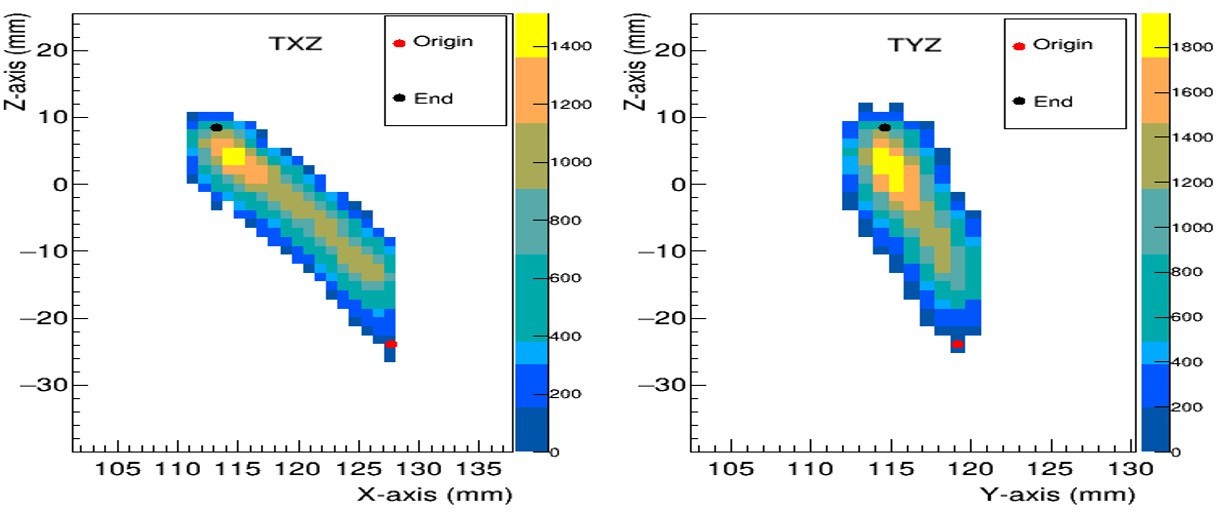}
    \caption{Projection of an event produced by an $\alpha$ particle traversing and ionizing the gas of a TPC with a Micromegas readout plane. The colors represent the integration of the signals in~XZ~(left) and in~YZ (right).}
    \label{fig:Introduction_EventReconstruction}
\end{figure}

The final analysis process, SignalToHits, utilizes the information previously introduced in the corresponding \textit{detector library} regarding the characteristics of the TPC (geometry, gas mixture, drift electric field, etc.), and particularly about the segmentation of the Micromegas (in pixels or \textit{strips}), to calculate observables that describe the event topology. For example, Figure~\ref{fig:Introduction_EventReconstruction} shows the reconstruction of the trajectory of an $\alpha$ particle during which it ionized the gas contained in the TPC.

\part{Scaling Micromega-based readout planes for the PandaX-III experiment}

\setlength{\baselineskip}{12pt} 
\selectlanguage{USenglish}
\graphicspath{{../Imagenes/}{../../Imagenes/}} 

\renewcommand{\listtablename}{Table of Contents}
\renewcommand{\tablename}{Table} 

\chapter{The PandaX-III Experiment and the Search for Neutrinoless Double Beta Decay}
\label{ch:IntroPandaX}
\chaptermark{The PandaX-III Experiment and the Search...}


\section{The Neutrino and Double Beta Decay}

In 2015, the Japanese Takaaki Kajita and the Canadian Arthur B. McDonald were awarded the Nobel Prize in Physics for the discovery of solar neutrino oscillations, made in the early 21st century at the Super-Kamiokande~\cite{Fukuda2001} and SNO~(Sudbury Neutrino Observatory)~\cite{Ahmad2002} neutrino observatories. Following them, other experiments were able to measure these oscillations, also at Super-Kamiokande~\cite{Jung2003} but for atmospheric neutrinos, and at KamLAND~\cite{Araki2005} for neutrinos from a nuclear reactor. This discovery provided evidence that neutrinos have a non-zero mass.

The study of the neutrino has a historically significant relationship with the beta decay process, which occurs when a nucleon in the atomic nucleus changes flavor and emits a (anti)electron and a (anti)neutrino. The study of this type of decay, already observed in the late 19th century, allowed Enrico Fermi to publish a theory in 1933 that described the interactions occurring in this decay through the weak force, including the neutrino in them. Subsequently, after Maria Goeppert Mayer proposed the idea of double beta decay in 1935 and Ettore Majorana suggested in 1937 the possibility that the neutrino could be its own antiparticle, Wendell H. Furry proposed in 1939 that double beta decay could occur without the emission of neutrinos if the neutrinos are of the Majorana type.

Double beta decay is a type of decay in which two neutrons simultaneously transform into two protons, or vice versa, allowing the atom to approach the optimal ratio of protons to neutrons, as in simple beta decay:

\begin{equation}
\begin{split}
&(A,Z)~{\rightarrow}~(A,Z+2)~+~2e^-~+~2\bar{\nu}_e \\
&(A,Z)~{\rightarrow}~(A,Z-2)~+~2e^+~+~2\nu_e
\label{eq:DobleBetaDecay}
\end{split}
\end{equation}

In practice, double beta decay can be observed when, following the law of conservation of energy, simple beta decay is forbidden. This occurs with the most stable elements, which have an even atomic number and an even number of neutrons. In these elements, the pairing force makes even-even nuclei much more bound, and simple beta decay is strongly suppressed, as it would cause a very large change in spin. Depending on the particles emitted in the decay, the process can be classified into several modes, of which two are the most well-known: double beta decay with neutrino emission ($2\nu\beta\beta$) and double beta decay without neutrinos ($0\nu\beta\beta$).

\begin{figure}[!ht]
    \centering
    \includegraphics[width=0.6\textwidth]{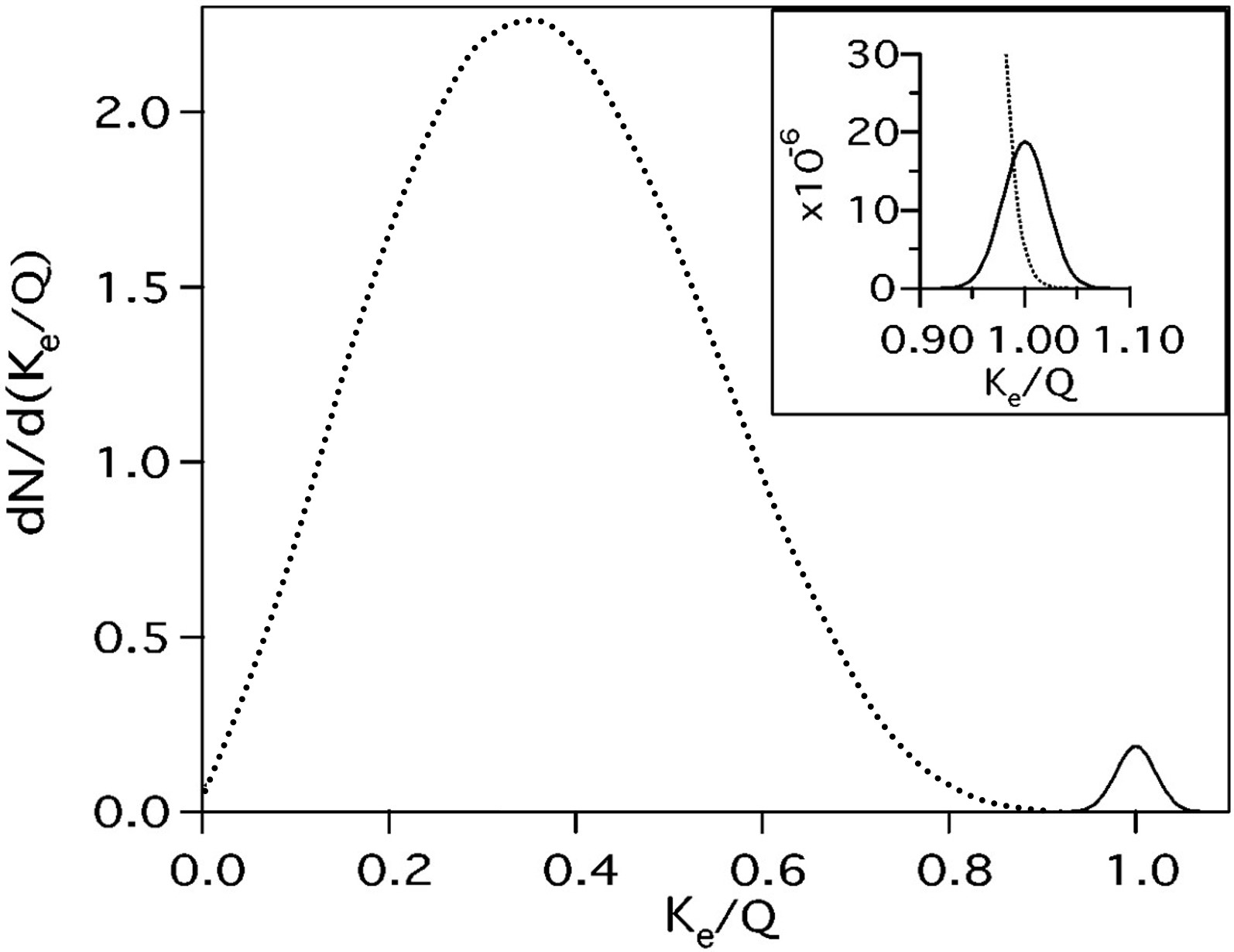}
    \caption{\cite{Elliott2003} Expected spectrum corresponding to the sum of the kinetic energy of the electrons $K_e$ ($Q=Q_{\beta\beta}$ is the endpoint) for the $2\nu\beta\beta$ decays normalized to 1 (dashed curve) and $0\nu\beta\beta$ (solid curve). The $0\nu\beta\beta$ spectrum is normalized to $10^{-2}$ (or $10^{-6}$ in the inset). The spectra are convoluted with a 5\% energy resolution, representative of several experiments. The inset illustrates in detail the expected spectra near the endpoint, where the decay $2\nu\beta\beta$, along with the natural background dependent on the transition energy $Q_{\beta\beta}$ of the measured isotope, represents the maximum possible background that allows the observation of the decay $0\nu\beta\beta$.}
    \label{fig:EspectroGeneralDobleBeta}
\end{figure}

For the decay $2\nu\beta\beta$, the lepton number is conserved; therefore, the process is allowed by the Standard Model. The transition energy $Q_{\beta\beta}$ is shared between the two emitted electrons and the two neutrinos. The experimental signal is the sum of the energies of the two emitted electrons, which is then continuous with an average value around $Q_{\beta\beta}/3$ and ends at the energy $Q_{\beta\beta}$, as shown in figure~\ref{fig:EspectroGeneralDobleBeta}. Several experiments, such as NEMO or CUPID, have been able to measure the half-lives of $2\nu\beta\beta$ for various isotopes.

In the decay $0\nu\beta\beta$, the lepton number is not conserved, which would imply physics beyond the Standard Model. The two electrons would carry all the available kinetic energy (the nuclear recoil is negligible for practical purposes), and the spectrum would show a single peak at the endpoint (figure~\ref{fig:EspectroGeneralDobleBeta}). If the decay $0\nu\beta\beta$ were observed, it would confirm that the neutrino is a massive Majorana particle ($\nu=\bar{\nu}$,~$m_{\nu}{\neq}0$), also allowing the determination of its effective mass.

Due to the low event rate in this process, with a predicted half-life exceeding $10^{26}$ years for the main isotopes, one of the greatest challenges for experiments is to efficiently discriminate events from the decay $0\nu\beta\beta$ from those of the radioactive background and from the events of the decay $2\nu\beta\beta$ itself. In recent years, efforts have been increased in the development of new and better low-background techniques, not only in the field of double beta experiments but also in other experiments searching for rare events, thus significantly reducing the radioactive background of the detectors. However, experiments searching for the decay $0\nu\beta\beta$ that use the detector itself as a container for the double beta emitting isotope are required to greatly improve discrimination techniques to differentiate those events from the decay $0\nu\beta\beta$ from those arising from the decay $2\nu\beta\beta$ itself.

\section{The T-REX Project and the PandaX-III Experiment}

The T-REX project (\textit{TPCs for Rare Event eXperiments})~\cite{Dafni2012} has aimed, over the past few years, to merge the most advanced reading techniques in gas TPCs and ultra-low background techniques, under the funding of the European Research Council, to develop practical ideas that could be used in the field of Physics focused on the search for rare events. The main component of these studies is the use of state-of-the-art Micromegas~\cite{Giomataris1996} readout planes and their subsequent development to meet the stringent requirements of this type of experiment.

The classic design of Micromegas has been utilized in various applications, including particle physics experiments such as ATLAS~\cite{Alexopoulos2010}, COMPASS~\cite{Alexeev2013}, or CAST~\cite{Anastassopoulos2017} at CERN, as well as in neutrino physics experiments like T2K~\cite{Abe2011}. Since its invention, significant efforts have been made to improve the performance of these detectors, starting with their construction technique. These efforts have resulted in two new types of Micromegas: \textit{bulk}~\cite{Giomataris2006} and \textit{microbulk}~\cite{Andriamonje2010} technologies.

Micromegas technology is a very promising readout for applications in the search for rare events with high-pressure TPCs, especially the microbulk type technology, due to its low radioactivity~\cite{Cebrian2011, Castel2019}. Therefore, the objective of the T-REX project was to explore this potential, and detectors of both types (bulk and microbulk) were constructed, tested in the laboratory, and also installed in operational experiments, such as the CAST experiment~\cite{Anastassopoulos2017}, achieving very good levels of background discrimination. Additionally, the commissioning and first data taking with the NEXT-MM prototype~\cite{Alvarez2014} were conducted for the search for 0$\nu\beta\beta$ decay, as one of the options to equip this experiment with a readout plane. This work included measurements with bulk and microbulk technologies in reference mixtures: Ar~+~2\%~iC$_4$H$_{10}$ (isobutane) and Xe~+~2\%~TMA (trimethylamine), demonstrating the topological reconstruction capabilities of the system. This work~\cite{Alvarez2014_2} is pioneering in the use of microbulk technology in a large high-pressure system, achieving an energy resolution of 10.6\% (1 bar) and 12.5\% (2.7 bar) at 30 keV.

The T-REX project also led to the construction and commissioning of the TREX-DM detector~\cite{Iguaz2016}, for the search for WIMPs~\cite{Irastorza2016a}, which is studied in the second part of this thesis.

In the T-REX project, it has been demonstrated that a TPC with Xenon and Micromegas readout planes is a highly competitive solution for the search for 0$\nu\beta\beta$ decay~\cite{Irastorza2016}. Microbulk Micromegas can be constructed and implemented with extremely low levels of radioactivity and show very good performance in Xe+TMA mixtures at 10 bar~\cite{Alvarez2014_2}, with favorable figures in terms of gain, spatial resolution, and energy resolution (3\% FWHM at the energy $Q_{\beta\beta}$). Furthermore, the Xe+TMA mixture exhibits a transverse diffusion of approximately 1 millimeter for a drift of 1 meter, a notable value that enhances the quality of the event's topological reconstruction, allowing for the adjustment of the pixel size in the design phase of the Micromegas to take advantage of this low diffusion. The capability to discriminate background in the region of interest, through the topological information of double beta events (a continuous track with two separated energy deposits at one end), is a potential to exploit for reducing the background in experiments.

One of the objectives of this thesis is the development of technical solutions to implement this readout concept at larger scales, maintaining the performance achieved in small-scale prototypes. These development efforts focused on the search for 0$\nu\beta\beta$ decay have been carried out thanks to the PandaX-III collaboration~\cite{Chen2017a}, which initiated the project for the construction of a first detector of 140 kg of Xenon enriched with $^{136}$Xe and Micromegas technology as the readout plane.

PandaX-III is a project for the search for 0$\nu\beta\beta$ decay at the CJPL (China Jinping Underground Laboratory). The first phase of the experiment consists of constructing a 140 kg TPC, with a sensitivity to the half-life of $^{136}$Xe in the 0$\nu\beta\beta$ decay of 10$^{26}$ years, after 3 years of operation. In a second phase, the construction of more TPCs is planned, with possible improvements in terms of energy resolution and background control, until a total of one ton of active mass is achieved. With a three-year lifespan, the complete experiment can reach a limit for $T^{0{\nu}}_{1/2}$ of 10$^{27}$ years.

\begin{figure}[!ht]
    \centering
    \includegraphics[width=0.7\textwidth]{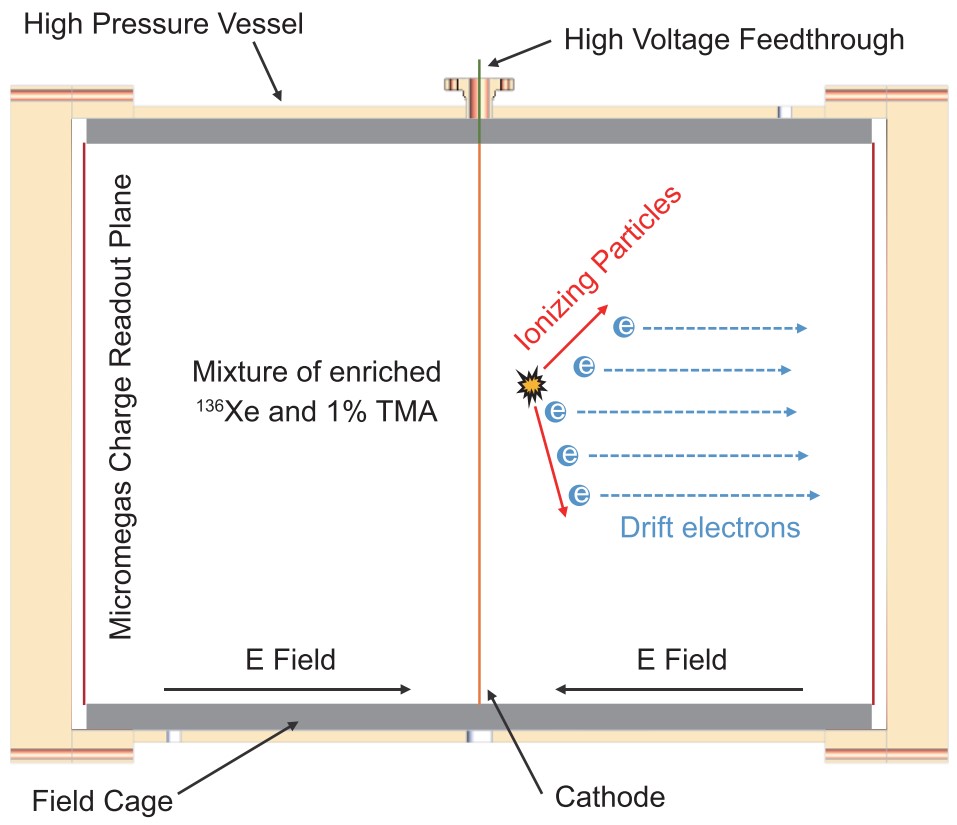}
    \caption{\cite{Chen2017a} Schematic of the PandaX-III TPC, with the main components.}
    \label{fig:IntroPandaX_EsquemaTPCpandaX}
\end{figure}

The central component of the PandaX-III experiment is a gas TPC at 10 bar pressure. The cylindrical TPC features a symmetrical design with a cathode in the middle and readout planes at both ends, as shown in figure~\ref{fig:IntroPandaX_EsquemaTPCpandaX}. For each half of the TPC, its active volume measures approximately 1 m in length and 1.5 m in diameter. By applying a negative voltage of up to 100 kV on the cathode, the generated electric field (drift field) displaces the primary electrons produced in the ionization of the gas by an ionizing event towards the two ends, where the readout planes are located. A field cage along the cylinder ensures a homogeneous drift field in the active volume.

From the beginning of the project, microbulk Micromegas technology was chosen as the main line for the development of the readout plane of the experiment's TPC. The construction of a 1.5 m diameter plane with this technology posed a challenge, as the maximum size that could be manufactured at that time was 20 x 20 cm$^2$, which required creating a tessellated readout plane with several Micromegas. Thus, the idea of developing a single Micromegas module that could be installed to form a larger readout plane emerged: the SR2M (\textit{Scalable Radiopure Readout Module}).

Each SR2M readout module, based on microbulk Micromegas technology, would cover 20 x 20 cm$^2$ of the total readout plane. Through the reading with the SR2M module, both the X-Y coordinates and the arrival time of the electrons at the module are measured. The spatial and temporal resolutions of the SR2M module are on the order of 1 mm and up to 10 ns, determined on one hand by the pixel size in the Micromegas (manufacturing parameter) and on the other hand by the maximum sampling frequency of the electronics used (100 MHz). The chamber that constitutes the TPC is made of radiopure copper and contains 140 kg of Xenon gas (90\% enriched with $^{136}$Xe) at 10 bar. The Xenon gas is mixed with 1\% TMA to improve the signal quality in the Micromegas in terms of energy resolution and electron diffusion.

A high-pressure gas system~\cite{Chen2017a} supplies and purifies the gas mixture used, with continuous recirculation. A prototype of this gas system has already been successfully commissioned as a subsystem of the TPC prototype being developed at the SJTU (Shanghai Jiao Tong University), prior to the construction of the final 140 kg TPC.

The readout electronics of PandaX-III is one of the key subsystems of the experiment. Among many of its requirements and desired characteristics, the readout electronics must provide low noise, high energy resolution, scalability to several tens of thousands of channels, and the lowest possible level of radioactivity. The electronics have been designed to integrate the charge in the Micromegas pixels and the signals from the cathode (mesh in the Micromegas), digitizing the signal after shaping it and sending the data packets to the acquisition computer. Each FEC (\textit{Front-End card}) contains 4 ASIC chips called AGET~\cite{AGETforTPCs}, each with 64 channels. This chip was developed in Saclay, IRFU CEA in France, as a generic chip to be used in a TPC with calorimetric and topological measurements.

So far, the PandaX-III project team has successfully constructed a TPC prototype with 7 SR2M modules to demonstrate the viability of the TPC system with Micromegas, achieving good results in terms of electron transmission, gain, and energy resolution with various radioactive sources ($^{241}$Am, $^{137}$Cs, and $^{55}$Fe). The construction of the full-size detector has already been completed, along with 52 SR2M modules based on Micromegas developed in China, and once the surface tests are concluded, it will be installed in the Jinping Underground Laboratory~\cite{Zhang2023}.

\setlength{\baselineskip}{12pt} 
\selectlanguage{USenglish}
\graphicspath{{../Imagenes/}{../../Imagenes/}} 

\renewcommand{\listtablename}{Table of contents}
\renewcommand{\tablename}{Table}

\chapter{The SR2M Module for Large Readout Planes}
\label{ch:TheSR2M}
\chaptermark{The SR2M Module for Large...}

\begin{figure}[!ht]
    \centering
    \includegraphics[width=0.8\textwidth]{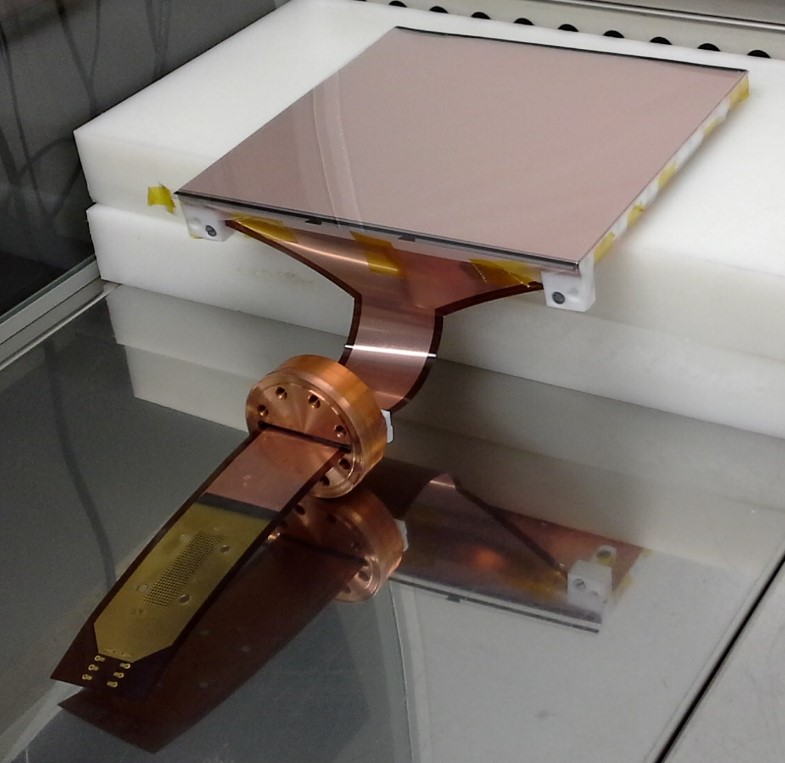}
    \caption*{}
    \label{fig:Photo_SR2M}
\end{figure}

\newpage

One of the objectives of this work is to achieve large readout planes through the installation of several microbulk-type Micromegas circuits~\cite{Andriamonje2010}, focusing on the design of a single module, referred to as the \textit{Scalable Radiopure Readout Module} (hereafter SR2M), which will allow the construction of different readout planes of varying sizes in the context of low-background and rare-event experiments. This module is composed not only of the Micromegas circuit but also of complementary pieces for its handling, assembly, and connection.

Microbulk-type Micromegas offer primarily two advantages over bulk types. The first is their low intrinsic radioactivity, as the materials from which they are made are primarily Copper and Polyimide (the generic name of the compound that makes up Kapton\textregistered). It is true that the lithographic methods with chemical baths during their manufacturing introduce a small contamination, especially in $^{238}$~U and $^{40}$~K \cite{Aznar2013}. However, there is room for improvement through the control of these lithographic processes to further reduce their radioactivity. The second advantage is their higher energy resolution, achieving 11\%~FWHM for 5.9~keV photons and 1.8\%~FWHM for 5.5~MeV alphas \cite{Andriamonje2010}. The ability to be below 1\%~FWHM in the double beta decay energy ($Q_{\beta\beta}$~=~2.457~MeV for the isotope $^{136}$Xe) makes this type of detector strong candidates for installation in experiments searching for neutrinoless double beta decay, such as the PandaX-III experiment.

This type of low-background and rare-event experiments defines the specifications for designing the SR2M module, which are listed below:

\begin{itemize}

\item \textbf{Scalability:} This is the key point and the main objective in the design of the SR2M module. Dark matter search experiments and neutrinoless double beta decay experiments increasingly require larger volumes of exposure/conversion. This entails larger readout planes capable of detecting events produced within the chamber, with spatial and energy resolution determined by each experiment. At the time the design of the SR2M module was proposed, the maximum dimensions for manufacturing a microbulk-type Micromegas were 20~x~20~cm$^2$, completely insufficient for large volume experiments that propose readout planes above 10$^4$~cm$^2$.

\item \textbf{Radiopurity:} Due to the need for a very low background in this type of experiments, the SR2M module must meet high radiopurity requirements. Microbulk-type Micromegas, as previously explained, have low intrinsic radioactivity~\cite{Aznar2013}, and are also potentially improvable. However, the module requires other pieces for its construction, limiting the materials that can be used. In \cite{Aznar2013}, a radiopurity study is conducted on a wide list of materials used in rare-event experiments with Micromegas, which will be an important reference when selecting materials for the SR2M module.

\item \textbf{Maximum efficiency in the collection of primary electrons:} In the construction of a large readout plane using modules, there may be adjustment gaps between them. These gaps, along with the distortion of the electric field in their vicinity, can cause the loss of primary electrons, reduce the event reconstruction capability, and worsen the energy resolution. Apart from the space between the modules, there may also be other processes that produce low efficiency or uniformity in the collection of primary electrons towards the amplification holes of the Micromegas (figure~\ref{fig:Introduction_MicromegasWorkingPrinciple}). All these processes occur during the transport of these electrons from the ionization region to the Micromegas (recombination, \textit{attachment}, electron transmission, etc.).

\item \textbf{Minimum number of connections to the electronics:} Intermediate connections in the connection chain from the Micromegas (from the corresponding pixel or \textit{strip}) to the readout electronics can lead to connectivity failures over time, changes in the impedance and capacitance of the channel due to weak connections or \textit{crosstalk}. Therefore, the design of the SR2M module must maximize the robustness of the connection chain.

\end{itemize}

\begin{figure}[!ht]
    \centering
    \includegraphics[width=\textwidth]{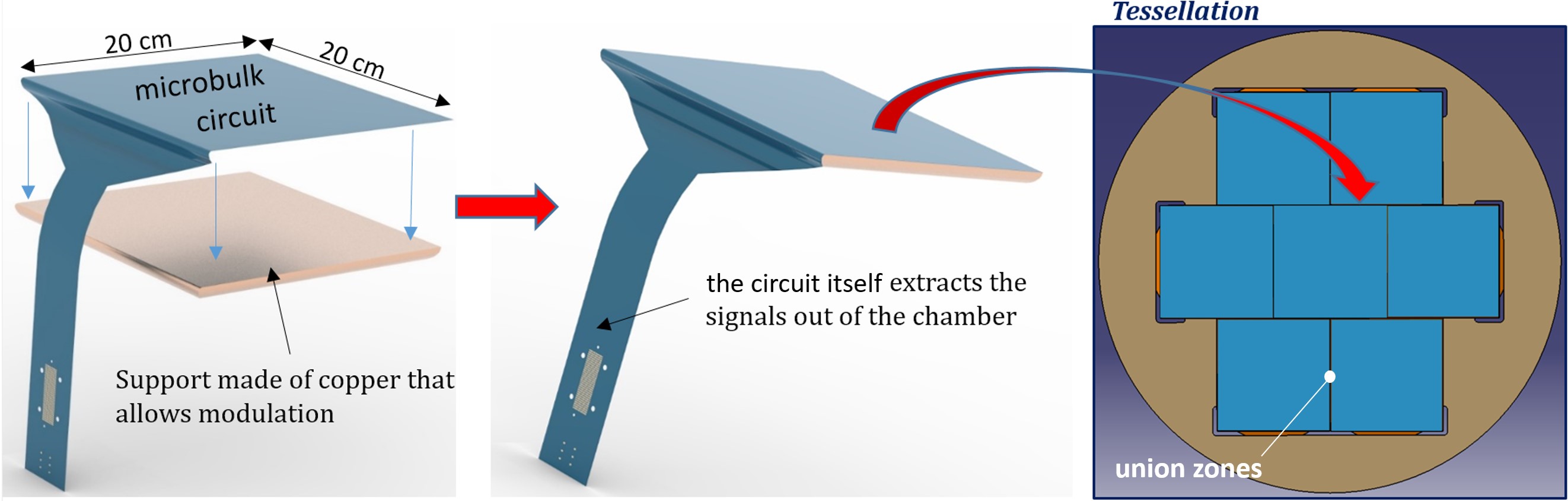}
    \caption{Illustration of the first conceptual design of the SR2M module. The Micromegas, with a tab for channel extraction, is attached to a rigid support for assembly in the readout plane. Microbulk-type Micromegas are sheets with a thickness on the order of tens of~$\mu$m, with a rigidity similar to that of a sheet of paper, so they must be handled with care.}
    \label{fig:SR2M_ConceptualDesign}
\end{figure}

With these specifications, a first conceptual design was launched (see figure~\ref{fig:SR2M_ConceptualDesign}) to iterate and advance in the process towards the final design. This first concept was based on the manufacture of a microbulk-type Micromegas with the largest possible active area at that time and with an extension of the circuit, or tab, used to extract the channels directly to the outside of the chamber. The Micromegas would be attached to a rigid support that would allow for manipulation and assembly into its final position in the readout plane.

\section{Micromegas Circuit}

\begin{figure} [htbp]
	\centering
	\begin{subfigure}[htbp]{\textwidth}
		\centering
		\includegraphics[width=0.85\textwidth]{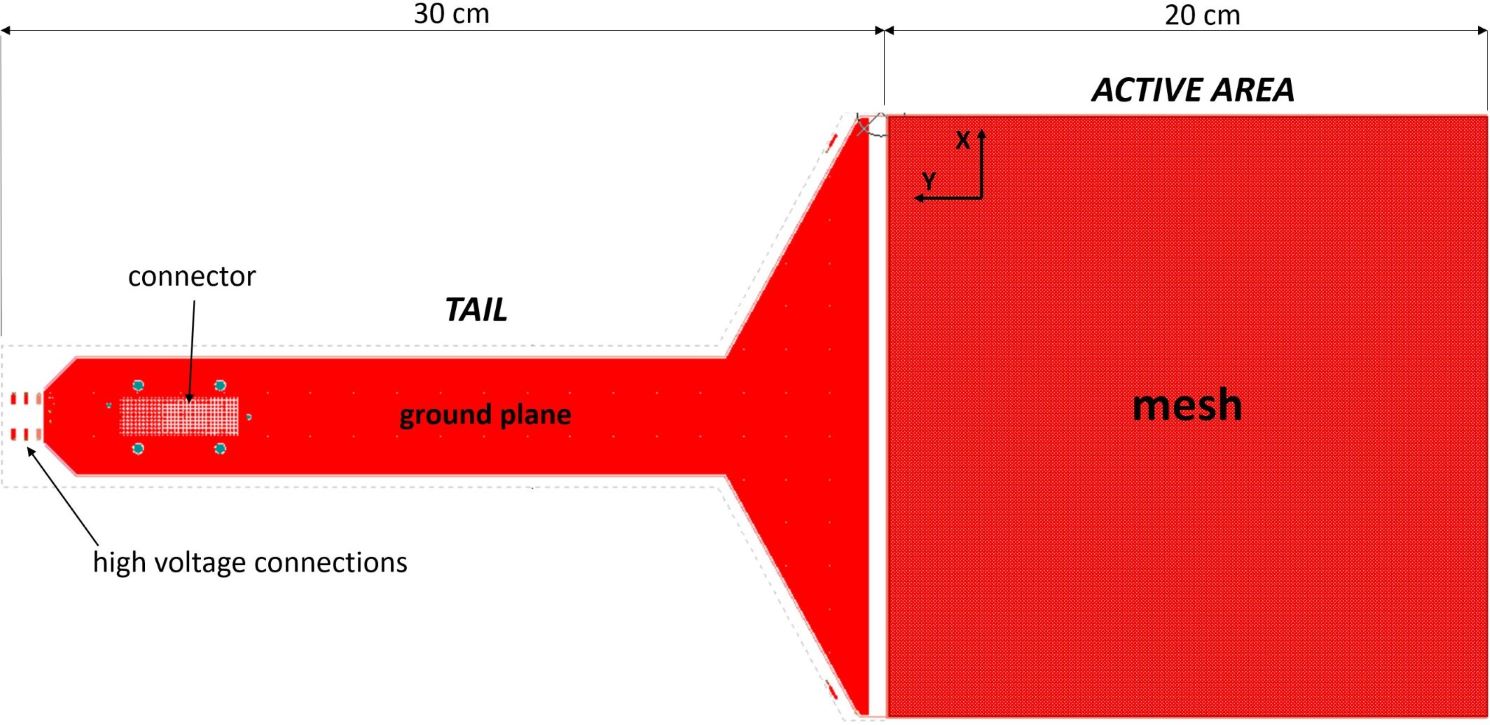}
		\caption{Top layer}\label{fig:SR2M_MM_Top}
	\end{subfigure}
	\begin{subfigure}[htbp]{\textwidth}
		\centering
		\includegraphics[width=0.85\textwidth]{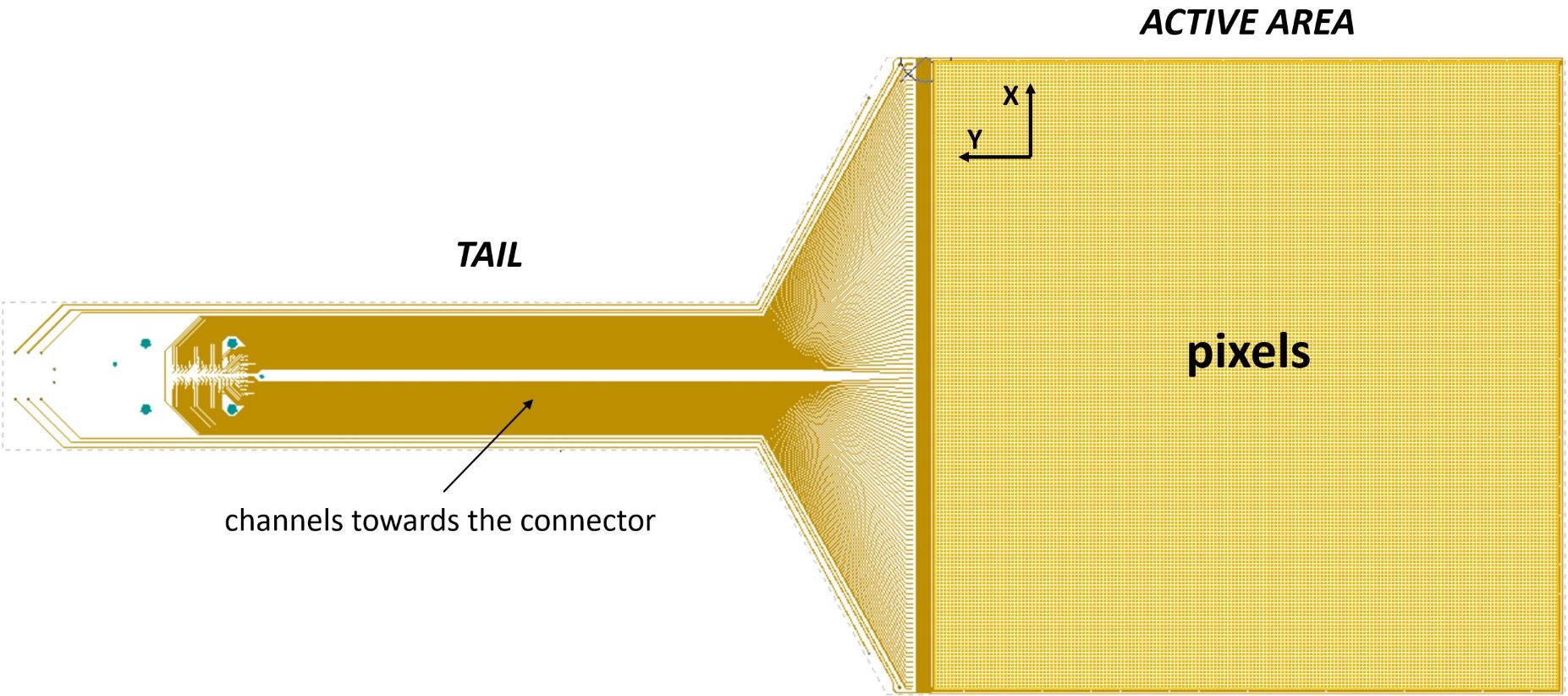}
		\caption{Intermediate layer}\label{fig:SR2M_MM_Inner}
	\end{subfigure}
	\begin{subfigure}[htbp]{\textwidth}
		\centering
		\includegraphics[width=0.85\textwidth]{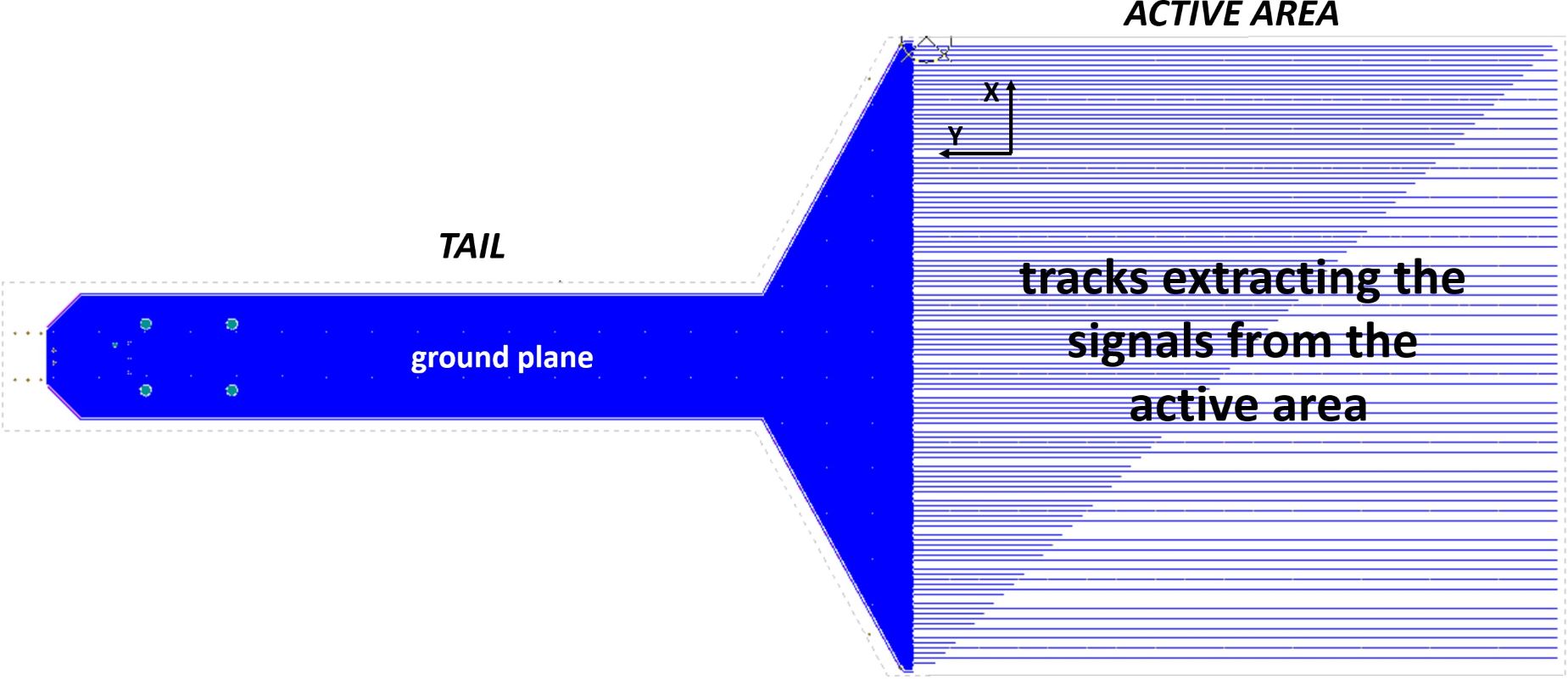}
		\caption{Bottom layer}\label{fig:SR2M_MM_Bottom}
	\end{subfigure}
	\caption{Layers of the Micromegas circuit designed for the SR2M module: \textbf{(\subref{fig:SR2M_MM_Top})~top layer} containing the mesh in the active area and a ground plane in the tab; \textbf{(\subref{fig:SR2M_MM_Inner})~intermediate layer} containing the pixels in the active area and the channel traces to the connector in the tab; \textbf{(\subref{fig:SR2M_MM_Bottom})~bottom layer} containing the traces that connect the pixels to form each of the strips along the X-axis and a ground plane in the tab (see Figure~\ref{fig:SR2M_XY_Pattern}).}
	\label{fig:SR2M_MM_Layers}
\end{figure}

\begin{figure}[!ht]
    \centering
    \includegraphics[width=\textwidth]{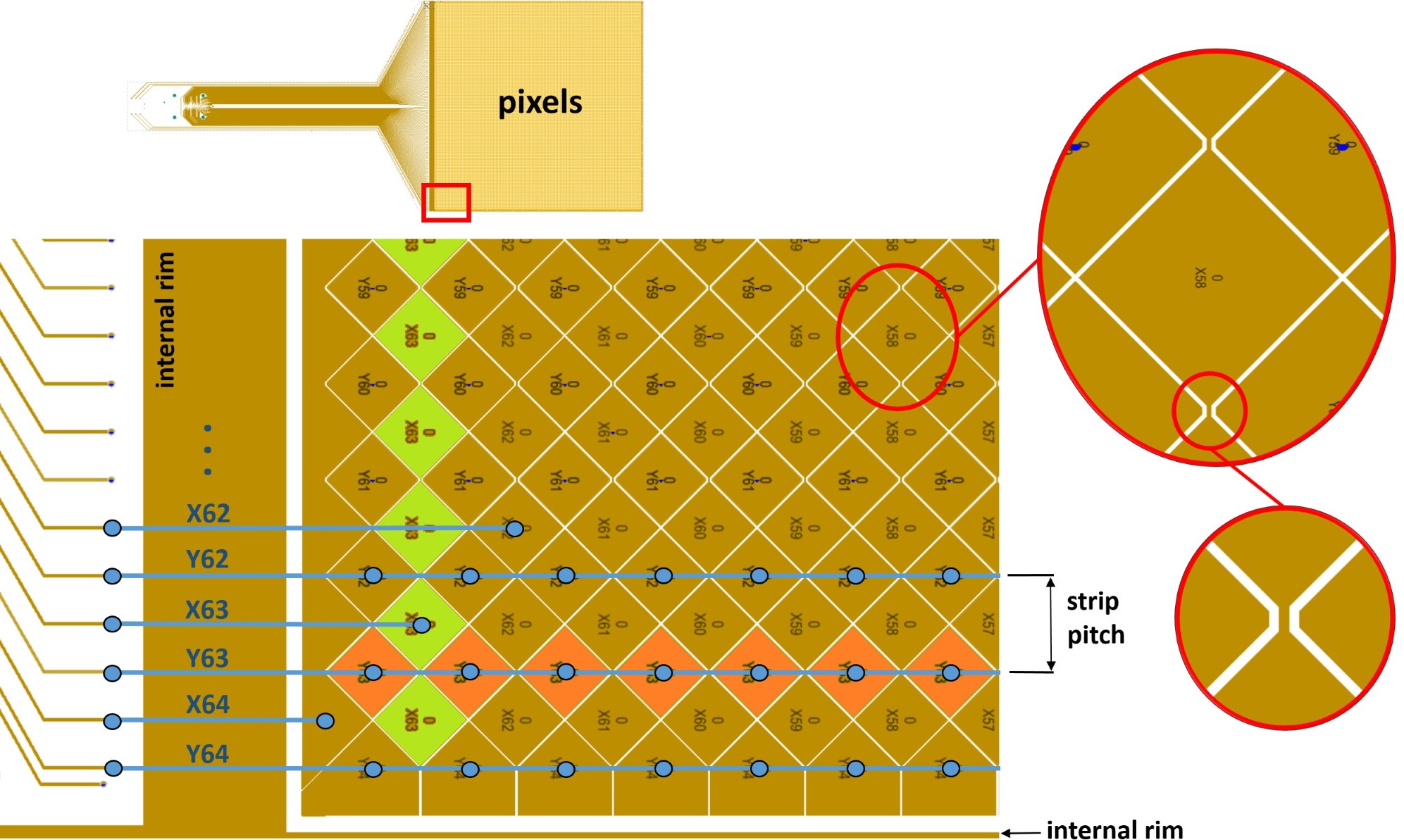}
    \caption{Intermediate layer in the lower left corner of the active area in the Micromegas circuit designed for the SR2M module. Half of the pixels are connected in the same layer (detail) forming the X-axis strips, while the other half connects through the bottom layer (blue lines) to form the Y-axis strips. A \textit{strip} is the surface formed by the sum of several pixels along one axis. The blue dots represent the holes (vias) that connect the intermediate layer with the bottom layer. The X63 strip (green) whose pixels are connected in the same layer is highlighted. The Y63 \textit{strip} (orange) whose pixels connect with a trace located in the bottom layer is also highlighted. Once all channels exit the active area through the bottom layer, they pass again, via vias, to the intermediate layer to reach the connector at the end of the tab (off the image). These traces running along the tab to the connector are shielded by respective ground planes in the top and bottom layers (Figure~\ref{fig:SR2M_MM_Layers}).}
    \label{fig:SR2M_XY_Pattern}
\end{figure}

For the development of the SR2M module, it was necessary to design a Micromegas circuit tailored to the module's specifications. This design can be seen in Figure \ref{fig:SR2M_MM_Layers}, which shows the three layers that make up the circuit. In Figure~\ref{fig:SR2M_XY_Pattern}, a superposition of the lower and intermediate layers corresponding to a corner of the active area can be observed.

The active area is 20~x~20~cm$^2$, the largest area that the manufacturing process allowed at the time of the SR2M module design. The only motivation for making a smaller area was better adaptation to any future geometry of the readout plane, as a smaller module can fill the space more optimally. However, a smaller area increases the total number of mechanical elements required, complicating assembly, and will also have more junction areas between the modules that are susceptible to causing the loss of primary electrons.

The maximum size of the copper-kapton-copper sheet that can be used for the manufacture of a microbulk-type Micromegas is 50~x~50~cm$^2$, on which a maximum active area of 20~x~20~cm$^2$ could be produced. These dimensions allowed for a tab in the Micromegas of 30~cm in length, sufficient to extract the channels outside the TPC if the wall/top is flat.

Figure \ref{fig:SR2M_XY_Pattern} shows the active area of the Micromegas according to the scheme in Figure~\ref{fig:SR2M_LayersActiveArea}, where an internal \textit{rim} will be introduced for the electron reintegration system (ERS), discussed in Section~\ref{sec:ERS}.

Some of the parameters considered during the design are shown in Table~\ref{tab:SR2M_ParametrosDiseñoMM}.

\begin{table}[!ht]
\centering
\begin{tabular}{|c|c|}
\hline
\textbf{Parameter} & \textbf{Value} \\ \hline
Active area dimensions (cm x cm) & 20 x 20 \\
Length of the extension (tab) (cm) & 30 \\
Minimum distance between channels in the tab area ($\mu$m) & 150 \\
Minimum distance between channels in the connector area ($\mu$m) & 80 \\
Width of the channel trace ($\mu$m) & 100 \\
Minimum distance between mesh traces and strips ($\mu$m) & 1000 \\
Minimum distance between mesh traces and internal \textit{rim} ($\mu$m) & 700 \\
Minimum distance between internal \textit{rim} and external \textit{rim} ($\mu$m) & 1100 \\
\hline
\end{tabular}
\caption{Design parameters of the Micromegas circuit for the SR2M module.}
\label{tab:SR2M_ParametrosDiseñoMM}
\end{table}

The parameters that define the active area and, therefore, the behavior of the Micromegas are: the amplification \textit{gap} (see Figure~\ref{fig:Introduction_MicromegasWorkingPrinciple}), which is the distance of separation between the upper plane or \textit{mesh} and the lower plane (pixel/strip); the diameter of the amplification hole, which together with the amplification \textit{gap} defines the amplification zone where the electron avalanche effect occurs; and the \textit{holes pitch}, the distance between the centers of the amplification holes distributed across the entire surface of the Micromegas. Based on the group's experience in operating different patterns, a D50P100G50 pattern (Diameter-Pitch-Gap, units in $\mu$m) was chosen for this case, a pattern for which the manufacturing methods used achieved high quality and that was yielding very good results in the CAST experiment~\cite{Anastassopoulos2017}.

\subsection{Strip Width and Pixel Size}
\label{sec:StripPitch_PixelSize}

When the lower plane of the Micromegas is segmented to form the pixels/strips, another parameter is defined: the strip width or strip pitch. The strip width is the distance between the centers of two consecutive strips along the same axis and is determined by the pixel size and the separation distance between them (strip pitch in figure~\ref{fig:SR2M_XY_Pattern}).

The most critical parameter, which can be strongly limited by the selection of the strip width, is the spatial resolution in the XY plane. The spatial resolution $\sigma_{xy_{~(MM)}}$ in the XY plane for a Micromegas, similar to that for a MWPC (Multi Wire Proportional Chamber) from which it inherits its operating principle, is defined as

\begin{equation}
\sigma_{xy_{~(MM)}}=\dfrac{p}{\sqrt{12}}
\label{eq:SpatialResolMM}
\end{equation}

\noindent where \( p \) is the strip width~\cite{WRLeo1994, Chen2017a}. The minimum spatial resolution that can be achieved with a Micromegas in the Z axis (drift direction) primarily depends on the precision with which the arrival time of the signals is measured and is not dependent on the strip width.

In general, when an event occurs within the active volume of a TPC and the primary electrons drift towards the readout plane, the lowest spatial resolution that can be achieved in the readout plane \( \sigma_{xy_{~(TPC)}} \) mainly depends on the transverse diffusion of the electrons, such that

\begin{equation}
\sigma_{xy_{~(TPC)}} = D_{T}\dfrac{\sqrt{z}}{\sqrt{P}}
\label{eq:SpatialResolTPC}
\end{equation}

\noindent where \( P \) is the pressure in the TPC, \( z \) is the drift length of the event, and \( D_{T} \) is the diffusion coefficient.

In any experiment, the ideal strip width will be the one that maximizes the intrinsic spatial resolution \( \sigma_{xy_{~(TPC)}} \) of the TPC. For this, the spatial resolution in XY in the readout plane (eq~\ref{eq:SpatialResolMM}) must be less than or equal to the spatial resolution in XY in the TPC (eq~\ref{eq:SpatialResolTPC}), such that

\begin{equation}
\sigma_{xy_{~(MM)}} = \dfrac{p}{\sqrt{12}} ~\leq~ \sigma_{xy_{~(TPC)}} = D_{T}\dfrac{\sqrt{z}}{\sqrt{P}}
\label{eq:SpatialResolMM-TPC}
\end{equation}

\begin{figure}[!ht]
    \centering
    \includegraphics[width=1.0\textwidth]{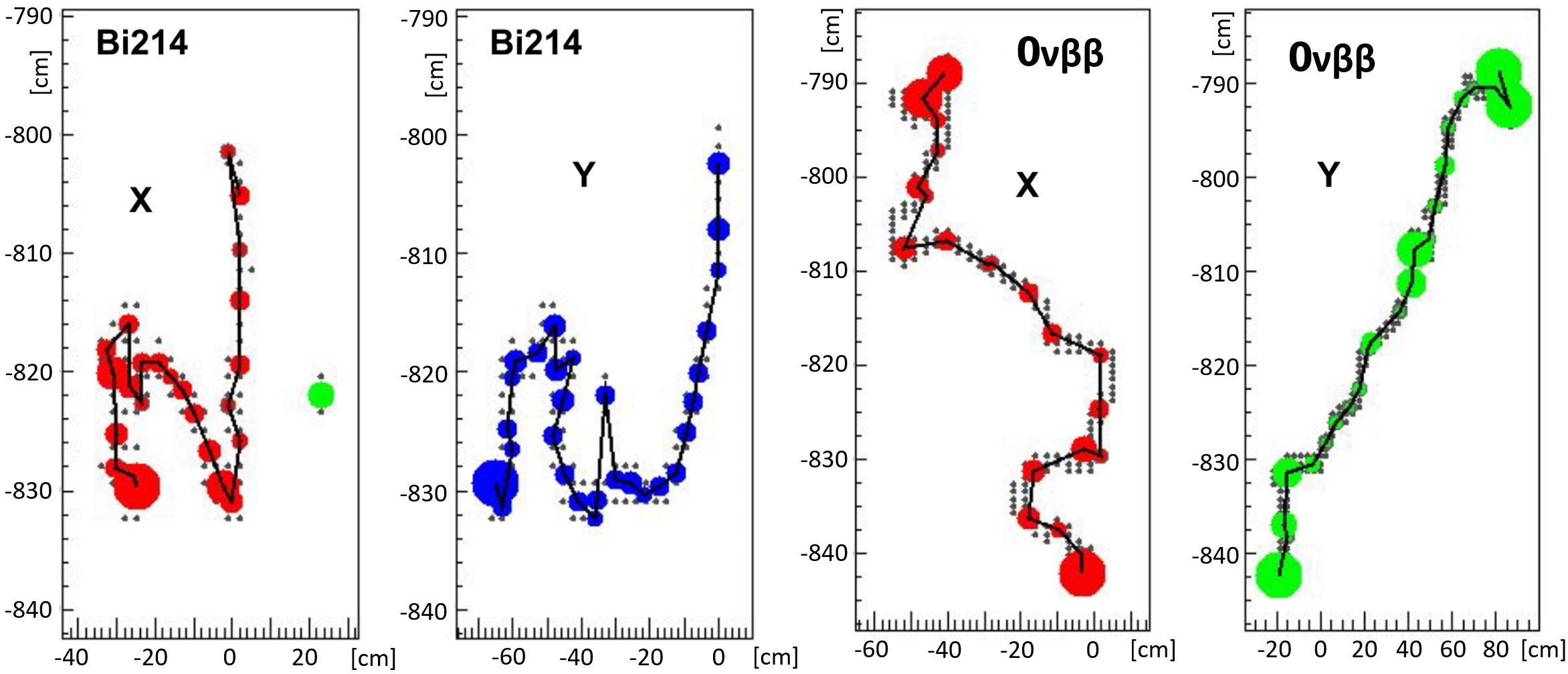}
    \caption{\cite{Chen2017a} Reconstruction of the projections in XZ (red) and YZ (blue) of the track of two Monte Carlo simulated events: a typical background event from the decay of \( ^{214} \)Bi (left) and another from the hypothetical 0$\nu\beta\beta$ decay of \( ^{136} \)Xe (right). It can be observed that the 0$\nu\beta\beta$ event contains an electron cloud at each end of the track, characteristic of double beta decay, whose identifications are key for discrimination and background reduction.}
    \label{fig:DobleBetaTrackVSbackground}
\end{figure}

The spatial resolution in an experiment like PandaX-III, which aims to reconstruct the tracks produced by the ionization of particles traversing the TPC, is a critical parameter. To define the necessary resolution in the experiment, several studies were conducted~\cite{Galan2016, Chen2017a, Lin2018a}, after which it was decided to use a strip width of 3.09 mm. This size would allow for a spatial resolution of \( \sigma_{xy_{~(MM)}} = 0.9 \) mm, sufficient to reconstruct the events of double beta decay of \( ^{136} \)Xe and to distinguish them from the background produced by gammas of the same energy (figure~\ref{fig:DobleBetaTrackVSbackground}). These events in the TPC at 10 bar have ionization tracks on the order of tens of centimeters and electron clouds at their ends (Bragg peak) of several millimeters.

\begin{figure}[!ht]
    \centering
    \includegraphics[height=8cm]{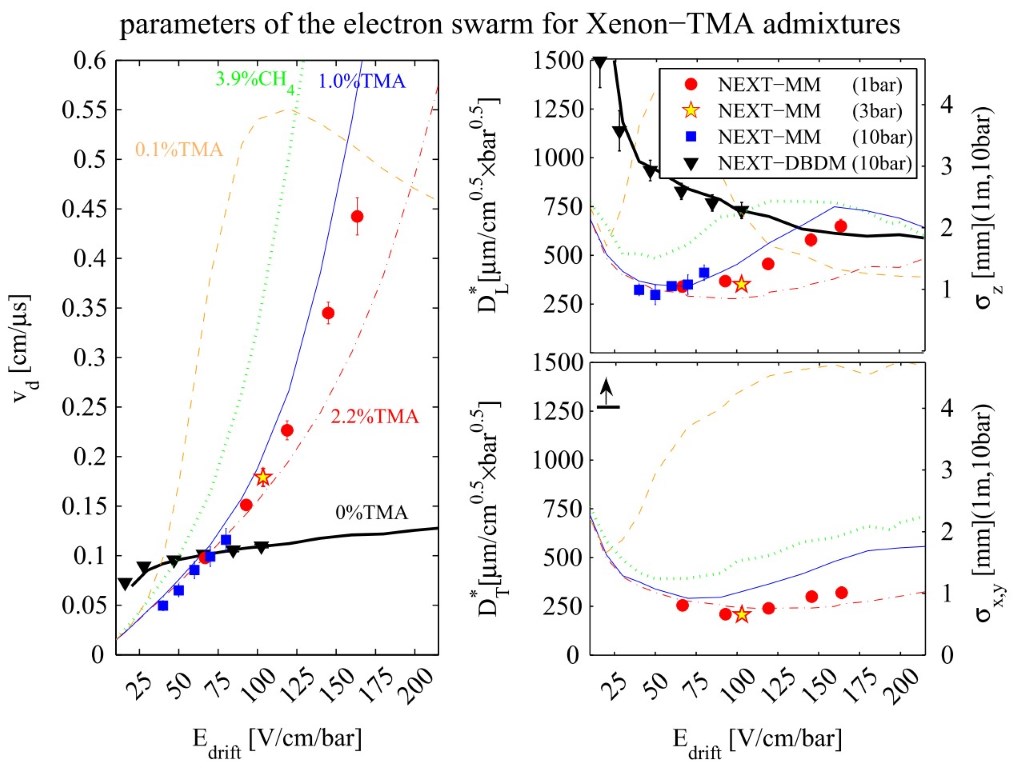}
    \caption{\cite{Gonzalez-Diaz2015} Representation of the main properties of various xenon mixtures in high-pressure TPCs for 1(1.0) 3(2.7) and 10(10.1) bar and different TMA mixtures, along with simulations, as a function of the reduced drift electric field. $\bullet$ Left: drift velocities. $\bullet$ Right: longitudinal and transverse diffusion coefficients. The simulation results for Xe-CH$_{4}$ (Gotthard) and Xe-TMA at 0.1\% mixture can be seen. For higher electric fields, it can be complemented with \cite{Alvarez2014_2}.}
    \label{fig:TranseverseDiffusionNEXT}
\end{figure}

\begin{figure}[!ht]
    \centering
    \includegraphics[width=0.8\textwidth]{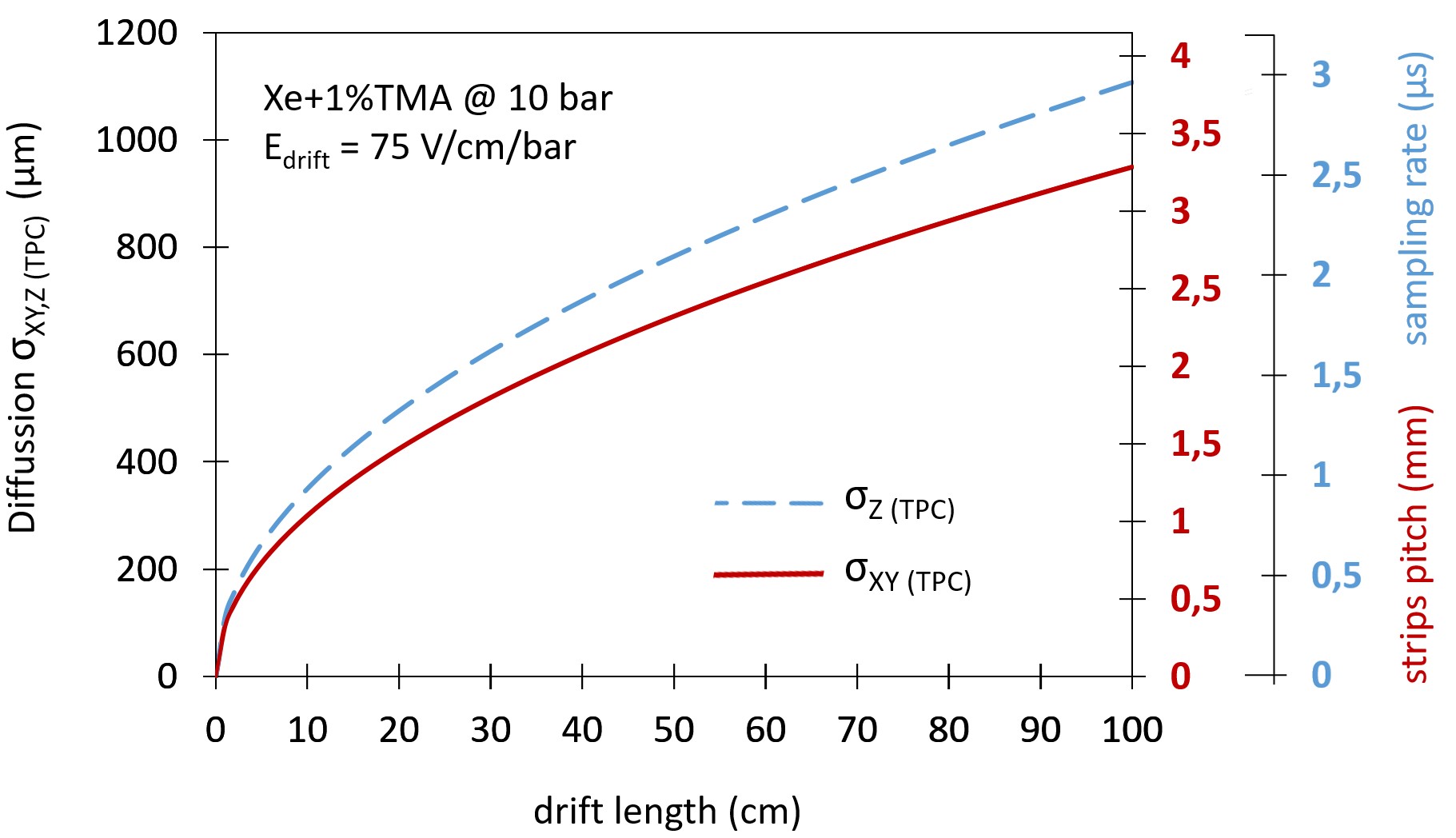}
    \caption{Transverse diffusion \( \sigma_{xy_{~(TPC)}} \) and longitudinal \( \sigma_{z_{~(TPC)}} \) in an event against its drift length, for a TPC with a xenon + 1\% TMA mixture, at 10 bar pressure and with a drift field of \( E_{drift} \approx 75 \) V cm\(^{-1}\) bar\(^{-1}\). The strip width of the Micromegas is added for each point, for which the resolution of the Micromegas is equal to the resolution of the TPC. The sampling rate of the electronics is also added for which the resolution of the electronics is equal to the resolution of the TPC.}
    \label{fig:SR2M_Plot_diffusionVSdriftLenght}
\end{figure}

The PandaX-III experiment consists of a TPC with 140 kg of \( ^{136} \)Xe + 1\% TMA at 10 bar pressure and with a maximum drift length of approximately 1 meter. In~\cite{Gonzalez-Diaz2015}, the diffusion of electrons in Xe+TMA mixtures is studied, from which figure~\ref{fig:TranseverseDiffusionNEXT} has been extracted. For the operating conditions of PandaX-III, the lowest transverse diffusion (\( D_{T} = 300 [{\mu}m/cm^{0.5} \times bar^{0.5}] \)) occurs for the drift field \( E_{drift} \approx 75 \) V cm\(^{-1}\) bar\(^{-1}\). Evaluating equation~\ref{eq:SpatialResolTPC} for the projected TPC of PandaX-III at 10 bar pressure, the spatial resolution \( \sigma_{xy_{~(TPC)}} \) of the TPC can be represented against the drift length of the event (figure~\ref{fig:SR2M_Plot_diffusionVSdriftLenght}). Additionally, using equation~\ref{eq:SpatialResolMM-TPC}, the maximum strip width can be represented in the same graph for which the event occurring at a distance \( z \) from the readout plane is read with the spatial resolution allowed by the TPC.

From figure~\ref{fig:SR2M_Plot_diffusionVSdriftLenght}, it can be inferred that with the strip width chosen for the first prototype of the SR2M (3.09 mm), the resolution of the TPC is lower than that achieved in the SR2M for any event produced throughout the volume. For example, if it were necessary to reduce the strip size, with a strip size of 2.4 mm or 1 mm, the resolution of the TPC would be maximally utilized for all events produced from \( z = 50 \) cm or \( z = 10 \) cm, respectively.

The decision to set the strip width of the Micromegas at 3.09 mm was a consensus and compromise. The commissioning and initial characterizations of the first prototypes of the SR2M module would allow for the evaluation of the impact of the chosen strip width on spatial resolution, in addition to other possible impacts such as energy resolution or the efficiency of the trigger electronics, and to make the decision to reduce or increase the strip width in future developments if necessary. Finally, after the characterizations and tests with the first prototypes of the SR2M, it would be demonstrated that the chosen strip width of 3.09 mm allowed for sufficient spatial resolution to perform topological reconstruction and thus discriminate those double beta events from background events~\cite{Galan2020}.

In conclusion, this section has presented the effects of transverse diffusion in the PandaX-III TPC on spatial resolution in the XY axes, a necessary study to set the strip width of the Micromegas. However, the effect of longitudinal diffusion, which limits the resolution of the TPC in the Z axis and has a direct impact on the choice of the sampling rate of the data acquisition (DAQ) electronics, has not been studied. Therefore, for completeness, we will examine the effects of longitudinal diffusion on the choice of the sampling rate.

Under the operating conditions of the PandaX-III TPC previously outlined, \( ^{136} \)Xe + 1\% TMA at 10 bar, the minimum longitudinal diffusion also occurs around \( E_{drift} \approx 75 \) V cm\(^{-1}\) bar\(^{-1}\) (figure~\ref{fig:TranseverseDiffusionNEXT}). As was done for transverse diffusion, the relationship between the distance traveled by an event during its drift, its longitudinal diffusion, and the sampling rate of the electronics can also be deduced (figure~\ref{fig:SR2M_Plot_diffusionVSdriftLenght}). Assuming that the temporal resolution of the TPC in the Z axis, the temporal resolution of the collection of primary electrons in the readout plane, is equal to \( \sigma_{t_{~(TPC)}} = \sigma_{z_{~(TPC)}}/{v_{d}} \), where \( v_{d} \) is the drift velocity, then

\begin{equation}
\sigma_{t_{~(DAQ)}} = \dfrac{S}{\sqrt{12}} ~\leq~ \sigma_{t_{~(TPC)}} = \dfrac{\sigma_{z_{~(TPC)}}}{v_{d}} = \dfrac{D_{L}\sqrt{z/P}}{v_{d}}
\label{eq:TimeResolMM-TPC}
\end{equation}

\noindent where \( z \) is the drift length of the event and \( S \) is the sampling rate of the readout electronics. The values \( v_{d} = 0.13 \) cm/\(\mu\)s and \( D_{L} = 350 \, {\mu}m/cm^{0.5} \times bar^{0.5} \) are extracted from figure~\ref{fig:TranseverseDiffusionNEXT} for \( E_{drift} \approx 75 \) V cm\(^{-1}\) bar\(^{-1}\).

To achieve the same spatial resolution in Z as in XY (\( \sigma_{xy_{~(MM)}} = 0.9 \) mm), previously set by the strip width of the Micromegas, the sampling rate must be set at 2.95 \(\mu\)s (0.33 MHz). The electronics planned for reading the SR2M modules is capable of operating up to 100 MHz, clearly above the minimum requirements.

\subsection{Connector Selection}

Within the T-REX project at the University of Zaragoza, an activity measurement campaign has been conducted on a wide range of materials and components, used or potentially usable in experiments involving rare events \cite{Aznar2013} \cite{Castel2019}. This measurement program is primarily carried out through gamma-ray spectroscopy with germanium, conducted at the Canfranc Underground Laboratory (LSC), and complemented by other techniques such as GDMS, ICPMS, or using the BiPo-3 detector.

These measurements, supported by others conducted in already published programs, have allowed for the generation of the background model for the experiment and, consequently, the design of the SR2M module.

One of the requirements for the design of the SR2M module was to avoid electronic chain connections within the chamber to reduce the likelihood of connectivity failures during operation, but mainly to avoid the use of commercial connectors whose activity is detrimental to the background. In \cite{Aznar2013}, two types of connectors were measured: the Panasonic P5K series connector (narrow pitch board-to-board with 120 channels) and the Samtec GFZ-30-03-G-10-AD connector (300 channels). Both showed activities of several mBq/piece in the isotopes $^{232}$Th, from the lower part of the $^{238}$U chain, and $^{40}$K. The NEXT experiment collaboration also analyzed the Hirose FFC/FCP connectors (Flexible Printed Circuit \& Flexible Flat Cable with 53 per piece) and the Molex thermoplastic connectors 503066-8011 \cite{Alvarez2013}, finding similar levels of activity.

\begin{figure}[h]
\begin{minipage}{17pc}
\includegraphics[width=0.95\textwidth]{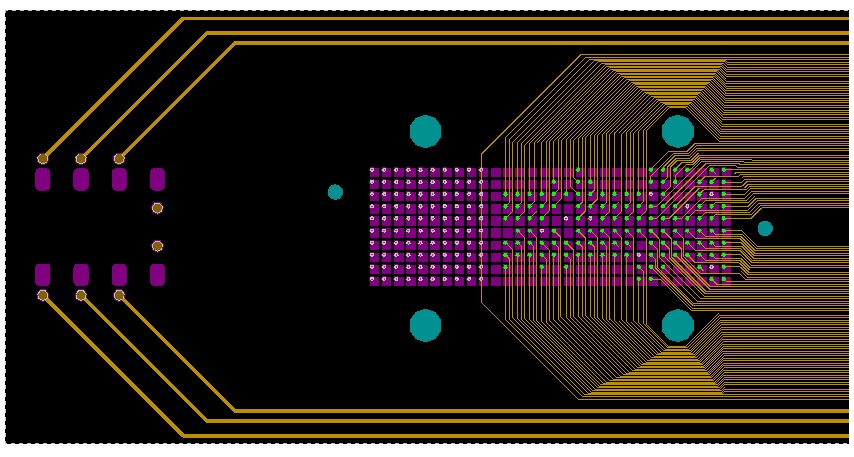}
\end{minipage}\hspace{0.1pc}%
\begin{minipage}{14pc}
\includegraphics[width=0.95\textwidth]{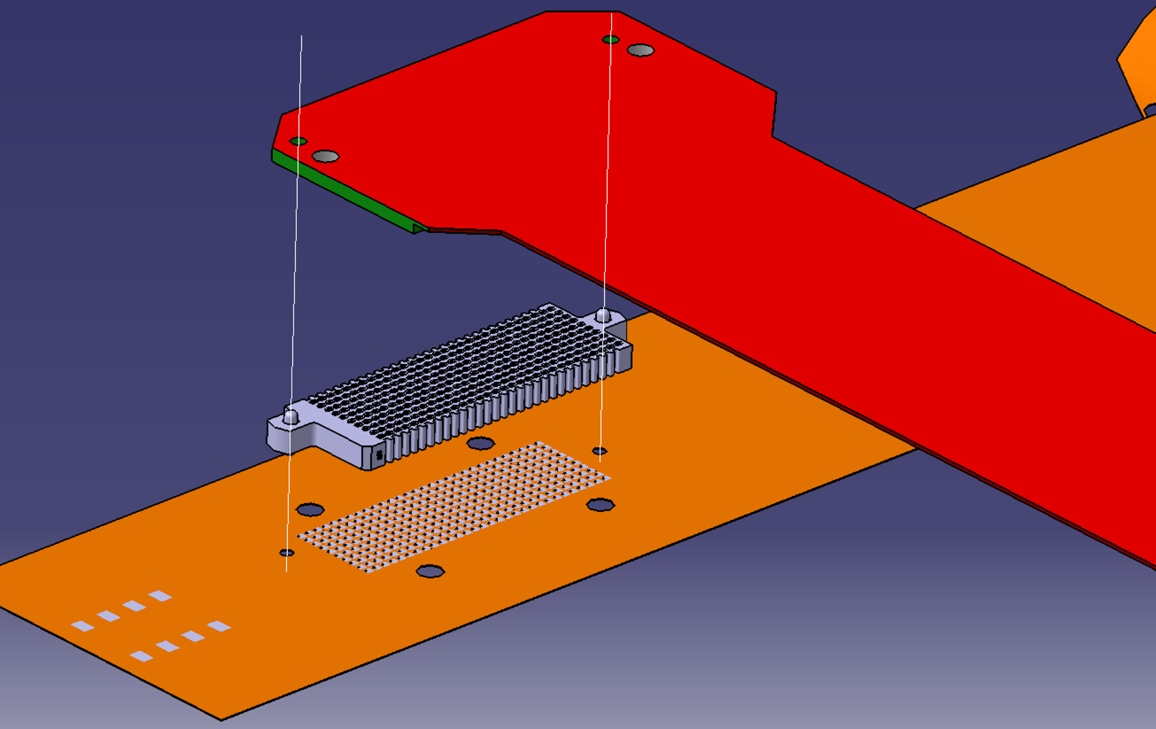}
\end{minipage}
\caption{$\bullet$~Left: routing of the channels in the Micromegas circuit of the SR2M module, in the area of the footprint of the Samtec GFZ-30-03-G-10-AD connector. $\bullet$~Right: exploded 3D view of the connection of the Samtec connector, the Micromegas, and the flat cable that carries the channels to the electronics.}
\label{fig:FootprintSR2M}
\end{figure}

Once the decision was made to avoid the installation of commercial connectors inside the chamber, it was necessary to decide on the connector to be installed outside the chamber, at the end of the Micromegas circuit's tab (figure \ref{fig:SR2M_Feedthrough_ChamberFlange}). Ultimately, it was decided to install the Samtec GFZ-30-03-G-10-AD connector because, although the total radioactivity level was similar to that of other connectors, the radioactivity per channel was lower, due to its high channel density. Additionally, the team had experience with its use (see figure~\ref{fig:FootprintSR2M}).

\section{Electron Reintegration System}
\label{sec:ERS}

In a Time Projection Chamber (TPC), the primary electrons resulting from ionization are directed towards the readout plane thanks to the application of an electric field (drift field~$E_{drift}$, see figure~\ref{fig:Introduction_MicromegasWorkingPrinciple}) generated by electrodes on both sides of the chamber. In the case of chambers with Micromegas readout planes, the mesh (the upper plane of the Micromegas) functions as the anode for generating this drift field.

In a readout plane composed of several SR2M modules, gaps may remain between the modules due to adjustment tolerances (see figure~\ref{fig:SR2M_ConceptualDesign}). As previously explained, these gaps, along with the distortion of the electric field in their vicinity, can lead to the loss of primary electrons, reduce the event reconstruction capability, and worsen the energy resolution.

\begin{figure}[!ht]
    \centering
    \includegraphics[width=0.9\textwidth]{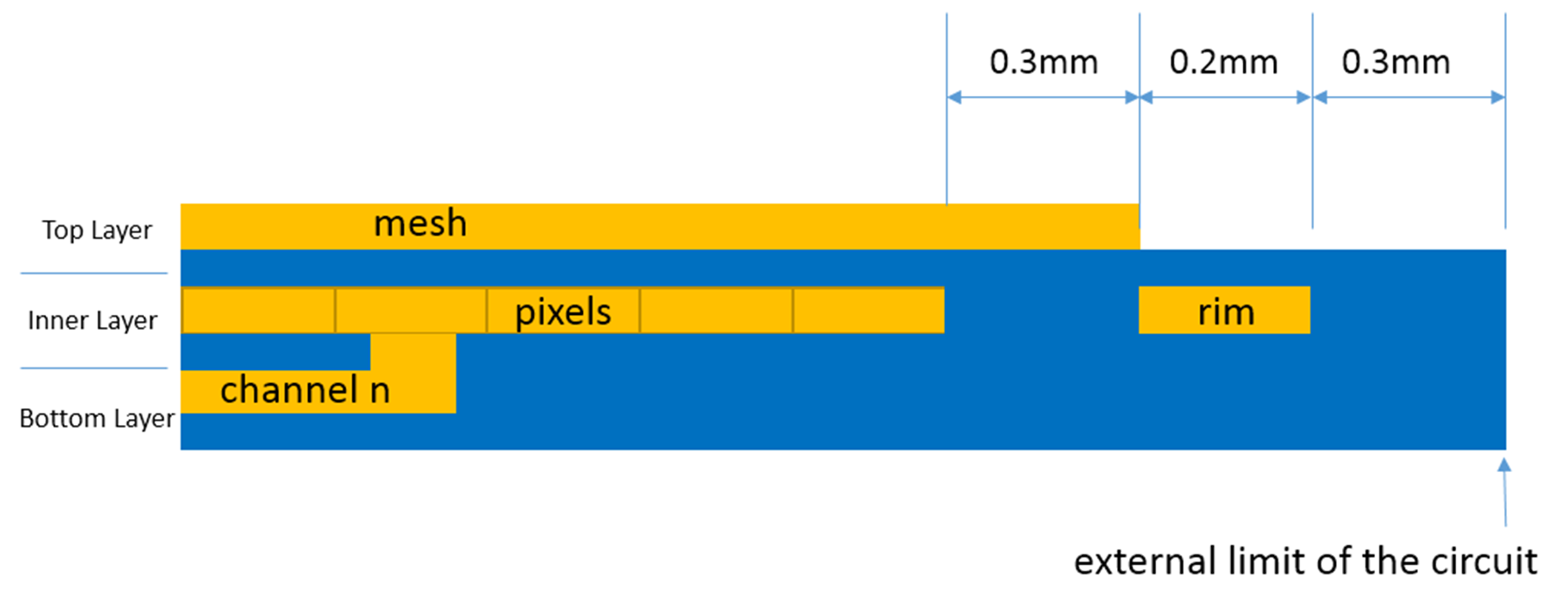}
    \caption{Schematic of the layers at the edge of the active area of the Micromegas, where the manufacturing process imposes minimum distances to be respected: the minimum distance between the edge of the mesh and the edge of the circuit is 0.5~mm~(0.2~+~0.3), and the minimum distance between the edge of the mesh and the nearest pixel/strip is 0.3~mm. The \textit{rim} electrode is added to force the direction of the drift electric field lines towards the mesh.}
    \label{fig:SR2M_LayersActiveArea}
\end{figure}

On the other hand, due to technical limitations during the manufacturing process, the mesh in a Micromegas cannot reach the very edge of the circuit (see figure~\ref{fig:SR2M_LayersActiveArea}). The minimum distance between the edge of the circuit and the mesh is 0.5~mm, while the minimum distance between the edge of the mesh and the last pixel/strip at the edge is 0.3~mm.

To mitigate these effects, produced by the gaps between modules and the distances at the edge of the Micromegas, the design of an Electron Reintegration System (ERS) was addressed. This involved the introduction of an extra electrode around the mesh, called the internal \textit{rim}, within the Micromegas circuit, as shown in figure~\ref{fig:SR2M_LayersActiveArea}. This electrode could be set to voltages above the voltage of the mesh to direct the lines of the drift electric field towards the mesh.

To verify the effect of this internal \textit{rim}, several electric field simulations were conducted at the junction of two SR2M modules using the open-source simulation software ELMER~\cite{ElmerSoftware}. In these simulations, the geometry of the module junction was generated along with the drift electric field typically used in TPCs (ranging from 100 to 300~V~cm$^{-1}$~bar$^{-1}$), and the voltage applied to the internal \textit{rim} was gradually increased until all the field lines fell within the active area. It was assumed that the primary electrons drifting towards the Micromegas follow the electric field lines. The results of the simulations indicated that this internal \textit{rim} should be set between 800 and 1000~V above the voltage of the mesh for all the field lines to fall within the active area.

\begin{figure}[!ht]
    \centering
    \includegraphics[width=\textwidth]{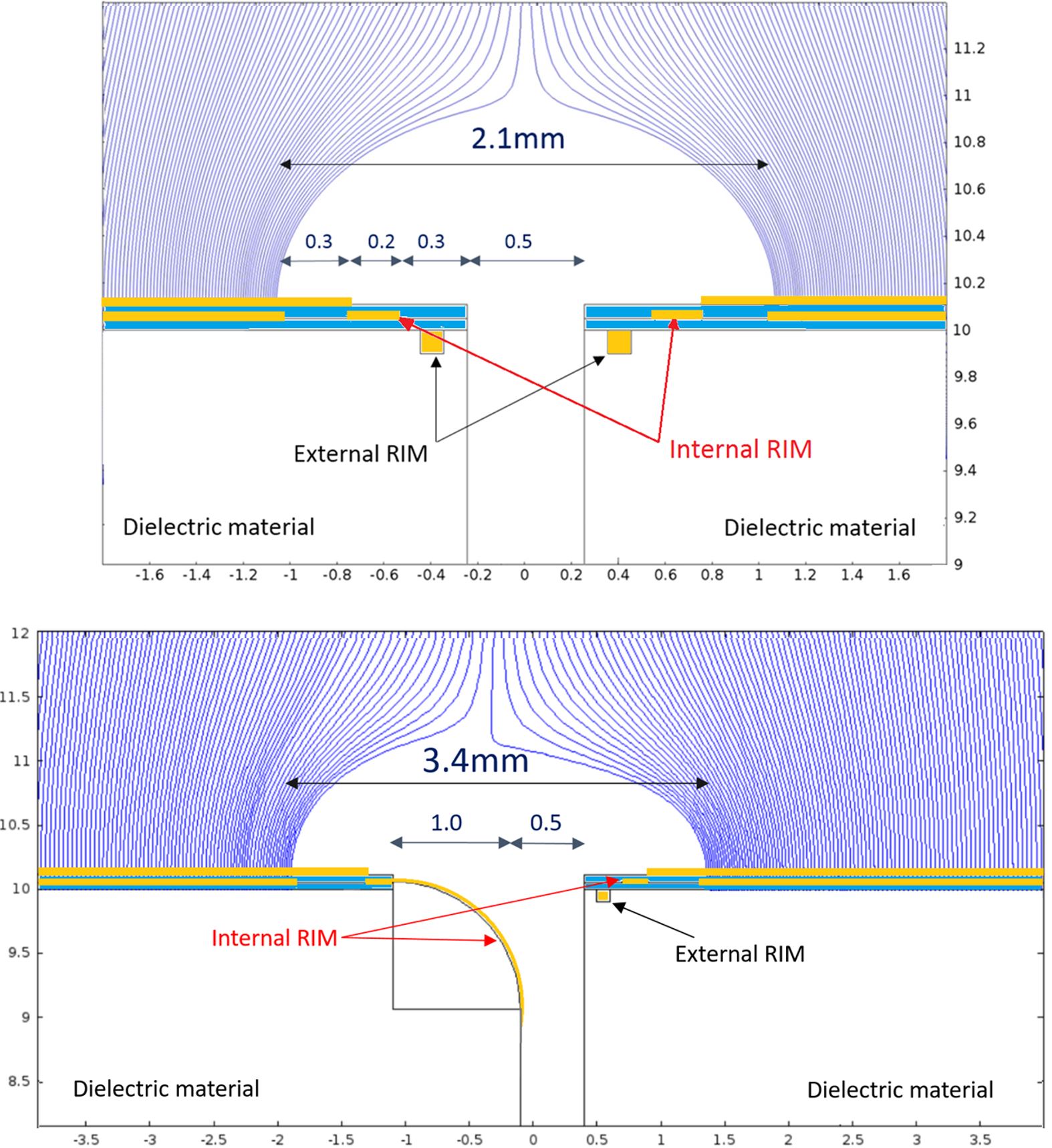}
    \caption{Simulation of the electric field in the junction area of two SR2M modules, in the cases of straight junction (top) and curved junction (bottom). Simulation values: Drift voltage 1000~V/cm (100~V~cm$^{-1}$~bar$^{-1}$); Voltage on the mesh 540~V; Voltage on the internal \textit{rim} 785~V; Voltage on the external \textit{rim} 1000~V.}
    \label{fig:SR2M_ElectricFieldUnion}
\end{figure}

Considering the situation within the Micromegas circuit, the internal \textit{rim} could easily withstand these voltage values without causing dielectric breakdown between layers, as the kapton with a thickness of 50~$\mu$m has a dielectric strength of 240~kV/mm~\cite{DuPont2017}. However, the distance of 0.3~mm between the internal \textit{rim} and the pixel/strip, both located in the same layer of the circuit, raised concerns about possible leakage currents between them if the potential difference was large, or about possible couplings that could increase noise in this and the adjacent \textit{strips}. Therefore, it was decided to add a second electrode, called the external \textit{rim}, embedded in the support of the Micromegas, which, being independent of the rest of the circuit, could be set to much higher voltages. With the addition of this second electrode, the same electric field simulations would be conducted, an example of which can be seen in figure~\ref{fig:SR2M_ElectricFieldUnion}, achieving the same results but with lower voltages on the internal \textit{rim}.

It is necessary to point out that the approximation used, in which the primary electrons in the conversion volume follow the electric field lines, introduces an error. This approximation is based on the mechanisms of electron transport in a gaseous medium in the presence of an electric field, where the electron experiences an acceleration, dependent on the field strength, and a deceleration, dependent on the mean free path of the electron in the gaseous medium, repeatedly throughout its trajectory \cite{Peisert}. The typical drift velocities (mean velocities) in noble gas mixtures are in the range of 2-3~cm/$\mu$s for Neon and Argon~\cite{Iguaz2016} and in the range of 0.05-0.6~cm/$\mu$s for Xenon~\cite{Gonzalez-Diaz2015}. The mean free paths of the electrons during their drift are in the range of 1-100~$\mu$m before being decelerated (found from the mean drift velocities and the acceleration experienced by an electron in a field of 100~V~cm$^{-1}$~bar$^{-1}$). Therefore, it is expected that the electrons in their trajectory do not deviate much from the field lines when the distances traveled are small, as in the case of the simulations performed, where the charge transport is simulated at a distance of 1.5~mm from the Micromegas. However, in areas with high field gradients, electrons will tend to deviate more easily from the field lines.

\begin{figure}[!ht]
    \centering
    \includegraphics[width=\textwidth]{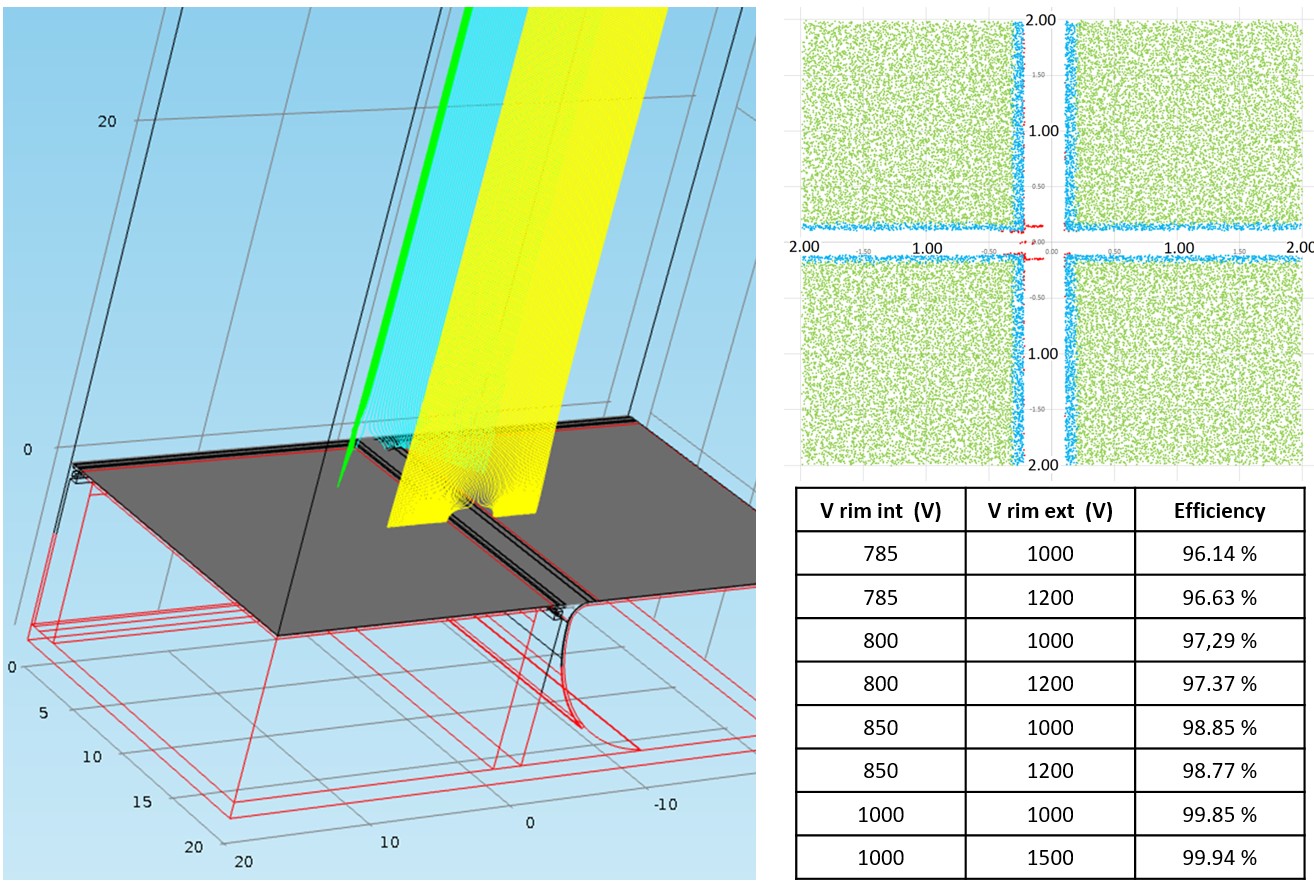}
    \caption{Simulation, using the Garfield++ software, of the drift of primary electrons in the junction area of four SR2M modules. Drift field of 1000~V/cm (100~V~cm$^{-1}$~bar$^{-1}$) and voltage on the mesh of 540~V, characteristic of the operation of a TPC at 10~bar. $\bullet$ Left: simulated geometry. $\bullet$ Upper right: top view of the readout plane with the final points of the electron trajectories, belonging to the simulation with internal \textit{V~rim~int} values of 785~V and external \textit{V~rim~ext} of 1000~V, with an efficiency of 96.14\%. In green, the electrons collected in the active area whose trajectory has not been affected by the ERS system. In blue, the electrons that have been affected by the ERS system and redirected towards the active area. In red, the electrons whose trajectory the ERS system has not been able to correct and that have ended up outside the active area. $\bullet$ Lower right: summary table of the results showing the efficiency of the ERS system for each group of internal-external \textit{rim} voltage values.}
    \label{fig:SR2M_SimulationGarfieldUnion}
\end{figure}

The electric field simulations helped determine the final geometry of the SR2M module at its edges, but to validate the method used and quantify the error made in the process, the Garfield++ software~\cite{GarfieldSoftware} was used to perform several simulations of the drift of primary electrons in an area of 4~x~4~cm$^2$, corresponding to the junction between four SR2M modules (see figure \ref{fig:SR2M_SimulationGarfieldUnion}). For each simulation, varying the voltages of the internal and external \textit{rim}, 10$^5$ electrons were generated, distributed homogeneously in a plane 20~cm above the readout plane. The trajectory for each of the electrons drifting towards the readout plane was calculated, resulting in the final point. The values of the drift field (100~V~cm$^{-1}$~bar$^{-1}$) and the voltage of the mesh (540~V) were the same for all simulations, coinciding with those applied in the previous electric field simulations (figure \ref{fig:SR2M_ElectricFieldUnion}), which are defined by the expected operating point in a TPC with Xe+TMA at 10~bar.

The efficiency of the ERS system is defined as the percentage of electrons that, with the application of certain voltage values on the internal and external \textit{rim}, end up falling into an active area of the readout plane, but which, if these voltages were not applied, would fall into the 'dead' zone between two SR2M modules. This efficiency, for each group of internal-external \textit{rim} values, can be seen in the table in figure \ref{fig:SR2M_SimulationGarfieldUnion}. It can be observed that as the voltages on both electrodes are increased, the losses of primary electrons are reduced, reaching efficiencies of 99.94\% with very reasonable voltages of 1000~V on the internal \textit{rim} (460~V above the mesh) and 1500~V on the external \textit{rim}.

\section{Feedthrough}
\label{sec:feedthrough}

The first conceptual design (figure \ref{fig:SR2M_ConceptualDesign}) proposed the extraction of the channels to the exterior of the TPC through the Micromegas circuit itself. The other proposed alternative involved extracting the channels through a flat cable, manufactured in layers of kapton-copper like the Micromegas circuit, which would require a connection inside the TPC between the Micromegas circuit and this flat cable. The chosen solution in figure \ref{fig:SR2M_ConceptualDesign}, eliminating the connection inside the TPC, offered mainly two advantages and one disadvantage compared to its alternative:

\begin{itemize}
\item \textbf{Radiopurity}: Commercial connectors with a high density of channels used in this type of connection have non-negligible radioactivity and can significantly contribute to the background of the detector. Avoiding any commercial connector inside the TPC would help control the background of the experiment.
\item \textbf{Robustness}: Connections over time can suffer relaxation in their materials, or movements due to various causes, or other processes such as oxidation that can weaken the connections and lead to channel loss or malfunction. Having the connection outside the TPC allows for easier intervention if necessary.
\item \textbf{Less manipulable}: To achieve a high level of tightness in the feedthrough with the Micromegas tab in the conceptual design (figure~\ref{fig:SR2M_ConceptualDesign}), the use of a radiopure epoxy (Epoxy Hysol measured in~\cite{Aznar2013}) was proposed to bond and seal the tab to a flange. This is the most robust solution proposed, but it implied having a flange, with the weight that entails, permanently attached to the SR2M module (see photograph on the cover of this chapter), which would make handling the SR2M module more cumbersome.
\end{itemize}

\begin{figure}[!ht]
    \centering
    \includegraphics[width=\textwidth]{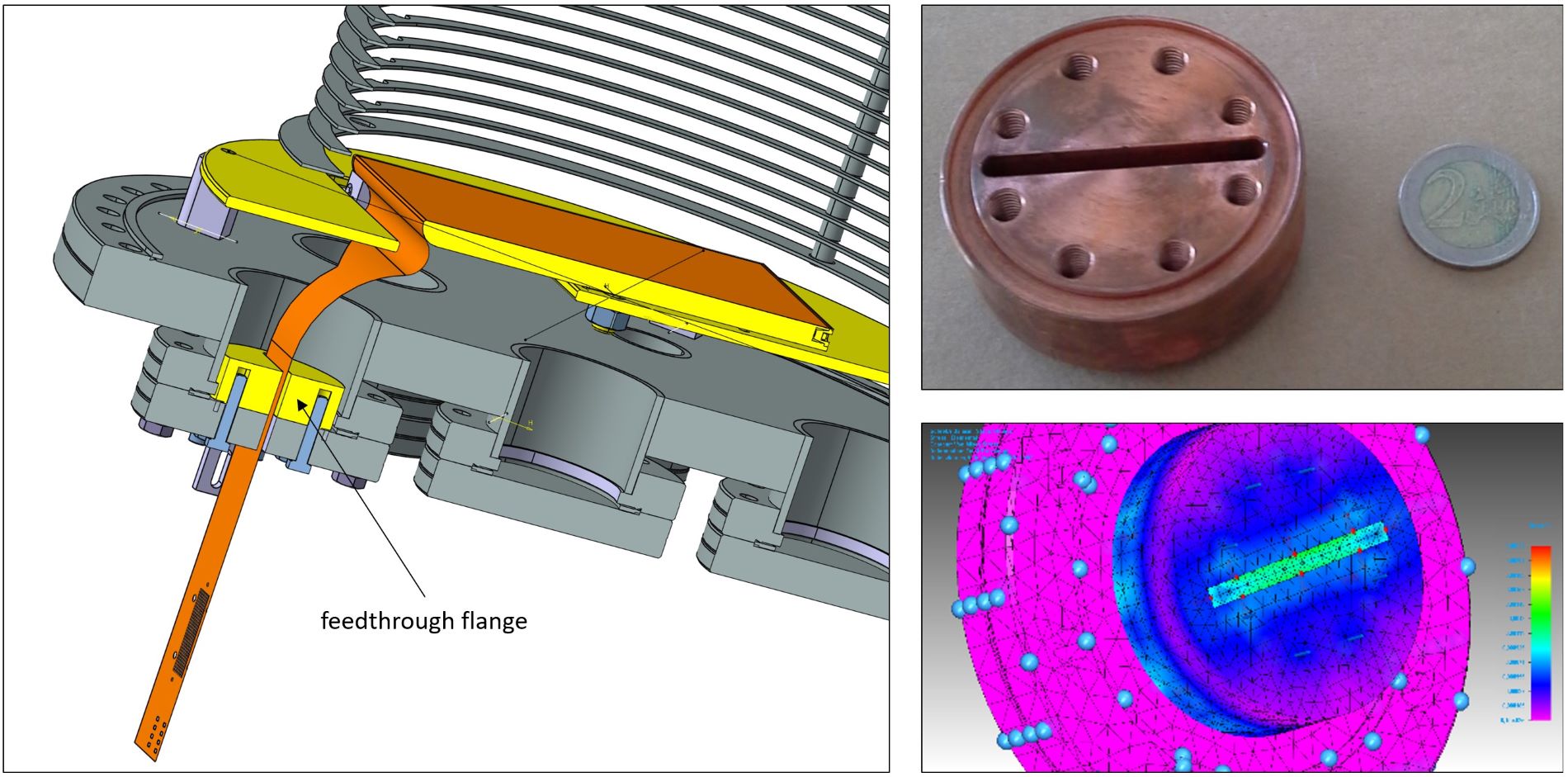}
    \caption{$\bullet$ Left: section of a 3D view of the SR2M module installed in the testing chamber at the University of Zaragoza. The Micromegas tab exits directly to the exterior of the chamber. $\bullet$ Upper right: photograph of the copper flange that adheres to the Micromegas tab. $\bullet$ Lower right: Study of stresses and deformations in the copper flange, in its final position.}
    \label{fig:SR2M_Feedthrough_ChamberFlange}
\end{figure}

Figure \ref{fig:SR2M_Feedthrough_ChamberFlange} presents the proposed solution for the feedthrough of the SR2M module: a copper flange with a central slot bonded to the Micromegas tab using epoxy, mounted on the interior of the chamber with a PTFE gasket. Both copper and PTFE are low-background materials. The dimensions of the flange were determined by the width of the Micromegas tab, the threaded holes for tightening the gasket, the diameter of the gasket, and the mechanical properties of the copper and epoxy.

With these parameters, a finite element simulation was conducted to verify the proper functioning and sizing of the system. The conditions during this simulation were as follows: 10~MPa of pressure on the surface of the PTFE gasket (according to the standard~EN13445-3 'Unfired pressure vessels~$-$~Part~3:~Design'); a tightening torque on the screws adequate to achieve 10~MPa of pressure on the PTFE gasket; 15~bar of pressure in the TPC (1.5 times the working pressure).

This simulation confirmed that the equivalent Von Mises stress in all parts of the system was below the elastic limit of the copper used (303~MPa) and the epoxy used (58~MPa). It was also verified that the shear stresses on the copper-epoxy bonding surface were below the maximum shear stresses for this material (17~MPa). Furthermore, this epoxy has a Young's modulus of 4.3~GPa, much lower than that of copper (110~GPa), which translates to lower stresses endured. Thus, this epoxy, which provides the sealing in the system, is capable of absorbing the deformations that occur in the copper flange that contains it and adapting to them, ensuring that it does not deteriorate, at least through mechanical means.

\begin{figure}[!ht]
    \centering
    \includegraphics[width=\textwidth]{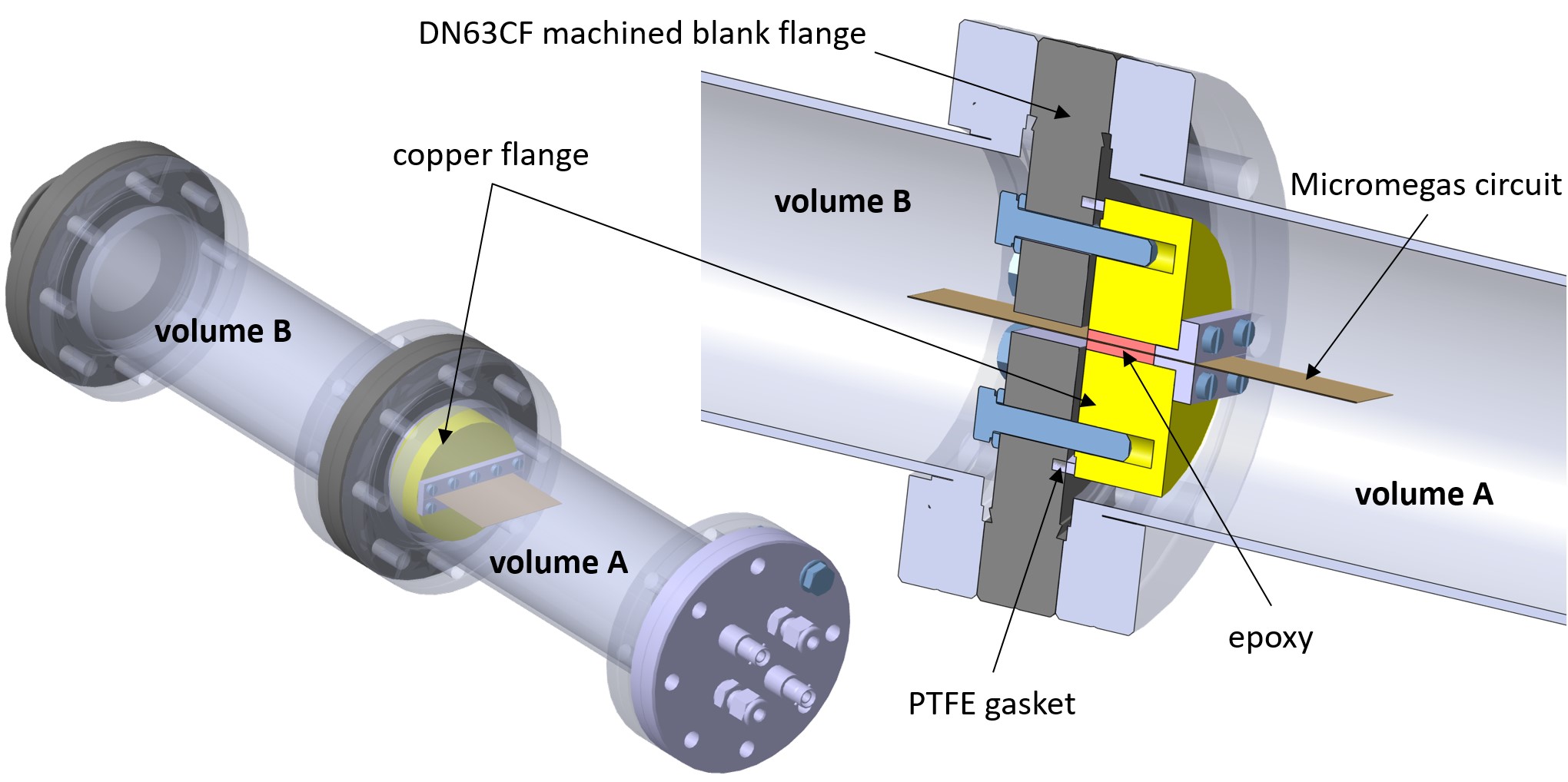}
    \caption{3D view and section of the test bench for measuring the tightness of the feedthrough system.}
    \label{fig:SR2M_FeedthroughTestBench}
\end{figure}

\begin{figure}[!ht]
    \centering
    \includegraphics[width=0.75\textwidth]{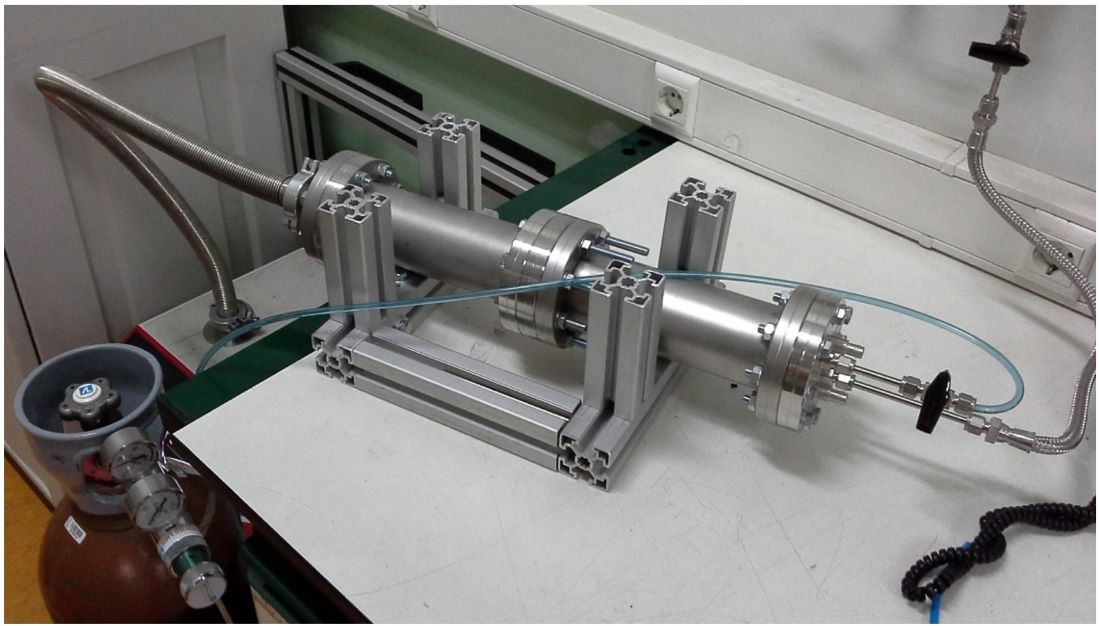}
    \caption{Photograph of the test bench for measuring the tightness of the feedthrough system, during a measurement session, where the different components indicated in figure~\ref{fig:SR2M_FeedthroughTestBench} can be distinguished.}
    \label{fig:SR2M_FeedthroughTestBenchFoto}
\end{figure}

\begin{figure}[!ht]
    \centering
    \includegraphics[width=\textwidth]{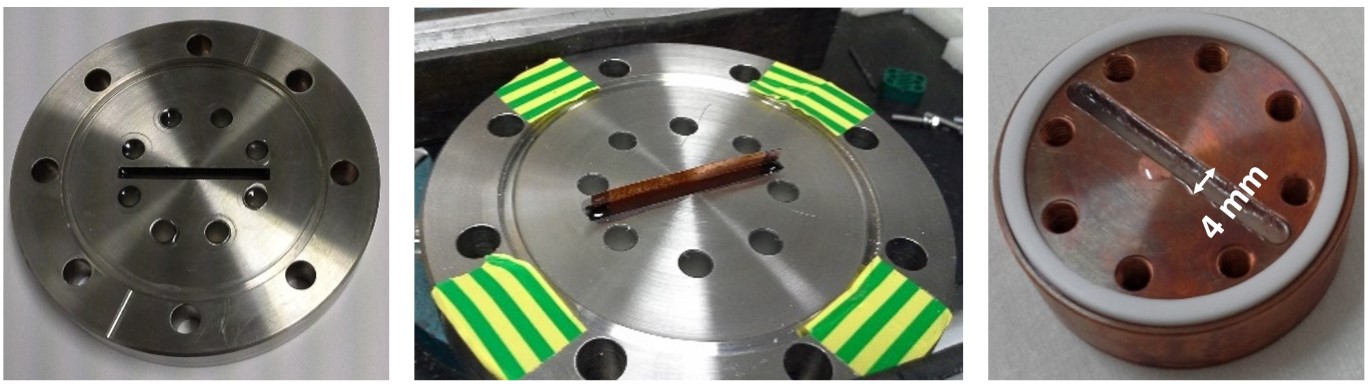}
    \caption{$\bullet$~Left: Standard DN63DF stainless steel flange used for measuring leaks due to the permeability of the epoxy. $\bullet$~Center: Standard DN63DF stainless steel flange used for measuring leaks due to the Micromegas circuit plus the epoxy. $\bullet$~Right: Copper knife flange used during the tightness measurements of the complete feedthrough system.}
    \label{fig:SR2M_EpoxyFlanges}
\end{figure}

After sizing the copper flange, the design and assembly of a test bench to measure the leak rate that could be achieved with the proposed feedthrough system was carried out (figures~\ref{fig:SR2M_FeedthroughTestBench}~and~\ref{fig:SR2M_FeedthroughTestBenchFoto}). This test bench consisted of two standard DN63CF stainless steel tubes (volumes A and B) separated by a DN63CF flange, initially blind, which would be machined to allow the passage of the Micromegas circuit (figure \ref{fig:SR2M_EpoxyFlanges}), simulating the slot in the final TPC. Volume A would be pressurized with a helium mixture up to a maximum of 10~bar, while volume B would be connected to a helium detector (model Pfeiffer Smart Test HLT 560), with a leak detection sensitivity on the order of 10$^{-12}$~mbar~l~s$^{-1}$, allowing for the quantification of the total leak from the feedthrough system.

During the measurements of the entire feedthrough system, the following procedure was followed:

\begin{enumerate}
\item The copper flange is attached, using screws, to the DN63CF flange with a tightening torque that ensures a minimum pressure of 10~MPa on the PTFE gasket.
\item The assembly to be measured (either just the DN63CF flange or the entire DN63CF and copper flange assembly) is mounted in the steel tubes (volumes A and B) with elastomeric gaskets.
\item The helium meter is connected to volume B, and a vacuum value of around 1.3·10$^{-3}$~mbar is achieved using the meter's own pump. The baseline of the meter was around 1.2·10$^{-9}$~mbar~l~s$^{-1}$ under these conditions, marking the minimum leak value that can be measured.
\item Volume A is filled with 1 bar of helium and 9 bar of nitrogen. Thus, volume A contains a helium proportion of 1/10.
\item The measurement begins with the helium detector, which initially usually indicates the baseline value, and sufficient time is allowed (several hours) until the measurement stabilizes.
\item The leak measurement is multiplied by a factor of 10 (the proportion of helium contained in volume A). This correction is an approximation and defines an upper limit.
\end{enumerate}

\begin{figure}[!ht]
    \centering
    \includegraphics[width=\textwidth]{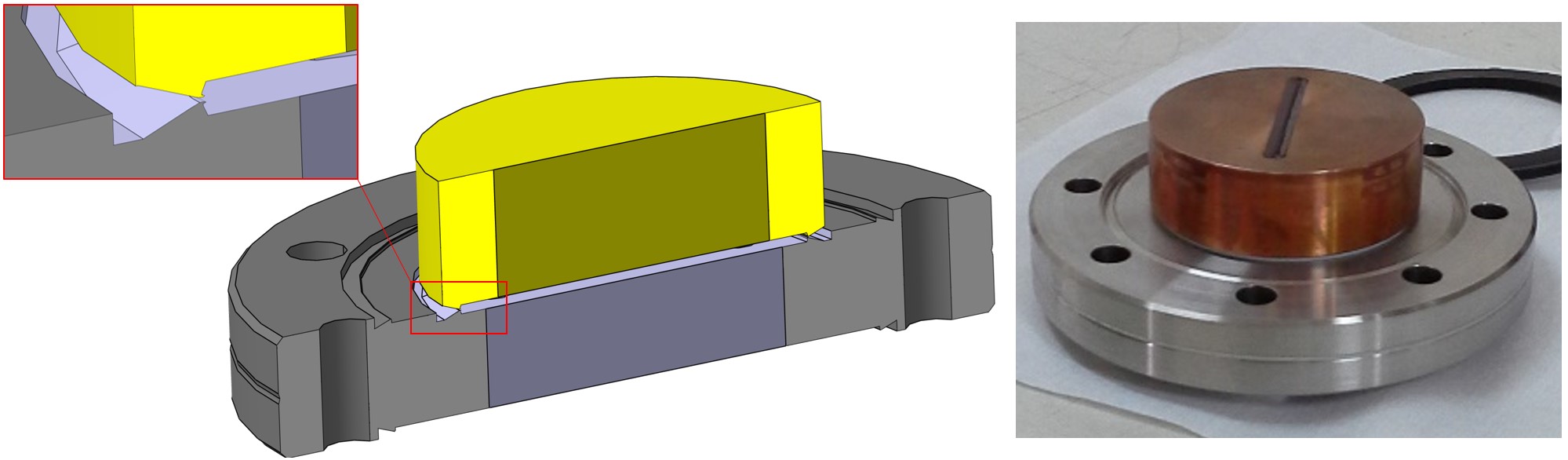}
    \caption{3D section view and photograph of the feedthrough system with knife edge.}
    \label{fig:SR2M_KnifeFlangeFeedthrough}
\end{figure}

With the measurements in the test bench, the leak rate for the feedthrough system was established at 2.5·10$^{-5}$~mbar~l~s$^{-1}$. This leak rate is very high for high vacuum systems, where it is usually below 10$^{-9}$~mbar~l~s$^{-1}$. The most serious consequence of having a high leak rate is the loss of gas in the TPC, and the measured level corresponds to losing 4~g of xenon per year for each of the feedthroughs installed. This is not a relatively high loss (the detector will operate with 140~kg of enriched xenon), but prolonged exposure over time (5~years) combined with a significant number of installed feedthroughs (52 modules planned) could lead to a more substantial gas loss (1~kg in 5~years).

This leak rate is dominated by the permeability of the PTFE gasket used. If it were desired to further reduce this leak rate, it would be necessary to study the installation of another type of gasket, while also respecting the requirement for radiopurity. For example, in high vacuum systems, copper gaskets with knife-edge flanges that deform the gasket to achieve tightness are commonly used. In the case of the designed feedthrough, a direct replacement of the PTFE gasket with a copper one is not possible, as the knife edge is also made of copper and would therefore also deform. One possible solution would be to use cold-worked electrolytic copper for the copper flange, which provides greater hardness, and to use non-preformed electrolytic copper for the gasket. Another possible solution would be to manufacture a piece in stainless steel, with the minimum material possible, that would contain the knife edge and could be joined, either by welding or epoxy, to the copper flange. The use of ASI~316L stainless steel could be the best option, as there are measurements confirming that its radioactivity is relatively low~\cite{Lawson2011}.

In the case of replacing the PTFE gasket and thereby reducing the leak rate of the feedthrough system, the next contribution would be the permeability of the epoxy, which has been measured at 1.4·10$^{-6}$~mbar~l~s$^{-1}$. In this case, as has been done to reduce the flow in the PTFE gasket by reducing its section, the area of the section in the epoxy slot can be reduced (assuming that the permeability in this material follows Fick's law). The width of the slot during the tests was 4~mm (see figure~\ref{fig:SR2M_EpoxyFlanges} right), and the Micromegas circuit has a thickness of 200~$\mu$m. It seems feasible to reduce the width of the slot below~1~mm. This minimum width of the slot will be determined by the filling process with the epoxy to ensure that it occupies all the space between the circuit and the walls of the slot. It seems feasible to achieve a reduction in the leak rate with this strategy of between a factor of 4 (slot width~1~mm) to a factor of 10 (slot width~400~$\mu$m). An increase in the diffusion length would also proportionally reduce the leak rate, but this would imply an increase in the total thickness of the copper flange, a more sensitive parameter as it affects the geometry around the feedthrough and the weight of the copper flange itself.

If these measures to reduce the leak rate were implemented, it would be necessary to verify that the effects on the circuit-epoxy surface, which may represent pathways for the transport of gas molecules, remain negligible. It would also be necessary to quantify the permeability of the Micromegas circuit itself.

As can be seen, there are several options that could be developed to reduce this leak rate, while also respecting the requirement for maximum radiopurity, but all of them would require further development.

\section{Copper Support and Tiling}

\begin{figure}[!ht]
    \centering
    \includegraphics[width=\textwidth]{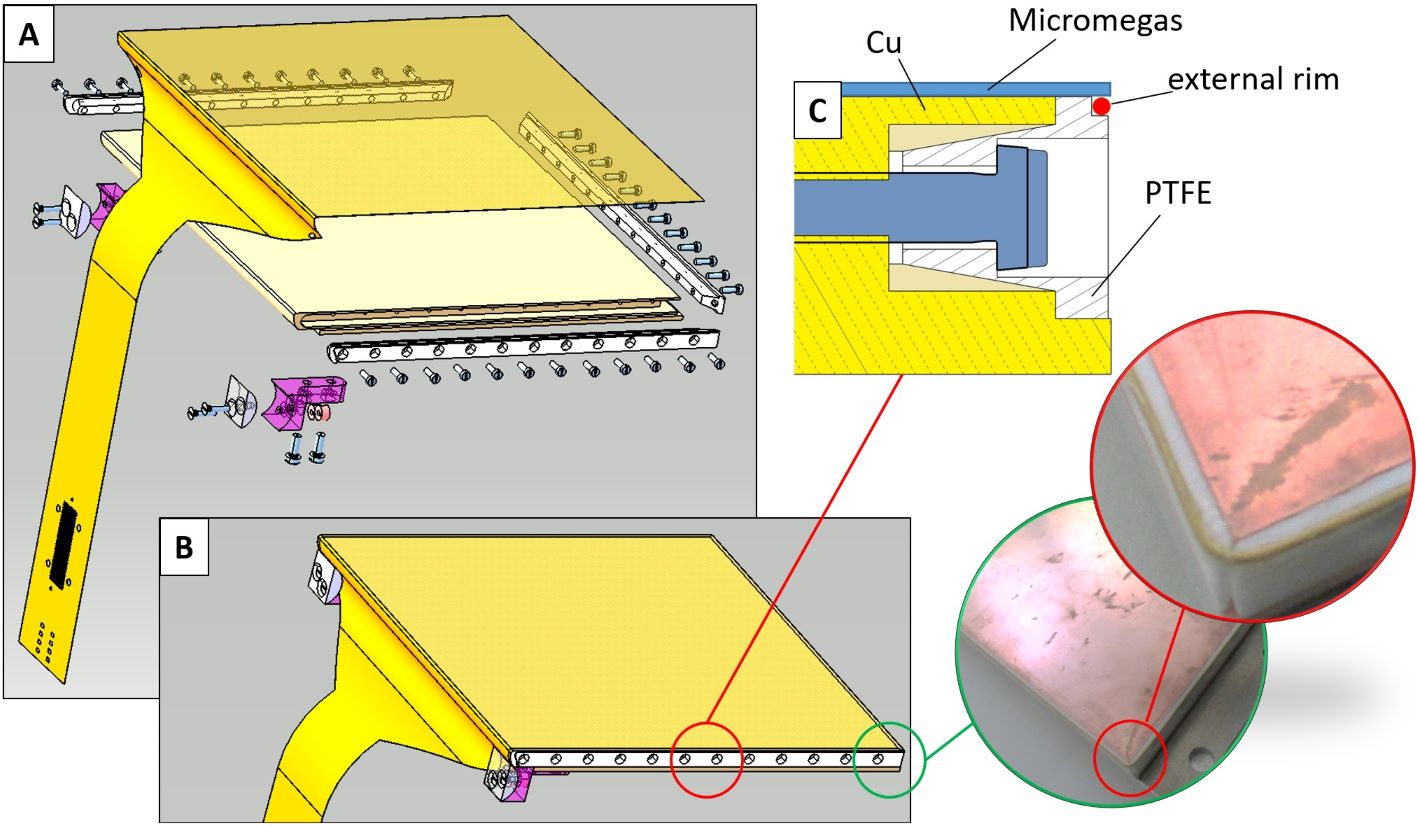}
    \caption{Exploded (A) and assembled (B) 3D view of the SR2M module. Detail section of the edge of the SR2M module (C), with the installation of a copper electrode (in red) that acts as an external rim.}
    \label{fig:SR2M_CatiaExplodedView}
\end{figure}

The microbulk Micromegas circuits are made of thin layers (50~$\mu$m) of kapton and copper, making them flexible and difficult to maintain their rigidity. To facilitate their handling and ensure a good fit between several circuits, as is the case for a module designed to generate a large readout plane, a set of mechanical components has been designed. Figure~\ref{fig:SR2M_ConceptualDesign} shows a conceptual illustration of the composition of an SR2M module and the installation of several of them to form a larger readout plane. Figure~\ref{fig:SR2M_CatiaExplodedView} presents a 3D view of all the components of an SR2M module. The Micromegas circuit is attached to the copper support using an acrylic-based adhesive commonly used in the manufacture of flexible circuits (Adhesive Transfer Tape - ATT. Ref: 3M F9460PC).

\begin{figure}[!ht]
    \centering
    \includegraphics[width=0.7\textwidth]{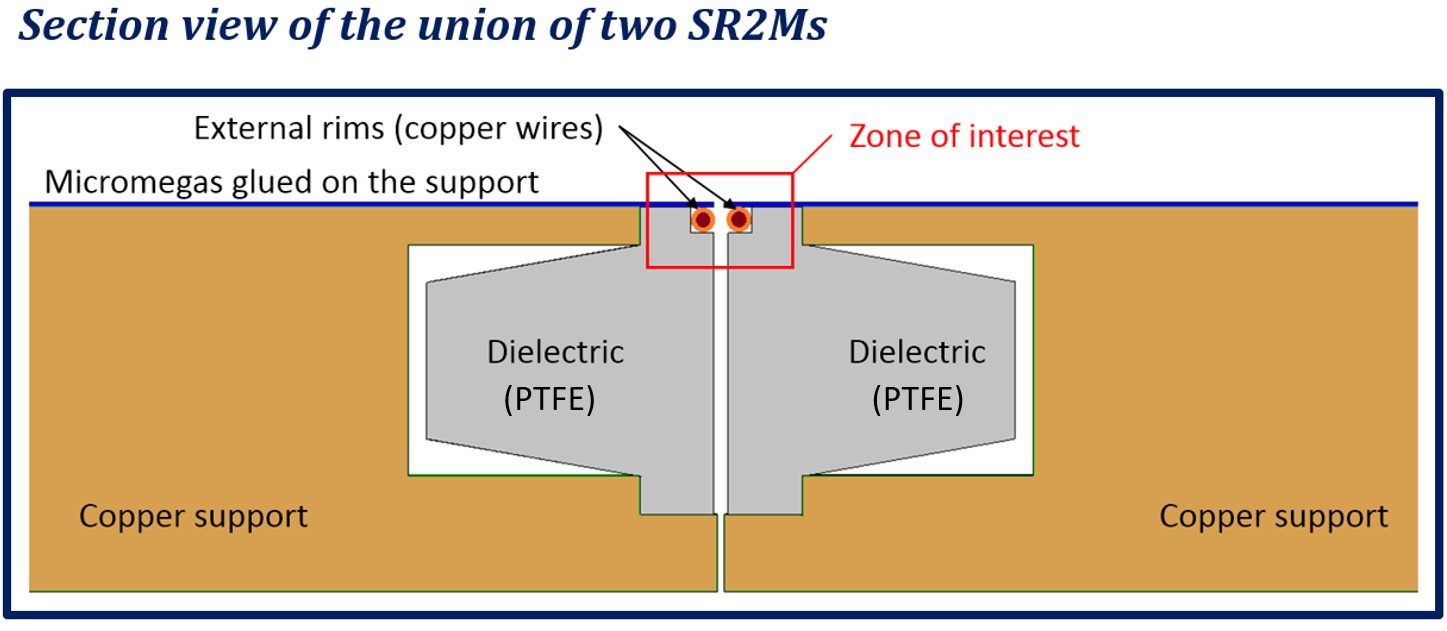}
    \caption{Section shown in figure~\ref{fig:SR2M_CatiaExplodedView} at the junction of two SR2M modules.}
    \label{fig:SR2M_SectionAndERS}
\end{figure}

Particular attention should be given to the solution adopted for integrating the external rim within the SR2M module. Figure~\ref{fig:SR2M_CatiaExplodedView} shows a detailed section of the edge of the SR2M module, where the external rim is installed. Figure~\ref{fig:SR2M_SectionAndERS} displays this same section at the junction of two modules, forming the geometry used for the study of the ERS in section~\ref{sec:ERS}.

\section{Installation of the SR2M Module}
\label{ch:InstallingTheSR2M}

After completing the design and fabrication of the first prototype of the SR2M module, it was installed in the TREX-$\beta\beta$ system, originally designed to house the NEXT-MM prototype \cite{Alvarez2014, DianaCarolinaHerreraMunoz}. This system consists of a stainless steel TPC, 28~cm in diameter and 35~cm in drift length, capable of reaching up to 10~bar of operating pressure, along with all the necessary subsystems and elements (gas system, \textit{slow control}, etc.).

The objective was to validate the first prototype of the SR2M module, prior to the fabrication of seven additional modules that would be installed in the laboratories of Shanghai Jiao Tong University, thus forming the largest readout plane ever executed with microbulk Micromegas detectors \cite{Chen2017a} \cite{Wang2020}.

\begin{figure}[!ht]
    \centering
    \includegraphics[width=\textwidth]{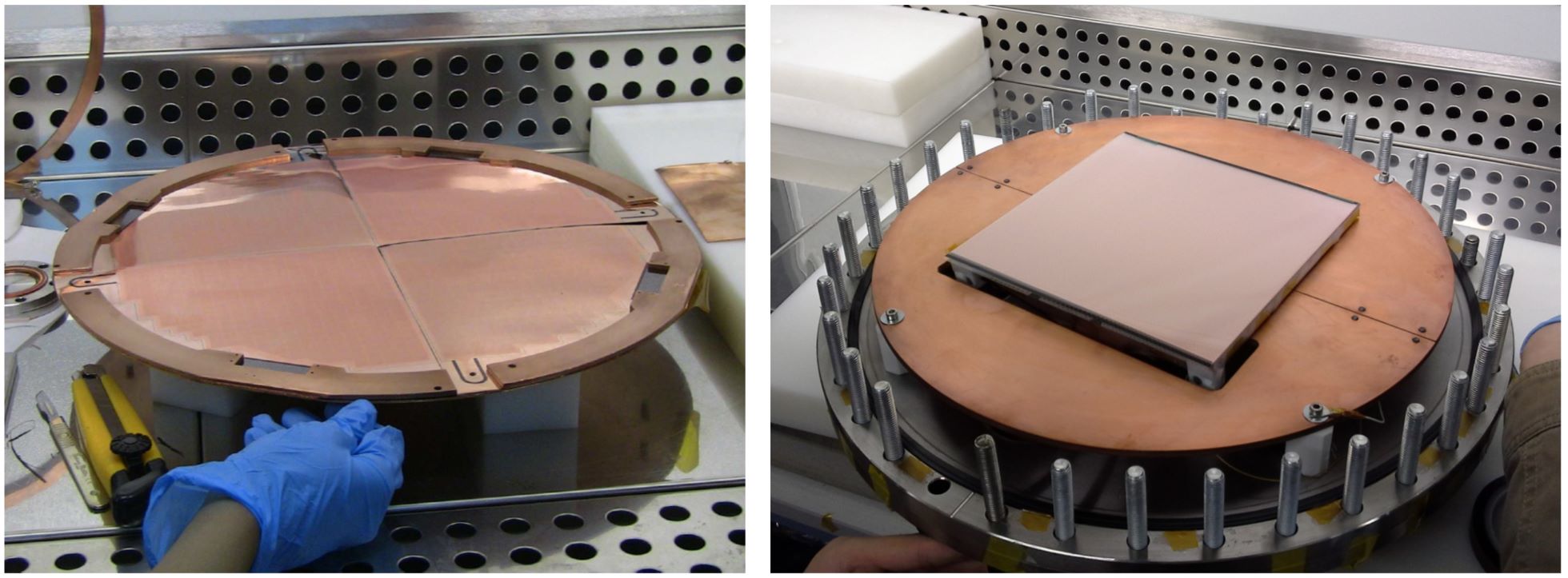}
    \caption{$\bullet$ Left: readout plane developed for the NEXT-MM prototype, consisting of four microbulk Micromegas \cite{Alvarez2014}. $\bullet$ Right: SR2M module in its final position after some modifications to the TPC for its positioning.}
    \label{fig:Before-After_NEXTMM_Plate}
\end{figure}

For the NEXT-MM prototype, a circular readout plane was composed of four Micromegas (figure \ref{fig:Before-After_NEXTMM_Plate} left). These four microbulk Micromegas were installed directly onto a circular copper support, secured at their perimeter. Due to the flexibility of the Micromegas circuit itself, the readout plane had some irregularities that could affect the quality of the data collected (transparency, spatial resolution, or energy resolution). However, the SR2M is independent and, among other things, resolves the issue of dimensional flatness.

To install the SR2M module in this TPC, it was necessary to design and fabricate another copper support attached to the bottom cover of the TPC on which to mount the SR2M module (figure \ref{fig:Before-After_NEXTMM_Plate} right). This copper support was in direct contact with the copper of the SR2M module and could be set to the appropriate potential independently of the rest of the system.

\begin{figure}[!ht]
    \centering
    \includegraphics[width=0.85\textwidth]{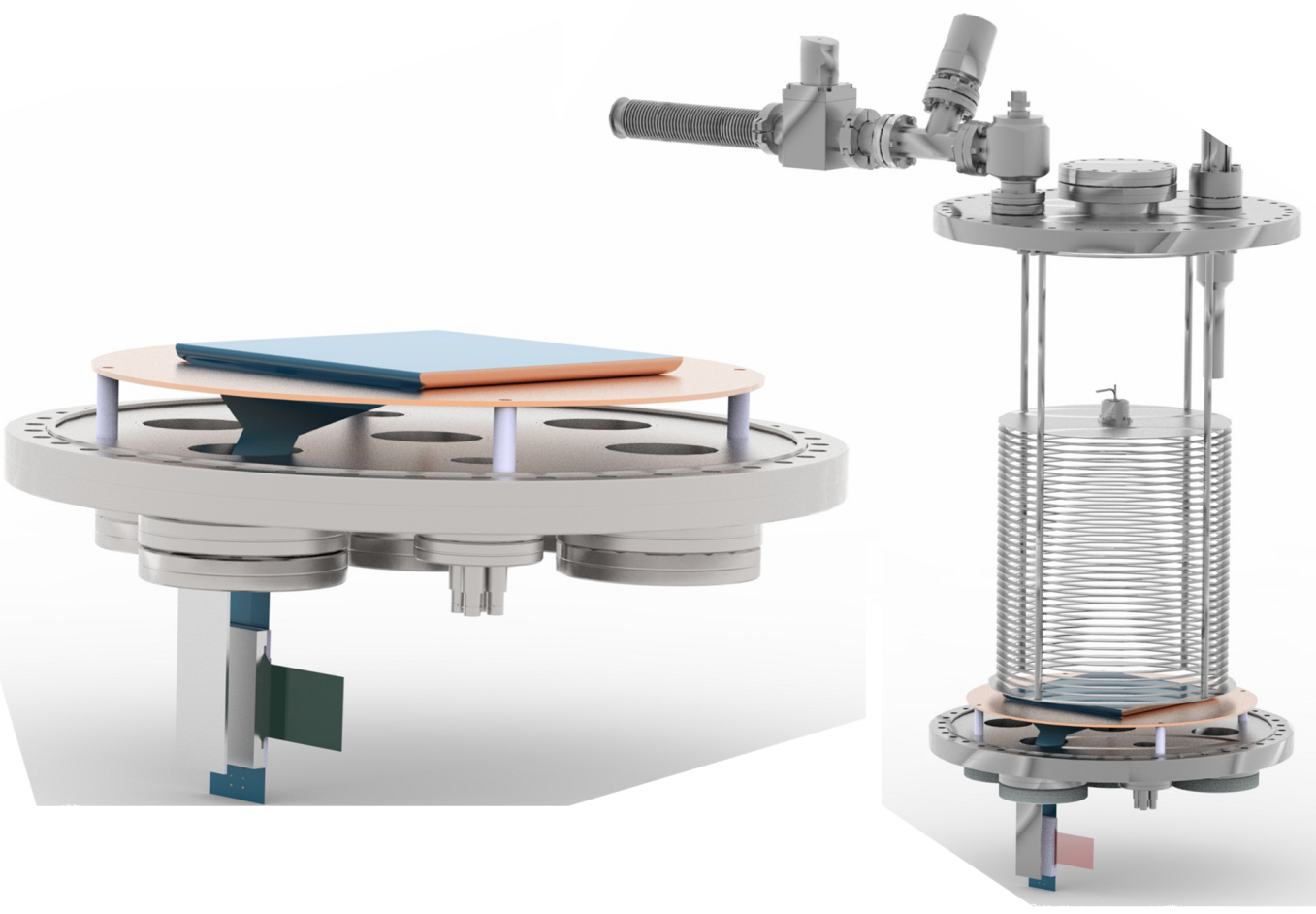}
    \caption{3D view of the SR2M module installed on the bottom cover of the TPC (left) and of the entire TPC (right).}
    \label{fig:Catia_SR2M_inNextChamber}
\end{figure}

Figure \ref{fig:Catia_SR2M_inNextChamber} left shows a 3D reconstruction of the SR2M module installed on the copper support, which has a galvanic separation from the bottom cover of the TPC via PTFE bushings. Figure \ref{fig:Catia_SR2M_inNextChamber} right shows a 3D reconstruction of the internal part of the TPC, where the drift rings and the cathode can be observed. Above the cathode, a small box can be distinguished, used to install the calibration source. For more details on the description of the components of the TPC, see \cite{Alvarez2014}.

\begin{figure}[!ht]
    \centering
    \includegraphics[width=0.70\textwidth]{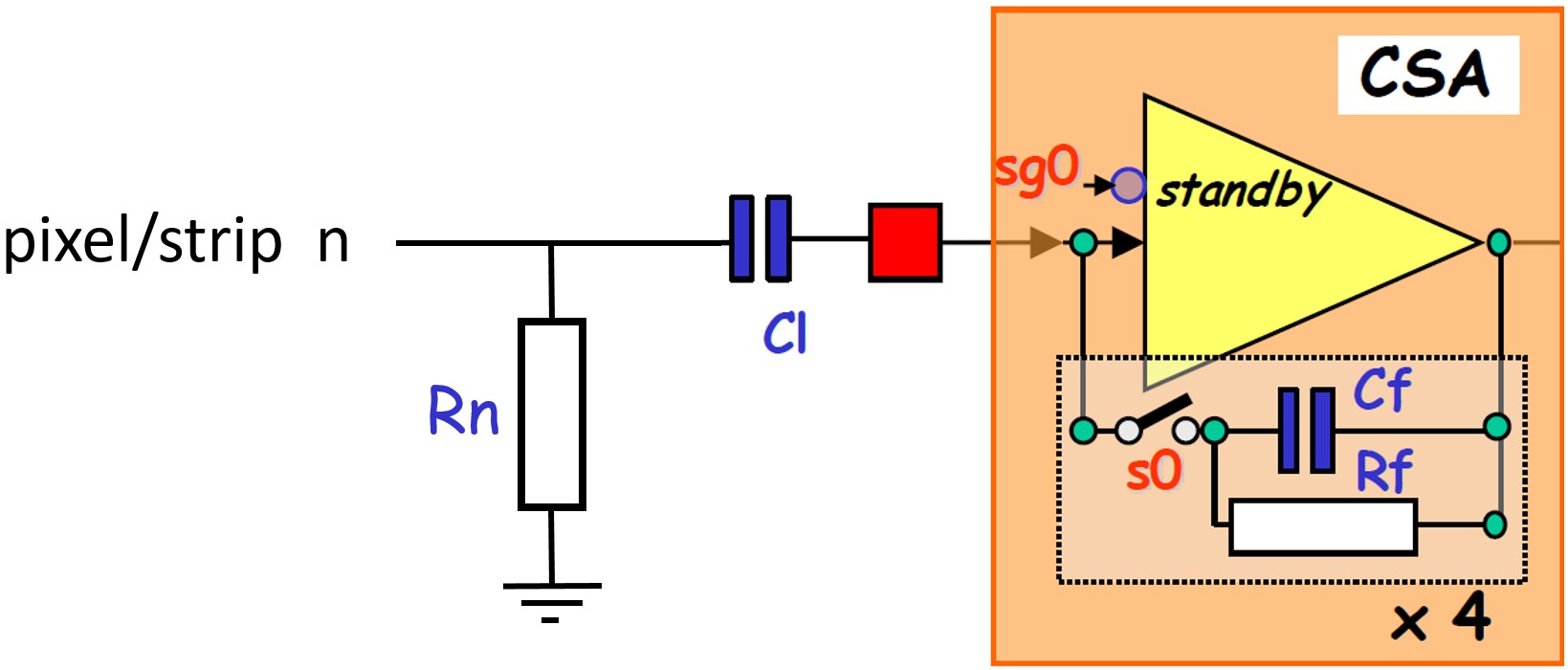}
    \caption{\cite{AGETforTPCs}~Connection scheme of each pixel/strip of the Micromegas with the acquisition electronics. The image shows the first stage of signal amplification via the programmable CSA (\textit{Charge Sensitive Amplifier}). Each pixel/strip is grounded through the resistor $Rn$, which in the case of the FEC-Feminos electronics has a value of~100~M$\Omega$.}
    \label{fig:SR2M_EsquemaConexionCanal}
\end{figure}

For the readout of the channels, the FEC-Feminos acquisition electronics were installed, recently acquired by the working group and used for the first time in this prototype. Through this electronics, and to achieve the necessary electric field in the amplification \textit{gap} to induce the electron avalanche effect, each channel or \textit{strip} is grounded through a 100~M$\Omega$ resistor (see figure~\ref{fig:SR2M_EsquemaConexionCanal}).

If a \textit{strip} is interconnected with the mesh or with a low resistance ($\ll$~100~G$\Omega$), due to a failure during the manufacturing process or during the operation of the detector, a leakage current occurs between the mesh and the \textit{strip}, complicating the proper control of the potential on the mesh. Additionally, it may also cause noise issues during data acquisition. During the connection work of the SR2M module, 10 channels with low resistance were discovered that could cause problems during the operation of the detector, so it was decided to disconnect them.

\begin{figure}[!ht]
    \centering
    \includegraphics[width=0.9\textwidth]{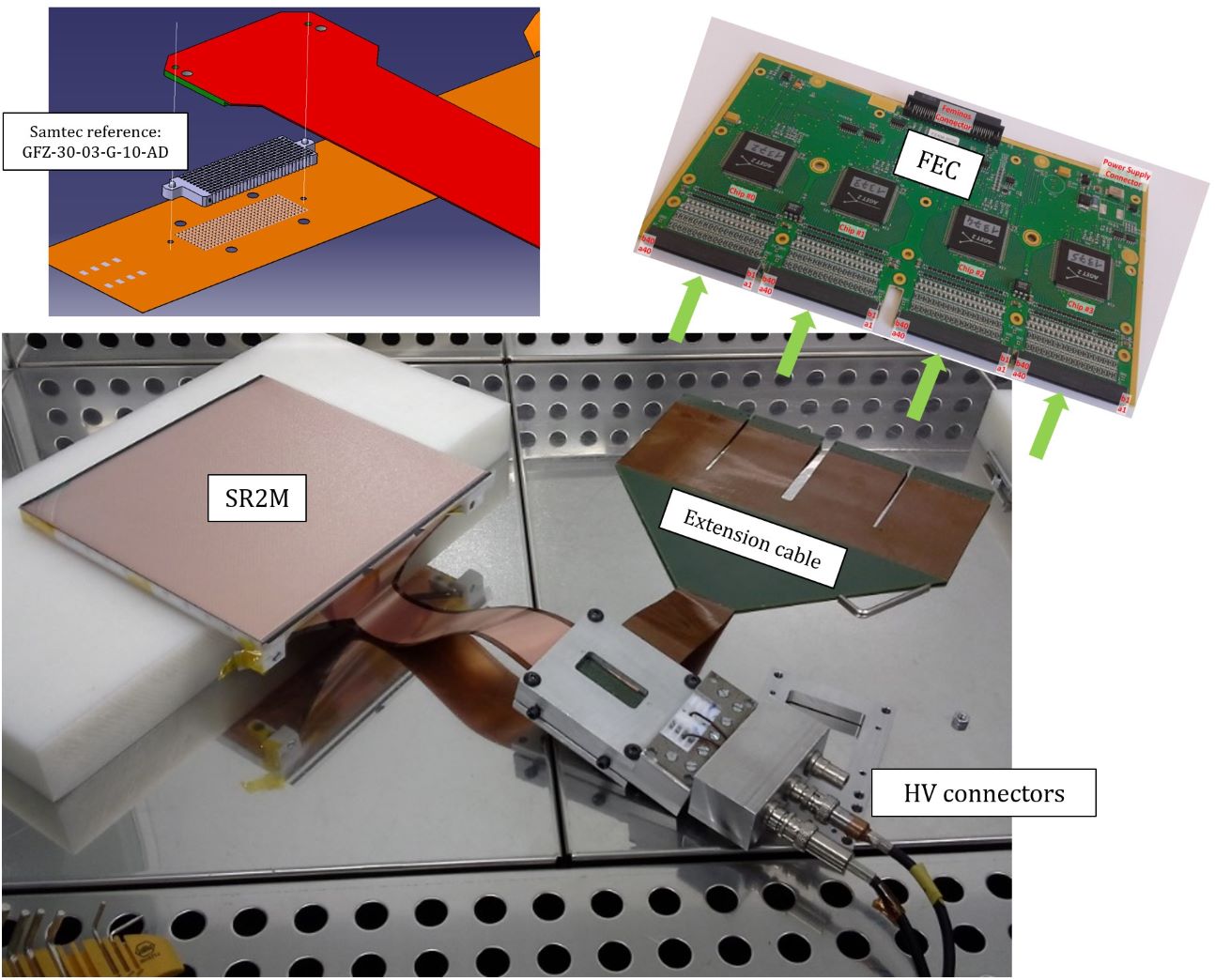}
    \caption{Photograph of the components in the connection chain from the Micromegas to the FEC-Feminos acquisition electronics.}
    \label{fig:TheSR2M_CadenaConexionElectronica}
\end{figure}

The connection chain (figure \ref{fig:TheSR2M_CadenaConexionElectronica}) consisted of the SR2M module itself, the Samtec connector (figure \ref{fig:FootprintSR2M}), and an extension cable that connected the channels to the FEC board. To supply voltages to both the mesh and the \textit{internal rim} and \textit{external rim} of the ERS, three SHV connectors were installed, connected via a cable soldered to the high voltage pads of the Micromegas circuit, which had previously undergone high voltage testing.

\begin{figure}[!ht]
    \centering
    \includegraphics[width=1.0\textwidth]{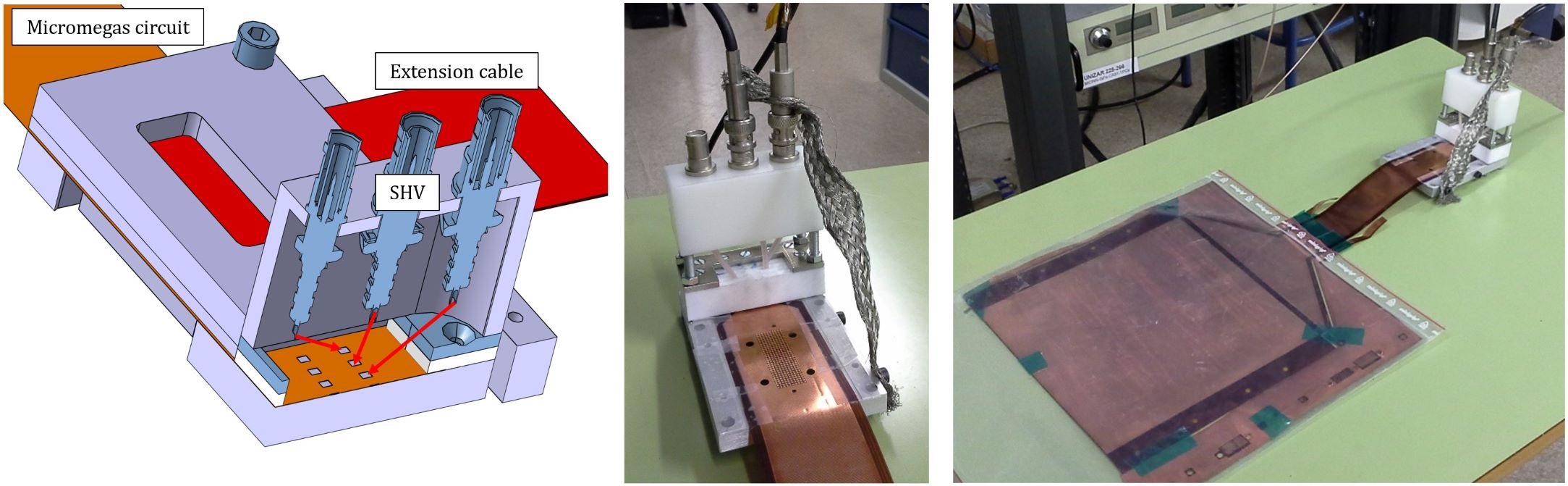}
    \caption{Setup of the high voltage connection for measuring leakage currents in the Micromegas circuit and in the connection area.  $\bullet$ Left: 3D section of the high voltage connectors. The red arrows from the SHV connectors to the pads of the Micromegas circuit represent the cables that were soldered for connection. $\bullet$ Center and right: photographs of the setup during the tests.}
    \label{fig:TheSR2M_PruebasHVconectorMM}
\end{figure}

High voltage tests on the Micromegas circuit and the connections were conducted 'in the air' with a relative humidity in the laboratory air of 62\% and a temperature of 21$^{o}$C. The objective of the tests was to ensure that no leakage currents occurred either within the Micromegas circuit or in the area of the connectors. To this end, the SHV connectors were mounted on the Micromegas circuit (figure \ref{fig:TheSR2M_PruebasHVconectorMM}), and voltages were applied according to the configurations reflected in table~\ref{tab:TheSR2M_PruebasHVconectorMM_Tabla}. As can be seen, none of the tests showed a detectable leakage current by the resolution of the source (0.5~nA), except for the test where the \textit{internal~rim} was set to 1000~V and the mesh to 500~V, where a leakage current of 1~nA was measured on the \textit{internal~rim}. This leakage was minimal and disappeared when hot air was applied to the connector area, indicating that the leakage likely occurred in that area and not within the Micromegas circuit itself.

This test highlighted the need to control humidity in high voltage connection areas. The different types of connections made during the tests were based on the area where leakage could occur (\textit{external~rim}-ground, \textit{internal~rim}-ground, mesh-ground, mesh-\textit{internal~rim}, connectors-ground).

\renewcommand{\tablename}{Table}
\begin{table}[!ht]
\centering
\begin{tabular}{l}
\includegraphics[width=0.9\textwidth]{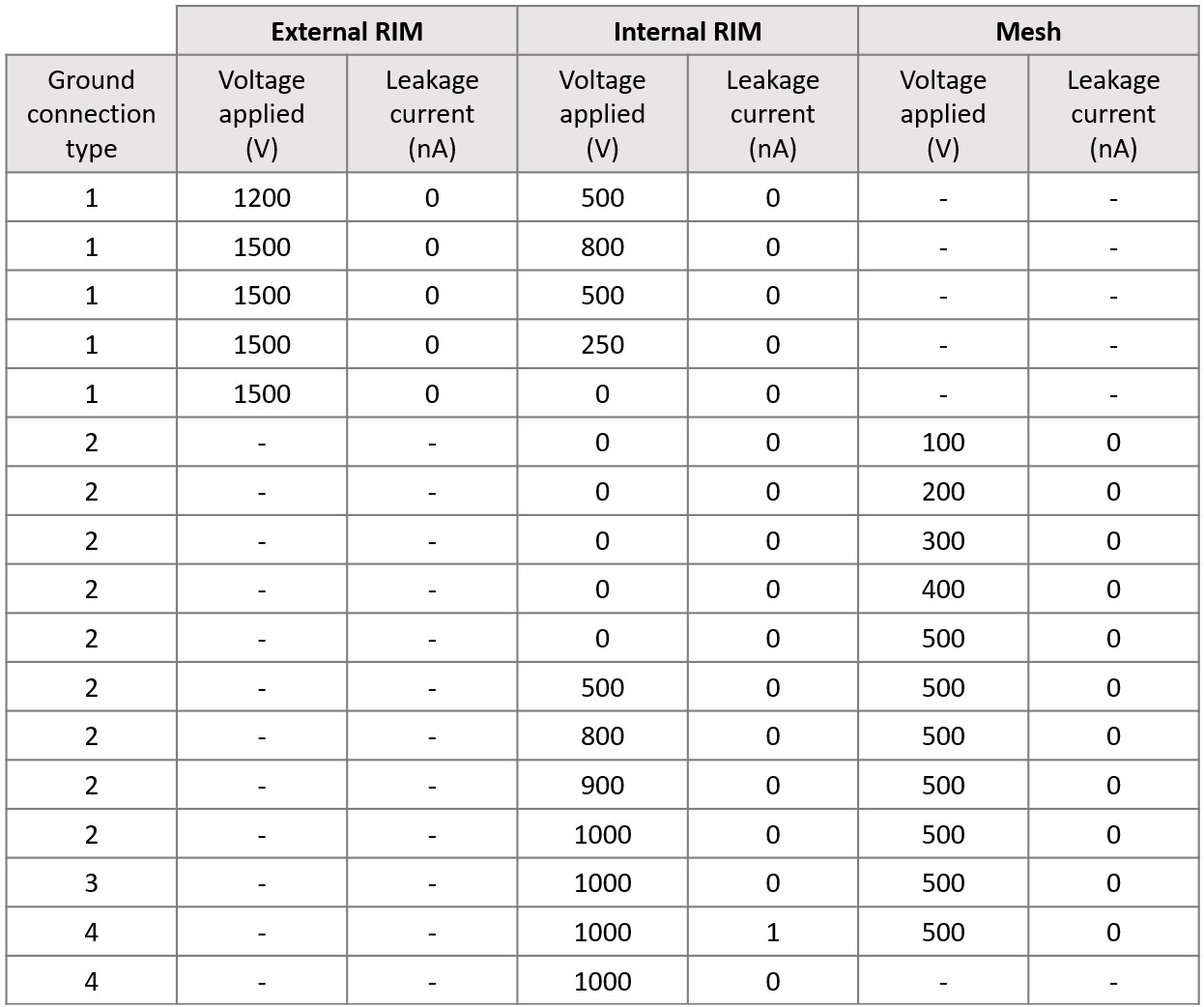}\\
\end{tabular}
\caption{Results of the leakage current measurements in the Micromegas circuit and in the connection area. Four types of grounding connections were made: (1)~Only the metal parts of the connector are grounded; (2)~Neither the metal parts of the connector nor the \textit{strips} are grounded; (3)~The metal parts of the connector are grounded but the strips are not; (4)~Both the metal parts of the connector and the strips (except for the 10 strips that had previously been identified as having an anomalous capacitance value) are grounded.}
\label{tab:TheSR2M_PruebasHVconectorMM_Tabla}
\end{table}

\setlength{\baselineskip}{12pt} 
\selectlanguage{USenglish}
\graphicspath{{../Imagenes/}{../../Imagenes/}} 

\renewcommand{\listtablename}{Table of Contents}
\renewcommand{\tablename}{Table}

\chapter{Commissioning of the SR2M Module}
\label{ch:DataTakingSR2M}

\begin{figure}[!ht]
  \centering
  \includegraphics[width=0.8\textwidth]{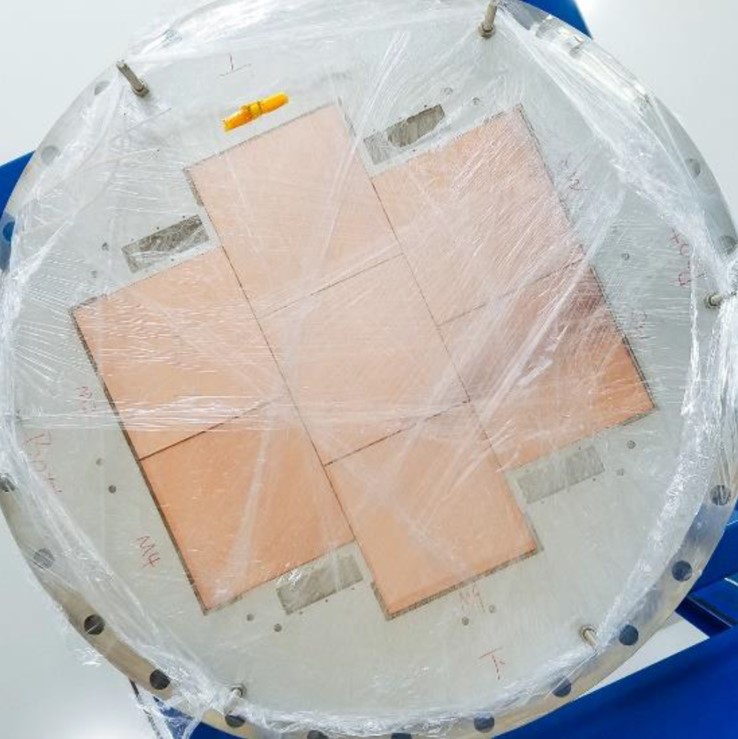}
  \caption*{}
  \label{fig:SR2MdataTaking_SR2MTessellationPictureSJTU_v2}
\end{figure}

\newpage

\section{First Spectrum}
\label{sec:PrimerEspectro}

The installation of the SR2M module in the TREX-$\beta\beta$ chamber took place on March 25, 2016. Prior to this, the chamber had been pumped for a week, reaching a vacuum value of 2.3·10$^{-4}$~mbar. The sum of the leak rate, outgassing, and permeability, verified in the days leading up to the installation, was 4·10$^{-4}$~mbar~l~s$^{-1}$. After the installation of the SR2M module on the bottom cover of the chamber and the $^{109}$Cd source on the cathode, pumping continued for 48~hours until the vacuum value prior to opening was achieved. At that moment, the injection of the Xenon~+~1\%~TMA mixture up to 1~bar of pressure was performed. This mixture had been kept in the bottle since it was used during the commissioning of the NEXT-MM prototype~\cite{Alvarez2014}. Subsequently, it was left in recirculation for 72~hours to allow the gas to be purified through oxygen and humidity filters. After the installation of the FEC-Feminos data acquisition electronics, and following 72~hours in recirculation, the voltage in the Micromegas was gradually increased, reaching a stable 270~V.

To collect the signal from the \textit{strips}, all channels were connected to a multi-channel data acquisition system. In the case of the SR2M module, it was connected to a FEC-Feminos acquisition system (see figure~\ref{fig:TheSR2M_CadenaConexionElectronica}), recently acquired by our research group and used for the first time in this prototype within the collaboration. Through this electronics, and to achieve the electric field in the amplification \textit{gap} necessary to induce the electron avalanche effect, each channel or \textit{strip} is grounded through a 100~M$\Omega$ resistor. Previously, 10 channels with anomalous capacitance values were disconnected (see section~\ref{ch:InstallingTheSR2M}).

\begin{figure}[!ht]
    \centering
    \includegraphics[width=0.65\textwidth]{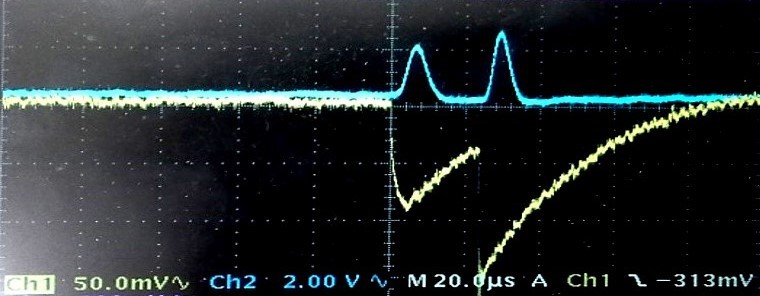}
    \caption{Two events generated by the $^{109}$Cd source and collected within the same acquisition time window (\textit{pile up}). Signals collected from the mesh of the SR2M module and read through a Canberra~2004 pre-amplifier plus a Canberra~2022 amplifier with a gain of x20 and a \textit{shaping time} of 4~$\mu$s. In channel 1, the signal from the pre-amplifier is visualized, and in channel 2, the signal from the amplifier.}
    \label{fig:TheSR2M_OsciloscopeFirstSignal}
\end{figure}

Before acquiring the signal from the channels with the FEC-Feminos electronics, it was decided to first acquire the signal from the mesh plane. The amplification chain necessary for its reading is simpler and serves as a way to verify, before launching longer and more complex data acquisitions with the channels, that the detector conditions (pressure, mixture quality, voltage on the mesh, drift field, etc.) are adequate for obtaining a good signal in the channels. To achieve the reading of the mesh signal, it was connected to a Canberra~2004 pre-amplifier followed by a Canberra~2022 amplifier. Figure \ref{fig:TheSR2M_OsciloscopeFirstSignal} presents one of the first signals collected in the prototype, which is a \textit{pile up} (accumulation of two events) within the acquisition time window, originating from the $^{109}$Cd source.

\begin{figure}[!ht]
    \centering
    \includegraphics[width=0.95\textwidth]{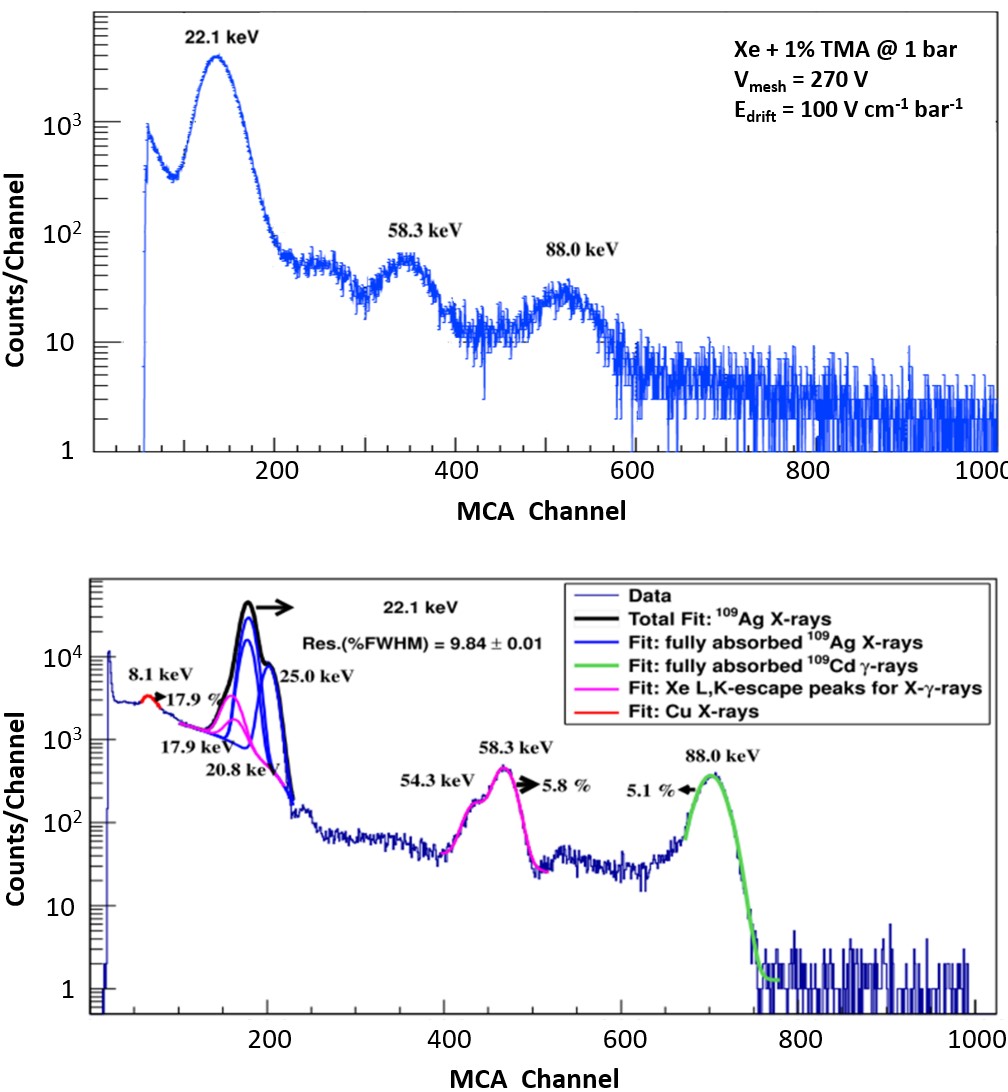}
    \caption{Comparison of the energy spectra generated by a $^{109}$Cd source in the SR2M module prototype and in the test TPC TREX-$\beta\beta$-0~\cite{ThesisDiana}. $\bullet$~Upper:~first spectrum of the SR2M module, generated from the mesh signal, with a mixture of Xe~+~1.0\%~TMA at 1~bar and a drift field of 100~V~cm$^{-1}$~bar$^{-1}$ over the 35~cm drift length. $\bullet$~Lower:~spectrum in TREX-$\beta\beta$-0, also generated from the mesh signal, with a mixture of Xe~+~0.8\%~TMA at 8~bar and a drift field of 245~V~cm$^{-1}$~bar$^{-1}$ over 1~cm drift length.}
    \label{fig:MCA_FirstSpectrum_SR2M}
\end{figure}

At the output of the amplifier, a multi-channel digital analyzer 'Amptek MCA 8000A' was installed to record the signal. The upper figure~\ref{fig:MCA_FirstSpectrum_SR2M} shows the resulting spectrum from collecting the mesh signal during a calibration with the $^{109}$Cd source and is compared with the spectrum collected in the TPC called TREX-$\beta\beta$-0~\cite{ThesisDiana} under similar conditions, shown in the lower figure~\ref{fig:MCA_FirstSpectrum_SR2M}. TREX-$\beta\beta$-0 is a 2.4~liter TPC used for the characterization of Micromegas with different mixtures of Xenon and TMA~\cite{ThesisDiana}, with a drift distance of 1~cm and a Micromegas with a single anode-cathode and $\sim$2~cm in diameter.

Comparing both spectra in figure \ref{fig:MCA_FirstSpectrum_SR2M}, it is observed that in the first spectrum collected from the SR2M module, the main peaks (22.1, 58.3, and 88.0~keV) appear; however, the resolution is not sufficient to distinguish the contributions from lower probability emissions near these peaks.

It should be noted that in the TREX-$\beta\beta$-0 chamber, the Micromegas has a diameter of 2~cm, which favors better energy resolution due to its size. A study of gains by zones on the surface of the SR2M would improve its energy resolution, as this would allow for a correction in the calculated energy of each event depending on the area where its charge was deposited.

It is necessary to remember that the TREX-$\beta\beta$-0 and TREX-$\beta\beta$ chambers have drift distances of 1~cm and 35~cm, respectively. This difference presumably causes less recombination and \textit{attachment} in the TREX-$\beta\beta$-0 chamber during the drift of electrons towards the readout plane than in the TREX-$\beta\beta$ chamber with the SR2M module, thus favoring its energy resolution.

Moreover, a smaller chamber with fewer materials inside that can produce \textit{outgassing}, combined with better sealing due to its design, enhances the gas quality in terms of impurities (Oxygen, humidity, etc.), reducing the \textit{attachment} that occurs during the drift of electrons towards the readout plane and also improving the performance of the Micromegas during the avalanche process.

\section{Gain and Electron Transmission}
\label{sec:GananciaTransparencia}

Typically, one of the first steps during the commissioning and characterization of a Micromegas is the realization of the electron transmission and gain curves. These curves help to find the optimal operating point and to verify the proper functioning of the Micromegas.

In a TPC, after the ionization of the gas due to the passage of a particle, the free electrons generated in the ionization move towards the readout plane thanks to the drift field. During this movement, some of these electrons may be lost before reaching the readout plane, primarily due to the processes of \textit{recombination} and \textit{attachment}. However, in the case of the Micromegas, where the amplification of the electrons occurs within the amplification hole, loss can also occur if a fraction does not enter the hole and falls directly onto the mesh plane.

\begin{figure}[!ht]
    \centering
    \includegraphics[width=0.9\textwidth]{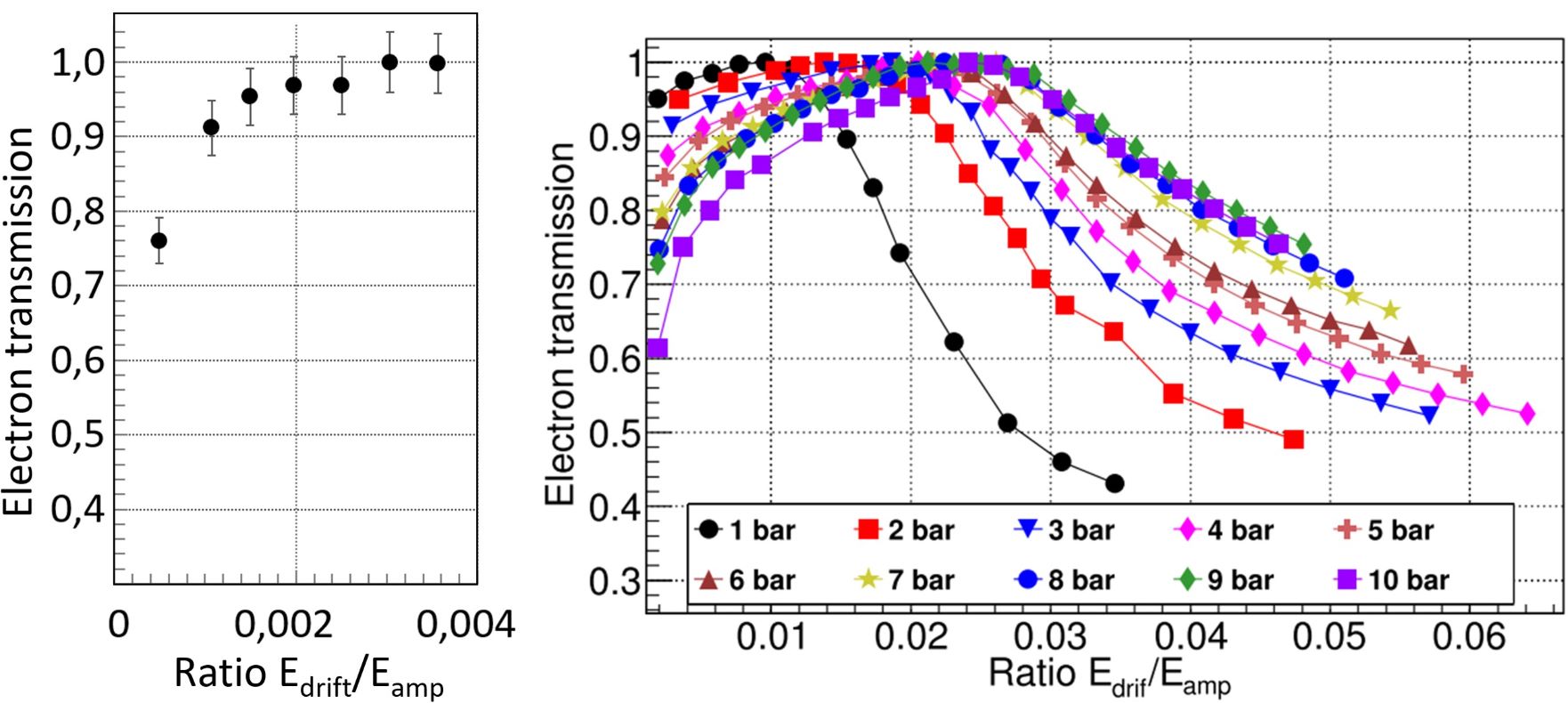}
    \caption{Comparison of electron transmission between the SR2M module prototype and the Micromegas of the TREX-$\beta\beta$-0 test TPC. $\bullet$~Left: electron transmission curve of the SR2M module versus the ratio between the drift field and the amplification field, with the mixture Xe+1\%~TMA at 1~bar pressure. $\bullet$~Right: electron transmission curve of the Micromegas of TREX-$\beta\beta$-0, for the mixture Xe+1.5\%~TMA and for different pressures~\cite{ThesisDiana}.}
    \label{fig:SR2MdataTaking_ElectronTransmision}
\end{figure}

The electron transmission curve, which evaluates the transmission of electrons from the ionization zone to the holes of the Micromegas itself~\cite{Ruiz-Choliz2015} (see figure~\ref{fig:Introduction_MicromegasWorkingPrinciple}), is typically represented as the variation of gain versus the ratio between the drift field and the amplification field. Figure~\ref{fig:SR2MdataTaking_ElectronTransmision} shows the curves developed in the prototype of the SR2M module and in the TREX-$\beta\beta$-0 chamber. The curve in the SR2M module prototype was obtained with V$_{mesh}$~=~270~V and only up to E$_{drift}$/E$_{amp}$~=~0.004, the point at which the cathode reached the maximum voltage (7~kV) before discharges occurred. Despite this limitation, the same upward trend as in TREX-$\beta\beta$-0 at 1~bar can be appreciated. However, the maximum electron transmission achieved in TREX-$\beta\beta$-0 occurs with E$_{drift}$/E$_{amp}$~=~0.01, which seems to indicate that the maximum electron transmission has not been reached in the SR2M module prototype, although it is likely close to the maximum. In any case, the objective of this first curve was to verify the proper functioning of the Micromegas and to choose a value (ultimately E$_{drift}$/E$_{amp}$~=~0.003) for the operation of the SR2M module, slightly lower than the maximum possible to avoid stressing the system. It is worth noting that the first spectrum obtained with the SR2M module prototype (figure \ref{fig:MCA_FirstSpectrum_SR2M}) was taken with a drift field of 100~V~cm$^{-1}$~bar$^{-1}$, which corresponds to the point E$_{drift}$/E$_{amp}$~=~0.002 in figure~\ref{fig:SR2MdataTaking_ElectronTransmision}, slightly below the maximum achievable electron transmission.

\begin{figure}[!ht]
    \centering
    \includegraphics[width=0.9\textwidth]{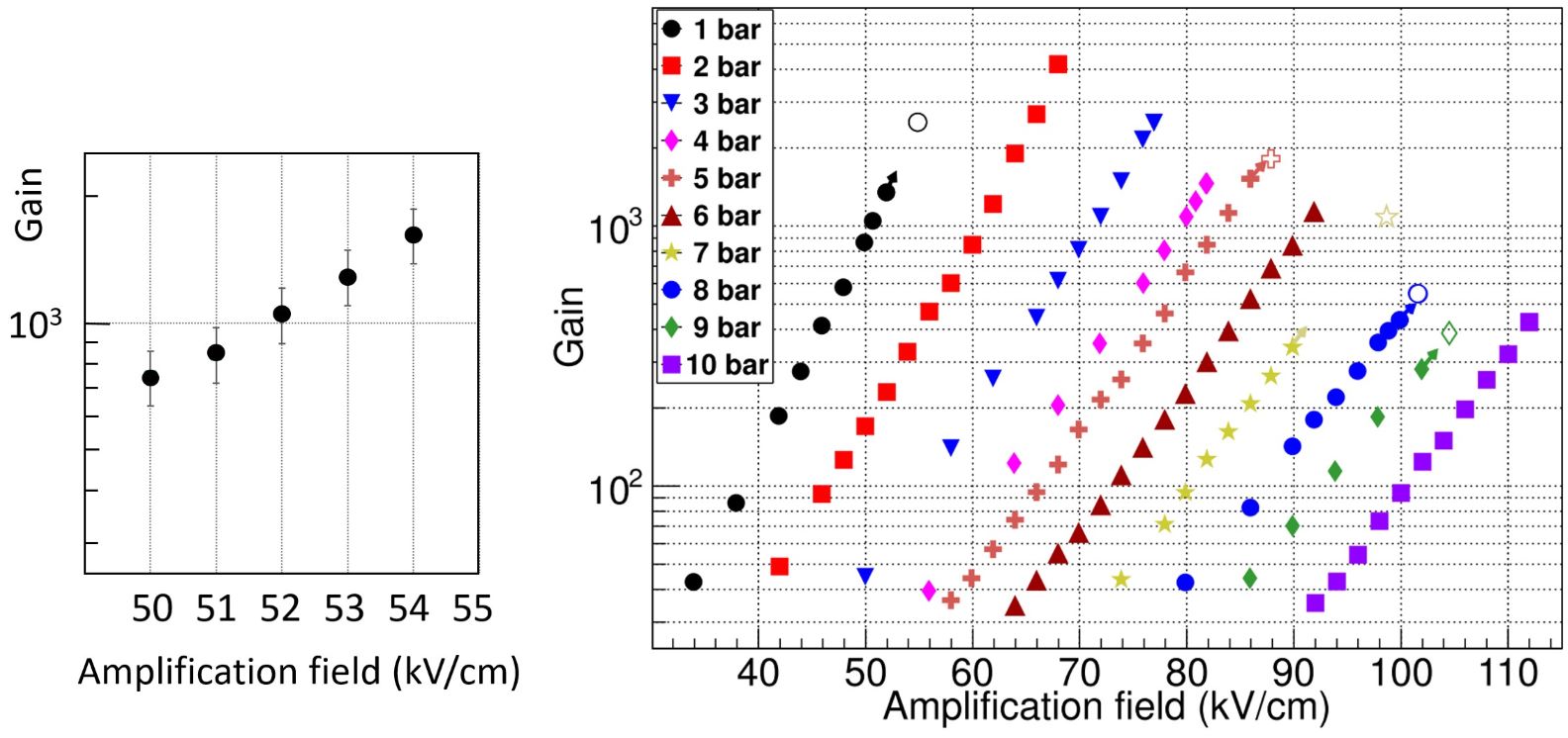}
    \caption{Comparison of gain between the SR2M module prototype and the Micromegas of the TREX-$\beta\beta$-0 test TPC. $\bullet$~Left: gain curve of the SR2M module versus the amplification field, with the mixture Xe+1\%~TMA at 1~bar pressure. $\bullet$~Right:~\cite{ThesisDiana} gain curve of the Micromegas of TREX-$\beta\beta$-0, with optimized TMA concentrations from 0.9\% to 1.2\% and different pressures from 1 to 10~bar.}
    \label{fig:SR2MdataTaking_GainPlot}
\end{figure}

The gain curve is typically represented as the absolute gain achieved by the Micromegas (multiplication factor of the primary electrons) versus the electric field of amplification applied in the holes where the avalanche effect occurs~\cite{Ruiz-Choliz2015}. Figure~\ref{fig:SR2MdataTaking_GainPlot} shows the curves developed in the prototype of the SR2M module and in the TREX-$\beta\beta$-0 chamber. The curve in the SR2M module prototype was obtained while maintaining E$_{drift}$/E$_{amp}$~=~0.003, the operating point chosen after the electron transmission curve was performed, and up to a maximum in the amplification field E$_{amp}$~=~54~kV, which corresponds to a voltage V$_{mesh}$~=~270~V. The curve with which it is compared (from 0.9\% to 1.2\%~TMA) reaches up to E$_{amp}$~=~55~kV, which corresponds to V$_{mesh}$~=~275~V. This difference is small considering that, under the same conditions, the maximum voltage achievable in two different Micromegas detectors can vary depending on the quality achieved in the manufacturing process and also on the wear to which it may have been subjected previously. Furthermore, it is expected that in the SR2M module prototype, with an active area of 20~x~20~cm$^{2}$, the probability of sparking increases compared to the Micromegas of TREX-$\beta\beta$-0, with an area of $\sim$2~cm$^{2}$. Aside from the aforementioned, the curve generated with the SR2M module prototype fits quite well with the curve it is compared to (1~bar), within the measurement errors.

\section{Reading of the Channels}
\label{sec:LecturaStrips}

To collect the signal from the channels, several parameters were adjusted in the acquisition electronics. The most important programmed parameters were as follows:

\begin{itemize}

\item \textit{Trigger delay}: 24~$\mu$s. This is the time that elapses from the moment a trigger occurs until the event recording stops. This parameter can be programmed according to the needs of each case. For example, it can be adjusted so that, for an event occurring very close to the surface of the Micromegas, the electronics can capture the signal from another event that occurred simultaneously in the area of the active volume farthest from the Micromegas, that is, near the surface of the cathode. Another example of adjustment is based on the length of the track that one wishes to reconstruct. For the commissioning, an intermediate value of 24~$\mu$s was chosen, which allowed for an acquisition window of about 11~cm (for Xe~+~1\%~TMA and a reduced drift field of 150~V~cm$^{-1}$~bar$^{-1}$, the drift velocity is about 0.45~cm~$\mu$s$^{-1}$, according to figure~\ref{fig:TranseverseDiffusionNEXT}) compared to the 35~cm drift length of the entire chamber. It should be noted that a high value of this parameter results in poorer resolutions along the Z-axis. It is also important to point out that the trigger of the electronics defines the initial time $t=0$, the moment when the first charge from the event is detected.

\item \textit{Sampling rate}: 100~ns. The value of the sampling rate multiplied by 512 (time bins or the number of values that the ADC has) will be the value of the time window collected in each event, in this case, 40.95~$\mu$s. This value is the sum of the previously selected \textit{trigger delay} and the time before the trigger that is desired to be recorded, on which the baseline and its fluctuation will be calculated during the event analysis.

\item \textit{Gain}: 240~fC. The gain of the amplifier in the AGET chip was adjusted to 240~fC, so that with the previously adjusted shaping time and the received charge, the signal remained within the dynamic range of the amplifier.

\item \textit{Shaping time}: 1014~ns. As can be seen in figure~\ref{fig:TheSR2M_OsciloscopeFirstSignal}, the response of the pre-amplifier is on the order of hundreds of nanoseconds. The response of the Micromegas to the arrival of the charge has a shorter time. The shaping time programmed in the electronics has been set to 1~$\mu$s, the maximum allowed by the electronics, as this value reduces high-frequency noise. Once work is done on noise reduction, the shaping time can be decreased.

\end{itemize}

\begin{figure}[!ht]
    \centering
    \includegraphics[width=1.0\textwidth]{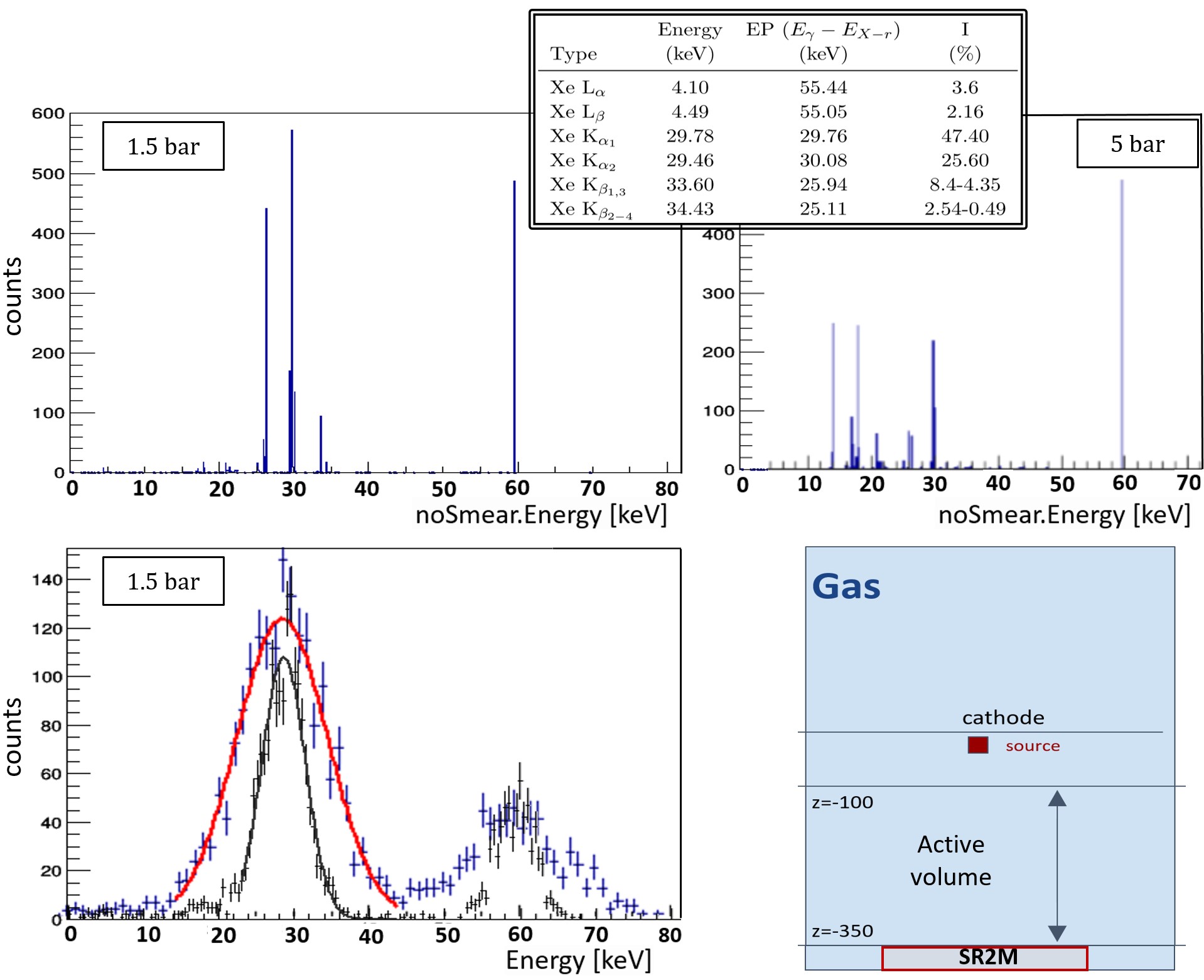}
    \caption{Simulation of the gamma emission of 59.54~keV from the $^{241}$Am source in the geometry (bottom right) of the TREX-$\beta\beta$ chamber, where the prototype of the SR2M module has been installed, for Xe~+~1\%~TMA at 1.5~bar (top left) and 5~bar (top right), showing the energy spectrum deposited within the active volume of the detector. The effects of scattering are shown for the simulation at 1.5~bar (bottom left) with two different resolutions: 49.7\%~FWHM (28.9\%) in red and 23.5\%~FWHM (13.7\%) in black for the peak at 28.49~keV (59.54~keV). A table with the escape peaks of a 59.54~keV photon in Xenon is also shown (top center). In addition to the peak at 59.54~keV from the gamma emission of the $^{241}$Am source and the corresponding escape peaks of this emission in Xenon, the simulations also show peaks corresponding to emissions from $^{237}$Np, a descendant of $^{241}$Am.}
    \label{fig:SR2MdataTaking_241AmSimulation}
\end{figure}

For data acquisition with the signal collected from the mesh plane, a source of $^{109}$Cd was installed. For data acquisition from the channels, a source of $^{241}$Am was installed, which, as observed in the simulation in figure \ref{fig:SR2MdataTaking_241AmSimulation}, has a well-defined peak at 59.54~keV. This source emits gamma rays of 59.54~keV (35.9\%) and 26.34~keV (2.4\%).

The simulation whose results are shown in figure~\ref{fig:SR2MdataTaking_241AmSimulation} was carried out within the PandaX-III collaboration, using the Geant~4 software~\cite{Agostinelli2003}, a software for simulating the passage of particles through matter, and the REST-for-Physics software~\cite{Altenmuller2022}, a framework for analysis and simulation in experiments searching for rare events, of which the Zaragoza group is the main developer.

\begin{figure}[!ht]
    \centering
    \includegraphics[width=0.60\textwidth]{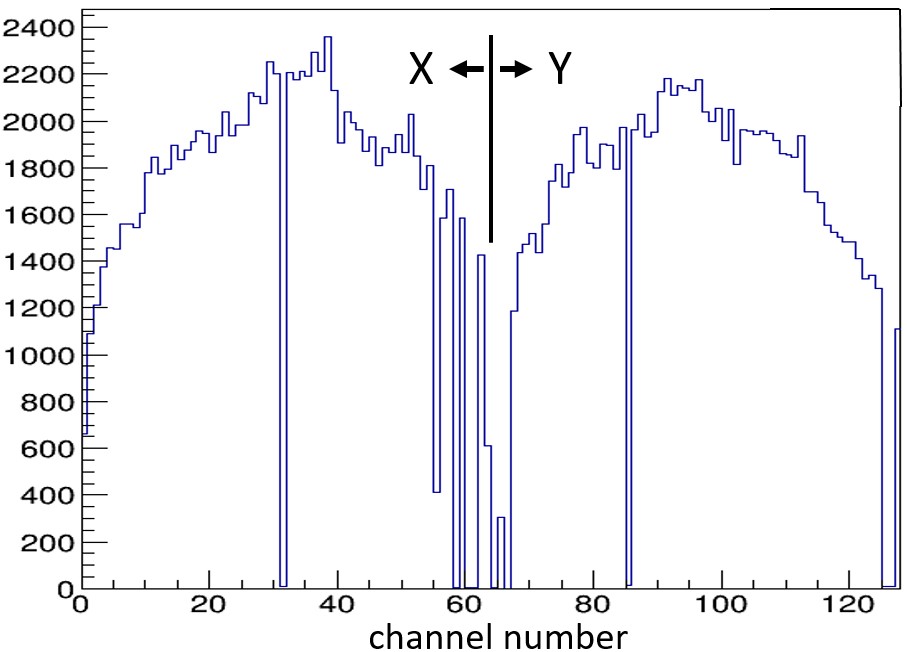}
    \caption{Activity map of the 128 channels of the SR2M module during a calibration with the $^{241}$Am source, for Xe~+~1\%~TMA at 3~bar. Ten defective channels can be observed: 31, 58, 60, 61, 66, 71, 85, 87, 125, 126. The SR2M module contains 64 \textit{strips} along the X-axis (1-64) and another 64 along the Y-axis (65-128).}
    \label{fig:SR2MdataTaking_DeadChannels}
\end{figure}

To take data with the new $^{241}$Am source, the pressure was increased to 3~bar, and the amplification field was adjusted to the maximum allowed before discharges occurred, E$_{amp}$~=~64~kV~cm$^{-1}$, which corresponds to a voltage in the mesh of 320~V. The drift field was similarly adjusted to the maximum allowed, E$_{drift}$~=~300~V~cm$^{-1}$~bar$^{-1}$, which corresponds to a voltage in the cathode of 31.5~kV. During data acquisition, the signals from the channels were collected as shown in figure~\ref{fig:SR2MdataTaking_DeadChannels}. It can be observed that there are 10 defective channels from which no signal is collected. These channels had been previously located and disconnected from the acquisition electronics.

\begin{figure}[!ht]
    \centering
    \includegraphics[width=1.0\textwidth]{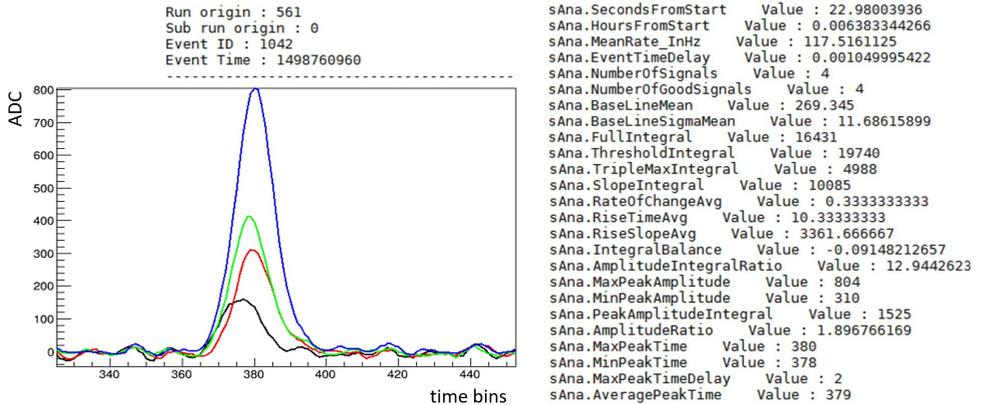}
    \caption{Event collected in the SR2M module during a calibration with the $^{241}$Am source, for Xe~+~1\%~TMA at 3~bar, and the observables calculated during the 'SignalAnalysis' process in REST-for-Physics.}
    \label{fig:SR2MdataTaking_SignalAndObservables}
\end{figure}

For data analysis, the REST-for-Physics software~\cite{Altenmuller2022} was used. In figure \ref{fig:SR2MdataTaking_RESTflow}, the processes in REST that comprise the analysis of each event collected by the detector are shown. Figure~\ref{fig:SR2MdataTaking_SignalAndObservables} displays an event collected and analyzed in the second process 'SignalAnalysis'. The first process 'MultiFEMINOSToSignal' transforms the data into \textit{root} format~\cite{ROOT}.

\begin{figure}[!ht]
    \centering
    \includegraphics[width=1.0\textwidth]{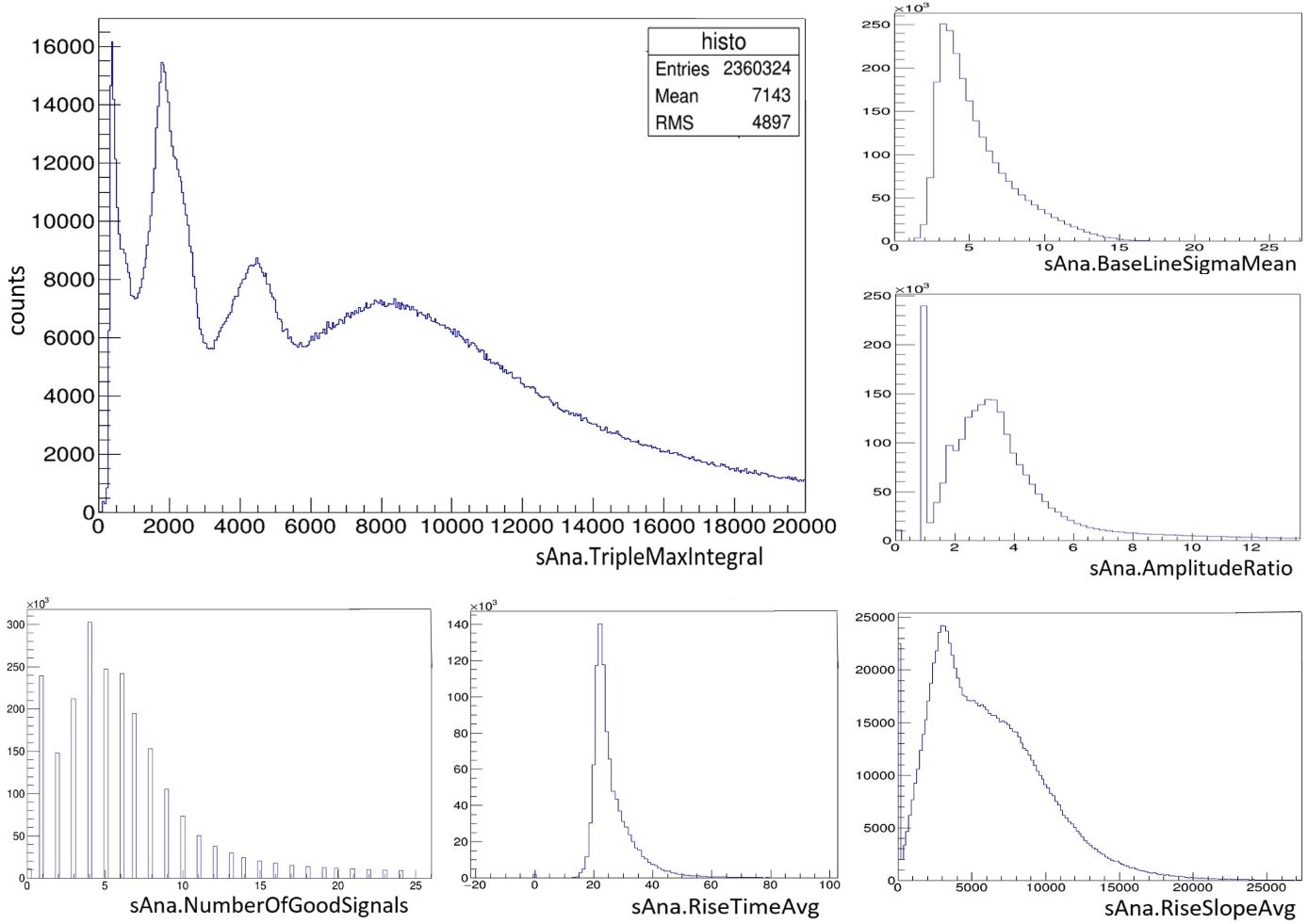}
    \caption{Histograms of some of the observables obtained after the first analysis process 'SignalAnalysis', based on the data collected in the SR2M module during a calibration with the $^{241}$Am source, for Xe~+~1\%~TMA at 3~bar. The observable 'sAna.TripleMaxIntegral' is used for estimating the deposited energy of the event.}
    \label{fig:SR2MdataTaking_HistogramasObservables}
\end{figure}

Figure \ref{fig:SR2MdataTaking_HistogramasObservables} shows the histograms of some observables. The energy spectrum, represented by the observable 'sAna.TripleMaxIntegral', shows three distinct peaks. These three peaks appear to correspond to the peaks at 59.54, 28.49, and 15-20~keV from the previously conducted simulation (see figure~\ref{fig:SR2MdataTaking_241AmSimulation}). The simulations were performed at 1.5 and 5~bar, while the data were taken at 3~bar.

\begin{figure}[!ht]
    \centering
    \includegraphics[width=0.75\textwidth]{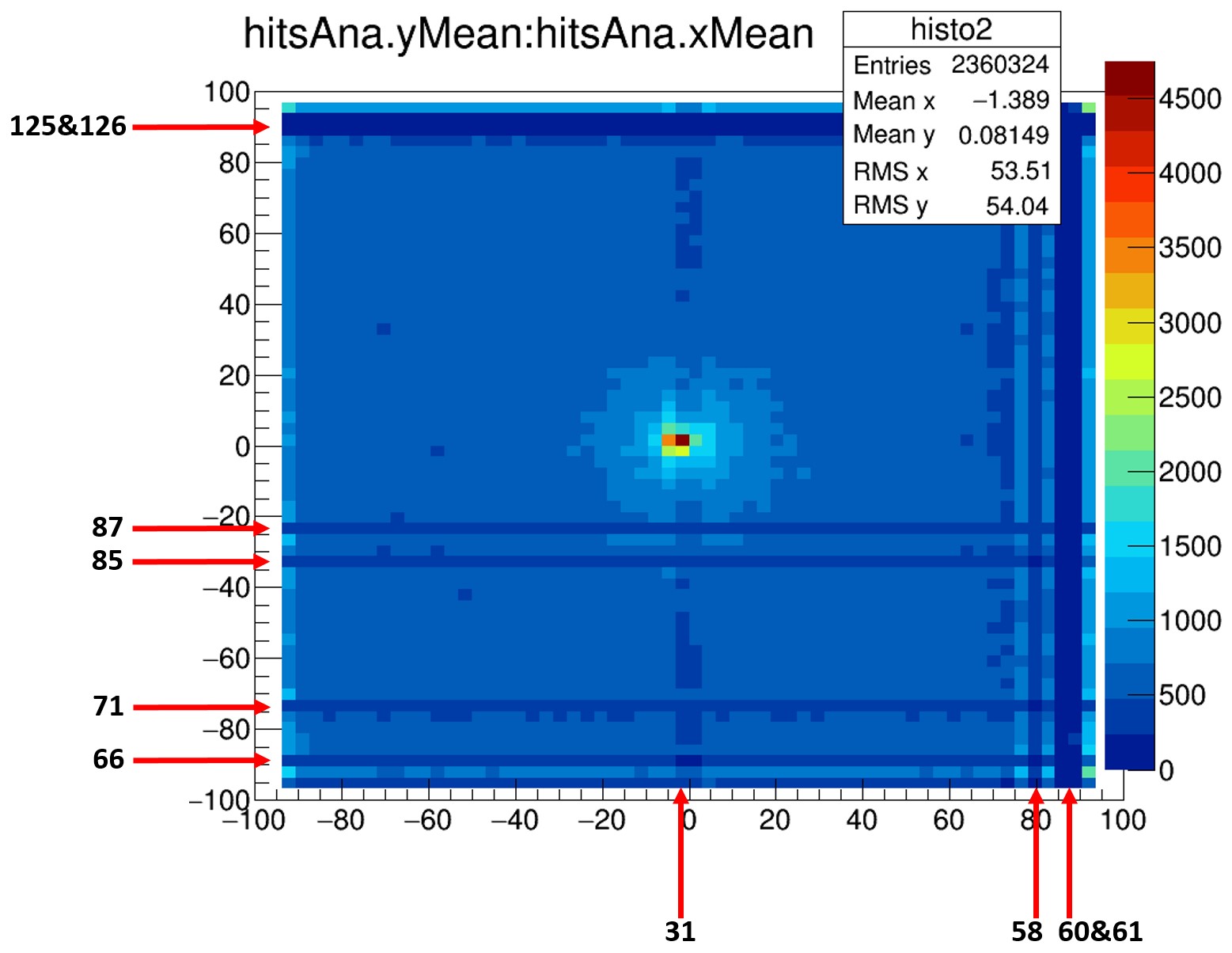}
    \caption{Activity map after the 'SignalToHits' analysis process, based on the data collected in the SR2M module during a calibration with the $^{241}$Am source, for Xe~+~1\%~TMA at 3~bar. The numbering of the defective \textit{strips} that do not collect signal is shown.}
    \label{fig:SR2MdataTaking_HitsMap}
\end{figure}

The analysis is completed with the 'SignalToHits' process, which uses the geometry of the detector to perform a topological analysis of the collected events (see figure~\ref{fig:SR2MdataTaking_HitsMap}).

\section{Reading Plane with 7 SR2M Modules}
\label{sec:ResultsFromTheSJTU}

Following the tests in Zaragoza with the first prototype of the SR2M module, the design was validated, and seven additional modules were manufactured to be assembled in the laboratories of the Shanghai Jiao Tong University (SJTU), in the 20~kg~(600~liters) TPC prototype~\cite{Lin2018a} that would serve as a testbed and demonstrator for the construction of the final 140~kg detector.

\begin{figure}[!ht]
    \centering
    \includegraphics[width=0.95\textwidth]{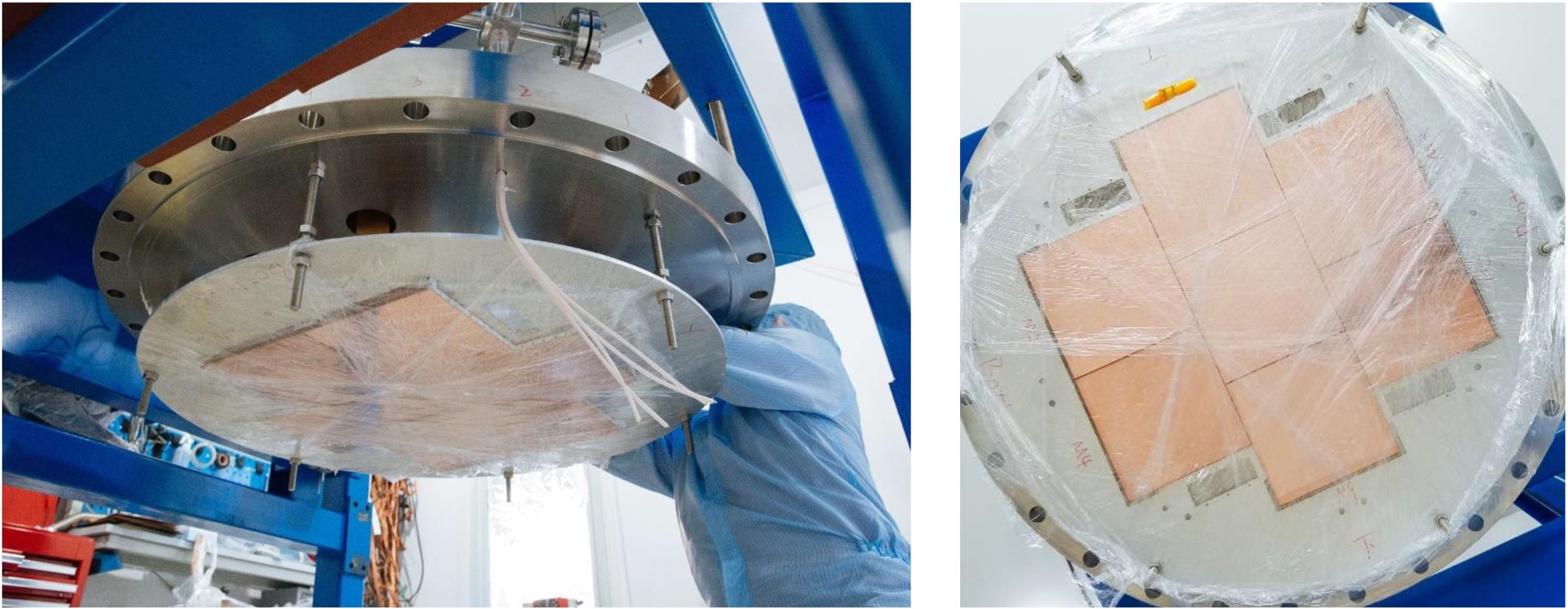}
    \caption{Photographs taken during the assembly of the 7 SR2M modules to form the reading plane in the SJTU test prototype.}
    \label{fig:SR2MdataTaking_SR2MTessellationPictureSJTU}
\end{figure}

Part of the installation and commissioning of the TPC, along with the reading plane formed by the 7 SR2M modules, was carried out throughout 2018. Figure~\ref{fig:SR2MdataTaking_SR2MTessellationPictureSJTU} shows two photographs taken during the assembly of the reading plane. The adjustment of the modules to each other was performed in such a way as to minimize the distance between them, within the errors of perpendicularity and parallelism inherited from the manufacturing process of the modules.

It is worth noting that the members of SJTU carried out all the necessary work to commission the 20~kg demonstrator. From Zaragoza, after validating the first prototype of the SR2M module, advisory support was provided for the installation and commissioning of the reading plane in the SJTU demonstrator, with on-site support only at specific times (figure~\ref{fig:SR2MdataTaking_SR2MTessellationPictureSJTU}).

\begin{figure}[!ht]
    \centering
    \includegraphics[width=1.0\textwidth]{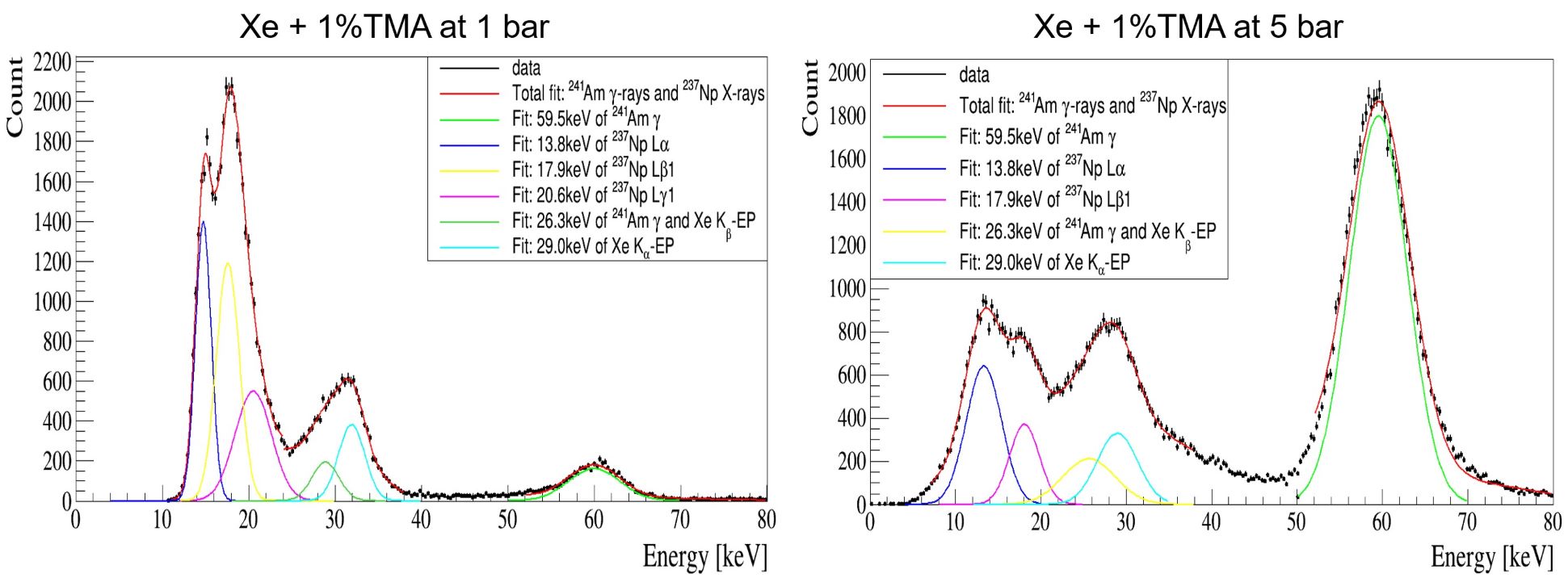}
    \caption{Spectra generated during data collection with a $^{241}$Am source in the SJTU test prototype, with Xe~+~1\%~TMA at 1 and 5~bar.}
    \label{fig:SR2MdataTaking_SJTU_Spectrum}
\end{figure}

The first data were collected using a mixture of Xe~+~1\%~TMA and a $^{241}$Am source located at the center of the cathode. Figure~\ref{fig:SR2MdataTaking_SJTU_Spectrum} shows the resulting spectra from the first data collected at 1 and 5~bar. The spectra match the simulations conducted under the conditions of the prototype.

\begin{figure}[!ht]
    \centering
    \includegraphics[width=1.0\textwidth]{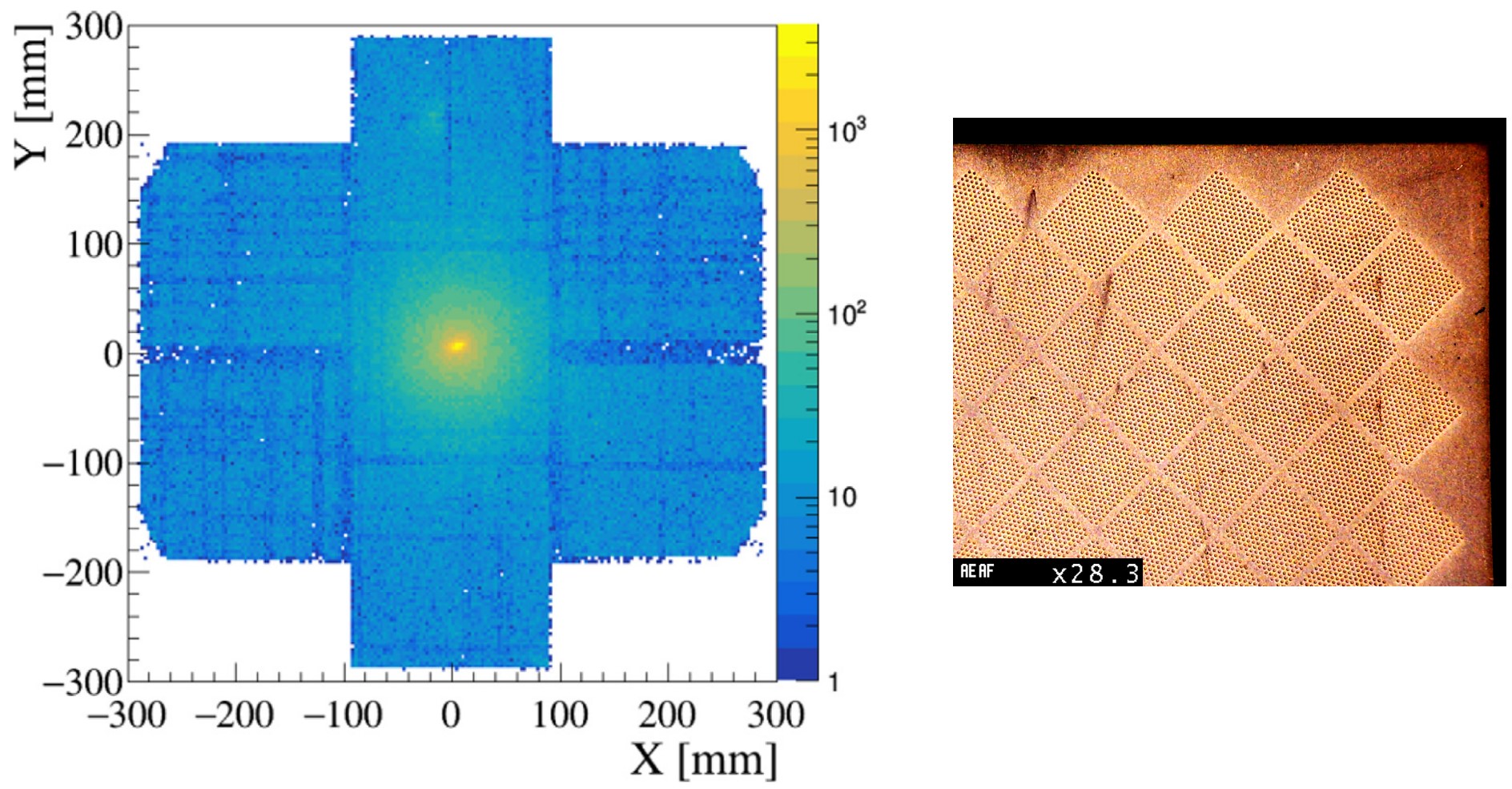}
    \caption{$\bullet$~Left: Activity map after the 'SignalToHits' analysis process, with 9.2\% defective channels, extracted from \cite{Wang2020}. $\bullet$~Right: Photograph with 28.3x magnification of the mesh surface in the SR2M module. It can be observed that there are no amplification holes at the edges.}
    \label{fig:SR2MdataTaking_SJTU_Hitmap}
\end{figure}

Figure~\ref{fig:SR2MdataTaking_SJTU_Hitmap} (left) shows the activity map across the entire reading plane formed by the 7 SR2M modules. It can be observed that there are many inactive channels, from which no signal is received, due to both connection failures in the chain leading to the acquisition electronics and defective strips in the Micromegas circuit itself. The spectra shown in figure~\ref{fig:SR2MdataTaking_SJTU_Spectrum} were obtained by strategically selecting areas with fewer defective channels.

A reduction in activity can also be observed at the junction edges between the modules. While this could be interpreted as a lack of efficiency in the electronic reinsertion system (ERS), presented in section~\ref{sec:ERS}, a more in-depth study revealed that, in fact, it was due to the absence of amplification holes in the strips at the edges of each module (figure~\ref{fig:SR2MdataTaking_SJTU_Hitmap} right). This was an error during the manufacturing of the Micromegas circuits, where the specifications reflected in the schematic of figure~\ref{fig:SR2M_LayersActiveArea} were overlooked. This manufacturing error had gone unnoticed during the testing and validation phase carried out in Zaragoza, as the efficiency study of the ERS needed to be conducted with at least two modules forming the same reading plane.

This manufacturing error did not allow for the experimental verification of the effectiveness of the external rim. It was also not possible to do so in subsequent versions of the SR2M module, as it was removed from the design, as described in chapter~\ref{ch:SR2Moptimization}. However, the internal rim was retained, whose effectiveness was demonstrated and is also presented in chapter~\ref{ch:SR2Moptimization}.

\setlength{\baselineskip}{12pt} 
\selectlanguage{USenglish}
\graphicspath{{../Imagenes/}{../../Imagenes/}} 

\renewcommand{\listtablename}{Table of contents}
\renewcommand{\tablename}{Table} 

\chapter{The FtF Connector and the Second Version of the SR2M Module}
\label{ch:SR2Moptimization}
\chaptermark{The FtF Connector and...}

\begin{figure}[!ht]
    \centering
    \includegraphics[width=0.7\textwidth]{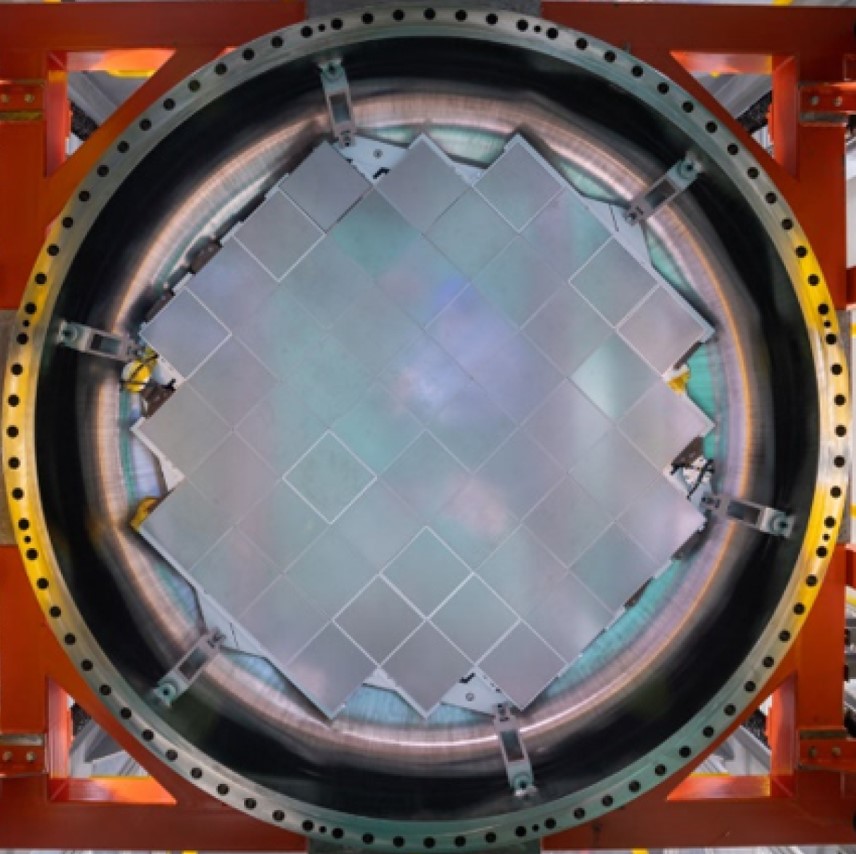}
    \caption*{}
    \label{fig:SR2Mv2_SJTU_1000kgReadoutPlane}
\end{figure}

\newpage


\section{The Face-to-Face (FtF) Connector}
\label{sec:FtFconnector}

Although the first prototype of the SR2M module was installed and tested satisfactorily, it soon became evident that there was a limitation in making the connection of the Micromegas outside the chamber, due to the limitation in the length of the tab for manufacturing reasons. This fact was critical when designing the TPC that would contain 140~kg of Xenon enriched to 90\% with the isotope~$^{136}$Xe at a pressure of 10~bar~\cite{Huang2020}. The cover of this chamber, on which the readout plane is installed, had to be flat if the first design of the SR2M module was used with the connection on the outside, to allow the tab to pass through the cover. However, the most optimal design for the cover was domed rather than flat, which was incompatible with the use of the first prototype of the SR2M module.

At this point, several alternatives were considered: the manufacture of a flat cover instead of a domed one; or the initiation of a set of activity measurements on a wide range of commercial high-density channel connectors available on the market at that time, with the hope of finding one with sufficiently low activity so as not to significantly increase the background of the experiment; or the design and manufacture of custom connectors with low-activity materials. In exploring this last option, a type of connector was proposed that would be the chosen alternative for the final solution, the so-called 'Face to Face' (FtF) connector.

\subsection{Principle of Operation of the FtF Connector}

\begin{figure}[!ht]
    \centering
    \includegraphics[width=0.95\textwidth]{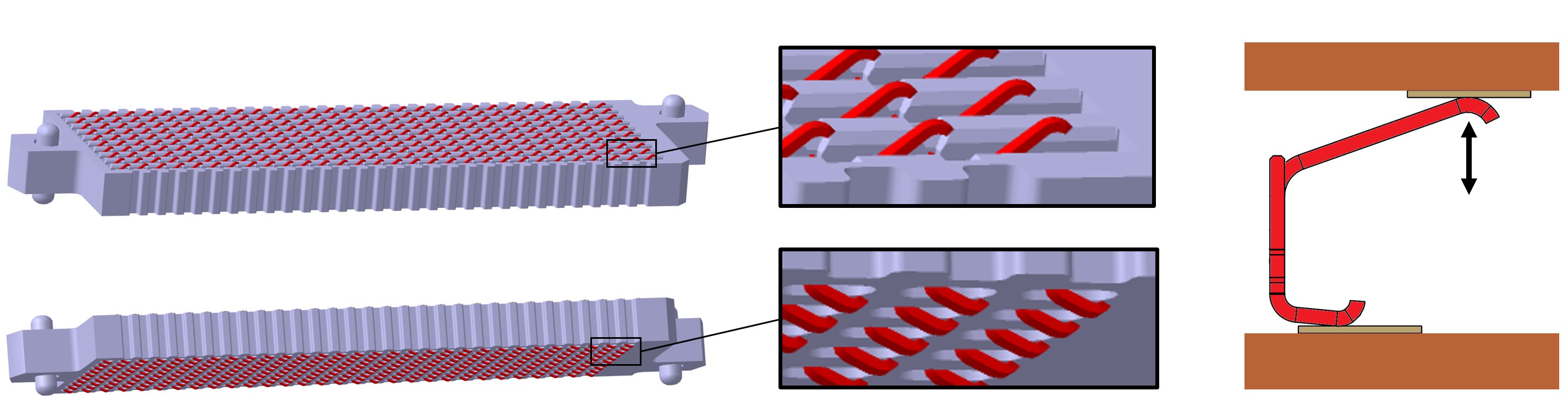}
    \caption{3D view of the Samtec GFZ-30-03-G-10-AD~\cite{SAMTEC} connector used in the first prototype of the SR2M module. On the right, a connection scheme of the Samtec between two opposing circuits is shown, through a U-shaped metal wire that allows for the absorption, via its elastic deformation, of certain errors in the position of the pads along the connection axis.}
    \label{fig:Samtec3D}
\end{figure}

Commercial high-density connectors, such as the Samtec GFZ-30-03-G-10-AD~\cite{SAMTEC} used in the first prototype of the SR2M module (figure~\ref{fig:FootprintSR2M}), connect the pad of a channel in one circuit with the pad of the channel in another opposing circuit (figure~\ref{fig:Samtec3D}). To achieve this, they utilize the elasticity of some of the materials they are composed of (in the case of Samtec, a metal wire) to absorb the error in the connection axis due to manufacturing tolerances (figure~\ref{fig:Samtec3D} right).

The circuits for which this type of connector is designed are rigid, so in the case of Micromegas circuits or flat cables, they must be joined to rigid supports to ensure proper connection.

\begin{figure}[!ht]
    \centering
    \includegraphics[width=0.7\textwidth]{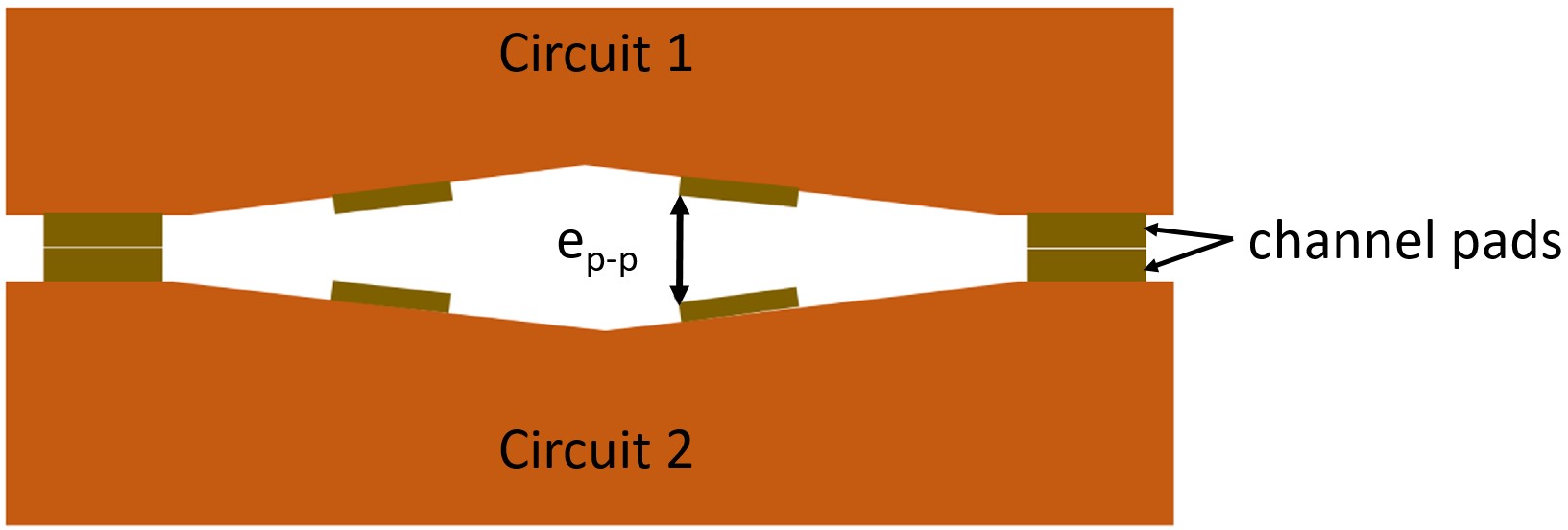}
    \caption{Schematic of the connection area between two opposing circuits, with e$_{p-p\max}$~=~32~$\mu$m being the sum of the dimensional tolerances of the circuits, imposed by the techniques generally used during the manufacturing processes of this type of circuits.}
    \label{fig:FtF_ToleranciasCircuitos}
\end{figure}

The FtF connector arises from the idea of establishing a direct connection between two circuits, facing their pads without any intermediate connector. However, facing two circuits in this manner does not guarantee the connection of all their pads, as there is an error in the position of the pads along the connection axis due to manufacturing tolerances (figure \ref{fig:FtF_ToleranciasCircuitos}).

Micromegas circuits and flat cables, which are manufactured from layers of kapton and copper, are flexible due to the small thickness of the layers that compose them. Taking advantage of this flexibility, the lack of connection in the scheme of figure \ref{fig:FtF_ToleranciasCircuitos} can be resolved by applying sufficient pressure between the circuits so that the resulting deformation ensures the connection of all their pads.

\begin{figure}[!ht]
    \centering
    \includegraphics[width=0.7\textwidth]{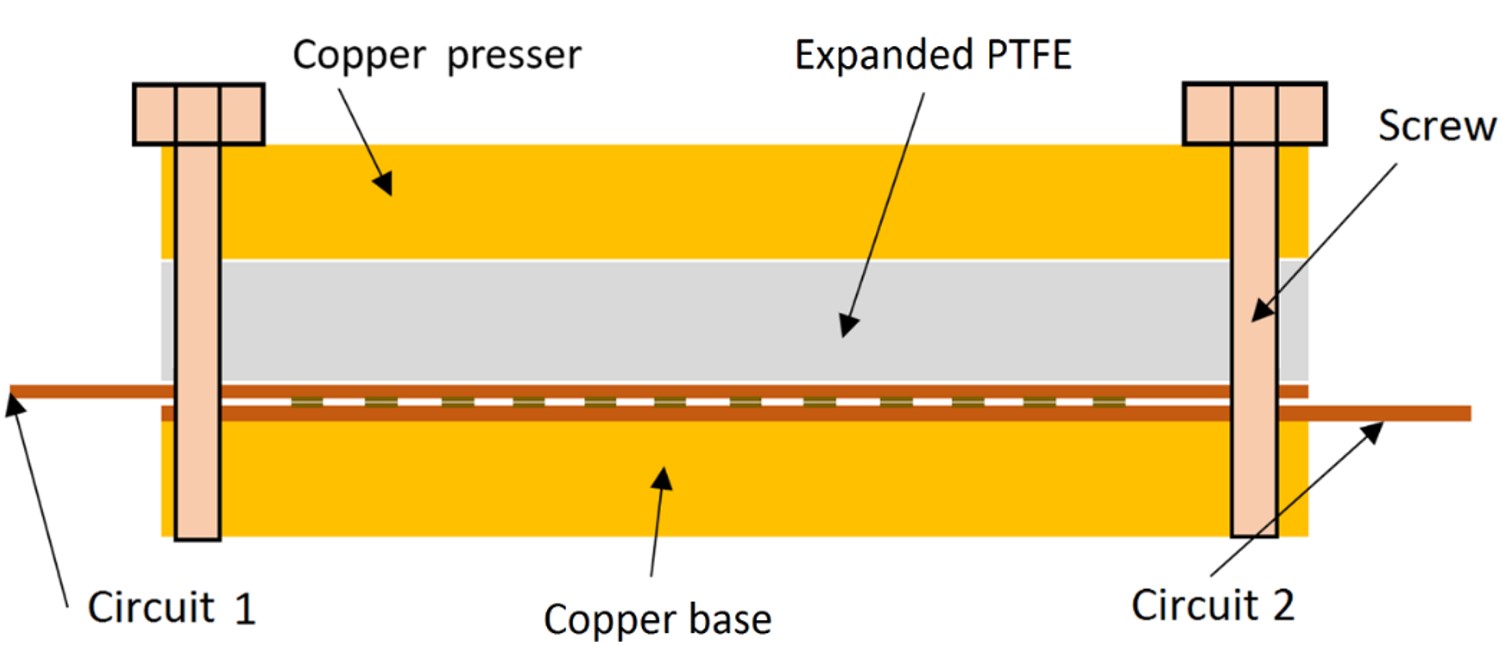}
    \caption{Connection scheme of the FtF connector. Two opposing circuits are pressed together by a rigid copper piece (Copper presser) and a piece of expanded PTFE that distributes the pressure homogeneously across the surfaces of the circuits.}
    \label{fig:FtF_EsquemaConector}
\end{figure}

The pressure on the circuits, to be effective, must be applied with a material whose elasticity allows for the absorption of the manufacturing tolerances of~32~$\mu$m (e$_{p-p\max}$~in~figure~\ref{fig:FtF_ToleranciasCircuitos}). The material proposed for this purpose was expanded PTFE, a material commonly used for sealing joints, but which, due to its properties, could distribute the pressure on the surfaces of the circuits homogeneously, thereby absorbing the dimensional tolerances. Furthermore, being composed of 100$\%$ PTFE, it should have low activity, as was later confirmed through measurements with ICPMS~\cite{Castel2019}. Figure~\ref{fig:FtF_EsquemaConector} shows the connection scheme of the FtF connector with all the necessary components.

\subsection{Analysis of Deformations Under Tension in the Contact Area}

Figure \ref{fig:FtF_ToleranciasCircuitos} shows the initial state of two facing circuits, where there are channel pads that are not in contact due to manufacturing tolerances. After the application of pressure, as shown in Figure \ref{fig:FtF_EsquemaConector}, the surface of circuit 1 must deform until all the channel pads come into contact.

\begin{figure}[!ht]
    \centering
    \includegraphics[width=\textwidth]{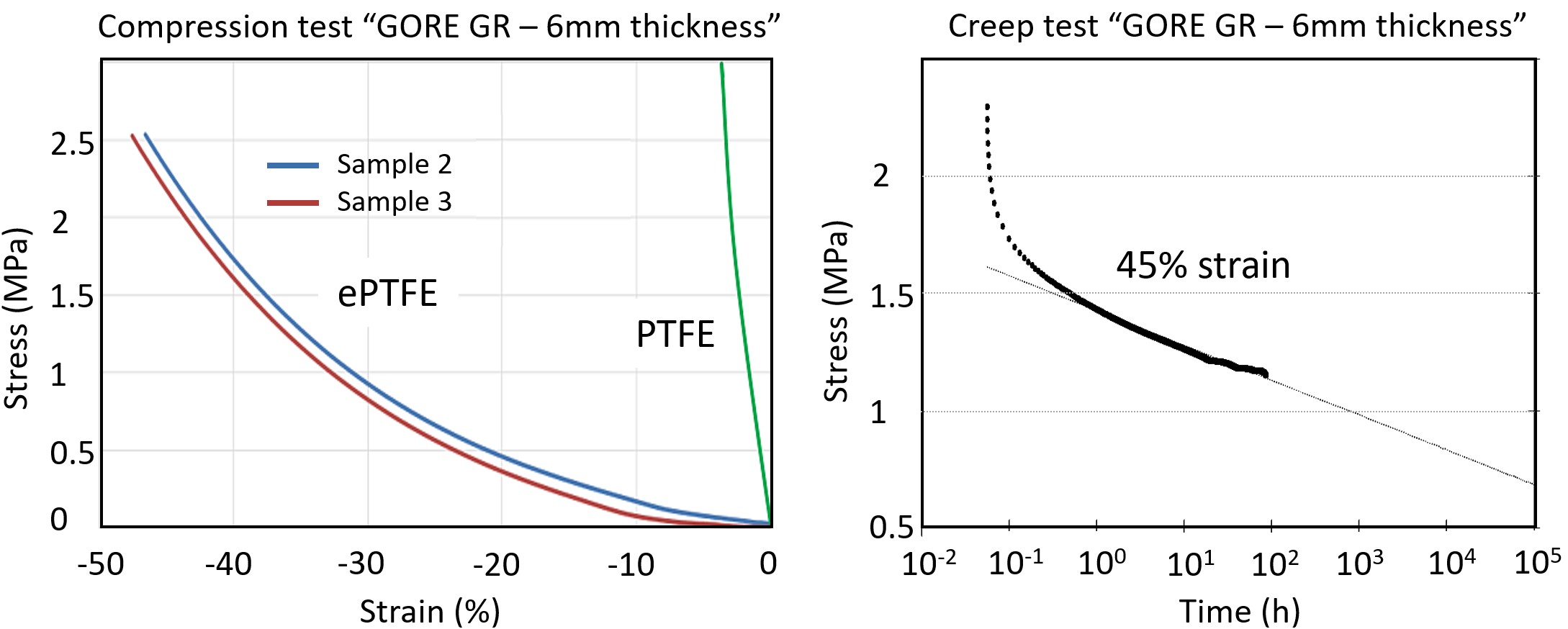}
    \caption{Results of the compression tests for expanded PTFE. $\bullet$~Left: stress-strain curve generated from tests performed on two samples. These are compared with the characteristic curve of standard PTFE (non-expanded), clearly showing the difference in elasticity. $\bullet$~Right: relaxation curve after the test in which a sample was maintained at 45$\%$ deformation for 100 hours (thick points) and the curve fit (thin points) to predict the long-term relaxation of expanded PTFE, which could be a cause of connection failure in the FtF connector.}
    \label{fig:FtF_EnsayosCompresion}
\end{figure}

To predict their behavior and adequately size the FtF connector solution, a stress and deformation analysis was conducted on the assembly. For this, it was necessary to know the mechanical properties of all the materials used: kapton, copper, and expanded PTFE. The mechanical properties of the first two were known, but some properties of the expanded PTFE were not available, so compression tests of this material had to be performed (figure~\ref{fig:FtF_EnsayosCompresion}), guided by the ASTM D695-10 'Standard Test Method for Compressive Properties of Rigid Plastics'~\cite{ASTM_D695}. 

These tests, apart from the stress-strain behavior (figure \ref{fig:FtF_EnsayosCompresion} left), also provided results on the relaxation effect of the material over time (figure \ref{fig:FtF_EnsayosCompresion} right), which is essential to ensure the long-term stability of the connection. The absence of a connection within the TPC had some important advantages, such as reducing potential connection issues during the operation of the detector that would require opening the chamber. Therefore, the FtF connector needed to have sufficient robustness to minimize the probability of failure during the long-term operation of the detector.

\subsubsection{FEM Analysis}

\begin{figure}[!ht]
    \centering
    \includegraphics[width=\textwidth]{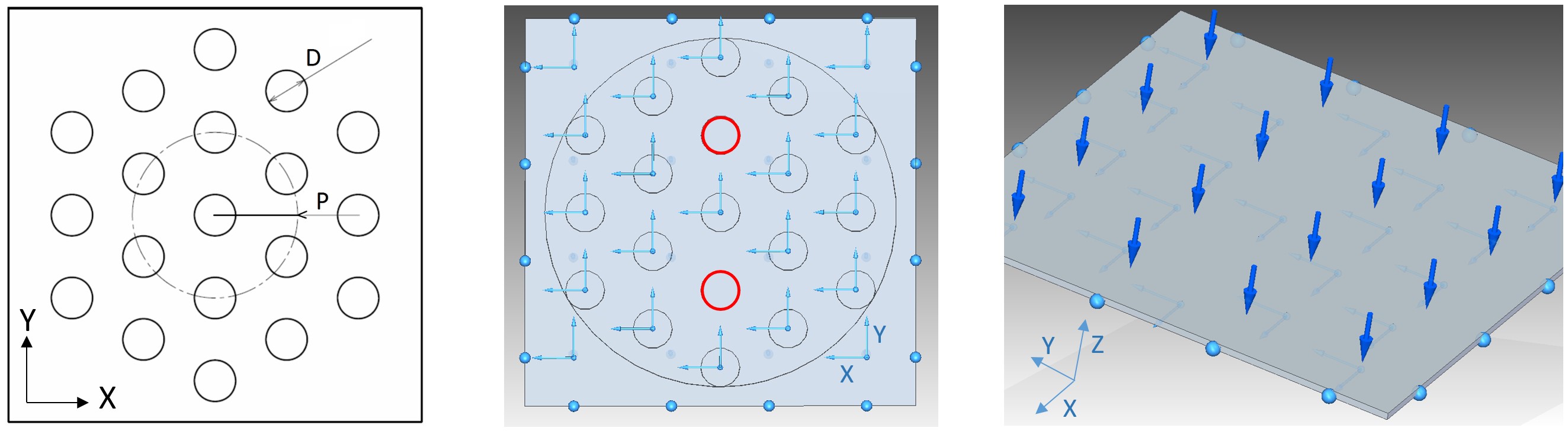}
    \caption{Finite element method simulation for the FtF connector area. $\bullet$~Left: simulated geometry with a hexagonal pattern, where D is the diameter of each channel's footprint and P is the distance between the centers of the footprints. $\bullet$~Center: boundary conditions imposed on the geometry: clamped along the entire contour; free displacement without restrictions on the footprints marked in red, which represent the situation in figure~\ref{fig:FtF_ToleranciasCircuitos}, where some footprints are not in contact; sliding allowed in the XY plane and restriction in the Z axis for the remaining footprints, as these footprints represent those that, according to figure~\ref{fig:FtF_ToleranciasCircuitos}, are indeed in contact. $\bullet$~Right: stress condition, in this case pressure on the surface of the circuit, representing the pressure that the expanded PTFE exerts on the surface of the circuit according to figure~\ref{fig:FtF_EsquemaConector}.}
    \label{fig:FtF_FEMconditions}
\end{figure}

To dimension the FtF connector, a stress and deformation analysis was performed on a flexible circuit that contained a hexagonal footprint pattern (figure~\ref{fig:FtF_FEMconditions}), optimal for covering flat surfaces. Appropriate boundary conditions were applied to simulate the case in which two opposing circuits had footprints that did not achieve contact, as shown in figure~\ref{fig:FtF_ToleranciasCircuitos}. Pressure was applied to the surface simulating the pressure exerted by the expanded PTFE, as illustrated in figure~\ref{fig:FtF_EsquemaConector}.

The simulations were conducted for a circuit that contained three layers of copper, each 17~$\mu$m thick, and two layers of kapton, separating the copper layers, each 275~$\mu$m thick. In total, the thickness was 601~$\mu$m, a quite reasonable thickness for the types of flat cables manufactured for these applications. A total of three simulations were launched for three different footprint patterns defined by D and P (figure~\ref{fig:FtF_FEMconditions} left) and with a total pressure distributed over the entire surface of 1~MPa (figure~\ref{fig:FtF_FEMconditions} right). The results are presented in table~\ref{tab:ResultadosSimulFtF}.

It is important to note that if in the FtF connection the pressure distributed by the expanded PTFE is applied to the Micromegas circuit, with a typical thickness of 250~$\mu$m with three layers of copper, the deformation will be much more favorable. That is, if in the simulation for the flat cable the minimum necessary deformation of 32~$\mu$m (figure~\ref{fig:FtF_ToleranciasCircuitos}) is reached, then it will be more than sufficient in the Micromegas.

\begin{table}[!h]
\centering
\begin{tabular}{|c|c|}
\hline
\textbf{Pattern Parameters} & \textbf{Displacement in z} \\ 
\hline
D = 1 mm, P = 2 mm (26 ch/cm$^2$) & 9.2 $\mu$m   \\
D = 0.7 mm, P = 3 mm (13 ch/cm$^2$) & 72.3 $\mu$m   \\
D = 0.7 mm, P = 3.5 mm (10 ch/cm$^2$) & 143.0 $\mu$m   \\
\hline
\end{tabular}
\caption{Results of the three simulations, using the finite element method, conducted for a kapton-copper circuit of 601~$\mu$m thickness and three different hexagonal patterns, defined by the footprint diameter D and the distance between footprint centers P, according to figure~\ref{fig:FtF_FEMconditions} left. Each pattern corresponds to a different channel density.}
\label{tab:ResultadosSimulFtF}
\end{table}

\begin{figure}[!ht]
    \centering
    \includegraphics[width=0.85\textwidth]{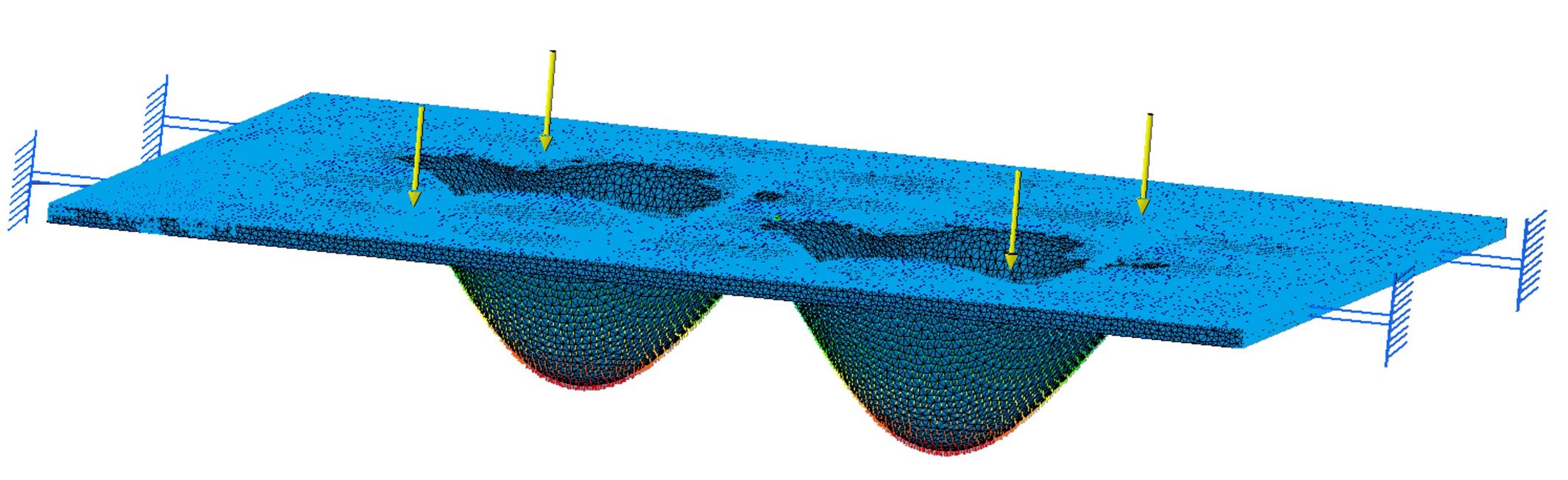}
    \caption{Deformations in the circuit after a simulation using the finite element method, whose geometry and conditions are shown in figure~\ref{fig:FtF_FEMconditions}.}
    \label{fig:FtF_FEMdeformation}
\end{figure}

From the configuration with D~=~0.7~mm and P~=~3~mm, and for larger P, the deformation in the circuit is such that it produces a displacement in the area of the unconnected footprints greater than 72.3~$\mu$m, a factor 2.3 greater than the minimum necessary displacement (32~$\mu$m). Therefore, future circuit designs that adopt the FtF solution for connections must respect these minimum values.


\subsubsection{Validation of the FEM Analysis}

To conclude the study of stresses and deformations, the method used was validated by comparing the results obtained with the finite element method and those obtained with the Navier analytical method for plates. In this case, the geometry used for validation was simpler than that of the simulated circuit, consisting of a square plate with dimensions similar to those of the circuit and subjected to a pressure of 1~MPa on the surface. The parameters used in both cases were:

\begin{itemize}
\item Length (a): 10.0 mm
\item Length (b): 10.0 mm
\item Thickness of the plate (h): 150 $\mu$m
\item Elastic modulus (E): 200 GPa
\item Poisson's ratio ($\nu$): 0.290
\item Surface pressure (P$_{0}$): 1 MPa
\end{itemize}

\begin{figure}[!ht]
    \centering
    \includegraphics[width=1.0\textwidth]{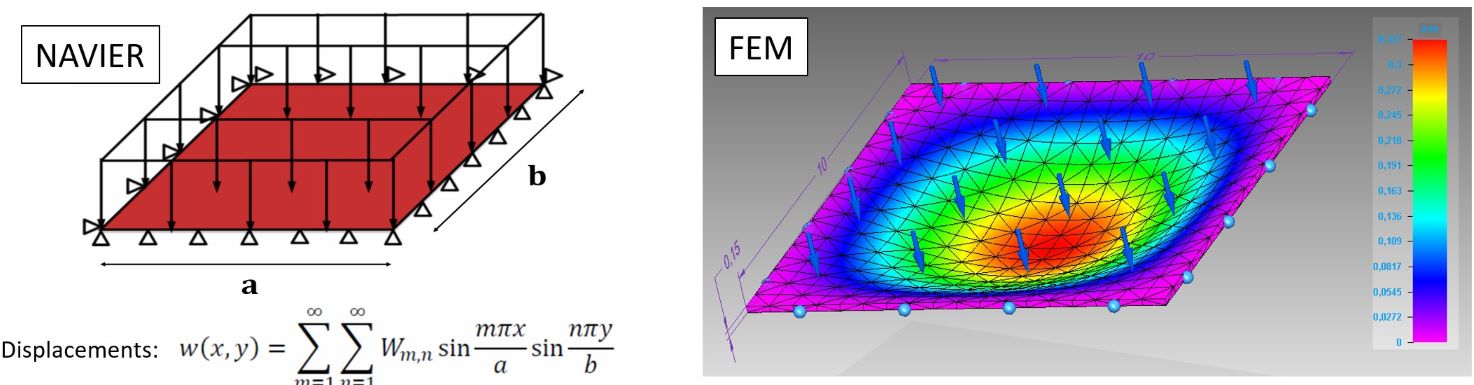}
    \caption{Navier method (left) for plates used to validate the finite element method (right) of the stress-deformation simulation of the circuit in the FtF zone. The displacement $w(x,y)$ in the Navier method is defined as the displacement experienced by a part of the plate in the direction of the Z-axis.}
    \label{fig:FtF_NavierValidation}
\end{figure}

The maximum displacement achieved with the finite element method was 327~$\mu$m, while with the Navier method it was 662~$\mu$m. The displacement with the Navier method was a factor of 2 greater than that with the finite element method, confirming that the latter is more restrictive. Therefore, the simulations of the circuits performed with this method were validated.

It is worth noting that the objective of the simulations was to achieve the minimum displacement necessary for two opposing pads without contact to make contact due to the deformation of the circuit on which pressure is applied. The chosen pad pattern (D~=~0.7~mm and P~=~3~mm) is established as the minimum that will be selected when designing this type of connection. It should be taken into account that this minimum pattern has been calculated for a circuit with a thickness of 601~$\mu$m. For circuits with greater thickness on which pressure is applied, this study will need to be redone. For thicknesses less than 601~$\mu$m, or patterns with P greater than 3 mm, the established pattern is entirely valid. Similarly, in the case of greater thicknesses, if for thicknesses less than 601~$\mu$m it is necessary to increase the channel density, which seems feasible a priori, this study will have to be redone.

\subsection{Test Bench for the FtF Connector}

\begin{figure}[!ht]
    \centering
    \includegraphics[width=\textwidth]{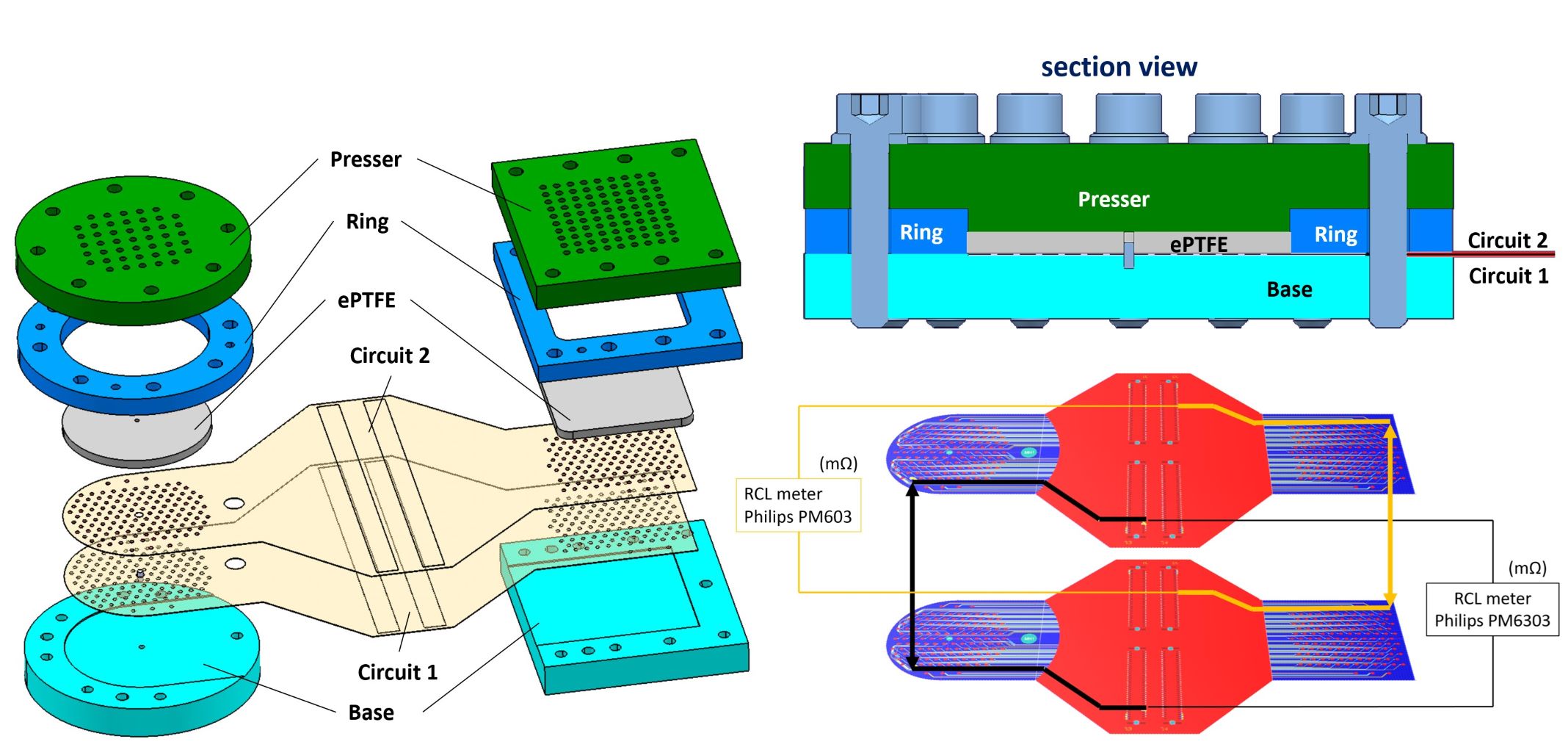}
    \caption{$\bullet$ Left: exploded 3D view of the test bench to test the FtF connector, formed by 2 circuits with 2 FtF connectors each (circular and square) facing each other, the expanded PTFE pieces that distribute the pressure in the circuits, and the mechanical pieces that hold and apply the initial pressure. $\bullet$~Right top: section of the test bench showing the system once assembled, with the PTFE deformed to its final thickness. $\bullet$~Right bottom: schematic of the facing circuits and the measurement system. An RCL meter measures the resistance between a point on one circuit and the corresponding point on the other circuit, connected through the FtF connector. This measurement is performed for both the 134 channels of the circular connector and those of the square connector.}
    \label{fig:FtF_DrawingsTestBed}
\end{figure}

\begin{figure}[!ht]
    \centering
    \includegraphics[width=0.9\textwidth]{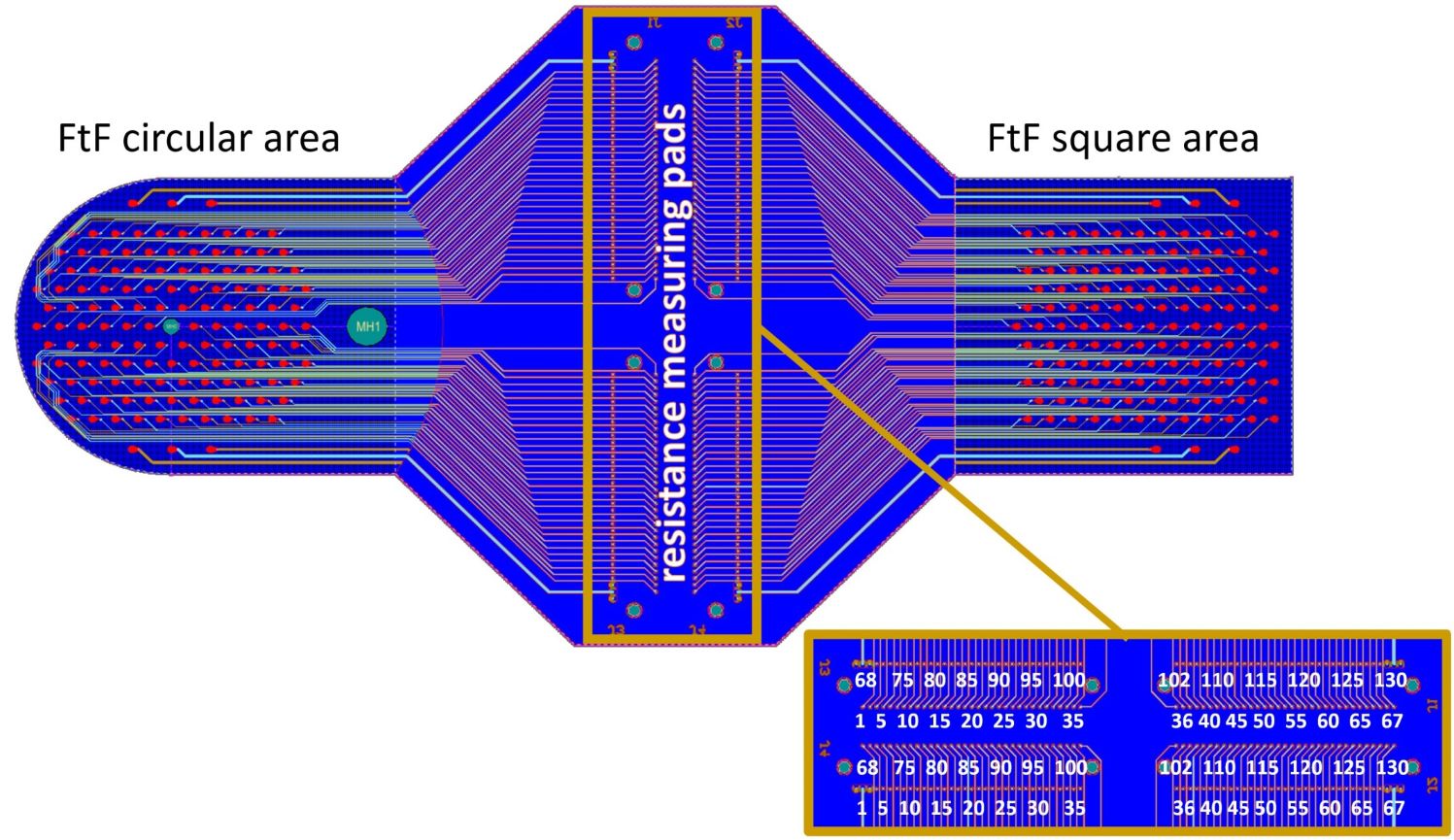}
    \caption{Schematic of the circuits used in the test benches. Both on the right (square) and on the left (circular), the areas with the footprints of the connectors are observed. Each footprint is connected through a trace to another footprint in the central area, where the resistance measurement is taken.}
    \label{fig:FtF_EsquemaCircuitoTestBed}
\end{figure}

To experimentally validate the FtF solution, a test circuit was designed and manufactured along with the necessary components for its assembly. Figure~\ref{fig:FtF_DrawingsTestBed} shows all the pieces that make up the assembly: two flexible kapton-copper circuits (figure~\ref{fig:FtF_EsquemaCircuitoTestBed}) with a thickness of 342~$\mu$m that are installed facing each other, with two FtF connectors at both ends (one with a circular area and the other square), a footprint pattern of D~=~1.7~mm and P~=~4~mm, and a total of 134 footprints in each FtF connector; two 'base' pieces, on which the circuits rest; two 'ring' pieces that hold the circuits to the 'base' pieces; two expanded PTFE pieces that distribute the pressure over the circuits and absorb manufacturing tolerances; and two 'presser' pieces that apply pressure to the expanded PTFE.

\begin{figure}[!ht]
    \centering
    \includegraphics[width=\textwidth]{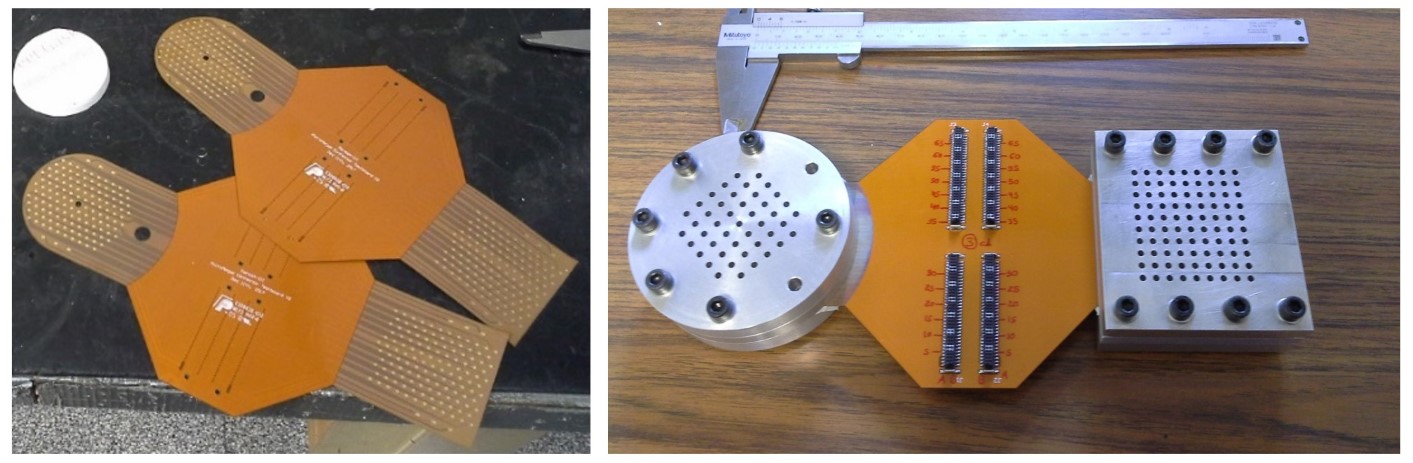}
    \caption{$\bullet$~Left: photograph of two circuits used in the tests. $\bullet$~Right: photograph of one of the test benches already assembled. In the central part, ERNI connectors were soldered to facilitate measurement with the RCL equipment through them.}
    \label{fig:FtF_TestBed}
\end{figure}

In the central part of the circuits, ERNI connectors are present (figures~\ref{fig:FtF_EsquemaCircuitoTestBed}~and~\ref{fig:FtF_TestBed}). Each channel of each ERNI connector corresponds to a footprint in the FtF connector, allowing the resistance between two facing channels to be measured using a Philips PM6303 RCL meter, which has a measurement sensitivity of 1~m$\Omega$. The measured resistance is the sum of the resistance of the two traces that run from the ERNI of each circuit to the footprint in the FtF plus the contact resistance between the two facing footprints. Assuming that the resistance of the traces does not change over time, as they are copper traces between two layers of kapton, a study was conducted on the variation of contact resistance in the FtF over time. A high variation in resistance or out of the meter's range could indicate that two facing footprints may have lost contact.

Three test benches were assembled with a total of 804 FtF connections: 134 footprints in each FtF connector, 2 FtF connectors (one circular and one square, to see if there was any difference between these two geometries) in each test bench. In two of the three test benches, the facing circuits were subjected to a pressure of 1~MPa, while in the third, the circuits were subjected to a pressure of 2~MPa. Additionally, to accelerate the relaxation and degradation processes of the materials, especially in the expanded PTFE, all test benches underwent 6 temperature cycles. The procedure for each of the 6 temperature cycles was as follows:

\begin{enumerate}
\item The 3 assemblies were placed in an oven at 60$^{o}$C (cycle~1), 65$^{o}$C (cycle~2), 73$^{o}$C (cycle~3), 78$^{o}$C (cycle~4), 88$^{o}$C (cycle~5), and 95$^{o}$C (cycle~6) for 6 hours in each cycle.
\item They were removed from the oven and left at room temperature for at least 12 hours.
\item They were placed in a refrigerator at 5$^{o}$C for 6 hours.
\item They were removed from the refrigerator and left at room temperature for at least 12 hours.
\item For each cycle, resistance measurements were taken for all channels.
\end{enumerate}

Using the Philips PM6303 RCL equipment, a series of measurements were conducted to measure the resistance of each connection at the start of the assembly and after each of the 6 temperature cycles. The first four temperature cycles and their corresponding measurements were conducted with a one-week interval between them, the fifth cycle was conducted three months after the fourth, and the sixth cycle was conducted seven months after the fifth.

\begin{figure}[!ht]
    \centering
    \includegraphics[width=0.9\textwidth]{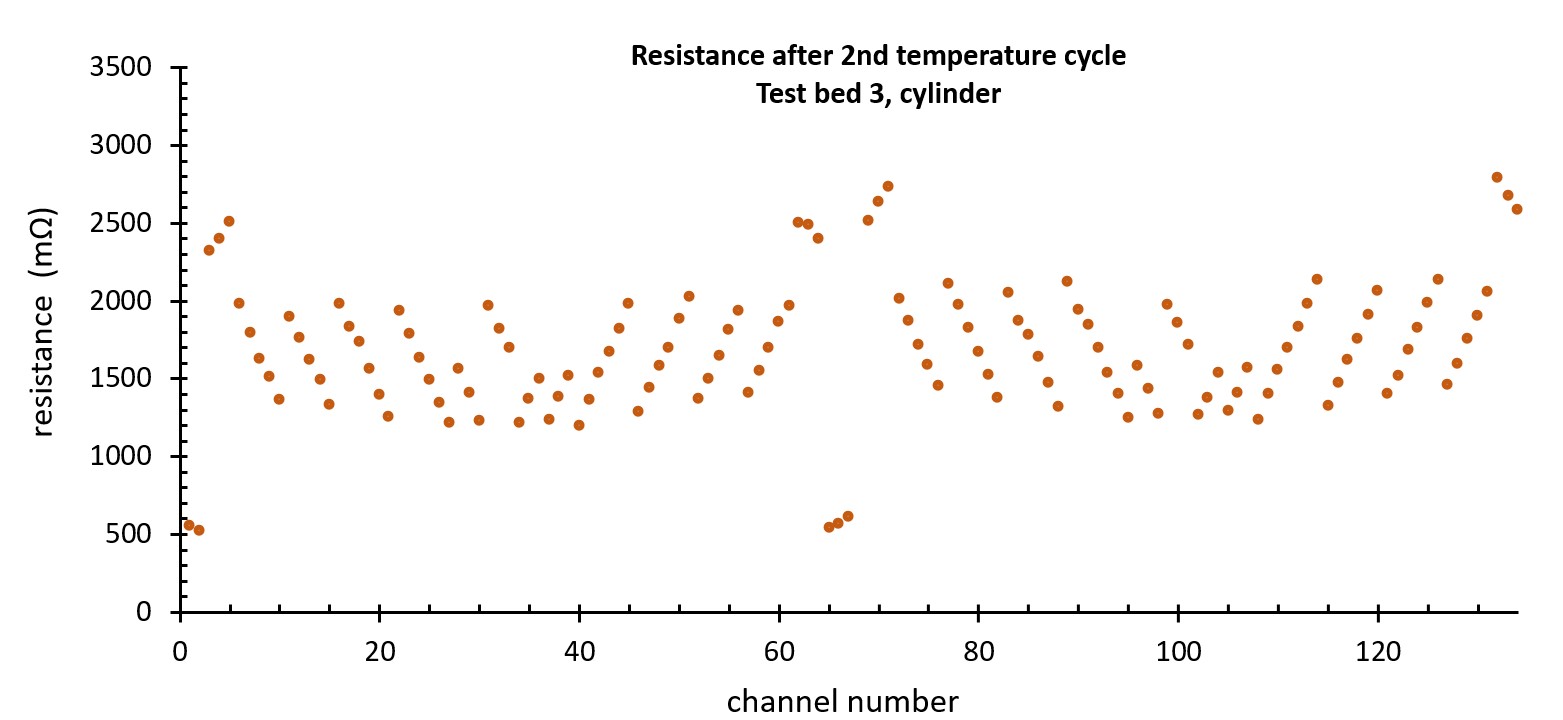}
    \caption{Measurements taken on the circular connector of test bench number~3, after the second temperature cycle at 65$^{o}$C and 5$^{o}$C.}
    \label{fig:FtF_ResistanceMeasurements}
\end{figure}

Figure~\ref{fig:FtF_ResistanceMeasurements} shows, as an example, the round of measurements taken on the circular connector of test bench number 3 after the second temperature cycle at 65$^{o}$C and 5$^{o}$C. The measurements for each channel (see the resistance measurement area in figure~\ref{fig:FtF_EsquemaCircuitoTestBed}) varied from 500~m$\Omega$ for the end channels, which had wider traces than the rest, to 2700~m$\Omega$ for the channels with narrower and longer traces. This effect of trace length on resistance value can be seen in the downward trend every 5 or 6 channels in figure~\ref{fig:FtF_ResistanceMeasurements}, corresponding to a row of footprints whose trace lengths to the measurement point decrease from greater to lesser. As can be seen in figure~\ref{fig:FtF_EsquemaCircuitoTestBed}, the footprints closest to the measurement area have shorter trace lengths, and vice versa.

\begin{figure}[!ht]
    \centering
    \includegraphics[width=0.95\textwidth]{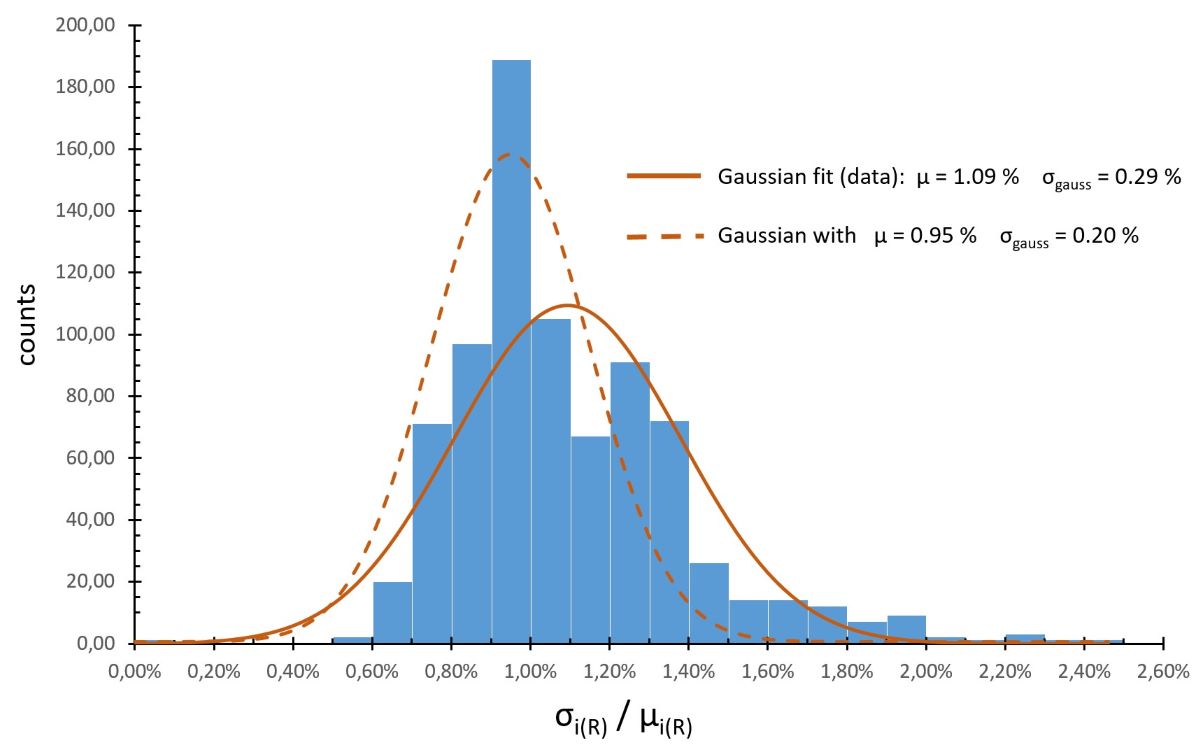}
    \caption{Histogram of the ratio between the standard deviation and the mean of the 7 measured resistance values for each of the 804 channels. Fit of the histogram to a normal distribution with $\mu$~=~1.09~$\%$ and $\sigma$~=~0.29~$\%$. Manual fit of the histogram to a normal distribution with $\mu$~= 0.95~$\%$ and $\sigma$~=~0.20~$\%$.}
    \label{fig:FtF_DesviacionMedidas}
\end{figure}

For each channel, corresponding to a contact between footprints of the circuit, 7 resistance values were taken, the first after the initial assembly of each test bench and the last after the sixth temperature cycle, 11 months after the first measurement. A total of 5628 resistance measurements were made for a total of 804 channels in the 3 assembled test benches. Figure~\ref{fig:FtF_DesviacionMedidas} shows the histogram of the ratio between the standard deviation and the mean of the 7 measured resistance values for each of the 804 channels.

\begin{equation}
\mu_{i_{(R)}} = \dfrac{1}{7}\sum_{i=1}^{7}R_{j}~~~~~~~~~~~~~~\sigma_{i_{(R)}} = \sqrt{\dfrac{1}{7}\sum_{i=1}^{7}(R_{j}-\mu_{i_{(R)}})^{2}}
\label{eq:GaussMedida}
\end{equation}

\begin{equation}
\mu = \dfrac{1}{804}\sum_{i=1}^{804}\sigma_{i_{(R)}}~~~~~~~~~~~~~~\sigma = \sqrt{\dfrac{1}{804}\sum_{i=1}^{804}(\sigma_{i_{(R)}}-\mu)^{2}}
\label{eq:GaussDistribucion}
\end{equation}

Equation \ref{eq:GaussMedida} represents the mean $\mu_{i_{(R)}}$ and the standard deviation $\sigma_{i_{(R)}}$ of the measurements of a channel over all cycles. Equation \ref{eq:GaussDistribucion} represents the mean $\mu$ and the standard deviation $\sigma$ of the standard deviations of the measurements of all channels over all cycles (fits to the normal distribution in figure \ref{fig:FtF_DesviacionMedidas}).

It would be expected that the distribution of the standard deviations of the measurements over time, assuming they are affected by a combination of random variables, would tend toward a normal or Gaussian distribution. The fit of the histogram to a Gaussian is also shown in figure~\ref{fig:FtF_DesviacionMedidas}. If the number of measurements per channel tended to infinity (7 measurements were taken per channel), the standard deviation of the measurement in all channels should tend toward a single value $\sigma_{i_{(R)}}=K$, and therefore the fit of the histogram would tend toward a distribution such that the mean would equal that single value $\mu=K$ and with a standard deviation equal to zero $\sigma=0$. The manual fit to a normal distribution centered on the value with the highest population of the histogram ($\mu$~=~0.95~$\%$), and a standard deviation adjusted to the population around that central value, is also presented in figure~\ref{fig:FtF_DesviacionMedidas} as an indication of the trend that one would expect in the data if the number of measurements per channel increased.

The objective of the measurements in the test benches was not to study the variations in the resistance of the channels, but rather the robustness of the connection, and the study is incomplete for such analysis. However, taking advantage of the accumulated data, a preliminary approximation can be made in the study of resistance variation. It can be said that the deviation of the data from the Gaussian fit around the central population may be due to statistical fluctuations resulting from the limited accumulated statistics, but they could also indicate that the measurements were affected by non-random variables.

After this study, and as an extension of it, ideas emerged to conduct a more in-depth study on the variation of resistance and capacitance, not only in the connector but throughout the connection chain to the electronics. The objective would be to understand how these variations could affect the signal of an event in the TPC recorded by the electronics.

The conclusion that can be drawn from these results, and the objective of this study, is that the FtF connector was robust and stable over time, as the 804 tested FtF contacts maintained their continuity, with none losing contact.

To conclude the findings of this study, which was a success and validated the operating principle of the FtF connector, it should be noted that there was no difference between the test benches whose circuits were subjected to different pressures (1 and 2~MPa). This allowed for the relaxation of the structural requirements of the mechanical parts of the connector, thereby reducing the amount of material used and its weight.

\section{The Second Iteration of the SR2M Module}
\label{sec:SR2Mv2_SJTU}

The manufacturing of the second version of the SR2M module, with some additional changes besides the introduction of the FtF connector, was carried out in 2019. The new modules were assembled in the laboratories of Shanghai Jiao Tong University (SJTU), in the 20 kg (600 liters) TPC prototype~\cite{Lin2018a} that would serve as a demonstrator. Ultimately, 52 more modules were manufactured, utilizing Micromegas technology developed directly at the University of Science and Technology in China~\cite{Feng2021}, to be installed in the final PandaX-III detector, which, after surface tests, will be installed in the Jinping Underground Laboratory~\cite{Zhang2023}.

\subsection{Modification of the ERS}
\label{sec:ModificacionERS}

To evaluate the impact of charge losses in defective channels and in the 'dead' zones between the modules, a study was conducted on the reconstruction of events from incomplete signals. These studies~\cite{Lobasenko2023}, based on the application of neural networks, indicated a high efficiency in the reconstruction of events. Furthermore, in preparation for the manufacturing of the new module, it was decided to simplify the design by eliminating the external \textit{rim}. 

The simulations carried out during the design of the first version of the SR2M demonstrated the efficiency of the external \textit{rim}, in combination with the internal \textit{rim}, forming the ERS system described in section~\ref{sec:ERS}. However, the system could not be experimentally tested due to a manufacturing error in the Micromegas, where the amplification holes in the mesh were not executed for those \textit{strips} located at the edge of the circuit (see figure~\ref{fig:SR2MdataTaking_SJTU_Hitmap}). The internal \textit{rim}, which was part of the Micromegas circuit (see figure~\ref{fig:SR2M_LayersActiveArea}), was retained in the second prototype, and its efficiency is shown in section~\ref{sec:PlanoSieteModulosv2}.

\subsection{Tensioning of the Micromegas Circuit}
\label{sec:TensingSolution}

In the first manufactured prototype of the SR2M module, the Micromegas circuit was joined to a copper support using an acrylic-based adhesive commonly used in the manufacture of flexible circuits. To achieve this adhesive bond, it was necessary to apply pressure to the surface of the Micromegas. This bonding process was carried out under specific measures following a specific protocol and did not pose any risk to the Micromegas. However, it was observed that the process did not have sufficient control to prevent some air bubbles from being trapped between the Micromegas circuit and the copper support. The trapped bubbles could not be detected visually, but during the operation of the TPC, when the chamber was evacuated, these bubbles caused localized deformation in some areas of the Micromegas. These deformations could lead to anomalous behavior in the area, and additionally, the trapped air could escape into the active volume of the TPC and contaminate the gas mixture.

\begin{figure}[!ht]
    \centering
    \includegraphics[width=\textwidth]{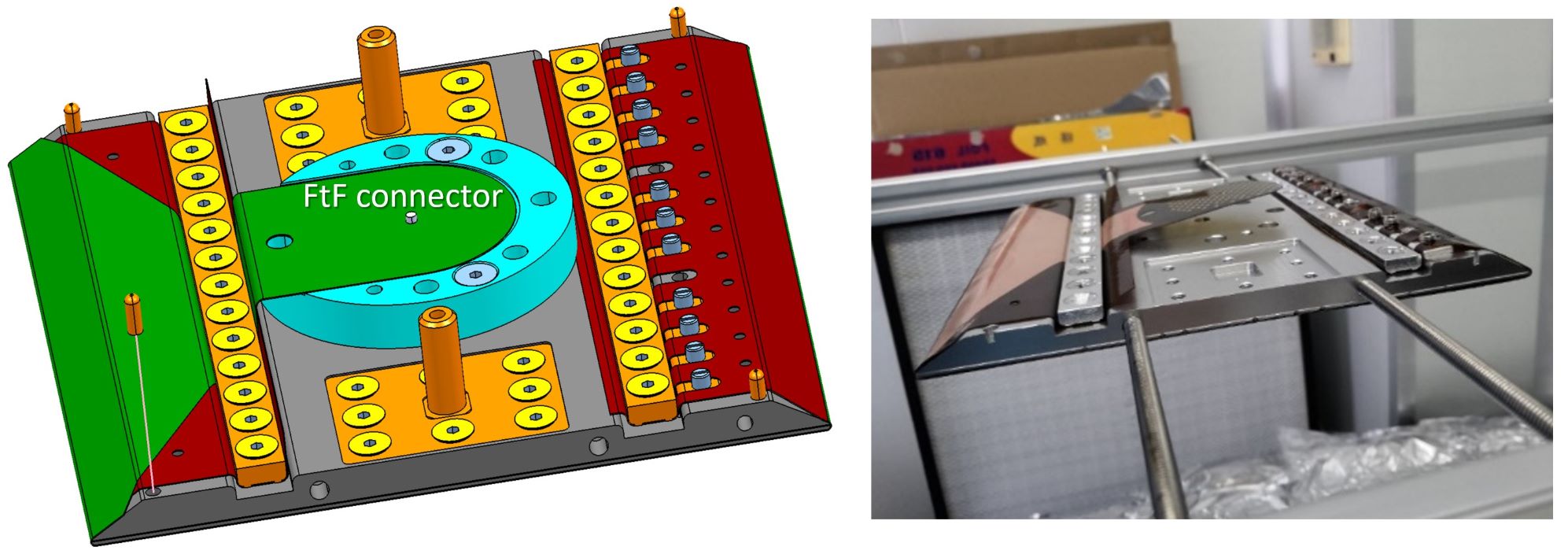}
    \caption{$\bullet$ Left: 3D view of SR2M module version 2 with the Micromegas circuit being tensioned onto the support. The back of the module, the active area is on the other side. The FtF connector is located in the center. $\bullet$~Right: Photograph of the SR2M module during assembly with the circuit being tensioned.}
    \label{fig:SR2M_v2_tensing}
\end{figure}

\begin{figure}[!ht]
    \centering
    \includegraphics[width=\textwidth]{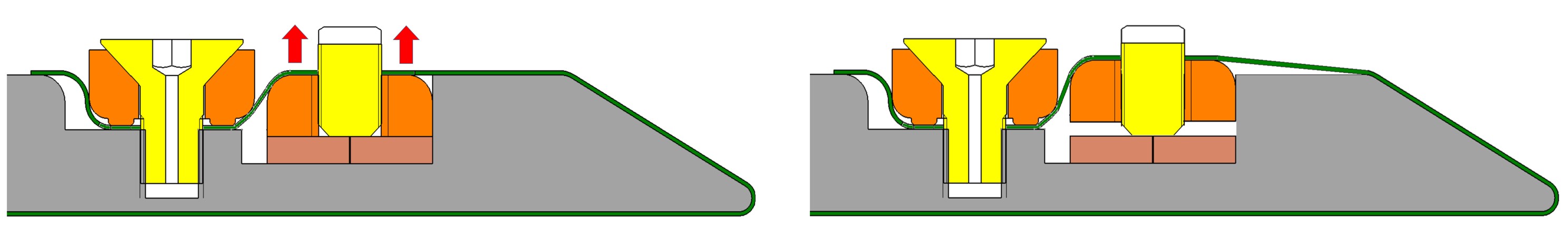}
    \caption{Section of the SR2M module showing the system used to perform circuit tension on the support.}
    \label{fig:SR2M_v2_tensingSection}
\end{figure}

To avoid this effect, and given that it was difficult to prevent bubble formation during the manual bonding process, the second prototype was designed and manufactured so that the Micromegas circuit could be tensioned over the support. Figure~\ref{fig:SR2M_v2_tensing} shows the second version of the SR2M module, in which the circuit is tensioned and the FtF connector is located in the center. Figure~\ref{fig:SR2M_v2_tensingSection} shows the system used so that, once the Micromegas circuit is positioned on the support, it can be tensioned.

\subsection{Readout Plane with 7 SR2M Modules}
\label{sec:PlanoSieteModulosv2}

\begin{figure}[!ht]
    \centering
    \includegraphics[width=1.0\textwidth]{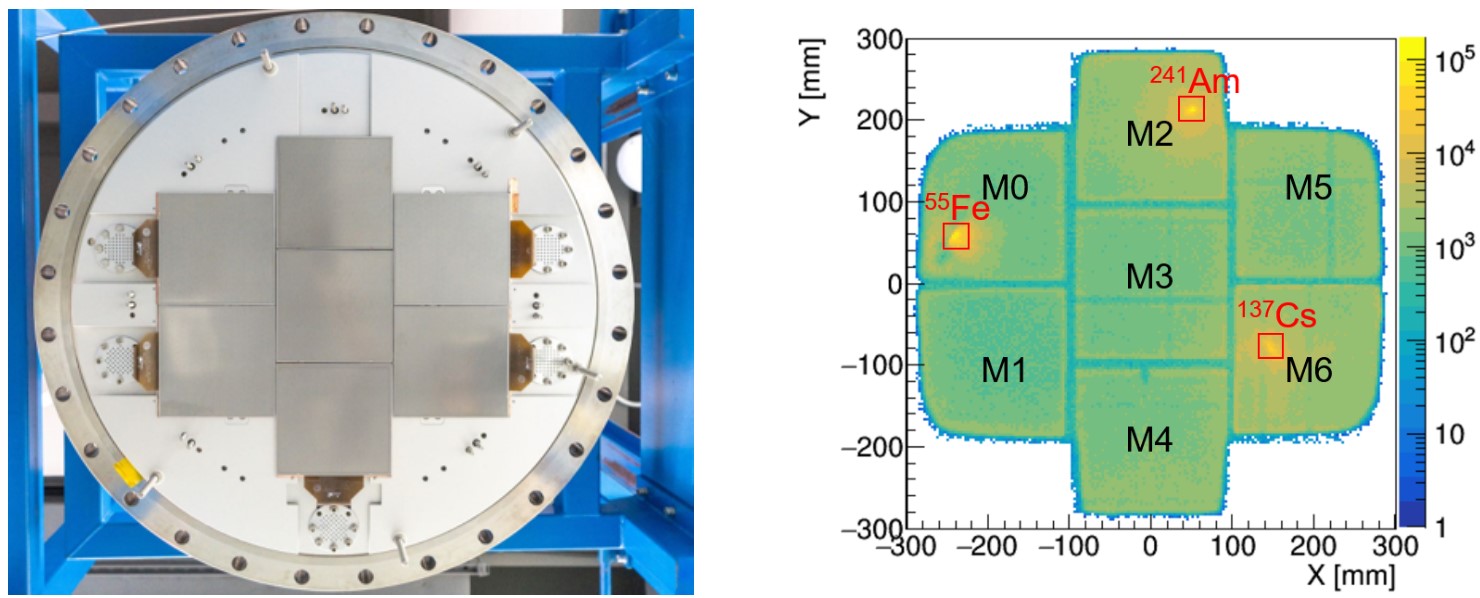}
    \caption{\cite{Zhang2023}~$\bullet$ Left: photograph of the readout plane formed by 7 modules of version~2 of the SR2M module. The connection inside the chamber and the use of the FtF connector to carry out the connection can be observed. $\bullet$~Right:~activity map of the readout plane formed by version~2 of the SR2M module with 1.1\% of defective channels.}
    \label{fig:SR2Mv2_SJTU_ReadoutPlane_Hitmap}
\end{figure}

Figure~\ref{fig:SR2Mv2_SJTU_ReadoutPlane_Hitmap}~on the left shows the new readout plane composed of version~2 of the SR2M module, with the connection inside the module's chamber to an extension cable that carries the signal from the interior to the exterior of the chamber, using the new FtF connector developed in this chapter. As can be seen in the image, the FtF connectors were mounted on one side of the module, rather than in the central and lower part. This was done to simplify the assembly in the demonstrator prototype. In the final readout plane, mounted in a TPC to contain 140~kg of Xenon enriched to 10~bar~\cite{Zhang2023}, the FtF connection is made at the back of the SR2M module, as designed.

Figure~\ref{fig:SR2Mv2_SJTU_ReadoutPlane_Hitmap}~on the right shows the activity map of this readout plane. Regardless of the evaluation of the response of the readout plane, formed by the second version of the SR2M module, to signals from ionizing events in the TPC, which is outside the scope of this thesis, there is a lower number of defective channels (1.1\%) compared to the previous readout plane (9.2\%) shown in figure~\ref{fig:SR2MdataTaking_SJTU_Hitmap}. This improvement cannot be attributed solely to the robustness of the FtF connector, as there were other factors that may have influenced it, such as the manufacturing process of the Micromegas or the increased knowledge in operating this type of detector, reducing damage to the channels. Considering these other factors, it can be said that much of the improvement was due to the robustness of the FtF connector, as in the previous readout plane (figure~\ref{fig:SR2MdataTaking_SJTU_Hitmap}) between 2\% and 4\% of the total channels were lost due to connection failures, whereas with the use of the FtF connector this loss due to connection failures has been completely eliminated.

A reduction in activity at the junction edges between the modules can also be observed. After eliminating the external rim in the second version of the SR2M module, the internal rim did not seem sufficient for the charge losses in the dead zones of the readout plane to be negligible. It should be noted that figure~\ref{fig:SR2Mv2_SJTU_ReadoutPlane_Hitmap}~on the right has been generated from the standard analysis (figure~\ref{fig:SR2MdataTaking_RESTflow}) performed with the REST-for-Physics software, without applying any event reconstruction methods from incomplete data, which could improve the reconstruction by up to 8\%~\cite{Lobasenko2023}.

\begin{figure}[!ht]
    \centering
    \includegraphics[width=0.65\textwidth]{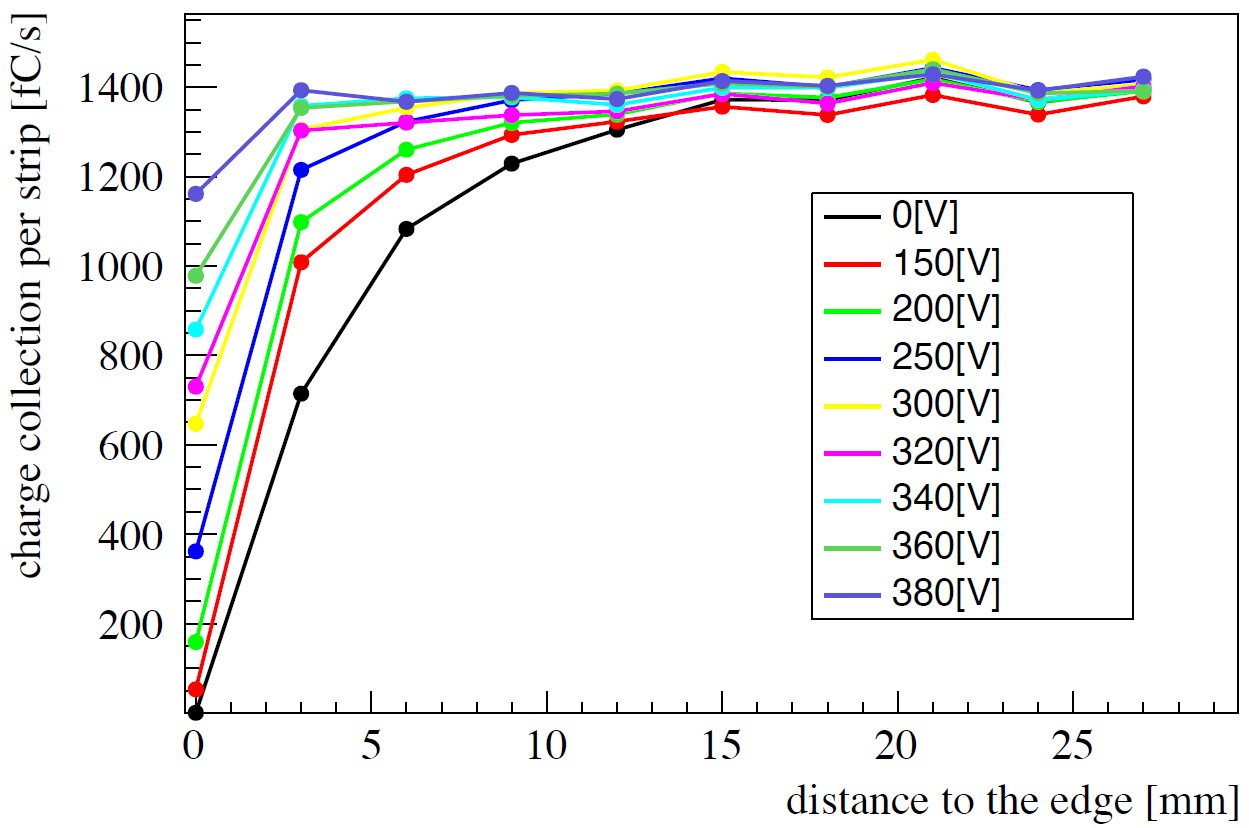}
    \caption{Charge collection at the edges of the SR2M-v2 module for different voltages applied to the internal rim, with a drift field of 100~V~cm$^{-1}$~bar$^{-1}$ and a voltage on the mesh of 300~V.}
    \label{fig:SR2MdataTaking_ChargeCollectionVsDistanceToEdge}
\end{figure}

Figure~\ref{fig:SR2MdataTaking_ChargeCollectionVsDistanceToEdge} shows the effectiveness of the internal rim. When the internal rim is disconnected, the charge loss in the last 10~mm of the module is significant, with a loss distribution ranging from 100\%~(distance to axis~=~0~mm) to 0\%~(distance to axis~=~15~mm). With an internal rim voltage of 380~V, 80~V above the mesh voltage (300~V), a small charge loss of about 15\% occurs in the last 2~mm.

It should be noted that the internal rim can be set to higher voltages, as the circuit design allows, so the small loss of 15\% observed in the last 2~mm could be reduced or even eliminated. This small loss was preferred to be accepted in order not to deform the electric field too much in that area, favoring a good spatial reconstruction of the event at the edge. Applying higher voltages to the rim would cause the field lines, which come perpendicular to the module (see figure~\ref{fig:SR2M_ElectricFieldUnion}), to deform inward, effectively recovering the charges in the edge area but, on the other hand, distorting the trace of the reconstructed event.

It seems that, after verifying the high effectiveness of the internal rim in the edge area of the module, the reduction in activity observed in figure~\ref{fig:SR2Mv2_SJTU_ReadoutPlane_Hitmap}~on the right is mainly due to the separation between the Micromegas circuits of adjacent modules. The tolerance in the positioning of the Micromegas with respect to the flat support to form the module, along with the adjustment tolerance when conforming the readout plane with all the modules, works against the collection of electrons, as was already anticipated in section~\ref{sec:ERS}, and in more detail in figure~\ref{fig:SR2M_ElectricFieldUnion}.

\subsection{Readout Plane with 52 Modules}

After validating the design of the SR2M-v2 module, the production of 52 modules was carried out using Micromegas technology developed directly at the University of Science and Technology in China~\cite{Feng2021}, in the particle detector and electronics laboratory. The development and testing of these modules are outside the scope of this thesis.

\begin{figure}[!ht]
    \centering
    \includegraphics[width=0.65\textwidth]{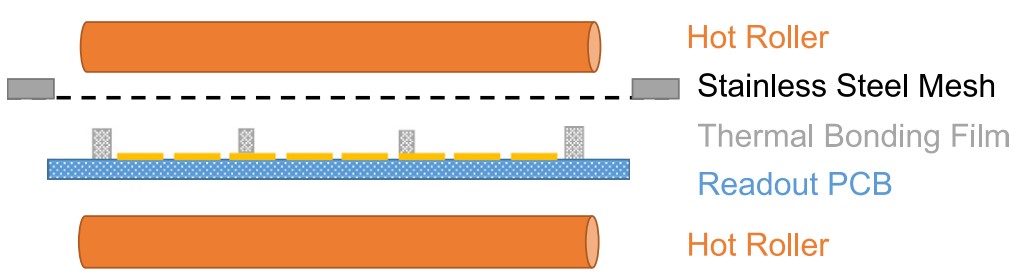}
    \caption{~\cite{Feng2021} Schematic of the thermal bonding process of the mesh with the readout circuit.}
    \label{fig:SR2Mv2_EsquemaThermalBonding}
\end{figure}

For over ten years~\cite{Guan2011, Zhang2014}, a new method based on a thermal bonding process has been developed for the fabrication of Micromegas detectors~\cite{Guan2011, Zhang2014}. A perforated mesh, which functions as a cathode for generating the avalanche effect, is thermally bonded to a printed circuit that contains the electron collection pixels/strips. The separation between the mesh and the pixels, the space where electron amplification occurs due to the avalanche effect, is achieved using pillars formed from the material itself with thermal adhesive (see figure~\ref{fig:SR2Mv2_EsquemaThermalBonding}). This method is similar to that used for manufacturing bulk-type Micromegas~\cite{Giomataris2006}, but the generation of pillars is not done through lithography, thus avoiding any use of chemical agents in the process. The thermal bonding films can have different thicknesses, allowing for the fabrication of detectors with an avalanche space ranging from 30~$\mu$m to several hundreds of~$\mu$m.

These detectors have achieved gains and energy resolutions of up to 10$^5$ and 16\%~(FWHM), respectively, in laboratory-optimized cases with Ar+7\%CO$_2$ mixtures and for X-rays of 5.9~keV from a $^{55}$Fe source~\cite{Feng2021}. Additionally, a new type of high-gain detector has been developed by installing a second mesh above the first, allowing for the avalanche effect to occur in two stages~\cite{Zhang2020, Zhang2018}. The high gain observed ($>$10$^6$), or even greater if the design is optimized, significantly reduces the energy thresholds of this type of detector.

Work is also underway on the development of resistive Micromegas, in which a thin layer of Germanium is deposited on the anode~\cite{Feng2022}. This layer protects the Micromegas in the event of discharges/sparks occurring in the amplification zone between the cathode and the anode, thus making them more robust. With these, high gains ($>$10$^5$) have been achieved, with a low spark rate and a high counting capacity ($>$100~kHz~cm$^{-2}$).

\begin{figure}[!ht]
    \centering
    \includegraphics[width=0.95\textwidth]{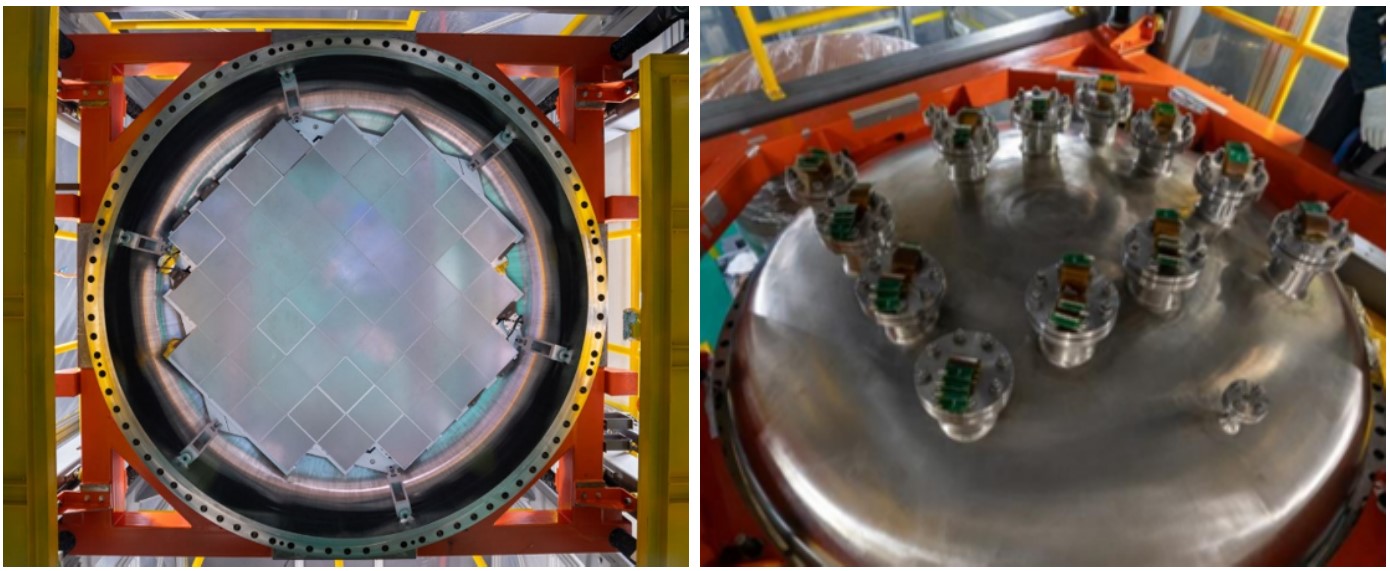}
    \caption{~\cite{Zhang2023} Overview images of the 140~kg PandaX-III detector. $\bullet$~Left:~readout plane assembled with 52 SR2M modules. $\bullet$~Right:~The upper cover has 12 DN-80 flanges that hold 52 extension cables.}
    \label{fig:SR2Mv2_SJTU_1000kgReadoutPlane_paper}
\end{figure}

These developments have been implemented in the new SR2M-v2 module, which was manufactured and installed in the PandaX-III chamber~\cite{Zhang2023} that will contain 140~kg of Xenon enriched with $^{136}$Xe, forming a readout plane of nearly 2 meters in diameter through the tiling of 52 SR2M-v2 modules (figure~\ref{fig:SR2Mv2_SJTU_1000kgReadoutPlane_paper}~left). Twelve DN-80 ports (figure~\ref{fig:SR2Mv2_SJTU_1000kgReadoutPlane_paper}~right) have been used to extract signals from the modules and to supply high voltage to them.

As explained, the PandaX-III project team has successfully constructed the TPC prototype with 7 SR2M modules to demonstrate the viability of the TPC system with Micromegas, achieving good results in terms of electron transmission, gain, and energy resolution with various radioactive sources ($^{241}$Am, $^{137}$Cs, and $^{55}$Fe). The construction of the full-size detector has already been completed, along with 52 SR2M-v2 modules based on Micromegas developed in China, and once surface tests are concluded, it will be installed in the Jinping Underground Laboratory~\cite{Zhang2023}.

\part{Optimization of the readout plane and subsystems in the TREX-DM experiment}

\setlength{\baselineskip}{12pt} 
\selectlanguage{USenglish}
\graphicspath{{../Imagenes/}{../../Imagenes/}} 

\renewcommand{\listtablename}{Table of contents}
\renewcommand{\tablename}{Table} 

\chapter{The search for the WIMP}
\label{ch:IntroTREXDM}

\begin{figure}[!ht]
    \centering
    \includegraphics[width=0.8\textwidth]{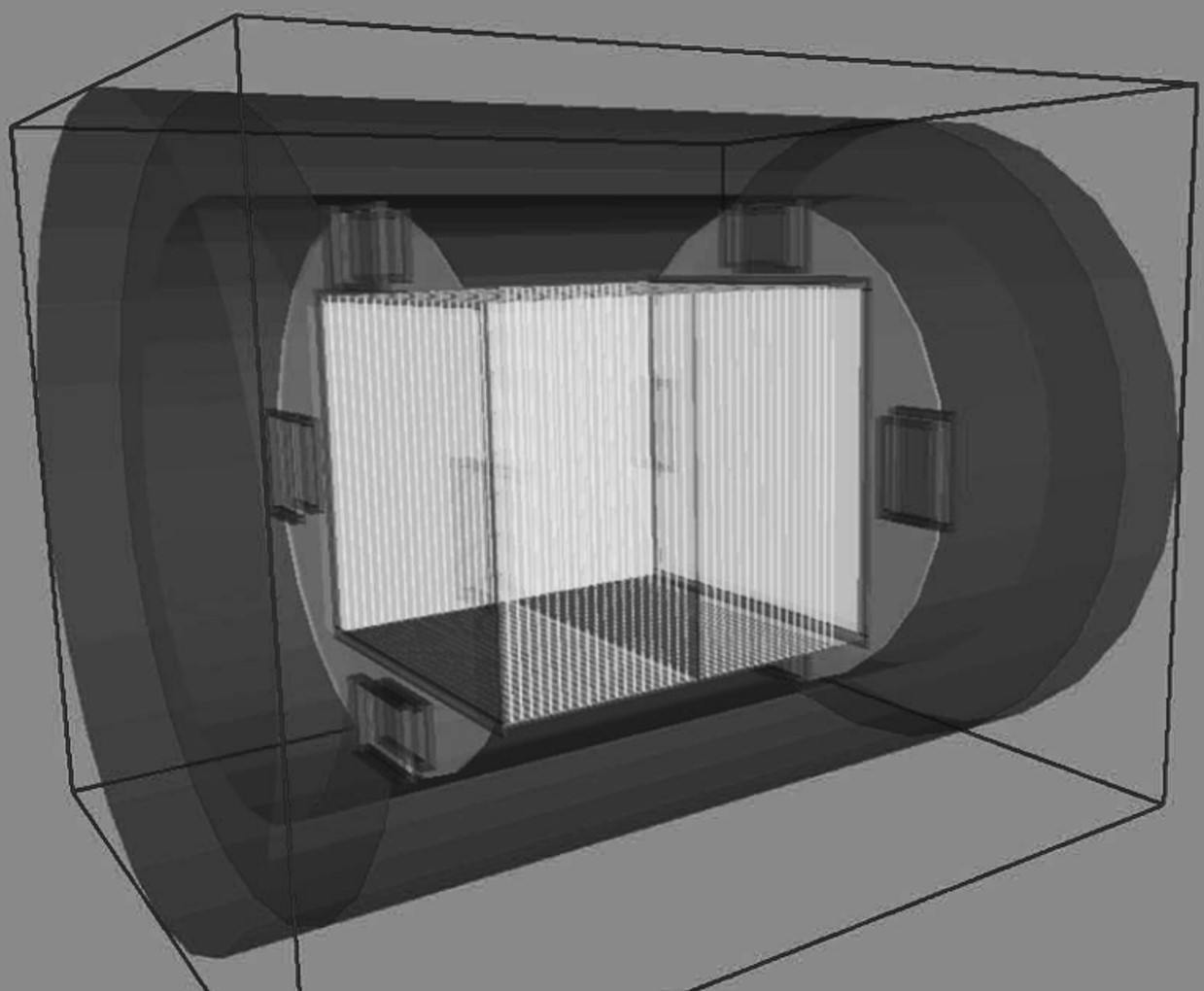}
    \caption*{}
    \label{fig:IntroTrexDM_GeometryForSimul}
\end{figure}

\newpage

The \mbox{TREX-DM} experiment (TPC for Rare Event eXperiments - Dark Matter) is a low-background TPC with readout planes formed by microbulk-type Micromegas detectors, the largest (25~x~25~cm$^2$) and the most radiopure ever constructed. The experiment has been developed for the search for WIMPs with a mass on the order of 10~GeV~c$^{-2}$, or even lower. It was installed and commissioned at the Canfranc Underground Laboratory (LSC) during the second half of 2018, and since then, up to the writing of this thesis, the experiment has undergone continuous improvement in operation and a significant reduction in both the background level and the initial energy threshold.


\section{The WIMP}

The observation that some astronomical objects move faster than expected, when compared to the value calculated by taking into account the gravitational attraction of visible objects, is one of the most convincing pieces of evidence for the existence of dark matter. On a galactic scale, the measurement of the velocity of stars and gas as a function of distance from the galactic center (the so-called galactic rotation curves) is the most prominent observation. The first hint of the existence of dark matter was found by F. Zwicky in 1933~\cite{Zwicky1933}, based on the measurement of the velocity of galaxies in the Coma cluster.

Another observational piece of evidence for the existence of dark matter is the observation of the dynamics of the system known as the Bullet Cluster, a cluster of galaxies passing through another. X-ray observations show that most of its baryonic mass decelerates as a result of the electromagnetic interaction of gas particles. However, gravitational lensing observations show that most of the total mass of the clusters does not slow down and passes through one another without any significant change. This can be explained if it is assumed that there is dark matter in the system that does not undergo electromagnetic interactions, nor self-interactions among the dark matter itself, beyond gravitational ones.

There is a third very convincing and independent piece of evidence from the previous ones, thanks to the measurements of anisotropy in the cosmic microwave background conducted by the Planck mission~\cite{Planck2015}, which indicate that the Universe is composed of approximately 4.2\% baryonic matter, 27.2\% dark matter, and 68.6\% dark energy.

Among the most attractive candidates for constituting dark matter~\cite{Bertone2005} is the WIMP (Weakly Interacting Massive Particle), a massive particle that interacts weakly. This particle appears in supersymmetry models and is sought in various experiments installed in underground laboratories. Its direct detection is based on calculating the rate of interactions expected in a given detector, so the comparison of results among all detectors requires the adoption of general hypotheses. The uncertainties of these hypotheses determine the precision of the expected interaction rate and, therefore, also determine the precision of the limits for the WIMP parameters.

The direct detection interaction rate of the WIMP depends on astrophysical, particle physics, and nuclear physics parameters. From the astrophysical point of view, the parameters to consider are the local density of the WIMP and the velocity distribution in the frame of the detector. From nuclear and particle physics, the parameters are the nuclear form factors and the interaction cross-sections, which depend on the theoretical framework in which the WIMP candidate arises. Additionally, experimental properties such as the \textit{quenching factor} and the detector's own characteristics, such as resolution or energy threshold, must be taken into account.

The extremely low velocity and energy of nuclear recoils induced by WIMPs pose formidable challenges in terms of background levels, threshold, and mass as a target for the WIMP. Over the past few years, detection techniques have been developed that have achieved increasingly larger target masses for the WIMP and lower background levels, thereby increasing sensitivity to WIMPs.

The effective cross-section $\sigma{_0}$ of the WIMP-nucleus interaction for zero momentum transfer is defined as

\begin{equation}
\sigma{_0}=\left( \dfrac{1+m_{\chi}/m_{n}}{1+m_{\chi}/m_{N}}\right) ^{2}A^{2}\sigma^{0}_{\chi^{n}}~~,
\label{eq:ElasticCrossSection}
\end{equation}

\noindent where $A$ is the mass number of the material used as a target for the WIMP, $\sigma^{0}_{\chi^{n}}$ is the WIMP-nucleon cross-section, $m_{\chi}$ is the mass of the WIMP, $m_{N}$ is the mass of the nucleus of the element used as a target for the WIMP, and $m_{n}$ is the mass of the nucleon, assuming the simplification that the neutron and proton have the same mass.

The leading experiments are those that use relatively heavy nuclei (Xenon or Germanium) as targets for the WIMP to exploit the $A^{2}$ term in equation~\ref{eq:ElasticCrossSection}.

\begin{figure}[!ht]
    \centering
    \includegraphics[width=0.95\textwidth]{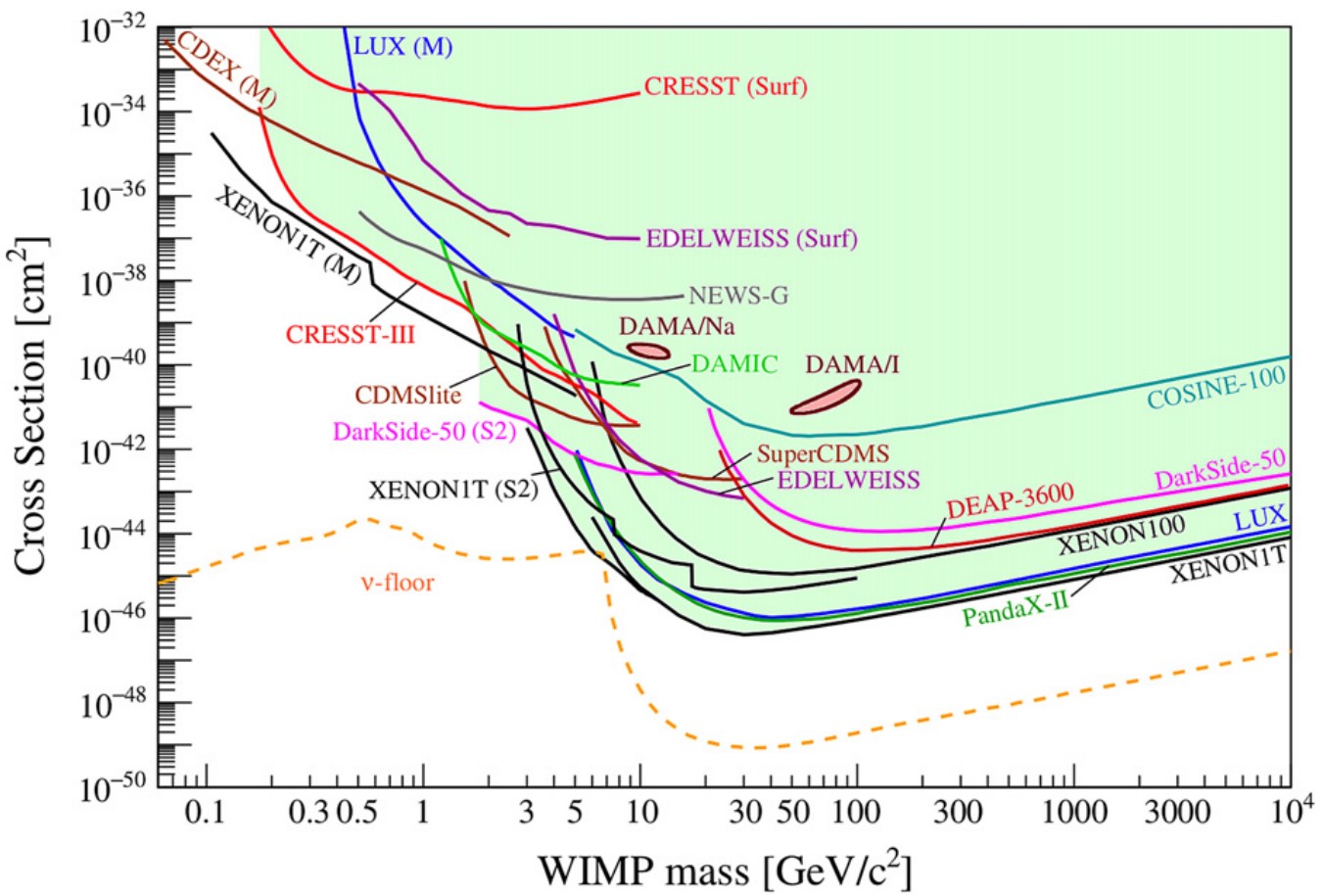}
    \caption{\cite{Billard2022} WIMP-nucleon cross-section versus WIMP mass, with the current limits from the most important experiments. Standard parameters for an isothermal WIMP halo have been assumed: $\rho_0$~=~0.3~GeV~cm$^{-3}$, v$_0$~=~220~km~s$^{-1}$, v$_{esc}$~=~544~km~s$^{-1}$. The results labeled 'M' have been obtained assuming the Migdal effect. The results labeled 'Surf' come from surface experiments. The lower limit imposed by neutrinos ($\nu$-\textit{floor}) for a Germanium target is a limit defined as the cross-section $\sigma_d$ with which a given experiment has a 90\% probability of detecting a WIMP, with a scattering cross-section $\sigma>\sigma_d$ greater than 3~sigma.}
    \label{fig:IntroTrexDM_ExclusionPlotPresente}
\end{figure}

Conventionally, WIMP searches are represented in the two-dimensional effective parameter space ($\sigma_{0}$, $m_{\chi}$), as shown in figure~\ref{fig:IntroTrexDM_ExclusionPlotPresente}. This representation is usually accompanied by a series of additional simplifying assumptions, for example, that the velocity distribution of WIMPs in the galactic halo follows a Maxwell-Boltzmann distribution, or that WIMPs interact exclusively (or primarily) with nuclei through coherent elastic scattering independent of spin. Although this conventional scenario is attractive for establishing a common ground for comparing the sensitivity of experiments, these assumptions must be taken into account. It should also be noted that there will come a time when detectors will be so sensitive that they will begin to detect solar neutrinos and the background caused by atmospheric neutrinos, through elastic neutrino-electron ($\nu$–e) and neutrino-nucleus (CE$\nu$NS for \textit{Coherent elastic neutrino-nucleus scattering}) scattering, generating electronic and nuclear recoils, respectively. This barrier is known as the \textit{neutrino floor} or \textit{neutrino fog}~\cite{Carew2024} and is also represented in the sensitivity diagram of figure~\ref{fig:IntroTrexDM_ExclusionPlotPresente}.

Most of the most relevant experiments are summarized, with basic information, in table~\ref{tab:IntroTrexDM_ListaExperimentos}. The low mass region (from $\sim$100~MeV~c$^{-2}$ to $\sim$5~MeV~c$^{-2}$) can be more easily explored by cryogenic bolometers (CRESST, SuperCDMS, EDELWEISS), with extremely low energy thresholds. Also, DAMIC-M, based on CCD, and the gas TPC of TREX-DM may become sensitive in this mass range. Exploring the medium-high mass range requires very large exposures (DarkSide-20k, ARGO, PandaX-4T, XENONnT, LZ, DARWIN).

\begin{table}[!ht]
\centering
\begin{tabular}{l}
\includegraphics[width=0.99\textwidth]{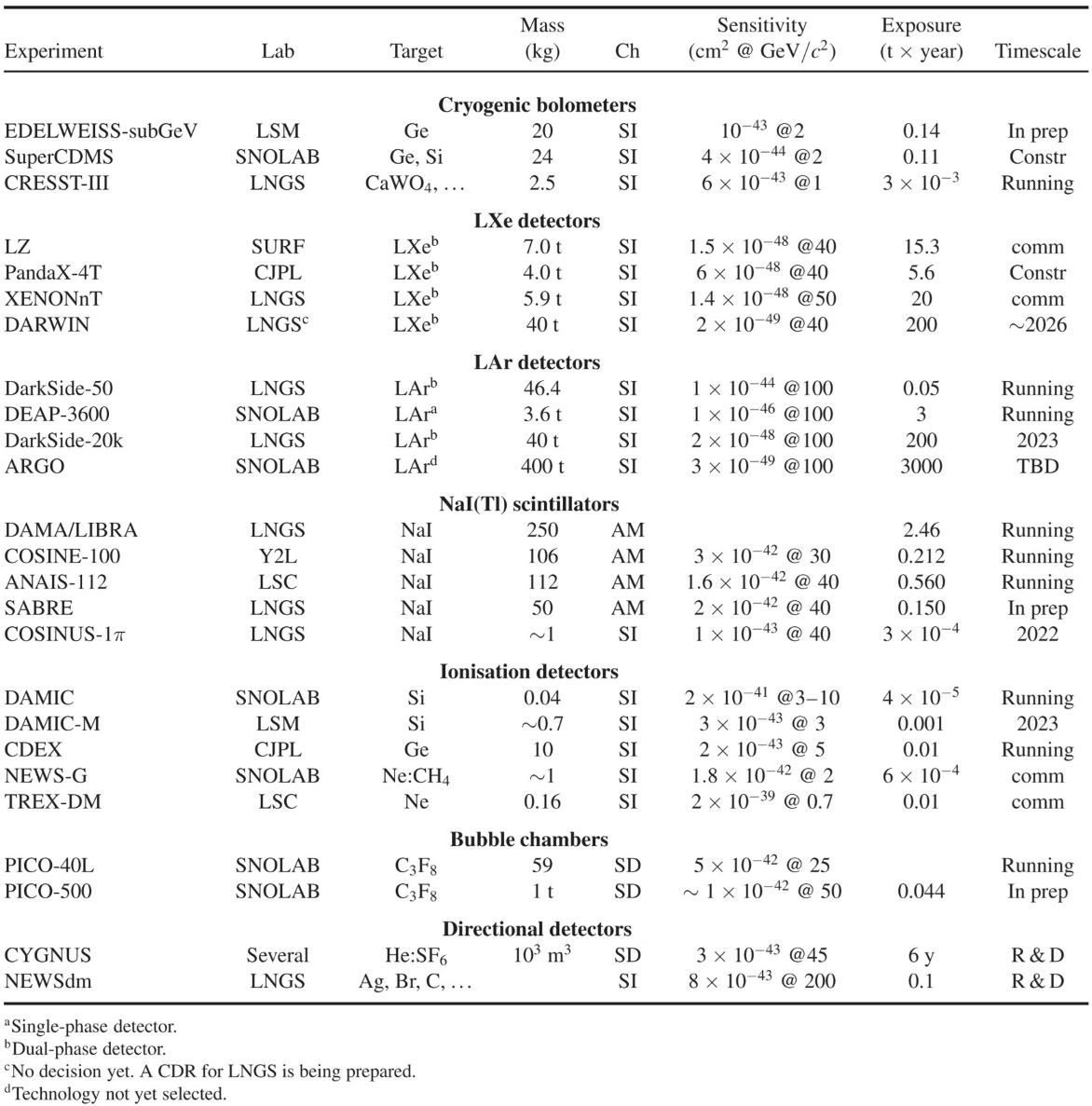}\\
\end{tabular}
\caption{\cite{Billard2022} Present and projected experiments for the direct detection of the WIMP. The main detection channel (Ch) of the experiments is abbreviated as: SI (spin-independent WIMP-nucleon interactions), SD (spin-dependent), AM (annual modulation). Sensitivity is given for this channel with the exposure time reflected in the table. It should be noted that many projects have multiple detection channels. 'comm.' means experiment in commissioning phase.}
\label{tab:IntroTrexDM_ListaExperimentos}
\end{table}

Leading experiments also utilize detection techniques that have the capability of nuclear recoil discrimination. This is the case, for example, of dual-phase liquid Xenon detectors (LUX-ZEPLIN~\cite{Akerib2020}, XENONnT~\cite{Aprile2023}), Germanium (SuperCDMS~\cite{Agnese2014}), or, more recently, through the use of liquid Argon (DarkSide~\cite{Agnes2023}). These experiments currently operate targeting WIMP masses that are already on the order of tons, with background levels of a few dru~(counts~keV$^{-1}$~kg$^{-1}$~day$^{-1}$) in the region of interest. Such impressive figures are achieved thanks to the availability of discrimination techniques that allow for distinguishing, with some efficiency, electron recoils (produced, for example, by gammas) from nuclear recoils.

In general, if a discovery of dark matter were to occur in any experiment, confirmation would be required using another target mass and possibly another technology, both to rule out misidentified systematic errors and to begin testing the most relevant properties of the discovered particle.

\section[Motivations for the search for low-mass WIMPs]{Motivations for the search for low-mass WIMPs \sectionmark{Motivations for the search for low-mass WIMPs...}}
\sectionmark{Motivations for the search for low-mass WIMPs...}

As previously explained, the detection of dark matter is one of the major challenges in the field of particle and astroparticle physics for the coming years. The evidence for the existence of dark matter is well-founded in various observations, such as the anisotropies in the cosmic microwave background, the distribution of matter in our galaxy, or its gravitational effect on visible matter. The nature of dark matter is still unknown, but it could be composed of new particles with masses and cross-sections characteristic of the electroweak scale.

Despite the enormous progress in WIMP search experiments in recent years, no convincing positive signal of WIMPs has been observed so far. Supersymmetry has also not been observed at the LHC (Large Hadron Collider) at the energies at which this collider has been operating, affecting the search for WIMPs, as there is a theoretical framework that identifies it as the neutralino of some supersymmetric extensions of the Standard Model. These facts have led to a reevaluation of general hypotheses, such as the different velocity distributions of WIMPs, and to the exploration of other theoretical and phenomenological frameworks, including the study of less conventional supersymmetry models or non-supersymmetric WIMP models.

Most of the effort made so far has focused on the search for WIMPs of relatively high masses (~$\gtrsim$~50~GeV~c$^{-2}$). This is partly because the search has been guided by theoretical considerations within the supersymmetric models established initially, along with the fact that the WIMP detection techniques available at that time were well-suited to this mass range. For lower masses, sensitivity decreases drastically due to the energy threshold effect.

As part of the strategy to go beyond conventional models, some recent experimental and phenomenological efforts have focused on studying WIMPs in the low-mass range ($<$~10~GeV~c$^{-2}$). Interest in this region of parameter space, traditionally beyond the reach of conventional experiments, increased with the emergence of a series of hints that could be interpreted as low-mass WIMPs, although these interpretations seem to have weakened over time. The well-known and controversial signal from DAMA/LIBRA~\cite{Bernabei2014} could only be understood with models outside the Standard Model, some of which include low-mass WIMPs. However, the DAMA/LIBRA signal has been excluded by many other experiments, including those that have used the same technology, such as the recent ANAIS experiment~\cite{Amare2021}. Regardless of the weight given to these suggestions or the theoretical motivation for low-mass WIMPs, it is clear that in the current situation, it is important to extend sensitivity to lower values of $\sigma_{0}$, but also to lower values of $m_{\chi}$.

In more experimental terms, sensitivity to low-mass WIMPs poses particular challenges. Conventional experiments are heavily limited at low masses due to the threshold effect and the limitations in nuclear recoil discrimination capabilities. Low-mass WIMPs leave energy deposits typically below the energy threshold of experiments based on heavy nuclei as targets for WIMPs, such as Xenon or Germanium. This means that the exclusion limits for low-mass WIMPs from these experiments are based on a very small fraction (1\% or less) of the WIMP interactions in the detector, those corresponding to the high-velocity tail of the distribution, with sufficient kinetic energies to produce a visible nuclear recoil in the detector. However, this part of the distribution is precisely the most uncertain, and in some galactic halo models, it may disappear completely, that is, those with a lower maximum WIMP velocity.

It is evident that to address the low-mass WIMP region, specific experiments optimized for this mass range are needed. A solid detection or exclusion must imply that a substantial fraction of the WIMP spectrum is above the experimental threshold. To achieve this, it is preferable to use light nuclei as targets for WIMPs, so that higher recoil energies can be kinematically reached, as well as techniques with a low energy detection threshold. These requirements are incompatible with the discrimination between nuclear and electron recoils, whose observable characteristics become blurred at low energies. Nevertheless, despite the described difficulties, experimental efforts are being made in this direction, such as DarkSide~\cite{Agnes2023}, XENON100~\cite{Aprile2016}, CDMSlite~\cite{Agnese2018}, DAMIC~\cite{Aguilar-Arevalo2016}, or CREST-III~\cite{Mancuso2020}, among others.

\section{The TREX-DM Experiment}
\label{sec:IntroTREXDM}

The \mbox{TREX-DM} experiment is a low-background TPC with reading planes formed by microbulk Micromegas detectors and has been developed for the search for WIMPs with a mass on the order of 10~GeV~c$^{-2}$, or even lower. It was installed and commissioned at the Canfranc Underground Laboratory (LSC) during the second half of 2018.

\begin{figure}[!ht]
    \centering
    \includegraphics[width=1.0\textwidth]{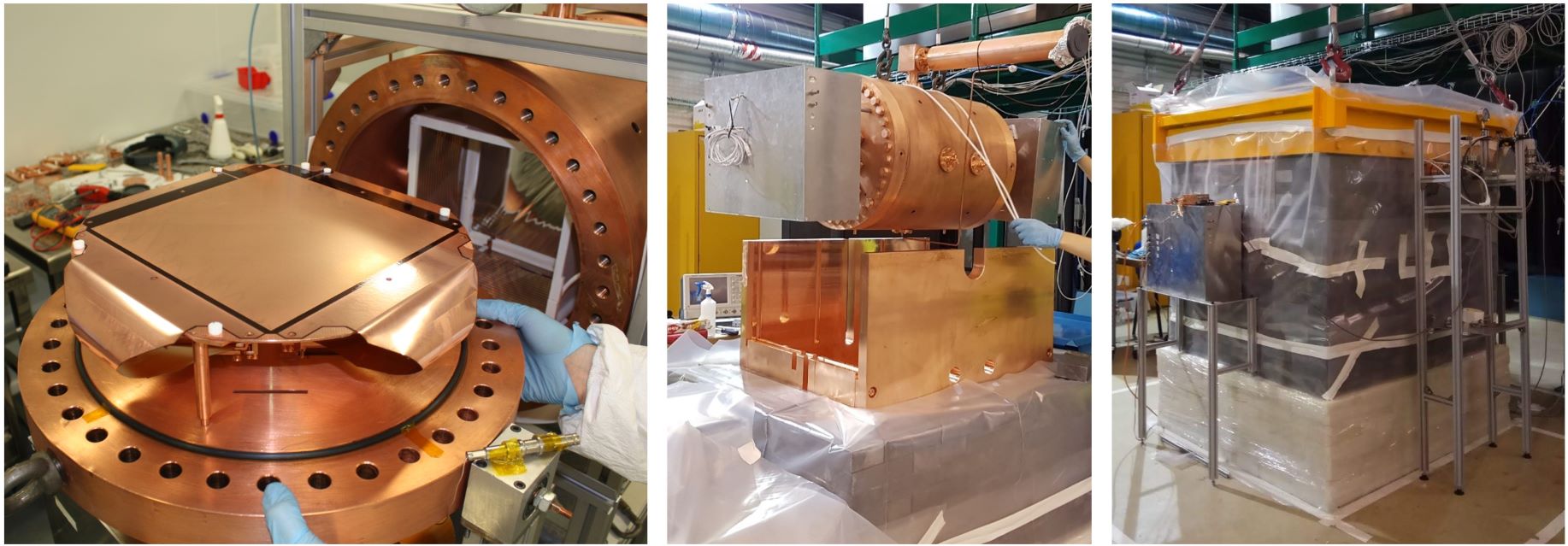}
    \caption{Photographs of the TREX-DM experiment. $\bullet$ Left: Microbulk Micromegas installed on one of the chamber lids. The detector has 2 reading planes, one on each lid. $\bullet$ Center: image taken during the installation of the detector in Hall~A of the LSC. $\bullet$ Right: the experiment installed and operational in Hall~A of the LSC.}
    \label{fig:IntroTrexDM_Fotos_MM_Camara_HallA}
\end{figure}

The \mbox{TREX-DM} detector (figure~\ref{fig:IntroTrexDM_Fotos_MM_Camara_HallA}) is designed to house 0.3~kg of Argon, or 0.16~kg of Neon at 10 bar, as a target for the WIMP. In some respects, the detector is an enlarged version of the low-background X-ray detectors with microbulk Micromegas developed for axion research in the CAST experiment~\cite{Anastassopoulos2017}, but with an active mass 100 times greater. The detector was constructed with the most stringent radiopurity specifications in mind, for which a specific campaign for material identification and measurements has been carried out~\cite{Castel2019}.

\begin{figure}[!ht]
    \centering
    \includegraphics[width=1.0\textwidth]{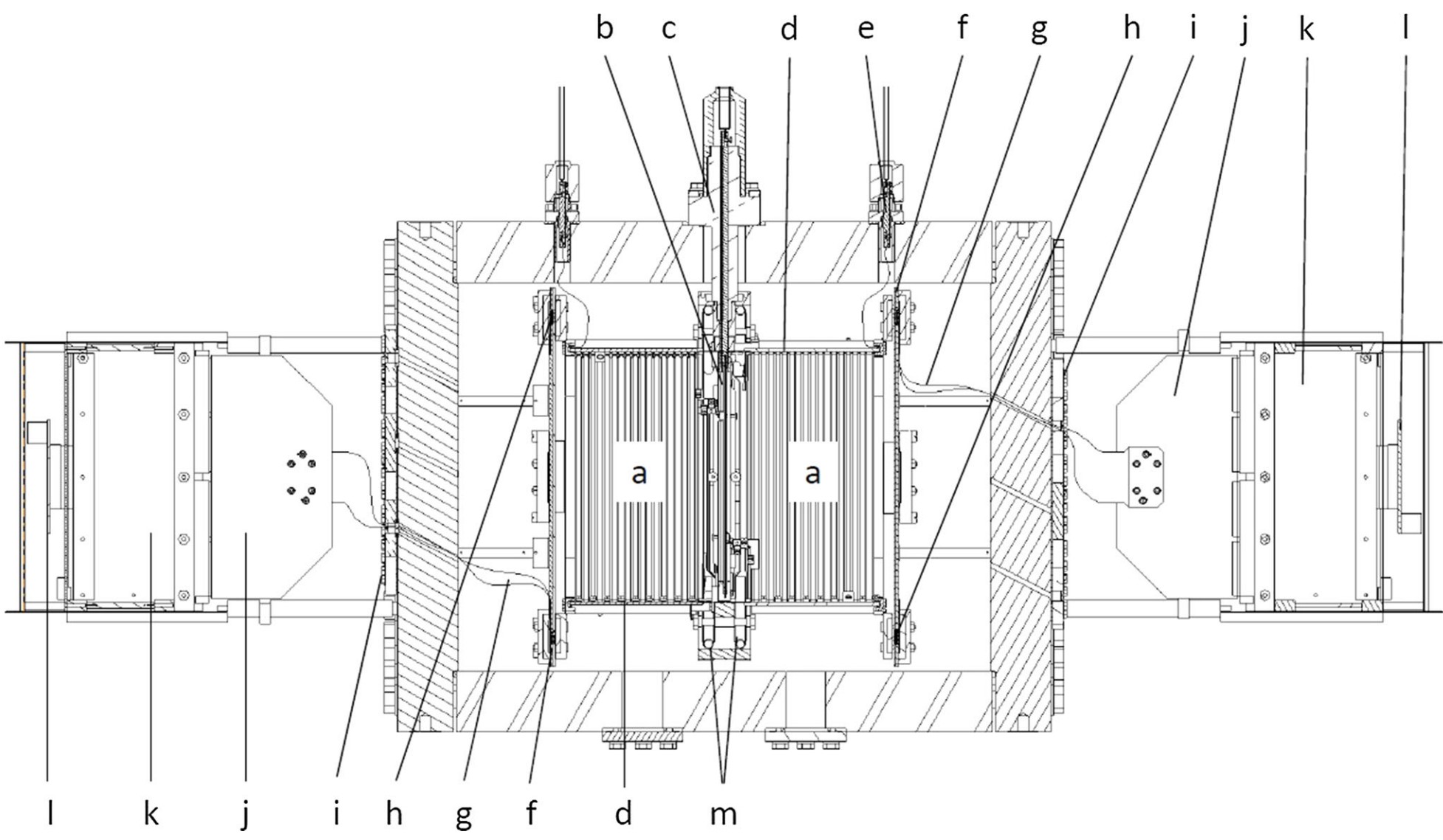}
    \caption{Longitudinal section of the \mbox{TREX-DM} detector~\cite{Iguaz2016}, with the different components: active volume~(a), central cathode~(b), high voltage feedthrough~(c), field cage~(d), feedthrough for the last ring of the field cage~(e), Micromegas planes~(f), flat extension cables~(g), Micromegas interior connectors$-$extension cables~(h), feedthrough for the signal carried by the extension cables~(i), intermediate board~(j), acquisition electronics~(k)(l), and calibration feedthrough~(m).}
    \label{fig:IntroTrexDM_PlanoCorteLongitudinal}
\end{figure}

Figure~\ref{fig:IntroTrexDM_PlanoCorteLongitudinal} shows a longitudinal section of the \mbox{TREX-DM} detector with all its components~\cite{Iguaz2016}, among which the reading planes themselves stand out, installed at both ends of the active volume defined by the field cage. In the center is the cathode, to which an appropriate voltage is applied to achieve the electric field (drift field) in the active volume. To achieve a homogeneous drift field, the field cage is used, formed by several rings that surround the active volume and are connected to each other and to the cathode through 100~M$\Omega$ resistors. The channels of the Micromegas are connected to the acquisition electronics via flat extension cables.

The chamber is made of forged and machined electrolytic copper, with a diameter of 0.5~m and a length of 0.6~m. The end caps are flat and also machined from oxygen-free electronic copper, 6~cm thick, sufficient to withstand pressures of up to 12~bar and to form part of the passive shielding.

\mbox{TREX-DM} operates at a pressure of up to 10~bar with Ar~+~1\%~Isobutane or Ne~+~2\%~Isobutane. The amount of quencher (Isobutane: iC$_4$H$_{10}$) in each mixture is calculated to not exceed the flammability limit, a fact that facilitates the implementation of the experiment underground. To evaluate the performance of the detector with these mixtures, a careful characterization of the microbulk Micromegas was performed with these mixtures and at different pressures~\cite{Iguaz2022}.

Before and during the design of the experiment, a thorough program of material measurements was carried out~\cite{Iguaz2016, Irastorza2016, Aznar2013}. This program was primarily based on gamma-ray spectrometry with germanium, conducted at great depth in Canfranc, but also complemented with other techniques such as GDMS, ICPMS, and measurements using the BiPo-3 detector at the LSC. This has allowed, on one hand, the design and construction of the detector and shielding according to the radiopurity specifications, and on the other, to provide information for constructing the background model of the experiment~\cite{Castel2019}.

One of the largest contributions to the radioactive background, according to the background model~\cite{Castel2019}, comes from the copper vessel, which underwent cosmogenic activation after being on the surface for several years, in the laboratories of the University of Zaragoza, during which the detector served as a demonstrator before its installation at the LSC. This contribution to the radioactive background could be reduced by constructing a new copper vessel.

Another of the most significant contributions to the background is produced by the activity of $^{40}$K measured in the Micromegas, and therefore a study was initiated to reduce this isotope introduced during the manufacturing process, which is described in section~\ref{subsec:MMreduccionFondo}.

\subsection{Operational History of the TREX-DM Experiment}

The commissioning of TREX-DM at the LSC, at the end of 2018, was carried out with an Argon~+~1\%~Isobutane mixture at a pressure of 1.5~bar. In early 2019, the mixture was changed to Neon~+~2\%~Isobutane to avoid background contributions from the $^{39}$Ar isotope contained in atmospheric Argon, which is present in a proportion of (8.0~$\pm$~0.6)·10$^{-16}$~g($^{39}$Ar)~/~g($^{nat}$Ar)~\cite{Benetti2007}, and could contribute to the detector's background level with a rate of 219~dru~\cite{Castel2019} in the region of interest (0.4$-$7~keV$_{ee}$). Furthermore, the latest low-threshold results achieved by the DarkSide experiment~\cite{Agnes2018} exhausted many of the opportunities for TREX-DM with the non-flammable Argon mixture.

The background model \cite{Castel2019}, developed after a comprehensive materials measurement program, indicated levels on the order of 1$-$10~dru in the region of interest (0.4$-$7~keV$_{ee}$). The operation of the detector during 2019, with the Neon~+~2\%~Isobutane gas mixture at 4~bar, allowed for the quantification of the low-energy background level in the range of 1 to 7~keV$_{ee}$, which was two orders of magnitude (1000~dru) higher than expected according to the background model.

This background level was dominated by Radon contamination in the gas system, primarily introduced by the oxygen and humidity filters. In this regard, a new line of research and development was initiated to construct Radon-free filters, the results of which can be consulted in~\cite{Altenmuller2021}. Ultimately, the Radon contamination was reduced by operating the detector in an open loop, directly from the mixing bottle without passing through the filters, with a very low flow rate of $\sim$~1~l/h. This open-loop operation mode led to a reduction of the low-energy background level from 1000 to 80~dru.

The reduction of Radon concentration in the gas system, thanks to the open-loop operation of the detector, revealed alpha particle activity on the surfaces of the detector, which had previously been masked by the volumetric Radon contamination. The activity of these surfaces, currently present in the detector, is likely caused by the progeny of Radon that have deposited on the surfaces, having been previously exposed to Radon-contaminated gas or to laboratory air during the time the TPC was open. The current low-energy background level in TREX-DM (80~dru) is likely dominated by these high-energy events emitted from the surfaces in contact with the active volume.

On the other hand, during the data acquisition campaigns carried out between 2019 and 2022, the energy threshold achieved has been between 1 and 1.5~keV$_{ee}$, with a high dependence on noise conditions and the gain achieved in the Micromegas, being even less than 1~keV$_{ee}$ in the last campaign of 2022, after installing new Micromegas (see chapter~\ref{ch:TREXDMoptimization}). The initial sensitivity estimates made for TREX-DM~\cite{Iguaz2016} were based on an energy threshold of 0.4~keV$_{ee}$, and although the current threshold is close to this value, several actions have been planned to reduce this threshold to 50~eV$_{ee}$.

In October 2022, when surface contamination of the internal components had been identified and was to be replaced, the experiment was completely dismantled at the request of the LSC management and stored in a safe place, awaiting a new location. In the spring of 2023, the experiment was reinstalled in the Lab2500 laboratory of the LSC, and the detector has been taking data since July 2023, with the aim of verifying that all systems are functioning correctly and that the background level remains at the same levels as before dismantling. Both points have been verified.

Once a clean room is installed by the LSC in the new location, as part of its services to the experiment, the chamber can be opened, and work to reduce the alpha particle background level can be carried out, along with other detector improvement tasks such as reducing the energy threshold (chapter~\ref{ch:GEM-MM}) or optimizing the gas mixture.

\setlength{\baselineskip}{12pt} 
\selectlanguage{USenglish}
\graphicspath{{../Imagenes/}{../../Imagenes/}{../../../Imagenes/}} 

\renewcommand{\listtablename}{Table of contents}
\renewcommand{\tablename}{Table} 

\chapter[TREX-DM en el LSC]{TREX-DM in the Canfranc Underground Laboratory}
\label{ch:TREXDMatLSC}

\begin{figure}[!ht]
    \centering
    \includegraphics[width=1.0\textwidth]{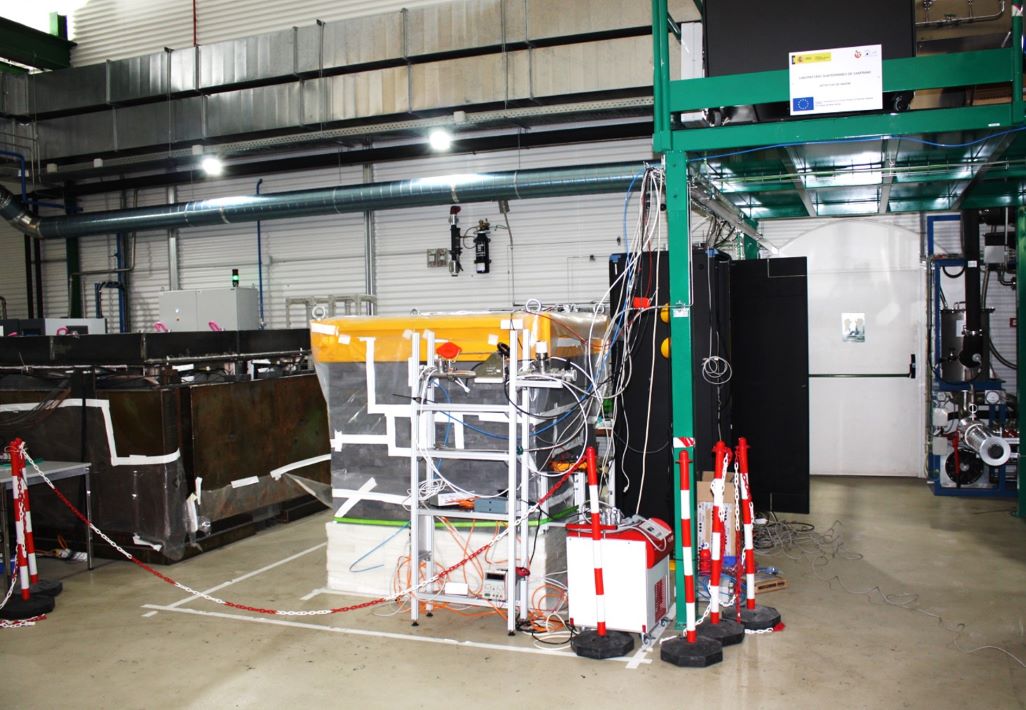}
    \caption*{}
    \label{fig:TREXDMatLSC_SiteIn2400}
\end{figure}

\newpage

In the spring of 2018, the TREX-DM experiment was relocated to the LSC, following approval from its scientific committee. Until then, the TPC had been installed and operated in its demonstrator phase in the laboratories of the University of Zaragoza, without shielding, using bulk-type Micromegas readout planes and FEC-FEM acquisition electronics with AFTER chip~\cite{Baron2008}. 

For the operation of the experiment at the LSC, a copper and lead shielding was assembled, the bulk-type Micromegas were replaced with new microbulk-type Micromegas (which are more radiopure), and a new FEC-Feminos electronics system with AGET chip was implemented~\cite{AGETforTPCs}.

\section{Installation at the LSC}
\label{sec:CommisioningAtLSC}

For the transfer of the experiment from the University of Zaragoza and its installation at the Canfranc Underground Laboratory, all components of the TPC were disassembled. Once all the components were in Canfranc, a cleaning process was carried out on all parts (acid baths for the copper pieces and alcohol washing for the surfaces of the other parts), and all internal components of the TPC were assembled in a clean room (class 7 in the ISO 14644-1 classification), including the new microbulk-type Micromegas. After that, the TPC was placed in Hall A of the LSC, and a copper and lead shielding was installed around it (see figure~\ref{fig:IntroTrexDM_Fotos_MM_Camara_HallA}). Subsequently, the gas system was activated, including the closed-loop system with a recirculation pump, along with the general control system or \textit{slow-control}. Finally, the data acquisition system ('DAQ system') was installed and put into operation, including the new electronics based on the AGET chip.

\begin{figure}[h]
\begin{minipage}{19pc}
\includegraphics[width=1\textwidth]{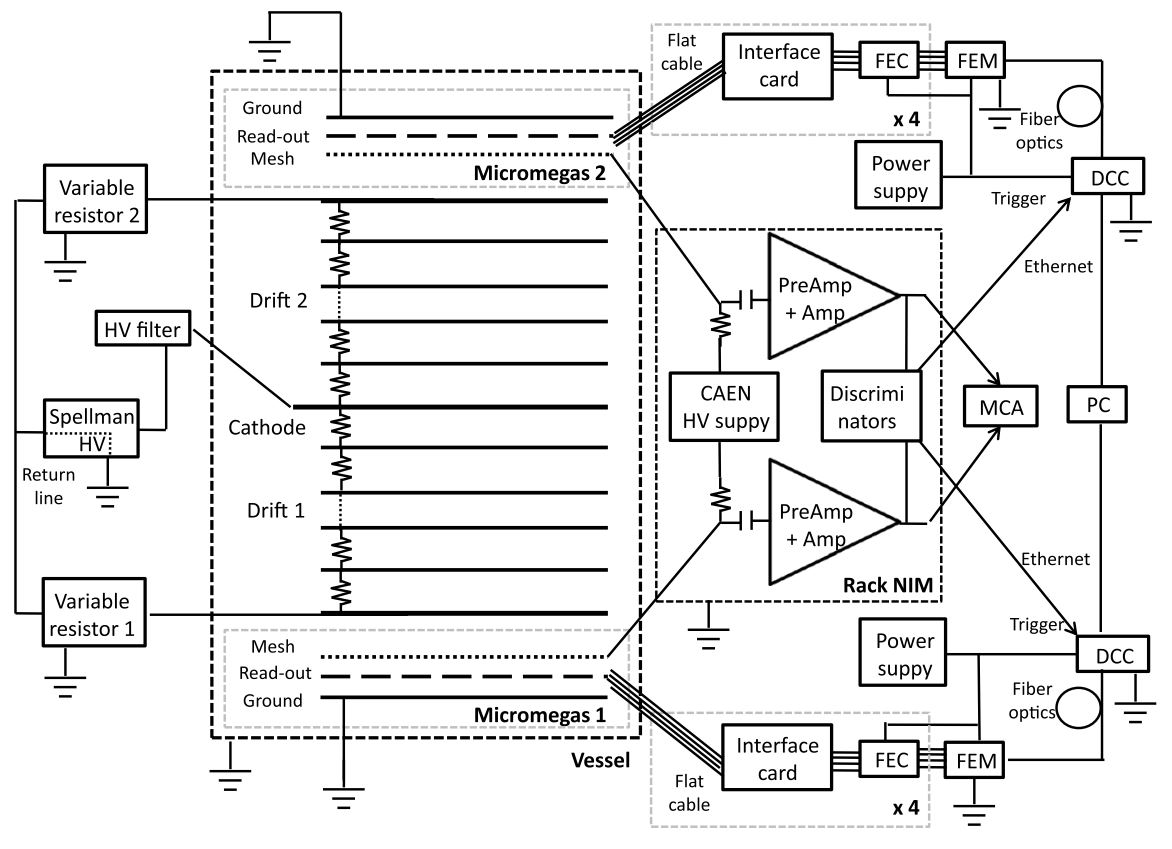}
\end{minipage}\hspace{2pc}%
\begin{minipage}{14pc}
\includegraphics[width=0.63\textwidth]{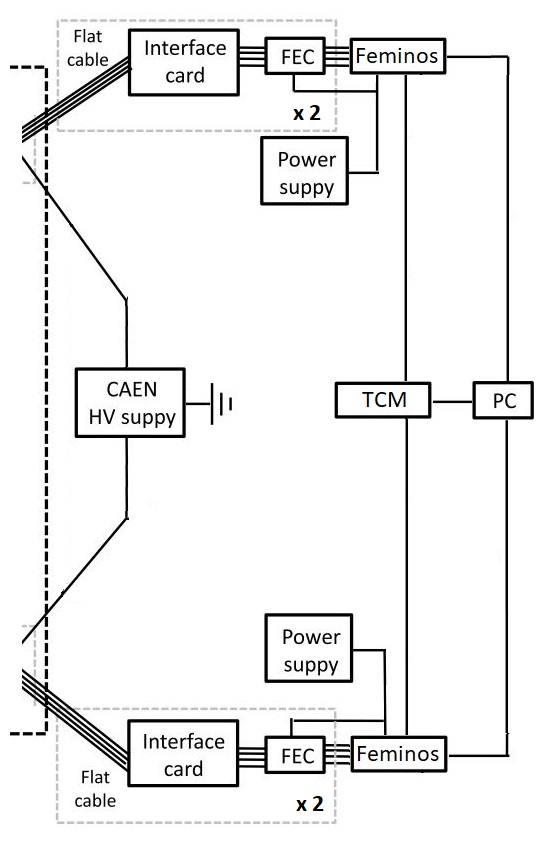}
\end{minipage}
\caption{$\bullet$~Left:~diagram of the FEC-FEM data acquisition electronics chain with the AFTER chip, mounted in the laboratories of the University of Zaragoza \cite{Iguaz2016}. $\bullet$~Right:~diagram of the FEC-Feminos data acquisition electronics chain with the AGET chip, mounted at the LSC, which replaces the previous one and in which the mesh reading chain has been eliminated.}
\label{fig:TREXDMElectronicsChain}
\end{figure}

During the phase when the TPC was installed in the laboratories of the University of Zaragoza, it operated with the FEC-FEM data acquisition electronics with the AFTER chip. This electronics requires reading the signal through the mesh of the Micromegas to trigger the acquisition of each event: when a particle traverses the TPC and is able to ionize the gas, the free electrons produced in this ionization are collected and amplified in the \textit{gap} of the Micromegas, inducing a signal both in the anode, which is segmented with pixels/strips, and in the mesh~\cite{Giomataris2006}. The signal generated in the mesh is used to trigger the acquisition of the FEC-FEM electronics with the AFTER chip, thus collecting the signal from the channels (figure~\ref{fig:TREXDMElectronicsChain}~left).

However, for the installation of the experiment at the LSC, it was decided to replace the FEC-FEM electronics with the AFTER chip with the FEC-Feminos electronics with the AGET chip. The main novelty of the AGET chip compared to the AFTER is its ability to generate the trigger with the signal from the channels themselves, which theoretically leads to a reduction in the energy threshold, among other advantages, by improving the signal-to-noise ratio~\cite{Iguaz2016}.

In the FEC-Feminos system with the AGET chip, the signal reading chain from the mesh is no longer necessary, and the trigger is performed with the data acquisition electronics itself (figure~\ref{fig:TREXDMElectronicsChain}~right). The use of several FEC-Feminos cards, as in the case of the TREX-DM experiment, with a total of 1024 channels, requires the installation of another card for synchronization and trigger management, called the TCM (\textit{Trigger Clock Module}).

\begin{figure}[!ht]
    \centering
    \includegraphics[width=0.85\textwidth]{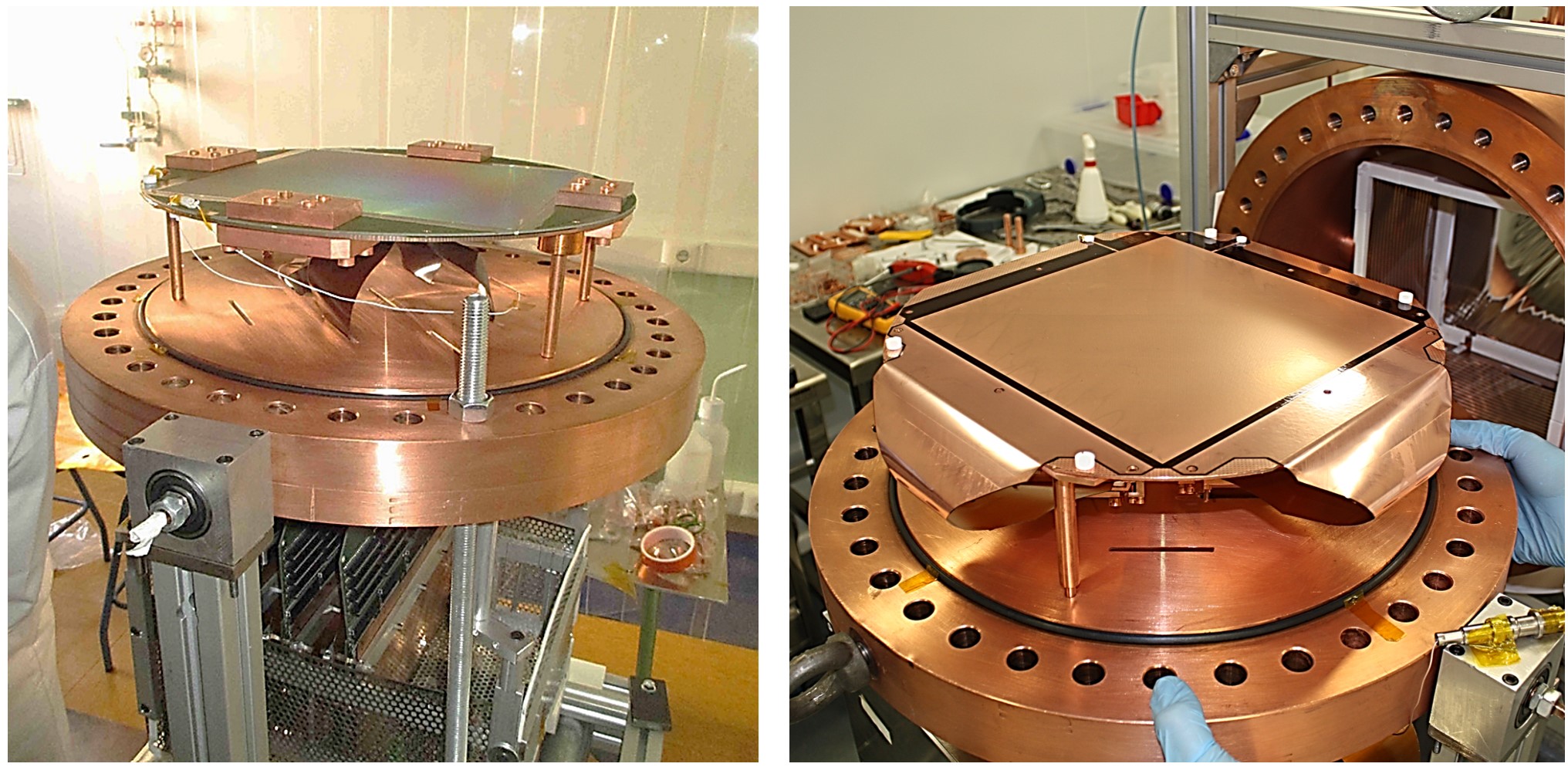}
    \caption{$\bullet$~Left:~photograph of the Micromegas bulk type reading plane, installed and used in the laboratories of the University of Zaragoza. $\bullet$~Right:~photograph of the Micromegas microbulk type reading plane, installed after the experiment was moved to the LSC.}
    \label{fig:TREXatLSC_MicromegasBulkmicroBulk}
\end{figure}

During the phase when the TPC was installed in the laboratories of the University of Zaragoza, a bulk-type Micromegas was used, which was also replaced by a microbulk-type one, mainly due to its higher radiopurity, for the commissioning at the LSC (figure~\ref{fig:TREXatLSC_MicromegasBulkmicroBulk}). During the measurement program conducted to evaluate the radiopurity of the experiment components~\cite{Iguaz2016}, it was confirmed that the raw materials (kapton and copper, mainly) used to manufacture the microbulk-type Micromegas were very clean, with only very low levels ($\sim$0.1~$\mu$Bq~cm$^{-2}$) quantified for the isotopes $^{208}$Ti and $^{214}$Bi, coming from the natural decay chains of $^{232}$Th and $^{238}$U.

It is worth noting that subsequent measurements of microbulk-type Micromegas~\cite{Castel2019}, after undergoing all manufacturing processes, confirmed an activity of $^{40}$K of 3.45~$\pm$~0.40~uBq~cm$^{-2}$, introduced during the mesh drilling or \textit{etching} process using chemical baths containing Potassium. This activity was subsequently reduced during the design and construction of the new Micromegas for TREX-DM (see section~\ref{subsec:MMreduccionFondo}).

\section{Startup with Ar~+~1\%~Isobutane}
\label{sec:TREXatLSC_PuestaENMArcha}

The work for the installation of the experiment extended until the spring of 2018, at which point the first calibration of the detector was performed with Ar~+~1\%~Isobutane at a pressure of 1.5 bar.

As part of the T-REX project program and prior to the installation of the microbulk Micromegas in the TREX-DM TPC, systematic measurements were conducted in the laboratory to study the response of this type of Micromegas in mixtures with Ar~+~1\%~Isobutane and Ne~+~2\%~Isobutane~\cite{Iguaz2022}. The Micromegas characterized in the laboratory had an active area of 2~cm in diameter with a single pixel and an amplification hole diameter of 55~$\mu$m, while the Micromegas installed in TREX-DM were pixelated (with \textit{strips}) and had an amplification hole diameter of 50~$\mu$m. The other two main parameters that typically define this type of Micromegas, the amplification \textit{gap} and the \textit{pitch} between holes, are the same in both cases, 50~$\mu$m and 100~$\mu$m, respectively.

\begin{figure}[!ht]
    \centering
    \includegraphics[height=5cm]{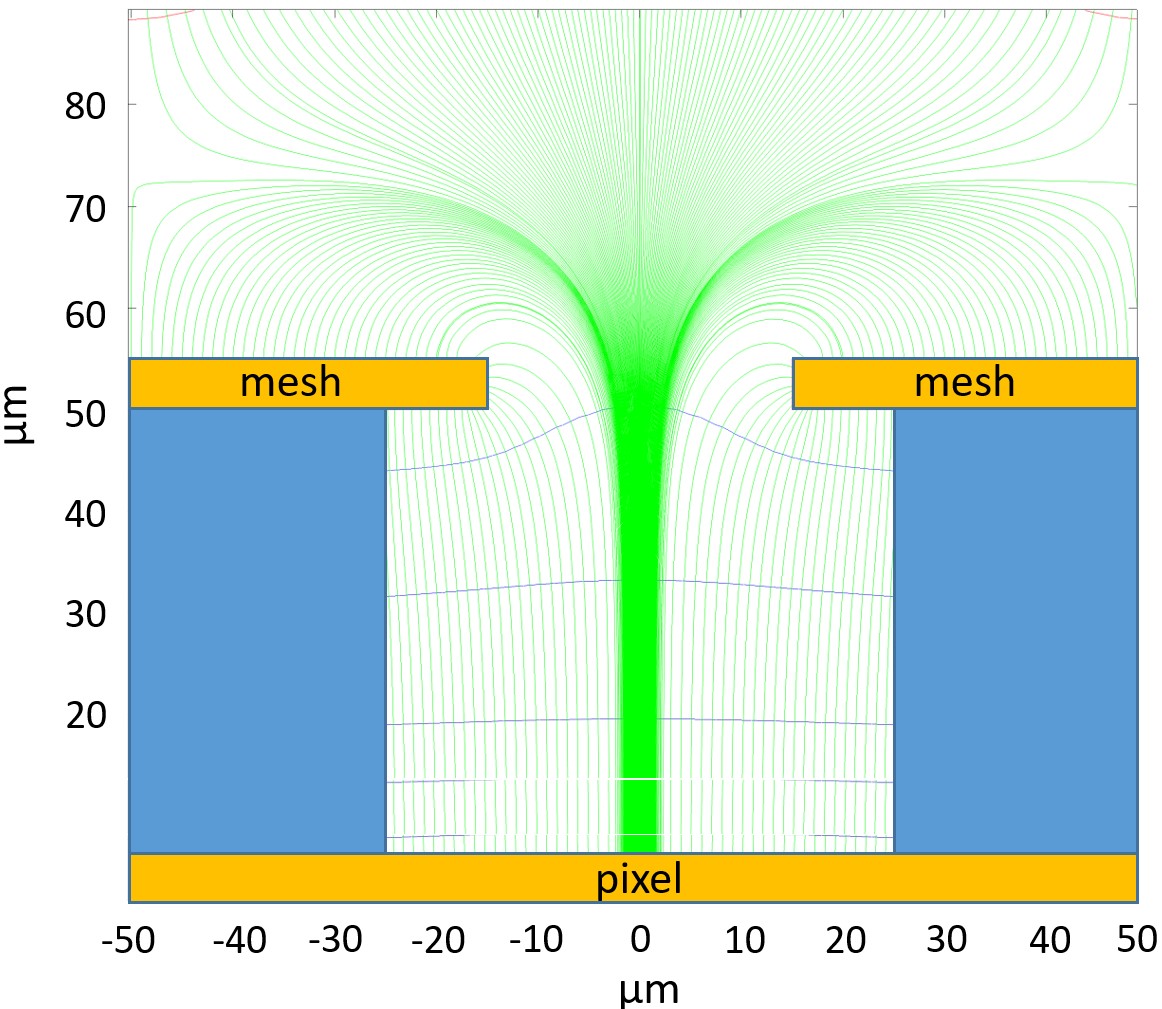}
    \caption{Simulation of the funnel effect produced by the electric field in the area of the amplification hole, in a Micromegas with a hole diameter of 30~$\mu$m and an amplification \textit{gap} of 50~$\mu$m}
    \label{fig:FunnelEffect}
\end{figure}

This difference in hole sizes between the Micromegas characterized in the laboratory and the one installed in TREX-DM is small, and it can be assumed that the response and behavior of both are similar. Even the smaller hole size in the one installed in TREX-DM favors the funnel effect (figure~\ref{fig:FunnelEffect}), although there are systematic measurements with Micromegas in Argon where this effect has a small impact on the response~\cite{Attie2014}. 

\begin{figure}[!ht]
    \centering
    \includegraphics[width=1.0\textwidth]{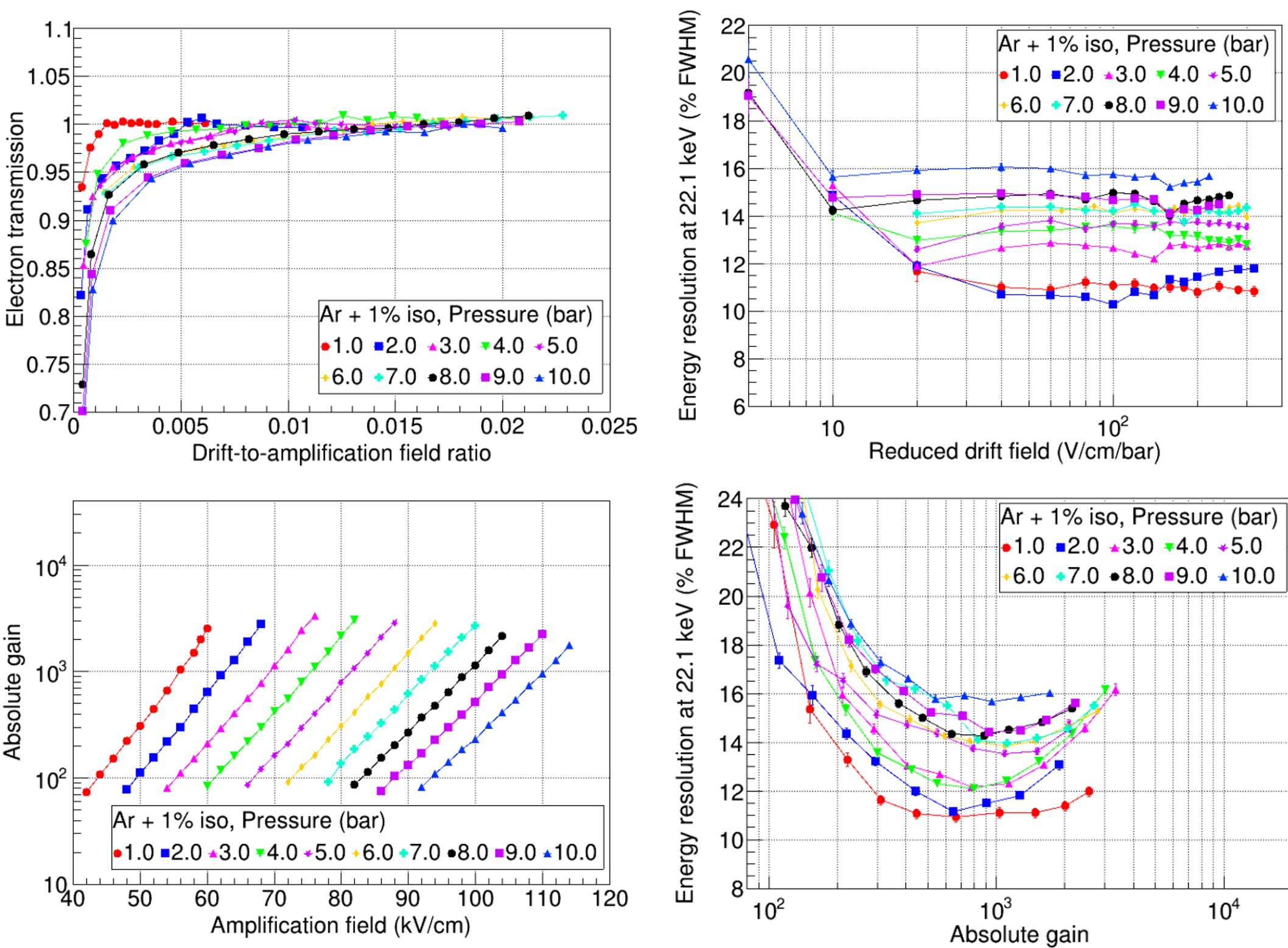}
    \caption{\cite{Iguaz2022} Different characterization curves of a Micromegas type microbulk (50~$\mu$m of \textit{gap} of amplification, 55~$\mu$m of hole diameter and 100~$\mu$m of \textit{pitch} between holes) in a mixture of Ar~+~1\%~Isobutane for different pressures from 1 to 10~bar.}
    \label{fig:CurvasArticuloIguaz2022}
\end{figure}

These systematic measurements~\cite{Iguaz2022} have been used to verify the correct behavior of the Micromegas installed in the TREX-DM TPC during the startup at the LSC. They have also been used to select the optimal operational points under the different pressure states and gas mixtures that have occurred in the experiment. The operational point is primarily defined by three parameters:

\begin{itemize}
 \item \textbf{Pressure}: The TREX-DM TPC was designed to operate and withstand a pressure of up to 10 bar. In the phase with Ar~+~1\%~Isobutane, the goal was to verify that the gas system and the response of the Micromegas were as expected before moving on to the mixture with Ne~+~2\%~Isobutane and increasing the pressure. Therefore, a conservative pressure of 1.5~bar was chosen. During the previous characterizations of the Micromegas~\cite{Iguaz2022}, several measurements were made at different pressures; however, none were made at the pressure of 1.5~bar, so an interpolation between the curves at 1 and 2~bar from figure~\ref{fig:CurvasArticuloIguaz2022} is assumed to approximate the operational point to 1.5~bar.
 
 \item \textbf{Drift field}: The chosen value for the drift field has been 200~V~cm$^{-1}$~bar$^{-1}$. With this value, we find ourselves at the maximum electron transmission and minimum energy resolution, which corresponds to the points in the curves of figure \ref{fig:CurvasArticuloIguaz2022} with a 'Drift-to-amplification field ratio' of 0.005 (assuming that the amplification field will be around 65~kV~cm$^{-1}$).  
 
 \item \textbf{Amplification field}: In TREX-DM, it is essential to reduce the energy threshold as much as possible to be sensitive to lower mass WIMPs. Therefore, throughout the experiment, the goal will always be to operate the detector with maximum gain, which directly translates into trying to achieve the highest possible amplification field. According to figure~\ref{fig:CurvasArticuloIguaz2022}, the maximum amplification fields reached in the systematic measurements were 60~kV~cm$^{-1}$ for 1~bar and 70~kV~cm$^{-1}$ for 2~bar, so the objective during the operation of the pixelated Micromegas installed in TREX-DM at 1.5~bar will be to reach around 65~kV~cm$^{-1}$. Considering that the amplification \textit{gap} is 50~$\mu$m, this implies a voltage value in the mesh of 325~V. It should be noted that the operational point with the maximum gain achieved ($\sim$2·10$^4$ and 12.5\%) in figure~\ref{fig:CurvasArticuloIguaz2022} is a factor of $\sim$2 better in gain and $\sim$1\% worse in energy resolution than the operational point with the best energy resolution ($\sim$1·10$^4$ and 11.5\%). It is preferable to sacrifice a little resolution, as resolution is not as critical as the energy threshold.
 
\end{itemize}

After selecting the operational point at 1.5~bar with Ar~+~1\%~Isobutane, defined by the three parameters described, the voltage in the mesh was gradually increased, at a rate of 1~V/s from the start and at a rate of 1~V/min from 260~V (80\% of the expected maximum voltage) to preserve the good condition of the Micromegas.

\begin{figure}[!ht]
    \centering
    \includegraphics[width=1.0\textwidth]{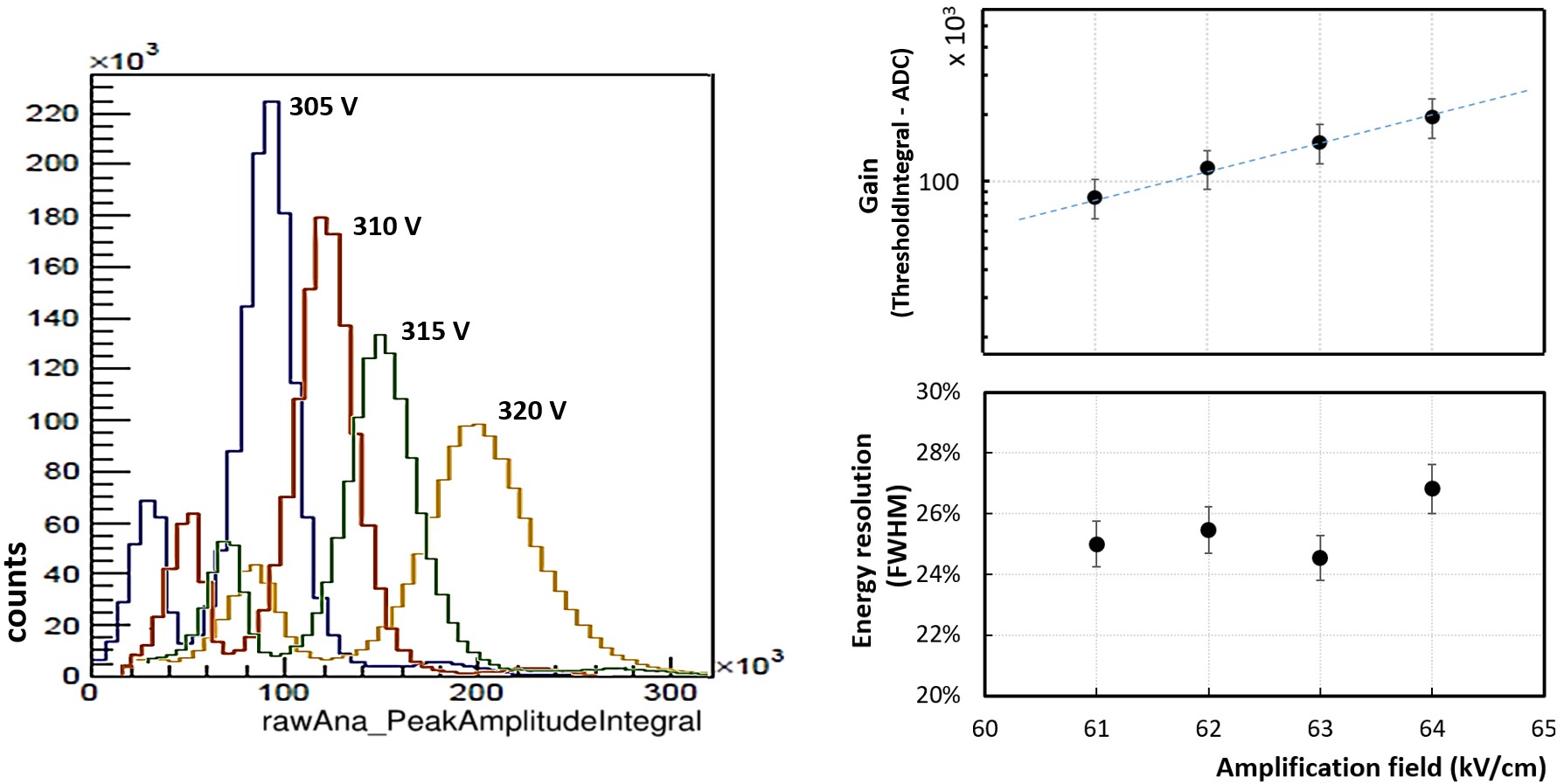}
    \caption{Results of the first calibrations with $^{109}$Cd in TREX-DM at 1.5~bar and Ar~+~1\%~Isobutano, with a drift field of 200~V~cm$^{-1}$~bar$^{-1}$ and different voltages in the mesh: 305~V~(61~kV~cm$^{-1}$), 310~V~(62~kV~cm$^{-1}$), 315~V~(63~kV~cm$^{-1}$), and 320~V~(64~kV~cm$^{-1}$). $\bullet$~Left:~Energy spectra for each of the calibrations, with peaks at 8~and~$\sim$22~keV. $\bullet$~Right: Energy resolution and gain curves constructed from the calibrations.}
    \label{fig:TREXDM_4calibraciones_Ar1Iso_1.5bar}
\end{figure}

Once 305~V was reached in the mesh, a 15-minute calibration was launched with a $^{109}$Cd source that could be manually introduced into the chamber. Following this, three more calibrations were conducted with an increase of 5~V in the mesh between each, until reaching 320~V, at which point sparks began to occur in the Micromegas. The results of these four calibrations can be seen in figure \ref{fig:TREXDM_4calibraciones_Ar1Iso_1.5bar}. The observable PeakAmplitudeIntegral, in units of ADCs, is representative of the energy deposited in each event. The gain curve resulting from these calibrations, expressed in ADCs, cannot be directly compared with the curve in figure~\ref{fig:CurvasArticuloIguaz2022}, as the latter depicts the absolute gain. To compare them, the electronics would have needed to be calibrated beforehand with the configuration used for the calibrations (\textit{Sampling rate, Gain, and Shaping time}), so that a correspondence between the signal collected in ADC and the charge injected in each channel could be achieved. However, it can be observed that the gain curve obtained from the calibrations has the expected behavior, with a slope similar to that of figure~\ref{fig:CurvasArticuloIguaz2022}. 

On the other hand, regarding the energy resolution curve constructed from the four calibrations, the fitting errors and the few points plotted do not allow us to determine if it behaves the same as that in figure~\ref{fig:CurvasArticuloIguaz2022}, where it is observed that the resolution reaches a minimum and then slightly increases up to the maximum gain. However, it is seen that the energy resolution in these calibrations (23$-$29~\%) is well above those achieved during the characterizations with the small Micromegas (11$-$13~

The voltage of 320~V, at which sparks began to occur more frequently in the Micromegas, is very close to the 325~V established as the expected maximum voltage derived from previous studies~\cite{Iguaz2022}. This is a good sign of the good condition and manufacturing quality of the Micromegas, as the one installed in TREX-DM has an active area of 625~cm$^2$ (4.5·10$^6$~holes) while the single pixel Micromegas from previous measurements has 4.5~cm$^2$ (3.2·10$^4$~holes). This difference of two orders of magnitude in the number of holes increases the probability of a spark occurring in one of them, between the mesh and a \textit{strip}. The fact that it was possible to reach 98.4\% of the expected maximum voltage seems to indicate the good construction quality of the Micromegas.

\begin{figure}[!ht]
    \centering
    \includegraphics[height=9cm]{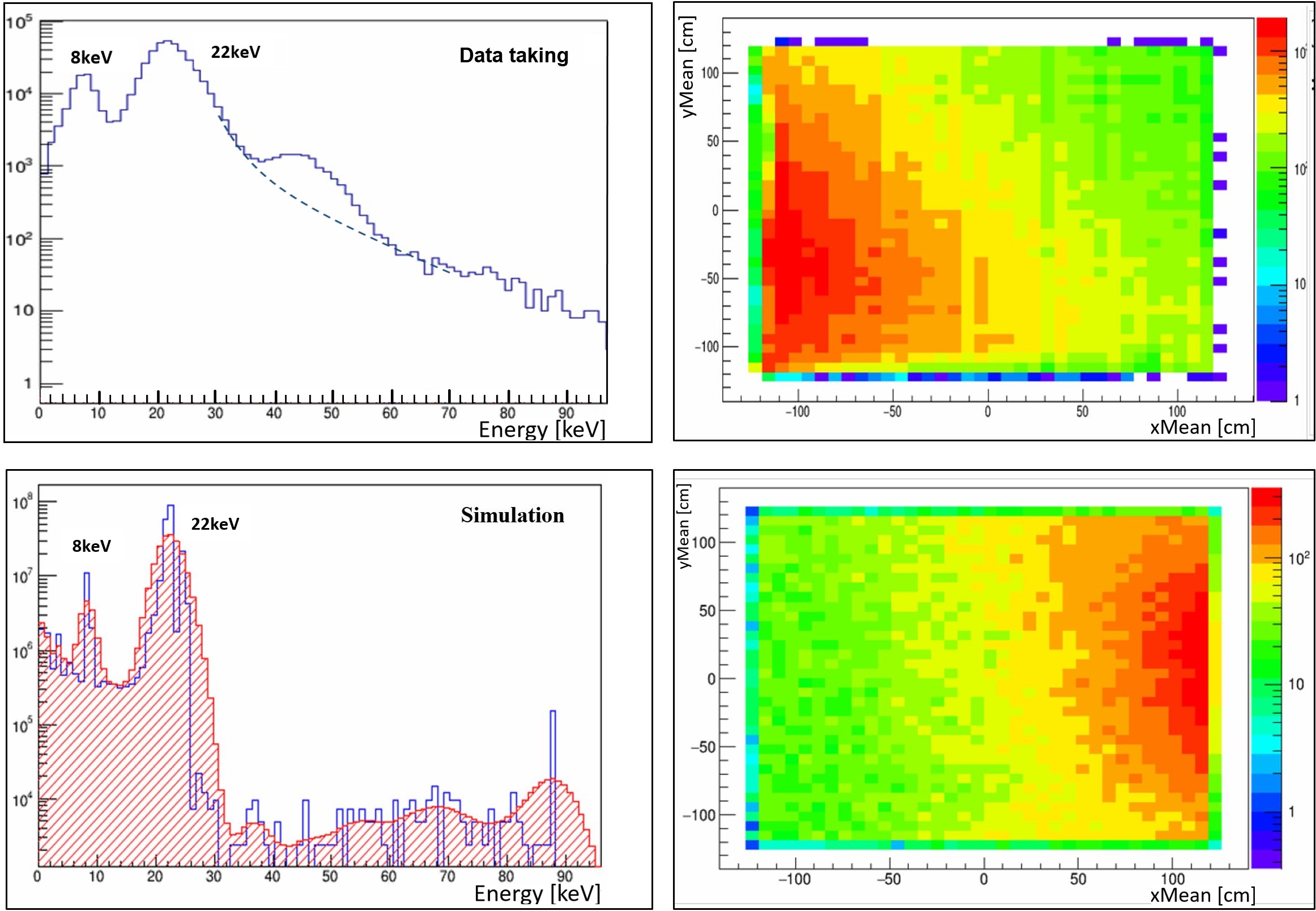}
    \caption{$\bullet$~Top: first calibration with $^{109}$Cd of one of the Micromegas in Ar~+~1\%~Isobutane at 1.5~bar: energy spectrum (left) and activity map in the Micromegas (right). Approximate manual adjustment (dashed line) for filtering events from \textit{pile up}. $\bullet$~Bottom: simulation of a calibration under the same gas mixture and pressure conditions: energy spectrum (left) and activity map in the Micromegas (right). The activity map, from which it can be determined that the $^{109}$Cd source is located on the right side, is flipped with respect to the X and Y axes, compared to the activity map of the experimental calibration where it is observed that the source is located on the left side.}
    \label{fig:TREDMPrimeraCalibracion}
\end{figure}

After these first calibrations to verify the correct response of the detector, a one-hour exposure calibration was taken, and its energy spectrum can be seen in figure~\ref{fig:TREDMPrimeraCalibracion}, distinguishing the peak at 22~keV from the $^{109}$Cd source (21.99, 22.16, 24.91, and 24.94~keV) and the peak at 8~keV corresponding to the fluorescence of copper. Another peak around 44~keV can also be seen, due to the accumulation of 22~keV events that are emitted with a temporal difference smaller than the time window programmed for acquisition (\textit{pile up}). A greater refinement in the analysis should identify these events and separate them into independent events that contribute independently to the spectrum. This should eliminate the peak at 44~keV and obtain a spectrum like the resulting one after the manual adjustment (in red) performed in figure~\ref{fig:TREDMPrimeraCalibracion}. The objective of the calibration is to verify the correct functioning of the detector, and the resulting spectrum corresponds to what was expected.

In figure~\ref{fig:TREDMPrimeraCalibracion}, a simulation performed with REST-for-Physics and Geant4 of a calibration in TREX-DM under the same conditions as that performed in the detector~\cite{ParisTrex2019} is also shown for comparison. The exposure time is greater in the simulation to accumulate more statistics.

In the low-energy region of the experimental spectrum, we see that there is no accumulation of events because during this calibration, a phase of startup of the experiment, the signal-to-noise ratio was not optimized, and a high energy threshold was established. Similarly, the analysis processes were not yet adjusted and optimized for this detector, so those signals coming from the $^{109}$Cd source that were significantly above the noise could be classified as noise, due to one of the first processes in the analysis chain (see figure~\ref{fig:SR2MdataTaking_RESTflow}).

In the central part, the two peaks of 8 and 22 keV appear, both in the simulation and in the experimental spectrum, in approximately the same proportion, while in the higher energy range, the simulation spectrum presents a population of events between 30 and 80 keV that is more populated in the case of the experimental spectrum. Even with a more refined analysis, in which the \textit{pile up} events are correctly identified and classified, it does not seem that at least the events above $\sim$60 keV could be filtered out. Additionally, in the simulation, a peak at 88 keV is observed, corresponding to a gamma emission from the $^{109}$Cd source; however, this peak is not observed in the experimental spectrum, likely hidden by a combination of background events and poorer energy resolution.

\begin{figure}[!ht]
    \centering
    \includegraphics[height=9.7cm]{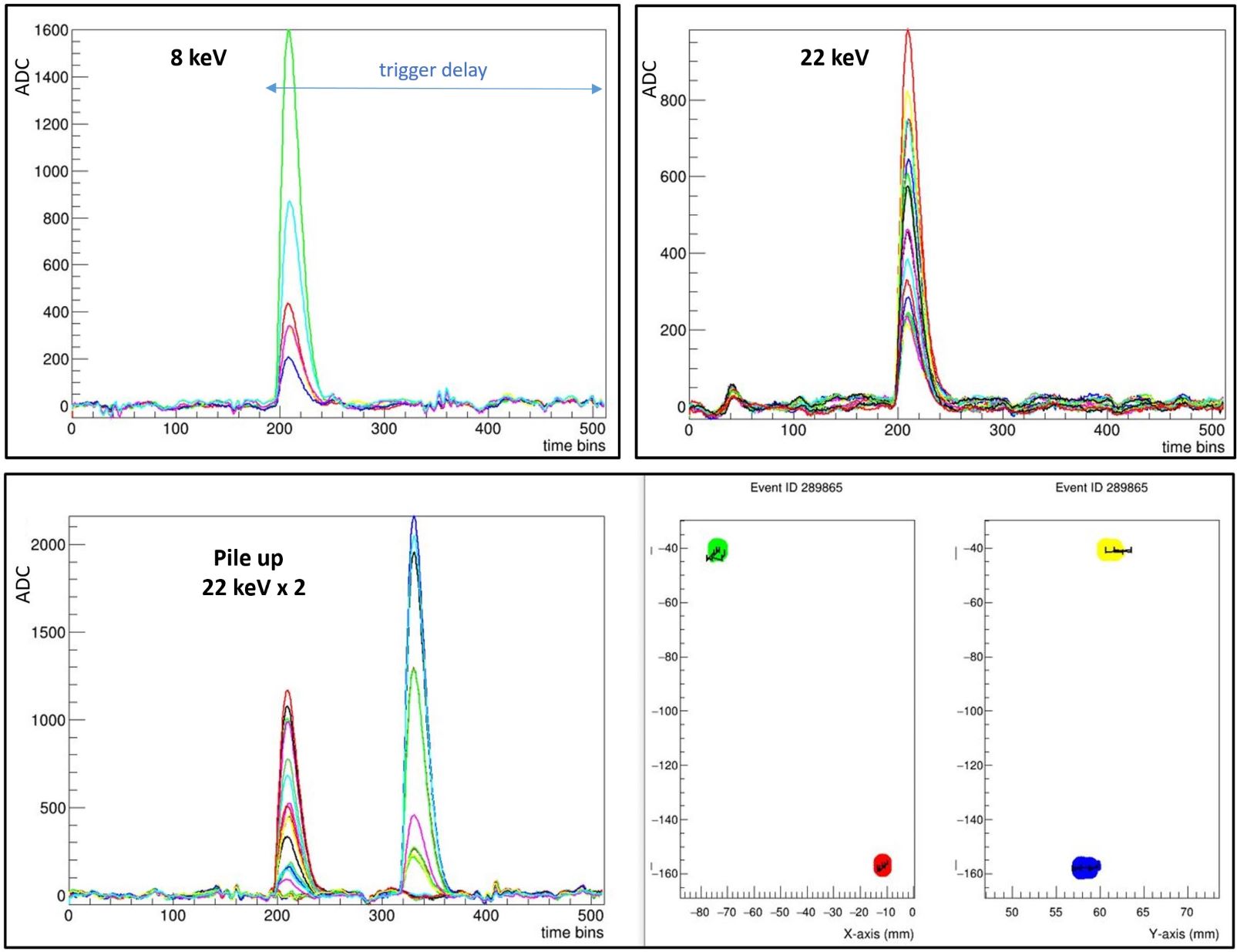}
    \caption{Raw signal from the channels in the first calibration with $^{109}$Cd of one of the two Micromegas in Ar~+~1\%~Isobutane at 1.5~bar, corresponding to figure \ref{fig:TREDMPrimeraCalibracion}. $\bullet$~Top left: event corresponding to the 22~keV peak from the $^{109}$Cd source (21.99, 22.16, 24.91, and 24.94). $\bullet$~Top right: event corresponding to the 8~keV peak from the fluorescence of copper. $\bullet$~Bottom: \textit{pile up} of two simultaneous 22~keV events and spatial reconstruction of the event.}
    \label{fig:TREDMPrimeraCalibracionSeñales}
\end{figure}

Figure \ref{fig:TREDMPrimeraCalibracionSeñales} shows three events directly collected by the electronics (\textit{raw data}) during the calibration whose spectrum is shown in figure~\ref{fig:TREDMPrimeraCalibracion}. The main parameters programmed in the electronics for reading and amplifying the signal, similarly to those programmed in the case of reading the SR2M in section~\ref{sec:LecturaStrips}, were as follows:

\begin{itemize}
\item \textit{Trigger delay}: 24~$\mu$s. This parameter was adjusted so that, for an event occurring very close to the surface of the Micromegas, the electronics could collect the signal from another event that occurred simultaneously in the area of the active volume farthest from the Micromegas, that is, near the surface of the cathode. To achieve this, it is essential to know the value of the drift velocity, which in this case for Ar~+~1\%~Isobutane and a reduced drift field of 200~V~cm$^{-1}$~bar$^{-1}$ is about 4~cm~$\mu$s$^{-1}$ (figure~13.11 of~\cite{JavierGraciaGarza}). The minimum \textit{trigger delay} in this case is about 5~$\mu$s, considering that there is a 19~cm drift distance between the cathode and the Micromegas. However, during the startup, and since the drift velocity has a strong dependence on experimental conditions (pressure variations, homogeneity of the drift field, gas quality), it was decided to extend this value to 24~$\mu$s, the same value previously programmed for reading the SR2M module, thus allowing sufficient margin for error.

\item \textit{Sampling rate}: 100~ns.

\item \textit{Gain}: 240~fC. The gain in the amplification stage is selected so that no event of interest can saturate the analog-to-digital converter (ADC). In this case, events of interest could be defined as those included in the energy range between the energy threshold and the end of the tail to the right of the 22~keV peak from the calibration source, up to 40-50~keV. Within this range, those of higher energy (40-50~keV) and occurring close to the Micromegas (due to having less diffusion) are those that have a higher charge per channel, and therefore a greater risk of saturation. Finally, being in the startup phase and to be able to compare with simulations with deposits up to 100~keV, a lower gain was decided upon.

\item \textit{Shaping time}: 1014~ns. The \textit{shaping time} of the amplification stage was selected to be as high as possible during the startup to filter out high-frequency noise as much as possible. However, a high integration time entails lower temporal resolution in the study of the avalanches caused by the arrival of primary electrons at the amplification \textit{gap} in the Micromegas, and reduces the possibilities of studying diffusion, among others. Since these diffusion studies had already been conducted for the gas mixture Ar~+~1\%~Isobutane in \cite{JavierGraciaGarza}, the strategy of noise reduction and thus the energy threshold was opted for.
\end{itemize}

Regarding the energy threshold achieved, on one hand, we have the 'energy threshold per channel', determined by the lowest amplitude signal that the electronics can collect, which will be equal to or greater than the baseline fluctuation. On the other hand, we have the 'trigger energy threshold', which is determined by the efficiency of the trigger in the electronics.

\begin{figure}[!ht]
    \centering
    \includegraphics[width=1.0\textwidth]{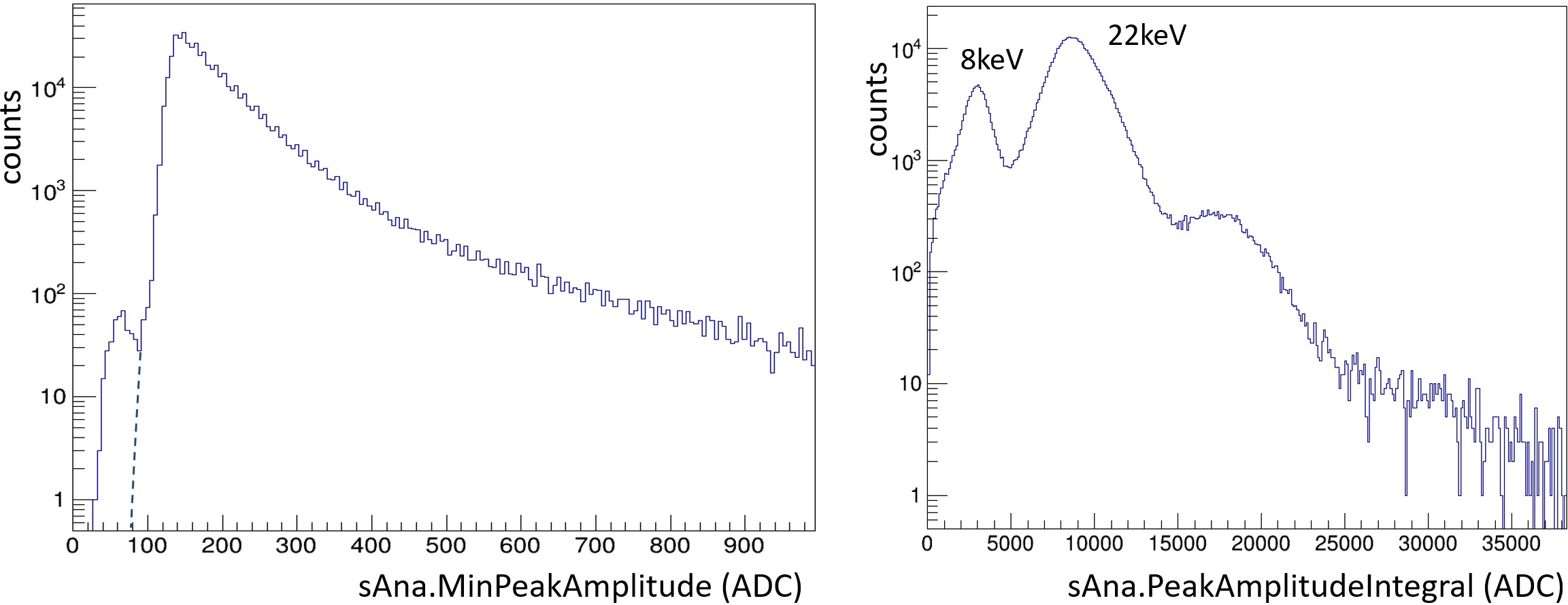}
    \caption{Estimation of the minimum energy threshold under the given conditions during the calibration corresponding to the spectrum in figure~\ref{fig:TREDMPrimeraCalibracion}, in Ar~+~1\%~Isobutane at 1.5~bar. $\bullet$~Left: histogram of the amplitude of the channel with the lowest amplitude in each event, among all channels, represented by the observable 'sAna.MinPeakAmplitude'. Manual adjustment (dashed line) to discard the small population on the left corresponding to noise signals that have passed the first cut in the analysis. $\bullet$~Right: histogram of the sum of the amplitudes in all channels for each event, represented by the observable 'sAna.PeakAmplitudeIntegral', which is representative of the energy deposited by the detected particle.}
    \label{fig:UmbralEnergia_TREXDM_Ar1Iso}
\end{figure}

Using the histogram from figure~\ref{fig:UmbralEnergia_TREXDM_Ar1Iso}, the 'energy threshold per channel' reached in the detector during the calibration with $^{109}$Cd corresponding to the spectrum in figure~\ref{fig:TREDMPrimeraCalibracion} can be estimated. This threshold is around 150~ADC, which corresponds to 400~eV if we use the 8~keV peak at 3000~ADC as energy calibration and assume that the energy deposited by a particle that ionizes the gas follows a linear relationship with the amplitude of the signal it generates.

This energy threshold of 400~eV coincides with the total threshold, or trigger energy threshold, expected for the experiment (region of interest 0.4-7~keV) but is higher than expected according to studies conducted in the CAST/IAXO experiment~\cite{Garza2018} with the new electronics based on the AGET chip, where an energy threshold per channel of 100~eV was measured. However, as previously mentioned, during this calibration, a phase of startup of the experiment, the signal-to-noise ratio was not optimized, so it was expected that with subsequent work to reduce noise in the experiment, this threshold would improve.

Regarding the 'trigger energy threshold', as mentioned earlier and as can be seen in the spectra of figures~\ref{fig:TREXDM_4calibraciones_Ar1Iso_1.5bar} and \ref{fig:TREDMPrimeraCalibracion}, it was very high, around 4~keV, due to the fact that the signal-to-noise ratio was not optimized, but also because it was affected by the trigger configuration in the electronics.

\begin{figure}[!ht]
    \centering
    \includegraphics[width=1.0\textwidth]{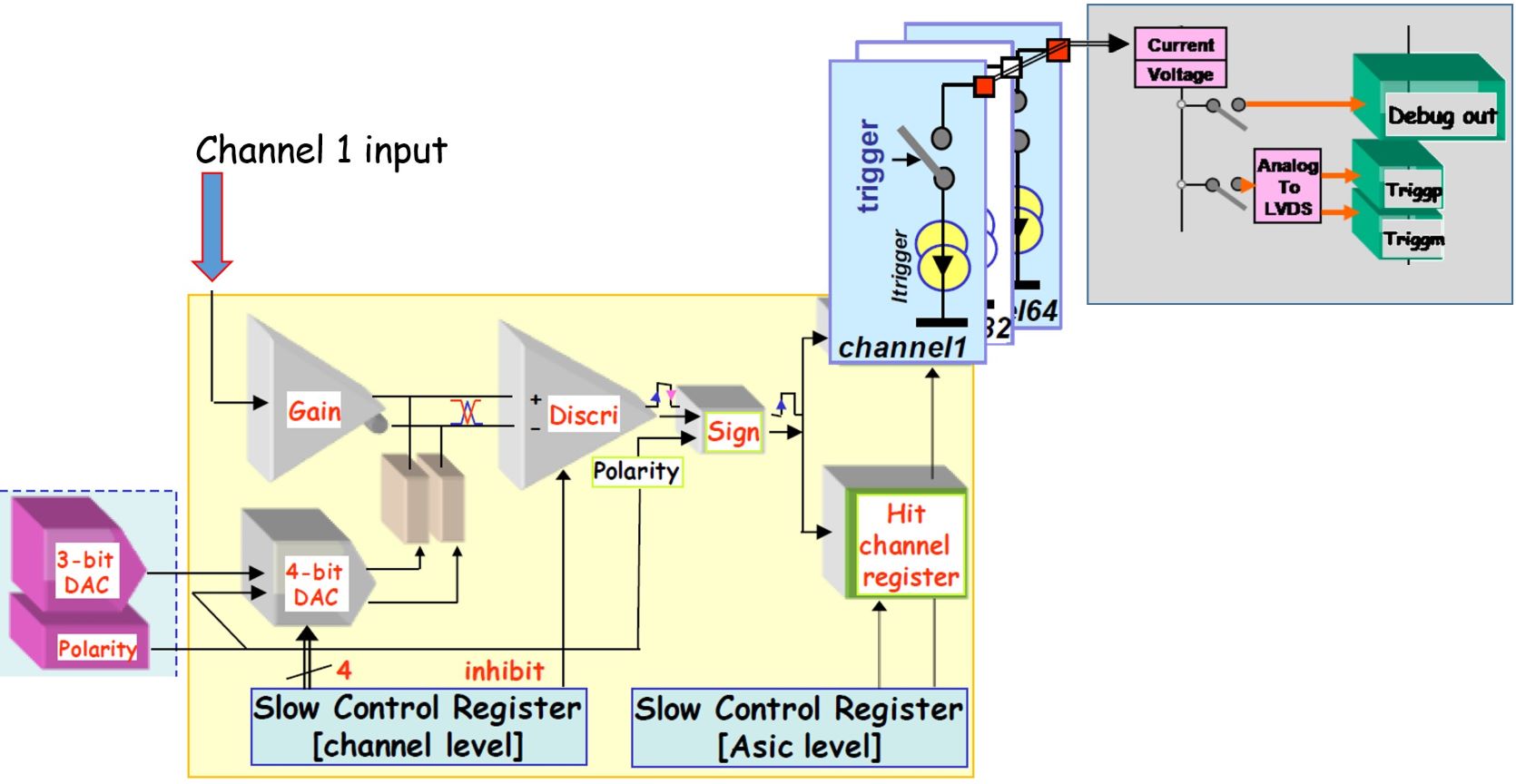}
    \caption{\cite{AGETforTPCs}~Block diagram for multiplicity management within the AGET chip. The signal from a channel ('Channel~1~input'), connected to the corresponding \textit{strip} of the detector, is amplified ('Gain'). Once amplified, it enters a discriminator ('Discri'), whose output threshold is programmable ('4-bit~DAC'), so that when the signal exceeds this threshold, the output of the discriminator activates the trigger contact ('trigger'), injecting a fixed intensity from the corresponding DC current source for that channel ('Itrigger'). The current-to-voltage converter ('Current/Voltage') collects from the analog adder the total current contributed by as many sources as channels have exceeded the programmed threshold in the discriminator. In this way, a 'multiplicity' signal of channels is generated, which must be processed and to which a threshold (multiplicity threshold) is applied, above which the trigger or acquisition trigger is launched, at which point the acquisition of the detector signals occurs.}
    \label{fig:EsquemaMultiplicity}
\end{figure}

The AGET chip~\cite{AGETforTPCs} performs the trigger depending on the programmed multiplicity, that is, the number of channels that exceed the programmed threshold at the same time. Although it would be desirable to trigger if a single channel exceeds the programmed threshold, which would have been adjusted significantly above the noise, in practice, this is difficult to achieve, as the chip uses an analog adder for multiplicity management that is also affected by noise (figure~\ref{fig:EsquemaMultiplicity}). This means that it is necessary for several channels to exceed the threshold for the trigger to occur, and this effective number of channels depends on several factors, such as the noise itself. Additionally, the distribution of the charge of an event among several channels from different chips is an unfavorable case for trigger management, as the multiplicity is programmed for each chip, meaning that the trigger will occur if the number of channels exceeding the threshold is above the multiplicity value in one of the chips.

In section~\ref{subsec:umbralNeon}, the study of the energy threshold for different pressures with the Ne~+~2\%~Isobutane mixture is addressed, and figure~\ref{fig:TREXDMatLSC_Ne_CurvasUmbralEnergia} shows a histogram with the number of channels per event, where the effect of multiplicity on trigger management is appreciated.

As will be seen later, the efforts directed at noise reduction and improving trigger efficiency in the electronics will have a direct impact on the improvement of the 'trigger energy threshold', which in practice defines the energy threshold of the experiment, being around 1~keV in many phases of data taking with Ne~+~2\%~Isobutane, and even below in some cases. Furthermore, the development of the GEM-Micromegas system (chapter~\ref{ch:GEM-MM}) will significantly reduce this threshold.

\section{Commissioning with Ne~+~2\%~Isobutane}

After confirming that the detector with Ar~+~1\%~Isobutane at 1.5~bar exhibited good performance, at the beginning of 2019, the mixture of Ne~+~2\%~Isobutane was adopted, and the pressure in the chamber was gradually increased while studying the detector's response at each operating point. The latest low-threshold results achieved by the DarkSide experiment~\cite{Agnes2018} exhausted many of the opportunities for TREX-DM with the non-flammable Argon mixture. The Neon mixture, compared to Argon, has a higher nuclear recoil energy in its interaction with the WIMP and a greater maximum gain achievable in the Micromegas. Both advantages allowed for a reduction in the energy threshold for TREX-DM, thereby increasing sensitivity to low-mass WIMPs.

However, as the pressure in the chamber increased, and consequently the voltage applied to the Micromegas, leakage currents began to appear in the connection chain from the Micromegas to the electronics, which often hindered the operation of the detector, eventually leading to more than 20\% of the channels being damaged. These problems and their consequences are explained in greater detail in section~\ref{subsec:CorrientesDeFuga}.

Despite this, and with significant effort to mitigate this issue, calibration and background data were successfully obtained at several operating points at different pressures. The studies of the detector's response at all these operating points are addressed in this section, while the results of the background data acquisitions are discussed in section~\ref{sec:BackgroundFisrtMeasurements}.

\begin{figure}[!ht]
    \centering
    \includegraphics[width=0.9\textwidth]{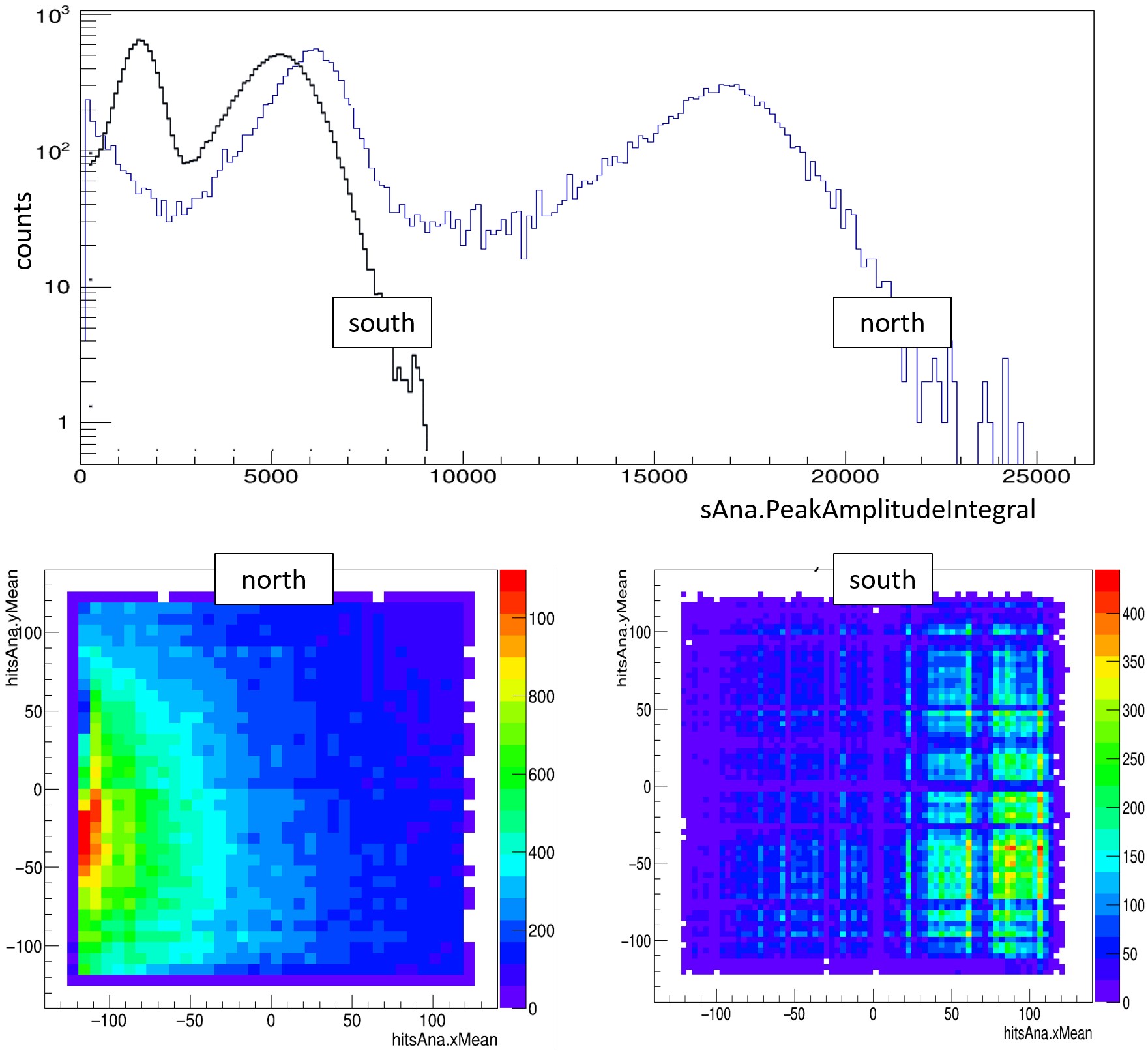}
    \caption{Results of the first calibrations in TREX-DM with Ne~+~2\%~Isobutane at 1.5~bar, with voltages on the mesh of 390~V~(North side) and 365~V~(South side) and with a drift field of 200~V~cm$^{-1}$~bar$^{-1}$. $\bullet$~Top:~Energy spectrum for each of the detectors, represented by the observable 'sAna.PeakAmplitudeIntegral' (sum of the maximum amplitude of all signals from the event). $\bullet$~Bottom left:~Activity map in the North detector. $\bullet$~Bottom right:~Activity map in the South detector, where channels with low or no activity due to connectivity failures or directly damaged channels are observed.}
    \label{fig:TREXDMatLSC_NeCalib1.5bar}
\end{figure}

The first calibration data with Ne~+~2\%~Isobutane were taken at 1.5~bar. Figure~\ref{fig:TREXDMatLSC_NeCalib1.5bar} shows the results of these calibrations for both detectors (North and South). In the North detector, a very high voltage could be applied to the mesh (390~V). The effective voltage applied was likely lower, as leakage currents on the order of tens or even hundreds of nA, as occurred during these data acquisitions, cause a voltage division between the mesh and ground, which can invalidate the reading of the voltage in the channel of the voltage source as the actual voltage applied to the mesh.

These leakage currents also occurred in the South detector, but in this case, they were unstable at lower voltages and only allowed reaching up to 365~V. The activity map of the South detector shows a significant number of damaged channels or poor connectivity. The damaged channels that are short-circuited with the mesh, or have low resistance to it, are the source of these leaks and the current division effect that obscures the actual voltage reading in the Micromegas.

The difference in gain between the two detectors observed in the spectrum of figure~\ref{fig:TREXDMatLSC_NeCalib1.5bar} is attributable to the difference in the voltages applied to the mesh, although due to the leakage currents, it is difficult to quantify the effective voltage difference at that moment.

After these initial calibrations, the pressure was progressively increased to 3, 4, 6, and 8~bar. The gains achieved at higher pressures were lower than that obtained at 1.5~bar: 50\% lower at 3~bar (380~V), 60\% lower at 6~bar (435~V), and 75\% lower at 8~bar (465~V). Although the limitation at each pressure for achieving the maximum voltage on the mesh, and therefore the maximum gain, was due to the leakage currents and not the occurrence of sparks in the Micromegas, as is usually the case.

\begin{figure}[!ht]
    \centering
    \includegraphics[width=0.99\textwidth]{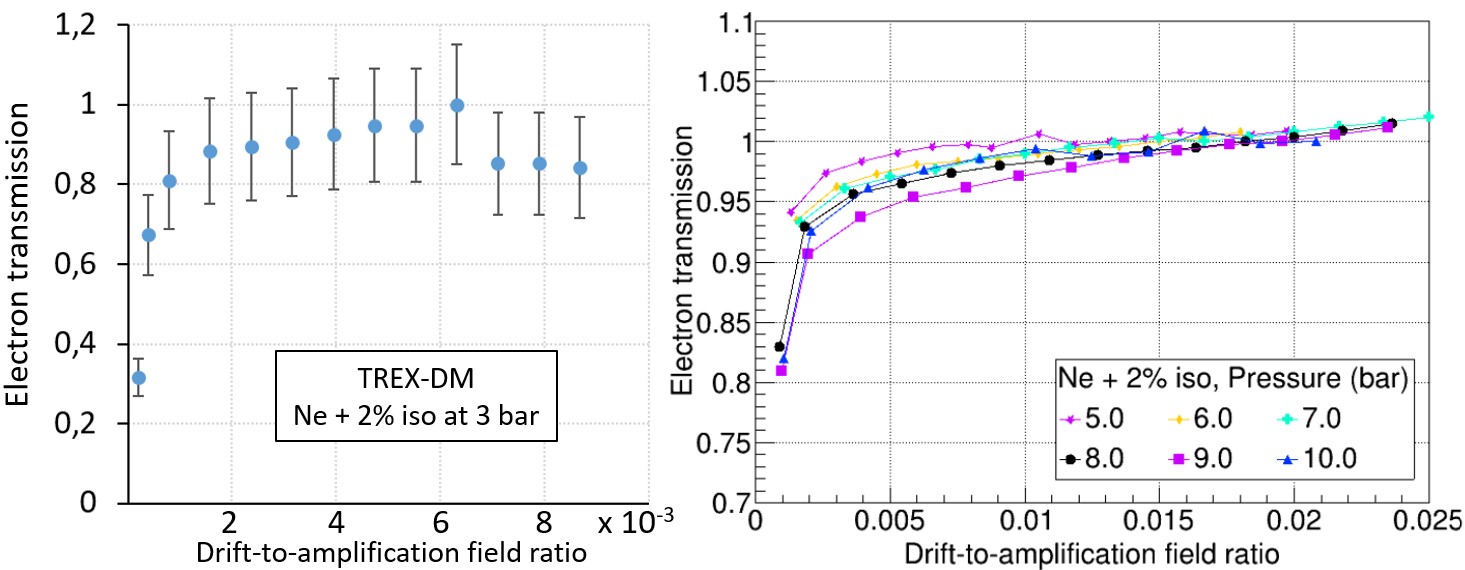}
    \caption{Electron transmission curves with Ne~+~2\%~Isobutane in TREX-DM at 3~bar~(left) and in the previous study~\cite{Iguaz2022} for different pressures~(right). The range of data in the case of TREX-DM is limited by the maximum voltage applicable to the cathode, defined as that immediately prior to the generation of a spark, which in this case at 3~bar was 12~kV~(V$_{drift}$/V$_{amp}$~=~8.7·10$^{-3}$).}
    \label{fig:TREXDMatLSC_Ne_CurvasTransp}
\end{figure}

\begin{figure}[!ht]
    \centering
    \includegraphics[width=0.99\textwidth]{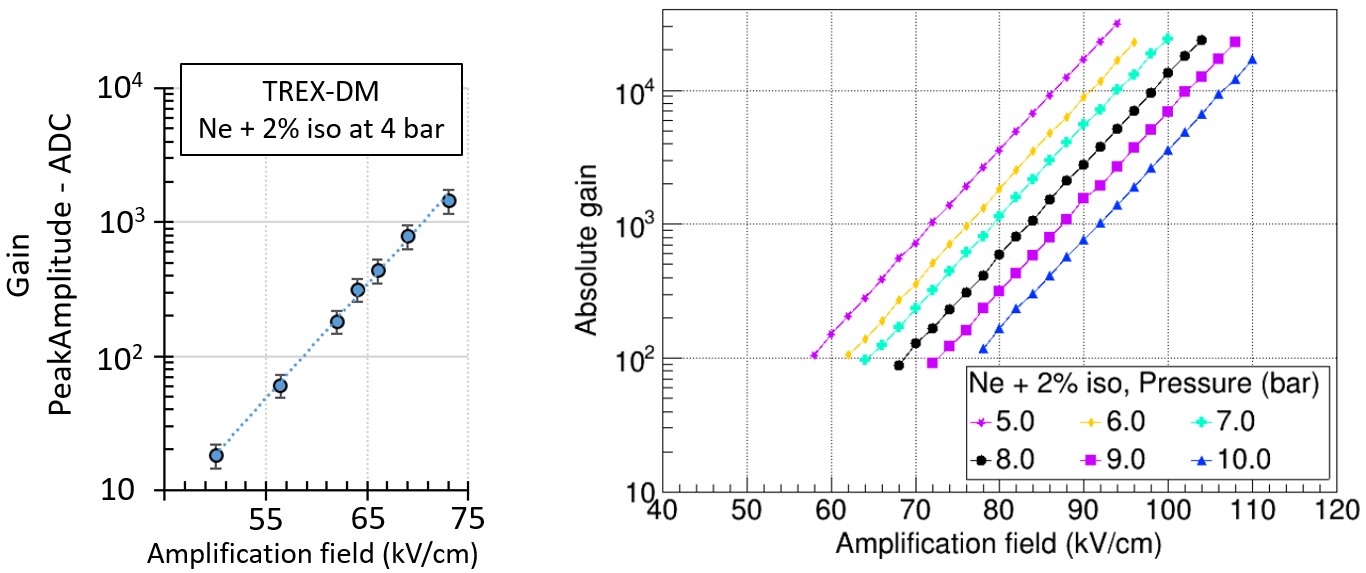}
    \caption{Gain curves with Ne~+~2\%~Isobutane in TREX-DM at 4~bar~(left) and in the previous study~\cite{Iguaz2022} for different pressures~(right). For a direct comparison of both curves and normalization on the Y-axis, it would be necessary to characterize and calibrate the acquisition electronics in TREX-DM, so that the correspondence between ADCs and charge received in the channels could be obtained, allowing the calculation of the absolute gain in the Micromegas of TREX-DM.}
    \label{fig:TREXDMatLSC_Ne_CurvasGain}
\end{figure}

\begin{figure}[!ht]
    \centering
    \includegraphics[width=0.99\textwidth]{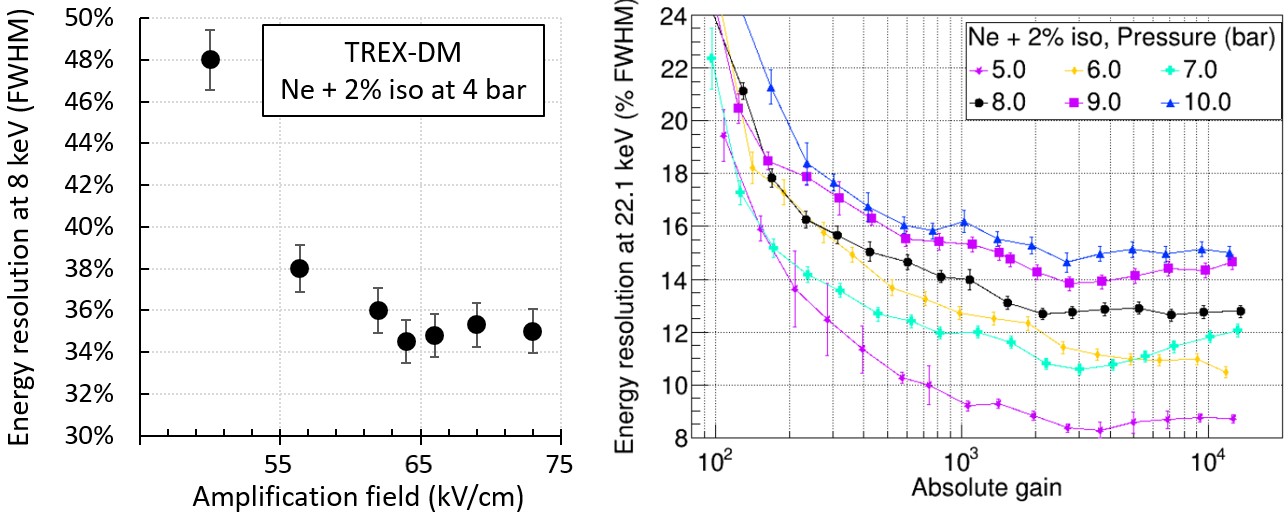}
    \caption{Energy resolution curves with Ne~+~2\%~Isobutane in TREX-DM at 4~bar~(left) and in the previous study~\cite{Iguaz2022} for different pressures~(right). The operating points and data acquisition for generating the curve in TREX-DM are the same as those used for generating the gain curve~(figure~\ref{fig:TREXDMatLSC_Ne_CurvasGain}). As in the gain curve, for a direct comparison of both curves, it would be necessary to characterize and calibrate the acquisition electronics in TREX-DM, so that the correspondence between ADCs and charge received in the channels could be obtained, allowing the calculation of the absolute gain in the Micromegas of TREX-DM.}
    \label{fig:TREXDMatLSC_Ne_CurvasResol}
\end{figure}

At 3~bar of pressure, an electron transmission curve was performed (figure~\ref{fig:TREXDMatLSC_Ne_CurvasTransp}) to compare the results with those obtained in previous studies with a smaller Micromegas~\cite{Iguaz2022} and also to define the optimal drift field for the TREX-DM TPC. After transitioning from 3 to 4~bar, a gain curve (figure~\ref{fig:TREXDMatLSC_Ne_CurvasGain}) and an energy resolution curve (figure~\ref{fig:TREXDMatLSC_Ne_CurvasResol}) were generated to make the same comparison with previous studies. It is worth noting that a more rigorous study would require, for each pressure, taking calibrations with different amplification fields combined with different drift fields, to develop the gain, transparency, energy resolution, and energy threshold curves. However, since these studies on the detector's response had previously been conducted in the laboratory for the same type of Micromegas~\cite{Iguaz2022}, the results from this study would be used to define most of the operating points throughout the series without needing to perform a complete characterization for each pressure in the experiment, which would entail additional time and effort.

Figure~\ref{fig:TREXDMatLSC_Ne_CurvasTransp} shows the electron transmission curve generated for TREX-DM compared to that generated in previous studies. The curve generated in the previous studies does not include the one corresponding to the pressure of 3~bar, although its shape can be estimated from the trend observed at the other pressures, indicating that the maximum transmission zone at 3~bar would be reached at approximately~V$_{drift}$/V$_{amp}$~=~5·10$^{-3}$. With this estimation, it can be said that the curve in TREX-DM seems to have already reached this maximum transmission zone. However, although the errors in this curve are large due to the high energy resolution achieved ($>$30\%), there is a sharp loss of transmission in the last three points, indicating an anomalous and unexpected behavior, likely due to geometric effects or distortions of the electric field at the edges of the Micromegas.

Figure~\ref{fig:TREXDMatLSC_Ne_CurvasGain} shows the gain curve generated for TREX-DM compared to that generated in previous studies. As with the transparency curves at 3~bar, the previous studies do not include the one corresponding to the pressure of 4~bar, but it can also be approximately extrapolated, estimating that the maximum gain that would be reached at 4~bar in the previous studies would be around 90~kV/cm, 17~kV/cm above the 73~kV/cm achieved in TREX-DM. It is important to reiterate that when the TREX-DM curve was generated, the leakage currents were what limited the detector's gain, not the generation of sparks. Aside from this, the slope of the curve is similar in both cases, indicating appropriate behavior of the Micromegas.

Using the same data used to generate the gain curve, the energy resolution curve shown in figure~\ref{fig:TREXDMatLSC_Ne_CurvasResol} has been generated, where it is compared with that generated in previous studies. The range of 50~$-$~75~kV/cm for which the curve in TREX-DM is drawn corresponds approximately to the range of absolute gain of 1·10$^2$~$-$~3·10$^3$ in the curve from the previous studies, if this correspondence is made with the support of the gain curve from these previous studies. The trend of the curve in TREX-DM is similar; however, the resolution values are much higher in TREX-DM. This could be due to several factors: greater diffusion due to the longer drift distance in TREX-DM, 19~cm compared to 1~cm in the previous studies, which causes the charge to be distributed among a greater number of channels, and if this charge is not sufficient for the signal to exceed the noise threshold, then this charge will not be read and will be lost; recombination and attachment processes also produced by the longer drift distance and possibly also by the quality of the gas (oxygen molecules, with high electronegativity, capture low-energy electrons during the collision~\cite{Gutierrez2011}); edge effects, which can cause part of the charge to be lost, so that through a more detailed analysis, applying cuts in the central area of the detector, the resolution values would improve; damaged channels where the charge they collect is lost.

\subsection{Leakage Currents and Stability Issues}
\label{subsec:CorrientesDeFuga}

\begin{figure}[!ht]
    \centering
    \includegraphics[height=3.2cm]{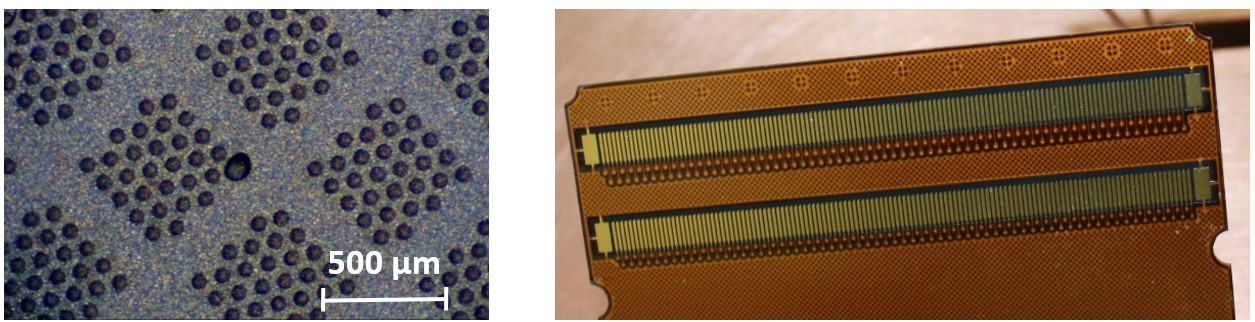}
    \caption{$\bullet$~Left:~photograph of an area of the mesh with a damaged hole. $\bullet$~Right:~connection footprint of the Fujipoly connector with a separation between channels of 150~$\mu$m.}
    \label{fig:AgujerosHuellaFujipolly}
\end{figure}

As the pressure in the chamber increased, and with it the amplification field, leakage currents began to appear in the connection chain from the Micromegas to the electronics, which often hindered the operation of the detector. These leakage currents originated from the \textit{strips} that, due to sparks produced in some of their amplification holes, had become short-circuited with the mesh or had a low resistance to it (see figure~\ref{fig:AgujerosHuellaFujipolly}). This degradation in some of the \textit{strips} while operating a Micromegas during the commissioning phase, until a stable operating point is reached, is common and increases the likelihood of its occurrence as the amplification field is increased. To minimize these damages, it is essential to control the gas conditions and not to overdrive the Micromegas.

To prevent a channel short-circuited with the mesh, or with low resistance between the mesh and a \textit{strip}, from causing leakage currents through the electronics that connects each \textit{strip} to ground (figure~\ref{fig:SR2M_EsquemaConexionCanal}), an interface card was designed to be installed in the connection chain between the Micromegas and the acquisition electronics, allowing these damaged channels to be disconnected, that is, to leave them floating. The number of damaged channels increased as the pressure in the TPC rose and with it the voltage in the Micromegas, reaching over 20\% of the channels damaged.

However, despite identifying the damaged channels and disconnecting them from the electronics with the help of the interface card, leakage currents continued to occur in the fujipoly connectors used to connect the Micromegas to the flat cables and also to connect the flat cables to the interface card. These connectors, which were introduced into the design due to their low radioactivity~\cite{Castel2019}, had a pitch of 150~$\mu$m (see figure~\ref{fig:AgujerosHuellaFujipolly}). When a channel was at the voltage of the mesh, and despite being disconnected from the electronics thanks to the interface card, and therefore disconnected from ground, it continued to cause current leaks to the adjacent channels in the area of the fujipoly connector.

These leakage currents were 1 or 2 orders of magnitude above the current injection value in the mesh (in nanoamperes), causing problems in monitoring the injection current in the mesh and, given their random variation, issues in stabilizing the detector. Furthermore, these leakage currents often caused significant noise fluctuations during data acquisition, triggering the acquisition rate and making it very difficult to control.

\begin{figure}[!ht]
    \centering
    \includegraphics[width=0.99\textwidth]{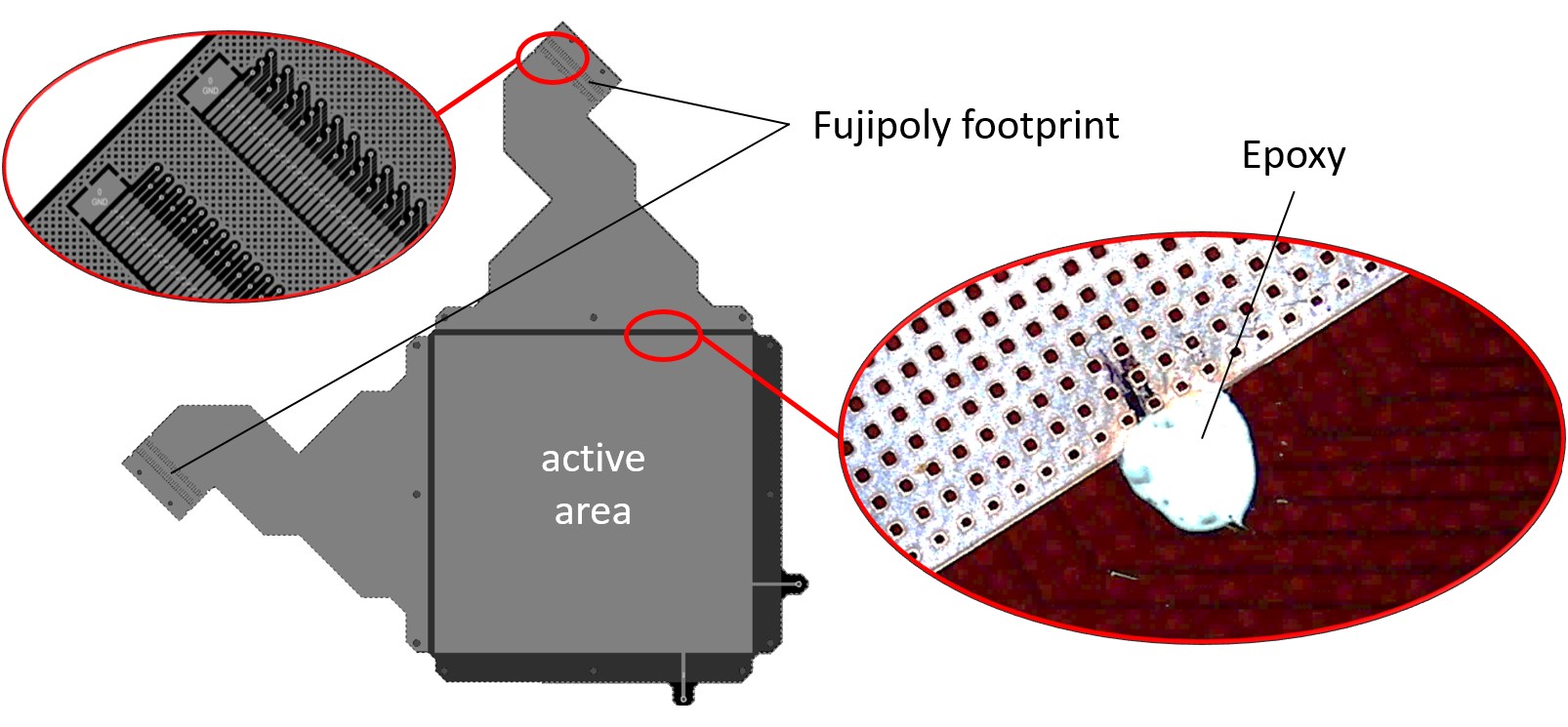}
    \caption{Micromegas installed in TREX-DM. $\bullet$~Upper left:~enlarged detail of the Fujipoly connector footprint. The small distance between the channel footprints (150~$\mu$m) and between the vias and the ground plane (200~$\mu$m) can be observed. $\bullet$~Right:~enlarged detail of the area where a channel was repaired. The central spot is a drop of epoxy applied in the area after cutting the problematic channel with a scalpel. The channels located in the intermediate layer can be recognized. The area with a square hole pattern is the ground plane that extends along the Micromegas tab to the connector. In this tab, the channels run through the intermediate layer with two ground planes, one upper and one lower.}
    \label{fig:TREXDM_ReparacionCanalMicromegas}
\end{figure}

The problems with the leakage currents necessitated two interventions throughout the year to attempt to eliminate them. The interventions involved stopping the detector, emptying the chamber, disconnecting the gas system and other subsystems, opening the upper lead shielding, transporting the chamber to the clean room of the LSC, opening the chamber, and performing work inside the chamber. Once the work in the clean room was completed, all operations were executed again in reverse order to restart the detector. These interventions lasted a total of about 4 to 5 weeks. Below is a chronological outline of the detector's operation periods and interventions carried out during 2019:

\begin{itemize}
	\item January-April: Change from the Ar~+~1\%~Isobutane mixture to Ne~+~2\%~Isobutane. First calibrations at 1.5~bar. Pressure increase to 3~bar and systematic measurements for the characterization of the Micromegas (gain, transparency, energy resolution curves). Pressure increase to 4~bar and calibration measurements. At this point, problematic leakage currents begin to appear.
	\item May-June: Intervention to attempt to eliminate the leakage currents in the fujipoly connectors. Additionally, the $^{109}$Cd source had to be reinstalled, as it had moved from its position and could not be locked for background data acquisition. Apart from the leakage currents in the fujipoly connectors, a leak was located from a channel in the Micromegas directly to ground, in the Micromegas circuit, which had to be repaired by cutting this channel once it was outside the active area (figure~\ref{fig:TREXDM_ReparacionCanalMicromegas}). This operation eliminated the leak, confirming that the channel-ground connection was located at some point along the path from the cut to the connector.
	\item July-September: Calibrations and background measurements at different pressures (1.5, 3, 4, 6, and 8~bar). Leakage currents began to appear from 4~bar of pressure. Despite this, pressure was continued to be increased to 6 and 8~bar, while attempts were made to keep these leakage currents controlled without affecting the stability of the detector. However, at a pressure of 8~bar, they began to be too high ($\sim$1000~nA), making it impossible to maintain the detector stable and in a safe operating mode.
	\item October-November: Intervention to solve the leakage currents in the fujipoly connectors and cut another channel short-circuited with ground in the Micromegas circuit.
\end{itemize}

\section{Energy Threshold}
\label{subsec:umbralNeon}

Special mention should be made of the study of the energy threshold in the series of data obtained at different pressures, as it is one of the keys to the experiment in the search for low-mass WIMPs.

\begin{figure}[!ht]
    \centering
    \includegraphics[width=0.9\textwidth]{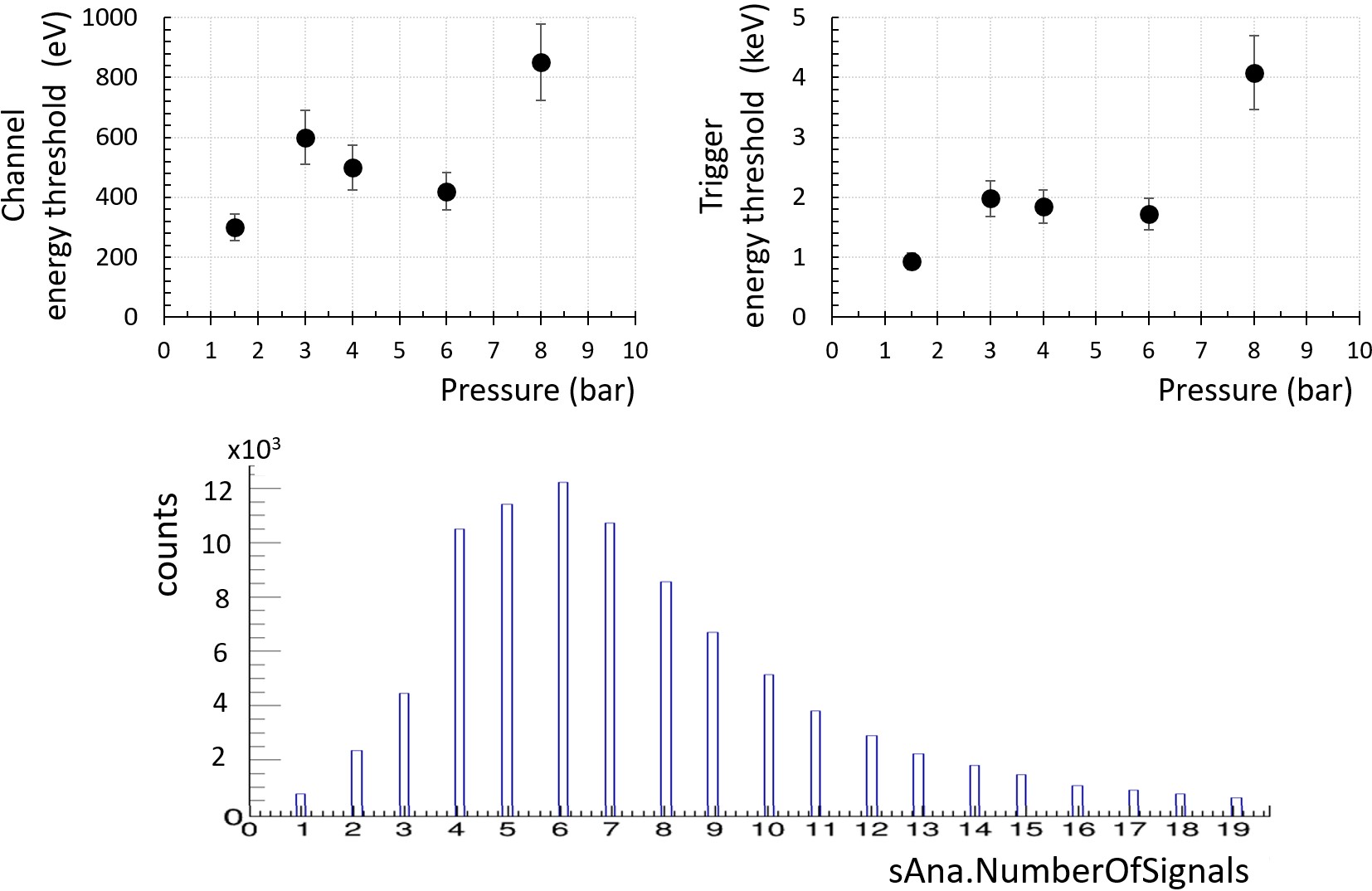}
    \caption{Curves of the 'energy threshold per channel' (upper left) and 'trigger energy threshold' (upper right) for the different pressures in the first data series with Ne~+~2\%~Isobutane, along with the histogram of the number of channels per event with signal above the noise during the calibration at 1.5~bar (lower).}
    \label{fig:TREXDMatLSC_Ne_CurvasUmbralEnergia}
\end{figure}

As explained in the commissioning with Ar~+~1\%~Isobutane, we can define an 'energy threshold per channel' and a 'trigger energy threshold.' Figure~\ref{fig:TREXDMatLSC_Ne_CurvasUmbralEnergia} shows the values of both thresholds for the five operating points reached, at five different pressure values, in Ne~+~2\%~Isobutane. There is a clear relationship between the energy threshold that can be reached in a channel and the trigger energy threshold, or effective threshold. This relationship is determined by the trigger efficiency in the acquisition electronics, which in turn has a strong dependence on multiplicity, that is, the number of channels that exceed the programmed threshold simultaneously (figure~\ref{fig:EsquemaMultiplicity}). This dependence of the trigger on multiplicity can also be seen in figure~\ref{fig:TREXDMatLSC_Ne_CurvasUmbralEnergia}, where the number of channels per event that have exceeded the noise level is presented, showing how the trigger efficiency decreases when the multiplicity is below 6 channels.

On the other hand, there is a high variability in the energy threshold with pressure. It would be expected that the energy threshold would be the same for each operating point, as the maximum gain achieved in a Micromegas~\cite{Iguaz2022} and the noise level in the electronics do not depend on pressure. However, this variability was largely due to the difficulty in finding the maximum gain and maintaining stable operation of the detector at each operating point, especially due to leakage currents. This meant that, particularly at higher pressures, where the applied voltage is greater and therefore the leakage currents as well, the detector was operated at a gain below the maximum. In addition, for the case of the trigger energy threshold, the number of channels over which the charge is distributed decreases as pressure increases, due to reduced diffusion, which also reduces the trigger efficiency.

After conducting the series of measurements at different pressures, and in an attempt to keep the detector stable, it was decided to operate at a single point at 4~bar, until the Micromegas were replaced with newly designed ones (chapter~\ref{ch:TREXDMoptimization}). During the operation of the detector at 4~bar, and thanks to efforts in noise reduction, energy thresholds of around 1.0~keV were achieved.

\begin{figure}[!ht]
    \centering
    \includegraphics[width=0.55\textwidth]{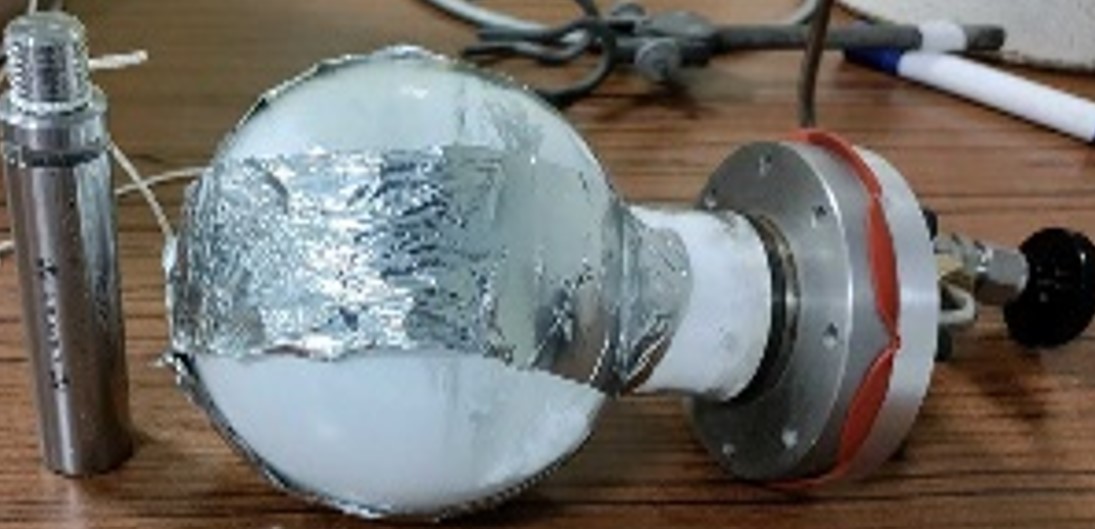}
    \caption{Photograph of the container with Calcium powder during irradiation tests.}
    \label{fig:TREXatLSC_37ArPolvo}
\end{figure}

As part of the strategy to reduce the energy threshold in TREX-DM, a new calibration system for the experiment has been developed over the last two years within the group, based on the emanation of the isotope $^{37}$Ar from a small amount of Calcium powder (figure~\ref{fig:TREXatLSC_37ArPolvo}), which has previously been subjected to a neutron flux for activation. This method has already been successfully used in low-background experiments such as XENON-1T~\cite{Aprile2023} or NEWS-G~\cite{Durnford2021}.

The isotope $^{37}$Ar has a half-life of 35.04 days and decays via electron capture, emitting photons of 2.82~keV (K shell with a probability of 90\%) and 0.27~keV (L shell with a probability of 9\%) in its decay. The objective of this development is to calibrate TREX-DM at increasingly lower energies, as part of the strategy to reduce the energy threshold.

The successful implementation of this calibration system will allow for a more precise determination of the effective energy threshold of TREX-DM.

\section{First Background Measurements}
\label{sec:BackgroundFisrtMeasurements}

The operation of the detector at different pressures during 2019, which allowed for the series of measurements described earlier, occasionally enabled stable background data collection over short to medium exposure times (days), despite the issues with leakage currents and their consequences in terms of gain, noise, and stability.

\begin{figure}[!ht]
    \centering
    \includegraphics[width=0.9\textwidth]{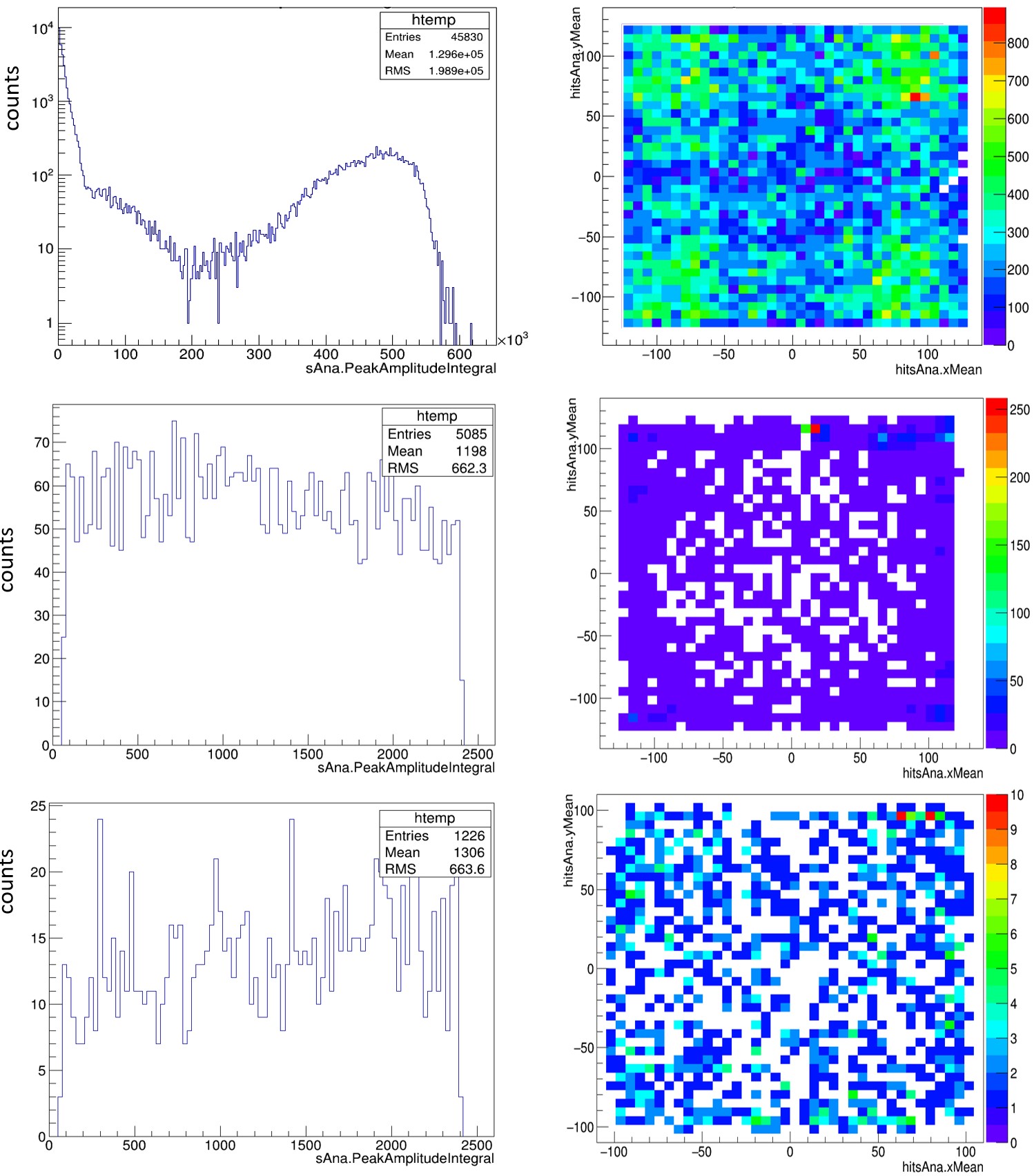}
    \caption{Results of the background measurements over 7 days with Ne~+~2\%~Isobutane at 3~bar. On the left, the energy spectra are presented, represented by the observable 'sAna.PeakAmplitudeIntegral', and on the right of each spectrum, the corresponding activity map is shown. $\bullet$~Top:~complete spectrum from the data collection over 7 days. $\bullet$~Center:~spectrum in the low-energy range ($\sim$1.2-7~keV). $\bullet$~Bottom:~spectrum in the low-energy range ($\sim$1.2-7~keV) and in the centered area of the detector of 20~x~20~cm$^2$ (the total area of the detector is 25~x~25~cm$^2$).}
    \label{fig:TREXDMatLSC_FirstBackground_Feb2019}
\end{figure}

The results of one of the first background measurements can be seen in figure~\ref{fig:TREXDMatLSC_FirstBackground_Feb2019}. When observing the spectrum over the full energy range, three populations can be differentiated: in the low-energy region up to 50·10$^3$ADC, the background decreases exponentially; between 50·10$^3$ and 200·10$^3$ADC, the slope of the background decline reduces; starting from 200·10$^3$ADC, the background begins to rise again, only to drop drastically around 600·10$^3$ADC.

In the lower energy region (figure~\ref{fig:TREXDMatLSC_FirstBackground_Feb2019}~center), which includes part of the target interest region of TREX-DM for the search for WIMPs (0.4-7~keV), the background is flat and yields an initial value of 2.9·10$^3$~dru (counts~keV${^{-1}}$~kg${^{-1}}$~d${^{-1}}$), nearly three orders of magnitude greater than predicted by the background model~\cite{Castel2019}. It is worth noting that the energy threshold during these measurements was around 1.2~keV, above the target threshold of the experiment of 0.4~keV.

If only the events collected in the central area of 20~x~20~cm$^2$ of the detector are considered for the lower energy region, the background value reduces to 1.1·10$^3$~dru. This suggests that there may be surface contamination on the internal walls of the active area.

This background value has been calculated from a basic analysis; however, while it is expected that a more in-depth and sophisticated analysis could reduce it by some factor, it is sufficient to conclude that there is some contamination in the detector causing this unexpected activity.

Although the first stable background measurements occurred at an operating point of 3~bar pressure, there were some operating points at 4, 6, and 8~bar that also allowed for background data collection, although typically the exposures were short due to the instability of the detector. Nevertheless, background values could be quantified at different pressures, all of which were consistent with the first background value measured at 3~bar. It is noteworthy that with the increase in pressure, the background events that appeared to originate from the walls had a shorter mean path, which allowed for a slight increase in the area of 20~x~20~cm$^2$ from the previous analysis at 3~bar, reducing the background value to 0.98·10$^3$~dru.

\section{Radon-Dominated Background Reduction}
\label{sec:RadonDominatedBackgroundReduction}

In the second intervention, in October-November 2019, the leakage currents occurring in the fujipoly connectors were reduced, but not completely eliminated. After this intervention, it became very difficult to maintain the detector stable with a voltage on the mesh above 350-400~V, as these leaks, which were strongly dependent on the voltage applied to the mesh, increased considerably above those voltage values. With these voltages of 350-400~V on the mesh and to achieve an acceptable amplification that would result in an energy threshold around 1~keV, since these leakage currents also affected the noise level in the experiment, it was decided to operate at 4~bar, which would allow for the study of the origin of the background that had been quantified in this first phase.

In parallel to these studies, a new design for the Micromegas and a new connection chain would be initiated to avoid these leakage currents (section~\ref{ch:TREXDMoptimization}). Furthermore, during the design of the new Micromegas, efforts would be made in the manufacturing process to attempt to reduce the contamination of $^{40}$K measured~\cite{Castel2019}.

\begin{figure}[!ht]
    \centering
    \includegraphics[width=0.8\textwidth]{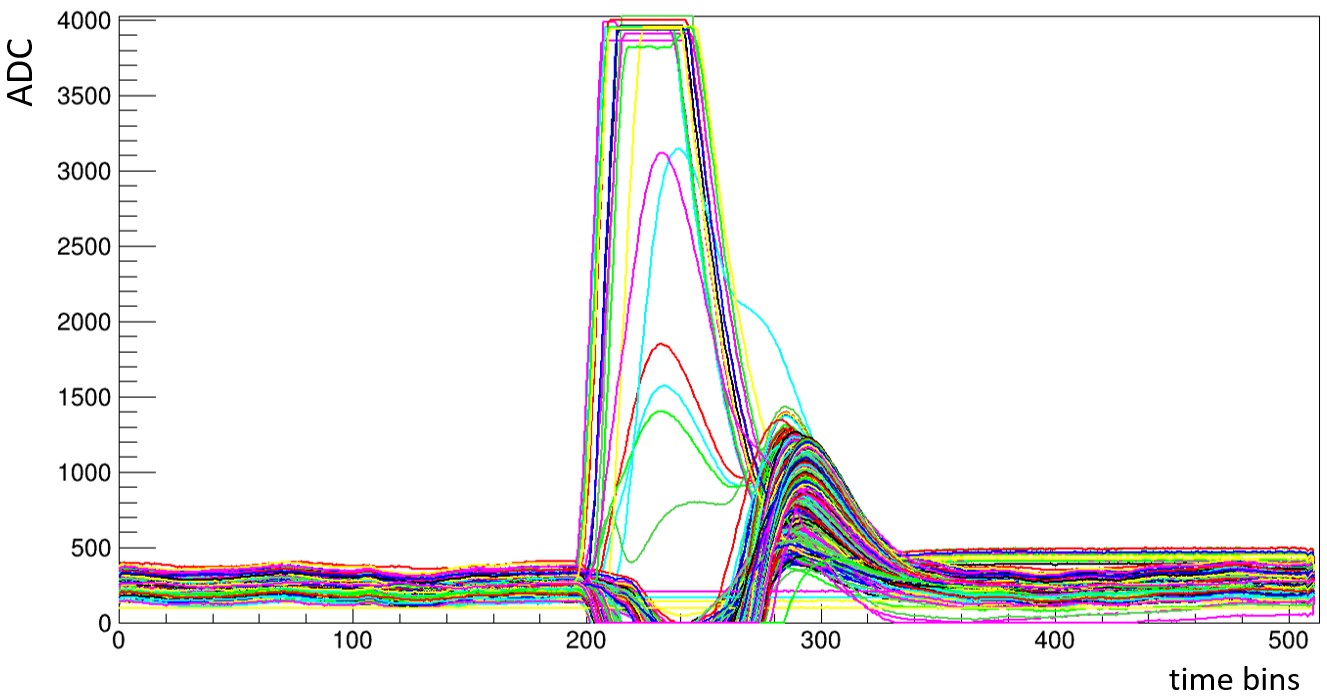}
    \caption{Event with saturation in some channels during a background data acquisition whose spectrum can be seen in figure~\ref{fig:TREXDMatLSC_FirstBackground_Feb2019}. This event has a PeakAmplitudeIntegral of 554·10$^3$~ADC, at the high-energy end of the spectrum.}
    \label{fig:TREXDMatLSC_EventoSaturadoHE}
\end{figure}

In figure~\ref{fig:TREXDMatLSC_FirstBackground_Feb2019}, it can be seen that the energy spectrum, represented by the observable PeakAmplitudeIntegral, contains a significant population of events on the right. Many of these events had energies such that many of the channels exceeded the dynamic range of the electronics and reached saturation (see figure~\ref{fig:TREXDMatLSC_EventoSaturadoHE}). This fact raised suspicions of contamination in the chamber with high-energy events, likely alpha particles, which led to the decision to proportionally reduce the gain in the Micromegas to be able to detect and collect these high-energy events without saturating the electronics.

\begin{figure}[!ht]
    \centering
    \includegraphics[width=0.7\textwidth]{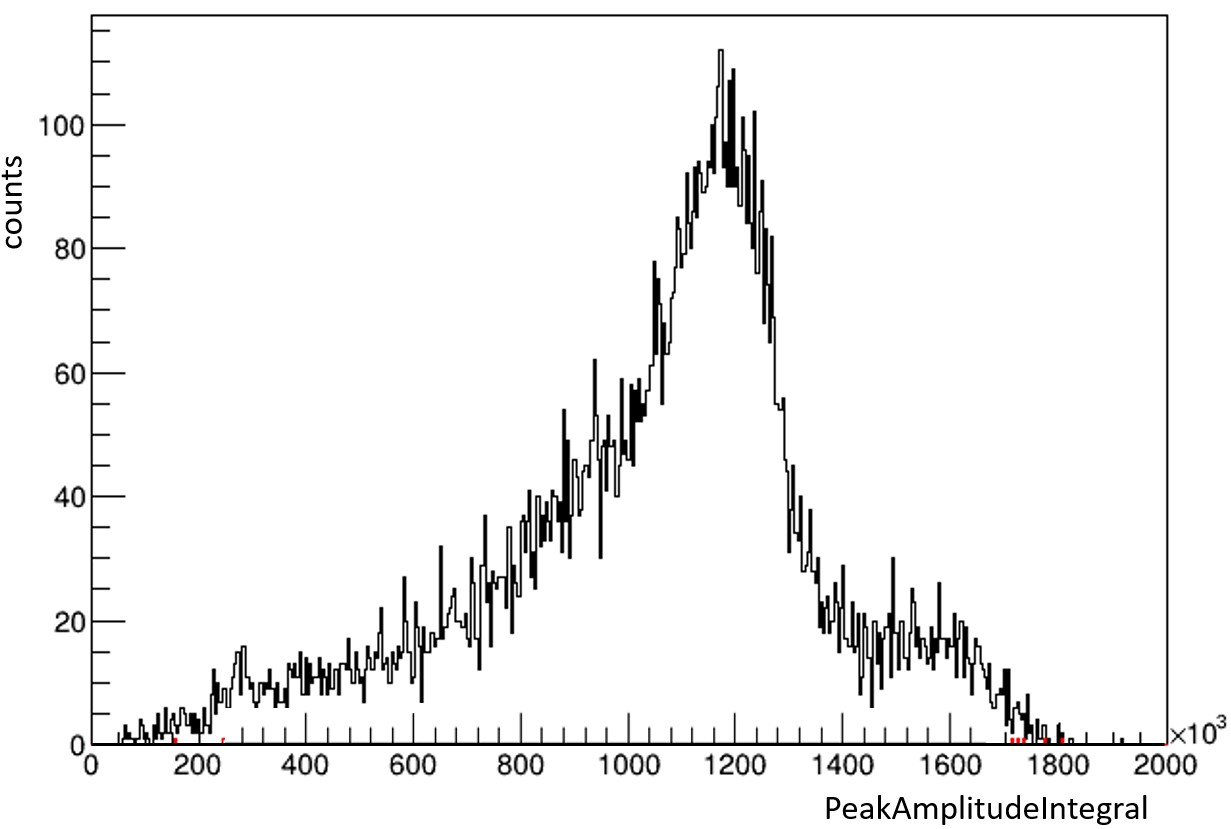}
    \caption{(Figure presented in the semiannual report for the scientific committee of the LSC in 2020) Energy spectrum during a background data acquisition (42 hours) at low gain, for high-energy events, with $V_{mesh}$~=~250~V. The most probable hypothesis is that the two distinguished peaks correspond to alphas of 5.3~MeV and 7.7~MeV from ${^{210}}$Pb and ${^{214}}$Po respectively.}
    \label{fig:TREXDMatLSC_HEespectro}
\end{figure}

Figure~\ref{fig:TREXDMatLSC_HEespectro} shows the result after a background data acquisition with low gain ($V_{mesh}=250V$). The events from the $^{109}$Cd source fell below the energy threshold in this data acquisition, making energy calibration impossible. The two peaks that are distinguished likely correspond to alpha particles of 5.3~MeV and 7.7~MeV from ${^{210}}$Pb and ${^{214}}$Po respectively, originating from the decay of the most common radon isotope, ${^{222}}$Rn.

If there was radon contamination in the TPC, it was expected that the low-energy electron emissions associated with its decay chain (see figure~\ref{fig:AlphaCAMM_NaturalChains}) would contribute to the background in the region of interest (0.4$-$7~keV). Moreover, these electron emissions were not taken into account in the background model, as radon contamination was not anticipated, which could explain all or part of the difference found between the measured background value and that predicted by the background model. This was later confirmed when, searching for a possible radon emanator within the gas system, the oxygen filter was isolated, resulting in a factor of~10 reduction in the activity of the previously detected high-energy events. Most importantly, this test also yielded almost the same factor of reduction in the low-energy events in the region of interest.

After discovering that the oxygen filter was a significant radon emanator within the gas system, other types of filters were tested, although none yielded satisfactory results. Ultimately, it was decided to operate the detector in an open loop, allowing the gas to flow directly from the mixing bottle to the TPC, bypassing the oxygen filter. A prior check was made of the minimum flow rate necessary to ensure that the degradation of the gas within the chamber would not negatively impact the gain and stability of the detector, concluding that with a small flow rate of~$\sim$1~l/h, the gas remained in good condition.

With this new mode of operation in an open loop, background measurements were taken that yielded background values in the low-energy region ($\sim$1$-$7~keV) of~$\sim$100~dru, an order of magnitude below the first measured value.

\setlength{\baselineskip}{12pt} 
\selectlanguage{USenglish}
\graphicspath{{../Imagenes/}{../../Imagenes/}} 

\renewcommand{\listtablename}{Table of contents}
\renewcommand{\tablename}{Table} 

\chapter{New Readout Plane for TREX-DM}
\label{ch:TREXDMoptimization}

\begin{figure}[!ht]
    \centering
    \includegraphics[width=0.9\textwidth]{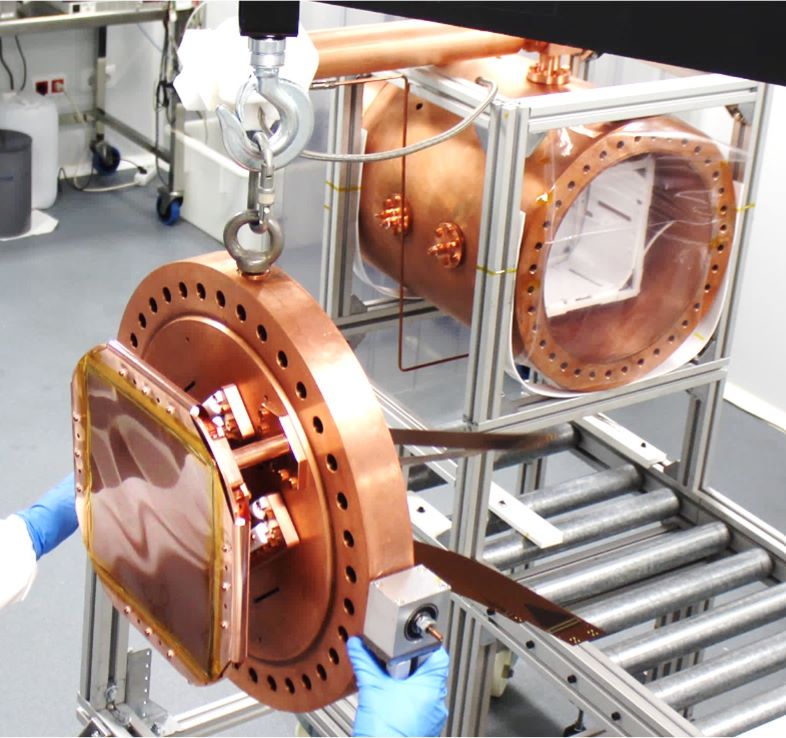}
    \caption*{}
    \label{fig:TREXDMoptimization_EnSalaBlanca}
\end{figure}

\newpage

In Chapter \ref{ch:TREXDMatLSC}, the commissioning process of the TREX-DM detector at the LSC during the first half of 2018, the initial calibrations with Ar~+~1\%~Isobutane during the second half of 2018, and the phase with Ne~+~2\%~Isobutane throughout 2019 have been described. In this last phase with Ne~+~2\%~Isobutane, as the pressure was increased, and consequently the amplification field in the Micromegas, leak currents began to appear in the fujipoly connectors that were installed at four points in the connection chain from the Micromegas to the acquisition electronics: two connectors located inside the chamber that connected the Micromegas circuit with the flat cables and two other connectors outside the chamber that connected the flat cables with the acquisition electronics.

Despite the efforts made to eliminate or reduce these leak currents, it was not possible to maintain the detector stable above 4~bar, so it was decided to keep the detector at this pressure and work on identifying the source of the background that had already been quantified. In parallel to this work, the design of new Micromegas and a new connection system that would eliminate these leak currents began. Furthermore, and no less importantly, during the design of these new Micromegas, efforts would be made in the manufacturing process to attempt to reduce the contamination of $^{40}$K measured up to that point~\cite{Castel2019}.

\begin{figure}
    \centering
    \includegraphics[width=0.85\textwidth]{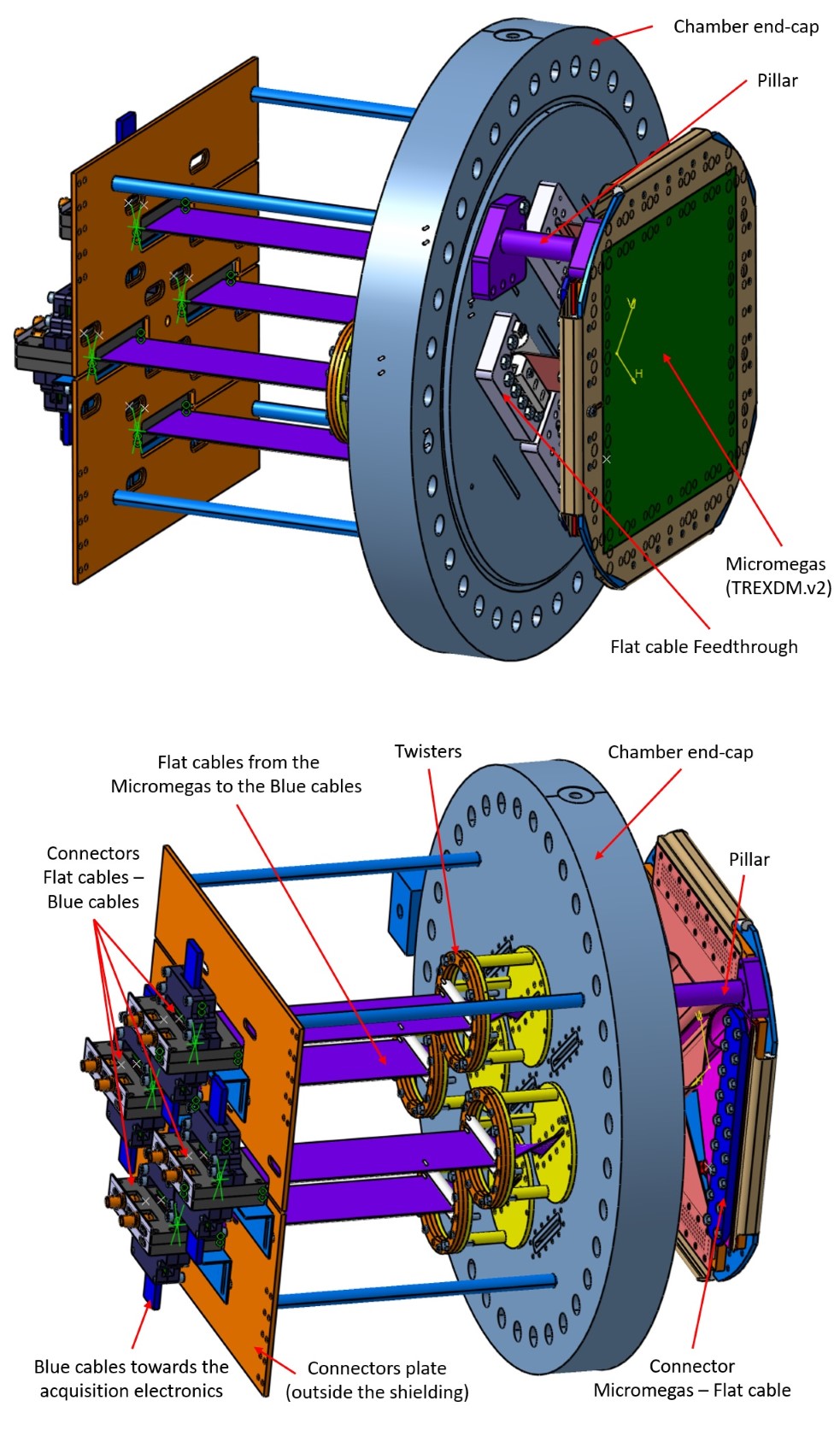}
    \caption{3D view of the new readout plane for TREX-DM with the new Micromegas, including the connection chain to the electronics.}
    \label{fig:TREXDM_Readout_ElectronicsChain_Catia}
\end{figure}

The installation of the new elements had to be done on the same chamber already certified for a working pressure of 10~bar, with minimal modifications possible. With these premises, the system designed can be seen in Figure \ref{fig:TREXDM_Readout_ElectronicsChain_Catia}. The new Micromegas connects via four 'Face to Face' (FtF) connectors (see Section~\ref{sec:FtFconnector}) to four radiopure flat cables that exit the chamber to the external wall of the lead shielding. There, the ends of the flat cables connect to the electronics through flexible and shielded cables already used in the previous version (blue cables).

This chapter describes in detail the design phase of the new Micromegas, the flat cables, and the feedthroughs. After this phase and its subsequent manufacturing, the installation and commissioning of the new readout plane are described. Finally, the first data taken with this readout plane in Ne~+~2\%~Isobutane are analyzed.


\section{New Micromegas}
\label{sec:NewReadouts}

The leakage currents that had been observed originated from the \textit{strips} which, due to sparks or the geometric variability of the amplification holes during manufacturing, had become short-circuited with the mesh or had a low resistance to it (see defect in a hole of the mesh in figure \ref{fig:AgujerosHuellaFujipolly}). This degradation in some holes is common during the startup phase of a Micromegas, until a stable operating point is reached, and the probability of degradation increases as the amplification field is raised in search of maximum gain. This degradation, along with the consequences of having some channels with a voltage equal to or very close to that of the mesh, was not taken into account in the design of the first Micromegas installed in TREX-DM.

To account for this degradation effect, during the design of the new Micromegas, the circuit areas were meticulously studied to avoid leakage from a channel to ground or from one channel to another (the channels are grounded through the electronics, figure~\ref{fig:SR2M_EsquemaConexionCanal}). This study was conducted assuming that a channel could reach a potential of 575 V, which corresponds to the voltage at maximum gain achieved in Ar~+~1\%~Isobutane at 10~bar according to figure~\ref{fig:CurvasArticuloIguaz2022} (115~kV/cm in a gap of 50·10$^{-4}$~cm). Similar voltages are reached in Ne~+~2\%~Isobutane.

With the experience accumulated during the operation of the detector, along with the detailed study of the areas in the circuit susceptible to current leakage, the design parameters for the next Micromegas were determined and can be seen in table~\ref{tab:TREXDM_MM_ParametrosDiseño}.

\begin{table}[!h]
\centering
\begin{tabular}{|c|c|c|}
\hline
\textbf{Parameter} & \textbf{MM v1} & \textbf{MM v2} \\ \hline
Minimum distance between channels ($\mu$m) & 75 & 500  \\
Minimum distance channel vias - ground ($\mu$m) & 200 & 4000  \\
Number of pads (units) & 2 & 4  \\
Distance between pads in the connector ($\mu$m) & 150 (fujipoly) & 4000 (FtF)  \\
Amplification distance ($\mu$m) & 50 & 50  \\
Distance between \textit{strips} in the active area ($\mu$m) & 50 & 100  \\
Hole pattern (Diameter-Pitch) ($\mu$m) & 50 - 100 & 60 - 110  \\
\hline
\end{tabular}
\caption{Comparison of design parameters between Micromegas~v1 (installed in TREXDM in 2018) and Micromegas~v2 (installed in TREXDM in 2022).}
\label{tab:TREXDM_MM_ParametrosDiseño}
\end{table}

\subsection{Study by Circuit Zones}

During this study, for the circuit elements (channels and grounds) that can have potential differences of up to 575~V between them, several cases were considered based on the location of these elements in the circuit: situated in the same layer, situated in different layers, and the particular case of the vias that connect multiple layers.

\begin{figure}
    \centering
    \includegraphics[width=\textwidth]{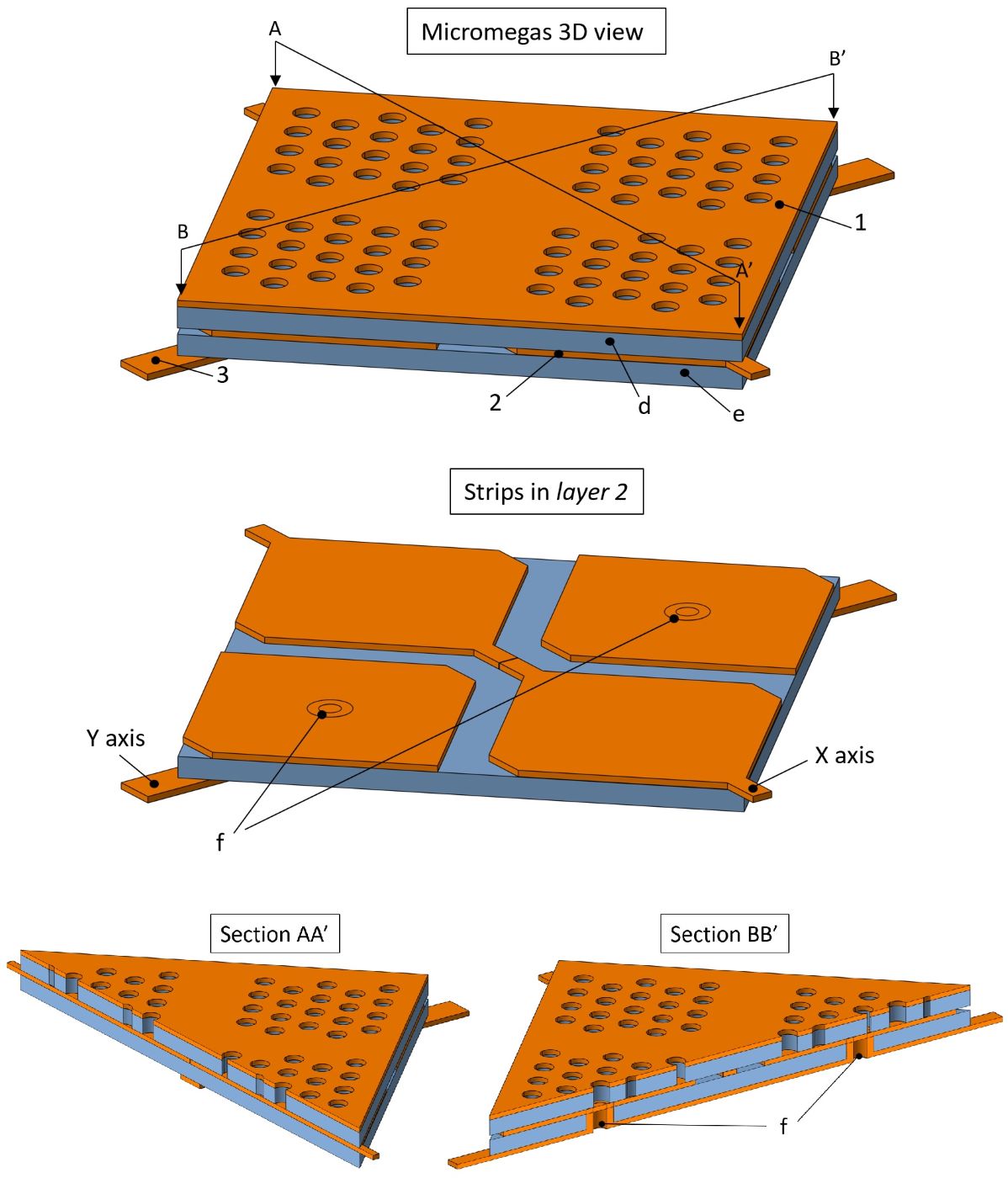}
    \caption{$\bullet$ Top:~3D view of a small area of the active region in a microbulk Micromegas, with the holes where electron amplification occurs through the avalanche effect. $\bullet$~Middle:~geometry of the strips in layer two. $\bullet$~Bottom:~sections AA' and BB' of the upper image. 
    In these images, the elements can be differentiated with the following labels: (1)~layer one with the mesh (cathode) and the amplification holes; (2)~layer two with pixels connected to each other in the same layer forming the X-axis strips and pixels connected to each other through layer three forming the Y-axis strips; (3)~layer three where the tracks connecting the Y-axis pixels are located; (d)~kapton layer separating layers one and two; (e)~kapton layer separating layers two and three; (f)~vias connecting the pixels of layer two with the tracks of layer three to form the Y-axis strips.}
    \label{fig:TREXDM_MM_Layers_Catia}
\end{figure}

The active area of a microbulk Micromegas~\cite{Andriamonje2010} has at least three copper layers separated by two kapton layers (figure \ref{fig:TREXDM_MM_Layers_Catia}). In addition to these three copper layers, other layers may be added at the bottom of the circuit as needed, but it is common to add another kapton layer to insulate layer three, which contains the Y-axis tracks. Outside the active area, in the routing of the channels to the connectors, all necessary layers can be used, but it is typical to use the same number of layers as in the active area, routing the channels through layer two and leaving layers one and three as ground planes for channel shielding.

In the case of TREX-DM, the thickness of the copper layers is 17~$\mu$m, which is the thickness commonly used in the manufacture of GEMs and Micromegas, and it is not usually changed unless specific needs require it. The thickness of the kapton layers in the active area is 50~$\mu$m, defined by the amplification gap.

\subsubsection{Elements Located in Different Layers of the Circuit}

In this case, no special measures were undertaken since the separation between layers of 50~$\mu$m of kapton in the Micromegas circuit, a distance set by the amplification \textit{gap}, ensures more than sufficient insulation. The breakdown of the dielectric, even in the most unfavorable cases where degradation occurs due to temperature and humidity \cite{Li2011}, is greater than 70~kV/mm. This field corresponds to a maximum potential difference between layers of the Micromegas circuit of over 3500~V. Assuming that the maximum voltage that will be reached between layers is 575~V, there is ample margin to absorb imperfections in the material and tolerances in the thickness of the separating kapton layer.

\begin{figure}
    \centering
    \includegraphics[width=8cm]{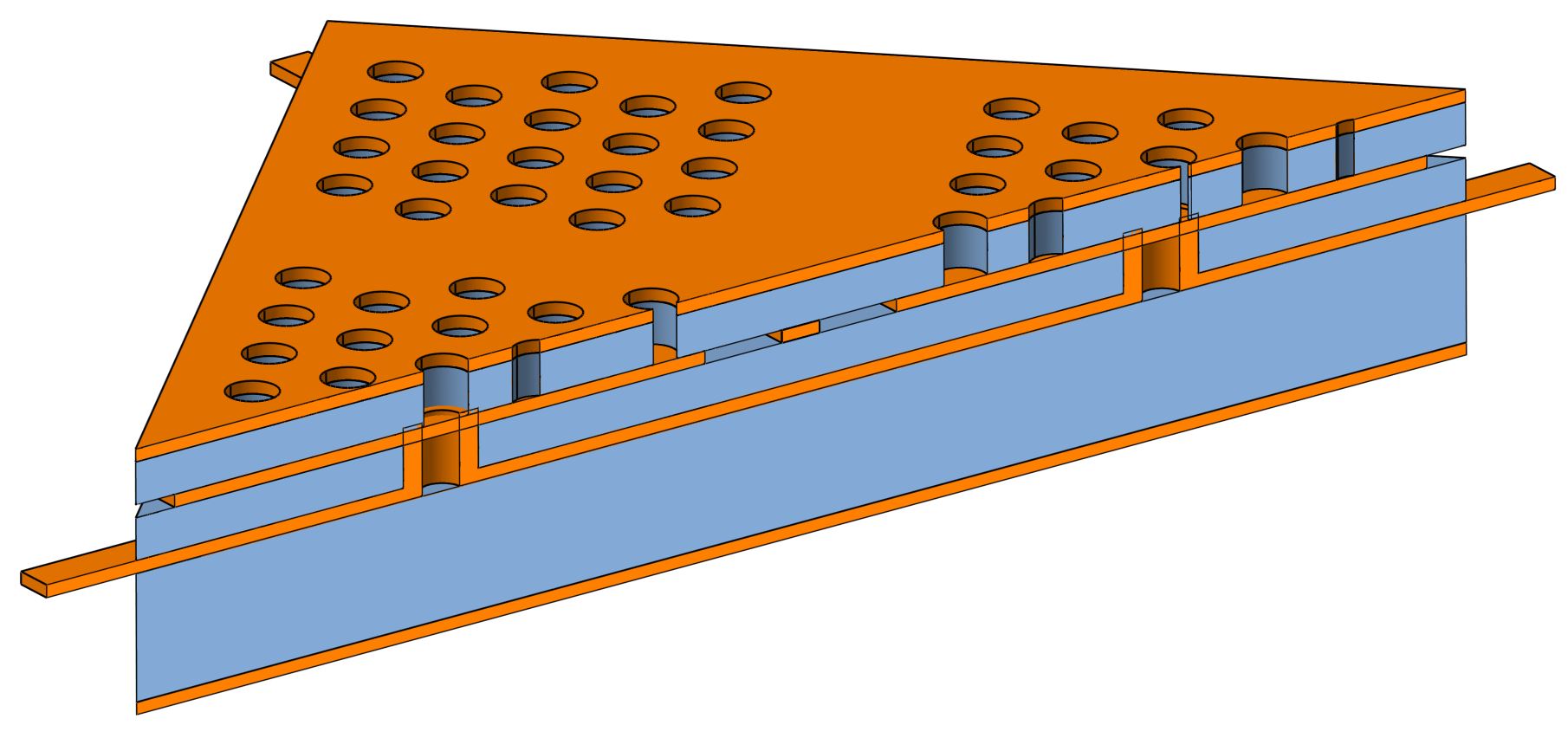}
    \caption{Section of the active area of the Micromegas with the added ground plane at the bottom, with a separation of 150~$\mu$m from layer~3 of the Micromegas circuit.}
    \label{fig:TREXDM_BottomGroundLayer}
\end{figure}

On the other hand, the Micromegas are usually glued or attached for their operation in a TPC to a surface that serves as a mechanical support. The relationship between the \textit{strips} and this support, separated by 50~$\mu$m (the lower coating of the Micromegas with 50~$\mu$m thick kapton), introduces a capacitance in the system of the same order as that which occurs between the \textit{strips} and the mesh, which are also separated by 50~$\mu$m. To reduce the \textit{strips}-ground capacitance, it was decided to add a ground plane at the bottom of the active area, 150~$\mu$m from layer~3 (figure \ref{fig:TREXDM_BottomGroundLayer}).

\subsubsection{Elements Located in the Same (Superficial) Layer of the Circuit}

In the case of a layer whose surface is in contact with the environment and has no protective coating, surface currents can occur between nearby circuit elements situated in this layer.

Current leaks in the area of the fujipoly connectors in the first Micromegas installed in TREX-DM occurred between the pads of adjacent channels or between the pads of the channels and the nearest ground plane, always due to one or more channels short-circuited with the mesh. These elements were located in the upper layer of the circuit, where the footprints of the connectors are situated.

During the operation of the first Micromegas installed in TREX-DM, it was observed that, in the connectors on the exterior of the chamber, the leakage currents were reduced when a flow of hot air was applied to the connector area. The airflow seemed to reduce the humidity in the connector area, thereby decreasing the leakage current, a relationship known in this type of material \cite{Amiour2021}.

To prevent this damage in the new Micromegas, the separation between the pads was increased from 150~$\mu$m in the previous version to 4000~$\mu$m in this new version.

\subsubsection{Elements Located in the Same (Internal) Layer of the Circuit}

In the case of an intermediate layer, where its surface is coated with the material used for bonding the layers of the circuit, and this material is uniformly distributed and adhered, the free path through which charges are conducted across the surface is eliminated.

However, having two elements in the layer that are very close to each other may result in the adhesive not adhering homogeneously and uniformly to the surface between them, leaving free paths for charge conduction.

To avoid or reduce the possibility of failure for this reason, the distance between the elements or channels in the same layer was increased from 75~$\mu$m in the previous version to 500~$\mu$m in this new version.

Furthermore, although this work does not address the study of crosstalk between channels in the Micromegas circuit, increasing the distance between channels favors the reduction of this effect.

\subsubsection{Vias in the Circuit}

Apart from the leakage currents that occurred on the surface of the fujipoly connectors, during two interventions carried out in the clean room, two leaks were located from the channels of the Micromegas to the ground planes, within the Micromegas circuit itself. These two leaks had to be repaired by cutting the two channels once they were out of the mesh (figure~\ref{fig:TREXDM_ReparacionCanalMicromegas}). This operation eliminated the leaks caused by these channels, confirming that the channel-ground connection that caused the leak was located somewhere along the path from the point where the channel was cut to the connector. The area most likely to harbor this short circuit was the one corresponding to the vias used to pass the channels from the intermediate layer to the upper layer, next to the footprint of the fujipoly connector (figures~\ref{fig:AgujerosHuellaFujipolly}~and~\ref{fig:TREXDM_ReparacionCanalMicromegas}), as this was where the distances between the channels and the ground planes were smallest. This could not be confirmed, as it would have been necessary to cut other areas of the circuit to do so, which posed the risk of damaging the Micromegas that needed to be installed as soon as possible in the chamber.

To prevent this damage in the new Micromegas, the separation between the vias (of channels or grounds) and the nearest elements was increased from 200~$\mu$m in the previous version to 4000~$\mu$m in this new version.

\subsection{Review of the Active Area}
\label{subsec:RevisionAreaActiva}

All relevant parameters were also reviewed, even if they did not have a direct implication in the cause of the leaks, including the parameters that define the active area.

There was no reason to modify the number of channels (512 channels) or the total dimensions of the active area (25 x 25 cm$^2$), but there were two incentives to review the hole pattern and the size of the pixels: on one hand, to attempt to improve the quality during manufacturing, and on the other, to try to increase the distance between pixels, so as to minimize potential leakage currents between them (elements in the same layer susceptible to being at different potentials).

Regarding the improved quality of the new active area compared to the previous one, the MPT (Micro-Pattern Technologies) group at CERN, where the microbulk Micromegas for TREX-DM are manufactured, recommended increasing the diameter of the amplification holes from 50 to 60~$\mu$m, or even more, after presenting the results obtained with the previous Micromegas. The main reason was that the drilling process of the holes or etching during manufacturing would be better controlled for larger hole sizes, thus enhancing the quality of the active area in terms of geometry. This could favor the maximum achievable voltage and therefore a greater achievable gain. However, there was also concern that an increase in the hole diameter could worsen the funnel effect in the area of the holes (figure \ref{fig:FunnelEffect}), thereby deteriorating the gain. Nevertheless, studies conducted~\cite{Attie2014} on Micromegas with different diameters and gaps indicated that the difference was small and could therefore be counteracted by the higher quality of the active area and greater achievable gains.

\begin{figure}
    \centering
    \includegraphics[width=\textwidth]{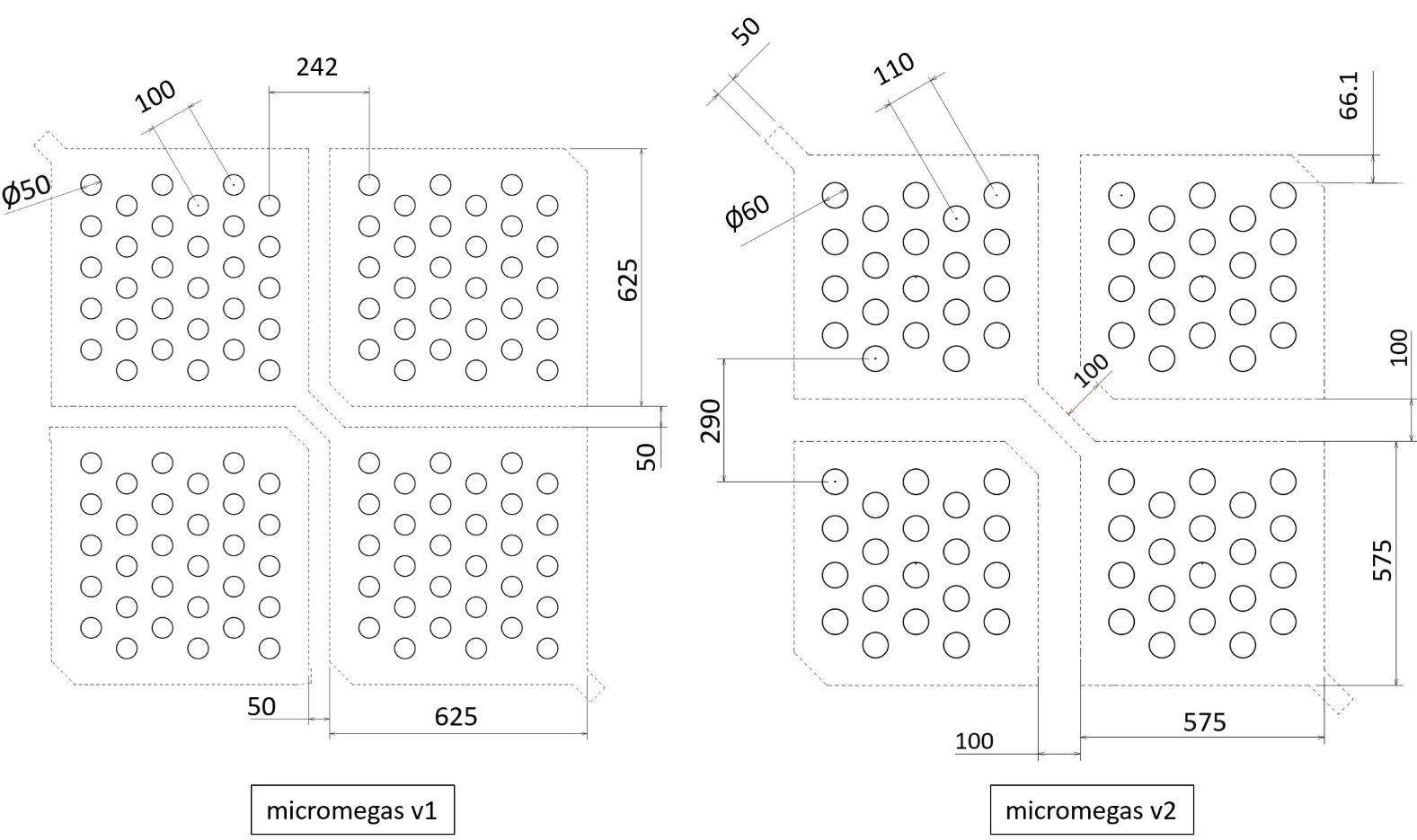}
    \caption{Hole pattern in the mesh superimposed on the geometry of the pixels/strips. $\bullet$ Left: first Micromegas installed in TREX-DM with a hole diameter of 50~$\mu$m, 100~$\mu$m center-to-center separation, and 50~$\mu$m separation between pixels. $\bullet$ Right: new Micromegas with a hole diameter of 60~$\mu$m, 110~$\mu$m center-to-center separation, and 100~$\mu$m separation between pixels.}
    \label{fig:TREXDM_PatronAgujeros}
\end{figure}

The distance between pixels was also reviewed. In the previous Micromegas, no leakage was observed in the active area, with a separation of 50~$\mu$m between the pixels (figure~\ref{fig:TREXDM_PatronAgujeros}). However, the high voltages expected in the Micromegas due to the operation of the detector at 10~bar and the potential future use of gas mixtures with a higher concentration of Isobutane, which would require even higher voltages, would increase the risk of leakage currents appearing in the area between pixels.

\begin{figure}
    \centering
    \includegraphics[width=\textwidth]{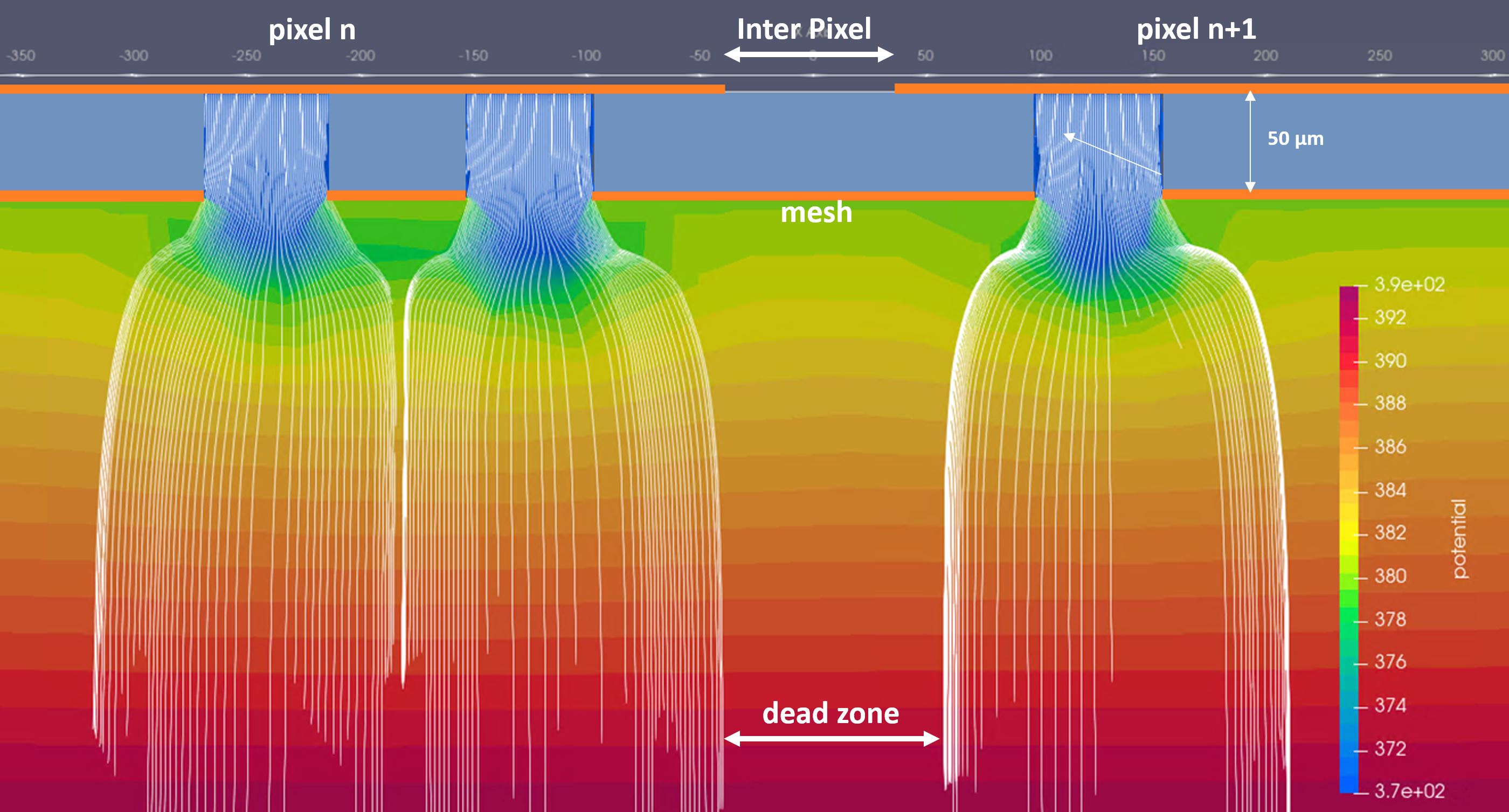}
    \caption{Simulation of the electric field near the mesh for the Micromegas D60P120IP260. Simulation conditions: 3~bar, 380~V on the mesh, and a drift field of 200~V~cm$^{-1}$~bar$^{-1}$. The dead zone between pixels is observed. It should be clarified that for simplification, no additional amplification holes have been added to the right or left of the image, as the objective was to quantify the loss of primary electrons in the area between pixels.}
    \label{fig:TREXDM_SimulacionPatron}
\end{figure}

With these incentives for reviewing the active area, the impact that these modifications could have on the collection of primary electrons from the conversion volume to the amplification holes was analyzed. For this purpose, three different patterns were chosen, and electric field simulations were performed for each of them, in the area closest to the mesh and at different pressures ranging from 1.5 to 10~bar (figure~\ref{fig:TREXDM_SimulacionPatron}), with a reduced drift field of 200~V~cm$^{-1}$~bar$^{-1}$. Applying the approximation that primary electrons in the conversion volume follow the electric field lines, for each simulation, the length of the area between pixels whose field lines did not end in the amplification hole, that is, ended in the mesh, was measured. This length of 'dead zone' for each simulation was applied to the entire perimeter of the pixel, resulting in the total 'dead area' for each pixel. The ratio between the 'dead area' and the total area of influence of the pixel (size of the pixel plus the separation between pixels) is the collection inefficiency for each pattern.

It is necessary to note that the approximation that primary electrons in the conversion volume follow the electric field lines introduces an error in the method and has not been quantified. This approximation is based on the mechanisms of electron transport within an electric field, where the electron experiences an acceleration, dependent on the field strength, and a deceleration, dependent on the mean free path of the electron in the gaseous medium, repeatedly throughout its trajectory~\cite{Peisert}. The typical drift velocities (mean velocities) in Argon or Neon mixtures are in the range of 2-3~cm/$\mu$s (table~3 of~\cite{Iguaz2016}), while the mean free paths of the electrons during their drift are around 1-100~$\mu$m (found from the mean drift velocities and the acceleration experienced by an electron in a field of 200~V~cm$^{-1}$~bar$^{-1}$) before being decelerated. It is expected that the electrons in their trajectory do not deviate much from the field lines, although in areas close to the hole, where the funnel effect occurs with a high field gradient, the electrons will tend to stray more easily from the field lines. Additionally, the effect of diffusion is not taken into account, although this effect tends to be counteracted if one assumes the simplification that there is the same number of electrons that can exit to the 'dead zone' as those that can enter from it to the amplification zone. In any case, the errors produced by these effects will be equivalent in each of the cases studied, so the comparative objective of the analysis remains valid without the need to account for all of them.

In the new design, it was intended to maintain, compared to the previous design, the same number of channels/strips and the same surface area of the active area, thus the same width of strips had to be maintained. The parameters susceptible to change would be the diameter of the holes (D), the distance between hole centers within the pixel (P), and the distance between hole centers between contiguous pixels (IP). This last parameter (IP) also determines the distance between pixels, as for manufacturing reasons there must be at least 65~$\mu$m (66.17~$\mu$m in figure~\ref{fig:TREXDM_PatronAgujeros}) between the edge of a hole and the edge of the pixel in which it is contained.

With these conditions, three different patterns were proposed for the new design, for which the field simulations and efficiency calculations were performed as described earlier. The results of these simulations can be seen in table~\ref{tab:TREXDM_EficienciaPatrones}. The different patterns and their results are discussed below:

\begin{itemize}
\item D50P100IP240: This is the same pattern used in the previous version installed in TREX-DM, as can be seen in figure~\ref{fig:TREXDM_PatronAgujeros}. However, this pattern still allowed for a slight increase in the distance between pixels from 50~$\mu$m to 60~$\mu$m. It is the pattern that shows the highest efficiency.
\item D60P120IP260: This pattern is proposed by the MPT (Micro-Pattern Technologies) group at CERN, arguing for its higher manufacturing quality. This pattern allowed for a separation between pixels of 70~$\mu$m. It has a collection efficiency approximately 3-4\% lower than the D50P100IP240 pattern.
\item D60P110IP290: This pattern arises from the idea of increasing the distance between pixels in the D60P120IP260 pattern, allowing for a separation between pixels of 100~$\mu$m. It has a collection efficiency approximately 10-12\% lower than the D50P100IP240 pattern.
\end{itemize}

\begin{table}[!h]
\centering
\begin{tabular}{l}
\includegraphics[width=0.8\textwidth]{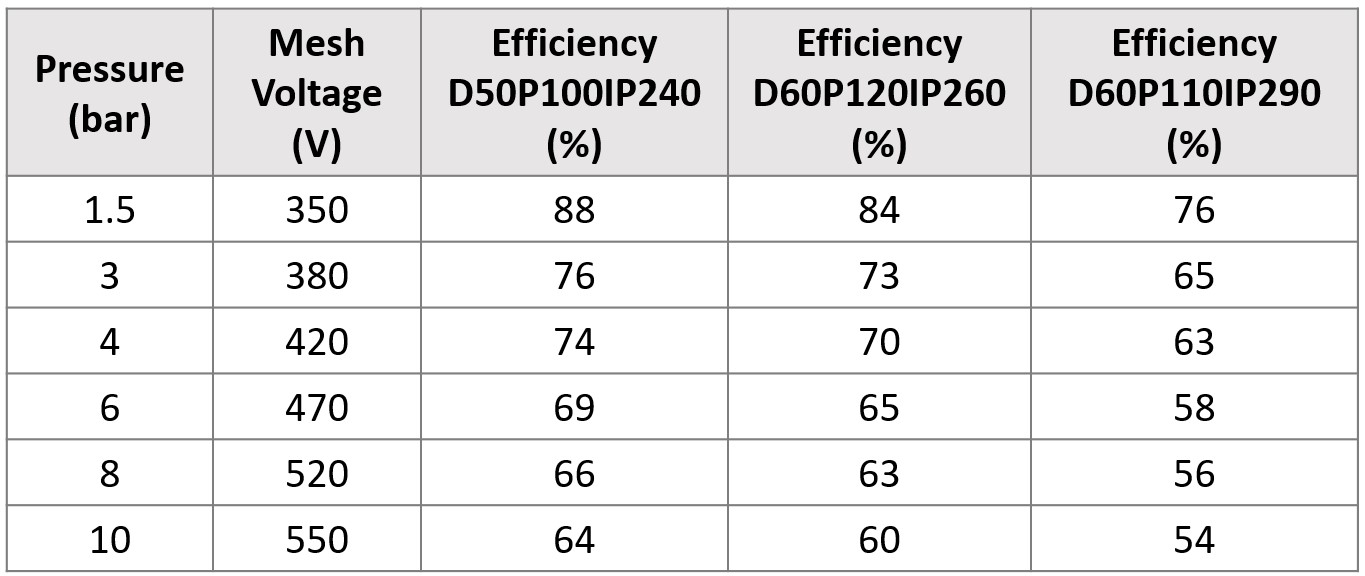}\\
\end{tabular}
\caption{Efficiency in the collection of primary electrons for three different patterns with different distances between holes of contiguous pixels (IP) of 240~$\mu$m, 260~$\mu$m, and 290~$\mu$m. In figure \ref{fig:TREXDM_PatronAgujeros}, two of these three patterns can be seen, with IP of 240 and 290~$\mu$m.}
\label{tab:TREXDM_EficienciaPatrones}
\end{table}

The D50P100IP240 pattern from the previous version of the Micromegas installed in TREX-DM was discarded, given the guarantee of being able to have a higher quality active area, the limited possible increase in the separation of the pixels (from 50 to 60~$\mu$m), and the verification of the small loss of efficiency (3-4\%) of the subsequent D60P120IP260 pattern. Among the two patterns with a hole diameter of D60, there was an advantage for the D60P120IP260 pattern in terms of efficiency. However, the advantage was reversed regarding the distance between pixels, with 100~$\mu$m for the D60P110IP290 pattern compared to 70~$\mu$m for the D60P120IP260 pattern. Finally, the mitigation of the leakage current problem was prioritized, with the D60P110IP290 pattern being chosen for the new Micromegas (figure~\ref{fig:TREXDM_PatronAgujeros}).

It is necessary to point out that the reduced drift field of 200~V~cm$^{-1}$~bar$^{-1}$ used in the simulations is the field applied during the operation of the previous Micromegas installed in TREX-DM, after having performed the transmission curve at 3~bar in Ne~+~2\%~Isobutane. This drift field in the plateau of the transmission curve coincides with the plateau (transmissions of 95-100\%) of the transmission curves performed in the previous study with small Micromegas~\cite{Iguaz2022} for all pressures. The greater loss of efficiency at higher pressures in the inter-pixel areas could be mitigated if lower drift fields are used at high pressures. However, on the other hand, these lower fields may lead to efficiency losses due to attachment and recombination during the drift of the primary electrons towards the readout plane. Finally, the adjustment of the drift field for the new Micromegas will be determined by the realization and study of the new transmission curves, with the detector in operation.

\begin{figure}
    \centering
    \includegraphics[width=7cm]{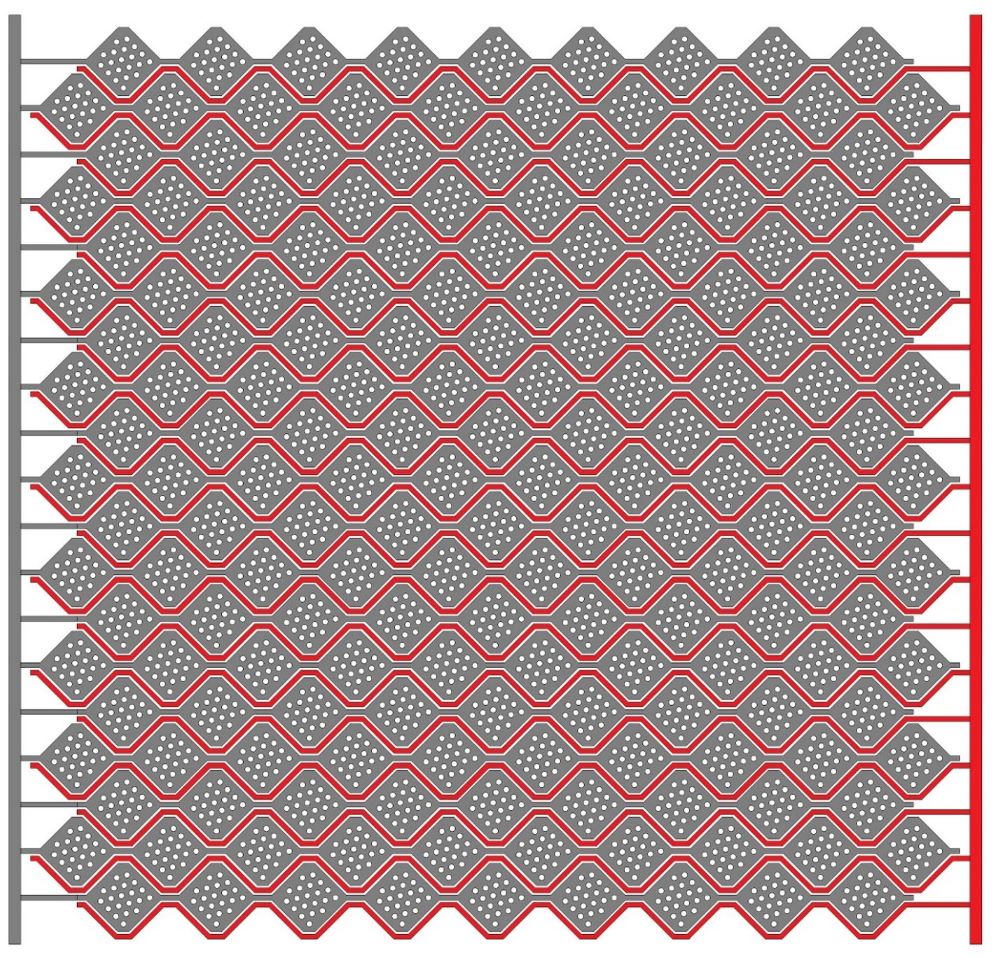}
    \caption{Possible development of a Micromegas in which the areas of the pixels in the mesh are surrounded by an electrode (in red) to which a higher potential than that of the mesh can be applied.}
    \label{fig:MeshConElectrodos}
\end{figure}

A possible solution to increase the efficiency of electron collection in the inter-pixel areas can be seen in figure \ref{fig:MeshConElectrodos}. This is a potential development of a Micromegas in which the areas of the pixels in the mesh are surrounded by an electrode to which a higher potential than that of the mesh can be applied. This solution may reduce the loss of efficiency in the inter-pixel area by diverting the field lines towards the holes. From a manufacturing perspective, this design is feasible, but the main problem is that if the electrode is too thin, it could break in some areas, leaving some parts of the electrode floating, which would produce the opposite effect to what is desired.

\subsection{Reduction of Background from the Micromegas}
\label{subsec:MMreduccionFondo}

The previous Micromegas installed in TREX-DM represented the main contribution to the background model~\cite{Castel2019}, with 2.30~dru~(counts~keV${^{-1}}$~kg${^{-1}}$~d${^{-1}}$) in Argon and 2.68~dru in Neon. This contribution was due to the activity of the isotope ${^{40}}$K contained in the Micromegas, which was quantified through measurements with germanium at 3.45~$\pm$~0.40~$\mu$Bq/cm${^2}$.

The strategy for measuring intrinsic radioactivity in a wide variety of materials, developed over several years within the T-REX project, allowed for the identification of the cause of this ${^{40}}$K contamination. During the manufacturing of the Micromegas by the MPT group at CERN, the drilling or etching process of the holes in the mesh is performed using Potassium Hydroxide (KOH). This chemical compound deposits as a residue on the walls of the holes, thus contributing to the contamination by the isotope ${^{40}}$K. Consequently, the MPT proposed a deionized water cleaning process that could be applied as the final step in the production process to reduce this contamination.

In the usual production of the Micromegas, after applying the bath with Potassium Hydroxide for drilling the holes in the mesh, a final bath of Potassium Permanganate (KMnO$_4$) is applied, aimed at cleaning impurities from the active area of the Micromegas. The new deionized water cleaning process would be applied after this bath, but it would require a new final bath with Potassium Permanganate. Therefore, the new cleaning process involved two additional baths: one with deionized water and one with Potassium Permanganate. The bath with Potassium Permanganate is less aggressive than the bath with Potassium Hydroxide applied during the drilling of the holes, so a reduction in ${^{40}}$K contamination was anticipated.

\begin{figure}
    \centering
    \includegraphics[width=\textwidth]{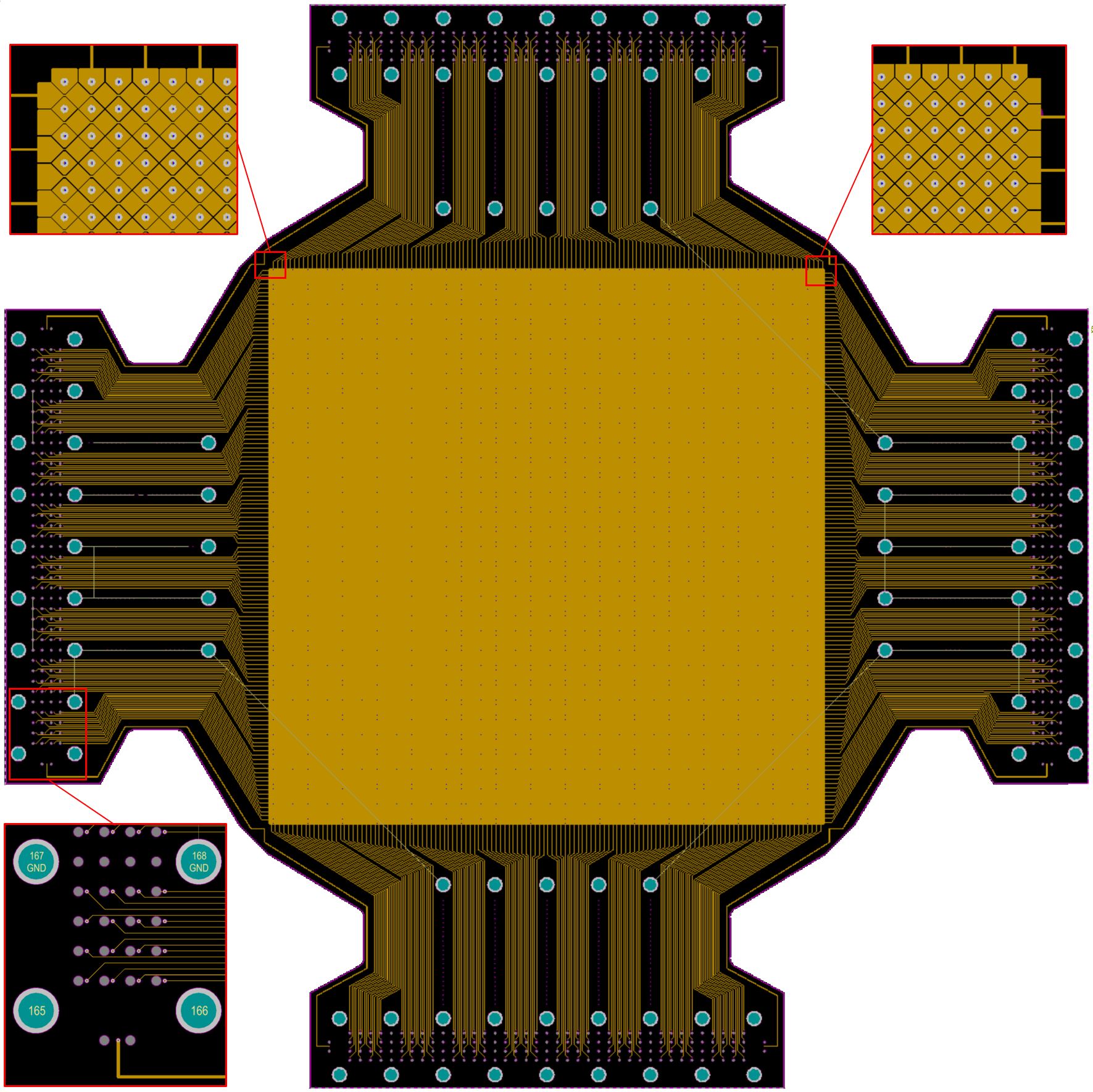}
    \caption{Layer~2 of the new Micromegas, with an active area of 25~x~25 cm${^2}$ and 512~channels. The upper details show the channels of each axis exiting the active area on both sides, alternately. The lower detail shows a small area of the FtF connector with a separation between pads of 4~mm. In the lower area of the lower detail, the pad and the wider trace that feeds the mesh can also be seen, with a minimum separation from the channels and the ground planes of 12~mm.}
    \label{fig:TREXDM_MMv2_Altium_completo}
\end{figure}

\begin{figure}
    \centering
    \includegraphics[width=6cm]{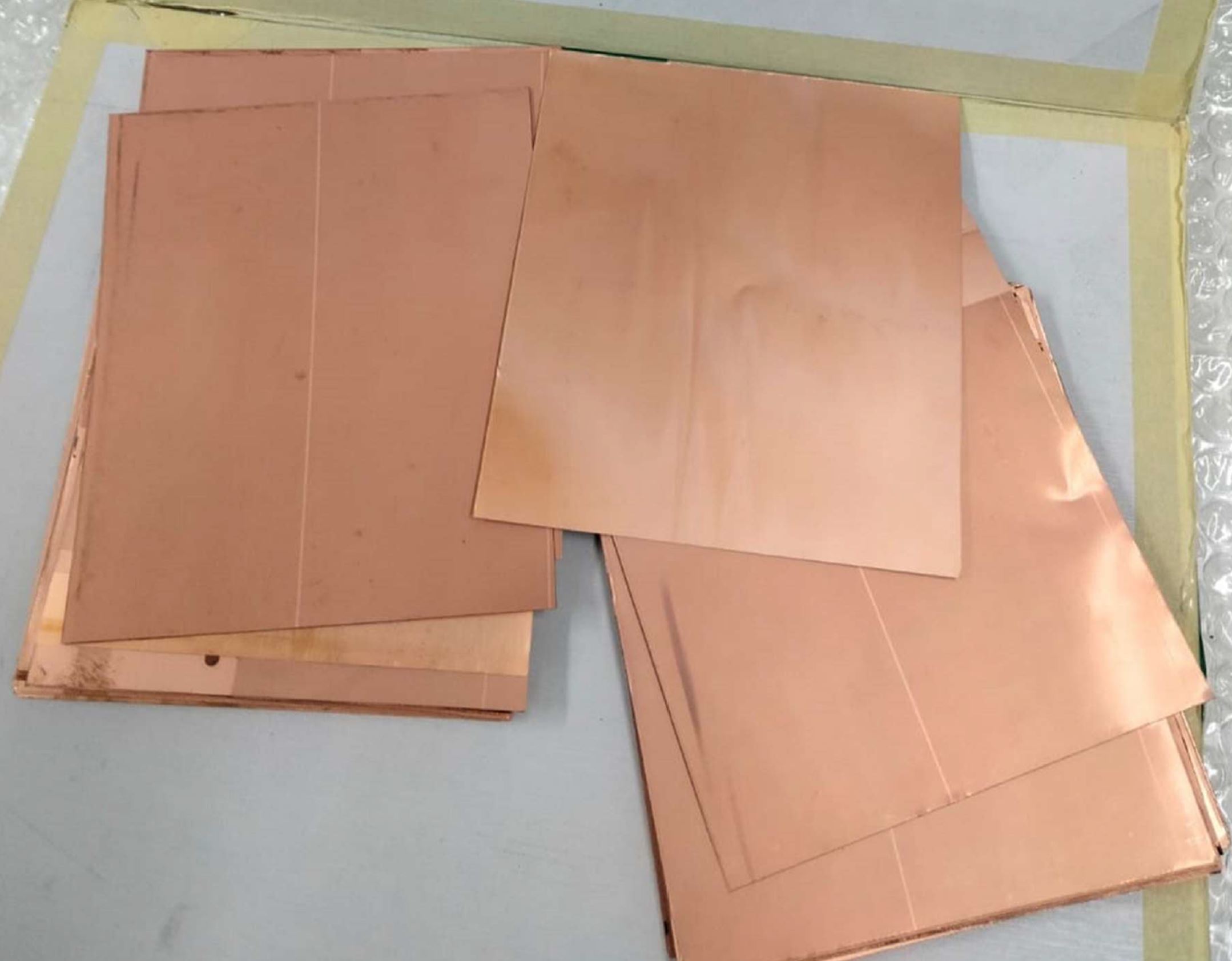}
    \caption{Witness circuits manufactured in parallel to the new Micromegas of TREX-DM and measured with the germanium 'GeAnayet' from the radiopurity service of the LSC for each cleaning process performed.}
    \label{fig:TREXDM_WitnessSamples}
\end{figure}

After several discussions, it was decided to apply the new cleaning process to the new Micromegas, and to quantify the degree of success of this process, it was proposed to measure the Micromegas using HPGe in Canfranc before and after this new cleaning process. To achieve a sensitivity below the previously measured 3.45~$\mu$Bq/cm${^2}$, a sample of at least 1~m${^2}$ was necessary. Thus, it was decided to manufacture, in parallel and applying exactly the same manufacturing processes simultaneously, several witness samples with a total area of 1~m${^2}$ (figure~\ref{fig:TREXDM_WitnessSamples}), and the results can be seen in table~\ref{tab:TREXDM_MM_radiopurity}. The steps followed during this process were as follows:

\begin{enumerate}
\item In the spring of 2020, the manufacturing order for 4~Micromegas with the new design (as shown in figure \ref{fig:TREXDM_MMv2_Altium_completo}) and the witness samples was launched.
\item In the spring of 2021, the manufacturing of the new Micromegas along with the witness samples was completed. The first measurement of the samples began with the germanium 'GeAnayet' from the radiopurity service of the Canfranc Underground Laboratory, with a total exposure of 29.66~days and a result of 1.07~$\pm$~0.23~$\mu$Bq/cm${^2}$. In this first step, the contamination had already been reduced by a factor of 3 compared to the previously measured value in~\cite{Castel2019}; however, the extra deionized water bath had not yet been applied. Although the exact reason for this reduction is not known, one of the most supported hypotheses is that the change in pattern in the active area affected the manufacturing process, reducing the area attacked by the KOH bath, the most aggressive of all, by 25\%, while also favoring the subsequent cleaning with larger diameter holes with~KMnO$_4$. Finally, since the activity value for ${^{40}}$K was already 3 times lower than that applied in the background model, it was decided to reserve and conclude two of the four Micromegas (MMv2-2 and MMv2-3).
\item In the summer of 2021, the new deionized water cleaning process was applied to the witness samples and the other two manufactured Micromegas (MMv2-4 and MMv2-5). The samples measured after this cleaning, this time with an exposure time of 58.66 days, yielded an activity of~\textless~0.63~$\mu$Bq/cm${^2}$, as an upper limit. However, although the activity had been significantly reduced, after this bath the Micromegas were not operational and required a final bath with~KMnO$_4$.
\item In the autumn-winter of 2021, the last bath with~KMnO$_4$ was applied to the samples and the two Micromegas that remained in the process (MMv2-4 and MMv2-5). The previous bath with deionized water produced several short circuits between the strips and the mesh that needed to be eliminated with the application of this last bath with~KMnO$_4$. However, in one of the Micromegas, many of the short circuits were not eliminated, and the MPT laboratory staff decided unilaterally to apply at least two more baths with~KMnO$_4$ only to the damaged Micromegas (MMv2-5). This decision disconnected this Micromegas from the radiopurity control process being conducted, so that only one Micromegas that had followed exactly the same processes as the samples remained, MMv2-4. The measurement of the samples with the germanium 'GeAnayet' had an exposure of 30.39~days, yielding an activity of~\textless~1.2~$\mu$Bq/cm${^2}$, as an upper limit, valid only for the Micromegas MMv2-4. 
\end{enumerate}

\begin{table}[!h]
\centering
\begin{tabular}{c c c c}
\textbf{MMv2-2} & \textbf{MMv2-3} & \textbf{MMv2-4} & \textbf{MMv2-5} \\ \hline \hline
\\
1.07 $\pm$ 0.23 & 1.07 $\pm$ 0.23 & \textless 1.2 & \textbf{?} \\
\end{tabular}
\caption{Results of the activity measurements ($\mu$Bq/cm${^2}$) of the isotope ${^{40}}$K for the witness samples and their correspondence with the manufactured Micromegas. Measurements were performed with the germanium 'GeAnayet' from the radiopurity service of the Canfranc Underground Laboratory. The activity of Micromegas MMv2-5 is unknown, as at one point in the process it was disconnected from the witness samples.}
\label{tab:TREXDM_MM_radiopurity}
\end{table}

\section{New Extension Cables}

In parallel with the design process of the new Micromegas, the design of the flat cables was carried out, which needed to connect the 512 channels of each Micromegas and extract them from the chamber to the acquisition electronics, located outside the lead shielding.

The decision was made to manufacture the cables in the MPT workshop, the same workshop where the Micromegas are produced, following the same processes used for them, except for the process of drilling the holes in the mesh, which is also the one that introduces contamination from ${^{40}}$K, in order to obtain cables with the lowest possible activity. The intrinsic radioactivity of the kapton-copper sheets used to manufacture the Micromegas, the same ones used for the flat cables, was measured~\cite{Castel2019} and shows very low radioactivity values (upper limits).

\begin{figure}
	\centering
	\begin{subfigure}{\textwidth}
		\centering
		\includegraphics[width=\textwidth]{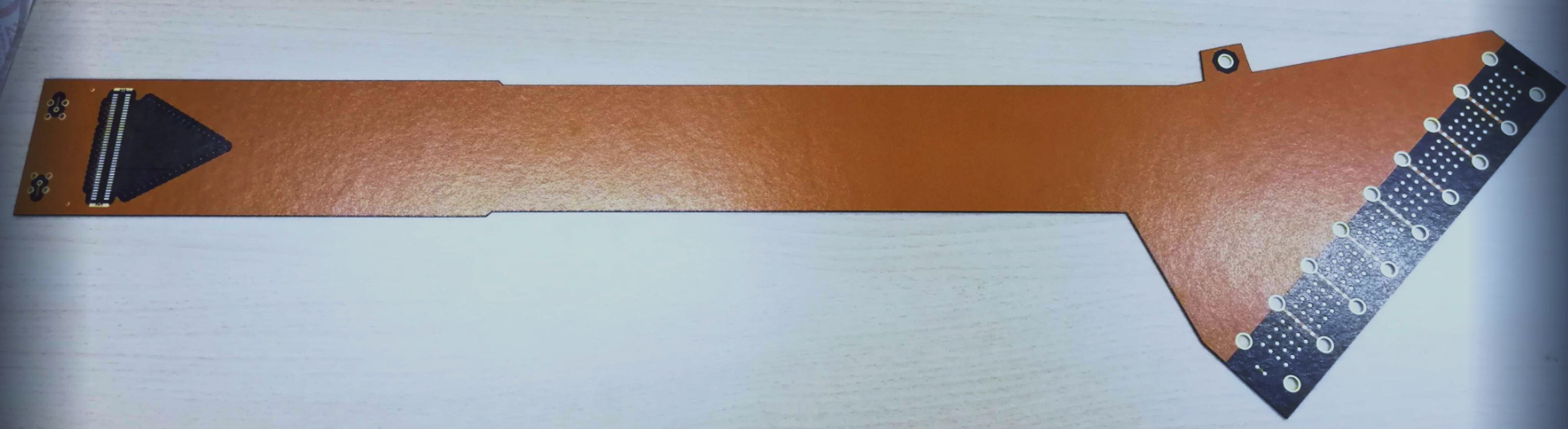}	
	\end{subfigure}
	\begin{subfigure}{\textwidth}
		\centering
		\includegraphics[width=\textwidth]{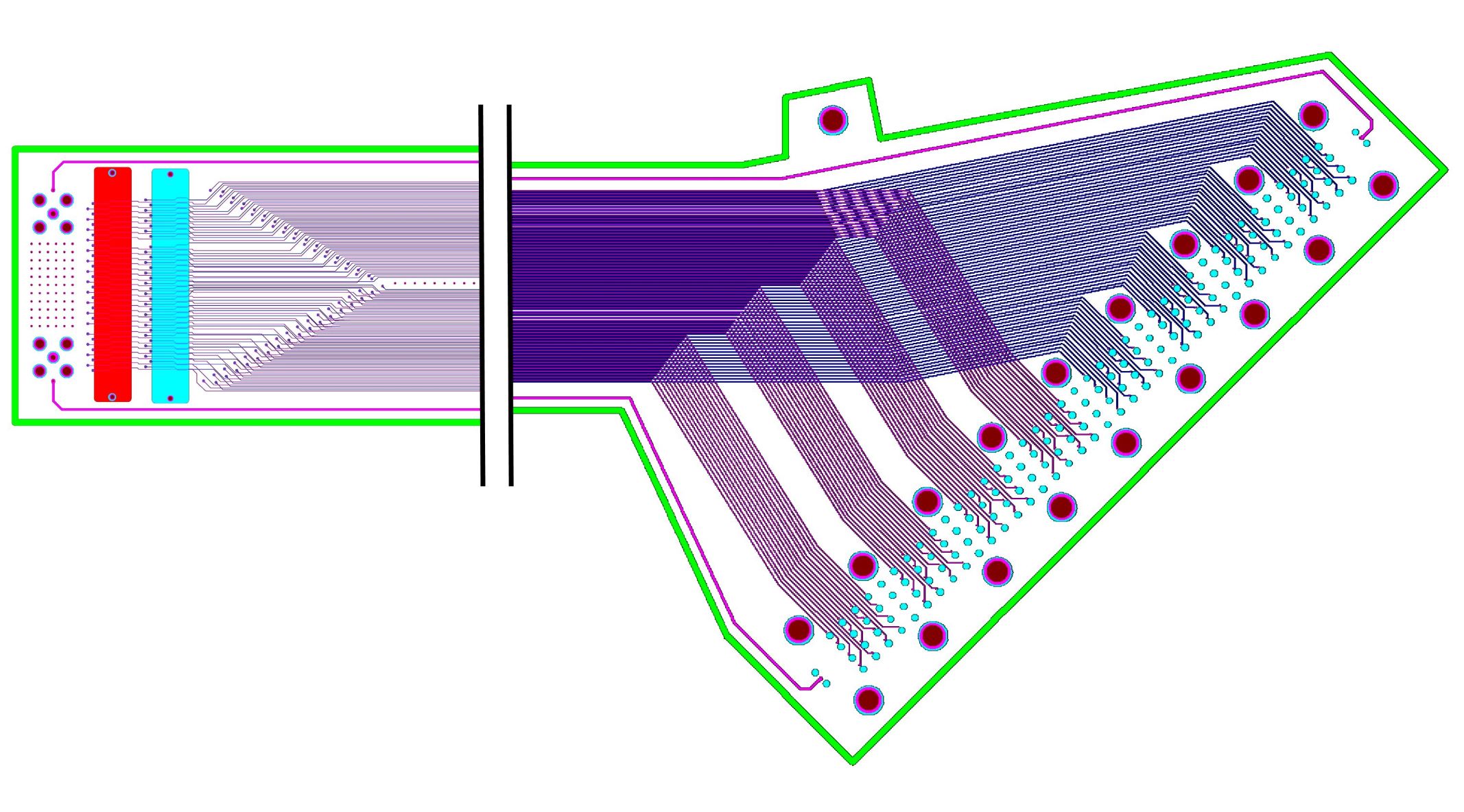}
	\end{subfigure}
	\caption{Flat cables designed in parallel to the new version of the Micromegas. They are composed of 4 layers of copper and separating layers of kapton between them.}
	\label{fig:TREXDMFlatCable}
\end{figure}

After several proposals, the cables were designed as shown in figure~\ref{fig:TREXDMFlatCable}, with a total of 128 channels each. At one end of the cable, the 'Face to Face' (FtF) connector was introduced, which connects to the Micromegas, while at the other end, two ERNI connectors were placed, the same type of connector mounted on the acquisition electronics board. The particular shape of the cable is due to the twist that must be made to pass through the cover via the already machined slots in it, used during the demonstration phase in the laboratories of the University of Zaragoza, prior to its transfer to the LSC.

The circuit consists of 4 layers of copper: layers 1 and 4 are ground planes for shielding the channels (obviously, these layers also contain the footprints of the connectors), while layers 2 and 3 contain the routing of the channels. The same parameters as those used for the design of the Micromegas were applied in terms of distances between elements (table \ref{tab:TREXDM_MM_ParametrosDiseño}).

The kapton layer that serves as separation between layers 2 and 3 is 50~$\mu$m thick, sufficient to prevent dielectric breakdown with the potential differences that may exist between channels (550~V). However, for the separation of layers 1 and 2 and layers 3 and 4, the thickness of the kapton was increased to 350~$\mu$m, to reduce the capacitance between the channels and the ground planes. The channel-ground capacitance of the flat cable with 350~$\mu$m of kapton has a value 3 times lower than the channel-ground capacitance of the Micromegas. In theory, this reduction in capacitance favors the charge collection in the reading of the Micromegas.

\section{New FtF Connectors}

Following the success of the 'Face to Face' connectors in the PandaX-III project (see section~\ref{sec:FtFconnector}), and after the negative experience with leakage currents in the connectors during the operation of the previous detector, it was decided to introduce this type of connector in the new Micromegas.

Tests with this type of connection in the PandaX-III project were conducted using flexible circuits with a pad distance of 4~mm. The objective of these tests was to verify the robustness of the system and the maintenance of connectivity over time. Leakage current tests were also performed, although they were not systematic, reaching up to 600~V due to limitations in the circuit itself; however, no leakage currents were observed at the order of the power supply resolution (nA).

During the tests of the 'Face to Face' connector, no measurements were taken against variations in ambient humidity, temperature, or pressure. However, within the TREX-DM chamber, the humidity is very low, and the temperature is kept constant in the range of 18-23$^{\circ}$~C, which justified the decision to include these connectors while maintaining the same distance of 4~mm between \textit{pads} (see figure~\ref{fig:TREXDM_MMv2_Altium_completo}).

\section{New Feedthroughs}

In parallel with the design process of the new Micromegas and the flat cables, the design of the new feedthroughs was carried out.

In the previous version of the readout system for TREX-DM (see section~\ref{sec:IntroTREXDM}), each side of the chamber had two flat cables with 256 channels each, connecting the Micromegas to the electronics located outside the lead shielding through the problematic fujipoly connectors. These cables were attached to a rectangular flange using low-activity epoxy, allowing each cable to be threaded from the outside into a slot in the cover during installation, and tightening this flange with a PTFE gasket to ensure sealing.

The new flat cables (figure~\ref{fig:TREXDMFlatCable}) feature a 'Face to Face' connector at one end. These connectors are too large to pass through the slot in the cover, so they had to be installed from the inside of the chamber or the slots in the cover had to be extended to allow the entire cable to pass through, which would weaken the cover and thus the chamber's ability to withstand high pressures. Therefore, a rectangular flange similar to the one in the previous version was designed, and threaded holes were drilled from the inside of the cover so that the flange could be mounted from within. This change also favored the sealing of the system because the operating pressure of the chamber worked in favor of the pressure between the flange and the PTFE gasket.

These modifications to the cover, minimal in nature, needed to be tested, as did the connection of the flat cable to the rectangular flange. To carry out these tests, a test bench was designed and built, using the same operating principle employed to test the feedthroughs of the SR2M module in the PandaX-III experiment (see section~\ref{sec:feedthrough} and figure~\ref{fig:SR2M_FeedthroughTestBench}).

\begin{figure}
    \centering
    \includegraphics[width=\textwidth]{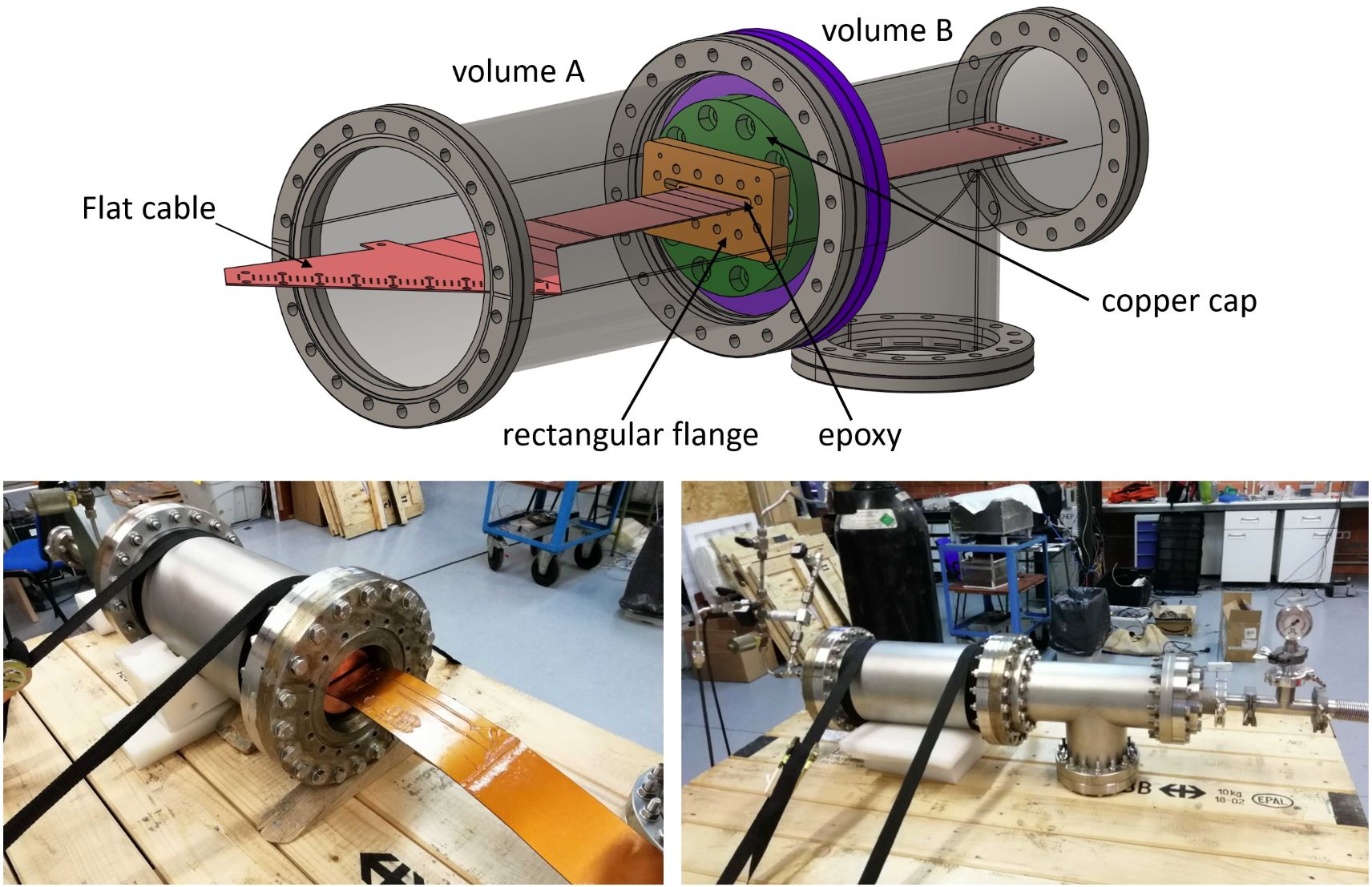}
    \caption{Test bench to validate the sealing of the 8~feedthroughs. The flat cable is joined to the rectangular flange with a low-activity epoxy. The flange-cable assembly is bolted to the copper cover, simulating the cover of the TREX-DM chamber, using copper screws. The sealing between the flange and the copper cover is achieved with a PTFE gasket. After assembly, volume~A is filled with a mixture of Nitrogen~+~10\%~Helium, and in volume~B, a helium meter is connected to measure the existing leak rate between the two volumes.}
    \label{fig:TREXDM_FeedthroughTestbench}
\end{figure}

Figure~\ref{fig:TREXDM_FeedthroughTestbench} shows some photos and the schematic of this test bench. Volume~A represents the internal volume of the TREX-DM chamber, subjected to a pressure of up to 10~bar with a mixture of Nitrogen~+~10\%~Helium. In this volume, the rectangular flange (previously glued with epoxy to the flat cable) is bolted onto a copper piece that represents the copper cover of TREX-DM. The sealing between the rectangular flange and the copper cover is achieved using a PTFE gasket. Volume~B represents the external part of the chamber, to which a leak detector (model Pfeiffer Smart Test HLT 560) is connected, with a leak detection sensitivity on the order of 1·10$^{-12}$~mbar~l~s$^{-1}$, allowing for the quantification of the total leak of the feedthrough system.

Using this test bench, the sealing of each of the eight manufactured feedthroughs was measured. Each feedthrough was mounted in the system, volume~A was set to 10~bar, and measurements were taken with the leak detector in volume B. Although intermediate measurements were taken as a check, the measurement representing the total leak, including the leak due to permeability in the gasket, was performed 24 hours after filling volume~A to 10~bar.

The measured leak values in the eight feedthroughs ranged from 9·10$^{-6}$ to 1.4·10$^{-5}$~mbar~l~s$^{-1}$, having applied the correction due to the proportion of Helium in volume~A~(10\%). These values are very close to the values measured (2.5·10$^{-5}$~mbar~l~s$^{-1}$) during the design of the SR2M in the PandaX-III experiment, which was expected given that they are similar flange and gasket systems.

\section{First Tests of the New Micromegas}
\label{sec:PruebasPrevias}

\begin{figure}
    \centering
    \includegraphics[width=\textwidth]{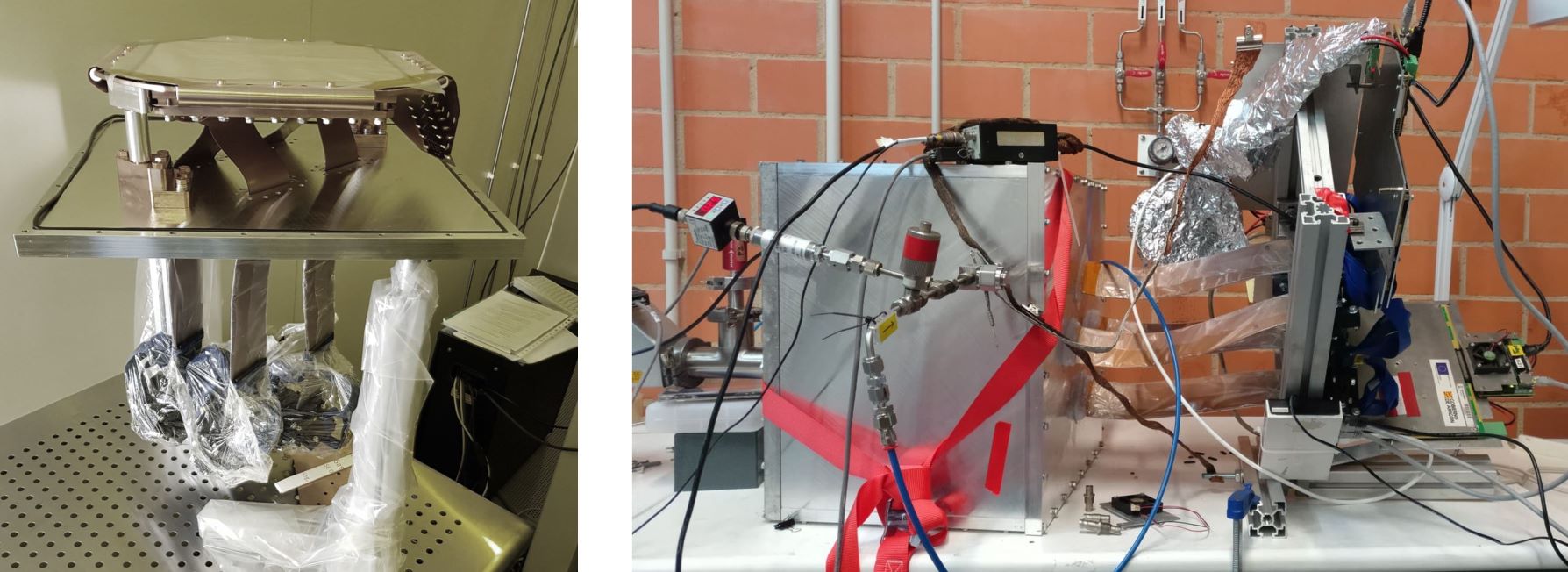}
    \caption{Test chamber for the new TREX-DM Micromegas, assembled in the laboratories of the University of Zaragoza. $\bullet$~Left:~photograph of one of the new Micromegas mounted on the lid of the test chamber, with the flat cables passing through the lid. $\bullet$~Right:~photograph of the final assembly of the test chamber, including the acquisition electronics.}
    \label{fig:TREXDM_TestChamber}
\end{figure}

After their fabrication, and prior to their definitive installation in the LSC, the four Micromegas were tested in the laboratories of the University of Zaragoza. The objective was to evaluate the quality of each Micromegas and the validity of the new connection chain, including mechanical elements. For this purpose, an aluminum chamber was manufactured to allow for pressures of up to 1.1~bar (figure~\ref{fig:TREXDM_TestChamber}). The installation of each Micromegas in this chamber was carried out in a Class 10,000 clean room.

After receiving the Micromegas, and as part of the quality control prior to their installation in the test chamber, the 'Philips RCL meter PM6303' was used to measure the capacitance and resistance between each of the channels and the mesh in each Micromegas.

The capacitance between a channel and the mesh is approximately that of two parallel planes of the same area as the \textit{strip} corresponding to the channel, separated by 50~$\mu$m of dielectric, in this case separated by kapton~\cite{DuPont2017}. The measured values, considered normal according to this approximation, ranged from 90 to 110~pF per channel. Capacitance values below this range indicate some connectivity issue between the measurement point, in this case the connector footprint, and the corresponding \textit{strip}. Capacitance values above this value, typically double, indicate that one channel is interconnected with another channel, usually from the opposite axis, due to a manufacturing error.

The resistance between a channel and the mesh is approximated, similar to capacitance, using the parallel plate model. The theoretical calculation with this model yields strip-mesh resistances on the order of hundreds of~T$\Omega$ for temperatures around 23~$^{\circ}$C~\cite{DuPont2017}. The measurement range of the equipment used reaches up to 200~M$\Omega$. Strip-mesh resistance values below this will cause leakage currents during the operation of the detector on the order of tens or units of~$\mu$A. The maximum leakage current values that allow for stable operation of the detectors are on the order of tens of~nA, so the resistance measurements performed in this quality control may be insufficient to detect all such issues. To complete this quality control, a voltage of 100~V is applied to the mesh, without the need to do so within a chamber with the gas mixture, and the current injected by the power supply is observed. The power supply used is a CAEN~1471, with a resolution of 0.5~nA, capable of detecting resistance values of up to 200~G$\Omega$, which serves as the definitive test to validate the insulation between \textit{strips} and mesh at the level necessary for the operation of the detectors.

During the checks with the 'Philips RCL meter PM6303', 512 measurements were made on each Micromegas, totaling 2048 for the four fabricated. The acceptable capacitance and resistance values were 90-110~pF and $>$200~M$\Omega$, respectively, and 28 of the 2048 measured had an anomalous value (table~\ref{tab:TREXDM_TestsResults}). During the load test, where 100~V was applied to the mesh, no channel showed any anomalies, apart from the 28 channels already identified in the previous measurements.

After the quality control, each Micromegas was installed in the test chamber (figure~\ref{fig:TREXDM_TestChamber}) and operated with Ar~+~1\%~Isobutane at 1.1~bar. One of the objectives in this chamber was to check the maximum voltage that could be stably reached on the mesh (table~\ref{tab:TREXDM_TestsResults}), under the same mixture and pressure conditions, which would directly determine the maximum gain that could be achieved. The Micromegas MMv2-2 and MMv2-3, which had not undergone the deionized water cleaning for background reduction, showed maximum voltage values between 10~V and 25~V higher than the Micromegas that had undergone cleaning. These differences in the maximum achievable voltage have a direct impact on the gain and can lead to gain differences ranging from 50\% (10 V) to 125\% (25 V)~\cite{Iguaz2022}.

The criterion for determining the maximum achievable voltage suffers from some ambiguity. It could be defined as the voltage applied 1~V below that which causes a spark between the mesh and the channel. This would allow for forcing the Micromegas until this spark occurs, as many times as necessary to accumulate the desired statistics. However, this criterion may not be sufficient for stable data acquisition over time, as it does not take into account the frequency with which this spark occurs. A high frequency may be unfeasible for stable data acquisition; therefore, more precisely, the maximum achievable voltage could be defined as the operating voltage at which a certain number of sparks occur per unit of time. To determine this value, it is necessary to do so during the operation of the detector, in a long exposure data acquisition, so it is usually done once the Micromegas is installed in its final configuration, in this case in the TPC of TREX-DM. It is important to note that sparks pose a high risk to the Micromegas, as during the discharge, part of the copper from the mesh may melt and fall onto the hole, causing a short circuit between the mesh and the corresponding channel, rendering this channel completely unusable. Therefore, in the search for the maximum achievable voltage, care must be taken with the detector to provoke as few sparks as possible.

However, to compare the maximum achievable gain among several Micromegas, in this case the four fabricated for TREX-DM, the first criterion described is valid, although it is expected that during the operation of the detector in long exposure data acquisitions, the maximum voltage reached by this criterion will be reduced. Finally, applying the first criterion, a test was conducted for each of the Micromegas, where the voltage on the mesh was gradually increased until discharge occurred. This was repeated two or three times to avoid risking the Micromegas, and the maximum value reached was recorded (see table~\ref{tab:TREXDM_TestsResults}).

\begin{table}[!h]
\centering
\begin{tabular}{ l  c  c  c  c }
\hline
\textbf{} & \textbf{MMv2-2} & \textbf{MMv2-3} & \textbf{MMv2-4} & \textbf{MMv2-5} \\ \hline
Ch. with normal capacity & 510 & 506 & 503 & 500 \\
Ch. with small capacity & 1 & 0 & 3 & 2 \\
Ch. with high capacity & 1 & 6 & 6 & 8 \\
Ch. with low resistivity & 0 & 0 & 0 & 2 \\
Maximum mesh voltage & 295 V & 305 V & 280 V & 285 V \\
\hline
\end{tabular}
\caption{Results of the tests on the four new Micromegas for TREX-DM in the laboratories of the University of Zaragoza. The number of channels found with any issues and the maximum voltage reached in Ar~+~1\%~Isobutane at 1.1~bar of pressure are represented.}
\label{tab:TREXDM_TestsResults}
\end{table}

\begin{figure}
    \centering
    \includegraphics[width=\textwidth]{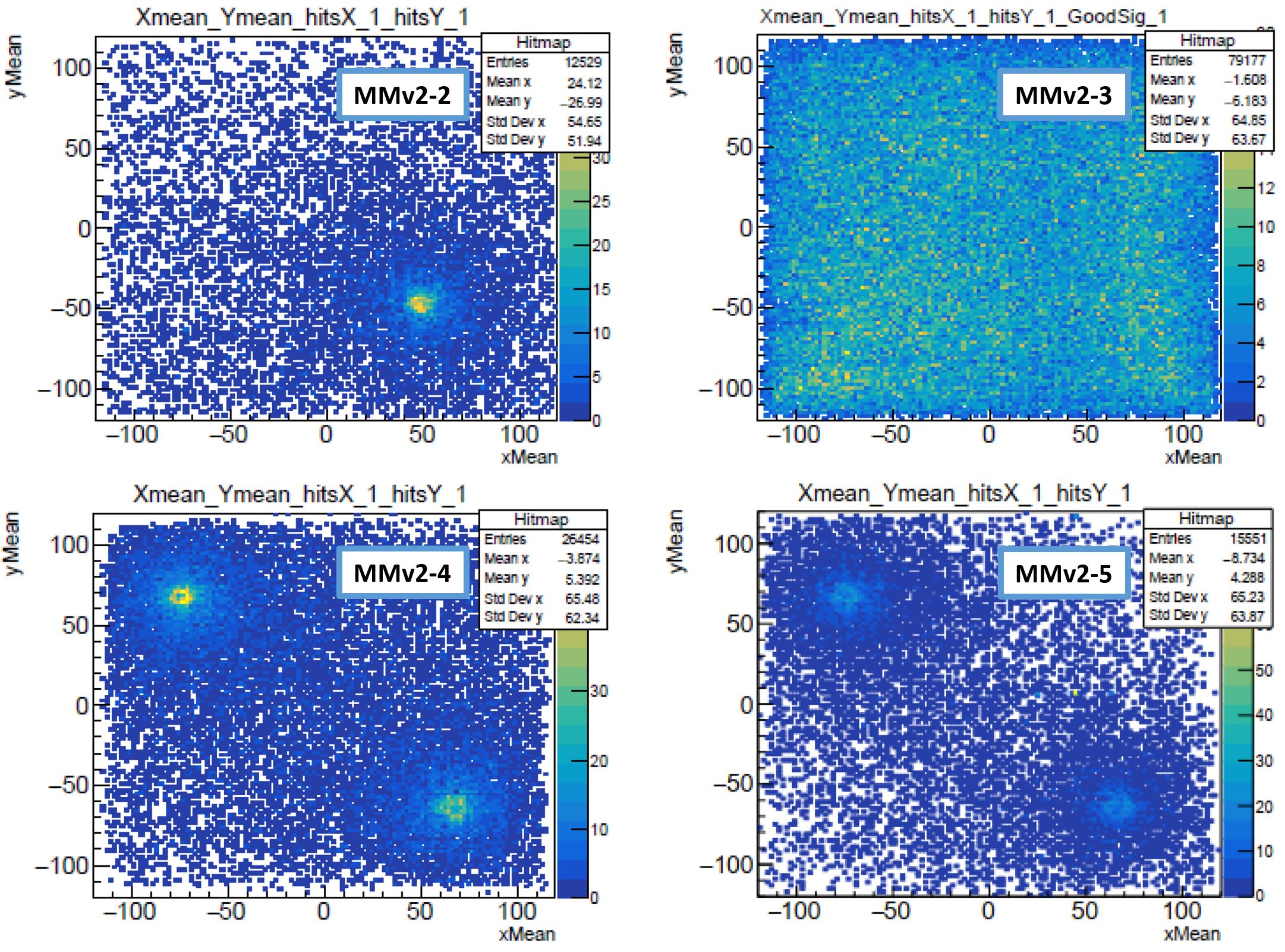}
    \caption{Activity maps of the channels for the four new TREX-DM Micromegas in the laboratories of the University of Zaragoza, during calibrations with the $^{109}$Cd source in Ar~+~1\%~Isobutane at 1.1~bar of pressure.}
    \label{fig:TREXDM_HitsMapsCamaraPruebas}
\end{figure}

For each Micromegas installed in the test chamber, and after determining the maximum achievable voltage, data were taken with one and two sources of $^{109}$Cd installed inside the chamber. Figure~\ref{fig:TREXDM_HitsMapsCamaraPruebas} shows the activity maps in the channels for the four Micromegas, where the sources of $^{109}$Cd can be located. It is worth noting that during the operation of the Micromegas MMv2-3, the first one installed in the test chamber, the $^{109}$Cd source was located several centimeters behind the perforated cathode; however, this was not sufficient to see its emission focus. During the installation of the second Micromegas, MMv2-2, the $^{109}$Cd source was installed directly on the cathode. For the Micromegas MMv2-4 and MMv2-5, an additional $^{109}$Cd source was added.

The objective of the data acquisition sessions with $^{109}$Cd was to validate the response of the Micromegas and rule out anomalies. In this regard, all four Micromegas showed an acceptable response.

After all these tests in the laboratories of the University of Zaragoza, the decision was made to install the Micromegas MMv2-2 and MMv2-3 in TREX-DM. These two Micromegas had fewer defective channels, achieved higher gains, and their radioactivity had been measured until the end of the manufacturing process, resulting in 1.07 $\pm$ 0.23~$\mu$Bq/cm${^2}$, a factor of 3 lower than that used in the background model.

\section[Installation and Data Acquisition with the New Micromegas in TREX-DM]{Installation and Data Acquisition with the New Micromegas in TREX-DM \sectionmark{Installation and Data Acquisition...}}
\sectionmark{Installation and Data Acquisition...}
\label{sec:TREXDMopt_TomaDatos}

\begin{figure} [htbp]
	\centering
	\begin{subfigure}[htbp]{\textwidth}
		\centering
		\includegraphics[width=\textwidth]{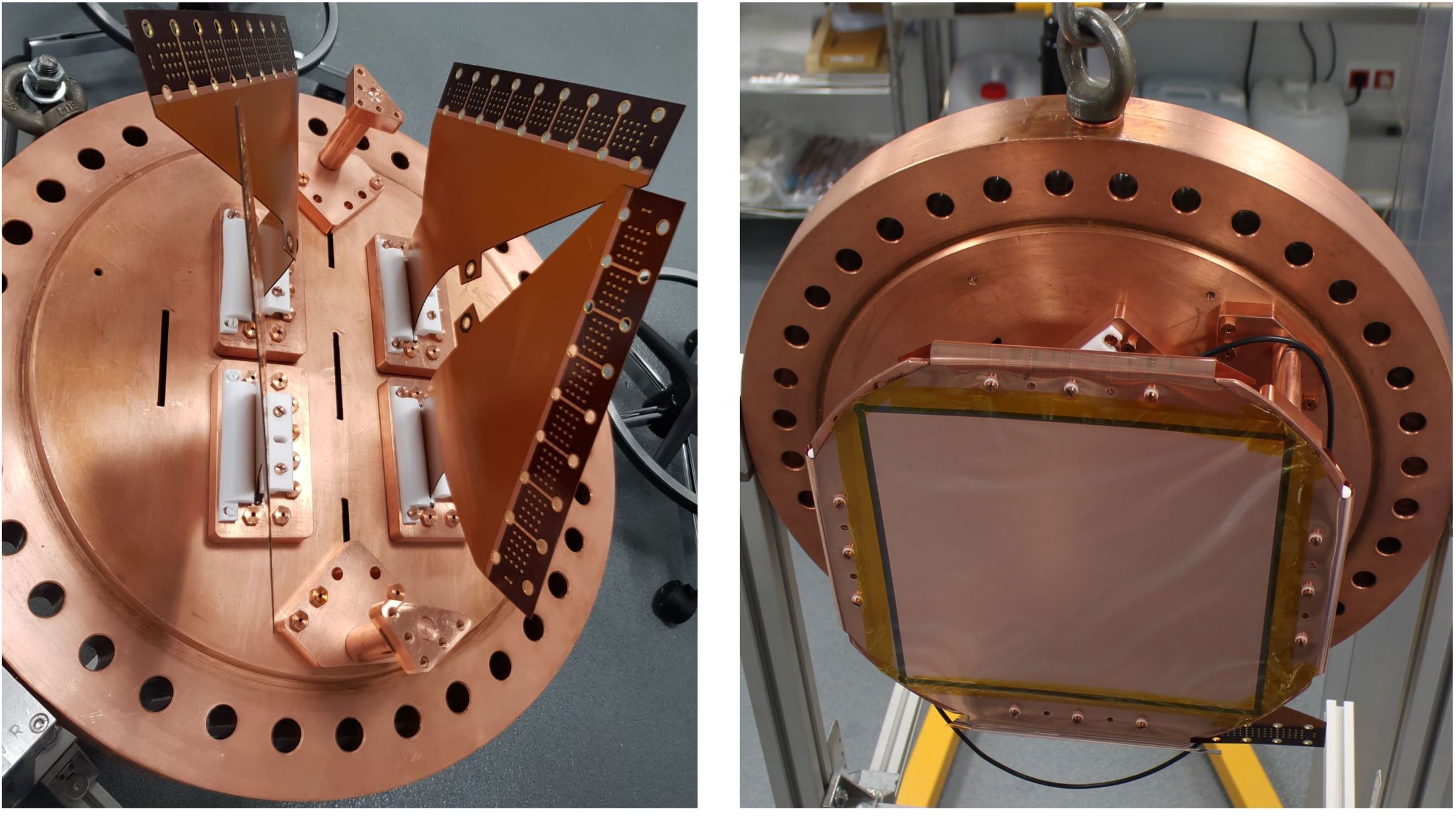}	
	\end{subfigure}
	\begin{subfigure}[htbp]{\textwidth}
		\centering
		\includegraphics[width=\textwidth]{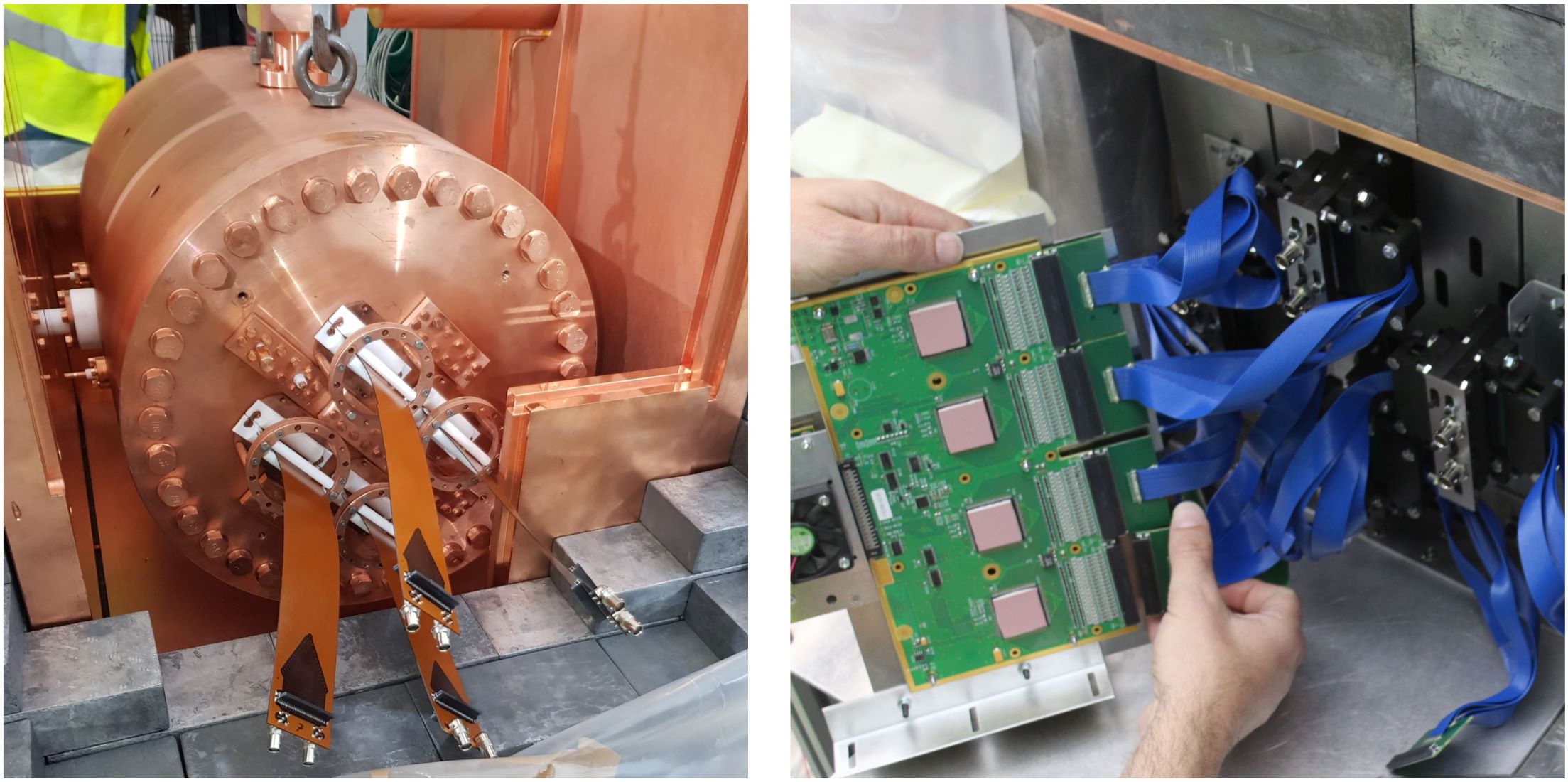}
	\end{subfigure}
	\caption{Photos taken during the installation of the new Micromegas. $\bullet$~Upper left:~view of the interior of the cover, with the flat cables and feedthroughs installed, a step prior to the installation of the Micromegas. $\bullet$~Upper right:~new Micromegas installed and connected to the flat cables, the active area is protected by a plastic sheet during the intervention. $\bullet$~Lower left:~view of the chamber installed in its final position within the shielding, a step prior to the assembly of the rest of the shielding. $\bullet$~Lower right:~photograph taken during the installation of the acquisition electronics. The blue cables connect the channels from the flat cables to the acquisition electronics FEC-Feminos.}
	\label{fig:TREXDM_FotosIntervencion}
\end{figure}

In the spring of 2022, an intervention was carried out at the Canfranc Underground Laboratory to dismantle the previous Micromegas and install the two newly selected ones (see figure~\ref{fig:TREXDM_FotosIntervencion}).

\begin{figure}
    \centering
    \includegraphics[width=0.9\textwidth]{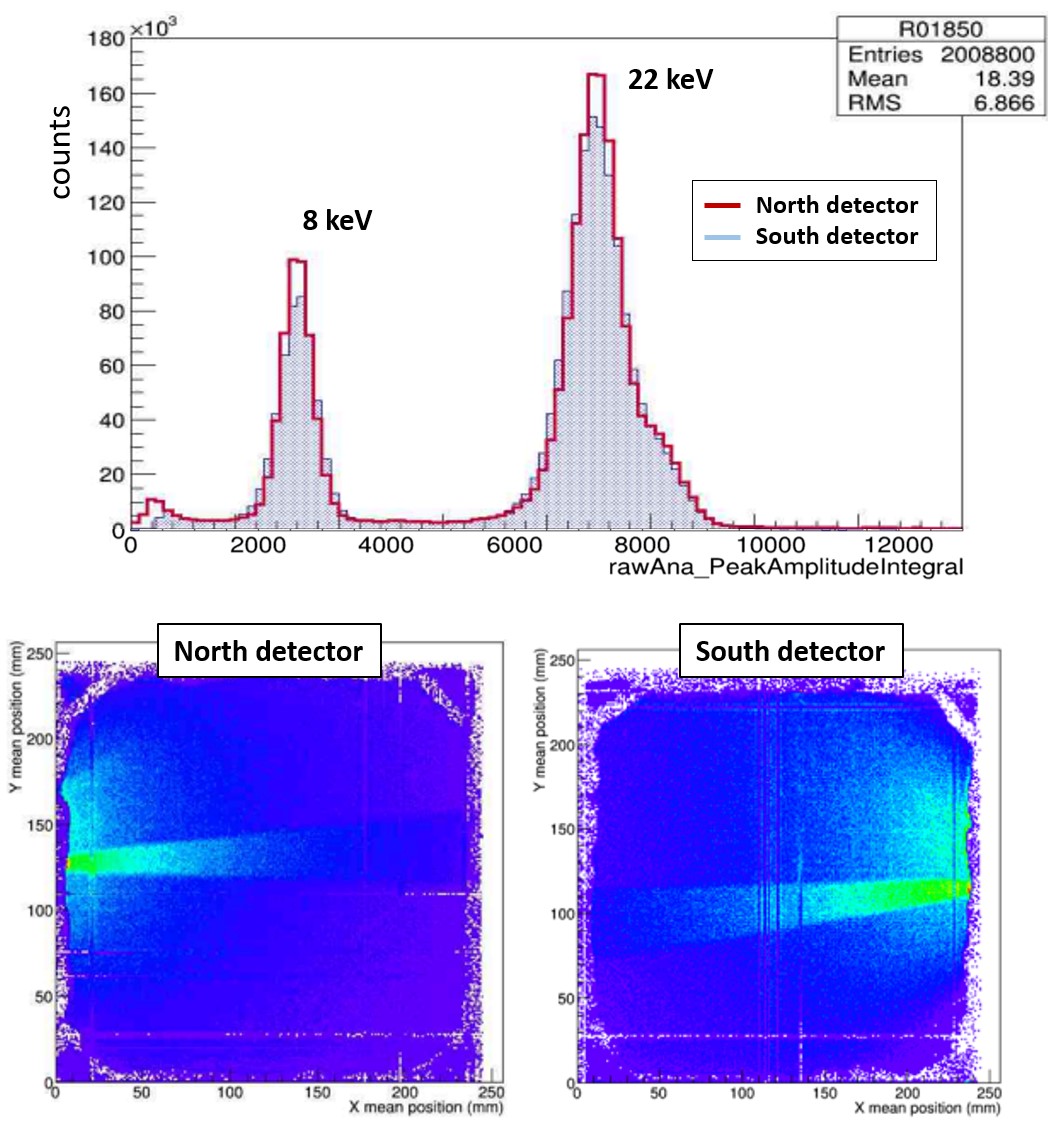}
    \caption{Results of the first calibration with the new Micromegas installed in TREX-DM, with Ne~+~2\%~Isobutane at 4~bar pressure. $\bullet$~Upper:~energy spectrum for the two readout planes at both ends of the chamber. $\bullet$~Lower:~activity maps for the two readout planes, where the beams from each of the $^{109}$Cd sources located on one side of each Micromegas can be distinguished.}
    \label{fig:TREXDM_CalibrationNewMM}
\end{figure}

Following this intervention, and to enable a preliminary comparison, the TPC was brought back to the same operating point prior to the intervention, with Ne~+~2\%~Isobutane at 4~bar pressure, with the same gain in the Micromegas (365~V) and the same drift field (200~V~cm$^{-1}$~bar$^{-1}$). The results of the first calibration can be seen in figure~\ref{fig:TREXDM_CalibrationNewMM}. The spectrum shows two peaks at 8 and 22~keV, with a resolution of 20\%~FWHM, better than the resolution achieved with the previous Micromegas (see figure~\ref{fig:TREXDMatLSC_Ne_CurvasResol}). The maps of the two installed Micromegas show that all channels are active, except for the anomalous channels identified in the previous tests of the detectors (see table~\ref{tab:TREXDM_TestsResults}).

The energy threshold in this calibration was 0.8~keV, a lower value than that achieved with the previous Micromegas ($\geq$1keV). This is likely due to a higher trigger efficiency of the acquisition electronics, as there are fewer damaged channels, combined with a better signal-to-noise ratio, due to reduced noise with the new connection chain. Furthermore, this threshold of 0.8~keV could be improved if the gain in the Micromegas were increased, as there is considerable room for improvement, theoretically allowing it to rise above 400~V (see figure~\ref{fig:TREXDMatLSC_Ne_CurvasTransp}). It should be noted that the voltage of 365~V used in this calibration was adjusted to allow comparison with the previous Micromegas under the same conditions. However, in the previous detectors, this voltage was defined conservatively due to stability issues caused by leakage currents.

Following the initial calibrations of the detector, a background data acquisition was initiated, yielding a background level approximately 20\% higher than in previous measurements, around ~$\sim$120~dru~(counts~keV${^{-1}}$~kg${^{-1}}$~d${^{-1}}$). This was likely due to the reduction in the energy threshold, combined with the greater efficiency of the acquisition electronics.

Since the background value was successfully reduced from 1.1·10$^3$ to ~100~dru, by changing the operating mode to open loop (see section~\ref{sec:RadonDominatedBackgroundReduction}), several efforts were made to further reduce it to the levels established by the background model: during interventions with the chamber open in the clean room, PTFE plates were placed on the walls of the active volume to block possible emissions from their surfaces; the cathode was also replaced on one occasion to eliminate potential contamination from radon decay nuclides that may have adhered to its surface.

\begin{figure}
    \centering
    \includegraphics[width=1.0\textwidth]{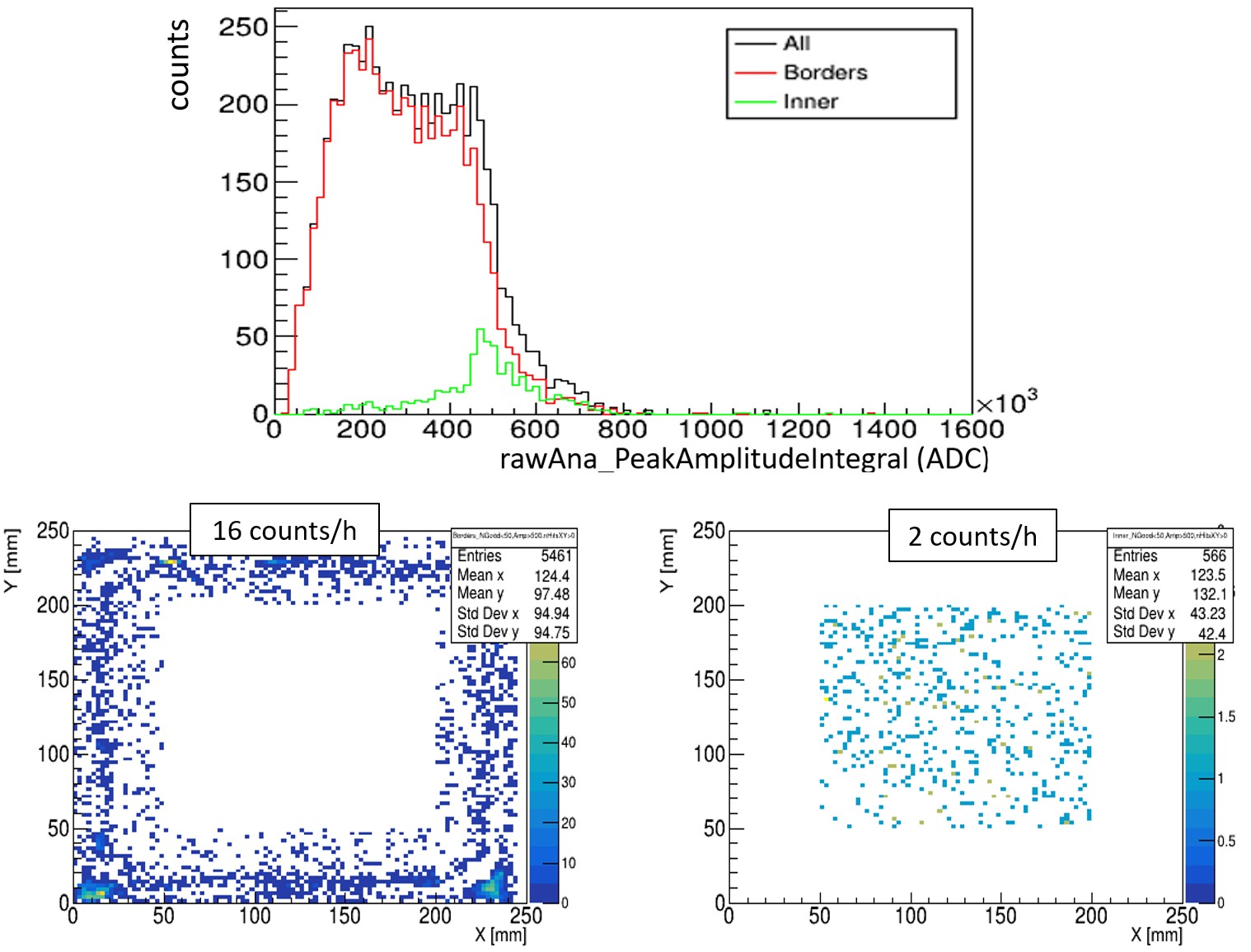}
    \caption{Results of the background data acquisition with the new Micromegas at low gain (210~V) for the detection of high-energy events, with Ne~+~2\%~Isobutane at 4~bar and 11 days of exposure. The spectrum (top) shows the total population of events collected across the entire detection plane (black line); along with the events collected in a centered area of 15~x~15~cm$^2$ (green line), whose activity map is shown in the lower right; and the events collected in the rest of the detection plane area (red line), whose activity map is shown in the lower left.}
    \label{fig:TREXDM_AlphasNewMicromegas}
\end{figure}

During the clean room intervention to install the new Micromegas, the opportunity was taken to replace the cathode for a second time; however, the contamination of its surface by alpha-emitting isotopes did not decrease (figure~\ref{fig:TREXDM_AlphasNewMicromegas}) compared to the measurements taken before the intervention. Everything seemed to indicate that the aluminized mylar used as the cathode was already contaminated before being installed in the chamber. Following this, in order to replace this mylar, a 16~$\mu$m thick kapton-copper sheet was prepared, the same material used for the fabrication of the Micromegas and therefore of low activity.

Within the series of activity measurements in materials carried out as part of the T-REX project~\cite{Castel2019}, both the aluminized mylar and the kapton-copper sheets used to construct the Micromegas were included. The results of both measurements were upper limits for the sensitive isotopes ($^{238}$U, $^{226}$Ra, $^{232}$Th, $^{228}$Th, $^{235}$U, $^{40}$K, $^{60}$Co, and $^{137}$Cs), as in both cases they were thin sheets with little mass and with an activity below the sensitivity of the detector used.

This highlighted the need to perform surface alpha particle measurements in the materials, leading to the design of the AlphaCAMM (Part III of this thesis). Until this detector is constructed and operational, the preparation of the kapton-copper sheet was decided upon, which, although like the mylar it had shown upper limits in its activity measurements, was made of the same material but of lesser thickness than the installed Micromegas. Additionally, it had not been subjected to the chemical attacks typical of their fabrication, so it was expected that its activity would be equal to or lower than that of the installed Micromegas.

However, the replacement of the mylar cathode with a kapton-copper sheet could not be carried out, as in October 2022 the experiment had to be dismantled and stored in a safe place, by order of the LSC management, awaiting the preparation of a new space for its installation. In May 2023, the experiment was relocated and installed in the new space, referred to as LAB 2500, within the underground facilities of the LSC and 100~m from the previous site, traversing the railway tunnel.

Once the installation and commissioning in the new location were completed, calibrations and background data were taken again, under the same previous conditions of mixture (Ne~+~2\%~Isobutane), pressure (4~bar), and gain (365~V), with results similar to those of the previous location.

After the initial checks (calibration, background, detector response), it was necessary to replace the contaminated cathode, for which a clean room was needed in the new location. This clean room was to be installed by the LSC as a service to the experiment; however, at the time of writing, the room has not yet been installed.

\chapter{Pre-amplification Stage with GEM}
\label{ch:GEM-MM}

\begin{figure}[!ht]
    \centering
    \includegraphics[width=1.0\textwidth]{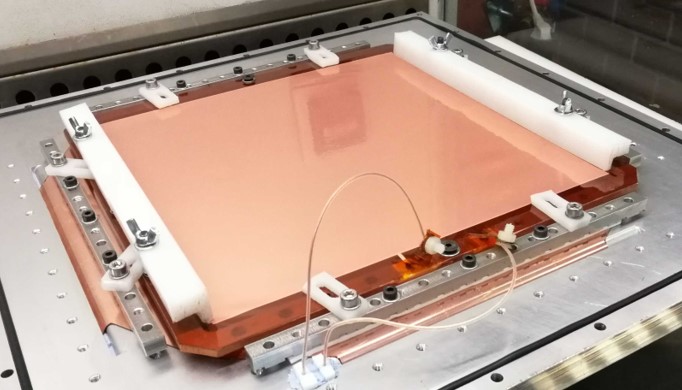}
    \caption*{}
    \label{fig:GEM_MM_PlanoLectura}
\end{figure}
\newpage

\section{Motivations}

\begin{figure}[!ht]
    \centering
    \includegraphics[width=1.0\textwidth]{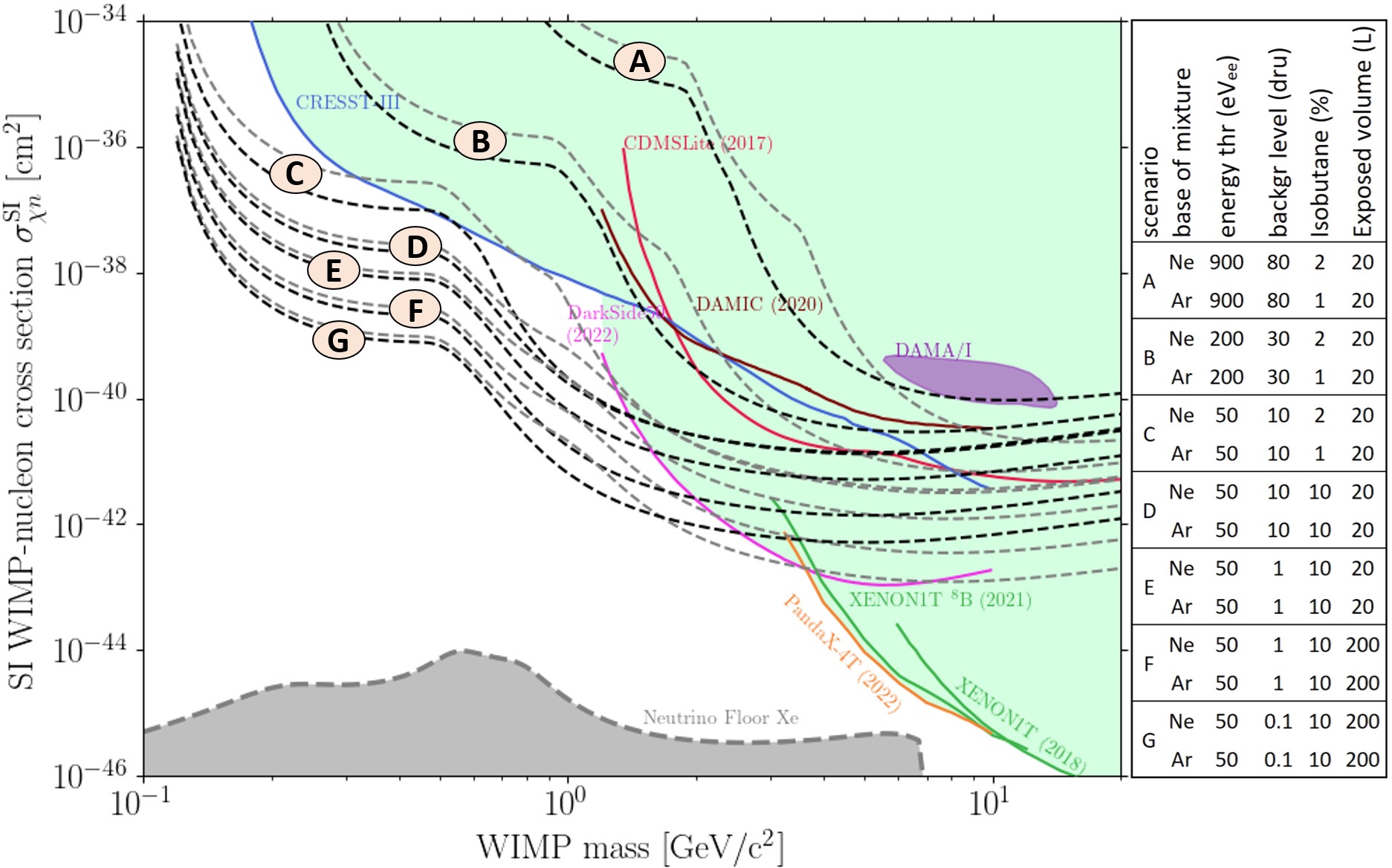}
    \caption{WIMP-nucleon cross-section versus WIMP mass, with the limits reached by the most important experiments and the limits that TREX-DM can achieve with one year of exposure, under the conditions outlined in the table on the right. Each scenario is represented with both a Neon mixture (black) and an Argon mixture (gray).}
    \label{fig:TREXDM_WIMP_ExclussionPlot_Ene24_v1}
\end{figure}

The initial sensitivity prospects for low-mass WIMPs of the TREX-DM experiment~\cite{Iguaz2016} were reviewed in November 2022, after 4 years of commissioning and data taking at the Canfranc Underground Laboratory, taking into account new results published by other experiments in the search for WIMPs (DarkSide50, Xenon1T, PandaX-4T). Following this review, the exclusion plot shown in figure~\ref{fig:TREXDM_WIMP_ExclussionPlot_Ene24_v1} was generated for mixtures with Isobutane of both Argon and Neon, with several possible scenarios for the future of the experiment, depending on the values achieved regarding energy threshold, background level, amount of Isobutane in the mixture, and exposure time. Scenarios F and G have been included under the assumption of a scaling of the experiment by a factor of 10 in terms of exposure volume.

The originally anticipated sensitivity for TREX-DM assumed an energy threshold of 400~eV$_{ee}$. However, due to the significant impact that reducing the threshold has on sensitivity, along with the growing need in the community to explore regions with lower WIMP mass, the necessity arose to seek new alternatives for reducing the energy threshold below the initial forecast.

The lowest possible energy threshold, which represents a limit in TPCs, is one that allows the detection of a single electron resulting from the ionization of the gas due to a particle passing through it. This limit depends on the ionization potential of the nuclei that make up the gas. For Argon and Neon, this energy is 15.76 and 21.56~eV, respectively~\cite{Smirnov2005}, although with the addition of elements that favor the Penning effect, this energy can be reduced to 10.55~eV, as in the case of adding Isobutane (iC$_4$H$_{10}$). Therefore, following the review of sensitivity in TREX-DM, the objective was set to achieve energy threshold values close to the limit imposed by physics, that is, significantly above these ionization potentials, around $\sim$50~eV$_{ee}$.

With the new Micromegas installed in TREX-DM, the lowest energy threshold achieved was 0.8~keV$_{ee}$ (see section~\ref{sec:TREXDMopt_TomaDatos}). At that time, the Micromegas were not operating at maximum gain, according to the gain curves from previous studies (see figure~\ref{fig:TREXDMatLSC_Ne_CurvasTransp}), so it seemed that there was room to further reduce the energy threshold. Additionally, more thorough work on reducing the noise level would help move in that direction. However, it would be difficult to reach the target threshold of $\sim$50~eV$_{ee}$.

Among the options to reduce the energy threshold was to increase the multiplication of electrons in the readout plane by installing a GEM (Gas Electron Multiplier)~\cite{Bondar1998} above the Micromegas, so that pre-amplification of the primary electrons would occur before reaching the Micromegas.

However, according to previous studies~\cite{Bondar1998}, the maximum gain achieved in a GEM worsens with pressure, at least in the Xe-CO$_2$ mixture and up to the 5 bar pressure reached in this study. This trend was confirmed for pure Argon and Xenon~\cite{Bondar2002}. But, in this same study, this trend seems to reverse for pure Neon, achieving higher gains at higher pressures. In any case, the need to reduce the energy threshold to a value close to the ionization potential implied the necessity to achieve factors between 20 and 30 of pre-amplification with the GEM. These pre-amplification factors seemed feasible, even at high pressures, but tests needed to be conducted with the Argon and Neon-based mixtures used in TREX-DM and at the nominal operating pressure (10 bar).

\subsection{Previous Studies}

To date, several studies have been conducted in which one or more GEMs have been installed above a Micromegas (generally bulk-type Micromegas). These studies have been diverse and aimed at different objectives, such as the study of GEM degradation in combination with a Micromegas \cite{Kane2003}, the study with a high rate of ionizing particles \cite{KANE2002}, the study of the ion back-flow returning to the drift field \cite{Aiola2016} \cite{Zhang2017}, or the mechanism of avalanche at low pressures for light noble gases \cite{Cortesi2018}. 

Generally, these studies were conducted in the field of high-energy physics and/or high ionization rates. Experiments involving rare events, particularly those focused on the search for low-mass WIMPs, such as TREX-DM, operate at low energies and ionization rates. However, there are some results and conclusions that can be drawn from these studies and applied in this case. 

For example, in \cite{Kane2003}, the degradation effect in the combined GEM-MM system due to charge accumulation during exposure to a photon source is studied. Following the measurements in this study, and after a visual inspection of the detectors, it is concluded that the GEM-MM system does not suffer any appreciable degradation. This study considers accumulated charges in the detectors on the order of mC/mm$^2$. In TREX-DM, the greatest charge accumulation could occur during long exposure calibrations with the $^{109}$Cd source at a rate of 200~Hz, where charges on the order of nC could accumulate distributed over the entire surface of the detectors (25~x~25~cm$^2$), well below the mC/mm$^2$ range considered in the referenced study. Therefore, it is expected that the GEM-MM system will not suffer degradation of this type.

In \cite{KANE2002}, the behavior of a GEM-MM system under high ionization rates and for various gas mixtures is studied. Among its results, it presents a comparison of the energy resolution between a calibration with photons from a $^{55}$Fe source, with amplification only in the Micromegas, and another calibration in which the GEM also provides pre-amplification, both with the same total gain. This result demonstrates that the energy resolution does not change when adding the GEM, at least under the conditions in which the measurements were made (90\%Ar+10\%DME). In this same study, it is also shown that the spark rate in the Ne-CF4-Isobutane (79:10:11) mixture is reduced by several orders of magnitude due to the addition of the GEM.

In \cite{Colas2004}, it is shown how the ion back-flow (IBF, for its acronym in English \textit{Ion Back-Flow}) in a TPC with Micromegas is proportional to the ratio between the drift field and the amplification field. Therefore, the value of the drift field in high-energy and/or high ionization rate experiments, where this ion flow can cause modifications to the field itself, is kept as low as possible. In the case of TPCs for rare events, it is not expected that this ion flow will produce sufficient distortion of the field to affect event reconstruction.

In addition to the described GEM-MM systems, recent developments of Micromegas \cite{Feng2021} have also been made, which are equipped with a double mesh, containing two amplification regions, similar to the GEM-MM system, but without a transfer region.

\section{Experimental Design}

\begin{figure}[!ht]
    \centering
    \includegraphics[width=0.8\textwidth]{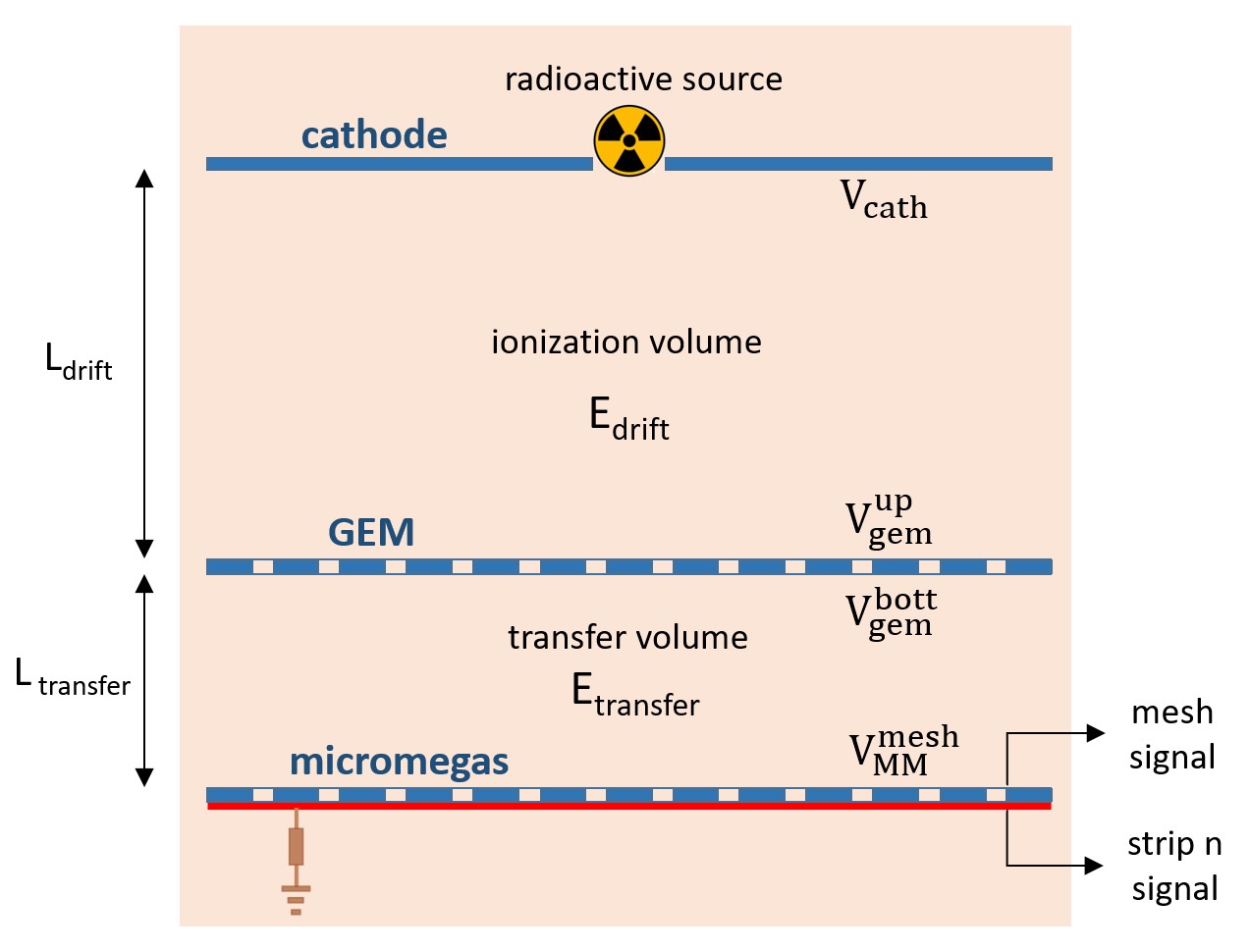}
    \caption{Schematic of the GEM-MM test system, with two differentiated zones: ionization volume and transfer volume.}
    \label{fig:GEM_MM_Esquema}
\end{figure}

To study the response of the TREX-DM detector with the introduction of the GEM pre-amplification stage, the scheme shown in figure \ref{fig:GEM_MM_Esquema} was proposed. A radioactive source of $^{55}$Fe or $^{109}$Cd is placed in a hole in the cathode, irradiating towards the ionization volume. Most of the ionizations due to the radioactive source occur in this volume, but some can also occur in the transfer volume, depending on the gas, pressure, and the mean path of the photons. The electrons produced in the ionization volume drift towards the GEM due to the electric field E$_{drift}$, where the first amplification occurs through the avalanche effect due to the amplification field \mbox{$E_{gem}^{amp}=(V_{gem}^{up}-V_{gem}^{bott})~/~50{\mu}m$}. The electrons resulting from the first amplification in the GEM enter the transfer volume and, thanks to the electric field E$_{transfer}$, drift towards the Micromegas, where the final amplification occurs due to the amplification field in the Micromegas \mbox{$E_{MM}^{amp}=(V_{MM}^{mesh}-0)~/~50{\mu}m$}. After this final amplification, the signal is read from the anode (\textit{strip}~n) or from the mesh.

\begin{figure}[!ht]
    \centering
    \includegraphics[width=0.9\textwidth]{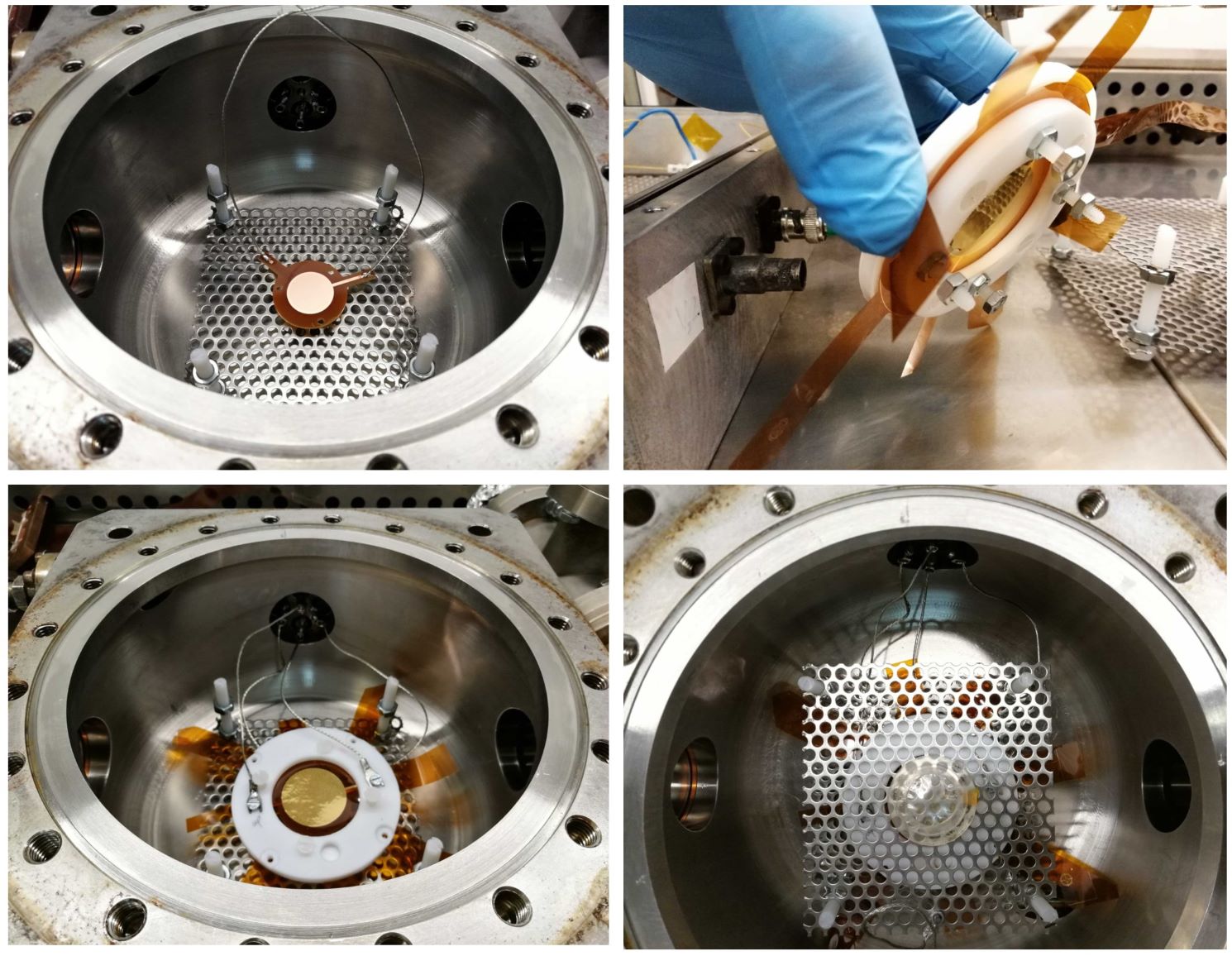}
    \caption{Photographs during the assembly in the 2-liter chamber with Micromegas and GEM of 20 mm diameter. $\bullet$~Top left:~top view of the interior of the chamber with the Micromegas. $\bullet$~Top right:~GEM mounted on a polyethylene frame. $\bullet$~Bottom left:~top view of the interior of the chamber once the GEM was placed on the Micromegas. $\bullet$~Bottom right:~top view of the interior of the chamber after placing the perforated cathode and the radioactive source on it.}
    \label{fig:GEM_MM_FotoCamaraPequena}
\end{figure}

To carry out the tests, two different setups were proposed. The first (figure \ref{fig:GEM_MM_FotoCamaraPequena}) consisted of a small 2-liter chamber in which a circular Micromegas with a diameter of 20 mm and a single channel was installed. The GEM, of the same size, was placed at L$_{transfer}$~=~10 mm, and the cathode was placed at L$_{drift}$~=~10 mm. This setup would allow measurements with Ar~+~1\%~Isobutane and Ne~+~2\%~Isobutane up to 10 bar of pressure.

\begin{figure}[!ht]
    \centering
    \includegraphics[width=\textwidth]{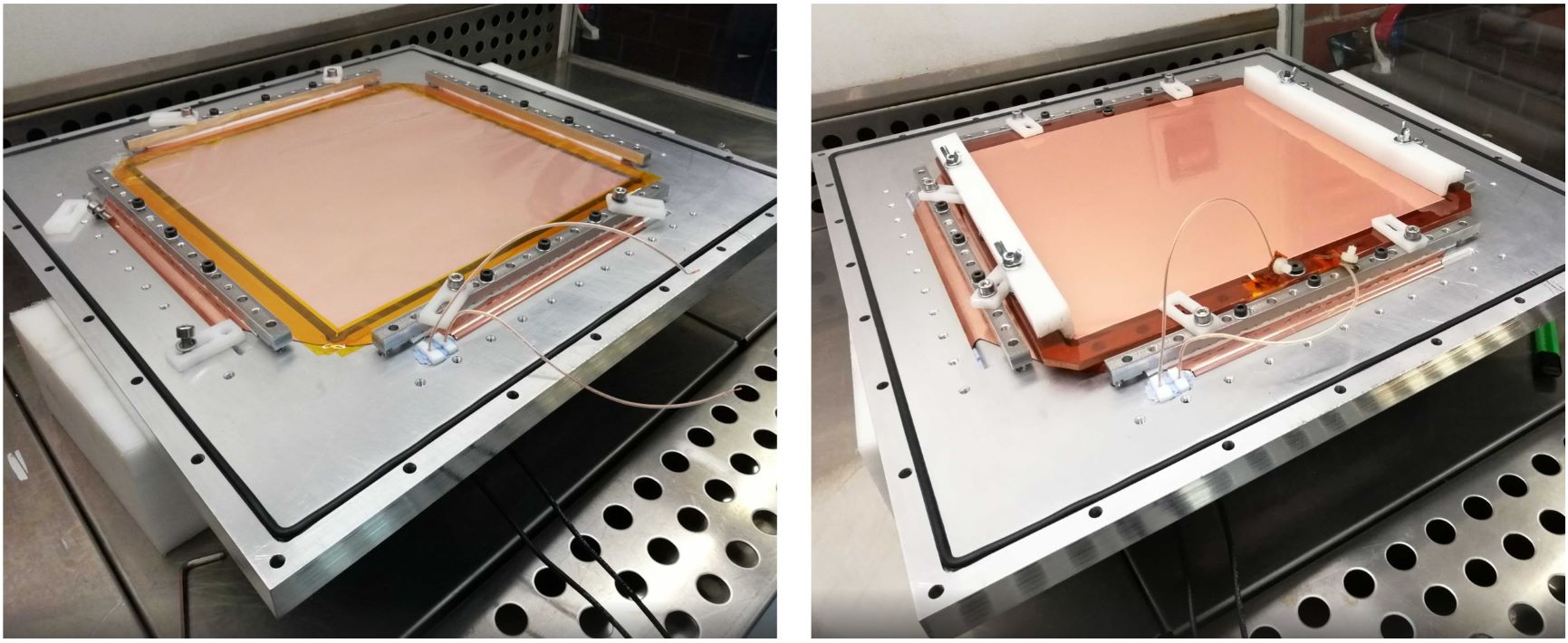}
    \caption{Photographs of the readout plane during the assembly in the 50-liter chamber. $\bullet$~Left:~Micromegas MMv2-5 installed on the lid of the chamber. $\bullet$~Right:~GEM of the same size as the Micromegas (25~x~25~cm$^2$) installed on top of it, at a distance of 10~mm.}
    \label{fig:GEM_MM_FotoReadoutCamaraGrande}
\end{figure}

The second setup (figure~\ref{fig:GEM_MM_FotoReadoutCamaraGrande}) involved using the same 50-liter chamber used for evaluating the new Micromegas of TREX-DM (section~\ref{sec:PruebasPrevias}). In it, the Micromegas MMv2-5 (table~\ref{tab:TREXDM_TestsResults}) was installed, and at a distance of L$_{transfer}$~=~10~mm, a GEM with the same active area (25~x~25~cm$^2$) was placed. At a distance of L$_{drift}$~=~100~mm, a perforated cathode was placed to allow irradiation from the source. This setup would allow measurements with Ar~+~1\%~Isobutane up to 1 bar of pressure and study how the introduction of a GEM affects the topological reconstruction of the events.

The GEMs used in both setups had the same characteristics and were manufactured by the MPT (Micro-Pattern Technologies) group at CERN: a kapton (50~$\mu$m) and copper (5~$\mu$m) foil, with a copper hole diameter of 70~$\mu$m, a kapton hole diameter of 60~$\mu$m, and a center-to-center hole distance of 140~$\mu$m. The only difference between them was the size of their active area.

In the case of the 2-liter chamber, the signal (figure~\ref{fig:GEM_MM_Esquema}) was read from the mesh and collected by an MCA (MultiChannel Analyzer) after being amplified. In the case of the 50-liter chamber, the signal was read from the \textit{strips} through the FEC-Feminos acquisition electronics, as is done in the TREX-DM experiment.

\section{Measurements in a 2-Liter TPC}
\label{sec:GEMMM_medidasDosLitros}

\subsection{Experimental Technique}

To supply the voltages to the Micromegas mesh (V$_{MM}^{mesh}$), to the GEM ($V_{gem}^{up}$ and $V_{gem}^{bott}$), and to the cathode (V$_{cath}$), a CAEN model N1471H power supply with 4 channels was used, with a voltage and current resolution of 1\% and 0.5~nA, respectively.

A radioactive source of $^{55}$Fe (X-rays of 5.9~keV) was installed in the cathode (figure~\ref{fig:GEM_MM_FotoCamaraPequena}), and Ne~+~2\%~Isobutane was used as the gas mixture, the same one that was being used at that time in TREX-DM.

\begin{figure}[!ht]
    \centering
    \includegraphics[width=\textwidth]{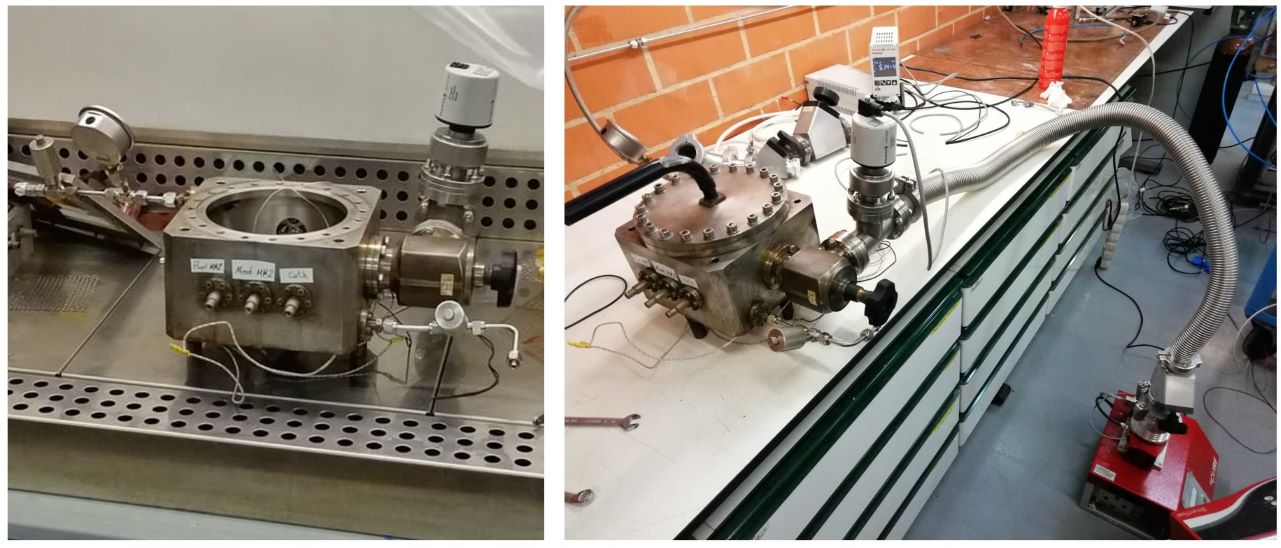}
    \caption{Photographs of the 50-liter chamber. $\bullet$~Left:~ Photograph during the assembly of the Micromegas and the GEM, inside a laminar flow cabinet. $\bullet$~Right:~Final installation situation, photo taken during the pumping of the chamber with a turbo pump (bottom right).}
    \label{fig:GEM_MM_FotoCamaraPequena2}
\end{figure}

The chamber had two ports with 6~mm tubing, valves for the inlet and outlet of the gas mixture, and a port with 40~mm tubing with a vacuum valve for pumping (figure~\ref{fig:GEM_MM_FotoCamaraPequena2}). The chamber had a leak rate below 10$^{-6}$~mbar~l~s$^{-1}$. The inlet port was connected to a take-off with a hand-reducer from the laboratory gas installation, and the outlet port was closed with the corresponding valve. The pumping port was connected to a turbo pump.

The procedure for operation, once all elements were installed and the chamber was closed, was as follows:

\begin{itemize}
\item Pumping with the turbo pump until 1·10$^{-4}$~mbar.
\item Closing the vacuum valve and filling the chamber with Ne~+~2\%~Isobutane to the operating pressure.
\item Gradually increasing the voltages in the Micromegas, the GEM, and the cathode to the operating point.
\item After 4 hours of measurements, and to ensure the quality of the gas in the chamber, the procedure was restarted from the first step.
\end{itemize}

For each operating point, defined by $V_{cath}$, $V_{gem}^{up}$, $V_{gem}^{bott}$, and $V_{MM}^{mesh}$, data were acquired for 5 minutes with approximately 30,000 events in each run. For the data runs where only amplification was present in the Micromegas, with the GEM functioning as a cathode and $E_{gem}^{amp}$~=~0~V, the aim was to collect events from the source that produced ionization solely in the transfer volume, between the GEM and the Micromegas; therefore, longer data runs were conducted to achieve the same number of events.

\subsection{Results}

\begin{figure}[!ht]
    \centering
    \includegraphics[width=0.9\textwidth]{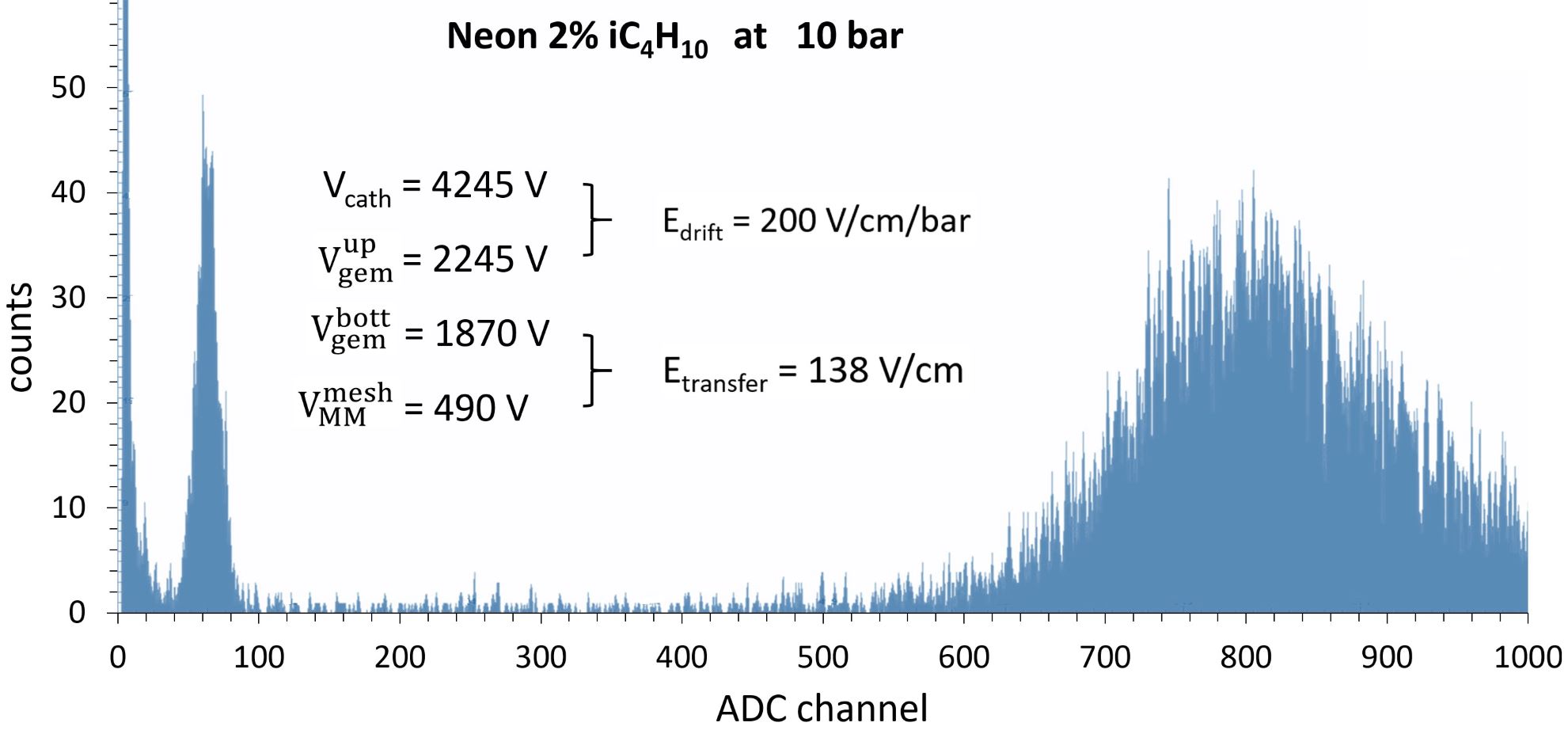}
    \caption{Energy spectrum generated during the initial tests in the GEM-Micromegas system with Ne~+~2\%~Isobutane at 10~bar. The first peak on the left, in channel 60 of the ADC, corresponds to events from the $^{55}$Fe source that produced ionization in the region between the GEM and the Micromegas. In this case, the GEM only acted as a cathode, without amplification. The peak on the right, in channel 825 of the ADC, corresponds to events from the source that produced ionization in the region between the cathode and the GEM. In this case, the GEM provides a pre-amplification of the primary electrons by a factor of 12.7.}
    \label{fig:GEM_MM_small_FirstSpectrum_Neon2Iso_10bar}
\end{figure}

For the initial measurements with the 2-liter chamber, it was filled with a mixture of Ne~+~2\%~Isobutane at 10~bar. After several tests adjusting both the amplification voltages of the GEM and the Micromegas, as well as the drift voltages between the Cathode-GEM and GEM-MM, the spectrum shown in figure~\ref{fig:GEM_MM_small_FirstSpectrum_Neon2Iso_10bar} was obtained. By comparing the peak on the left, corresponding to the ionization in the region between the GEM and the Micromegas, with the peak on the right, corresponding to the ionization in the region between the cathode and the GEM, a pre-amplification factor of 12.7 was calculated due to the installation of the GEM.

It should be noted that, at this operating point, the voltage of the Micromegas was well below the maximum value (550~V) reached in previous studies for this mixture~\cite{Iguaz2022}. Furthermore, GEMs can withstand higher voltages than Micromegas, so presumably, the GEM was also well below its maximum value. However, increasing these voltages caused saturation in the amplification chain, even though the amplifier was set to the minimum gain and maximum shaping time.

To estimate approximately what pre-amplification values could be achieved in these initial measurements, and knowing that systematic measurements would be taken later with an appropriate amplification chain, the amplifier was removed and an oscilloscope was used instead. In this way, and in a somewhat imprecise manner, the pre-amplification factor was estimated as follows:

\begin{itemize}
\item At the same operating point where a pre-amplification factor of 12.7 was achieved (figure~\ref{fig:GEM_MM_small_FirstSpectrum_Neon2Iso_10bar}), the trigger threshold of the oscilloscope was adjusted to have an event rate around 1~Hz. This threshold turned out to be 3.4~V.
\item The amplification field in the GEM was increased by raising $V_{gem}^{up}$ in steps of 10~V, adjusting the cathode voltage at each step to keep the drift field in the ionization volume constant at $E_{drift}$~=~200~V~cm$^{-1}$~bar$^{-1}$. For each increase of 10~V in the GEM, the event rate increased, so the oscilloscope threshold was adjusted again to maintain an approximate rate of 1~Hz. With this method, $V_{gem}^{up}$ was raised to 2285~V (40~V above the previously applied value, figure~\ref{fig:GEM_MM_small_FirstSpectrum_Neon2Iso_10bar}), with a trigger threshold of 10.1~V.
\item The ratio between the trigger threshold of the last measurement (10.1~V) and the trigger threshold of the first measurement (3.4~V), multiplied by the pre-amplification factor of the first measurement, results in the total pre-amplification factor (37.7). The values for all measured operating points can be seen in table~\ref{tab:GEM_MM_small_Tabla_Neon2Iso_10bar}.
\end{itemize}

\begin{table}[!h]
\centering
\begin{tabular}{l}
\includegraphics[width=0.90\textwidth]{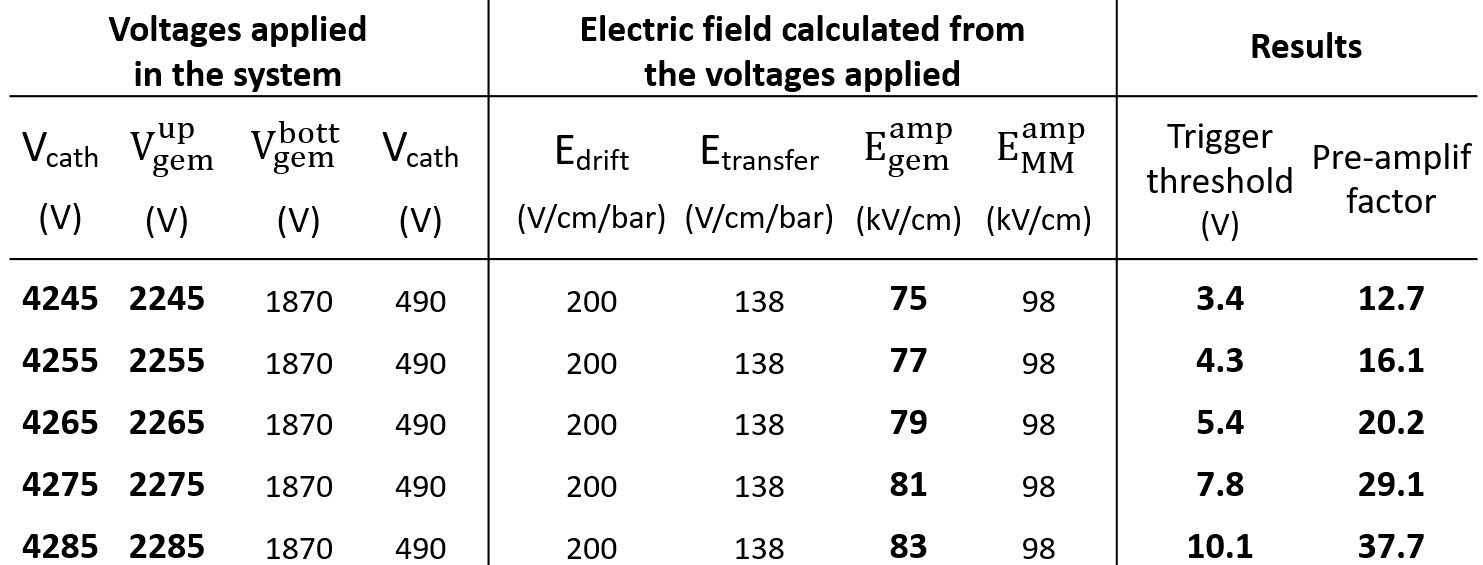}\\
\end{tabular}
\caption{Parameters of the initial measurements in the 2-liter chamber with Ne~+~2\%~Isobutane at 10~bar. Combined readout chain: MCA in the first measurement and oscilloscope in the remaining measurements.}
\label{tab:GEM_MM_small_Tabla_Neon2Iso_10bar}
\end{table}

In the last measurements, with $V_{gem}^{up}$~=~2285~V, increases in current began to occur, on the order of hundreds of nanoamperes, in the power supply channel of the Micromegas. There were several hypotheses to explain these currents (sparks, ion backflow, etc.), although it would be necessary to carry out a series of systematic measurements, and perhaps a review of the experimental design, to study the origin of these currents.

\subsection{Conclusions}

The method used to obtain the pre-amplification factor is imprecise, relying on the adjustment of the trigger threshold with the oscilloscope. Furthermore, no prior electron transmission curves have been performed, which would have allowed for the adjustment of the fields to achieve maximum transmission. However, it is sufficient to provide a first approximation and to understand what pre-amplification factors can be expected in TREX-DM with this system.

The energy threshold achieved by TREX-DM during its operation at the LSC has been 1~keV$_{ee}$, even lower with the new Micromegas and the new connection chain. In simple terms, if this threshold is reduced by a factor of 37.7 as the measurements suggest, the energy threshold of TREX-DM could reach 26~eV$_{ee}$. Therefore, it seems reasonable that the new prospects for TREX-DM, through the installation of a GEM as a pre-amplification stage, aim for an energy threshold of 50~eV$_{ee}$ (figure~\ref{fig:TREXDM_WIMP_ExclussionPlot_Ene24_v1}), a conservative reduction by a factor of 20 compared to the current threshold.

The initial measurements presented here with the 2-liter chamber, and the achievement of promising pre-amplification factors, would motivate further systematic measurements with this same chamber at various pressures and with different mixtures of Argon and Neon with Isobutane. This would also signify the starting point for the design of a complete GEM-MM readout plane for TREX-DM, utilizing one of the new Micromegas (chapter~\ref{ch:TREXDMoptimization}) and a GEM of the same size, whose initial tests were conducted in a 50-liter chamber and are described below.

\section{Measurements in 50-Liter TPC}

\subsection{Experimental Technique}

As in the case of the 2-liter chamber, a CAEN model N1471H power supply with 4 channels was used in the 50-liter chamber to supply the voltages to the Micromegas mesh (V$_{MM}^{mesh}$), to the GEM ($V_{gem}^{up}$ and $V_{gem}^{bott}$), and to the cathode (V$_{cath}$), with a voltage and current resolution of 1\% and 0.5~nA, respectively.

On this occasion, to study a greater energy range, a radioactive source of $^{109}$Cd was installed in the cathode, as shown in figure~\ref{fig:GEM_MM_Esquema}. With this source, it was expected to differentiate a peak at 22~keV (21.99, 22.16, 24.91, and 24.94) and a mono-energetic peak of 8~keV, the latter caused by the fluorescence of copper on the surface of the Micromegas.

\begin{figure}[!ht]
    \centering
    \includegraphics[width=\textwidth]{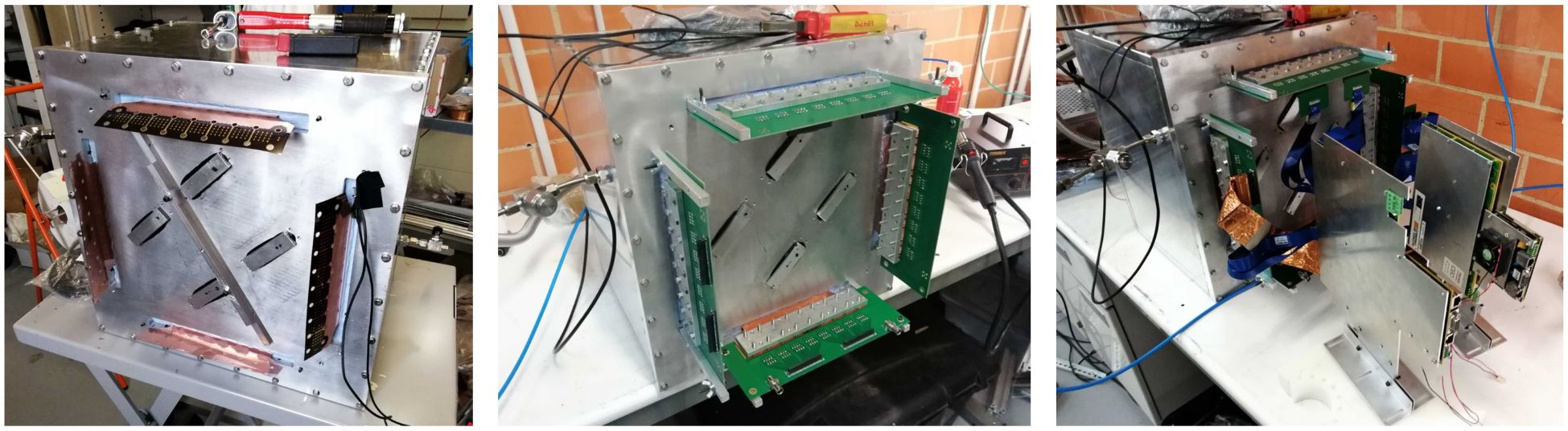}
    \caption{Photos of the 50-liter chamber. $\bullet$~Left:~lid with the 4 tabs of the Micromegas extending outside. $\bullet$~Center:~assembly of the four PCBs connecting the Micromegas to the FEC-Feminos acquisition electronics. $\bullet$~Right:~complete assembly with the two FEC-Feminos boards used and connected to the PCBs via 8 flexible cables.}
    \label{fig:GEM_MM_FotosExteriorCamaraGrande}
\end{figure}

The Micromegas MMv2-5 was chosen, and this time it was installed directly on the lid of the chamber (figure~\ref{fig:GEM_MM_FotoReadoutCamaraGrande}), differently from the previous testing phase before the installation in TREX-DM (section~\ref{sec:PruebasPrevias}, figure~\ref{fig:TREXDM_TestChamber}), where the Micromegas was positioned at a certain distance from the lid and the extraction of the channels was done using flat cables. This time, to simplify the assembly, it was decided to extract the four tabs of the Micromegas from the chamber through four slots (figure~\ref{fig:GEM_MM_FotosExteriorCamaraGrande}). The use of these slots facilitated the installation of the Micromegas; however, this resulted in a poor vacuum being achieved (10~mbar), so it was necessary to use a flow rate of 8~l/h and wait for several renewals of the gas volume to carry out the measurements.

The four tabs of the Micromegas, each with 128 channels, were connected to the FEC-Feminos electronics through a rigid PCB (Printed Circuit Board), designed and manufactured for this assembly, replacing the flexible cables designed and manufactured for its installation in TREX-DM (figure~\ref{fig:TREXDMFlatCable}). Like the flexible cables, these PCBs had an FtF connector on one side, which connected to the Micromegas, and on the other side, two ERNI connectors to connect to the FEC-Feminos.

The chamber had two ports with 6~mm tubing, valves for the inlet and outlet of the gas mixture, and a port with 40~mm tubing with a vacuum valve for pumping. The inlet port was connected to a take-off with a hand-reducer from the laboratory gas installation, and the outlet was connected to a flow meter. The pumping port was connected to a primary pump.

The operational procedure, once all elements were installed and the chamber was closed, was as follows:

\begin{itemize}
\item Pumping with a primary pump until 10~mbar.
\item Closing the vacuum valve and filling the chamber with Ar~+~1\%~Isobutane up to 1.05 bar.
\item Opening the outlet valve with regulation of the flow meter to 8~l/h.
\item Maintaining the flow for 72 hours. With a total of 11.5 renewals, the amount of air remaining inside the chamber is 3.4~ppm of the volume.
\item Gradually increasing the voltages in the Micromegas, the GEM, and the cathode until the operating point.
\end{itemize}

\begin{figure}[!ht]
    \centering
    \includegraphics[width=\textwidth]{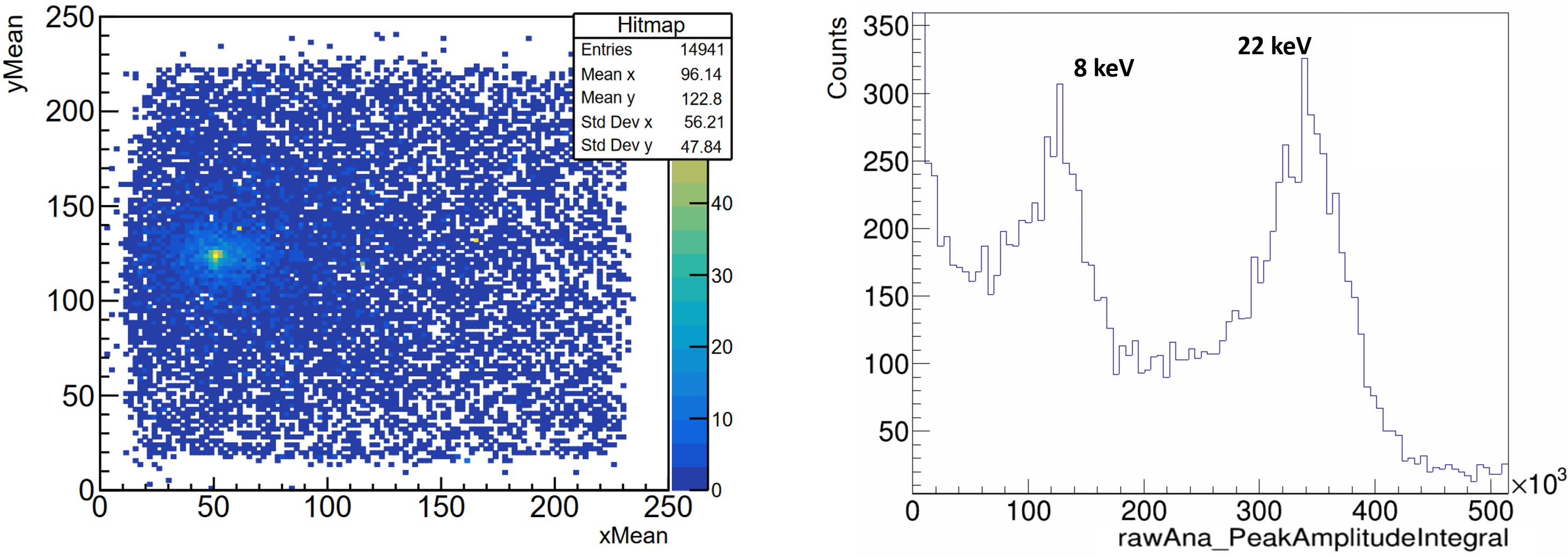}
    \caption{Representation of the unfiltered (raw) data from a 5-minute data run, for Ar~+~1\%~Isobutane at 1~bar, with $V_{MM}^{mesh}$~=~290~V and $V_{gem}^{up}$~-~$V_{gem}^{bott}$~=~270~V. $\bullet$~Left:~activity map in the Micromegas (25~x~25~cm$^2$), where the position of the $^{109}$Cd source is located. $\bullet$~Right:~energy spectrum represented by the observable PeakAmplitudeIntegral (sum of the maximum signal amplitude over all channels of the event), where the 8~keV peak, corresponding to the fluorescence of the copper from which the Micromegas is made, and the composite peak at $\sim$22~keV (21.99, 22.16, 24.91, and 24.94) from the $^{109}$Cd source are distinguished.}
    \label{fig:GEM_MM_grande_Spectrum_Vmm290_Vgem270}
\end{figure}

For each operating point, defined by $V_{cath}$, $V_{gem}^{up}$, $V_{gem}^{bott}$, and $V_{MM}^{mesh}$, data were acquired for 5 minutes with approximately 15,000 events in each run (figure \ref{fig:GEM_MM_grande_Spectrum_Vmm290_Vgem270}). For the data runs where amplification was only in the Micromegas, with the GEM functioning as a cathode and $E_{gem}^{amp}$~=~0~V, the aim was to collect events from the source that produced ionization only in the gas volume between the GEM and the Micromegas; therefore, longer data runs were conducted to achieve the same number of events.

For each data run, corresponding to a different operating point, the events collected in the area with the highest event density, corresponding to the area of the $^{109}$Cd source, were selected, in 0~$<$~xMean~$<$~100 and 75~$<$~yMean~$<$~175. Subsequently, the spectrum was fitted to an exponential for the background, plus a Gaussian for the 8~keV peak and several Gaussians for the peaks of the $^{109}$Cd source at $\sim$22~keV, thus allowing the calculation of the position and resolution of each peak.

\subsection{Results}

\subsubsection{Electron Transmission}

\begin{figure}[!ht]
    \centering
    \includegraphics[width=\textwidth]{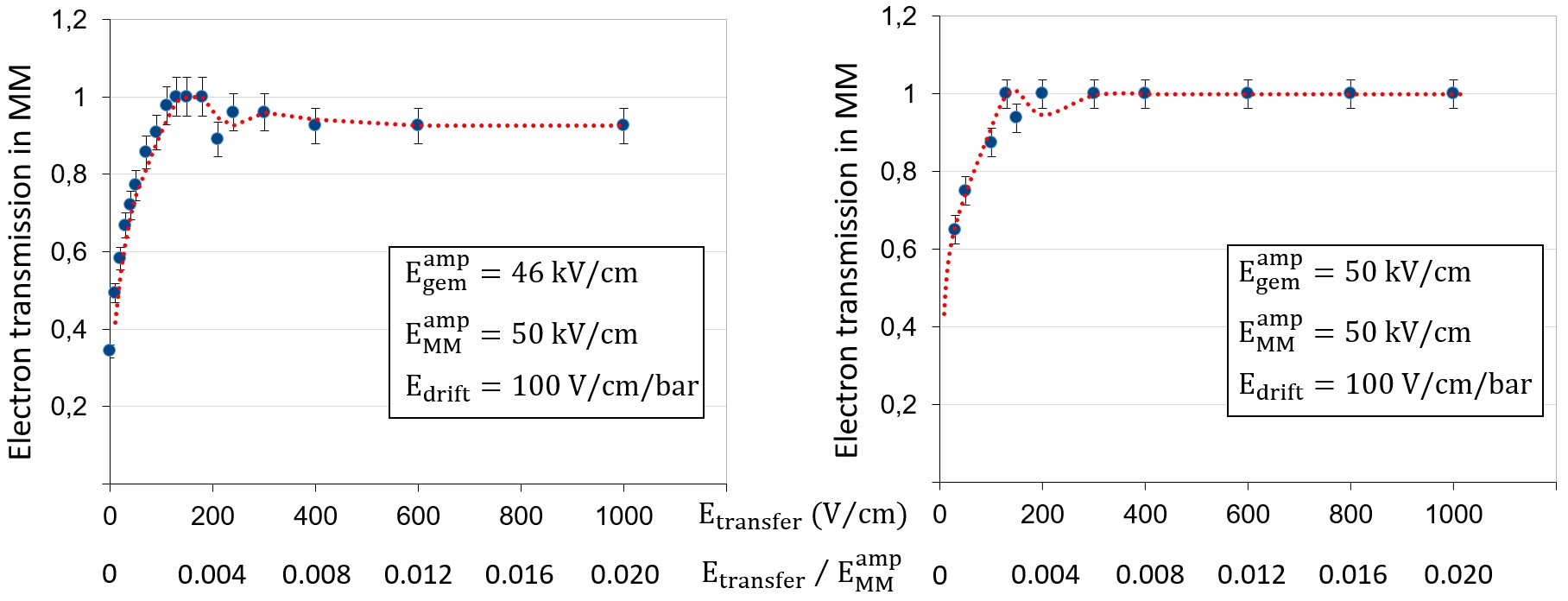}
    \caption{Electron transmission curves in the Micromegas, with Ar~+~1\%~Isobutane at 1~bar, for two different amplification fields in the GEM: E$_{gem}^{amp}$~=~46~kV/cm~(left) and 50~kV/cm~(right).}
    \label{fig:GEM_MM_CurvasTransparenciaMM_CamaraGrande}
\end{figure}

\begin{figure}[!ht]
    \centering
    \includegraphics[width=0.65\textwidth]{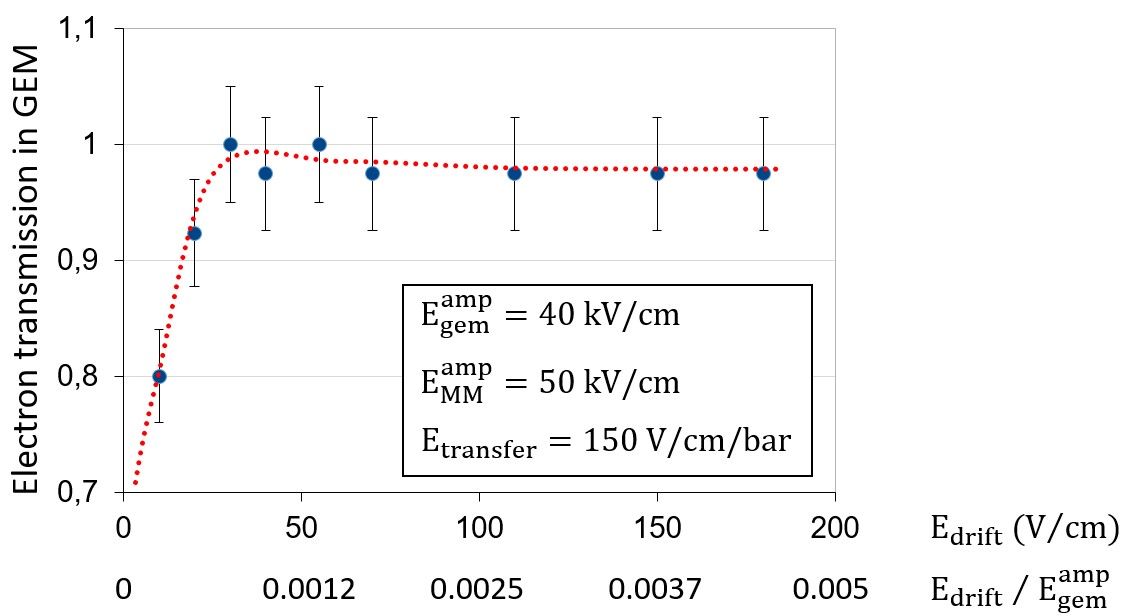}
    \caption{Electron transmission curves in the GEM installed in the GEM-MM system, with Ar~+~1\%~Isobutane at 1~bar.}
    \label{fig:GEM_MM_CurvasTransparenciaGEM_CamaraGrande}
\end{figure}

First, data were collected at different operating points to generate the electron transmission curves for both the Micromegas (figure~\ref{fig:GEM_MM_CurvasTransparenciaMM_CamaraGrande}) and the GEM (figure~\ref{fig:GEM_MM_CurvasTransparenciaGEM_CamaraGrande}).

The efficiency of electron transmission in a MPGD (Micro Pattern Gaseous Detector) primarily depends on the relationship between the electric fields in the region close to the detector surface and inside the amplification hole~\cite{Giomataris1996} (see also section~\ref{subsec:RevisionAreaActiva} and figure~\ref{fig:TREXDM_SimulacionPatron}).

The first transmission curve generated was for the Micromegas. The usual way to do this is by fixing the amplification field E$_{MM}^{amp}$ and varying the drift field above the Micromegas. In the case of the GEM-MM system, the drift field above the Micromegas is the transfer field E$_{transfer}$. To perform the measurements in this manner, it is necessary to ensure that the pre-amplification stage introduced by the GEM remains constant. In this sense, despite keeping the amplification field E$_{gem}^{amp}$ in the GEM and the drift field E$_{drift}$ above constant, the variation of E$_{transfer}$ for generating the transmission curve in the Micromegas affects the field configuration in the region close to the lower surface of the GEM, likely affecting the number of electrons generated in the avalanche that are expelled from the amplification hole into the transfer volume.

The generated curve (figure~\ref{fig:GEM_MM_CurvasTransparenciaMM_CamaraGrande}) has the expected shape: a loss of transmission for small drift fields, due to recombination and attachment processes during the drift of the electrons, and a flat region or plateau after a certain field value. For high drift fields, no loss of transmission is observed for primary electrons that end up in the mesh plane instead of in the amplification holes, although this was also not observed in previous studies~\cite{Iguaz2022} for the same gas mixture (figure~\ref{fig:CurvasArticuloIguaz2022}).

The transmission curve for the GEM was generated by keeping the values of $V_{MM}^{mesh}$, E$_{transfer}$, $V_{gem}^{up}$, and $V_{gem}^{bott}$ constant while varying the field E$_{drift}$. With this variation of E$_{drift}$, and therefore the variation of E$_{drift}$~/~E$_{gem}^{amp}$, the transmission curve was generated as it has historically been represented~\cite{Giomataris1996}~\cite{Iguaz2012}~\cite{Iguaz2022}, referenced to the maximum gain achieved, and assuming that the combined amplification stage (GEM / transfer zone / Micromegas) remains constant, as the parameters defining it are not modified.

In the transmission curve for the GEM (figure~\ref{fig:GEM_MM_CurvasTransparenciaGEM_CamaraGrande}), it can be seen that at low values of E$_{drift}$~/~E$_{gem}^{amp}$ there is a loss of transmission until reaching a flat region or plateau where the transmission remains constant, coinciding with this trend also observed in previous measurements for Micromegas~\cite{Iguaz2022} and the same gas mixture (figure~\ref{fig:CurvasArticuloIguaz2022}). However, due to the drift length and the maximum voltage that the cathode can withstand, measurements with E$_{drift}$ greater than 200~V/cm could not be taken. This limitation made it impossible to verify whether the plateau continued to be maintained.

\begin{figure}[!ht]
    \centering
    \includegraphics[width=1.0\textwidth]{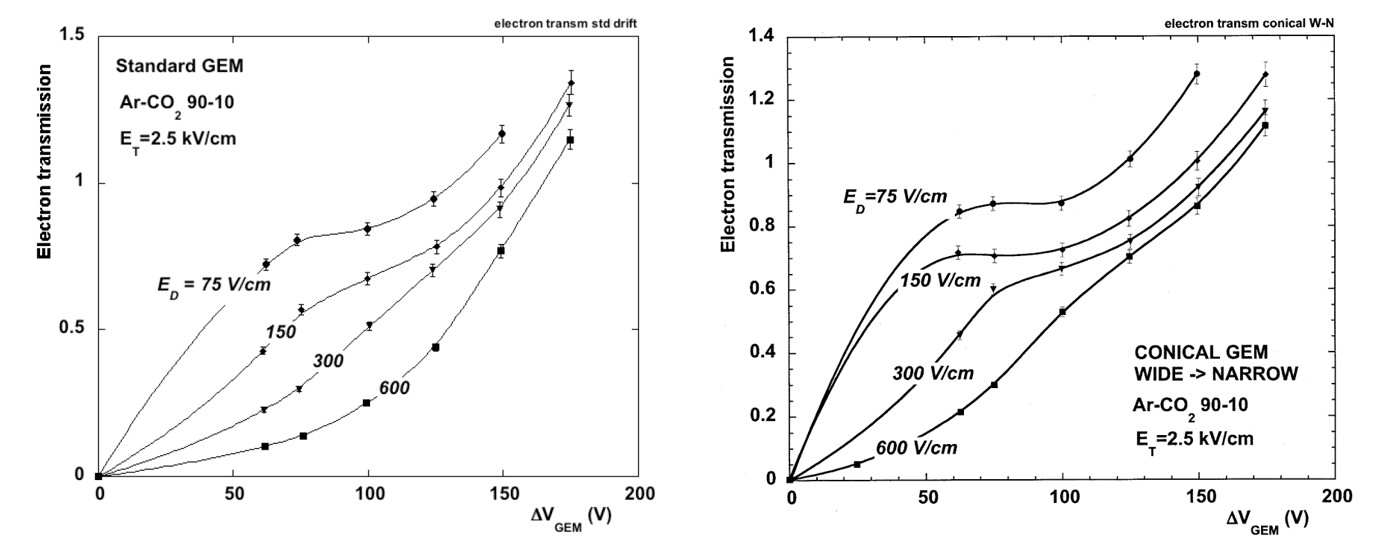}
    \caption{\cite{Sauli2003}~Electron transmission curves, for Ar~+~10\%~CO$_2$ at 1~bar.}
    \label{fig:GEM_MM_PaperSauliaEtAl_ElectronTransmission}
\end{figure}

In previous studies on electron transmission in a GEM~\cite{Sauli2003} (figure~\ref{fig:GEM_MM_PaperSauliaEtAl_ElectronTransmission}), it is shown that, for the same amplification field E$_{gem}^{amp}$ in the GEM ($\Delta$V in the figure), there is a loss of transmission as the drift field E$_{drift}$ (E$_D$ in the figure) increases, coinciding with the behavior observed in some studies for Micromegas~\cite{Iguaz2012}. Furthermore, this study also shows how transmission increases as the amplification field E$_{gem}^{amp}$ increases, tending to converge.

Generally, in the case of GEMs, the transfer fields after pre-amplification are usually much greater than the drift fields prior to it~\cite{Sauli2003}, precisely to favor the extraction of electrons from the hole into the transfer volume. On the other hand, high values of the transfer field have the opposite effect in the Micromegas, attracting electrons towards the mesh instead of towards the holes. The ideal transfer field will therefore be one that, by summing the two effects described in the GEM and the Micromegas, achieves higher transmission rates.

With the transmission curves generated for both the Micromegas and the GEM, it was possible to select the field ratios to achieve maximum electron transmission from the ionization (and transfer) volume to the amplification holes in the GEM (and to the holes in the Micromegas). This relationship of fields would be maintained for all subsequent operating points: $E_{transfer}/{E_{MM}^{amp}}=$0.008 and $E_{drift}/{E_{gem}^{amp}}=$0.0025.

\subsubsection{Gain}

The gain in a gaseous detector, such as a Micromegas or a GEM, can be defined as the multiplication factor that occurs on the primary electrons, originating from the ionization volume and entering the amplification holes, due to the electron avalanche process. The multiplication factor of the electrons through the electron avalanche process strongly depends on the electric field present in the amplification holes, which in turn is determined by the voltage difference between the upper and lower planes of the detector and the distance separating them.

Typically, to carry out the gain study of a Micromegas or a GEM, a TPC (Time Projection Chamber) is used with a radioactive source emitting in the energy range to be studied. Each gain curve is generated for a specific gas mixture and pressure. The variation of the voltage applied between the upper and lower planes, and therefore the variation of the electric field that precedes the electron avalanche process, will cause a variation in the multiplication factor of the electrons.

\begin{figure}[!ht]
    \centering
    \includegraphics[width=\textwidth]{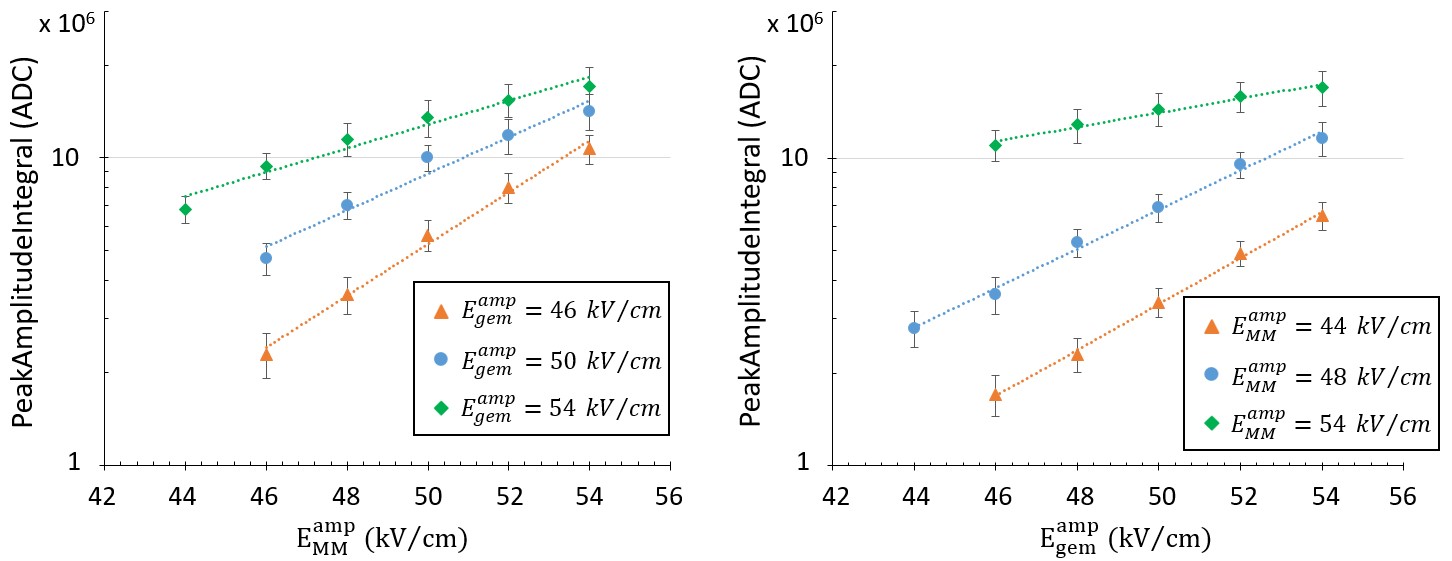}
    \caption{Total gain curves of the GEM-MM system in Ar~+~1\%~Isobutane at 1~bar, with the field ratios for transmission of 0.008 in the Micromegas and 0.0025 in the GEM. The gain is represented by the observable PeakAmplitudeIntegral. $\bullet$~Left:~Total gain curve as a function of the amplification field in the Micromegas, for various values of the amplification field in the GEM. $\bullet$~Right:~Total gain curve as a function of the amplification field in the GEM, for various values of the amplification field in the Micromegas.}
    \label{fig:GEM_MM_gainCurves_CamaraGrande}
\end{figure}

In the case of the developing GEM-MM system, there are two amplification stages whose respective gains depend on the applied amplification fields. The signal is read after the second amplification stage, in the Micromegas. The study of the total gain of the system must take into account both amplification fields, so several curves were developed for both cases, as shown in figure~\ref{fig:GEM_MM_gainCurves_CamaraGrande}.

The first curves generated (figure~\ref{fig:GEM_MM_gainCurves_CamaraGrande}~left) show the total gain as a function of the amplification field in the Micromegas. For each operating point, where the amplification field in the Micromegas increases compared to the previous one, the other voltages must be adjusted to keep the amplification field in the GEM constant (46, 50, and 54~kV/cm) and also to maintain constant the field ratios that ensure maximum electron transmission in both the Micromegas and the GEM.

The second series of curves generated (figure~\ref{fig:GEM_MM_gainCurves_CamaraGrande}~right) shows the gain as a function of the amplification field in the GEM. In this case, for each operating point, where the amplification field in the GEM increases compared to the previous one, only the voltage at the cathode needs to be adjusted to maintain the field ratio in the GEM area.

In both series of curves, the behavior of the gain is as expected; the gain increases exponentially as the amplification field increases. However, it is noteworthy that the different curves generated from the same series have different slopes, tending to converge. Intuitively, one would expect that a specific increase in the amplification field in one of the two stages would cause a specific jump in the total gain of the system, regardless of the initial gain from which it starts. However, for example, it is observed that the jump in gain that occurs when moving from the operating point E$_{MM}^{amp}$~=~46~kV and E$_{gem}^{amp}$~=~46~kV to the operating point E$_{MM}^{amp}$~=~46~kV and E$_{gem}^{amp}$~=~54~kV is by a factor of $\sim$7, while when moving from E$_{MM}^{amp}$~=~54~kV and E$_{gem}^{amp}$~=~46~kV to the operating point E$_{MM}^{amp}$~=~54~kV and E$_{gem}^{amp}$~=~54~kV, it is by a factor of $\sim$3.

\begin{figure}[!ht]
    \centering
    \includegraphics[width=\textwidth]{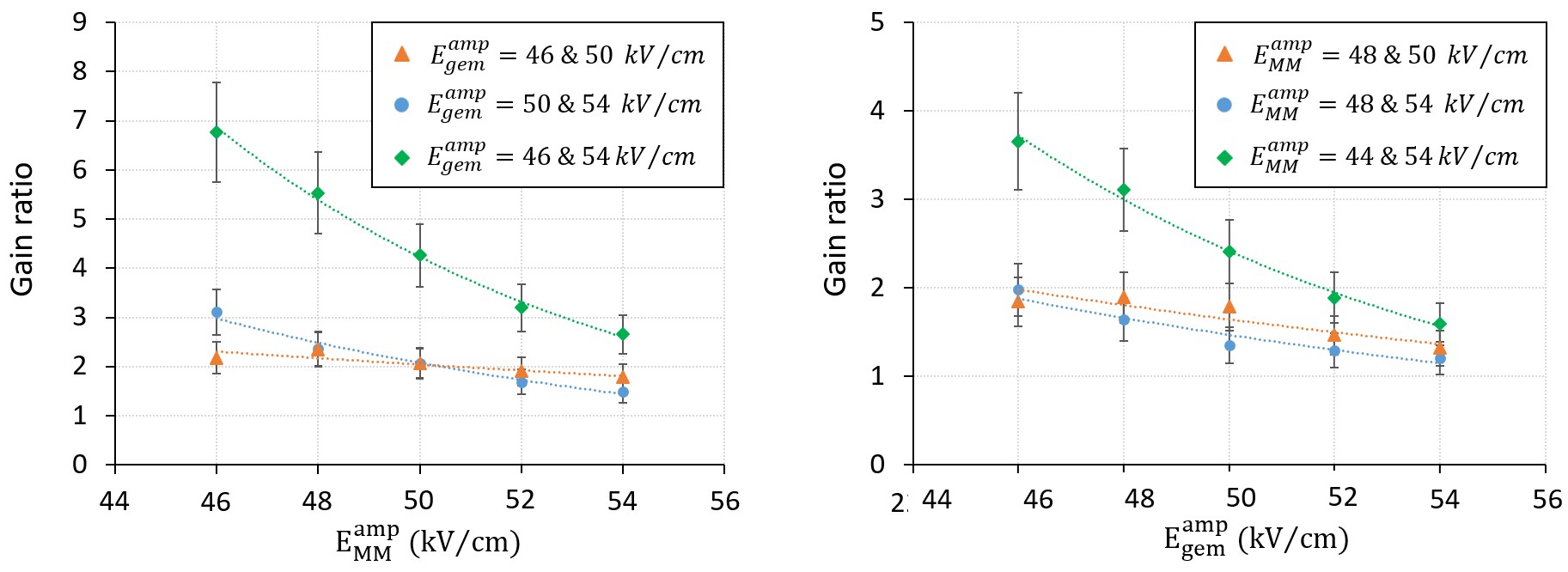}
    \caption{Gain ratios. $\bullet$~Left:~Ratio between the gains of two operating points with the same E$_{MM}^{amp}$ and different E$_{gem}^{amp}$. $\bullet$~Right:~Ratio between the gains of two operating points with the same E$_{gem}^{amp}$ and different E$_{MM}^{amp}$.}
    \label{fig:GEM_MM_RelacionGanancias_CamaraGrande}
\end{figure}

This effect is extensively shown in figure~\ref{fig:GEM_MM_RelacionGanancias_CamaraGrande}. Each curve is the comparison between two of the gain curves shown in figure~\ref{fig:GEM_MM_gainCurves_CamaraGrande}; for each amplification field on the X-axis, the ratio between the gains of the two compared curves is plotted. In all generated curves, a drop in the ratio is observed as the amplification field increases, with a steeper slope in the curves that compare amplification fields with a greater difference between them (E$_{gem}^{amp}$~=~46$\&$54~kV and E$_{MM}^{amp}$~=~44$\&$54~kV).

To explain this effect, it is necessary to study the factors involved in the processes of electron transport and electron avalanche. It is important to understand the factors that influence the transmission of electrons from the ionization volume to the amplification hole~\cite{Nikolopoulos2011}. Additionally, there may be polarization effects on the surfaces of kapton in the amplification holes that can cause variations in the electric field~\cite{Correia2014}, such that these variations in the topology of the electric field in the hole influence the electron avalanche process~\cite{Attie2014}. In general, a study and understanding of the electron avalanche process is necessary~\cite{Zerguerras2015,Ruiz-Choliz2015}. The study of the behavior of detectors under high particle fluxes~\cite{Giomataris1996} can also help understand cases with high charge density.

On the other hand, to complete the gain curve of the Micromegas and to determine the pre-amplification factors achieved through the introduction of the GEM, further measurements should be conducted over a wide range of amplification fields in the GEM, from 0~kV/cm to the maximum allowed before sparking. In this regard, to perform measurements where the GEM does not produce pre-amplification, the exposure time increases considerably because the ionization from the source events only occurs in the transfer volume. Therefore, only one data acquisition was launched with the maximum gain in the Micromegas (60~kV/cm) and without pre-amplification in the GEM. The comparison between the gain achieved in this data acquisition (the maximum gain that can be achieved in a TPC under these conditions and with a Micromegas readout plane) and the maximum gain achieved during the generation of the curves (E$_{gem}^{amp}$~=~54~kV/cm and E$_{MM}^{amp}$~=~54~kV/cm) yielded a factor of 50 for pre-amplification. Additionally, it should be noted that for the gain curves, neither of the two detectors was pushed to their limits, so it is expected that there is room to achieve even greater pre-amplification factors.

After confirming a factor of 37.7 for pre-amplification in this thesis, factors around 100 were observed during systematic measurements in the 2-liter chamber, under the same mixture and pressure conditions as in the initial measurements described in section~\ref{sec:GEMMM_medidasDosLitros}. These studies are not included in this report and will soon be the subject of a publication.

\subsubsection{Energy Resolution}

\begin{figure}[!ht]
    \centering
    \includegraphics[width=\textwidth]{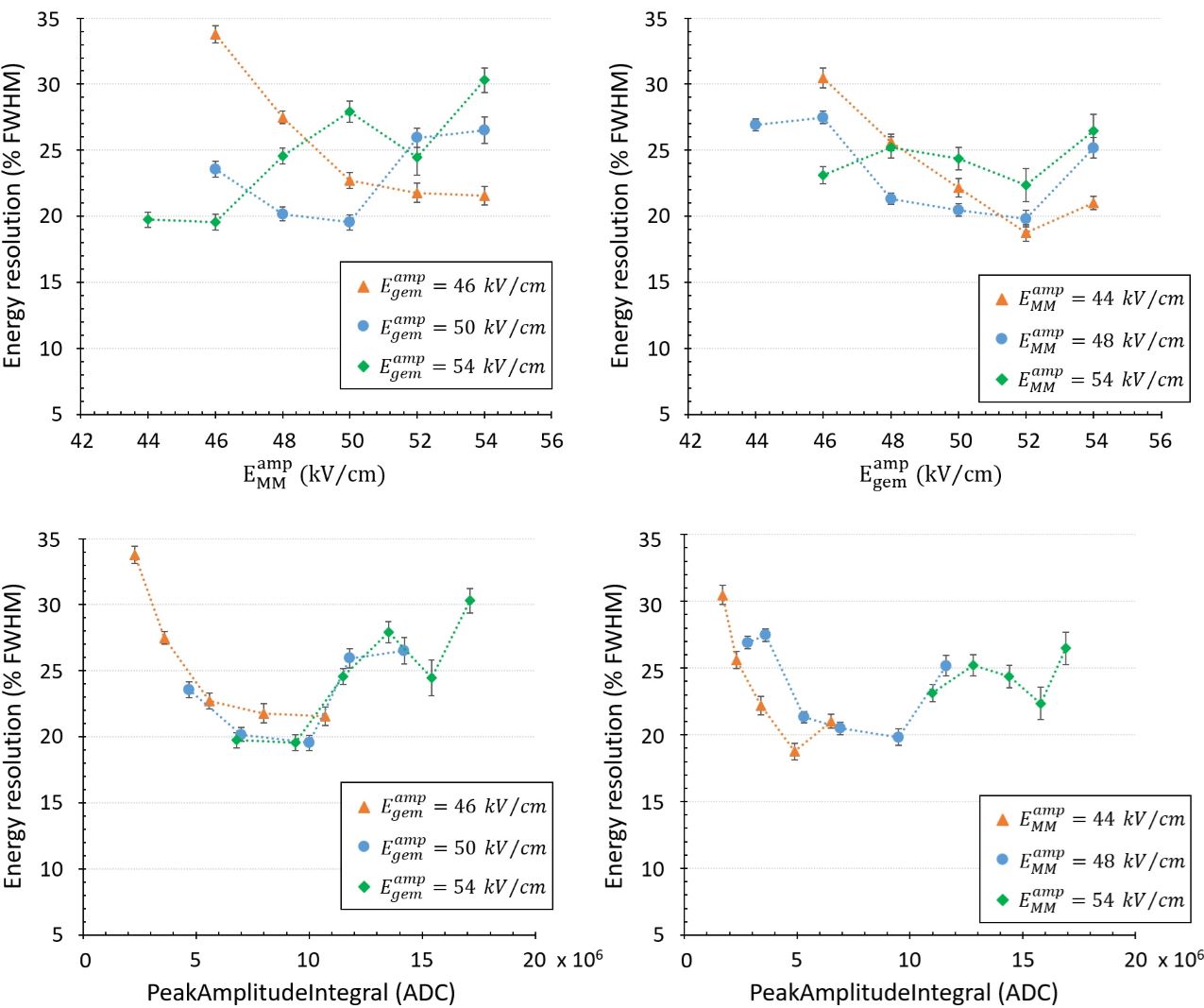}
    \caption{Energy resolution curves of the GEM-MM system in Ar~+~1\%~Isobutane at 1~bar, with the field ratios for transmission of 0.008 in the Micromegas and 0.0025 in the GEM. The gain is represented by the observable PeakAmplitudeIntegral. $\bullet$~Left:~resolution versus the amplification field in the Micromegas (top) and versus the total gain (bottom), for various values of the amplification field in the GEM. $\bullet$~Right:~resolution versus the amplification field in the GEM (top) and versus the total gain (bottom), for various values of the amplification field in the Micromegas.}
    \label{fig:GEM_MM_ResolutionCurves_CamaraGrande}
\end{figure}

In figure~\ref{fig:CurvasArticuloIguaz2022}, a curve of energy resolution versus gain is shown for Ar~+~1\%~Isobutane at 1~bar, conducted in previous studies with single-channel Micromegas~\cite{Iguaz2022}. It is observed that the curve demonstrates an improvement in resolution with increasing gain until reaching an optimal zone, after which it deteriorates in the final segment for the highest gains. In the GEM-MM system (figure~\ref{fig:GEM_MM_ResolutionCurves_CamaraGrande}), the trend of improvement in resolution up to an optimal zone appears to occur in those curves where the Micromegas or the GEM has low amplification (E$_{gem}^{amp}$~=~46~kV/cm and E$_{MM}^{amp}$~=~44~kV/cm), while in the other curves, there are no data for the low gain region.

The best energy resolution achieved across several operating points among all curves is around 20\%, while the optimal resolution reached in figure~\ref{fig:CurvasArticuloIguaz2022} is 12\%. It should be noted that for all generated curves, a basic data analysis has been used, without applying any topological cuts or gain corrections. The generation of a gain map of the detector, characterized by zones, allows for an energy correction of each event and an improvement in the energy resolution in the spectrum. Furthermore, it is important to remember that previous studies with single-channel Micromegas~\cite{Iguaz2022} were conducted in a more sealed TPC with a drift length of 1~cm, compared to the lower sealing and the 10~cm drift of the 50-liter test chamber. These conditions favor recombination and attachment processes in the 50-liter chamber, affecting both gain and energy resolution, although this could not be quantified.

\subsection{Conclusions}

Following the measurements in the 2-liter chamber, where a pre-amplification factor of 37.7 was achieved for Ne + 2\% Isobutane at 10 bar, it was necessary to verify the behavior of the GEM-MM system in larger and pixelated readout planes, in order to validate the system for its implementation in TREX-DM.

The results obtained from the measurements in the 50-liter chamber are satisfactory and indicate good performance of the system. From a topological perspective, the measurements allow for the correct identification of the event topology (figure~\ref{fig:GEM_MM_grande_Spectrum_Vmm290_Vgem270}). The generated electron transmission curves exhibit expected behavior (figures~\ref{fig:GEM_MM_CurvasTransparenciaMM_CamaraGrande} and \ref{fig:GEM_MM_CurvasTransparenciaGEM_CamaraGrande}). The generated gain curves also show the anticipated behavior (figure~\ref{fig:GEM_MM_gainCurves_CamaraGrande}), although it is observed that the gain jump that occurs when any of the amplification fields is increased (figure~\ref{fig:GEM_MM_RelacionGanancias_CamaraGrande}) is dependent on the total gain. Nevertheless, a pre-amplification factor of approximately 50 has been measured in Ar + 1\% Isobutane at 1 bar. Furthermore, the energy resolution curves show very promising values of around 20\%, which could be improved with the application of a more refined analysis utilizing a gain map for the energy correction of each event.

After these measurements in both chambers, the GEM system was validated, and preparations began for the installation of the GEM on the Micromegas in the TREX-DM detector. At the time of writing this thesis, everything is ready for installation, pending the establishment of a clean room at the location where TREX-DM has been relocated, Lab2500.

\part{Development of AlphaCAMM: a high-sensitivity alpha particle detector}

\setlength{\baselineskip}{12pt} 
\selectlanguage{USenglish}
\graphicspath{{../Imagenes/}{../../Imagenes/}} 

\renewcommand{\listtablename}{Table of contents}
\renewcommand{\tablename}{Table} 

\chapter{The first demonstrator as an alpha particle detector}
\label{ch:AlphaCAMM_Intro}
\chaptermark{The first demonstrator as a detector...}

\begin{figure}[!ht]
    \centering
    \includegraphics[width=0.8\textwidth]{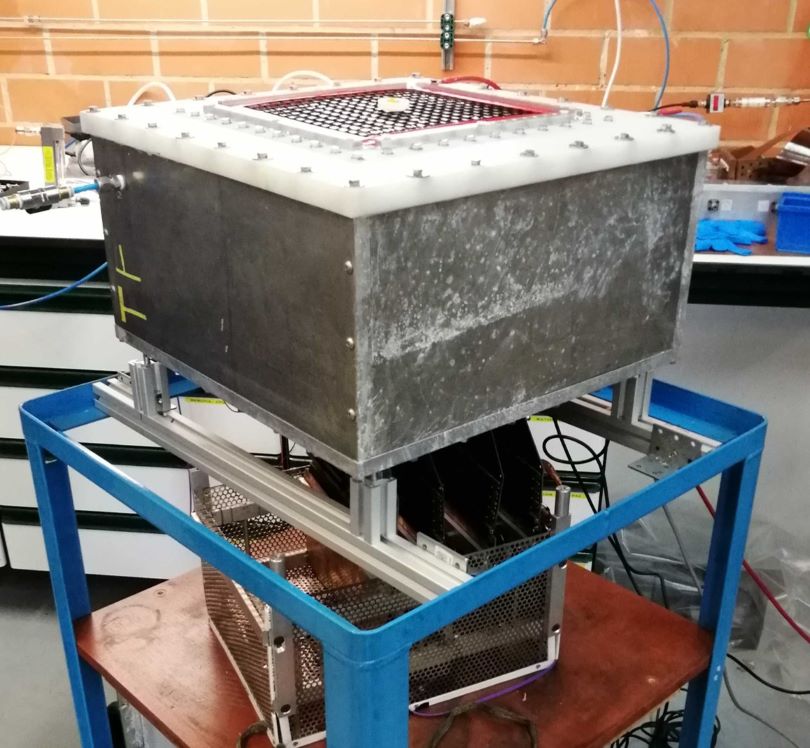}
    \caption*{}
    \label{fig:AlphaCAMM_FotoDemostradorPortada}
\end{figure}

\newpage

\section[Necesidad de un detector de partículas alfa superficiales]{Need for a Surface Alpha Particle Detector \sectionmark{Need for a Surface Alpha Particle Detector}}
\sectionmark{Need for a Surface Alpha Particle Detector}
\label{sec:NecesidadAlphaCAMM}

Experiments involving rare events aim to achieve the minimum possible background level, allowing for maximum sensitivity enhancement and exploitation. In the case of TREX-DM, a background model was developed~\cite{Castel2019} that took into account all possible contributions: radioactive isotopes in the materials used for the construction of the detector, Radon gas and its progeny outside the TPC, and activity from the environment within the LSC (gamma rays, muons, and neutrons).

From the developed model, it was concluded that the main contributions to the background, within the region of interest (RoI) of 0.4$-$7~keV$_{ee}$, were threefold: the Micromegas themselves ($\thicksim$ 2.5 dru), the field cage ($\thicksim$ 1.5 dru) made of kapton-copper sheets, and the copper chamber ($\thicksim$ 1.2 dru) that would have undergone cosmogenic activation in the laboratories of the University of Zaragoza, during the phase prior to its definitive installation in the LSC.

These three contributions could be reduced by manufacturing new Micromegas (see section \ref{sec:NewReadouts}), redesigning a new low-background field cage, and replacing the copper chamber with a new one that would reduce surface exposure.

\begin{figure}[!ht]
    \centering
    \includegraphics[width=\textwidth]{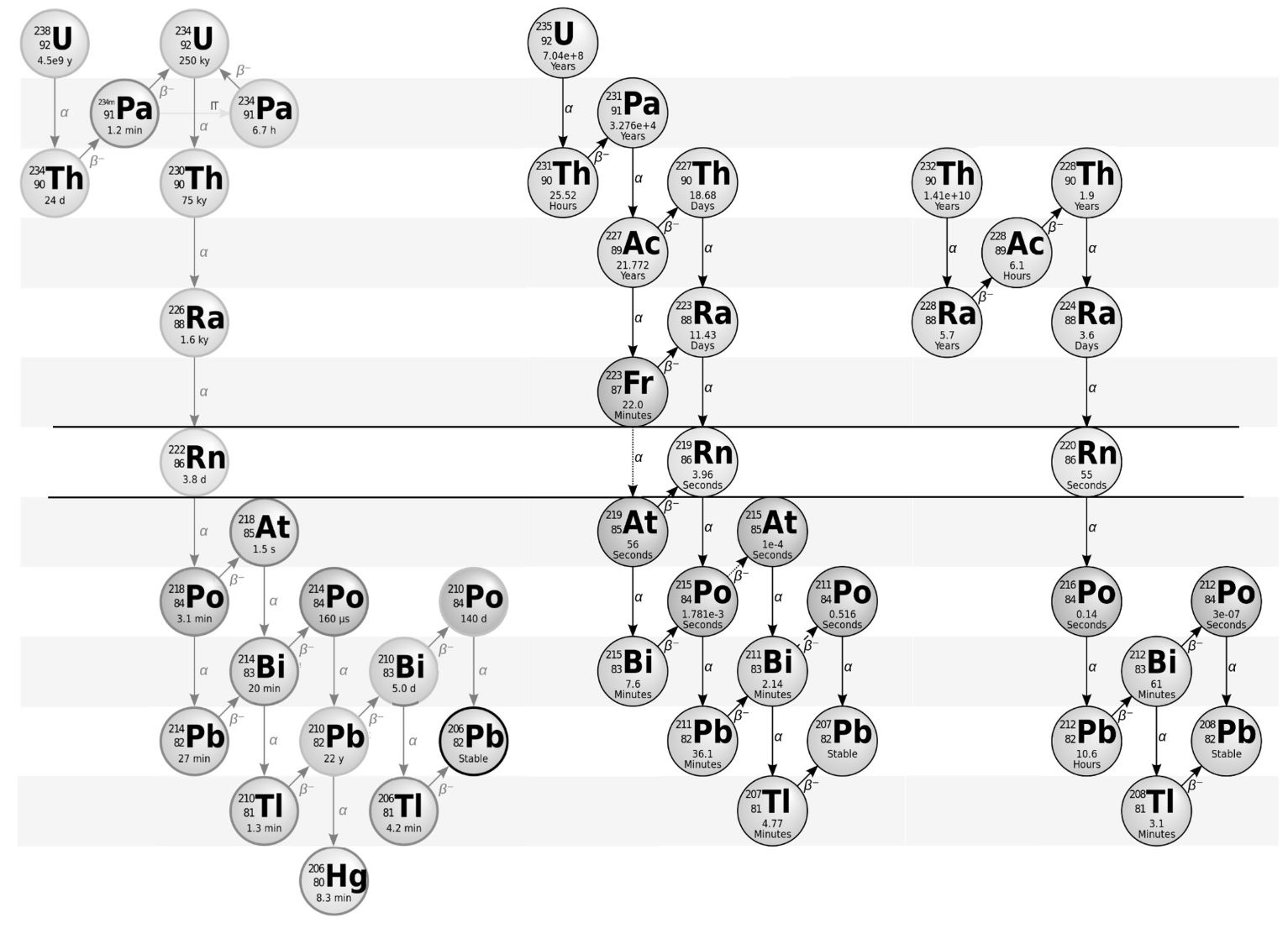}
    \caption{\cite{chains_wikipedia} Natural decay chains containing the nuclide Radon. When a radium isotope contained in a material decays into a radon isotope, a noble gas, it diffuses more easily out of the material. Once Radon has diffused from the material into the gas volume, the isotopes of Polonium and Lead, with very short half-lives, remain positively charged after the decay of their parents, making them susceptible to being attracted to charged surfaces. These processes favor the accumulation primarily of the long-lived isotope $^{210}$Pb on surfaces, significantly contributing to the background in low-background detectors.}
    \label{fig:AlphaCAMM_NaturalChains}
\end{figure}

However, after the initial data with Ne~+~2\%~Isobutane in TREX-DM, it was observed that the detector's background in the region of interest was two orders of magnitude greater than the level predicted by the background model (see section~\ref{sec:BackgroundFisrtMeasurements}). It was found that this elevated level was caused by Radon contamination in the gas system. To address this, an activity of high-energy events associated with alpha-emitting decays was first identified (figure~\ref{fig:TREXDMatLSC_HEespectro}), and it was subsequently confirmed that a reduction in this activity caused a corresponding reduction in the low-energy region, in the region of interest of 0.4$-$7~keV$_{ee}$, at an approximately 1:1 ratio. If there was Radon contamination in the TPC, it was expected that low-energy electron emissions associated with natural decay chains (see figure~\ref{fig:AlphaCAMM_NaturalChains}) would contribute to the background in the region of interest.

After discovering that the oxygen filter was a significant Radon emanator within the gas system, other types of filters were tested, although none yielded satisfactory results. Ultimately, it was decided to change the mode of operation and work with the detector in an open loop, passing the gas mixture directly from the bottle to the chamber and expelling it into the environment through the chamber outlet, with a low flow rate ($\thicksim$~1~l/h). This approach achieved a reduction of the background by one order of magnitude (see section \ref{sec:RadonDominatedBackgroundReduction}). Nevertheless, the background level remained one order of magnitude higher than that predicted by the background model, this time dominated by the emission of alpha particles from surfaces closest to the active volume. To mitigate its impact, several measures were implemented, such as installing a PTFE piece that internally covered the entire field cage to block emissions from its surface, or replacing the mylar of the cathode with a new one. However, the reduction in background due to these actions was minimal.

Therefore, the need arose to measure the concentrations of radioactive isotopes in natural decay chains. Specifically, the progeny of Radon, whose decay causes contamination of exposed surfaces and whose alpha emissions were the source of the background problem in TREX-DM and other low-background experiments. Thus, it became necessary to measure the surface contaminations derived from these isotopes.

For measuring the radiopurity of materials, in order to develop the background model for TREX-DM, various spectrometric techniques (HPGe, GDMS, ICPMS, BiPo-3) have been employed. However, these techniques are challenging to use for quantifying Radon progeny due to their low sensitivity to these isotopes. Additionally, the added difficulty of measuring surface contaminations would require these techniques to clearly separate the surface from the rest of the material, except for the BiPo-3 detector~\cite{Barabash2017}, which is specifically designed for measuring thin materials.

Gamma emission-based spectrometry techniques, such as HPGe (High Purity Germanium detectors), are generally not sensitive to low-energy lines from gamma emissions of isotopes further down the natural decay chains (figure~\ref{fig:AlphaCAMM_NaturalChains}), where the Radon progeny is found. There are exceptions, such as BEGe (Broad Energy Germanium) detectors, which can achieve sensitivities of a few $\mu$Bq/kg in the gamma emission of $^{210}$Pb at 46.5~keV, as in the case of the Roseberry detector \cite{Akerib2020radiopurity}, which reaches a sensitivity of 3.47~$\mu$Bq/kg. Even with the use of these BEGe-type germaniums, some pieces are difficult to measure, especially thin or flat pieces that have small mass and large surface area, such as mylar or kapton-copper sheets.

On the other hand, mass spectrometry techniques, such as GDMS (Glow Discharge Mass Spectrometry) or ICPMS (Inductively Coupled Plasma Mass Spectrometry), achieve sensitivities close to those of BEGe (units of $\mu$Bq/kg). These assays are destructive, although they typically require a small amount of mass for testing, which may be unfeasible in the case of measuring final pieces to be installed in the detector, which have large surface areas and low mass. Furthermore, although these techniques have been improved, they generally do not allow differentiation between isotopes of the same element.

For all these reasons, within the TREX-DM team, the need arose to develop a surface alpha particle detector with high sensitivity, capable of measuring contaminations in future pieces that would be installed within the TREX-DM chamber, especially for large surface pieces/sheets ($\sim$~25~x~25~cm$^{2}$) relative to the total size of the experiment. The sensitivity goal for this alpha particle detector was set at 100~nBq/cm$^{2}$, with a background level of 60~nBq/cm$^{2}$.

As previously noted, during the operation of TREX-DM, a 1:1 relationship of low-energy events to high-energy events was observed, meaning that for every detected alpha particle, there was a low-energy event in the region of interest of 0.4$-$7~keV$_{ee}$. Assuming this relationship, the projected sensitivity for the alpha detector (100~nBq/cm$^{2}$) would allow for the measurement of activities from pieces/sheets of 25~x~25~cm$^{2}$ that contribute to the TREX-DM background level of $\geq$~5~dru for Neon and $\geq$~2.5~dru for Argon.

The background model in the TREX-DM experiment~\cite{Castel2019} indicates a level on the order of 1$-$10~dru in the region of interest 0.4$-$7~keV$_{ee}$. In a second stage, by reducing the contribution from some elements of the detector (Micromegas, field cage, and copper chamber), the model points to a level below 1~dru. For this second stage, the sensitivity of the AlphaCAMM should be increased.

\section{First Demonstrator}
\label{sec:PrimerDemostrador}

\begin{figure}[!ht]
    \centering
    \includegraphics[width=0.7\textwidth]{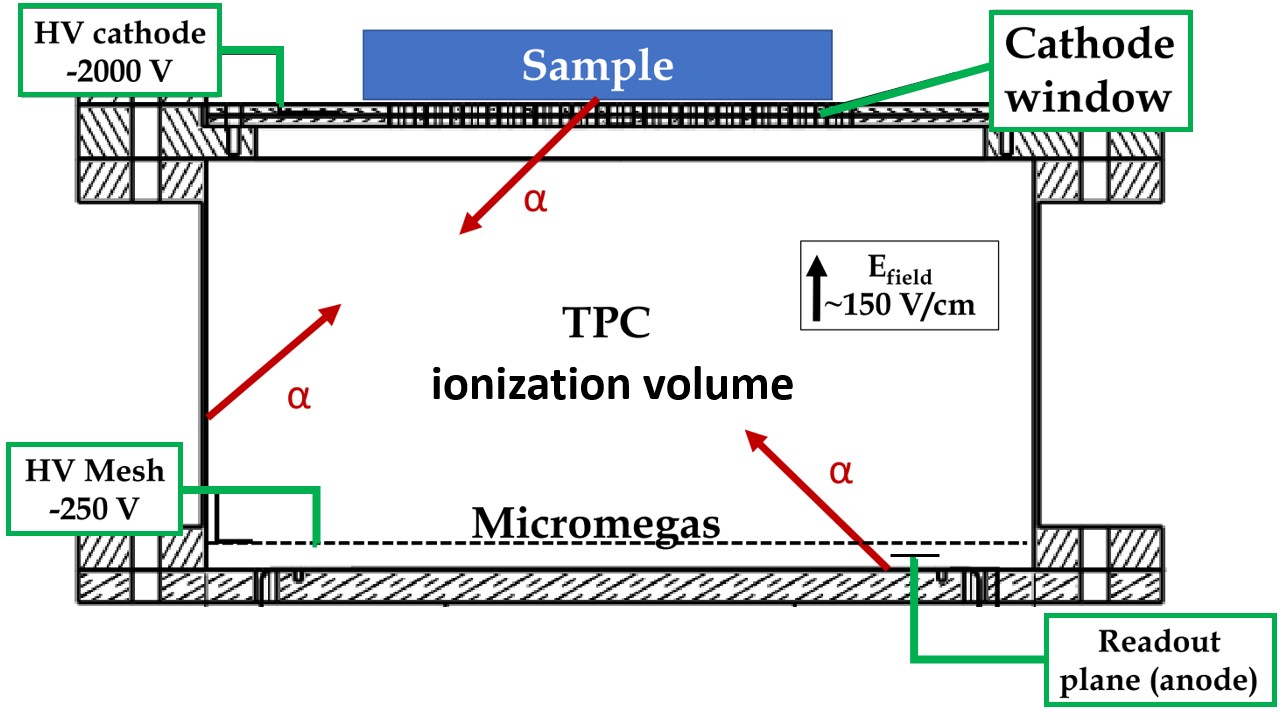}
    \caption{Schematic of the operating principle of the AlphaCAMM demonstrator. The sample is placed on a grid above the cathode, which is transparent to $\alpha$ particles. The reading plane based on a Micromegas (bulk type) provides topological information for the reconstruction of the tracks (direction and sense), so only $\alpha$ particles directed towards the Micromegas contribute to the measurement.}
    \label{fig:AlphaCAMM_EsquemaDemostrador}
\end{figure}

As a first step towards the design and construction of the AlphaCAMM, an initial prototype was developed~\cite{Altenmuller2022AlphaCAMM} as a demonstrator of the operating principle, which would serve as a starting point for defining the final design specifications. For this prototype, a TPC (Time Projection Chamber) with Micromegas was proposed, the schematic of which can be seen in Figure \ref{fig:AlphaCAMM_EsquemaDemostrador}, where a sample is placed with the surface to be measured facing the ionization volume, allowing the ionization in the gas produced by a particle emitted from the surface to be detected. The pixelation (\textit{strips}) of the Micromegas enables the reconstruction of the direction of the $\alpha$ particles. Furthermore, due to the Bragg peak, a known effect whereby an $\alpha$ particle deposits more energy at the end of its path while slowing down in a medium, the direction of its trajectory can also be determined.

\begin{figure}[!ht]
    \centering
    \includegraphics[width=0.95\textwidth]{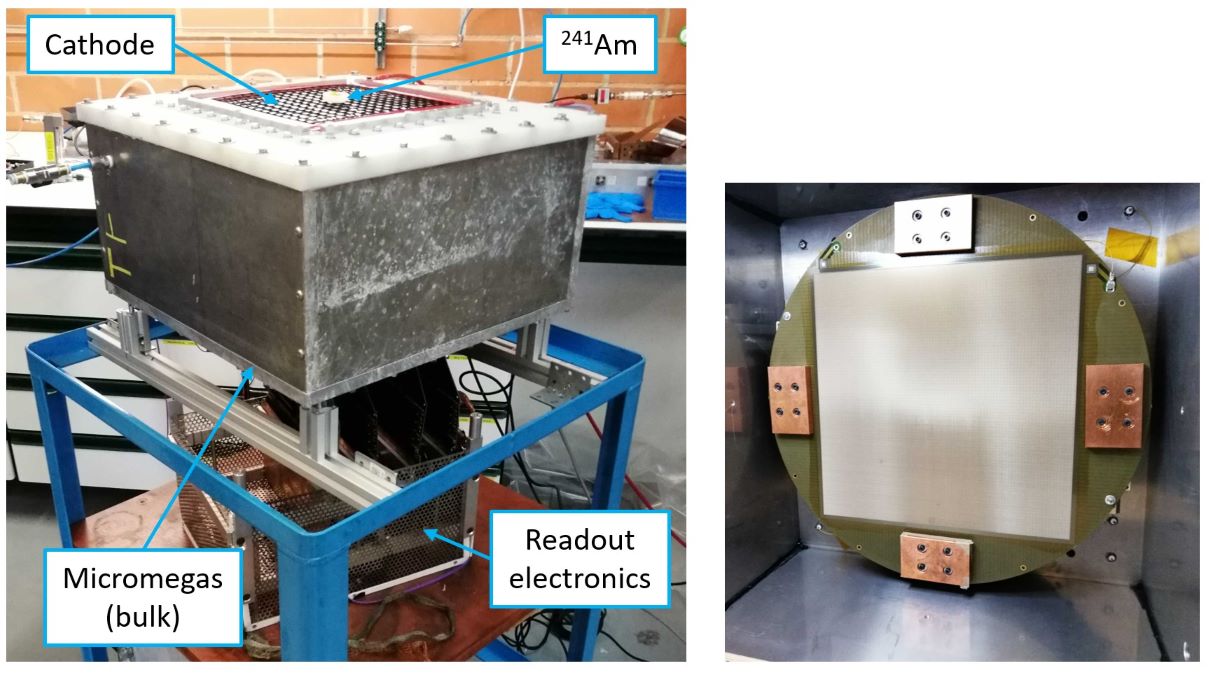}
    \caption{Photo of the AlphaCAMM demonstrator, not radiopure. The components can be seen: the cathode made of mylar, with a grid on top where samples to be measured and calibration sources can be placed, in this case a $^{241}$Am source; the Micromegas, in this case of bulk type, at the bottom and inside the chamber; and the acquisition electronics with which all channels are read. In the photo on the right, the bulk Micromegas of 26 x 26 cm$^{2}$ installed inside the chamber can be seen.}
    \label{fig:AlphaCAMM_FotoDemostrador}
\end{figure}

The Micromegas installed in the demonstrator (Figure~\ref{fig:AlphaCAMM_FotoDemostrador}) was the same as that installed in the TREX-DM chamber during the initial tests in the laboratories of the University of Zaragoza, prior to the transfer of the experiment to the LSC~\cite{Iguaz2016}. This Micromegas is of the bulk type, with an active area of 26~x~26~cm$^{2}$ and 432 channels per axis. The readout electronics are based on the AFTER chip~\cite{Baron2008}, an earlier version of the AGET chip used in the TREX-DM acquisition electronics. As previously explained, the AFTER chip requires, unlike its successor, an external signal to trigger the acquisition, so a readout-amplification-discrimination chain was set up to read the signal from the mesh and provide the trigger signal to the electronics, as was done in the initial tests of TREX-DM~\cite{Iguaz2016}.

After the demonstrator was set up, several measurements were conducted using a $^{241}$Am source, placed on a grid above the cathode and at its center (Figure \ref{fig:AlphaCAMM_FotoDemostrador}), emitting $\alpha$ particles with an energy of approximately 5.4~MeV.

\begin{figure}[!ht]
    \centering
    \includegraphics[width=0.95\textwidth]{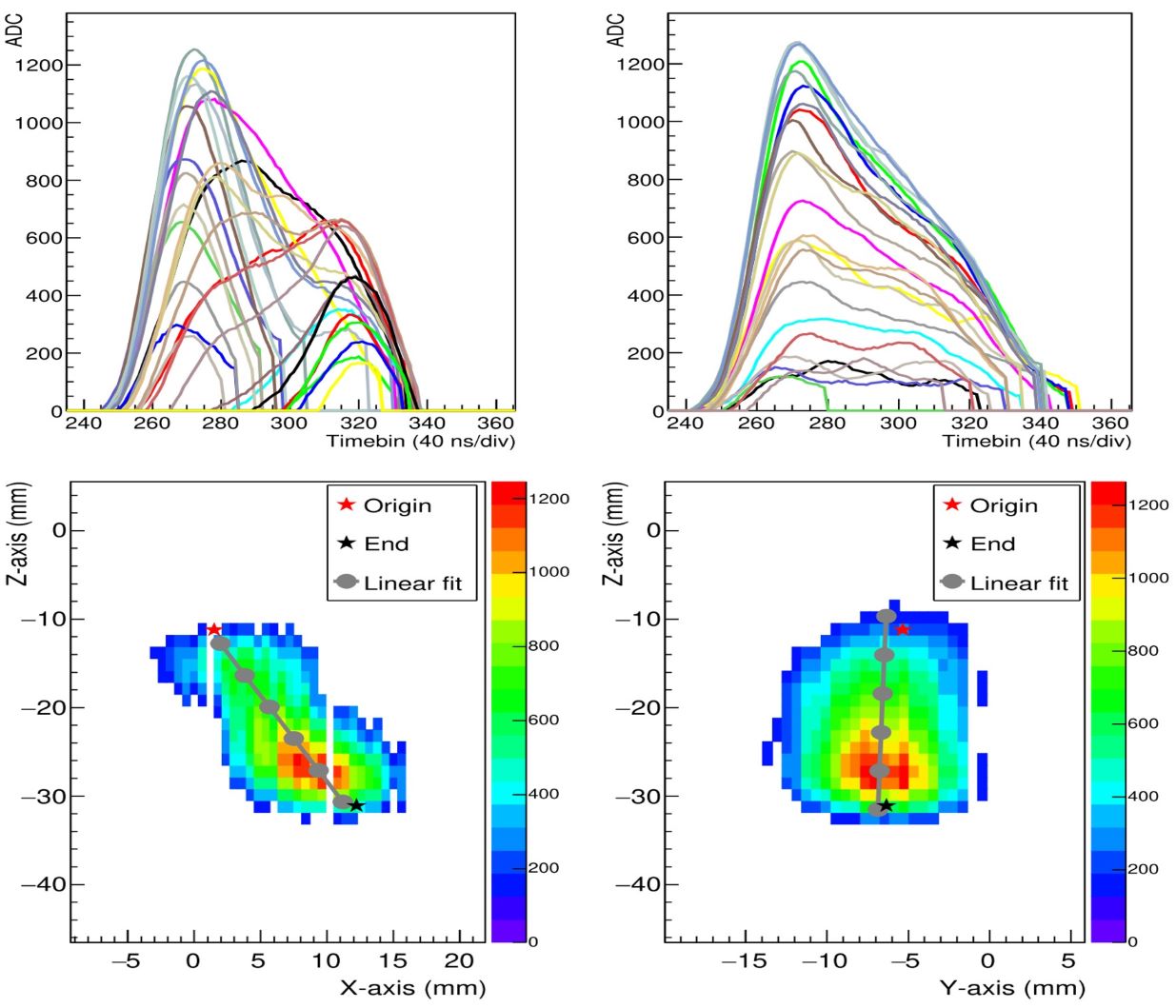}
    \caption{$\bullet$~Upper:~digitized signals from an event produced by the $^{241}$Am source and collected by the \textit{strips} of the X axis (left) and Y axis (right). Each color represents a strip/channel of the corresponding axis. $\bullet$~Lower:~projection of an event produced by the $^{241}$Am source, where the colors represent the integration of the pulses in XZ (left) and in YZ (right), while the gray lines represent the linear fit that defines the trajectory, with the start (red star) and end (black star) of the same \cite{Altenmuller2022AlphaCAMM}.}
    \label{fig:AlphaCAMM_SignalsDemostrador}
\end{figure}

For the reconstruction of the $\alpha$ particle tracks (Figure \ref{fig:AlphaCAMM_SignalsDemostrador}), the digitized pulses are integrated over time using the sampling times configured in the electronics (40~ns), thus reconstructing the topology of the energy transfer from the $\alpha$ particle to the gas along the Z-axis. This reconstruction in Z, along with the reconstruction in the XY plane of the Micromegas, allows for the definition of the direction and sense of the trajectory followed by the particle during its path through the ionization of the gas. This reconstruction method has been tested in this demonstrator~\cite{Altenmuller2022AlphaCAMM}, and the direction of 99.9\% of the $\alpha$ particles emitted from the source was correctly identified.

It is worth noting that, as is often the case in TPCs with a single readout plane based on a single detection technology, t$_0$ (the temporal origin of the event, which occurs when the first signal from the charge generated by an ionizing particle passing through the TPC is detected) reconstructs a relative Z coordinate, meaning that the absolute Z coordinate where the ionization occurred is unknown. In the case of gas TPCs, the diffusion of the primary electrons along the drift distance to the Micromegas can provide information about~t$_0$~\cite{Gonzalez-Diaz2015}. Therefore, future studies on diffusion under the same operating conditions as those of the AlphaCAMM can be conducted to increase the accuracy of spatial reconstruction along the Z-axis, which could significantly improve the sensitivity of the detector.

\setlength{\baselineskip}{12pt} 
\selectlanguage{USenglish}
\graphicspath{{../Imagenes/}{../../Imagenes/}} 

\renewcommand{\listtablename}{Table Index}
\renewcommand{\tablename}{Table} 

\chapter{Background Model by Emanation for AlphaCAMM}
\label{ch:AlphaCAMM_ModeloFondo}
\chaptermark{Background Model by Emanation...}

\begin{figure}[!ht]
    \centering
    \includegraphics[width=0.8\textwidth]{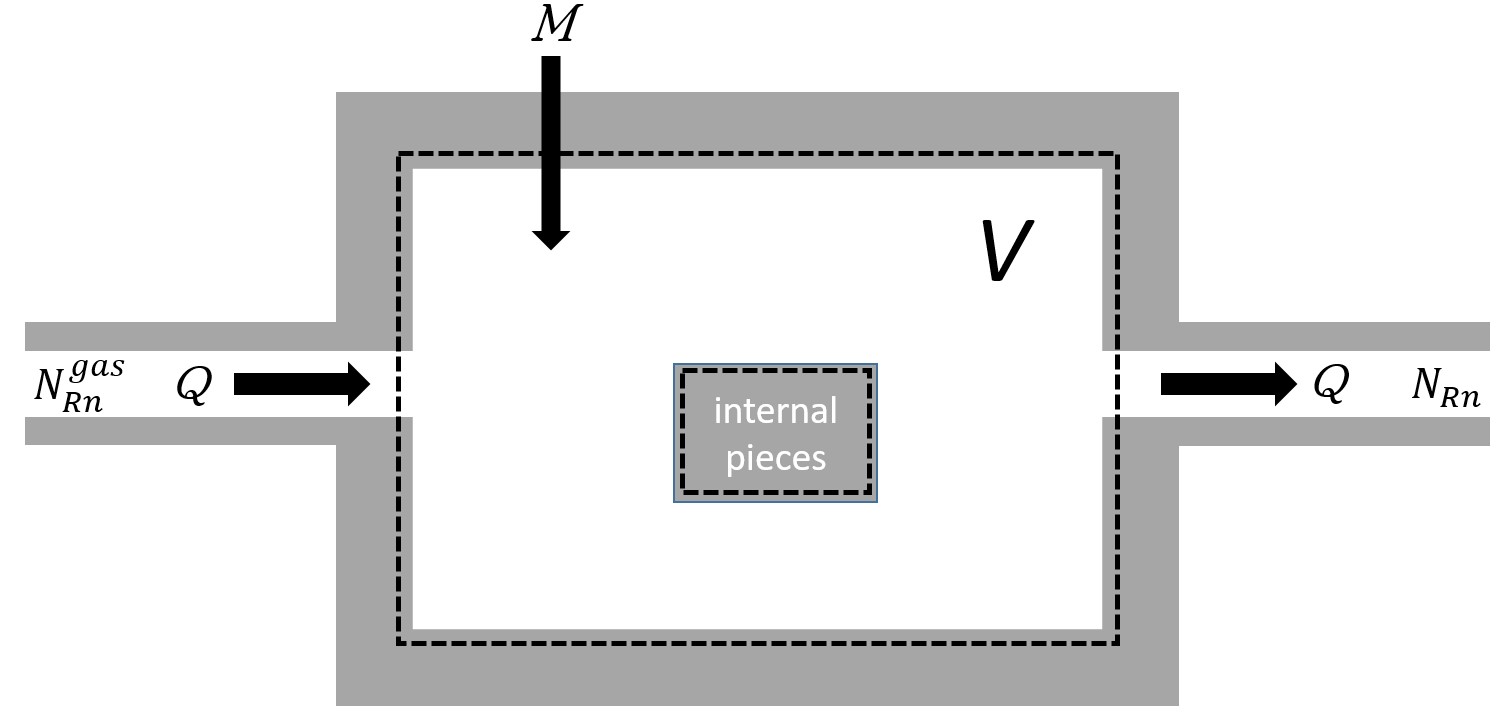}
    \caption*{}
    \label{fig:EsquemaVolumenEstudioPortada}
\end{figure}

\newpage

\section{Radon Emanation Background Model}
\label{sec:ModeloDeFondo}

With the first demonstrator of the AlphaCAMM, the ability to discriminate according to the direction and sense of the event has been demonstrated, allowing for the elimination of the background of $\alpha$ particles that originate from the walls of the chamber and from the Micromegas itself. Thus, the background of the detector in the energy range of interest (1-10~MeV) will be primarily dominated by the emissions of $\alpha$ particles from the decays occurring within the gas volume, whose trajectories have a component in the direction of the Micromegas, potentially confusing them with an event coming from the sample. Another significant contribution to the detector's background may come from the emissions of $\alpha$ particles resulting from decays occurring in the last $\sim$250~$\mu$m of the surface of the material from which the cathode/sample holder will be constructed.

The main nuclides that can contribute to the emission of $\alpha$ particles in the detection volume of the AlphaCAMM are those corresponding to the different isotopes of Radon. It is known that these isotopes belonging to the natural decay chains, due to their noble gas nature, easily diffuse in materials and are emanated to the outside. Therefore, we can expect that contamination from Radon will be introduced from the materials with which the TPC is constructed, from the laboratory air itself, or from the gas system used to supply the chamber.

The isotopes of Radon that can be introduced into the AlphaCAMM, and therefore contribute to the background of $\alpha$ particles in the detection volume, can mainly originate from three natural chains: from the $^{238}$U chain, where the isotope $^{226}$Ra decays into $^{222}$Rn; from the $^{235}$U chain, where the isotope $^{223}$Ra decays into $^{219}$Rn; and from the $^{232}$Th chain, where the isotope $^{224}$Ra decays into $^{220}$Rn (see figure \ref{fig:AlphaCAMM_NaturalChains}).

Once a Radon isotope decays within the gas volume, its descendants, although they can also decay via $\alpha$ emission, will primarily do so from the surfaces of the detector. This is because, once a Radon nuclide decays into the corresponding Polonium nuclide, it becomes charged and is therefore attracted to the surfaces of the detector, primarily to the cathode. During the journey of the Polonium nuclide towards the surfaces, for short-lived times such as that of $^{215}$Po~(1.781~ms), significant $\alpha$ emissions may occur from the descendants of Radon from the gas volume rather than from the surfaces.

\subsection{Radon Concentration in the Study Volume}

Any of the three isotopes of Radon, from the three described decay chains, can contribute to the background.

From the Bateman equations, which describe the temporal evolution of nuclide concentrations in a decay chain (eq~\ref{eq:BatemanEquation}), it is possible to deduce the equation that describes the temporal evolution of the concentration $N_2$ of a nuclide (i=2) considering an initial concentration of its parent $N_1(0)$ (eq~\ref{eq:BatemanEquationN2}).

\begin{equation}
\dfrac{dN_{i}(t)}{dt}=\lambda_{i-1}N_{i-1}(t)-\lambda_{i}N_{i}(t)
\label{eq:BatemanEquation}
\end{equation}

\begin{equation}
\dfrac{dN_{2}(t)}{dt}=-\lambda_{2}N_{2}(t)+\lambda_{1}N_{1}(0)e^{-\lambda_{1}t}
\label{eq:BatemanEquationN2}
\end{equation}

\begin{figure}[!ht]
    \centering
    \includegraphics[width=0.7\textwidth]{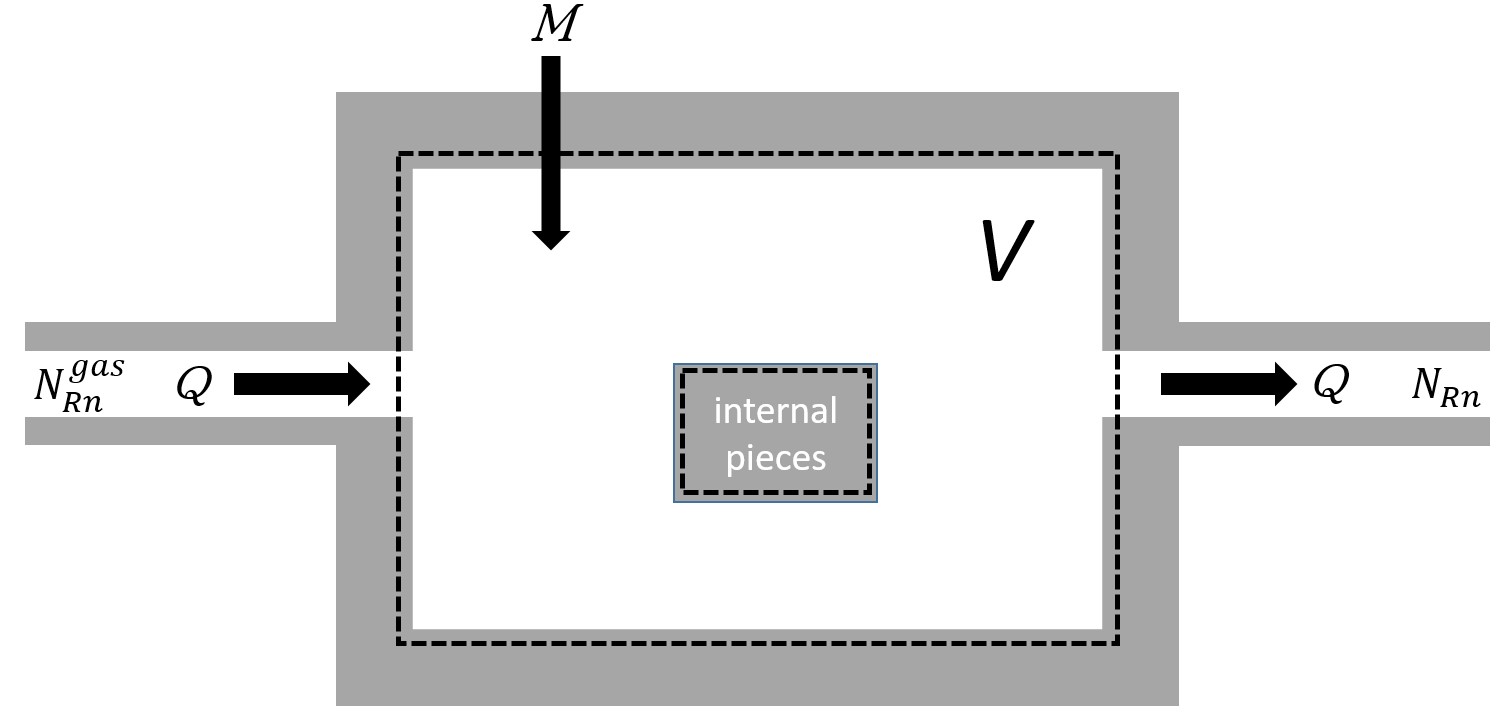}
    \caption{Definition of the study volume~$V$ (dashed line), formed by the gaseous volume inside the chamber plus a part of the volume within each material. Note that the dashed line encompasses part of the walls of the chamber and part of the internal pieces, representing the volume of material that, through the emanation of Radon, contributes to the total concentration in volume $V$.}
    \label{fig:AlphaCAMM_EsquemaVolumenEstudio}
\end{figure}

Equation~\ref{eq:BatemanEquationN2} can describe the temporal evolution of the concentration of Radon (number of nuclides per unit volume) if i=1 is the nuclide of Radium and i=2 is the nuclide of Radon. Furthermore, this equation can be extended for the case of a chamber with an internal volume $V$ and a constant flow rate $Q$ of input/output, the scheme of which is presented in figure~\ref{fig:AlphaCAMM_EsquemaVolumenEstudio}.

For its application to the study volume $V$, the terms that describe the transport of Radon through the flow $Q$ of input and output are added to equation~\ref{eq:BatemanEquationN2}, with $N_{Rn}^{gas}$ being the concentration of Radon in the input gas mixture and $N_{Rn}(t)$ the concentration in the volume of the chamber and, therefore, that of the output flow. Another term $M$ is also introduced, representing a constant flow of Radon into volume $V$ through processes of leaks, permeability, and outgassing. The introduction of these new terms results in equation~\ref{eq:AlphaCAMM_BatemanEqInitial}, which describes the evolution of the concentration of Radon in a chamber with an internal volume $V$ and a flow rate of input/output $Q$, in which constant contributions of Radon also occur due to leaks, permeability, and outgassing processes.

The emanation of Radon from the materials of which the chamber is constructed is a process included within the study volume $V$ and described by equation \ref{eq:AlphaCAMM_BatemanEqInitial} through the term $\lambda_{Ra}N_{Ra}$. Note that this term corresponds to the contribution of Radon through the decay of Radium, but the concentration of Radium will decrease due to the decay of this isotope. However, this case in which only Radium nuclides are found within the material, but no parent nuclides are present, is the least common and only occurs in the case of direct contamination with this element. It is more common for the existence of Radium nuclides to be due to the presence of parent nuclides from the corresponding natural decay chain, and these are also in secular equilibrium. Therefore, to reflect this reality in the model, it has been assumed that the contribution from Radium will be constant. Additionally, if the activity of some of the parent elements is known (direct measurement of materials, for example, using spectrometric techniques) and assuming secular equilibrium in the chain, the concentration of Radium in the material can be determined and directly introduced into the model.

For simplicity in notation, and from here on, the concentrations that appear without superscript refer to those contained in the gaseous volume $V$ of the chamber, the object of study, while those that appear with superscript, such as $N_{Rn}^{gas}$, refer to those contained outside the study volume. Furthermore, the application of equation~\ref{eq:AlphaCAMM_BatemanEqInitial} is valid for each of the three decay chains where a Radium nuclide decays into a Radon one (figure~\ref{fig:AlphaCAMM_NaturalChains}).

\begin{equation}
\dfrac{dN_{Rn}(t)}{dt}=-\lambda_{Rn}N_{Rn}(t)+\lambda_{Ra}N_{Ra}+(N_{Rn}^{gas}-N_{Rn}(t))\dfrac{Q}{V}+M
\label{eq:AlphaCAMM_BatemanEqInitial}
\end{equation}

It should be clarified that, for the completeness of equation \ref{eq:AlphaCAMM_BatemanEqInitial}, the terms that describe the transport of Radium nuclides and their parents between the volume $V$ of the chamber and the outside should be added, just as has been done with the Radon nuclides. However, it is assumed that the exchange of Radium and its parents with the outside through gaseous transport processes can be neglected, assuming that it is very unlikely to find molecules of these elements in suspension. Therefore, only the concentration of Radium $N_{Ra}$ that may be present within the materials of which the chamber is composed is taken into account, and also, as has been assumed in the model, in secular equilibrium with the corresponding decay chain. Any introduction of Radium or its parents into the system would be accidental and, therefore, would represent a source of unknown background.

It should also be clarified that this model is limited to the background produced by the Radon nuclides in the gaseous volume and that any direct contribution from alpha decay by its parents, their descendants, or any other isotope contained in the gaseous volume or in the materials has not been taken into account. In this regard, it should be noted that the discrimination of events by their topology reduces the background coming from the surfaces of the detector, except for those coming from the surface that contains the sample to be measured. In this case, special care should be taken in the cleaning or replacement of this surface with the necessary regularity, depending on the exposure produced.

The solution to equation \ref{eq:AlphaCAMM_BatemanEqInitial} is shown in \ref{eq:AlphaCAMM_BatemanEqSolution} and has two differentiated terms. The first represents the contribution from the initial concentration of Radon $N_{Rn}(0)$. The second represents the contribution from emanation from within the chamber due to the decay of Radium $N_{Ra}$ into Radon, plus the contribution of Radon that may be introduced into the chamber from the outside through the gas system ${\theta}N_{Rn}^{gas}$, plus the contribution $M$ from leaks, permeability, and outgassing. The renewal constant is defined as $\theta=Q/V$ such that $t_{ren}=P_{op}/\theta$ is the time it takes to renew all the gas contained in $V$ with a flow rate of input/output $Q$ and at an operating pressure $P_{op}$.

\begin{multline}
N_{Rn}(t)=N_{Rn}(0)e^{-(\lambda_{Rn}+\theta)t}+\dfrac{\lambda_{Ra}N_{Ra}+M+{\theta}N_{Rn}^{gas}}{(\lambda_{Rn}+\theta)}(1-e^{-(\lambda_{Rn}+\theta)t})\
\label{eq:AlphaCAMM_BatemanEqSolution}
\end{multline} 

Let us define here the initial conditions of the system. The final design of the AlphaCAMM detector will follow the same operating principles as the demonstrator described in section \ref{sec:PrimerDemostrador}. An operating mode of this detector is assumed in which, for the measurement of a sample, the following steps are performed:
\begin{enumerate}
\item The chamber is opened, the sample is placed, and a vacuum cycle is performed in the chamber.
\item Gas is injected into the chamber until the operating pressure is reached.
\item The output port of the chamber is opened, and a constant flow rate $Q$ of input and output is adjusted.
\item Once the steady-state pressure and flow rate of input/output are reached, data collection begins for the measurement of alpha particles from the sample.
\end{enumerate}

Equation \ref{eq:AlphaCAMM_BatemanEqSolution} is applied to a system with a flow rate of input/output $Q$, therefore the first moment when this condition is met and where $t=0$ is defined is when, once the operating pressure has been reached, the output port is opened and steady-state is achieved with a constant flow rate of input and output.

\subsection[Contribution from the Initial Concentration of Radon]{Contribution from the Initial Concentration of Radon~$N_{Rn}(0)$}
\label{subsec:ConcentInicialRadon}

The initial concentration of Radon that may exist within the gas volume of the detector and its evolution is represented by the first term of equation~\ref{eq:AlphaCAMM_BatemanEqSolution}. The moment $t=0$ is defined as the time when, after injecting gas into the chamber, a steady state with a flow rate $Q$ of input and output is reached. To calculate the initial concentration of Radon $N_{Rn}(0)$ at $t=0$, we will describe the system before reaching steady state, starting from the moment when the transition from the vacuum cycle to the filling of the chamber occurs. To do this, it is necessary to modify equation~\ref{eq:AlphaCAMM_BatemanEqInitial}, eliminating the transport of Radon in the outflow, and evaluate it at $t=t_{fill}$ (the filling time of the chamber), such that $N_{Rn}(0)=N_{Rn}^{0}(t=t_{fill})$.

In this case, the contribution from the concentration of Radium during the filling process in the chamber is $N_{Ra}$, just as it is for the system in operation with a flow rate of input and output $Q$, since this value depends on the materials from which the chamber is constructed. If we assume that the processes by which Radon is expelled from the materials, as described in section \ref{sec:ContribucionInicialRadio}, do not depend on either the pressure or the concentration of Radon in the chamber, then equation \ref{eq:AlphaCAMM_BatemanEqSolution_NRa} that defines $N_{Ra}$ is also valid in the case of the filling process of the chamber.

On the other hand, $N_{Rn}^{0}(0)$ is the concentration of Radon after the vacuum cycle, at the moment of starting the filling of the chamber. To correctly calculate $N_{Rn}^{0}(0)$, it would be necessary to rewrite equation \ref{eq:AlphaCAMM_BatemanEqInitial}, just as is being done for the filling process, to describe the system during the vacuum cycle. This would require including the transport of gas molecules through the vacuum pump, which entails studying and modeling the system with the pump. To simplify this step, as long as the vacuum cycle is not decisive in the final results of the Radon background, we can approximate by assuming that the concentration of Radon remaining in the volume after the vacuum cycle originates solely from the laboratory air that filled the volume when the chamber was opened, disregarding any Radon emanations that may have occurred from the materials during the vacuum cycle and that the pump was unable to extract. Therefore, the concentration of Radon $N_{Rn}^{0}(0)$ that remains in the detector after this vacuum cycle can be approximated by

\begin{equation} 
N_{Rn}^{0}(0)=\hspace{1mm}{P_{vac}}\cdot{N_{Rn}^{air}}~~,
\label{eq:AlphaCAMM_BatemanEqSolution_NRn0(0)}
\end{equation}

where $P_{vac}$ is the pressure reached during the vacuum cycle and $N_{Rn}^{air}$ is the concentration of Radon in the air of the location where the detector operates.

In the case of the contribution to the concentration of Radon from the processes of leaks, permeability, and outgassing during the filling cycle ($M_{fill}$), we can assume that the flow rate contributing to the volume of the chamber from these processes has a concentration of $N_{Rn}^{air}$, since the exchange of molecules in the processes of permeability and leaks is clearly between the exterior and interior volumes, while the outgassing process releases gas that has previously been trapped through the surfaces of the materials when exposed to a gas, in this case, the laboratory air. We can also assume, for simplification, that the flow into the volume of the chamber due to these processes is constant and does not depend on the pressure in the chamber. Similarly, as done in the simplification of the vacuum cycle, a more detailed study can be conducted if it is observed that the filling cycle is decisive in the final results of the Radon background.

We can define $r_{Q}=Q_{M_{fill}}/Q_{fill}$ as the ratio between the sum of the flow rates of leaks, permeability, and outgassing $Q_{M_{fill}}$ and the filling flow rate $Q_{fill}$. We can also define $r_{N}=N_{Rn}^{air}/N_{Rn}^{gas}$ as the ratio between the concentration of Radon in the air $N_{Rn}^{air}$ and the concentration of Radon in the input gas $N_{Rn}^{gas}$. Then $M_{fill}$ can be approximated as

\begin{equation} 
M_{fill}=\dfrac{Q_{M_{fill}}}{V}\hspace{0.4mm}N_{Rn}^{air}=r_{Q}r_{N}\theta_{fill}N_{Rn}^{gas}~~,
\label{eq:AlphaCAMM_M_fill}
\end{equation}

where $\theta_{fill}=Q_{fill}/V$ is the renewal constant during the filling, just as in equation \ref{eq:AlphaCAMM_BatemanEqSolution} when describing the system in operation with a flow rate of input and output. Thus, $t_{fill}=P_{op}/\theta_{fill}$ is the time it takes to fill the volume $V$ with an input flow rate $Q_{fill}$ up to an operating pressure $P_{op}$.

Additionally, $Q_{M_{fill}}$ is the total gas flow contributed by the processes of leaks, permeability, and outgassing and can be experimentally determined by direct measurement in the chamber. To do this, the chamber is subjected to a vacuum cycle, and after reaching the minimum pressure, the chamber is isolated and the evolution of the pressure over time is monitored. This evolution will directly report the flow $Q_{M_{fill}}$ that can be directly applied in equation \ref{eq:AlphaCAMM_M_fill}. 

With all this, we rewrite equation \ref{eq:AlphaCAMM_BatemanEqInitial} but applied to the system during the filling of the chamber

\begin{equation}
\dfrac{dN_{Rn}^{0}(t)}{dt}=-\lambda_{Rn}N_{Rn}^{0}(t)+\lambda_{Ra}N_{Ra}+N_{Rn}^{gas}\theta_{fill}+r_{Q}r_{N}N_{Rn}^{gas}\theta_{fill}
\label{eq:AlphaCAMM_BatemanEqInitialFilling}
\end{equation}

whose solution evaluated at $t=t_{fill}$ is

\begin{multline}
N_{Rn}^{0}(t_{fill})={P_{vac}}{N_{Rn}^{air}}\left[e^{-\lambda_{Rn}t_{fill}}\right]\\
+N_{Ra}\left[\dfrac{{\lambda_{Ra}}}{\lambda_{Rn}}(1-e^{-\lambda_{Rn}t_{fill}})\right]\\
+\dfrac{Q_{M_{fill}}}{V}N_{Rn}^{air}\left[\dfrac{1}{\lambda_{Rn}}(1-e^{-\lambda_{Rn}t_{fill}})\right]\\
+P_{op}N_{Rn}^{gas}\left[\dfrac{1}{\lambda_{Rn}t_{fill}}(1-e^{-\lambda_{Rn}t_{fill}})\right]\
\label{eq:AlphaCAMM_BatemanEqSolutionFilling}
\end{multline}

\begin{figure}[!ht]
    \centering
    \includegraphics[width=0.97\textwidth]{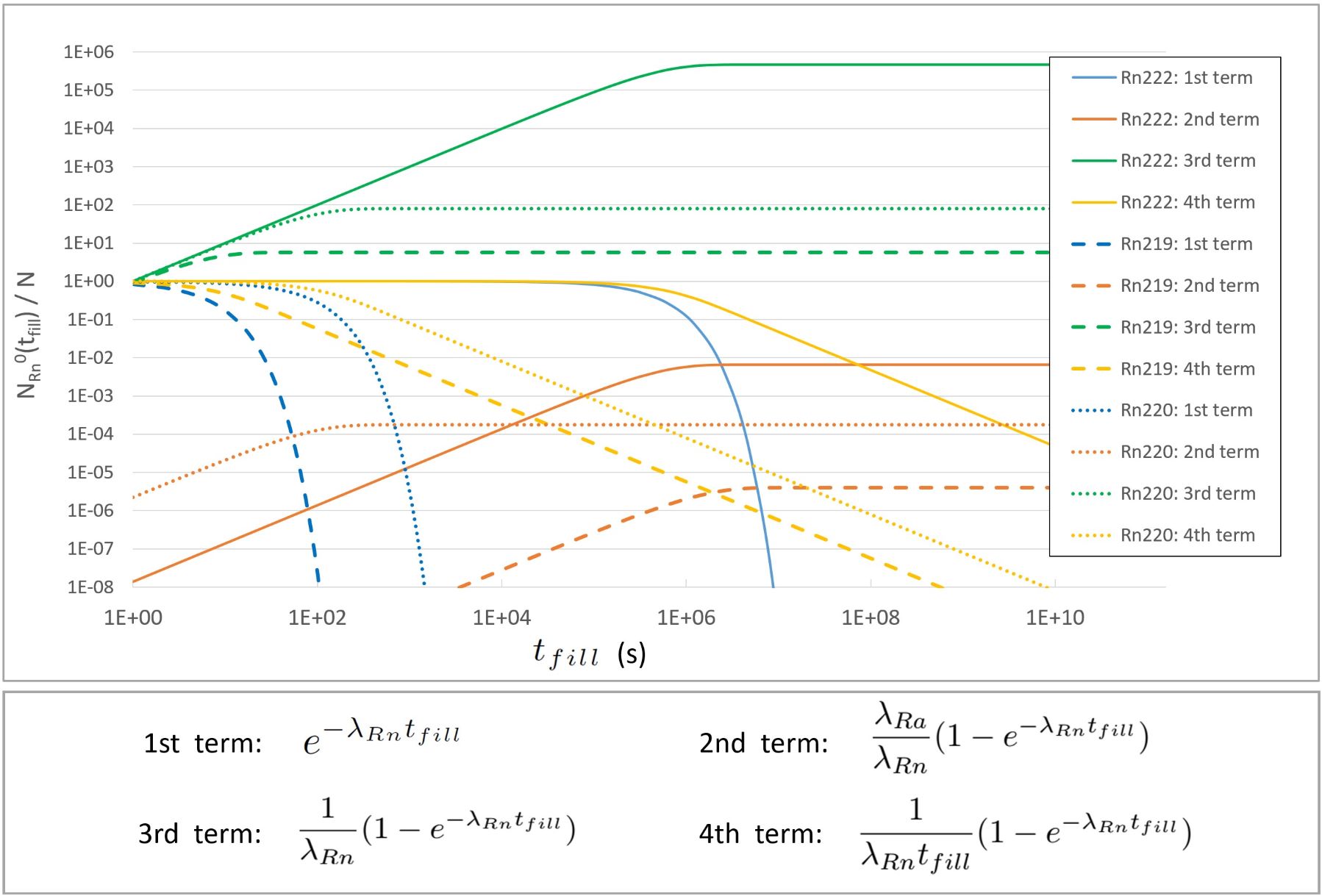}
    \caption{Evolution of each of the four terms of equation~\ref{eq:AlphaCAMM_BatemanEqSolutionFilling} that describes the filling process in the chamber, for the three natural decay chains. Solid lines represent the terms corresponding to the decay chain of the isotope $^{222}Rn$, dashed lines represent the terms corresponding to the decay chain of the isotope $^{219}Rn$, and dotted lines represent the terms corresponding to the decay chain of the isotope $^{220}Rn$. In blue, the first term corresponding to the contamination of Radon after the vacuum cycle; in orange, the second term corresponding to the contamination from Radon emanating from the decay of Radium; in green, the third corresponding to contamination from the processes of leaks, permeability, and outgassing; and in yellow, the fourth corresponding to contamination from Radon in the gas mixture.}
    \label{fig:AlphaCAMM_NRn0Evolution_allRange_allIsotopes}
\end{figure}

In figure \ref{fig:AlphaCAMM_NRn0Evolution_allRange_allIsotopes}, the evolutions over the filling time $t_{fill}$ of the four terms of equation~\ref{eq:AlphaCAMM_BatemanEqSolutionFilling} applied to each of the three natural decay chains are shown. A wide range of $t_{fill}$ is displayed to observe the evolutions, but in practice, the filling time will be on the order of seconds or minutes.

\subsection[Contribution from the concentration of Radium]{Contribution from the concentration of Radium~$N_{Ra}$}
\label{sec:ContribucionInicialRadio}

As previously justified, in equation \ref{eq:AlphaCAMM_BatemanEqInitial}, and therefore also in its solution shown in equation \ref{eq:AlphaCAMM_BatemanEqSolution}, the contamination from radium and its progenitors that may occur from the outside once the chamber is in operation has been disregarded. However, the contribution from radium that may be intrinsically present in the materials forming the chamber is taken into account, as represented in equation \ref{eq:AlphaCAMM_BatemanEqSolution}. Part of the radium from $N_{Ra}$ will decay into radon, releasing into the volume of the chamber if the necessary processes occur, thus contributing to the radon population in the volume.

Radon atoms located within the grains of a material can hardly be expelled to the surface of the material due to their low diffusion coefficients. However, if the radon atoms are located in the interstitial space of the grains, they can diffuse more easily to the surface. Therefore, the release of radon from the material occurs through the following processes \cite{Ishimori2013}:

\begin{enumerate}
\item Emanation: the radium atoms found within the grain of the material decay to form radon atoms, and due to the nuclear recoil caused by this decay, some of these atoms escape into the interstitial space between the grains. The emanation coefficient ($\varepsilon$) is defined as the fraction of radon atoms generated that escape from the grain, and several factors affect this coefficient: the distribution of radium within the grain, which, although uniform in most primary minerals, can be primarily at the surface in other types of secondary minerals; the grain size, such that if radium is uniformly distributed, the emanation coefficient would be inversely proportional to the grain diameter; the moisture content, which favors emanation from the grain (the emanation coefficients in saturated materials can be 2 to 6 times greater than in dry materials); mineralogical characteristics, such as structure, porosity, grain shape, etc.
\item Transport: through diffusion or flow drag, the radon atoms that have previously been emanated from the grains are transported to the surface of the material. Although flow drag can be significant in cases of transport in rock, in most cases the dominant transport mechanism is diffusion, thus quantified by the diffusion coefficient (D). Several factors can affect this coefficient: the distribution and size of the pore; the moisture content, where 1$\%$ humidity can reduce diffusion by up to four orders of magnitude compared to 0$\%$ humidity; temperature, which favors diffusion~\cite{Ishimori2017}. It is known \cite{Ishimori2017,Wojcik2017} that for a nuclide with a decay constant $\lambda$, diffusing in a material with a diffusion coefficient~D, the diffusion length $L_{D}$ (the distance at which, after diffusing within the material, the concentration of the corresponding nuclide is half of the initial concentration) is given by $L_{D}=\sqrt{D/\lambda}$.
\item Exhalation: the radon atoms that have reached the surface of the material, after the processes of emanation and transport (mainly diffusion), are released into the external atmosphere. Since radon is a noble gas, with very low chemical reactivity that prevents it from bonding with surface molecules, and also due to the low concentration of radon in the volume of the chamber, favoring a high concentration gradient between the material and the volume, we will simplify and assume that 100$\%$ of the radon atoms that reach the surface are exhaled into the volume of the chamber. In any case, this assumption is conservative, and the results derived from it will set an upper limit for radon contamination.
\end{enumerate}

In equations \ref{eq:AlphaCAMM_BatemanEqInitial} and \ref{eq:AlphaCAMM_BatemanEqSolution}, $N_{Ra}$ appears as the concentration of radium in the study volume V. As previously justified, it is very unlikely to find radium molecules suspended in a gas, and therefore the concentration of radium that may exist in the gaseous volume is disregarded. However, its concentration within all the materials that make up the detector and contribute to the radon population cannot be neglected. Taking into account, for each material, the emanation process through the emanation coefficient $\varepsilon$ and the diffusion process through the diffusion length $L_{D}$, $N_{Ra}$ can be defined as

\begin{equation} 
N_{Ra}=\dfrac{1}{V}\sum_{Materials}N_{Ra}^{'}\cdot\varepsilon\cdot[\hspace{1mm}L_{D}\cdot\rho\cdot{S}\hspace{1mm}]
\label{eq:AlphaCAMM_BatemanEqSolution_NRa}
\end{equation}

\noindent where $N_{Ra}^{'}$ is the amount of radium (number of nuclides) contained in the material per unit mass, $\rho$ is the density of the material, and S is the surface area of the material in contact with the volume of the chamber. Thus, if $N_{Ra}^{'}$ is measured in the materials that form the chamber, using various spectrometric techniques, the contribution of each part of the detector to the total radon concentration can be determined.

It should be noted that the boundaries defining the gaseous volume within the chamber extend into the materials by a length equal to the diffusion length $L_{D}$, thus forming the study volume~$V$ (see figure \ref{fig:AlphaCAMM_EsquemaVolumenEstudio}).

\subsection{Contribution from External Sources via Flow (Q) and by \textit{Outgassing}, Permeability, and Leaks (M)}

In equation~\ref{eq:AlphaCAMM_BatemanEqSolution}, the contribution of Radon that can be introduced into the chamber from the outside through the gas system (${\theta}N_{Rn}^{gas}$) is taken into account. This gas typically comes pre-mixed in bottles from the supplier's treatment plant, and it can be estimated that there are several sources from which Radon can be introduced into the gas mixture before entering the detector chamber: from the gas processing plant, emanation from the bottle, or emanation from the components of the gas system. In any case, the values of the concentrations in the gas mixture can be quantified once the AlphaCAMM is operational, through background data collection in \textit{sealed mode}, that is, without inflow/outflow. Thus, by measuring the decay times along with the associated energy, the origins of the contaminations can be identified, and therefore, necessary measures can be taken for their mitigation, if needed.

Equation~\ref{eq:AlphaCAMM_BatemanEqSolution} also considers the contribution to the concentration of Radon in volume~$V$ from \textit{outgassing}, permeability, and leaks during the measurement regime of the detector with inflow and outflow rate~$Q$. The term~$M$, which encompasses these contributions, can be broken down as follows:

\begin{equation} 
M=\dfrac{G+J+rR}{V}\hspace{0.4mm}N_{Rn}^{air}
\label{eq:AlphaCAMM_BatemanEqSolution_M}
\end{equation}

\noindent where $G$ is the gas flow introduced into $V$ through \textit{outgassing}, $J$ is the flow introduced through permeability, $R$ is the flow introduced through leaks, and $r$ is the back-diffusion factor. As in the case of the filling cycle, it is assumed that the flow introduced into the volume $V$ of the chamber from these processes has a concentration of $N_{Rn}^{air}$.

The total gas flow contributed by the processes of \textit{outgassing}, permeability, and leaks ($G+J+R$) can be experimentally determined through direct measurement in the chamber. To do this, the chamber is subjected to a vacuum cycle, and after reaching the minimum pressure, the chamber is isolated, and the evolution of the pressure over time is monitored. This evolution will directly report the sum of the flows due to the three processes, but the leak flow ($R$) can be experimentally determined through a helium leak test. The term $r$, which represents the proportion of Radon introduced due to the back-diffusion process in the leaks, is dependent on the geometry of the leak, so it will need to be measured independently.

\section{Evaluation of the Emanation Background Model}
\label{sec:EvaluacionModeloFondo}

Once the Radon emanation background model in the volume of the chamber has been established, it is necessary to evaluate the model under operational conditions that are closest to real conditions in order to define the specifications required for the design and construction of the detector.

To this end, several parameters have been defined, which are collected in table~\ref{tab:AlphaCAMM_TablaParametrosCamara}, taking into account practical reasons and serving as a starting point for defining the final design parameters. The operational pressure~$P_{op}$ is set slightly above atmospheric pressure, so that the mechanical demands on the chamber are minimized. The volume $V$ of the chamber is set at 17 liters, an approximate volume in which a Micromegas with an active area of 25~x~25~cm$^2$ and approximately 10~cm of drift distance can be installed. The inlet and outlet flow rate $Q$ is set at 5~l/h~(1.39·10$^{-6}$~m$^3$/s), a flow rate that is manageable from the perspective of gas consumption in open-loop operation. The filling time of the chamber t$_{fill}$, starting from vacuum pressure, is estimated at 100 seconds at a flow rate of 612~l/h, minimizing this time as much as possible to reduce the significant amount of Radon that leaks into the chamber. The leak level $R$, along with the level of \textit{outgassing} $G$ and permeability $J$, is set at a typical value for vacuum equipment of 1·10$^{-6}$~mbar~l~s$^{-1}$~(1·10$^{-12}$~m$^{3}$/s).

\begin{table}[!ht]
\centering
\begin{tabular}{l}
\includegraphics[width=0.33\textwidth]{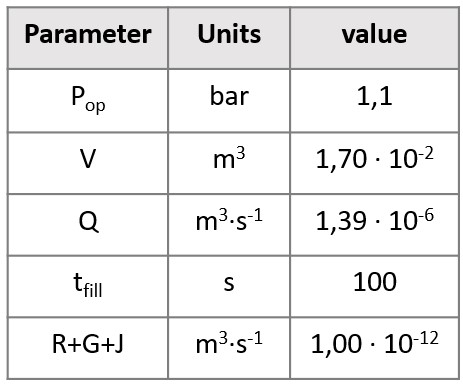}\\
\end{tabular}
\caption{Main design parameters of the chamber.}
\label{tab:AlphaCAMM_TablaParametrosCamara}
\end{table}

The equation \ref{eq:AlphaCAMM_BatemanEqSolution} describes the evolution of the Radon concentration $N_{Rn}(t)$ due to emanation within the study volume, inside the chamber volume, and therefore the evolution of the activity ($A_{Rn}(t)=N_{Rn}(t)\lambda_{Rn}$) due to the decays of the Radon nuclides contained within it. Since the Radon concentrations are volumetric, the activity will also be expressed volumetrically in Bq/m$^{3}$, so it will need to be multiplied by the volume~$V$ of the chamber to obtain the detectable activity. The terms used in equation~\ref{eq:AlphaCAMM_BatemanEqSolution} are defined in table~\ref{tab:AlphaCAMM_TablaParametrosModelo}, and are explained below.

\begin{table}[!ht]
\centering
\begin{tabular}{l}
\includegraphics[width=0.65\textwidth]{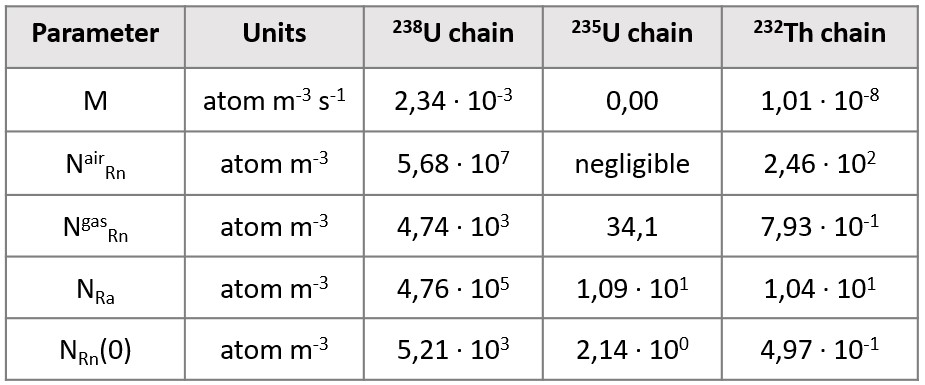}\\
\end{tabular}
\caption{Main parameters of the Radon emanation background model.}
\label{tab:AlphaCAMM_TablaParametrosModelo}
\end{table}

\subsection{Flow of Radon M due to leaks, permeability, and \textit{outgassing}}

During the operation of the detector above atmospheric pressure, there exists a flow of Radon M that enters the chamber due to leaks, permeability, and \textit{outgassing}. To calculate this flow, equation \ref{eq:AlphaCAMM_BatemanEqSolution_M} is applied, where r represents the retro-diffusion coefficient. This coefficient depends on several factors (geometry of the leak, pressure, concentrations of exterior/interior Radon, etc.) and is difficult to model and predict; thus, an arbitrary coefficient of r~=~0.1 has been chosen as a starting point. This coefficient can be empirically adjusted through the study of leaks and their corresponding retro-diffusion during the operation of the detector.

Therefore, taking the typical value of $R+G+J$ for vacuum equipment (table~\ref{tab:AlphaCAMM_TablaParametrosModelo}), approximating the contribution of each process $R$, $G$, and $J$ to this total value equally, and using a retro-diffusion coefficient of r~=~0.1, the contribution from leaks, permeability, and \textit{outgassing} can be evaluated as shown in table~\ref{tab:AlphaCAMM_TablaParametrosModelo}.

Note that in the $^{235}$U chain, the contribution is zero, as it has been assumed that due to the short half-life of $^{219}$Rn (3.96~s), the concentration of this isotope in the laboratory air is negligible ($N_{219Rn}^{gas}$~$\approx$~0).

\subsection[Concentration of Radon in the air]{Concentration of Radon in the air $N^{air}_{Rn}$}

The concentration of Radon in the ambient air of the Canfranc Underground Laboratory, where the AlphaCAMM will operate, has been monitored over the past few years \cite{Perez-Perez2022} \cite{Amare2022}, with an average activity of 69~Bq/m$^{3}$ and fluctuations throughout the year reaching up to 120~Bq/m$^{3}$.

These measurements do not distinguish between the different isotopes of Radon; however, in \cite{Amare2006}, a radiological analysis of the rock in the Canfranc Underground Laboratory is conducted, reporting a ratio in the activity of two of the isotopes A$^{rock}_{232Th}$~/~A$^{rock}_{238U}$~$\approx$~2. This ratio is close to the typical value of~1.1~\cite{Evans1997}.

The diffusion length $L_D$ of Radon within a solid, in this case, the rock of the laboratory, depends on the diffusion coefficient~$D$ and the decay constant~$\lambda$, such that~$L_D$~=~$(D/\lambda)^{1/2}$. Assuming secular equilibrium within the rock in the chains of $^{238}$U and $^{232}$Th, such that the same activity of $^{222}$Rn is found as that of $^{238}$U and the same activity of $^{220}$Rn as that of $^{232}$Th, it can be stated that the ratio of the activities of both Radon isotopes in the rock A$^{rock}_{220Rn}$~/~A$^{rock}_{222Rn}$~$\approx$~2 \cite{Amare2006}. Assuming also that the diffusion coefficients of both isotopes are similar D$_{r220Rn}\approx$~D$_{r222Rn}$, the concentrations of the two Radon isotopes in the ambient air of the laboratory can be related as follows:

\begin{equation} 
\dfrac{A^{air}_{220Rn}}{A^{air}_{222Rn}}=2\times\dfrac{L_{D220Rn}}{L_{D222Rn}}=2\times\sqrt{\dfrac{\lambda_{222Rn}}{\lambda_{220Rn}}}=2.59\cdot10^{-2}
\label{eq:AlphaCAMM_RelaciónConcentracionesAire}
\end{equation}

In this calculation, the transport of Radon molecules within the volume of the laboratory once they are emanated from the rock has not been taken into account, for which a detailed study of the air flows within the laboratory would be necessary. However, a conservative case can be assumed in which the AlphaCAMM operates in a location close to the rock, in which case the transport time of the nuclide $^{220}$Rn is small. Therefore, as reflected in table~\ref{tab:AlphaCAMM_TablaParametrosModelo}, the concentration of $^{222}$Rn has been calculated by applying the highest value from the measurements made in \cite{Amare2022} (120~Bq/m$^{3}$), resulting in a concentration of $N_{222Rn}$~=~5.68·10$^7$~atom/m$^{3}$. The concentration of $^{220}$Rn has been calculated by applying the ratio A$^{air}_{220Rn}$~/~$A^{air}_{222Rn}$~=~2.59·10$^{-2}$, resulting in $N^{air}_{220Rn}$~=~2.46·10$^2$~atom/m$^{3}$.

In the case of the concentration of the isotope $^{219}$Rn, from the $^{235}$U chain, it has such a short half-life (3.96 seconds) that its emanation from the rock is minimal. Furthermore, this difficulty is compounded by the transport from the rock to the volume of air in the vicinity of the detector. Therefore, the concentration of this isotope in the ambient air of the laboratory has been considered negligible ($N_{219Rn}^{gas}$~$\approx$~0).

\subsection[Radon Concentration in the Gas Mixture]{Radon Concentration in the Gas Mixture $N^{gas}_{Rn}$}

The different concentrations of the various isotopes of Radon in the gas mixture that will be used for the operation of the AlphaCAMM are more difficult to estimate. This gas typically comes pre-mixed in bottles from the supplier's treatment plant, and its quality control does not account for the minimization or quantification of Radon concentrations. It can be estimated that there are several sources from which Radon can be introduced into the gas mixture before entering the detector chamber:

\begin{itemize}
\item From the gas processing plant, where the mixing occurs and the gas is introduced into the bottle for transport. In this case, the storage times favor the reduction of the concentration due to decay, especially for isotopes with shorter half-lives. Therefore, as long as the initial concentrations are not excessive, the isotopes $^{220}$Rn and $^{219}$Rn will have a negligible concentration.

\item Emanation from the bottle. The content of Radium in the material from which the bottle is made will be decisive for the concentration of Radon emanating from it. The different bottles that may be supplied by the vendor, coming from various batches of material for their manufacture, make it difficult to control. During the operation of the detector, the level of contamination introduced from the bottle can be determined, and if necessary, a container made from radiopure material can be installed to serve as an intermediate storage, thus allowing for a reduction in concentration due to decay.

\item Emanation from the gas system. The components of the gas system (tubes, joints, valves, sensors, etc.) will cause an emanation that is difficult to estimate. As in the previous cases, these contributions can only be quantified during operation and mitigated if necessary. Additionally, in this case, contamination could also occur from the laboratory air into the gas system through leakage and permeability processes, but the tightness of such systems is at very low leakage values, around 10$^{-8}$~mbar~l~s$^{-1}$.
\end{itemize}

Given the difficulty in predicting the concentrations in the gas mixture, and with the aim of evaluating the background model prior to the design phase, a very conservative value for the activities of $^{222}$Rn and $^{220}$Rn of 10~mBq/m$^{3}$ has been chosen. In \cite{Simgen2009}, the activity of $^{222}$Rn in different Argon mixtures from various suppliers and batches is measured, obtaining values ranging from 0.007 to 8.4~mBq/m$^{3}$. An activity of 10~mBq/m$^{3}$ for these two isotopes yields concentrations of $N^{gas}_{222Rn}$~=~4.74·10$^3$~atom/m$^{3}$ and $N^{gas}_{220Rn}$~=~7.93·10$^{-1}$~atom/m$^{3}$ (table~\ref{tab:AlphaCAMM_TablaParametrosModelo}).

In the case of the concentration of $^{219}$Rn in the gas mixture, the time that elapses from the emanation of this isotope, in the bottle or the gas system, until it enters the detector chamber can be seconds if the flow rate is high or the distance is short. Therefore, although it could initially be disregarded just like the concentration in the laboratory air, it will be taken into account. To estimate its concentration for the background model, the known natural ratio between the isotopes $N_{235U}$~/~$N_{238U}$~=~0.0072 will be used, and secular equilibrium will be assumed within both chains, so that for a concentration of $N^{gas}_{222Rn}$~=~4.74·10$^3$~atom/m$^{3}$ (10~mBq/m$^{3}$), a concentration of $N^{gas}_{219Rn}$~=~34.1~atom/m$^{3}$ (5.97~Bq/m$^{3}$) corresponds.

In any case, the values of the concentrations in the gas mixture can be quantified once the AlphaCAMM is operational in \textit{sealed mode}. Thus, by measuring the decay times along with the associated energy, the origins of the contaminations can be identified, and therefore, the necessary measures can be taken for their mitigation.

During the design and construction phase of the AlphaCAMM, no Radon filter was introduced for the gas system, which could be installed at the entrance of the chamber, as done in \cite{Simgen2009}, where concentrations of $^{222}$Rn in Argon are reduced by a factor greater than 1000. If high concentrations of Radon are found in the gas mixture during the commissioning of the AlphaCAMM, the development and installation of such filters in the system may be considered.

\subsection[Concentration of Radium]{Concentration of Radium $N_{Ra}$}
\label{subsec:ConcentracionRadio}

To estimate the concentration of Radium within the chamber, the equation~\ref{eq:AlphaCAMM_BatemanEqSolution_NRa} must be used, which sums the concentrations of all the materials that contribute to the final concentration. Table~\ref{tab:AlphaCAMM_TablaRadioMateriales} shows three materials that are planned for the construction of the AlphaCAMM.

Based on the activity in the material of the constituent element of each chain measured with spectrometric techniques, the concentrations per unit mass for each Radium isotope ($N^{'}_{226Ra}$, $N^{'}_{223Ra}$, and $N^{'}_{224Ra}$) are calculated, assuming secular equilibrium. Using the material parameters $\varepsilon$ (emanation coefficient), $D_r$ (Radon diffusion coefficient), $\rho$ (density), and $S$ (surface area exposed to the internal volume of the chamber), the concentration in the volume $V$ of each Radium isotope ($N_{226Ra}$, $N_{223Ra}$, and $N_{224Ra}$) is calculated, whose decay will contribute to the Radon background in the detector.

Each material, with a mass concentration $N_{Ra}^{'}$ corrected by the emanation coefficient $\varepsilon$, will contribute to the concentration in the chamber only along the diffusion length $L_D$. The materials listed in Table~\ref{tab:AlphaCAMM_TablaRadioMateriales}, which are planned for the construction of the AlphaCAMM, have been chosen for their low background:

\begin{itemize}

\item Micromegas. The AlphaCAMM will have as its readout plane one of the four Micromegas designed and manufactured for installation in TREX-DM (see section~\ref{sec:NewReadouts}). Over time, the activities of several samples have been measured~\cite{Castel2019}, selecting the highest among them to develop the background model. A high emanation coefficient has been considered, assuming that the nuclear recoil produced during the decay of Radium within the material allows the corresponding Radon nuclide to diffuse towards the surface. The value of the diffusion coefficient $D$~=~6.10·10$^{-15}$ has been extracted from~\cite{Wojcik2017} (Supronyl Polyamide).

\item Stainless Steel AISI~316L. This is the material initially chosen for the construction of the body and lids of the chamber, as most of these stainless steels measured~\cite{Lawson2011} fall within an acceptable activity range (0.07$-$6.91~mBq/kg for $^{238}$U, 0.03$-$5~mBq/kg for $^{235}$U, and 0.07$-$8.7~mBq/kg for $^{232}$Th). For the background model calculation, the value measured with GDMS (Glow Discharge Mass Spectrometry) from the batch of stainless steel AISI~316L intended for the manufacture of the AlphaCAMM was chosen, where a Uranium concentration of 0.002~ppm (24.8~mBq/kg) and a Thorium concentration of 0.001~ppm (4.1~mBq/kg) were measured. This technique did not allow distinguishing between the different isotopes of Uranium and Thorium, so it was assumed that all Thorium was $^{232}$Th, and the known natural ratio between the isotopes $N_{235U}$~/~$N_{238U}$~=~0.0072 was applied to quantify the concentrations of both Uranium isotopes. The emanation coefficient has been determined considering the typical grain size of stainless steel AISI~316L (25$-$55~$\mu$m)~\cite{Wang2021} and the relationship of the emanation coefficient with grain size \cite{Ishimori2017}. The diffusion coefficient has not been found in the literature, as all measurements found have been for quantifying the emanation of Radon in various samples of stainless steels, so it was decided to approximate it to the same value as that of aluminum~\cite{Ruvira2022}.

\item Electrolytic Copper (ETP). This will be used in the manufacture of the internal parts necessary for the operation of the detector (cathode, sample holder, electric field rings, etc.). This material is commonly used in low-background experiments, and its activity has been measured~\cite{Castel2019}. Considering that the grain size of ETP copper can vary widely, from tens of nanometers to hundreds of micrometers, depending on various factors such as cold deformation or applied thermal processes, a value for the emanation coefficient typical of intermediate grain sizes (5$-$10~$\mu$m)~\cite{Ishimori2017} has been chosen. The diffusion coefficient, as in the case of copper, has not been found in the literature, so it was decided to approximate it to the same value as that of aluminum~\cite{Ruvira2022}.

\end{itemize}

The sum of the concentrations of Radium from all materials results in the total concentration of Radium for each isotope ($N_{226Ra}$, $N_{223Ra}$, and $N_{224Ra}$) in the volume of the chamber (Tables~\ref{tab:AlphaCAMM_TablaParametrosModelo}~and~\ref{tab:AlphaCAMM_TablaRadioMateriales}).

\begin{table}[!ht]
\centering
\begin{tabular}{l}
\includegraphics[width=0.98\textwidth]{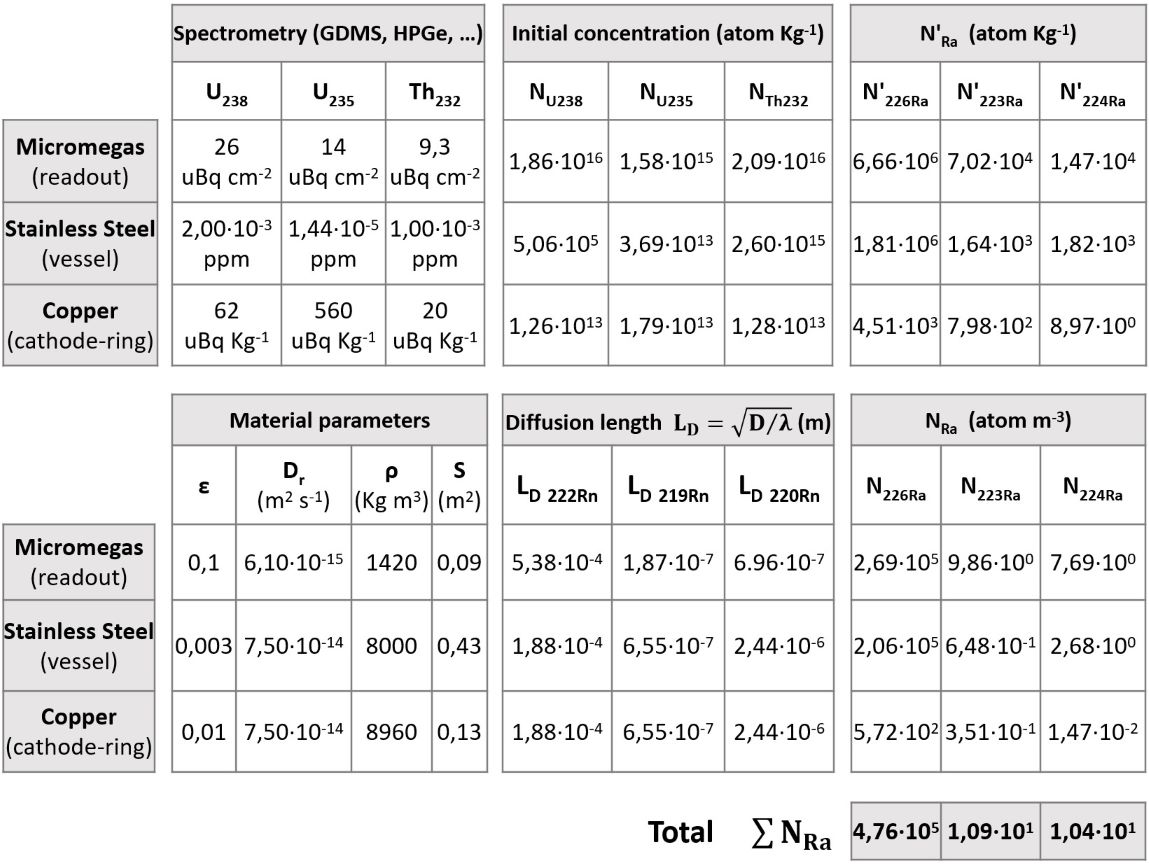}\\
\end{tabular}
\caption{Table of the materials that will be used in the construction of the AlphaCAMM, with the necessary parameters for calculating the total concentration of Radium according to equation \ref{eq:AlphaCAMM_BatemanEqSolution_NRa}. Note that the final concentration results for each isotope are also reflected in Table~\ref{tab:AlphaCAMM_TablaParametrosModelo}.}
\label{tab:AlphaCAMM_TablaRadioMateriales}
\end{table}

\subsection[Initial Concentration of Radon]{Initial Concentration of Radon $N_{Rn}(0)$}

The initial concentration of Radon $N_{Rn}(0)$, as described in section~\ref{subsec:ConcentInicialRadon}, will depend on the pumping process during the vacuum cycle and the subsequent filling of the chamber. It is calculated by evaluating equation~\ref{eq:AlphaCAMM_BatemanEqSolutionFilling} at the filling time, which for the background model has been set at t$_{fill}$~=~100~seconds.

All parameters of equation \ref{eq:AlphaCAMM_BatemanEqSolutionFilling} have been defined and are presented in tables~\ref{tab:AlphaCAMM_TablaParametrosCamara} and~\ref{tab:AlphaCAMM_TablaParametrosModelo}. Q$_{Mfill}$~=~$R+J+G$ is also included, being the sum of the flow rates of leaks, permeability, and outgassing during the filling process.

\begin{table}[!ht]
\centering
\begin{tabular}{l}
\includegraphics[width=0.75\textwidth]{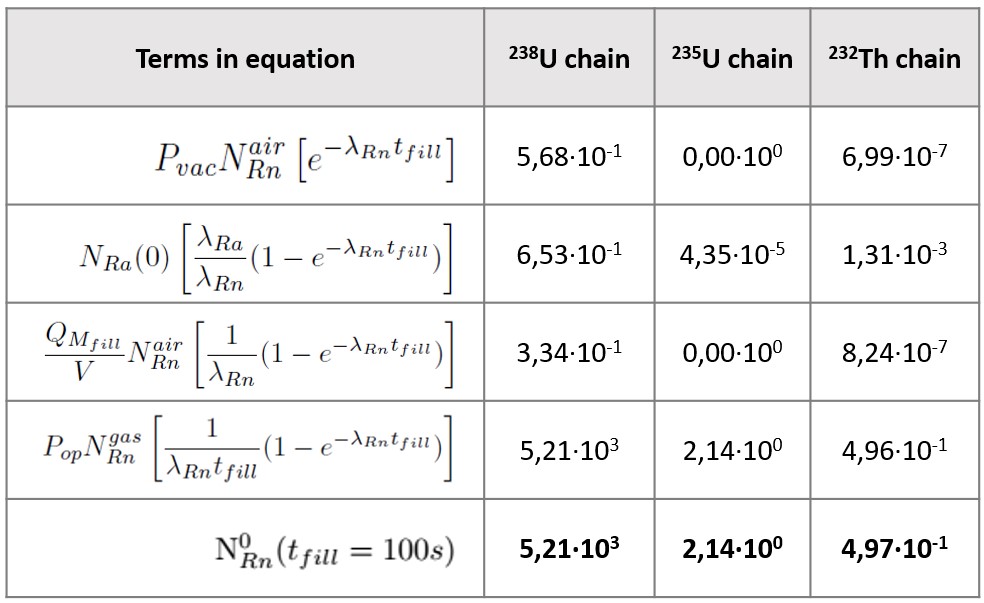}\\
\end{tabular}
\caption{Table with the four terms of equation \ref{eq:AlphaCAMM_BatemanEqSolutionFilling} and their evaluation at t$_{fill}$~=~100~seconds for each decay chain. The initial concentration of Radon $N^{0}_{Rn}(t_{fill}=100s)$ for each chain is the sum of all its terms.}
\label{tab:AlphaCAMM_TablaValoresRadonInicial}
\end{table}

Table~\ref{tab:AlphaCAMM_TablaValoresRadonInicial} shows the results obtained for each term of equation \ref{eq:AlphaCAMM_BatemanEqSolutionFilling} evaluated at the filling time t$_{fill}$~=~100~seconds. This way, the contribution of each element to the initial concentration of Radon can be observed.

\subsection{Results of the Emanation Background Model}

\begin{figure}[!ht]
    \centering
    \includegraphics[width=0.90\textwidth]{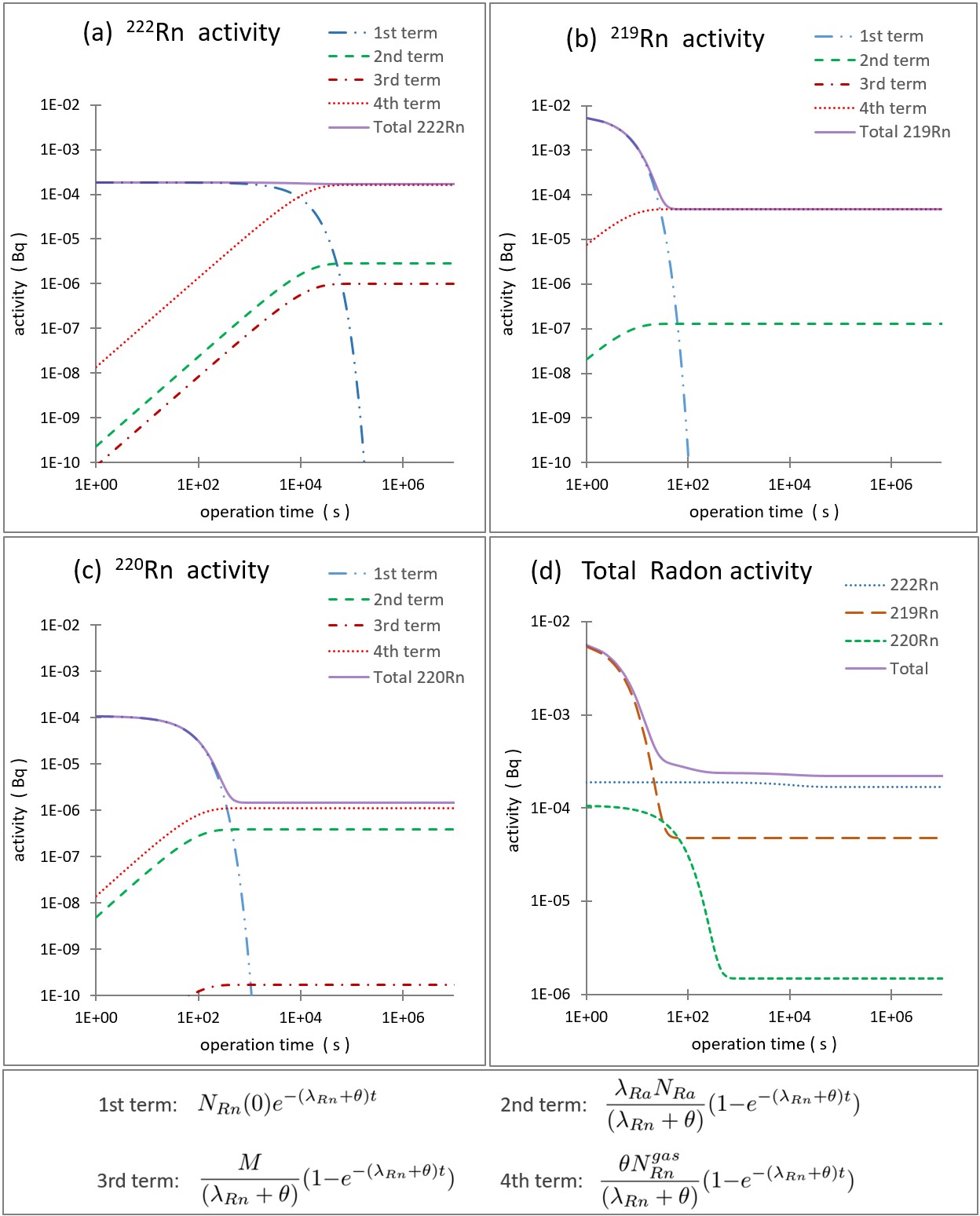}
    \caption{Results of the radon emanation background model. In the first three graphs (a-c), the evolution of the contribution to the activity from each radon isotope is shown, including the evolution of the contribution from each term of equation~\ref{eq:AlphaCAMM_BatemanEqSolution} separately: in blue, the first term regarding the initial radon contamination after the filling cycle; in green, the second term regarding contamination from radon emanating from the decay of radium; in red dashed, the third term regarding the processes of leaks, permeability, and outgassing; and in red dotted, the fourth term regarding contamination from radon in the gas mixture. The fourth graph (d) shows the evolution of the total contribution of each isotope to the total activity.}
    \label{fig:AlphaCAMM_GraficasModeloFondo}
\end{figure}

After defining and calculating the parameters of the model, the results obtained can be seen in Figure \ref{fig:AlphaCAMM_GraficasModeloFondo}. It represents the activity of each radon isotope over time and the origin of its activity: contamination from initial concentration (1$^{st}$ term); contamination from radium in the materials (2$^{nd}$ term), which after their decay emanate radon; contamination from leaks, permeability, and outgassing (3$^{rd}$ term); and contamination from the gas system (4$^{th}$ term).

As previously explained, the target background level for the AlphaCAMM is 60~nBq/cm$^{2}$. Scaling to the entire detection surface of the Micromegas (25~x~25~cm$^{2}$), the background level in the AlphaCAMM, to achieve that sensitivity, must be~$\leq$~3.7·10$^{-5}$ Bq. As can be seen in graph~\ref{fig:AlphaCAMM_GraficasModeloFondo}(d), the activity from radon emanated after 500 seconds of operation tends to 2·10$^{-4}$ Bq, an order of magnitude greater than necessary.

The main contribution comes from the isotope $^{222}$Rn, whose graph~\ref{fig:AlphaCAMM_GraficasModeloFondo}(a) shows that the 4$^{th}$ term is the primary contributor. Below the level of the isotope $^{222}$Rn, but still above the necessary background level, is the contribution from the isotope $^{219}$Rn (4.75·10$^{-5}$~Bq). This contribution, due to the concentration of this isotope in the gas, is likely overestimated as it has been assumed that all radon emanated from the bottle or the gas system can reach the detection volume, despite its short half-life (3.96 s).

Aside from the activity levels reached after evaluating the model with the previously calculated parameters, a trend is observed in the three graphs that describe the contribution of each isotope. In all of them, initially, the activity is dominated, as expected, by the radon $N_{Rn}$(0) that remains within the chamber after filling. As $N_{Rn}$(0) decreases due to the combination of its decay and the flow that carries it away, the activity becomes dominated by the largest of the other three terms, usually the 4$^{th}$ term corresponding to the concentration of radon in the gas mixture.

An adjustment of the flow rate to another value will modify the time constant $\lambda_{Rn}+\theta$, such that if the flow rate is increased, the steady state will be reached sooner, and vice versa.

Furthermore, the adjustment of the flow rate will also affect the constants of the last three terms. In the 2$^{nd}$ and 3$^{rd}$ terms, an increase in the flow rate will cause a decrease in the constant and, therefore, a decrease in their contribution to the activity, which is expected since higher flow rates result in greater carryover of radon emanated within the chamber (2$^{nd}$ term) and radon introduced by leaks, permeability, and outgassing (3$^{rd}$ term). In the 4$^{th}$ term, the variation of the flow rate will not affect the constant, as long as $\theta\gg\lambda_{Rn}$, meaning that the activity will only be affected when very low flow rates are applied such that the renewal time $t_{ren}=P_{op}/\theta$ is of the same order or greater than the half-life of the corresponding radon isotope. In that case, an example being the operation mode without flow, \textit{sealed mode}, the activity will be significantly reduced on the timescale of the half-lives of the involved isotopes.

\setlength{\baselineskip}{12pt} 
\selectlanguage{USenglish}
\graphicspath{{../Imagenes/}{../../Imagenes/}} 

\renewcommand{\listtablename}{Table of Contents}
\renewcommand{\tablename}{Table} 

\chapter{Design, Manufacturing and Commissioning of the AlphaCAMM}
\label{ch:AlphaCAMM_PuestaMarcha}
\chaptermark{Design, Manufacturing, and Commissioning...}

\begin{figure}[!ht]
    \centering
    \includegraphics[width=0.8\textwidth]{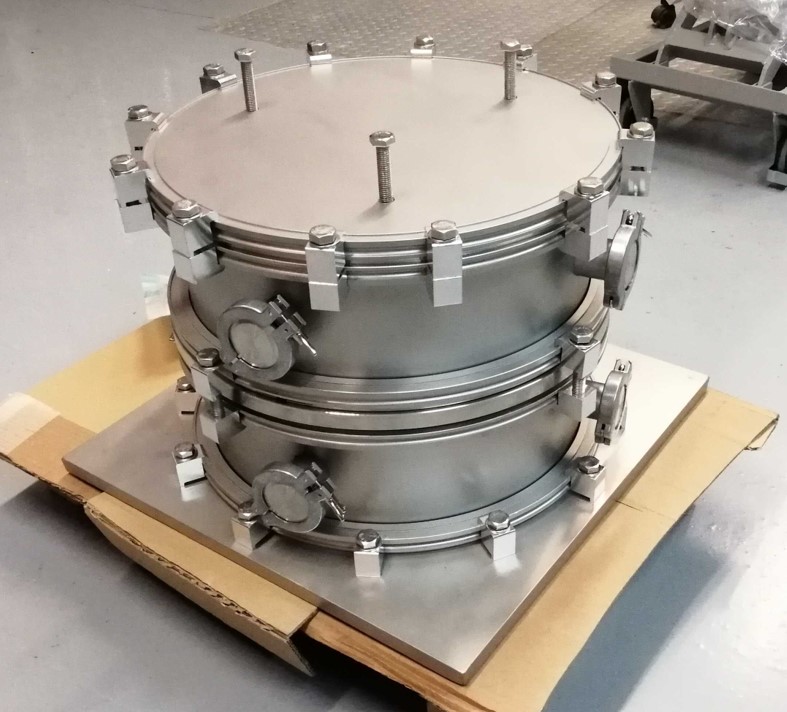}
    \caption*{}
    \label{fig:AlphaCAMM_CamaraInicial}
\end{figure}

\newpage


\section{Final Design and Manufacturing}
\label{sec:DiseñoFinal}

\begin{figure}[!ht]
    \centering
    \includegraphics[width=0.80\textwidth]{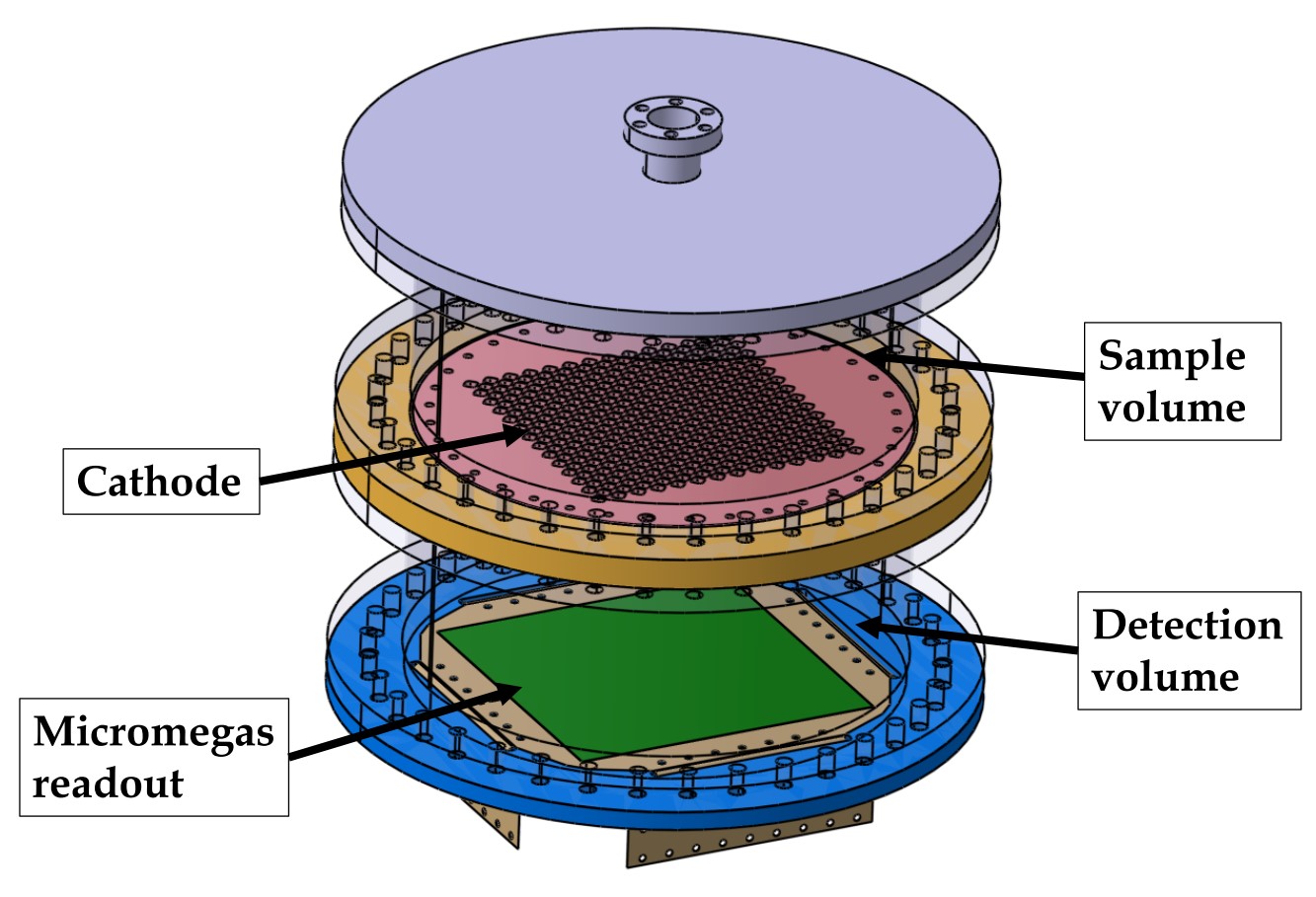}
    \caption{Conceptual design of the AlphaCAMM with two differentiated volumes, one for housing the sample and the other as a detection volume.}
    \label{fig:AlphaCAMM_ConceptualDesign}
\end{figure}

In order to carry out the final design for subsequent manufacturing, it was proposed to adopt a conceptual solution as shown in figure \ref{fig:AlphaCAMM_ConceptualDesign}. This involves two chamber bodies that form two independent volumes separated by a sample holder grid on which the sample to be measured is placed, similar to the first demonstrator described in section \ref{sec:PrimerDemostrador}, and whose schematic is shown in figure \ref{fig:AlphaCAMM_EsquemaDemostrador}. The separating grid functions as a cathode to generate the drift field in the detection volume.

\begin{figure}[!ht]
    \centering
    \includegraphics[width=0.95\textwidth]{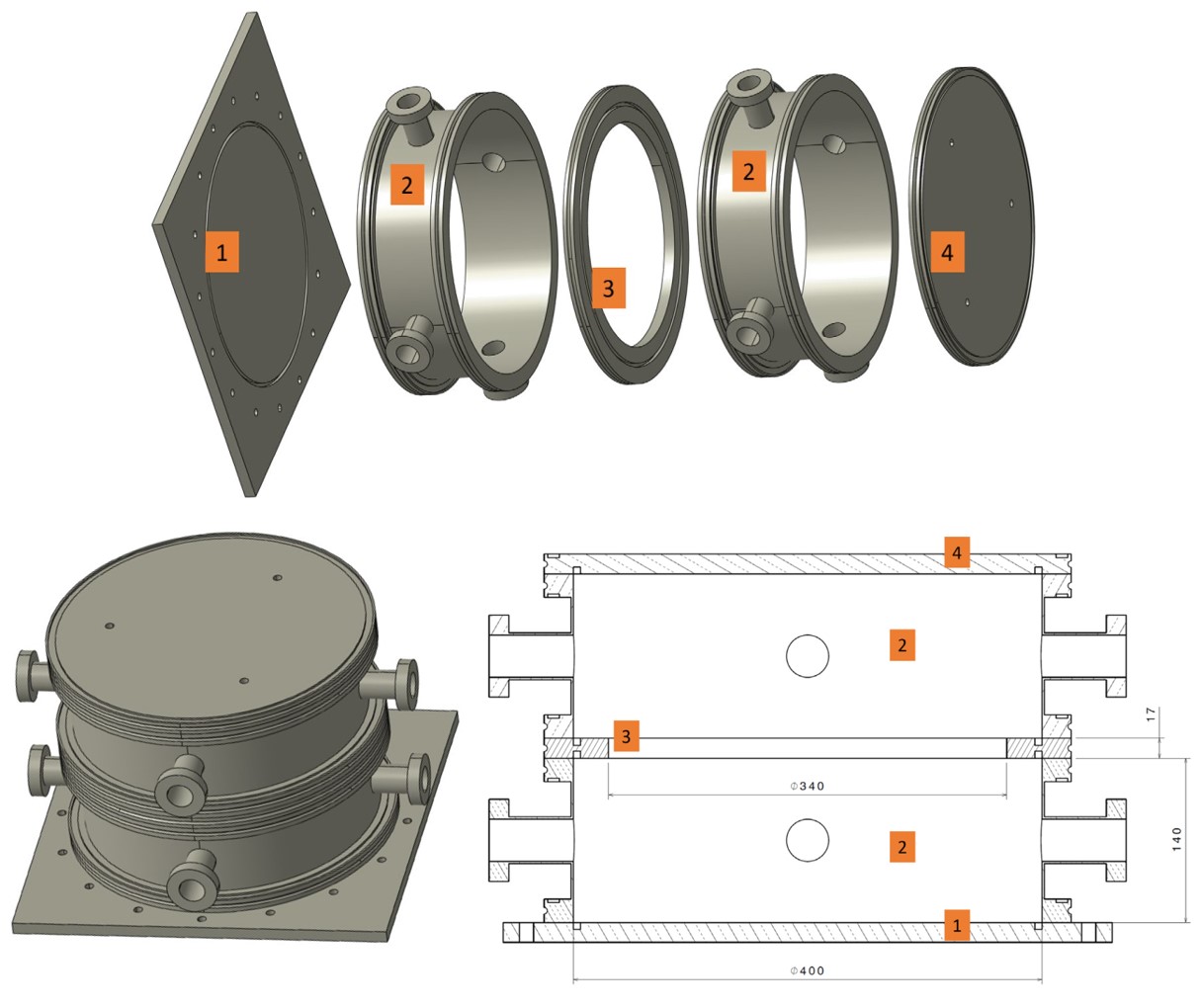}
    \caption{Different views of the final design of the AlphaCAMM chamber, where its parts can be distinguished: lower cover (1), body (2), inter-flange (3), and upper cover (4).}
    \label{fig:AlphaCAMM_PlanosProduccion}
\end{figure}

For the manufacturing of the chamber by an external company, the following technical specifications were defined:

\begin{itemize}
\item Material: AISI~316L stainless steel with limits on Uranium concentration ($\leq$~10~mBq/kg).
\item Working pressure~$\leq$~1.45~barA.
\item Leak level + Permeability + \textit{Outgassing}~$\leq$~1·10$^{-6}$~mbar~l~s$^{-1}$.
\item Working temperature: 15$^{o}$C$-$30$^{o}$C.
\item Geometry and dimensions according to figure \ref{fig:AlphaCAMM_PlanosProduccion}, with elements according to the DN-ISO-K standard.
\end{itemize}

\subsection{Operational Options}

With this conceptual design, several options were proposed for operating the detector. The detection volume is a TPC (Time Projection Chamber) in which the sample holder grid functions as the cathode, using a mixture of Argon/Isobutane and a working pressure between 1.00 and 1.45 bar. However, the sample volume could operate under different conditions:

\begin{enumerate}

\item Sample volume exposed to air: this is the simplest solution, where the volume is filled with laboratory air at atmospheric pressure. Under these conditions, the Radon contained in the laboratory air that decays near the surface of the cathode/sample holder could contribute to the background. On the other hand, the cathode/sample holder would have to withstand a pressure difference of 1 bar, only if vacuum cycles are performed in the detection volume.

\item Sample volume under vacuum: a pump connected to the volume would maintain it at a stable vacuum level. Under these conditions, Radon contamination in this volume would be reduced due to the laboratory air and possible emanation from the sample. Inevitably, the cathode/sample holder would have to withstand a pressure difference of 1 bar or more.

\item Sample volume with nitrogen flow: ensuring a Radon level in the nitrogen flow low enough not to contribute significantly to the detector's background. Similar to the previous case 'under vacuum', these conditions would reduce contamination from Radon contained in the laboratory air and that which could be emanated from the sample. And like in the first case 'in air', the cathode/sample holder would have to withstand a pressure difference of 1 bar, only if vacuum cycles are performed in the detection volume.

\item Sample volume with gas mixture flow: the outflow in the detection volume would be injected into the sample volume, so that both volumes would share the same gas mixture and maintain the same pressure. The Radon level in the gas mixture, necessarily under control to reduce the background level in the TPC, would presumably be sufficient not to contribute to it from the sample volume. In this case, being interconnected, the cathode/sample holder would only have to withstand differential pressures during pumping in the vacuum cycle, which would be performed for each measurement, on the order of tens to hundreds of millibar.

\end{enumerate}

In all the previous cases, the decay of a Radon nuclide in the sample volume would produce ions that would be attracted to the surface of the sample holder, which is at a high (negative) potential as it also functions as a cathode. The decay of the Radon daughters on this surface, many of them through $\alpha$ emissions, could affect the detector's background and thus reduce its sensitivity. To mitigate this effect, it was proposed to place a plate at the opposite end of the sample volume, on the upper cover (figure~\ref{fig:AlphaCAMM_PlanosProduccion}), with a certain potential that would allow the electric field in the sample volume to be inverted, thereby attracting the ions resulting from the decays of these nuclides. This trap for the Radon daughters could be implemented at any time during the commissioning of the detector.

\subsection{Design of the Cathode/Sample Holder}

The cathode/sample holder is a fundamental part of the detector as it serves several functions: supporting the weight of the samples while keeping their surfaces in contact, separating the detection and sample volumes in a hermetic manner, and functioning as a cathode in the TPC with a potential distributed across the entire surface.

To support the weight of the samples, a tensioned grid or a perforated plate was considered. To separate the detection and sample volumes, the installation of a mylar sheet, attached to the grid, was proposed. The only condition for fulfilling the function of a cathode is that the material used must be conductive.

\begin{figure}[!ht]
    \centering
    \includegraphics[width=1.0\textwidth]{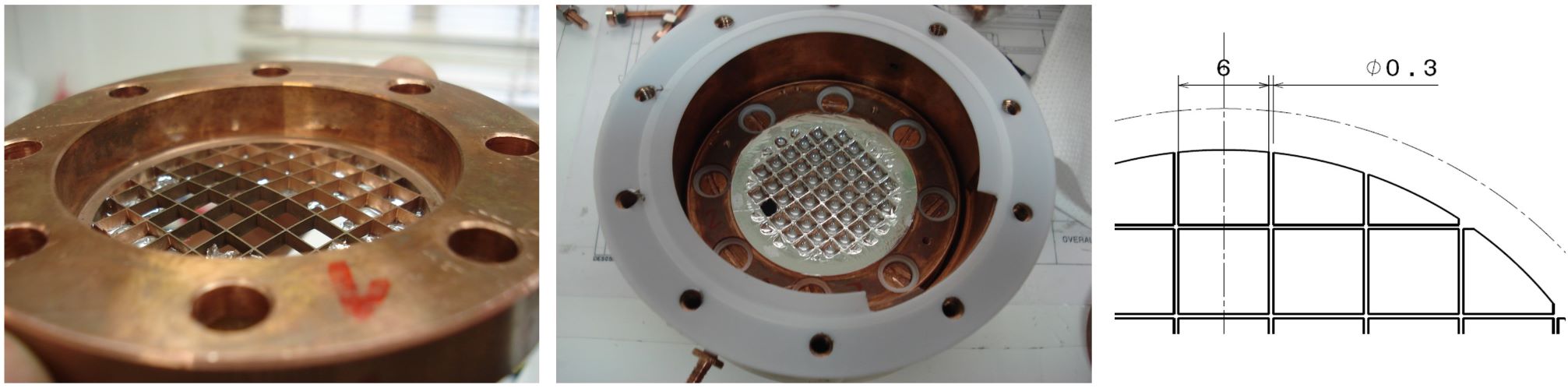}
    \caption{Photos taken during the destructive tests carried out on the copper-mylar grid of the cathode in the CAST experiment \cite{Aznar2015}. The chamber was subjected to increasing pressure, reaching 3.1~bar and 3.0~bar in two measured samples, just before the mylar ruptured. $\bullet$~Left:~the pressure caused plastic deformation of the copper, permanently deforming the grid. $\bullet$~Center:~rupture zone of the mylar upon reaching maximum pressure. $\bullet$~Right:~geometry of the copper grid.}
    \label{fig:AlphaCAMM_FotoCASTrupturaMylar}
\end{figure}

\begin{figure}[!ht]
    \centering
    \includegraphics[width=0.9\textwidth]{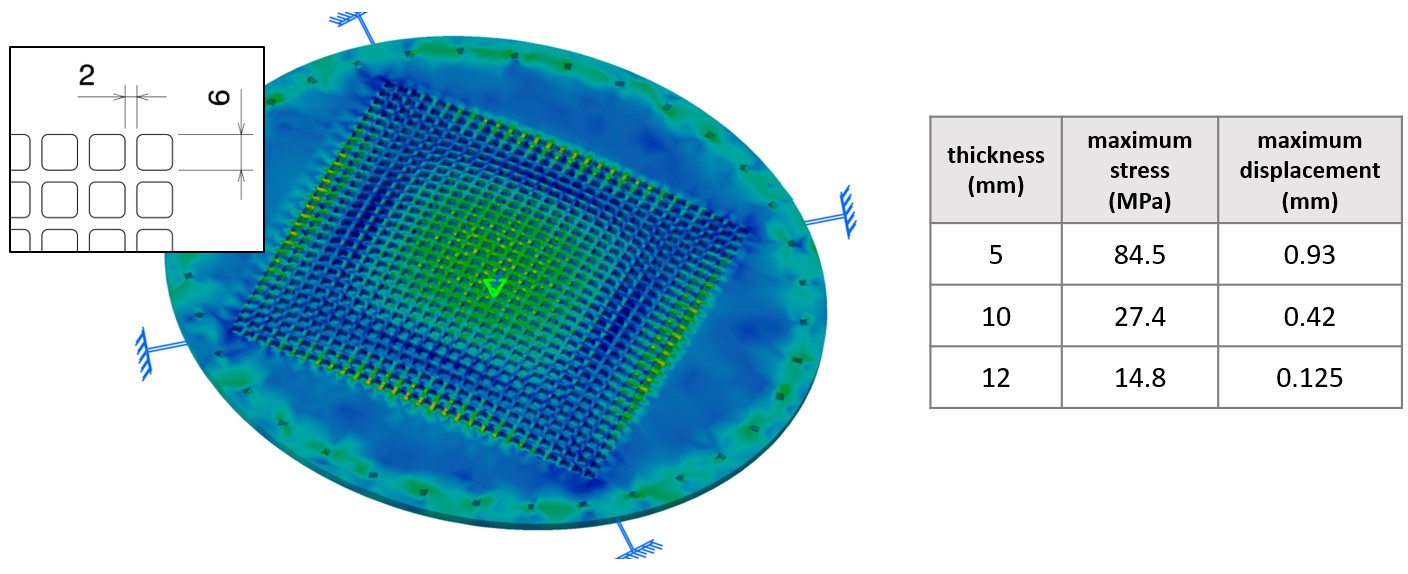}
    \caption{Image from the finite element study, showing the stress-strain responses of the cathode/sample holder to the application of 1~bar of pressure on its surface with a grid of 6~x~6~mm$^2$. The table shows the results for three different grid thicknesses.}
    \label{fig:AlphaCAMM_FEMrejilla}
\end{figure}

With these components, and for the most unfavorable case in which the cathode/sample holder must withstand a pressure difference of 1~bar, a stress-strain study was conducted using finite elements. This study was based on a copper grid with holes of 6~x~6~mm$^2$, to which aluminized mylar of 3.5~$\mu$m thickness is attached. This composition, during the development phase of the cathode in the CAST experiment~\cite{Aznar2015}, had withstood pressures of up to 3~bar (figure~\ref{fig:AlphaCAMM_FotoCASTrupturaMylar}). From the results of this simulation (figure~\ref{fig:AlphaCAMM_FEMrejilla}), and for copper with a yield strength of 33~MPa (a limit that could be increased through cold deformation), a grid thickness of at least 10~mm is necessary.

\begin{figure}[!ht]
    \centering
    \includegraphics[width=0.80\textwidth]{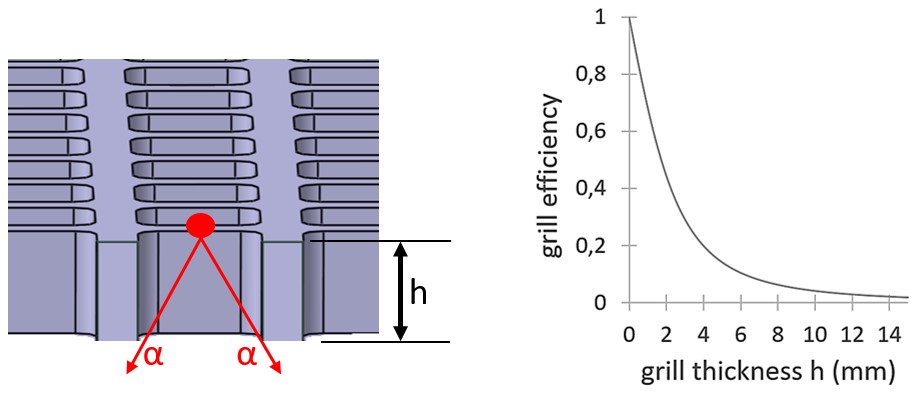}
    \caption{$\bullet$ Left: section of the cathode/sample holder grid with a hole size of 6~x~6~mm$^2$ and a thickness of~\textit{h}. The emission of an alpha particle from the upper surface where the sample rests has an entry efficiency into the detection volume (lower volume) that depends on~\textit{h}. $\bullet$ Right: entry efficiency of the alpha particles into the detection volume versus the thickness~\textit{h} of the grid.}
    \label{fig:AlphaCAMM_EficienciaRejilla}
\end{figure}

The formation of the cathode/sample holder through the combination of a grid and a mylar sheet will result in a loss of efficiency due to the opacity generated by this element when a sample is placed on top. This efficiency is very low, around 4$\%$ (figure~\ref{fig:AlphaCAMM_EficienciaRejilla}), when the grid thickness is 10~mm. Therefore, if one wishes to increase the transparency of the grid, alternative solutions must be sought: increasing the yield strength of the copper, using stainless steel, reducing the differential pressure between volumes, etc. Additionally, the electrons resulting from the ionization of the gas by the alpha particle, during their travel between the walls of the grid, may be lost because the drift electric field in that area is likely not sufficiently homogeneous.

On the other hand, the use of mylar in the cathode to ensure the hermeticity between both volumes causes a stopping of the alpha particles that pass through it. The stopping power for alpha particles in the range of 3 to 10~MeV is between 1.15 and 0.58~MeV~cm$^{2}$~g$^{-1}$ \cite{NIST-ASTAR}. For example, for the proposed mylar thickness of 3.5~$\mu$m and an alpha particle of 5~MeV passing through, there is an energy loss of 0.45~MeV.

\section{Assembly and Commissioning}

Following the fabrication of the chamber according to the design specifications, the detection volume was assembled first. The objective in this first stage was to measure the background level in the detection volume and, once quantified, to assemble and connect the second volume with the sample holder grid to perform material measurements.

\begin{figure}[!ht]
    \centering
    \includegraphics[width=1.0\textwidth]{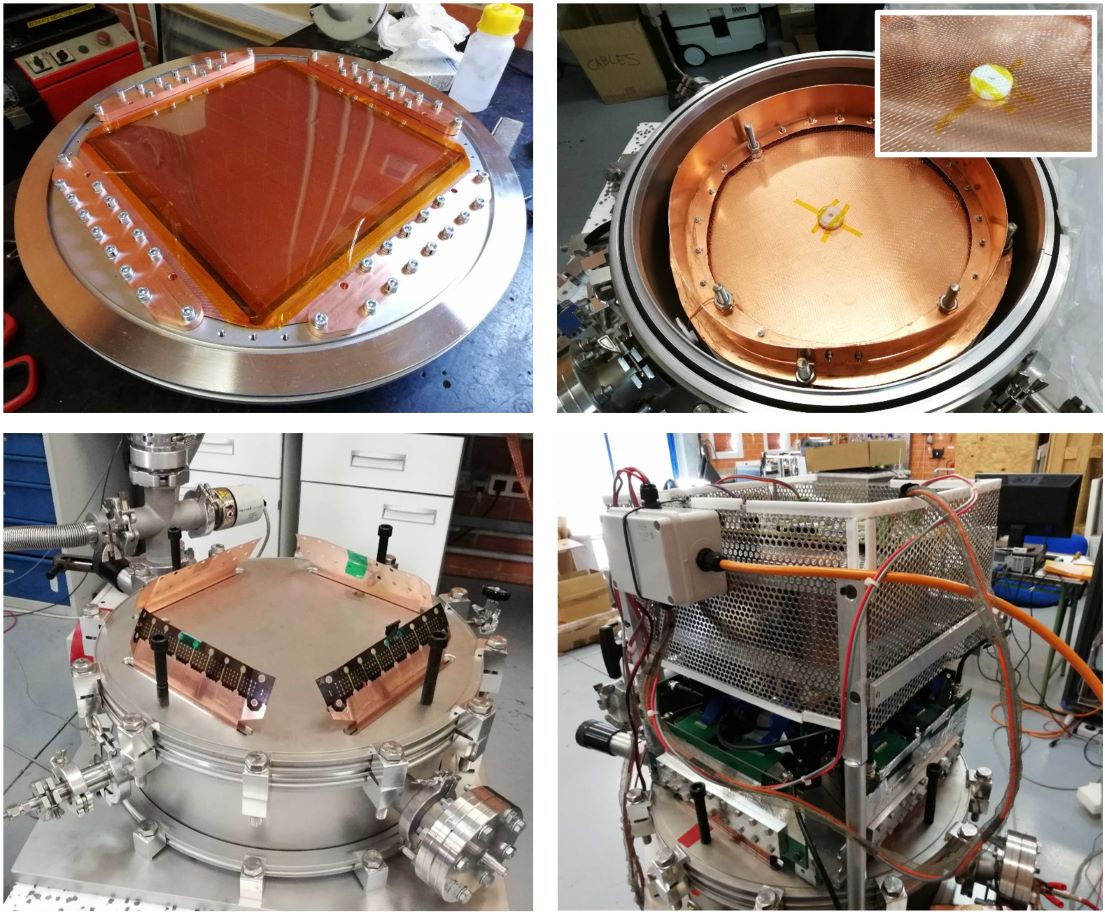}
    \caption{Photographs taken during the assembly of the AlphaCAMM (first stage). $\bullet$~Top~left:~Micromegas installed on the stainless steel cover. The sealing of the flaps that lead the channels to the outside is achieved using four copper clamps and stainless steel screws. $\bullet$~Top~right:~cathode with a $^{241}$Am source in the center. A copper ring was also installed on the cathode, which served to maintain a uniform electric field along the drift length. The detailed photo shows a copper mesh above the $^{241}$Am source, installed as a first test for the construction of the sample holder grid. $\bullet$~Bottom~left:~closed chamber, prior to the assembly of the acquisition electronics. The flaps of the Micromegas can be seen before they were connected to the electronics. $\bullet$~Bottom~right:~final situation of the detector during the first calibration.}
    \label{fig:AlphaCAMM_FotoCamaraFinal}
\end{figure}

Figure~\ref{fig:AlphaCAMM_FotoCamaraFinal} shows several photographs taken during the assembly in this first stage. The Micromegas was installed on the top cover. The cathode was installed on the bottom cover, along with a copper ring to maintain a uniform electric field along the drift length. The electronics that were installed were based on the AFTER chip~\cite{Baron2008}, which necessitated the assembly of a readout-amplification-discrimination chain to read the signal from the mesh and thus provide the trigger signal to the electronics, as was done in the first demonstrator~\cite{Altenmuller2022AlphaCAMM}.

\begin{figure}[!ht]
    \centering
    \includegraphics[width=1.0\textwidth]{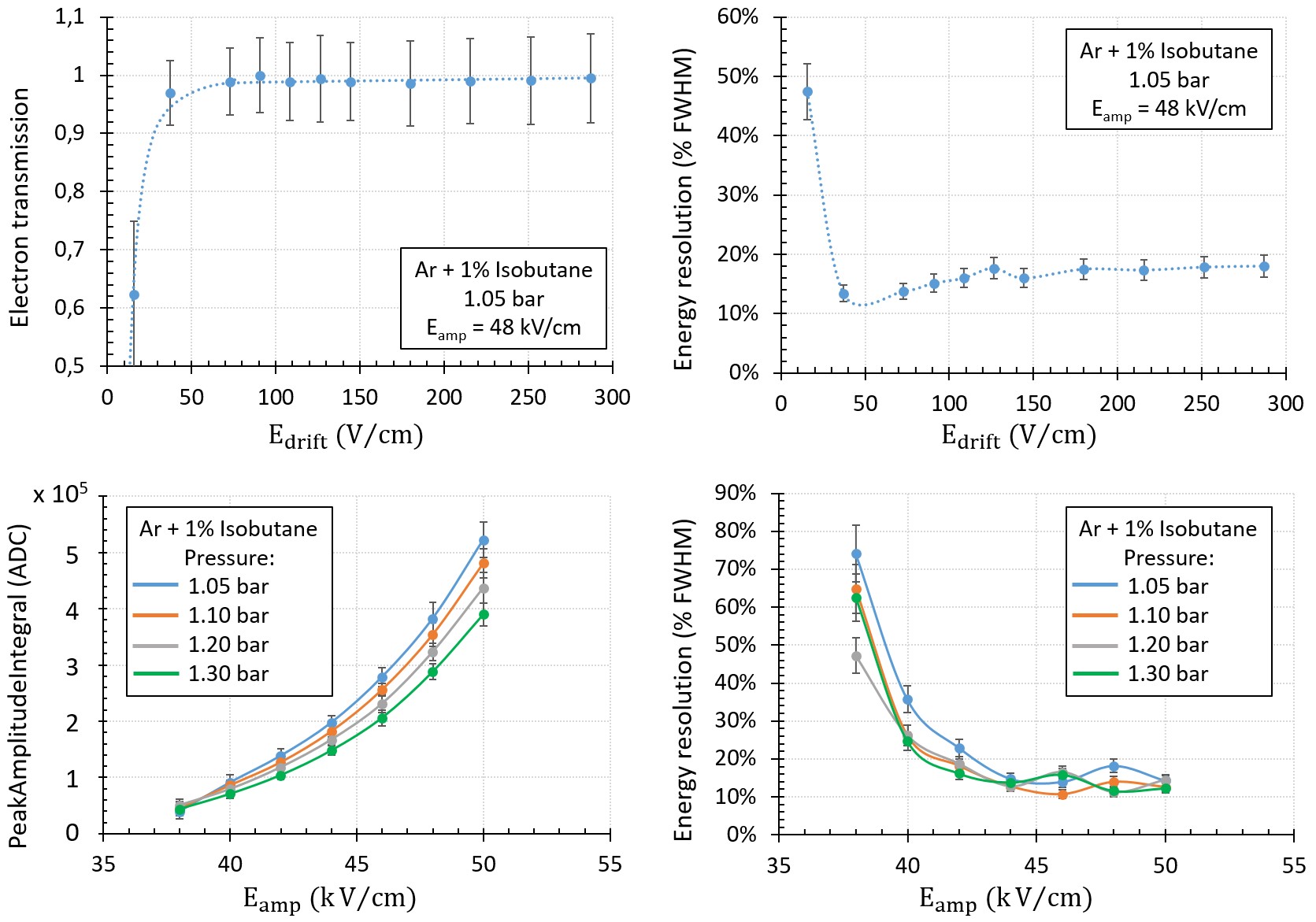}
    \caption{Characterization curves of the Micromegas in the AlphaCAMM, for Ar + 1\% Isobutane. $\bullet$~Top~left:~Electron transmission curve versus drift field, at 1.05 bar and E$_{amp}$ = 48 kV/cm. $\bullet$~Top~right:~Energy resolution curve versus drift field, at 1.05 bar and E$_{amp}$ = 48 kV/cm. $\bullet$~Bottom~left:~Gain curve versus amplification field in the Micromegas, for different pressures from 1.05 to 1.30 bar. $\bullet$~Bottom~right:~Energy resolution curve versus amplification field in the Micromegas, for different pressures from 1.05 to 1.30 bar.}
    \label{fig:AlphaCAMM_CaracterizationCurves}
\end{figure}

The first data acquisitions, with the $^{241}$Am source inside emitting $\alpha$ particles of 5.4 MeV and with a mixture of Ar + 1\% Isobutane, served to generate the electron transmission, gain, and energy resolution curves (figure~\ref{fig:AlphaCAMM_CaracterizationCurves}). This verified the proper functioning of the detector and allowed for the selection of an operating point.

Note that the maximum amplification field (E$_{amp}$) possible in a microbulk-type Micromegas, for the gas mixture used (Ar + 1\% Isobutane) and in the applied pressure range (1.05–1.30 bar), is around 60 kV/cm (figure~\ref{fig:CurvasArticuloIguaz2022}~\cite{Iguaz2022}). The gain levels achieved with the maximum amplification fields, as seen throughout this thesis, allow for energy thresholds with this gas mixture around 1 keV. For the detection of $\alpha$ particles, with an interest range between 1 and 10 MeV, the system gain needed to be reduced by approximately a factor of 1000, to avoid saturating the acquisition electronics. While theoretically this gain could be reduced solely through the acquisition electronics, by lowering the gain in its amplification stage, a combination of gain reductions in both the electronics and the Micromegas was preferred. In this way, a reduction in the amplification field in the Micromegas would allow operation with voltages below the maximums, thus ensuring a safe zone against discharges and potential damage to the channels. Furthermore, the amplification field ultimately chosen for the operation of the detector (E$_{amp}$ = 48 kV/cm, corresponding to 240 V on the mesh) offers the best energy resolution, according to previous studies conducted with small Micromegas~\cite{Iguaz2022}, and as can be seen in figure~\ref{fig:AlphaCAMM_CaracterizationCurves}, where resolutions between 15 and 20\% are achieved, even lower in some points.

\begin{figure}[!ht]
    \centering
    \includegraphics[width=1.0\textwidth]{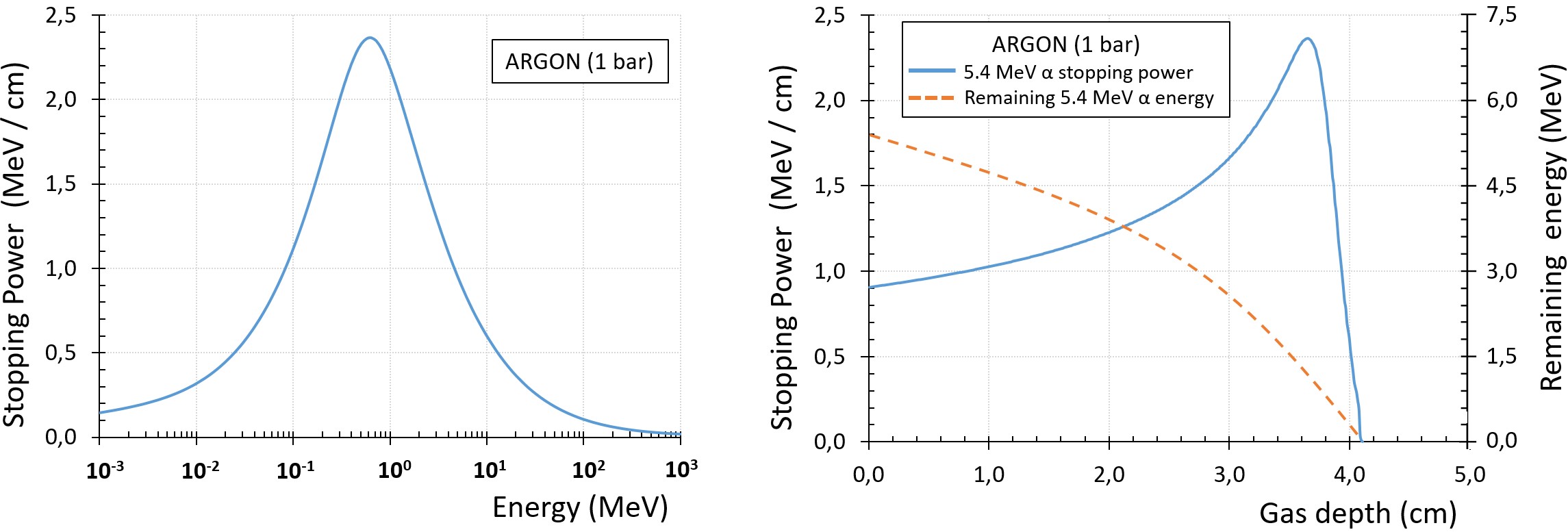}
    \caption{$\bullet$~Left:~Stopping power of $\alpha$ particles versus particle energy, for Argon at 1 bar pressure. Data to generate the curve collected from~\cite{NIST-ASTAR}. $\bullet$~Right:~Ionization density and remaining energy of the 5.4 MeV $\alpha$ particles (initial energy) versus their range in the gas, for Argon at 1 bar pressure, calculated from the stopping power.}
    \label{fig:AlphaCAMM_StoppingPower}
\end{figure}

The $\alpha$ particles emitted by the $^{241}$Am source have an energy of 5.4 MeV and, taking into account the stopping power of Argon at 1 bar pressure (figure~\ref{fig:AlphaCAMM_StoppingPower}~left), their range in the detection volume of the AlphaCAMM can be calculated, along which gas ionization occurs. Figure~\ref{fig:AlphaCAMM_StoppingPower}~right shows the ionization density (MeV/cm) of the 5.4 MeV $\alpha$ particles along their trajectory in Argon at 1 bar, where the Bragg peak can be appreciated at the end of this, with a higher ionization density. The remaining energy of the particle along its trajectory is also represented. The length of the track, according to the fit performed, is 4.11$\pm$0.06 cm.

\begin{figure}[!ht]
    \centering
    \includegraphics[width=0.7\textwidth]{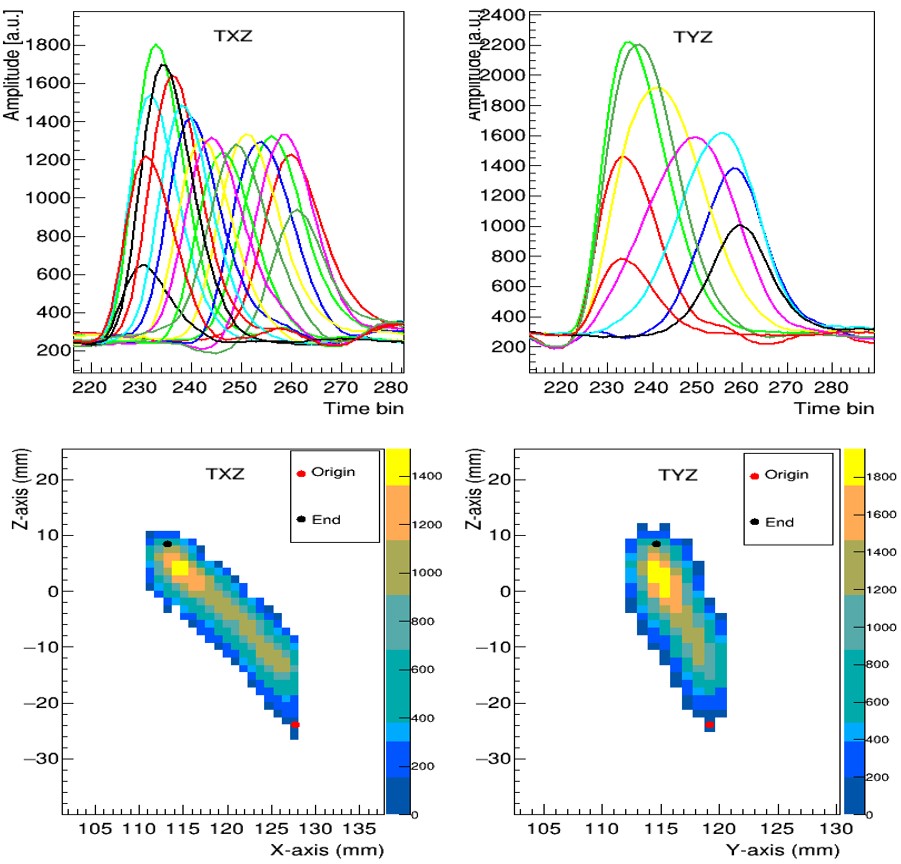}
    \caption{Event from the $^{241}$Am source, recorded in the AlphaCAMM during the first calibration. $\bullet$~Top:~Digitized signals of the event recorded by the X-axis (left) and Y-axis (right) strips. Each color represents a strip/channel of the corresponding axis. $\bullet$~Bottom:~Projection of an event produced by the $^{241}$Am source, where the colors represent the integration of the pulses in XZ (left) and YZ (right), with the start (red star) and end (black star) of the trajectory.}
    \label{fig:AlphaCAMM_FirstCalibrationSignals}
\end{figure}

Figure~\ref{fig:AlphaCAMM_FirstCalibrationSignals} shows an event from the $^{241}$Am source during one of the first calibrations. In the projections on the XZ and YZ planes, the part of the trajectory with the highest ionization density (Bragg peak) can be differentiated, from which the direction can be deduced. In the case of the event shown, it is emitted from the cathode, where the source is located, towards the Micromegas, situated at the top according to the graph. It should be noted that the code used for the topological analysis of the tracks has been developed by the research group in which this thesis is framed~\cite{Altenmuller2022AlphaCAMM}, but its development is not part of this work.

The spatial component Z is calculated from the integration of the pulses in the sampling time of the electronics. By definition, since it is a time projection chamber, the time difference between two pulses in the same event will determine, taking into account the drift velocity of the electrons in the gas, the spatial difference in Z. In this case, the source events, of known length, can also be used to validate the drift velocity used.

\begin{figure}[!ht]
    \centering
    \includegraphics[width=0.99\textwidth]{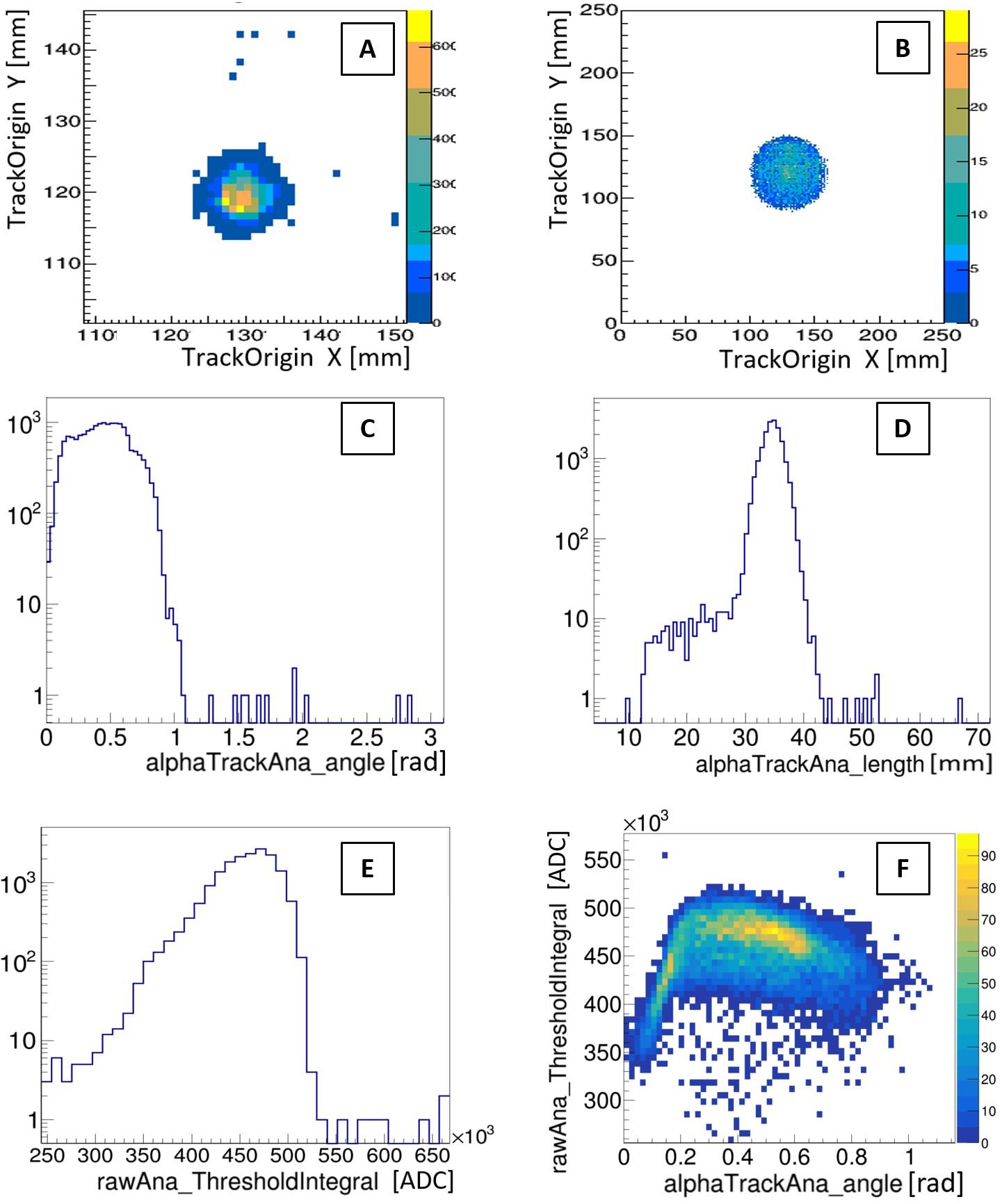}
    \caption{Results of the analysis for one of the calibrations with the $^{241}$Am source. $\bullet$~\textbf{A}:~XY projection of the origin of the trajectory of each recorded event. $\bullet$~\textbf{B}:~XY projection of the end of the trajectory of each recorded event. $\bullet$~\textbf{C}:~Histogram of the angle of the trajectory of each event with respect to the Z axis. From 0 to $\pi$/2 the direction is towards the Micromegas, from $\pi$/2 to $\pi$ the direction is towards the cathode. $\bullet$~\textbf{D}:~Histogram of the length of the trajectory. $\bullet$~\textbf{E}:~Histogram of the energy, represented by the observable ThresholdIntegral. $\bullet$~\textbf{F}:~Histogram of the energy of each event (represented by the observable ThresholdIntegral) versus the angle of the trajectory.}
    \label{fig:AlphaCAMM_FirstCalibrationAnalisis}
\end{figure}

Figure~\ref{fig:AlphaCAMM_FirstCalibrationAnalisis} shows some results from the analysis for one of the calibrations with the $^{241}$Am source. The source casing is made of polyethylene and has a diameter of 20 mm, with a central hole of 5 mm in diameter where the Americium material is deposited. This shape can be deduced from the XY projection of the origin of the trajectories, with a central area showing a higher density of events. The XY projection of the end of the trajectories shows a circular shape with a larger diameter than the source (approximately 5 cm) due to the isotropic emission of the source, although there is some collimation due to the thickness of the polyethylene casing that supports the source. This isotropic emission and the semi-collimation of the source are confirmed in the trajectory angle histogram, with a flat distribution between 0 and 1 rad that confirms the isotropy of the emission and with an absence of events between 1 and $\pi$/2 rad explained by the opacity provided by the thickness of the source casing, affecting the solid angle and partially collimating the emission.

\begin{figure}[!ht]
    \centering
    \includegraphics[width=1.0\textwidth]{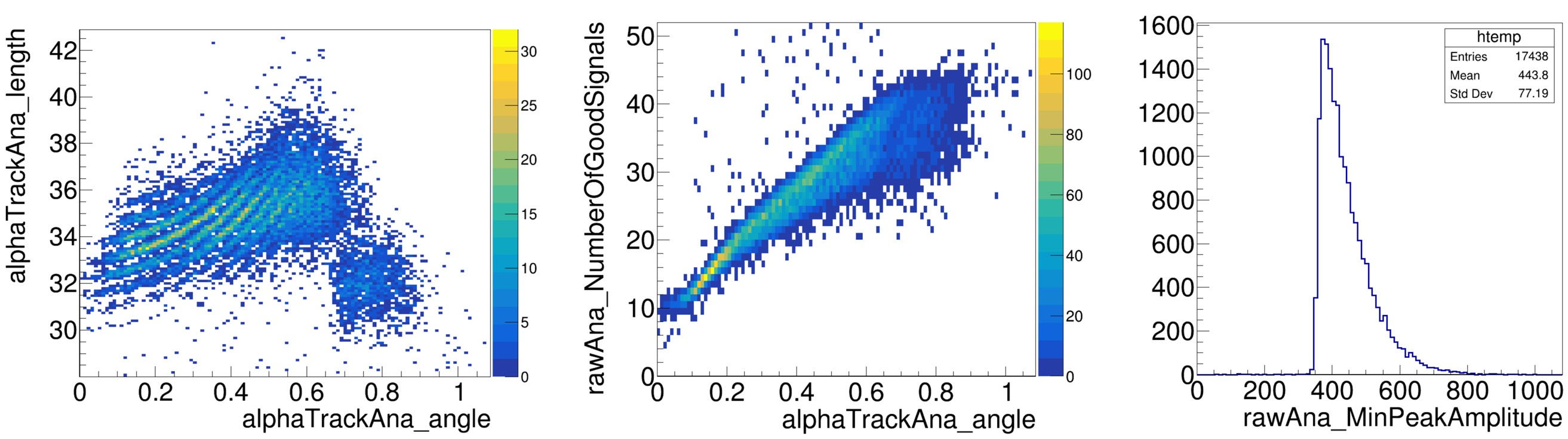}
    \caption{$\bullet$~Left:~Histogram of the length of the trajectory versus the angle of the trajectory with respect to the Z axis. $\bullet$~Center:~Histogram of the number of signals/channels in each event versus the angle of the trajectory with respect to the Z axis. $\bullet$~Right:~Histogram of the amplitude of the lowest amplitude signal in each event.}
    \label{fig:AlphaCAMM_ProblemaLongitudTraza}
\end{figure}

As seen in the trajectory length histogram, the majority of the collected events (99.73\%) are found in the population with a mean of 34.5 mm and a standard deviation of 1.65 mm. The difference between the measured length and the theoretical length found from the stopping power (4.11$\pm$0.06 cm) may be due to several factors:

\begin{itemize}
\item An error in the estimation of the drift velocity value used for the analysis, which would affect the Z component. In that case, the more horizontal trajectories, with angles close to $\pi$/2 rad, will have little Z component and mostly X and Y, so the measured length of the trajectory will have a greater dependence on the pixelation of the Micromegas. In the angle-length histogram of figure~\ref{fig:AlphaCAMM_ProblemaLongitudTraza}, it can be seen that the length has a slight dependence on the angle of the trajectory between 0 and 0.7 rad, which raises suspicion of a drift velocity in the analysis lower than that present in the detector. For angles greater than 0.7 rad, with a low component in Z and high in X and Y, one would expect the length to approach the theoretical value; however, this does not happen, and contrary to expectations, this population has a length shorter than the mean. This strange behavior points to the need to optimize the analysis. The number of signals read with respect to the angle of the trajectory, which is also shown in figure~\ref{fig:AlphaCAMM_ProblemaLongitudTraza}, has a dependence as expected.

\item Effect of the 'energy threshold per channel' (see section~\ref{sec:TREXatLSC_PuestaENMArcha}), which would cause the initial part of the trajectory, with lower ionization density, to be lost. In figure~\ref{fig:AlphaCAMM_ProblemaLongitudTraza}, the histogram of the amplitude of the lowest amplitude channel in each event can be seen, which shows a threshold per channel of 400 ADCs. If this effect is relevant for the length calculation, it would be expected that, since the ionization density increases progressively with the length of the track, there would be a progressive increase in the amplitude of the channels at the beginning of the trajectory. Thus, the signal from the channel that is below 400 ADCs is lost, which corresponds to the previous channel/strip before the start of the trajectory, while the next channel is already above this level and represents the first channel/strip of the trajectory. To delve deeper into this effect, a more sophisticated analysis would need to be conducted, focusing part of it on the initial area of the trajectory.

\item Thickness of the protective film of the $^{241}$Am source, which causes a loss of energy of the $\alpha$ particle as it passes through. The material on which the solid/liquid containing the radioactive isotope $^{241}$Am is deposited is unknown, nor is the depth at which it is deposited, nor if there is a protective film above the surface on which it is deposited to prevent the detachment of the radioactive material.
\end{itemize}

Continuing with the results of the analysis shown in figure~\ref{fig:AlphaCAMM_FirstCalibrationAnalisis}, the energy spectrum shows a clear peak at 462·10$^3$ ADCs with a standard deviation of 26·10$^3$ ADCs, which corresponds to the energy of 5.4 MeV of the $\alpha$ particles emitted by the source. However, there is a population to the left of the peak with lower energy, which corresponds to the population with an angle less than 0.25 rad. This population and its dependence on the angle can be clearly seen in the histogram ThresholdIntegral$-$TrackAnaAngle of the same figure~\ref{fig:AlphaCAMM_FirstCalibrationAnalisis}. For small angles where all the charge concentrates on a few strips/channels of the Micromegas, the non-linear behavior of the Micromegas amplification with charge density is accentuated. High charge densities affect the amplification field, reducing the gain. This effect could be taken into account in the analysis and a correction dependent on the charge density arriving per channel could be made.

\section{Background Measurements}
\label{sec:AlphaCAMM_MedeidasFondo}

The first background measurements conducted after the assembly and commissioning of the detection volume yielded a background level of 2.29~$\pm$~0.14~$\mu$Bq/cm$^2$, a factor of approximately 40 greater than the target background level (60~nBq/cm$^2$). For this background measurement, the volume was evacuated to 5·10$^{-5}$~mbar, and then the mixture of Ar~+~1\%~Isobutane was injected up to 1.05~bar. The filling cycle from the vacuum pressure to the operating pressure lasted 110 seconds, after which voltages were increased in the Micromegas and the cathode, and data acquisition was initiated with a total measurement duration of 165 hours.

\begin{figure}[!ht]
    \centering
    \includegraphics[width=0.85\textwidth]{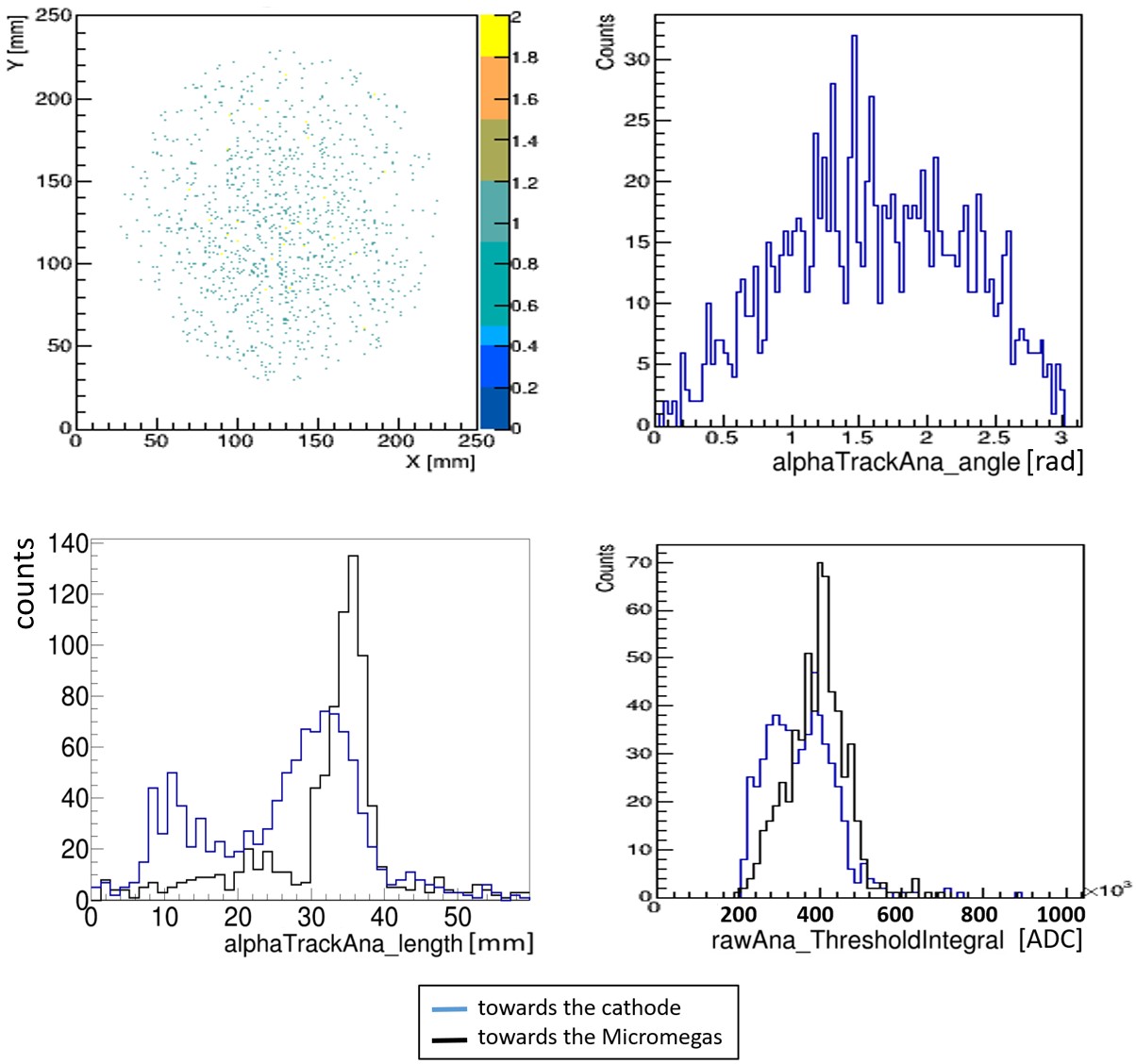}
    \caption{Results of the analysis for the first background data acquisition in the AlphaCAMM, during the first stage with the detection volume.}
    \label{fig:AlphaCAMM_AnalisisPrimerFondo}
\end{figure}

Figure~\ref{fig:AlphaCAMM_AnalisisPrimerFondo} shows some graphs from the analysis of this first background measurement. The origin of the trajectories is uniformly distributed across the detector area. To discriminate the events originating from the ring (Figure~\ref{fig:AlphaCAMM_FotoCamaraFinal}) installed to direct the drift field, a fiducial cut was made, and only those events whose origin is located 1 cm or more from the surface of the ring were considered.

The angular distribution shows a higher population of events centered around $\pi$/2 (horizontal trajectories, perpendicular to the Z-axis), which seems to indicate the ring as a source of background radiation. However, as explained, a fiducial cut was made with a volume that is a cylinder with a diameter of 23 cm, 2 cm smaller than the diameter of the ring (25 cm), so that an $\alpha$ particle emitted from the surface of the ring towards the detection volume immediately interacts with the atomic electrons of the gas mixture, ionizing it and defining the origin of the trajectory very close to the surface, thus remaining outside the fiducial volume.

It was ultimately determined that the cause of this angular distribution was the selection of the drift velocity during the analysis. The Z component has a strong dependence on this velocity, and an underestimation of it causes an overestimation of the angle of the trajectories, thus resulting in a distribution with a higher population at angles close to the horizontal.

The correction of the drift velocity introduced in the analysis would lead to a correction of the angular distribution, resulting in a flat distribution, indicating that the background was primarily dominated by decays occurring in the gas volume of the detector. However, while this was the most probable origin, it could not be ruled out that both exposed and opposing surfaces, that of the Micromegas and that of the cathode, were the origin or part of the origin of the background, contributing the same proportion of events.

Following the first background measurement, and with the aim of identifying its source for reduction, several actions were taken (flow variation, cleaning, removal of elements, variation of drift distance, ...) summarized in Table~\ref{tab:AlphaCAMM_TablaHistoricoFondo}.

As can be seen, no significant reduction in the background level was achieved despite several modifications and checks carried out over several weeks of measurements. The lowest background level measured in this series was 1.69~$\pm$~0.21~$\mu$Bq/cm$^2$, a factor of approximately 30 greater than the target background level (60~nBq/cm$^2$).

\begin{table}[!h]
\centering
\begin{tabular}{l}
\includegraphics[width=0.95\textwidth]{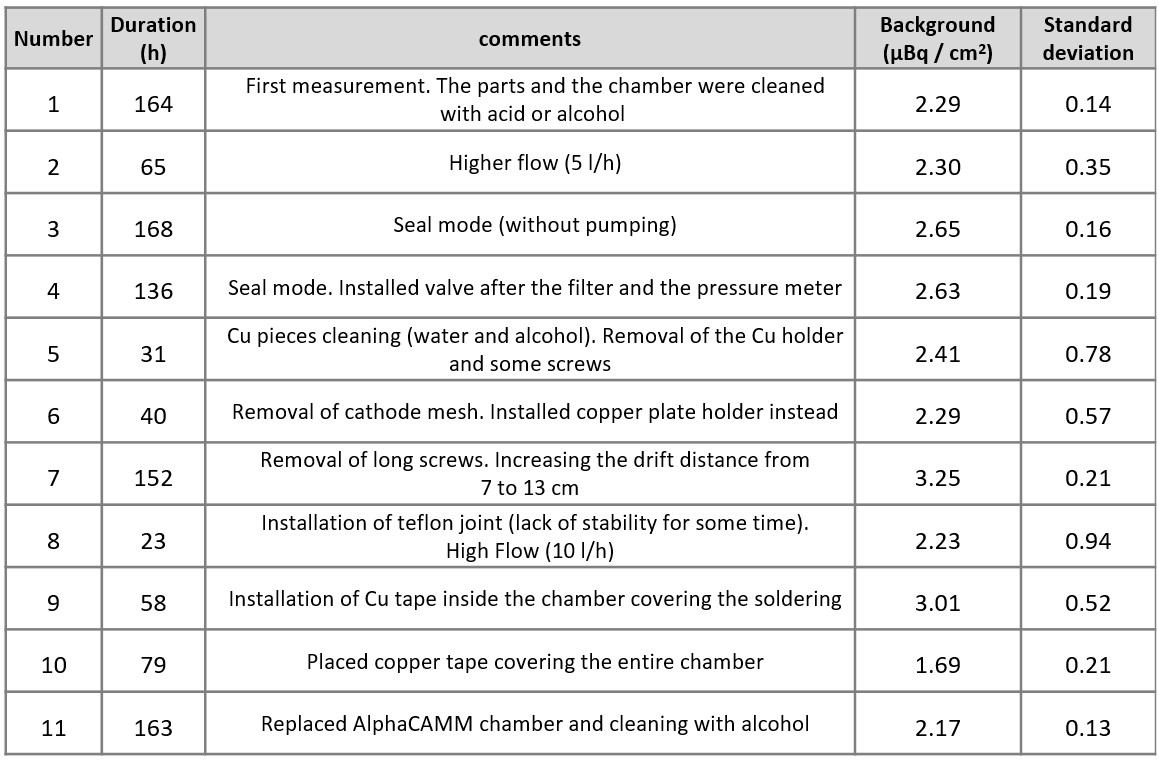}\\
\end{tabular}
\caption{List of data acquisitions under different conditions and modifications. Each time the camera was opened, or a new measurement cycle was started, the volume was pumped down to 5·10$^{-5}$~mbar. Subsequently, the mixture of Ar~+~1\%~Isobutane was injected up to 1.05~bar, with a process duration of approximately 110 seconds. Finally, the voltages in the Micromegas and the cathode were increased, and after about 100 seconds for this operation, data acquisition was launched. Unless explicitly stated, the measurements were conducted with a flow rate of 2~l/h.}
\label{tab:AlphaCAMM_TablaHistoricoFondo}
\end{table}

\subsection{Actions for Background Identification}

The first measurement conducted ('Number~1' in table~\ref{tab:AlphaCAMM_TablaHistoricoFondo}) was performed after cleaning with acid (internal parts) and alcohol (the stainless steel body of the chamber) and after assembly as shown in figure~\ref{fig:AlphaCAMM_FotoCamaraFinal}. The results have already been presented (figure~\ref{fig:AlphaCAMM_AnalisisPrimerFondo}), yielding a background of 2.29~$\pm$~0.14~$\mu$Bq/cm$^2$.

\begin{figure}[!ht]
    \centering
    \includegraphics[width=0.90\textwidth]{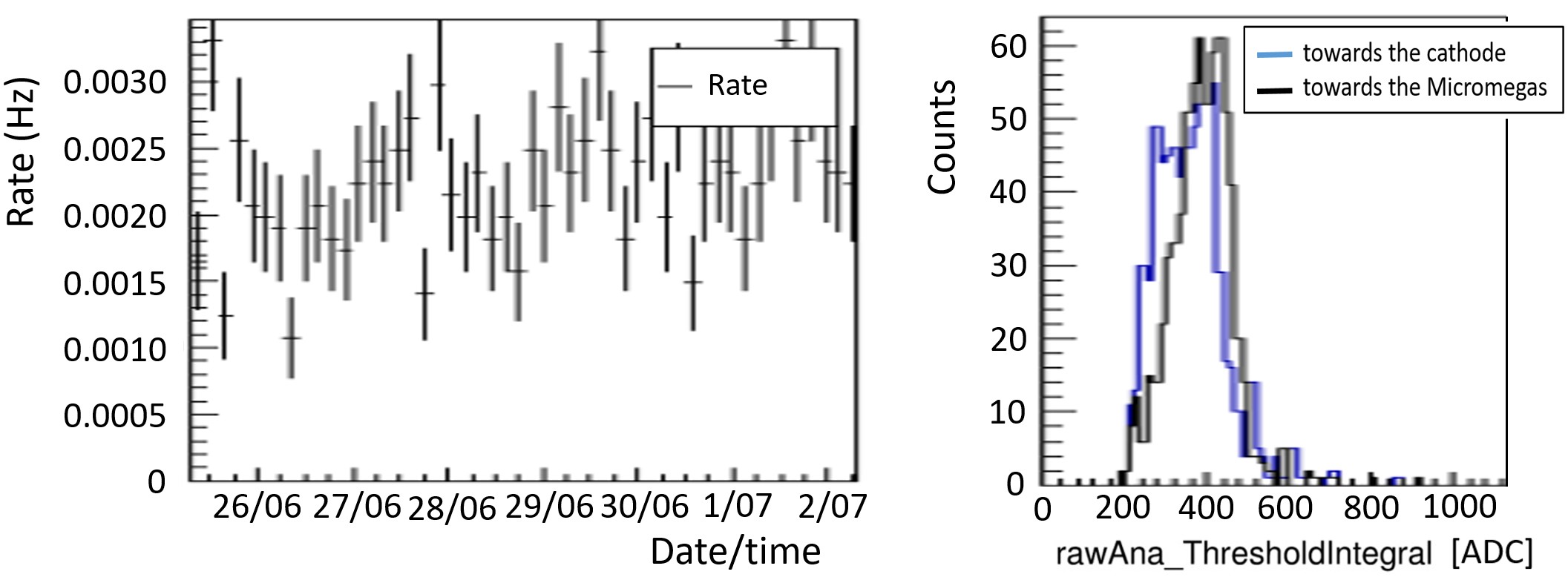}
    \caption{Evolution of the event acquisition rate over time (left) and energy spectrum (right) for the background measurement of 168 hours in \textit{sealed mode} ('Number 3').}
    \label{fig:AlphaCAMM_RateFondoSealModeR154}
\end{figure}

These initial measurements suggested that the source of the background was volumetric, so a background measurement was decided upon by increasing the flow rate from 2 to 5 l/h ('Number~2'), resulting in a similar value (2.30~$\pm$~0.35~$\mu$Bq/cm$^2$). Following this, and without performing a vacuum cycle in the chamber, but rather directly closing the inlet and outlet, another measurement was taken in \textit{sealed mode} ('Number~3'), recording a background measurement of 2.65~$\pm$~0.16~$\mu$Bq/cm$^2$. This background level, calculated by integrating the entire measurement time, is slightly higher than the previous ones. Although an increasing trend in the rate over the measurement time would be expected, given that the same conditions as the previous measurement were used, the data dispersion does not clearly show this trend (figure~\ref{fig:AlphaCAMM_RateFondoSealModeR154}).

The significant increase in the background level during data acquisition in \textit{sealed mode} raised suspicions about the existence of a component that produced Radon emanation within the measurement volume. In addition to the common components in the construction of the detector (stainless steel, copper, PTFE, Micromegas), there were two that remained within the volume when measurements were taken in \textit{sealed mode}: a dust particle filter and a pressure sensor at the chamber's entrance. The next measurement ('Number~4') would be conducted after installing a new valve between the measurement volume and these two components, ensuring they were outside the measurement volume, but the result was similar to the previous one (2.63~$\pm$~0.19~$\mu$Bq/cm$^2$).

For the next two measurements, the internal parts of the chamber were cleaned with distilled water and alcohol ('Number~5'), and the copper mesh was removed ('Number~6'), a commercial copper mesh that had not been previously measured with spectrometric techniques, and therefore its composition was unknown. Both results were compatible with the previous background levels (2.41~$\pm$~0.78~$\mu$Bq/cm$^2$ and 2.29~$\pm$~0.57~$\mu$Bq/cm$^2$, respectively).

To confirm the hypothesis that the disintegrations responsible for the background level occurred throughout the detector volume, and not on the surfaces of the cathode and the Micromegas, the drift distance was increased from 7 to 13 cm, expecting a proportional increase in the background level. The background level in this measurement ('Number~7') increased to 3.25~$\pm$~0.21~$\mu$Bq/cm$^2$. This increase by a factor of approximately 1.42 compared to the first measurement was less than the increase in drift distance (factor of approximately 1.86). A hypothesis that could explain this difference is that there is a second component that could be 1.17~$\pm$~0.66~$\mu$Bq/cm$^2$ (the error in measuring the position of the cathode, and therefore in the drift length, introduces a significant error in the calculation).

The next measurement ('Number~8') was made after replacing the rubber gasket (Viton) in one of the two chamber lids (the lid with the Micromegas) with an expanded PTFE gasket. This measurement aimed to verify whether the Viton gasket was producing Radon emanation. Since the chamber was designed and manufactured with the ISO-K system using a Viton gasket and aluminum clamps (figure~\ref{fig:AlphaCAMM_FotoCamaraFinal}), the installation of an expanded PTFE gasket, which requires greater pressure on its surface to ensure sealing, caused the leak rate in the system to increase. The result of the measurement (2.23~$\pm$~0.94~$\mu$Bq/cm$^2$), although lower than the previous one, cannot be considered valid as the high leak rate may degrade the gas quality.

The previous instability could have been resolved by manufacturing new stainless steel clamps, allowing for increased tightening torque and thus pressure on the expanded PTFE gasket, but due to the delay this would cause, it was preferred to reassemble the Viton gasket and proceed with other tests.

After conducting several tests, including cleaning the detector components, the possibility was raised that the stainless steel chamber itself was producing the Radon emanation that caused the observed background level. The batch of stainless steel AISI~316L used to manufacture the chamber had been measured with GDMS (see section~\ref{subsec:ConcentracionRadio}), yielding satisfactory results regarding Uranium and Thorium content. However, these measurements could not ensure an acceptable concentration of elements below the parent isotopes in the three natural decay chains. Furthermore, although the specifications for the chamber's manufacturing process required that the welds be made solely with the measured material and with extreme cleanliness, the manufacturer might have made an error in the process and contaminated the chamber.

\begin{figure}[!ht]
    \centering
    \includegraphics[width=1.0\textwidth]{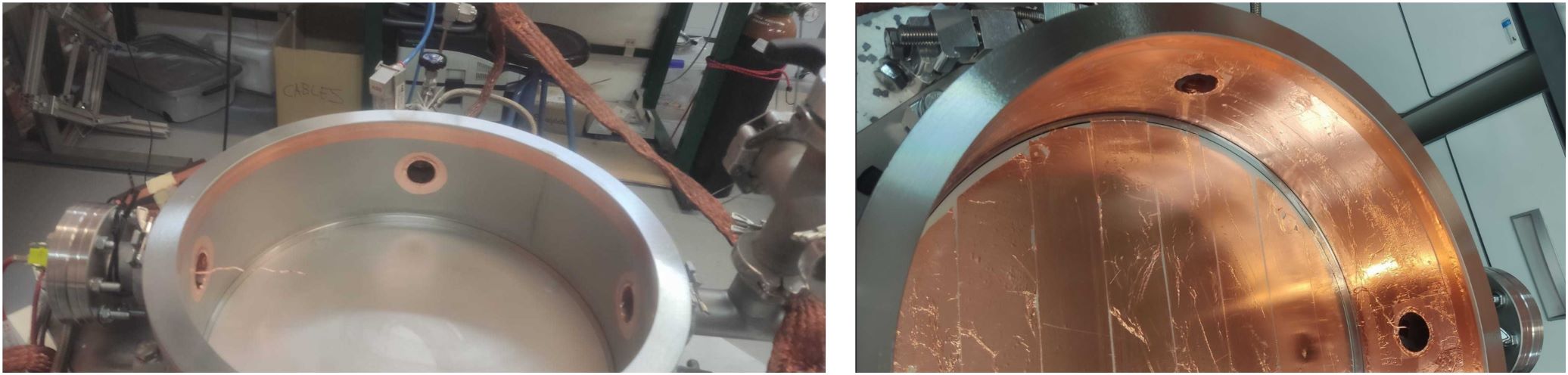}
    \caption{Photographs after covering the welds with copper tape (left) and 66\% of the chamber (right).}
    \label{fig:AlphaCAMM_CobreRecubreCamara}
\end{figure}

Therefore, the next two measurements aimed to determine whether the stainless steel of the chamber was producing Radon emanation. In the first intervention, the welds were covered with 35~$\mu$m thick copper tape ('Number~9'), and in the second intervention, the same tape covered 66\% of the chamber's surface ('Number~10'), that is, the bottom lid where the cathode is supported and the body or cylinder of the chamber. The lid where the Micromegas was installed was not covered. Photos of these two interventions can be seen in figure~\ref{fig:AlphaCAMM_CobreRecubreCamara}.

The background levels measured after these interventions were reduced to 3.01~$\pm$~0.52~$\mu$Bq/cm$^2$ in the case of the weld coverage, and to 1.69~$\pm$~0.21~$\mu$Bq/cm$^2$ in the case of the 66\% coverage of the chamber. This last measurement represents a reduction of the background level by 48\% compared to the previous one (3.25~Bq/cm$^2$ in 'Number~7'). However, this percentage of reduction does not match the percentage of surface covered with copper tape, reinforcing the hypothesis that part of the measured background level has a second component. In this case, assuming the total reduction of emanation from the surface covered by the copper tape (the composition and emanation level of the copper tape are unknown, thus introducing an error in this simplification), it has been calculated that the second component is 0.89~$\pm$~0.45~$\mu$Bq/cm$^2$. This value is compatible with the previously calculated value when checking the background level before and after increasing the drift distance (1.17~$\pm$~0.66~$\mu$Bq/cm$^2$).

\section{Measurement of Materials}

\begin{figure}[!ht]
    \centering
    \includegraphics[width=1.0\textwidth]{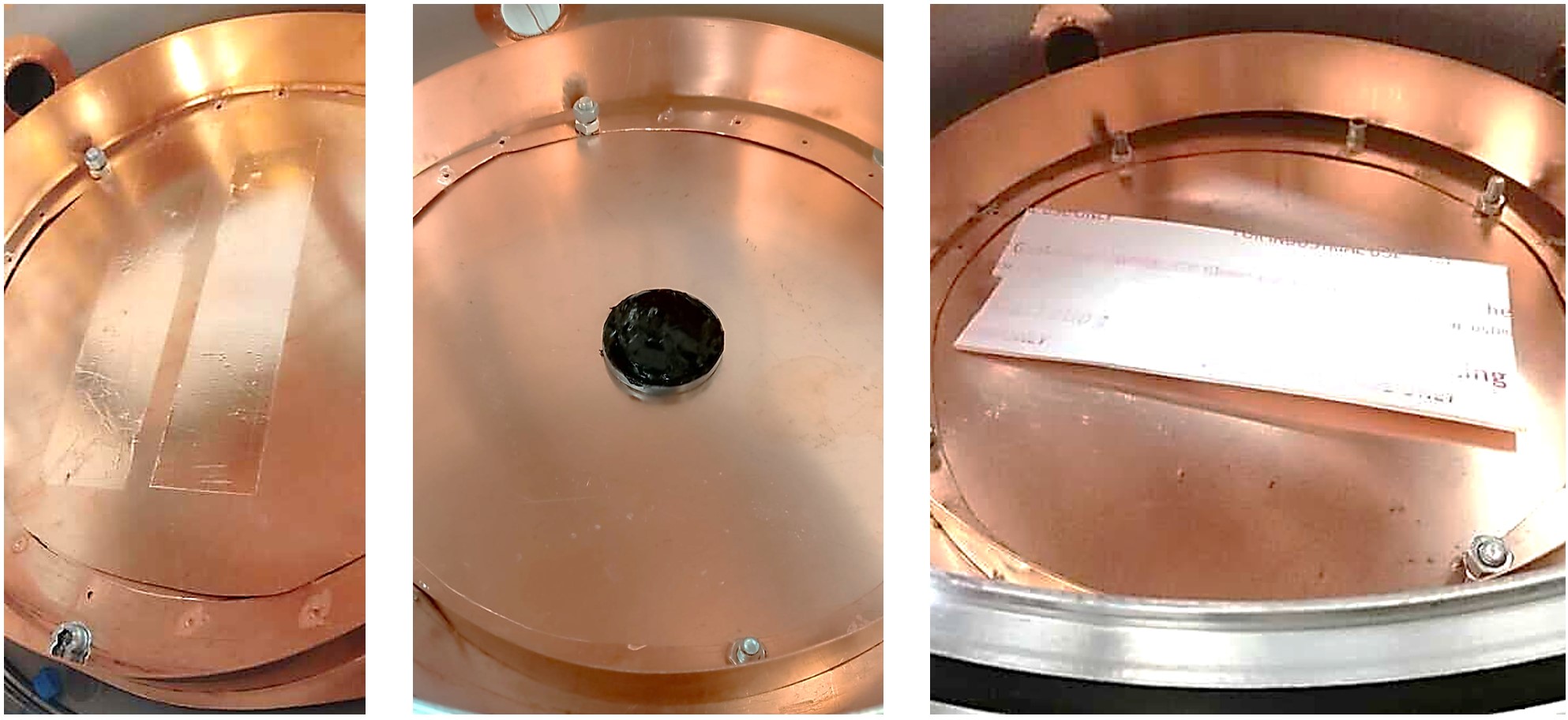}
    \caption{Photographs of the samples measured in the AlphaCAMM: 35~$\mu$m thick copper adhesive tape attached to the surface of the cathode (left), container with solidified silicone (center), and a 20~x~10~cm$^2$ piece of expanded PTFE (right).}
    \label{fig:AlphaCAMM_FotoMedidaMuestras}
\end{figure}

\begin{figure}[!ht]
    \centering
    \includegraphics[width=1.0\textwidth]{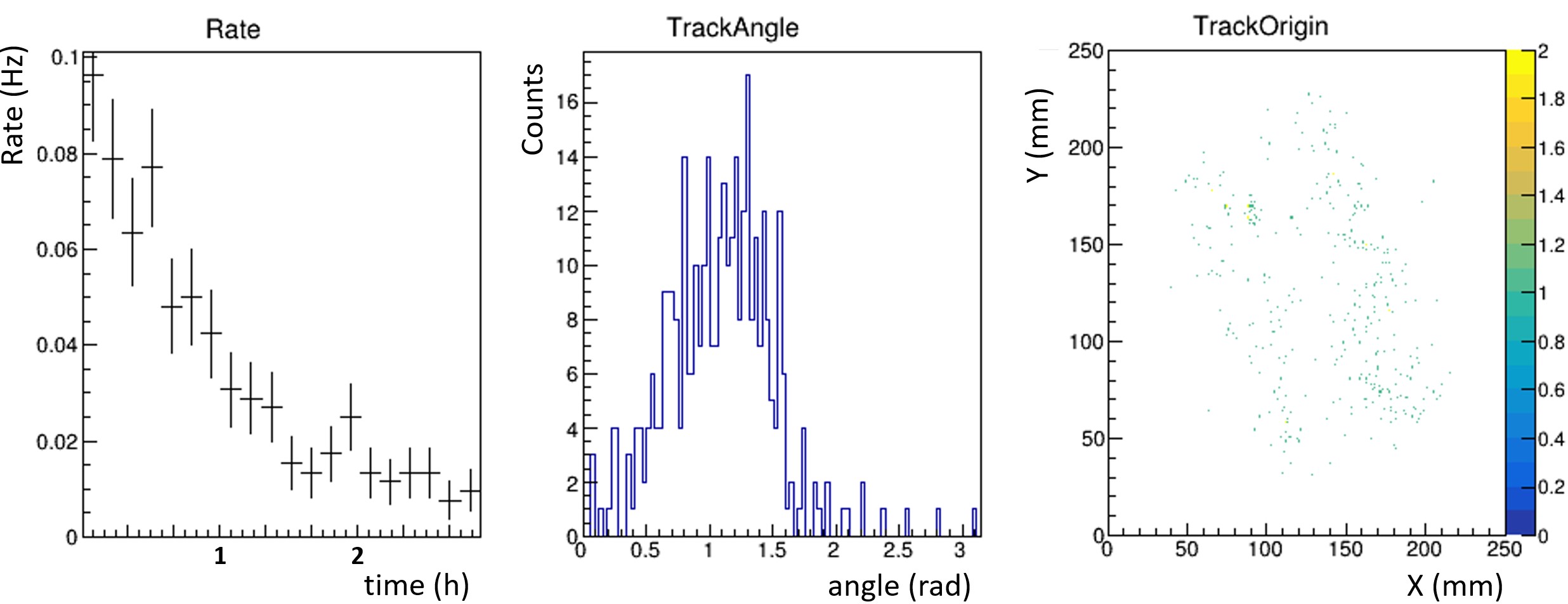}
    \caption{Results of the analysis for the measurement of the 20~x~10~cm$^2$ sample of expanded PTFE. $\bullet$ Left: evolution of the event acquisition rate over the measurement time (2.88 hours). $\bullet$ Center: histogram of the angular distribution of each trajectory. $\bullet$ Right: projection in the XY plane of the origin of each trajectory.}
    \label{fig:AlphaCAMM_MedidaePTFE}
\end{figure}

As initial tests, activity measurements were performed on 3 different samples (figure~\ref{fig:AlphaCAMM_FotoMedidaMuestras}):

\begin{itemize}

\item 35~$\mu$m thick copper tape: tape used to cover the welds and walls of the chamber during the work to identify the source of the background, as described in section~\ref{sec:AlphaCAMM_MedeidasFondo}. Two pieces measuring 25~x~5~cm$^2$ each were attached to the cathode, and a measurement was taken with a duration of 15.7 hours, yielding a result of (2.37~$\pm$~1.06~$\mu$Bq/cm$^2$) compatible with the background level at that time (table~\ref{tab:AlphaCAMM_TablaHistoricoFondo}, 'Number 9': 3.01~$\pm$~0.52~$\mu$Bq/cm$^2$).

\item Silicone 'SILKRON SPG Plus Krafft': a small amount (estimated to be around 1-3 grams) was used to seal small areas in the feedthrough of the Micromegas, leading to suspicions during the background identification work that Radon might be emanating into the AlphaCAMM. An emanation measurement was prepared in which a glass container was filled with 25 g of this silicone and placed on top of the cathode during a 2-hour measurement. The results (2.97~$\pm$~2.23~$\mu$Bq/cm$^2$) were compatible with the background at that time (table~\ref{tab:AlphaCAMM_TablaHistoricoFondo}, 'Number 9': 3.01~$\pm$~0.52~$\mu$Bq/cm$^2$). Although the error in the measurement is high due to the exposure time, the larger amount of material in the measured sample, compared to that used to seal the AlphaCAMM, helped to rule out this silicone as the main source of the background.

\item 20~x~10~cm$^2$ piece of expanded PTFE: as a first test of measurements of materials used in TREX-DM, a sample of expanded PTFE with a thickness of 3 mm was measured, which had previously been cleaned with acetone (removal of red lettering) and methanol. The results of the analysis can be seen in figure~\ref{fig:AlphaCAMM_MedidaePTFE}. The acquisition rate decayed over the measurement time (2.88 hours). The integration of the rate in the last 60 minutes yielded an activity in the sample of 37.5~$\pm$~15.7~$\mu$Bq/cm$^2$. The trajectory of 95.8\% of the events was classified as directed from the sample towards the readout plane ($\leq$~$\pi$/2~rad), as can be confirmed in the 'TrackAngle' histogram of the same figure. The activity coming from the sample can be clearly represented in the projection in the XY plane of the origin of each trajectory, showing an accumulation of events in an area whose geometry matches that of the measured sample.

\end{itemize}

Particular attention should be given to the observation of the decrease in the rate during the measurement time of the expanded PTFE sample. The data can be fitted to an exponential, and if we assume the hypothesis that this rate decrease is due to the decay of a single radioactive isotope, the decay constant according to the fit is \mbox{$\lambda=2.46(\pm0.6)\cdot10^{-4}$}. Therefore, the half-life of the isotope is \mbox{$ln2/\lambda=2817\pm551$~seconds} (46.90$\pm$9.2~minutes). The radioactive isotopes from the natural chains (figure~\ref{fig:AlphaCAMM_NaturalChains}) that are closest to this half-life value and emit an $\alpha$ particle in their decay, or have direct descendants with shorter lives that emit an $\alpha$ particle, are: $^{214}$Bi (20 min), $^{223}$Fr (22 min), $^{211}$Pb (36 min), and $^{212}$Bi (61 min).

Among the mentioned isotopes, $^{212}$Bi and $^{211}$Pb are the closest to the measured value. $^{212}$Bi emits an $\alpha$ particle in its decay, but $^{211}$Pb does not directly emit an $\alpha$ particle; instead, it has a direct descendant, $^{211}$Bi, with a shorter half-life (2.14 minutes) that does emit it. Considering the precision in the measurement of the sample and the fit performed, any of the mentioned isotopes could be responsible for the rate decrease, or even a combination of several isotopes.

One possible explanation given for this high activity is that during the cleaning of the sample, where its surface is rubbed with a cloth soaked in acetone or alcohol, the surface of the sample may become charged. The ions of the mentioned isotopes, or those of their closest progenitors in the chain leading to Radon, remain positively charged and suspended in the laboratory air after the decay of the corresponding Radon nuclide, thus being attracted to the surface of the sample. Surface contamination due to Radon progeny is a widespread problem in low-background experiments~\cite{Guiseppe2011}.

\begin{figure}[!ht]
    \centering
    \includegraphics[width=0.55\textwidth]{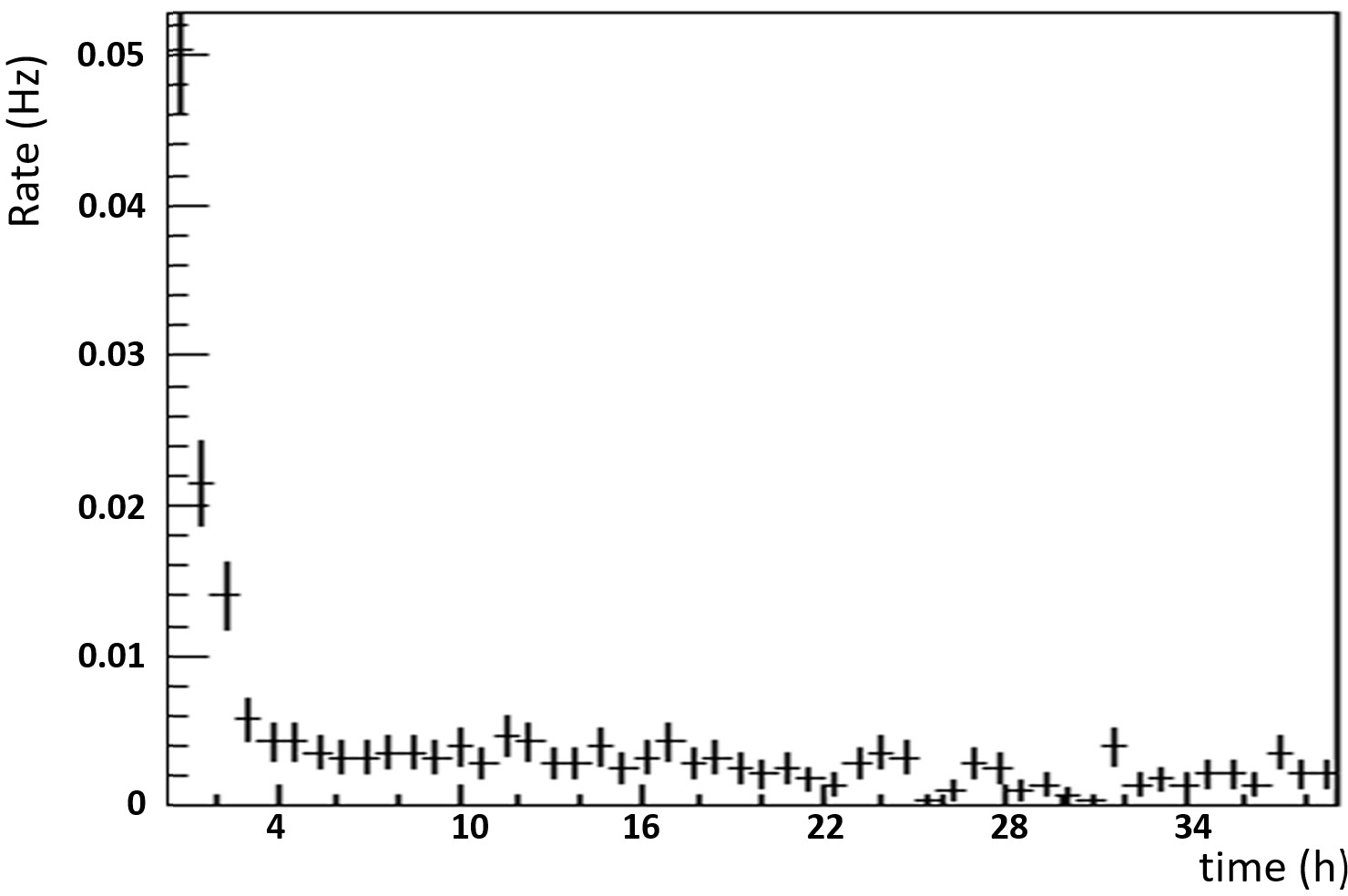}
    \caption{Results of the analysis for the second measurement of the 20~x~10~cm$^2$ sample of expanded PTFE. Evolution of the event acquisition rate over the measurement time (38.7 hours).}
    \label{fig:AlphaCAMM_MedidaePTFE_larga}
\end{figure}

This hypothesis was reinforced by observing, in a second measurement, that the effect continued to occur (figure~\ref{fig:AlphaCAMM_MedidaePTFE_larga}). In this second measurement, the detector was opened, the sample was cleaned again, and reintroduced. The time taken to close the detector, pump, inject the gas mixture, and launch the acquisition can vary between approximately 45 and 75 minutes. This second measurement, aside from reinforcing the hypothesis about the introduction of a contaminant during the cleaning of the sample, also allowed for the verification that the event rate after the first 4 hours of measurement was compatible with the detector's background.

\section{Conclusions}

Experiments involving rare events aim to achieve the minimum possible background level, allowing for maximum sensitivity enhancement and exploitation. During the operation of TREX-DM, where the background model~\cite{Castel2019} predicted a level between~1~and~10~dru, a level was reached that was an order of magnitude greater than predicted, primarily due to surface contamination from radioactive isotopes resulting from Radon decay. Consequently, the need arose to measure these surface contaminations with sufficient sensitivity (100~nBq/cm$^2$) to quantify activities of components/films contributing to the background level of the experiment in the same order as predicted by the background model ($\geq$~5~dru for Neon and $\geq$~2.5~dru for Argon), leading to the development of the AlphaCAMM.

Following the design, manufacture, and commissioning of the AlphaCAMM, as described in this chapter, a background level of~1.69~$\pm$~0.21~$\mu$Bq/cm$^2$ was achieved for this detector, placing the sensitivity of the AlphaCAMM at~3~$\mu$Bq/cm$^2$, a factor of~$\sim$30 worse than required (100~nBq/cm$^2$). With this sensitivity, it is only possible to measure contaminations from components contributing to the background of TREX-DM at the same order as the current background level (see sections~\ref{sec:BackgroundFisrtMeasurements}~and~\ref{sec:TREXDMopt_TomaDatos}), but not below it, which is the primary objective of this development.

According to the measurements shown in section~\ref{sec:AlphaCAMM_MedeidasFondo}, the emanation from the stainless steel walls appears to be the source of the main contributor to the background of the AlphaCAMM. Therefore, a plan has been made to execute a copper coating over the entire chamber, which is currently being carried out while writing these lines. The electrolytic copper (ETP) to be used for this coating has been measured~\cite{Castel2019} (table~\ref{tab:AlphaCAMM_TablaRadioMateriales}), and its activity for the parent isotopes in the natural decay chains is one to two orders of magnitude lower than that of the stainless steel used.

It is worth noting that, as demonstrated in the background model for the AlphaCAMM developed in chapter~\ref{ch:AlphaCAMM_ModeloFondo}, the activity measured with GDMS in the stainless steel used in the manufacture of the AlphaCAMM was sufficient to achieve the initially proposed sensitivity target. The evidence of emanation from this material observed during the commissioning of the detector necessitates a review of the model parameters that describe the processes of emanation, transport, and exhalation of Radon within this material (section~\ref{sec:ContribucionInicialRadio}). Other hypotheses are also proposed to explain this excess in emanation, such as contamination of the material during the manufacturing process or the initial concentration of radioactive isotopes from the natural decay chains that were not quantified in the GDMS measurement of the material.

In general, the parameters introduced in the Radon emanation model developed in chapter~\ref{ch:AlphaCAMM_ModeloFondo} need to be reviewed and adjusted according to experimental measurements.

Following the hypothetical reduction of the background level after the copper coating of the chamber, it will be possible to confirm or refute the hypothesis of a second component (superficial or volumetric) proposed earlier. In any case, after quantifying the new background level, the new situation will need to be studied to determine whether the sensitivity is sufficient to begin systematic measurements of materials or if it is necessary to continue with efforts to identify and reduce the background level.

\addcontentsline{toc}{part}{Summary and conclusions}

\setlength{\baselineskip}{12pt} 
\selectlanguage{spanish}
\graphicspath{{../Imagenes/}{../../Imagenes/}} 

\chapter*{Summary and Conclusions}
\label{ch:Conclusiones}

\pagestyle{myheadings}
\markboth{SUMMARY AND CONCLUSIONS}{SUMMARY AND CONCLUSIONS}

Throughout this thesis, and as its central axis, four large readout planes have been developed, the largest realized to date with microbulk Micromegas. The first, the SR2M module in the context of the PandaX-III experiment, allows for the tessellation of readout planes as large as necessary. The next two developments occur within the context of the TREX-DM experiment: one as an optimization of the previously installed readout plane in the experiment, and the other (the GEM-Micromegas system) as a fundamental part of the strategy to reduce the energy threshold of TREX-DM. For the development of the fourth readout plane, the design made during the optimization of TREX-DM has been utilized, resulting in the construction of a highly sensitive $\alpha$ particle detector that, although it originates in the context of the TREX-DM experiment, has the potential to grow independently by addressing one of the most ambitious challenges in many experiments: the reduction of the background level of detectors.

The number of experiments searching for unlikely and low-background events has increased in recent years, due to the interest from various disciplines in this type of experiment, and this number is expected to rise in the future. Therefore, the developments made here have significant projections, for which it will be necessary to continue working on the improvement of each of them. In this regard, new lines of research can be opened for the use of resistive Micromegas~\cite{Chefdeville2021, Feng2022} or Micromegas with segmented mesh~\cite{Diakaki2018}, which have not been addressed in this thesis but are a near-term objective within the group.

Although the central axis of this thesis has been the development of large readout planes, with a significant technical component, efforts have always been made to contextualize them within the fields in which they are deployed.

\section*{The SR2M Module in the PandaX-III Experiment}
\label{sec:ConclusionesPandaX}

A TPC with Xenon and Micromegas readout planes is a highly competitive solution for the search for the 0$\nu\beta\beta$ decay~\cite{Irastorza2016}. Microbulk Micromegas can be constructed and implemented with extremely low levels of radioactivity and show very good performance in Xe+TMA mixtures at 10 bar~\cite{Alvarez2014_2}, with good figures in terms of gain, spatial resolution, and energy resolution (3\% FWHM at the $Q_{\beta\beta}$ energy).

One of the objectives of this thesis was to develop technical solutions to implement this readout concept at larger scales while maintaining the performance obtained in small-scale prototypes. This was developed in the context of the PandaX-III experiment, where microbulk Micromegas technology was chosen as the main line for the development of the TPC readout plane of the experiment. The construction of a 1.5 m diameter plane with this technology posed a challenge, thus the idea of developing a single Micromegas module that could be installed to form a larger readout plane was born: the SR2M (Scalable Radiopure Readout Module).

During the design of the SR2M, the design of a Micromegas circuit was addressed for the first time in this thesis, which in this case had the particularity of containing a 30 cm long strip that allowed the signal to be extracted outside the TPC without using any connection inside it. In this circuit design, the width of the strips in the Micromegas was defined according to the needs for the topological reconstruction of events, but below the capabilities of the TPC itself allowed by diffusion processes, leaving room for future improvements if necessary. The electron reinsertion system (ERS) was also incorporated into this design, whose efficiency was demonstrated during the testing phase. Additionally, a feedthrough connected to the SR2M module itself and all the necessary mechanical components for its installation to form the final readout plane were added.

This prototype was tested and characterized individually in the laboratories of the University of Zaragoza, introducing for the first time in this thesis the FEC-Feminos acquisition electronics and the REST-for-Physics analysis framework. The results were satisfactory regarding the detector's response behavior, although some problematic points emerged that needed to be resolved in a second iteration of the module's design: connection issues, damaged channels, limitations in the length of the strip, or bubbles on the surface of the Micromegas.

After the testing and validation phase of the first prototype, 7 identical modules were manufactured, which were assembled in the laboratories of Shanghai Jiao Tong University (SJTU), forming the first tessellated readout plane for PandaX-III. This first readout plane served as a testing ground for the team at SJTU to acquire the necessary knowledge for its operation.

For the fabrication of a second prototype, some changes were made to the design: the Face-to-Face connector was introduced, the method of attaching the Micromegas circuit to the support was changed, and the external rim was eliminated. With the experience gained by the SJTU team in operating the previous prototype, another readout plane was formed with another 7 modules of the second version. Among other things, the study of the efficiency of the ERS was addressed, demonstrating its efficiency. Furthermore, the success of the FtF connector allowed its implementation in TREX-DM and, although not described in this thesis, also in the IAXO experiment.

For the fabrication of the 52 modules that would be installed to form the readout plane of the final chamber, Micromegas designed and manufactured at the University of Science and Technology in China were used, whose work was outside the scope of this thesis but adopted all the solutions developed up to the second version of the SR2M. The construction of the full-size detector has already been completed, along with 52 SR2M modules based on Micromegas developed in China, and once the surface tests are concluded, it will be installed in the Jinping Underground Laboratory~\cite{Zhang2023}.

\section*{Present and Future of the TREX-DM Experiment}
\label{sec:ConclusionesTREXDM}

TREX-DM has been developed for the search for WIMPs with a mass on the order of 10 GeV c$^{-2}$ or less, using Neon as the main gas or even using Argon with low concentration in $^{39}$Ar. Due to the lack of signal so far, it is interesting to extend the search towards lighter WIMPs, below 1 GeV c$^{-2}$. To have the opportunity to reach these regions for the WIMP mass, the current background levels (80 dru) and energy threshold (1 keV$_{ee}$), achieved in TREX-DM in 2022, must be improved. Additionally, the increase of Isobutane in the gas mixture leads to greater sensitivity towards lower WIMP masses below 1 GeV c$^{-2}$, primarily due to a lower mass of the target nuclei with which the WIMP interacts.

\begin{figure}[!ht]
    \centering
    \includegraphics[width=1.0\textwidth]{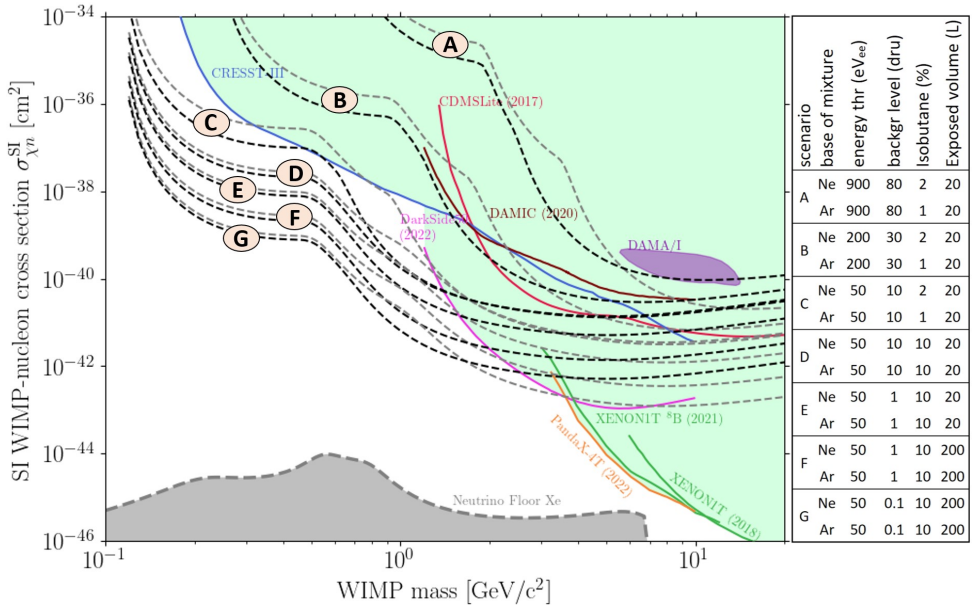}
    \caption{WIMP-nucleon cross-section versus WIMP mass, with the limits reached by the most important experiments and the limits that TREX-DM, under the conditions outlined in the table on the right, can reach with one year of exposure. Each scenario is represented both with a Neon mixture (black) and with an Argon mixture (gray).}
    \label{fig:TREXDM_WIMP_ExclussionPlot_Ene24_v1_2}
\end{figure}

Figure~\ref{fig:TREXDM_WIMP_ExclussionPlot_Ene24_v1_2} shows several future projections for the experiment's sensitivity, with a 1-year exposure in all of them, assuming spin-independent interaction, standard values of the WIMP model in the galactic halo, standard values of astrophysical parameters, and different detector parameters: energy threshold, background level, amount of Isobutane in the mixture, and volume of gas exposed to a pressure of 10 bar. The drawn curves represent the scenarios through which TREX-DM will progress as the mentioned parameters are gradually improved: scenario A has been reached during past campaigns, scenarios B-E are expected to be reached in the short to medium term, while scenarios F-G represent the case where the experiment is scaled by a factor of 10 in the long term.

\subsection*{Reduction of the Background Level}
\label{subsec:TREXDM_BackgroudReduction}

The components that come into direct contact with the active volume of the detector, and therefore possibly contribute to the measured background level, are the Micromegas readout planes, the field cage (walls that delimit the active volume and contain the rings to achieve the drift electric field), the cathode, and the gas mixture. The microbulk Micromegas installed in TREX-DM in 2018, when the experiment was installed at the LSC, was replaced by new Micromegas in 2022. The new design is more robust from the perspective of operational stability and is also a factor of 3 more radiopure than the previous version, which had been used to develop the background model~\cite{Castel2019}. This reduction contributes to the low-energy background level from the new Micromegas, on the order of 1 dru.

A detailed study has concluded that currently the main contribution to the low-energy background level is due to surface contamination from $\alpha$ particles. Furthermore, the study indicates that, among the contaminated surfaces in contact with the active volume, the cathode is the main contributor, likely due to the progeny of Radon that has been deposited on its surface. The cathode consists of a copper frame and a stretched aluminized mylar sheet over it. Replacing this element with a new one is one of the first actions that will be carried out once the clean area is installed and ready for use in the new location of the experiment.

By applying an 18 x 18 cm$^2$ cut over the area of the Micromegas (25 x 25 cm$^2$) during the analysis of the background data, so that only events in that region pass the cut, it allows for the quantification of events emitted from the cathode. However, there is also a significant population of events emitted from the walls of the field cage, which, as in the case of the cathode, is likely due to the progeny of Radon deposited on them. This population can be discriminated through topological studies and the application of the previously explained cut in the area of the Micromegas; however, these cuts considerably reduce the volume sensitive to WIMPs. Once the chamber can be reopened in a clean area, the walls of the field cage, made of kapton-copper sheets, will also be replaced.

For the development of the background model, an exhaustive program of material measurements has been carried out, mainly through gamma-ray spectrometry with germanium and completed with other techniques such as GDMS or ICPMS. However, these techniques are not the most suitable for measuring concentrations of particular isotopes, such as $^{210}$Pb or $^{210}$Po. For this purpose, a novel high-sensitivity $\alpha$ particle detector, called AlphaCAMM, is being developed, which is described in this thesis. The material measurements planned to be carried out with this detector will allow for an appropriate selection of those that will be installed within the TREX-DM chamber, thereby reducing the high-energy background observed so far and, as expected, also the low-energy background in the region of interest.

\subsection*{Reduction of the Energy Threshold}
\label{subsec:TREXDM_ThresholdReduction}

To explore the parameter space of low-mass WIMPs, it is essential to achieve the minimum possible energy threshold and background level. In figure~\ref{fig:TREXDM_WIMP_ExclussionPlot_Ene24_v1_2}, it can be observed that a reduction from 900 eV$_{ee}$ (minimum energy threshold achieved by TREX-DM in 2022) to 50 eV$_{ee}$ has a significant impact on sensitivity in the low-mass WIMP regions, below 1 GeV c$^{-2}$.

The energy threshold strongly depends on the electronic gain achieved in the Micromegas, which in turn depends on the electric field applied in its amplification holes~\cite{Giomataris1996}. With the aim of increasing the total electronic gain, a new readout plane has been developed in which a GEM has been installed above the Micromegas, achieving a stage of electronic pre-amplification prior to that produced in the Micromegas.

With this GEM-Micromegas combination, data have been taken in the laboratory with Ar~+~1\%~Isobutane and Ne~+~2\%~Isobutane mixtures, from 1 to 10 bar, achieving pre-amplification factors ranging from 12 for Ne~+~2\%~Isobutane at 10 bar to over 100 for Ar~+~1\%~Isobutane at 1 bar (these studies will soon be the subject of a publication in article form). The pre-amplification factor depends, among other parameters, on the gas pressure, but even in the worst-case scenario, a factor of 12 with Ne~+~2\%~Isobutane at 10 bar, the GEM-Micromegas system can potentially reach energy thresholds close to the energy deposited by a single electron resulting from the ionization of a particle passing through the active volume.

The GEM-based pre-amplification planes are already ready to be installed in TREX-DM and, once the chamber can be opened in a clean area, they will be installed along with the new cathode and the new field cage. Additionally, the implementation of the new calibration system based on the emanation of the isotope $^{37}$Ar (figure~\ref{fig:TREXatLSC_37ArPolvo}) will allow for a more precise determination of the effective energy threshold of TREX-DM.

\subsection*{Optimization of the Gas Mixture}
\label{subsec:TREXDM_MixtureOptimization}

The first sensitivity estimates made for TREX-DM~\cite{Iguaz2016} were conducted for a quantity of Isobutane in the gas mixture such that the resulting mixture could be classified as non-flammable (1\% in Argon and 2\% in Neon). At that time, the sensitivity estimates were better for Neon than for Argon (scenarios A-C in figure~\ref{fig:TREXDM_WIMP_ExclussionPlot_Ene24_v1_2}). However, a higher amount of Isobutane in both mixtures reduces the difference between them, as can be seen when comparing both mixtures in scenarios D-G. Most importantly, the increase of Isobutane in both mixtures enhances sensitivity to WIMPs for masses below 1 GeV c$^{-2}$, primarily due to the reduction of the mass of the nuclei in the mixture.

Given the evidence of improvement by adding more Isobutane, and taking advantage of the fact that the new Lab2500 space is further away from the other spaces in the underground facility, the LSC has accepted the proposal to increase the concentration of Isobutane, both for Neon and for Argon, up to 10\%. This concentration value is a compromise solution that significantly improves the sensitivity of the detector (see figure~\ref{fig:TREXDM_WIMP_ExclussionPlot_Ene24_v1_2}) but with an acceptable risk, considering that the jump in sensitivity from 10\% Isobutane is not as notable as the jump that occurs when moving from 1\% (Argon) or 2\% (Neon) to 10\%.

The use of flammable gases necessitates the application of new and extreme safety measures in the new Lab2500 space. These measures are being designed, and their implementation is planned for the medium term.

\subsection*{Conclusions}

Although the current background level of TREX-DM is almost an order of magnitude higher than predicted by the background model, the origin of much of this background has been identified, and the short-term actions described above will be taken to reduce it. Furthermore, the energy threshold achieved during the data-taking campaigns is approaching that of the initial projections; however, the new projections are more ambitious, and a new GEM-Micromegas combination has been developed so that the detector can approach single electron detection.

The good performance of TREX-DM, along with future plans to reduce the background, the energy threshold, and optimize the gas mixture, demonstrates that TREX-DM has the potential to explore unexplored regions of WIMP mass below 1 GeV c$^{-2}$~\cite{Castel2023}.

\section*{Present and Future of the AlphaCAMM}
\label{sec:ConclusionesAlphaCAMM}

The AlphaCAMM, although it originates in the context of the TREX-DM experiment, has the potential to grow independently by addressing one of the most ambitious challenges in many experiments: the reduction of the background level of detectors.

For the measurement of the radiopurity of materials, different spectrometric techniques (HPGe, GDMS, ICPMS, BiPo-3) are commonly used. However, these techniques are difficult to use for the quantification of Radon progeny due to their low sensitivity to these isotopes. Additionally, the added difficulty of measuring surface contaminations would require these techniques to clearly separate the surface from the rest of the material.

Within the TREX-DM team, the need arose to develop a high-sensitivity surface $\alpha$ particle detector capable of measuring contaminations in future pieces that would be installed within the TPC, especially for large surface pieces/sheets. The sensitivity objective for this $\alpha$ particle detector was set at 100 nBq/cm$^{2}$, with a background level of 60 nBq/cm$^{2}$.

As a first step towards the design and construction of the AlphaCAMM, a first prototype~\cite{Altenmuller2022AlphaCAMM} was developed as a demonstrator of the operating principle and would serve as a starting point for defining the final design specifications. Subsequently, a background model based on Radon emanation was developed, which helped in the appropriate selection of materials and the design of the corresponding TPC, and will also aid in interpreting the measurements and identifying the source of the detector's background. Finally, it has been manufactured, set up, and the first material measurements have been taken.

However, the current background level of the detector is 1.69 $\pm$ 0.21 $\mu$Bq/cm$^2$, placing the sensitivity of the AlphaCAMM at 3 $\mu$Bq/cm$^2$, a factor of $\sim$30 worse than necessary (100 nBq/cm$^2$). The emanation from the stainless steel walls appears to be the main contributor to the background of the AlphaCAMM. Therefore, the execution of a copper covering over the entire chamber has been planned, which is being carried out during the writing of these lines, and aims to significantly reduce this emanation.

In any case, after quantifying the new background level, the new situation will need to be studied to decide whether the sensitivity is sufficient to begin systematic material measurements or if it is necessary to continue with the identification and reduction of the background level.

If the AlphaCAMM demonstrates its potential, now limited by the background level, to measure samples with the projected sensitivity, it could have great prospects, not only within the TREX-DM experiment but also in other low-background experiments.

\newpage
\renewcommand{\chapter}[2]{} 
\bibliography{bibliografia}







\end{document}